\documentclass[3p,times]{elsarticle}
\usepackage{hyperref}  
\hypersetup{
  colorlinks,
  allcolors=blue
}

\usepackage{ecrc}
\volume{00}
\firstpage{1}
\journalname{Physics Reports}
\runauth{R. T. Clay, S. Mazumdar}
\usepackage{graphicx}
\usepackage{amssymb}
\usepackage{amsmath}
\usepackage[percent]{overpic}

\biboptions{numbers,sort&compress} 
\journal{Physics Reports}

\begin{document}
\begin{frontmatter}
\title{From charge- and spin-ordering to superconductivity in the organic charge-transfer
  solids}
\author{ R. T. Clay$^{\rm{a}}$, S. Mazumdar$^{\rm{b}}$}
\address{$^{\rm{a}}$Department of Physics and Astronomy and HPC$^2$ Center for Computational
Sciences, Mississippi State University, Mississippi State, MS 39762, USA.}
\address{$^{\rm{b}}$Department of Physics, University of Arizona, Tucson, AZ 85721, USA}
\date{\today}

\begin{abstract}
We review recent progress in understanding the different spatial
broken symmetries that occur in the normal states of the family of
charge-transfer solids (CTS) that exhibit superconductivity (SC), and
discuss how this knowledge gives insight to the mechanism of the
unconventional SC in these systems.  A great variety of spatial broken
symmetries occur in the semiconducting states proximate to SC in the
CTS, including charge ordering, antiferromagnetism and spin-density
wave, spin-Peierls state and the quantum spin liquid. We show that a
unified theory of the diverse broken symmetry states necessarily
requires explicit incorporation of strong electron-electron
interactions and lattice discreteness, and most importantly, the
correct bandfilling of one-quarter, as opposed to the effective
half-filled band picture that is often employed. Uniquely in the
quarter-filled band, there is a very strong tendency to form nearest
neighbor spin-singlets, in both one- and two-dimension.  The
spin-singlet in the quarter-filled band is necessarily
charge-disproportionated, with charge-rich pairs of nearest neighbor
sites separated by charge-poor pairs of sites in the insulating
state. Thus the tendency to spin-singlets, a quantum effect, drives a
commensurate charge-order in the correlated quarter-filled band. This
charge-ordered spin-singlet, which we label as a paired-electron
crystal (PEC), is different from and competes with both the
antiferromagnetic (AFM) state and the Wigner crystal (WC) of single
electrons. Further, unlike these classical broken symmetries, the PEC
is characterized by a spin gap. The tendency to the PEC in two
dimension is enhanced by lattice frustration. The concept of the PEC
mirrors parallel development of the idea of a density wave of Cooper
pairs in the superconducting high T$_{\rm c}$ cuprates, where also the
existence of a charge-ordered state in between the antiferromagnetic
and the superconducting phase has now been confirmed.  Following this
characterization of the spatial broken symmetries, we critically
reexamine spin-fluctuation and resonating valence bond theories of
frustration-driven SC within half-filled band Hubbard and
Hubbard-Heisenberg Hamiltonians for the superconducting CTS. We
present numerical evidence for the absence of SC within the
half-filled band correlated-electron Hamiltonians for any degree of
frustration. We then develop a valence-bond theory of SC within which
the superconducting state is reached by the destabilization of the PEC
by additional pressure-induced lattice frustration that makes the
spin-singlets mobile. We present limited but accurate numerical
evidence for the existence of such a charge order-SC duality. Our
proposed mechanism for SC is the same for CTS in which the proximate
semiconducting state is antiferromagnetic instead of charge-ordered,
with the only difference that SC in the former is generated via a
fluctuating spin-singlet state as opposed to static PEC.  In
\ref{othersc} we point out that several classes of unconventional
superconductors share the same band-filling of one-quarter with the
superconducting CTS. In many of these materials there are also
indications of similar intertwined charge order and SC. We discuss the
transferability of our valence-bond theory of SC to these systems.
\end{abstract}

\end{frontmatter}
\newpage
\tableofcontents
\newpage

\section{Introduction and Scope of the Review}
\label{intro}
Superconductivity (SC) is only one of many mysterious features of the
high critical temperature (high T$_{\rm c}$) superconductors. More
enigmatic is perhaps the violation of the Landau Fermi liquid theory
in the normal state, which had been a cornerstone of traditional
condensed matter theory until thirty years ago, when high $T_{\rm c}$
in the cuprate materials was first discovered \cite{Bednorz86a}. The
list of experiments that demonstrate the unusual characteristics in
the so-called normal state of the cuprates, especially in the
under-doped region, is now very long
\cite{Timusk99a,Norman05a,Keimer15a}. Similar unusual behavior have
now been observed in the iron-based superconductors
\cite{Stewart11a,Si16a,Kobayashi17a}. It is widely recognized that
these peculiarities, including in particular the proximity of SC to
multiple competing or coexisting broken symmetries separated by
quantum criticalities, are the common features of systems with
strongly correlated charge carriers \cite{Dagotto05a}.  Furthermore,
there is now widely-shared agreement that there exist many other
correlated-electron superconductors, albeit with relatively lower
T$_{\rm c}$ (due to perhaps smaller one-electron bandwidths)
\cite{Mazumdar89a,Uemura91a,Fukuyama87a,Fukuyama90a,McKenzie97a,Iwasa03a,Capone09a,Mazumdar12a,Baskaran16a}. The
present review is largely about one such family of materials with
strong electron-electron (e-e) interaction, low-dimensional organic
charge-transfer solids (CTS).  Even as we restrict our discussions to
this one family of materials, we believe that the key concept, viz.,
SC in the CTS evolves from a Wigner crystal of nearest neighbor
spin-singlet electron pairs that we term the Paired Electron Crystal,
may be widely transferable.  In order to avoid confusion, henceforth
we will use the term ``Wigner crystal'' (WC) to refer to the Wigner
crystal of single electrons only; the ordered state of spin pairs will
be referred to as the ``Paired Electron Crystal'' (PEC).  We show in
our work that the tendency to form such a charge-ordered spin-singlet
state, a quantum effect, is particularly strong in the
correlated-electron quarter-filled band, where this broken symmetry
competes with other more thoroughly researched broken symmetries such
as antiferromagnetism, quantum spin liquid and the WC of single
electrons.  A generalization of our theory is that there must exist
other unconventional correlated-electron superconductors that share
the same bandfilling of one-quarter with the CTS, and that exhibit
some of the same peculiar features. In particular, there should exist
evidence for the occurrence of the same PEC. We discuss several such
materials (families of materials) in \ref{othersc}. While clearly
further research on these other materials are warranted before firm
conclusions can be arrived at, the identification of any common theme
between these diverse families is encouraging.

SC in the CTS has been known since 1980
\cite{Jerome80a,Jerome80b}. While the first organic superconductors
were ``low T$_{\rm c}$'' and belonged to the weakly two-dimensional (2D)
(and are hence often referred to as ``quasi-one-dimensional'',
quasi-1D) TMTSF family (see \ref{molecules} for a compendium of the full names of
organic molecules referred to by their acronyms in the review), within
a few years the more strongly 2D ``high T$_{\rm c}$'' superconducting
compounds belonging to the BEDT-TTF (hereafter ET) family were
discovered \cite{Parkin83a}. The chemical formulas of the two series
of compounds are similar [(TMTSF)$_2$X versus (ET)$_2$X, where X is a
  closed-shell inorganic anion.]  While (TMTSF)$_2$X exhibit the same
basic crystal structure independent of X, (ET)$_2$X are obtained with
many different crystal structures, which are labeled $\alpha$, $\beta$,
$\theta$, $\kappa$, ...  etc., with no particular significance
attached to these Greek prefixes. The original anions X in the family
(ET)$_2$X were the same as in the (TMTSF)$_2$X. Subsequently, however,
compounds with more complicated polymeric anions such as Cu(NCS)$_2$,
Cu[N(CN)$_2]$Cl, Cu[N(CN)$_2$]Br etc. became more popular because of the
higher T$_{\rm c}$'s reached with these compounds. The compound
$\kappa$-(ET)$_2$Cu[N(CN)$_2$]Br is an ambient pressure superconductor below 11.6
K \cite{Kini90a}, in contrast to the others in which SC is reached
only under pressure.  Several other families of superconducting CTS,
including compounds of BEDT-TSF (hereafter BETS) as well as compounds
with asymmetric cations such as DMET are known.  We point the reader
to several excellent resources on the chemistry and physics of these
materials \cite{Ishiguro,Williams,Saito07a,Lebed}. While the charge
carriers in the above materials are holes, CTS superconductors with
electrons as charge carriers are also known \cite{Kato04a}. These last
compounds have the general chemical formula X[M(dmit)$_2$]$_2$, where
M = Ni, Pd or Pt, and X$^+$ is a closed-shell monovalent cation,
such as EtMe$_3$P$^+$, EtMe$_3$Sb$^+$ etc.

Two characteristic features are common to nearly all known
superconducting CTS. First, with very few exceptions
\cite{Lyubovskaya87a,Naito05a,Taniguchi07a}, all hole carriers have
2:1 cation:anion stoichiometry, implying a hole concentration of
$\frac{1}{2}$ per cationic molecule \cite{Ishiguro,Williams,Saito07a}
(the stoichiometry of CTS superconductors with large unit cells is
sometimes written as 4:2; the carrier concentration per cationic
molecule is the same in these.) We will refer to the carrier
concentration per molecule as $\rho$, which we believe is a more
fundamental quantity than apparent bandfillings in these strongly
correlated systems \cite{Torrance77a,Torrance78a,Mazumdar83a}.  The
exceptions mentioned above actually prove the ``$\rho=\frac{1}{2}$
rule'', in that these systems are complex structures in which the
deviation of $\rho$ from $\frac{1}{2}$ is very small.  Similarly, the
electron carriers have 1:2 cation:anion stoichiometry
\cite{Kato04a}. With the anions forming the active layer now, the
electron concentration $\rho$ is $\frac{1}{2}$ per anionic molecule.

The second characteristic common to superconducting CTS is that in
nearly every case, the ambient pressure state at low temperature T is
an unconventional semiconductor with spatial broken symmetry, and SC
is obtained by application of pressure at constant carrier
concentration. These two features, SC being limited to a very narrow
carrier concentration range or even fixed concentration, as well as an
apparent {\it semiconductor-superconductor transition}, are common to
many known correlated-electron superconductors beyond the CTS
\cite{Mazumdar12a}.  We discuss several such families of materials
materials in \ref{othersc}.  Interestingly, the electron-doped cuprate
superconductors share these characteristics \cite{Armitage10a}.

We report a two-pronged effort to understanding the mechanism of SC in
the CTS. On the one hand, we report recent experimental and
theoretical studies that have attempted to understand the spatial
broken symmetries in the semiconducting states proximate to SC. We
will be interested in the detailed {\it patterns of charge, bond and
  spin distortions in the insulating phases, beyond merely their
  periodicities.}  We will attempt to show that the patterns of
distortions, as well as the {\it phase relationships between any
  coexisting broken symmetries} are both relevant when trying to
arrive at a clear understanding of the complex phenomenon of
correlated-electron SC. (We will, for instance, show that the
bond-charge distortion pattern in the charge-ordered phase proximate
to SC is universal, while other patterns that are also observed do not
lead to SC \cite{Clay12a}.) Hence our discussions of SC will be
limited strictly to theoretical approaches that are based on {\it
  discrete lattice Hamiltonians.}  We refer the reader to recent
reviews for discussions of competing theories that start from
continuum electron-gas models
\cite{Bourbonnais08a,Bourbonnais11a}. Following the discussions of
spatial broken symmetries, we discuss various proposed theoretical
mechanisms of SC in the CTS, focusing on theories that emphasize the
correlated-electron nature of the CTS.  We show that spin-fluctuation
and resonating valence-bond (RVB) theories of SC that are based on
mean-field theories of e-e interaction or approximate approaches such
as the random phase approximation (RPA), fluctuation-exchange (FLEX)
or the dynamic mean-field theory (DMFT) fail to explain the SC in the
CTS, even as they might successfully explain select
non-superconducting spatial broken symmetry phases
\cite{Clay08a,Dayal12a}.  Our principal goal is to report the
collected theoretical and experimental results which indicate that
quantum fluctuations in systems with the specific carrier
concentration of $\frac{1}{2}$ per unit drive a strong tendency to
form charge-ordered local singlets. The strong quantum effects are not
limited to 1D, but will also occur in 2D and perhaps even 3D in the
presence of strong geometric lattice frustration (see \ref{spinelsc}
and \ref{c60}).  We discuss experiments that reveal these local
singlets in the dominant spatial broken symmetry states in the CTS at
the lowest T. A new viewpoint about correlated-electron SC, that has
conceptual overlaps with both the RVB theory
\cite{Anderson73a,Anderson87a} and bipolaron theories of SC
\cite{Alexandrov94a}, and is yet distinct, is shown to emerge. The
proposed mechanism of SC, while still incomplete, also has strong
overlaps with the emergent ideas of intertwined charge
order-superconductivity in the cuprates \cite{Fradkin15a} as well as
the proposition that the charge-ordered phase there is a density wave
of Cooper pairs \cite{Cai16a,Mesaros16a}.

This review is broken into the following parts. In Section \ref{early}
we introduce our theoretical model and describe early applications of
the theory to the family of CTS as a whole, both non-superconducting
and superconducting.  The purpose of this section is to show that many
of the later developments, in particular, the conclusion that the
specific carrier density per molecule $\rho=\frac{1}{2}$ is special,
emerges already from the early literature.  In Section \ref{broken-1d}
we develop a comprehensive theory of broken symmetries specifically
for $\rho=\frac{1}{2}$, in quasi-one dimension.  This is where we
first encounter the concepts of Bond-Charge-Density Wave (BCDW) and
Bond-Charge-Spin-Density Wave (BCSDW), which are simply PECs in one
dimension, or on weakly coupled 1D chains. The two most notable
conclusions of the theoretical work here are, (a) even within the
extended Hubbard model with nonzero nearest neighbor repulsion, the WC
is not the obvious ground state; rather, for a wide range of realistic
parameters, the ground states are the BCDW and BCSDW; (v) there exists
a broad range of realistic Coulomb parameters where even as the high
temperature phase is the WC, at the lowest temperatures the BCDW
dominates. In Section \ref{1d-expt} we discuss experimental
observations in the Fabre (TMTTF) and Bechgaard (TMTSF) salts, along
with non-traditional $\rho=\frac{1}{2}$ cationic CTS. The wide
applicability of the PEC concept in these quasi-one-dimensional
materials is pointed out. In Sections \ref{2d-section} and
\ref{2dtheory} we discuss experiments and theory in
quasi-two-dimension.  From explicit calculations we develop the PEC
concept in two dimensions. Manifestations of the PEC in both cationic
and anionic quasi-two-dimensional CTS are the occurrence of insulating
stripes, and nonzero spin gap. In Section \ref{pec2dcts} we present
detailed evidence that in every case where there occurs transition to
SC from a charge-ordered state, and the pattern of the charge-order is
known, this pattern corresponds to that of the PEC.  Following this we
present detailed numerical calculations of superconducting pair-pair
calculations and present our conclusions. We then point out in
\ref{othersc} that there exist many other unconventional
semiconductors characterized by strong electron-electron interactions,
frustrated lattice and $\rho=\frac{1}{2}$. ``Generic'' unconventional
behavior is observed in these cases, as we point out in \ref{othersc},
where we discuss briefly the experimental observations in the
individual systems.

\paragraph{Guide to the reader}

Because of the lengths of the individual sections and subsections that
follow, and the seemingly diverse topics that are discussed, we
present a summary paragraph at the end of each of Sections 2-6.
Sections \ref{1d-expt} and \ref{2d-section} review in detail the
experimental work on the quasi-one and quasi-dimensional materials and
are by necessity quite dense; we have therefore presented summary
paragraphs here after each subsection.  Readers who are short of time
are encouraged to read just the introduction and summary paragraphs of
these sections, and refer back to subsections on specific materials as
needed in the following sections.

\section{Electron-electron interaction effects - survey of early work}
\label{early}

Many of the concepts related to e-e interaction effects in
CTS have their roots in early work done through the 1960s to the
1980s. We report on these studies here, not merely for the sake of
completeness. The justification of many of the claims made in the
present review come from comparisons of early and recent theories
against experiments from this period as well as later work. It is
also conceivable that the astute reader may still find new directions
of research from this report on early work.
  
\subsection{Classification: cationic, anionic and two-chain conductors} 

Several early reviews on CTS emphasize the role of e-e interactions
\cite{Soos74a,Soos75a,Torrance79a}. The earliest CTS were mostly {\it
  mixed stack} systems in which planar aromatic $\pi$-electron donors
$D$ and acceptors $A$ alternated along quasi-1D stacks, with
intermolecular distances slightly smaller than the sum of the van der
Waals' radii of the $D$ and $A$ molecules.  Predominantly neutral
$\cdots DADA \cdots$ as well as predominantly ionic $\cdots
D^+A^-D^+A^- \cdots$ were both known. Later demonstrations of pressure
\cite{Torrance81a}, temperature \cite{Torrance81b} and light
\cite{Koshihara90a} induced neutral-ionic transitions in the CTS have
led to a wide field of research, but mixed stack CTS are outside the
scope of this review, as they are necessarily $\frac{1}{2}$-filled
band semiconductors.

Prototype one-chain {\it segregated stack} CTS are the anionic M-TCNQ,
in which the M are alkali metals that donate one complete electron to
the acceptor TCNQ molecules, and in which the TCNQ form acceptor-only
$\cdots A A A A \cdots$ segregated stacks. The semiconducting
behavior of M-TCNQ, along with their unusual optical and magnetic
behavior led to the Hubbard model description of these systems,
\begin{equation}
\label{Hubbard}
H=-\sum_{\langle ij \rangle \sigma}t_{ij} c^\dagger_{i\sigma}
c_{j\sigma} + U\sum_{i} n_{i,\uparrow}n_{i,\downarrow} 
\end{equation}
where $c^\dagger_{i\sigma}$ creates an electron or a hole with spin
$\sigma$ ($\uparrow$ or $\downarrow$) on the frontier orbital of
molecular site $i$, $\langle ... \rangle$ implies nearest neighbors
(NN), and all other terms have their usual meanings. The TCNQ$^-$
stacks with one electron per anion are 1D Mott-Hubbard semiconductors
with $U\geq 4|t|$ \cite{Torrance75a} [Very recently, Soos {\it et al.}
  have suggested that Na-TCNQ and K-TCNQ are charge-ordered, $\cdots$
  TCNQ$^{2-}$TCNQ$^0$TCNQ$^{2-}$TCNQ$^0$ $\cdots$ \cite{Kumar11a}. We
  believe that additional theoretical and experimental work are needed
  before this issue could be considered as settled.]

Many different groups pointed out that 1D chains with $\rho \neq 1$
are expected to be conductors in spite of large $U$
\cite{Soos74a,Soos75a,Torrance75a,Torrance79a}. Research on conducting
CTS began in earnest with the discovery of the peculiar T-dependent
transport in NMP-TCNQ, a two-chain segregated stack CTS. Initial work
had assumed that the degree of charge-transfer $\rho$ from NMP to TCNQ
was 1 \cite{Epstein71a}, but later more precise measurements
\cite{Epstein81a} indicated this to be $\frac{2}{3}$. Two-chain
conductors TTF-TCNQ \cite{Ferraris73a,Coleman73a}, TSF-TCNQ,
HMTTF-TCNQ and HMTSF-TCNQ \cite{Keller74a} with high conductivities
and with $\rho=$ 0.59, 0.63, 0.74 and 0.74, respectively, were
discovered around the same period, establishing firmly the concept of
conductors with incomplete charge-transfer ($\rho \neq 1$).  Two-chain
$\rho=1$ systems, such as HMTTF-TCNQF$_4$ and HMTSF-TCNQF$_4$, are
also Mott-Hubbard semiconductors \cite{Torrance80a,Hawley78a}. The
one-chain cationic systems of interest in the present Review were
discovered subsequently. We refer the readers to several conference
and workshop proceedings from this period that give excellent
discussions of the work done during this period
\cite{Keller74a,Keller77a,Devreese79a}.

\subsection{Charge-spin decoupling}
\label{decoupling}

Early theoretical works that introduced the concept of charge-spin
decoupling in the context of the quasi-1D CTS were by Bernasconi {\it
  et al.} \cite{Bernasconi75a} and Klein and Seitz \cite{Klein74a}.
Bernasconi {\it et al.} showed that the Peierls instability, which in
noninteracting 1D chains opens a gap at the Fermi wavenumbers $\pm$
k$_{\rm F}$, will for large $U$ occur in a band of {\it spinless}
Fermions at new wavenumbers $\pm$ 2k$_{\rm F}$, leading to a so-called
4k$_{\rm F}$ instability. This was one of the earliest discussions of
metal-insulator transitions leading to Bond Order Wave (BOW) and
periodic lattice distortion with wavevector other than the
conventional Peierls' 2k$_{\rm F}$.  In this review we will use BOW to
describe a periodic lattice distortion that leaves the site charge
densities homogeneous; in general however for $\rho\neq 1$ a BOW must
coexist with a simultaneous periodic distortion of the charge density
on each site (see Section \ref{1dlimit}).  Specifically for
$\rho=\frac{1}{2}$, the ordinary 2k$_{\rm F}$ BOW instability would
correspond to bond tetramerization (2k$_{\rm F}=\pi/2a$, where $a$ is
the lattice constant of the undistorted lattice) while the 4k$_{\rm
  F}$ BOW instability implies bond dimerization (4k$_{\rm F}=\pi/a$).

Klein and Seitz showed that the lowest excitations  of the large $U$
1D Hubbard model with nearest neighbor-only hoppings ($t_{ij} = t$) reduces 
in the $U >> |t|$ limit to
\begin{eqnarray}
H&=& N_2U -t\sum_{i=1}^N (a^\dagger_i a_{i+1} +H.c.) +  H_{\rm spin} \\
\label{klein}
H_{\rm spin} & = & \sum_{\theta,i=1}^{N_1} J_{\theta} 
\vec{S}_i\cdot\vec{S}_{i+1}
\label{hspin}
\end{eqnarray}
where $N$ is the total number of sites, $N_1$ and $N_2$ are the number of 
single and double occupancies, respectively ($N_2=0$ for $\rho
\leq 1$), $a^\dagger_i$ creates a spinless Fermion, and $H_{\rm spin}$ is
the Heisenberg spin Hamiltonian describing spin
excitations corresponding to each different occupancy $\theta$ of the 
spinless Fermion band. The authors gave a closed form expression for the
exchange integral within $H_{\rm spin}$ for the ground state occupancy 
($\theta = 0$) of
the spinless Fermions \cite{Klein74a}, 
\begin{equation}
  \label{exchange}
  J_0 = \frac{2t^2}{U}\rho[1 - \frac{\sin(2 \pi \rho)}{2 \pi \rho}]
\end{equation}

This work formed the basis for comparing the temperature-dependence of
the static magnetic susceptibilities $\chi(T)$ of 1D CTS with
arbitrary $\rho$ below the metal-insulator transition temperature to
the Bonner-Fisher $\chi(T)$ of the 1D Heisenberg chain
\cite{Torrance77a}. Coupling of the spins to the lattice can therefore
lead to the spin-Peierls transition even for $\rho \neq 1$.
Interestingly, the spin-Peierls (SP) transition was observed in the
$\rho=1$ CTS \cite{Bray75a,Bray83a} considerably before it would be
seen in inorganic systems, and the experimental observations led to
sophisticated theoretical developments of this phenomenon
\cite{Cross79a}.  Much of our current understanding of the SP
transitions in the $\rho=\frac{1}{2}$ cationic CTS (see next section)
come from these early papers.

\subsection{The 4k$_{\rm F}$ instability, Peierls CDW versus Wigner crystallization}
\label{4kf}

\begin{figure}[tb]
  \center{\resizebox{2.0in}{!}{\includegraphics{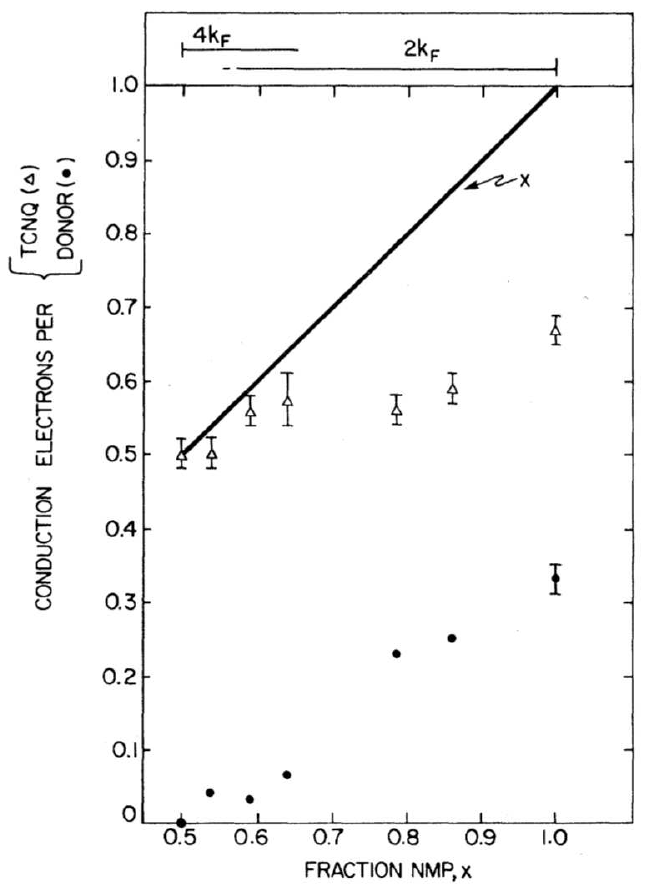}}}
  \caption{Number of conduction electrons per TCNQ (triangles) and per
    donor molecule (dots) in (NMP)$_x$(phen)$_{1-x}$TCNQ. The solid
    line is equal to $x$.  Reprinted with permission from
    Ref. \cite{Epstein81a}, $\copyright$ 1981 The American Physical
    Society.}
\label{Epstein}
\end{figure}
Diffuse X-ray scattering \cite{Pouget76a,Comes77a,Kagoshima76a} and
inelastic neutron scattering \cite{Mook76a,Shirane76a} of TTF-TCNQ
gave the first experimental detection of the 4k$_{\rm F}$ instability
in the donor stack of TTF-TCNQ.  Subsequent measurements detected the
4k$_{\rm F}$ instability in the one-chain compounds MEM(TCNQ)$_2$
\cite{Bosch77a,Huizinga79a,Huizinga82a} and
(NMP)$_x$(phen)$_{1-x}$TCNQ for $0.5\leq x \leq 0.54$
\cite{Epstein81a}.  MEM(TCNQ)$_2$ continues to enjoy the status of
being one of the most well-characterized materials, in that both the
4k$_{\rm F}$ and the 2k$_{\rm F}$ phases here were clearly identified
as bond dimerized and tetramerized, respectively
\cite{Huizinga79a}. The systems (NMP)$_x$(phen)$_{1-x}$TCNQ were
interesting as the carrier concentration in the TCNQ acceptor stack
could be varied almost continuously by changing the ratio between the
electron donor NMP and the neutral phenazine molecule. Strong diffuse
X-ray scattering at 4k$_{\rm F}$=$\pi/a$, where $a$ is the lattice
constant was found at $x$ exactly equal to 0.5, where there occurs
complete charge-transfer from each NMP molecule. As $x$ is increased
from 0.5, there is a concomitant but nonlinear increase in $\rho$,
with $\rho$ reaching the value $\frac{2}{3}$ at $x=1$. Interestingly,
while for low electron densities both 4k$_{\rm F}$ and 2k$_{\rm F}$
instabilities on the TCNQ stack are observed, for $x>\frac{2}{3}$
there is no 4k$_{\rm F}$ scattering at all and only 2k$_{\rm F}$
scattering is observed.  This specific observation was the first
experimental demonstration of the systematic carrier concentration
dependent behavior of correlated one-dimensional bands (see Section
\ref{rho-dependence}), and is summarized in Fig.~\ref{Epstein}. Very
interestingly, the 4k$_{\rm F}$ scattering vector is unshifted between
$x=0.5$ and $x=0.54$ \cite{Epstein81a}. This was interpreted as
evidence for the formation of fractionally charged $e/2$ solitons
\cite{Epstein82a,Conwell83a}, which had been theoretically suggested
earlier by Rice and Mele within the $U \to \infty$ Peierls-Hubbard
model \cite{Rice82a}.

\begin{figure}[tb]
  \begin{center}
    \resizebox{2.5in}{!}{\includegraphics{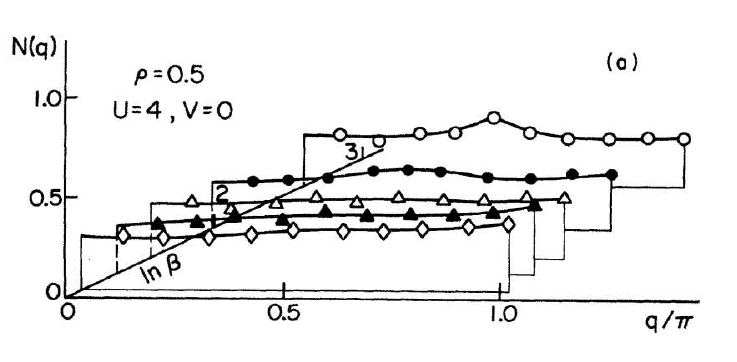}}\hspace{0.1in}%
    \resizebox{2.3in}{!}{\includegraphics{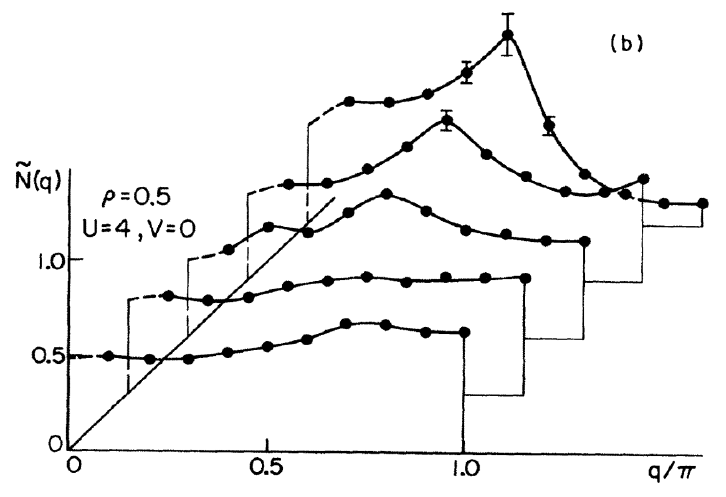}}\hspace{0.1in}%
  \end{center}
\caption{(a) Charge susceptibility of the 1D Hubbard model as a
  function of wavevector and inverse temperature $\beta$ for a 20-site
  lattice with $U=4$ and $V=0$ at $\rho=\frac{1}{2}$. Here $q=\pi$
  ($\frac{\pi}{2}$) corresponds to 4k$_{\rm F}$ (2k$_{\rm F}$).  (b)
  bond-order susceptibility for the same parameters.  Reprinted with
  permission from Ref.~\cite{Hirsch84a}, $\copyright$ 1984 The
  American Physical Society.}
\label{HirschU}
\end{figure}
Identification of the ground state as a manifestation of the 4k$_{\rm
  F}$ Peierls charge-density wave (CDW) (see above) appeared
reasonable, but it was suggested by Torrance \cite{Torrance78a} that
the mechanism of the instability was in fact different.  Torrance was
of the opinion that the 4k$_{\rm F}$ instability was driven by
intersite Coulomb interactions and resulted in a gain in {\it
  potential energy}, as opposed to the gain in kinetic energy that
drives the Peierls instability.  This was the first foray into the
concept of {\it charge ordering} (CO).  Very soon after this Hubbard
in a seminal paper developed the concept of Wigner crystallization
driven by long range intersite Coulomb interaction $V_{ij}$, within
the extended Hubbard Hamiltonian (EHM),
\begin{equation}
H=-\sum_{\langle ij \rangle \sigma}t_{ij}c^\dagger_{i\sigma}
c_{j\sigma} + U\sum_{i} n_{i,\uparrow}n_{i,\downarrow} 
+ \frac{1}{2} \sum_{i,j} V_{ij} n_{i} n_{j}
\label{EHM}
\end{equation}
where $n_i = \sum_{\sigma}n_{i,\sigma}$ is the total number of
electrons on site $i$.  Note that the intersite interaction in the
last term is not limited to NN.  Hubbard's actual work was for the
limits of $U=\infty$ and $t_{ij}=0$, in which case Eq.~\ref{EHM}
reduces to a classical Ising-like Hamiltonian $H_{\rm red}=\frac{1}{2}
\sum_{i,j} V_{ij} n_{i} n_{j}$, with $n_i=0,1$.  Hubbard was able to
derive the WC site occupancies by electrons in 1D for all $\rho$ for
arbitrary but downward convex $V_{ij}$.  Interestingly, for the
special case of $\rho=\frac{1}{2}$, Hubbard found two competing Wigner
crystals, denoted by $\cdots 1010 \cdots$ and $\cdots 1100 \cdots$,
where 1 and 0 refer to sites occupied and unoccupied by charge
carriers, respectively. Subsequent theoretical work showed that
intersite Coulomb interactions beyond NN play insignificant role in
the CTS \cite{Bloch83a}.  We shall hereafter retain only the NN
intersite Coulomb interaction and refer to this as $V$. We will refer
to the Hamiltonian with $U$, $V$ and $|t|$ as the extended Hubbard
Hamiltonian. The configuration $\cdots 1100 \cdots$ plays no important
role in the absence of long range interaction, within this {\it
  classical picture} of lattice instability. We will refer to the
configuration $\cdots 1010 \cdots$ only as the WC for
$\rho=\frac{1}{2}$.
\begin{figure}[tb]
  \begin{center}
    \resizebox{1.8in}{!}{\includegraphics{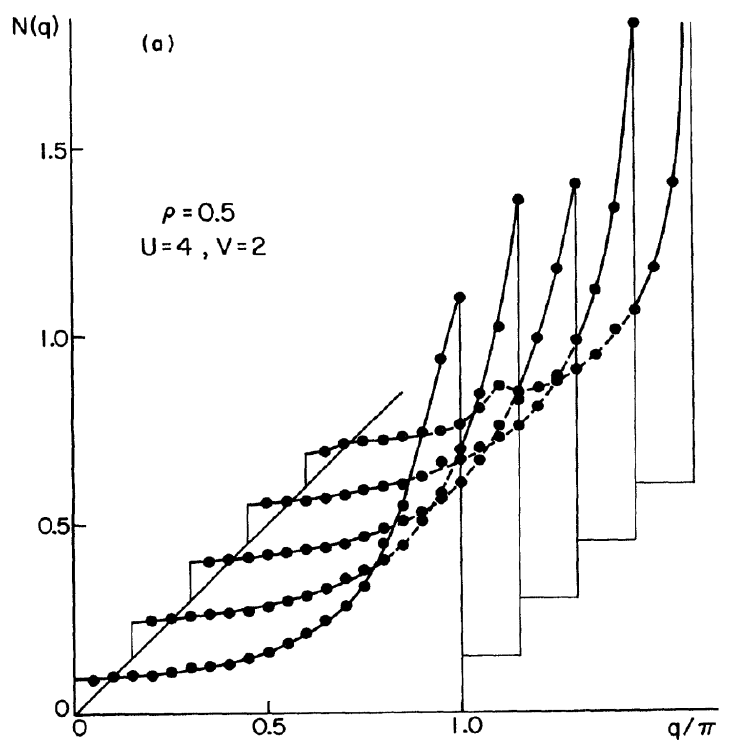}}\hspace{0.2in}%
    \resizebox{2.0in}{!}{\includegraphics{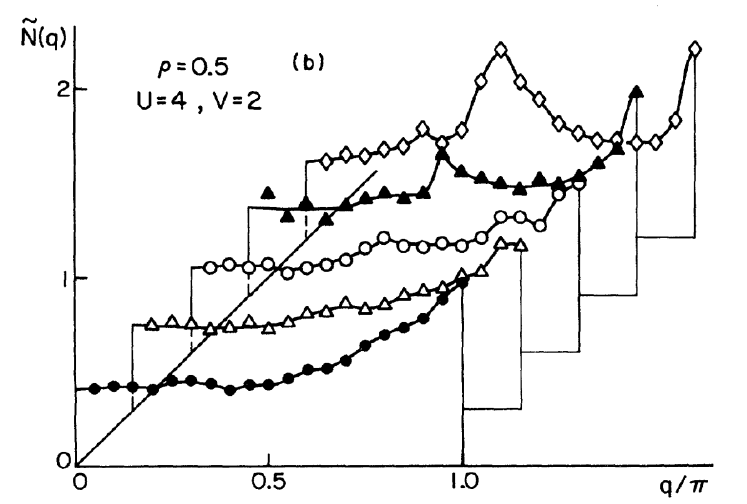}}
  \end{center}
\caption{(a) Charge susceptibility of the 1D extended Hubbard model as
  a function of wavevector and inverse temperature $\beta$ for a
  40-site lattice with $U=4$ and $V+2$ at $\rho=\frac{1}{2}$.  Here
  $q=\pi$ ($\frac{\pi}{2}$) corresponds to 4k$_{\rm F}$ (2k$_{\rm
    F}$). (b) bond-order susceptibility for the same parameters.
  Reprinted with permission from Ref.~\cite{Hirsch84a}, $\copyright$
  1984 The American Physical Society.}
\label{HirschUV}
\end{figure}

\begin{figure}[tb]
  \begin{center}
    \begin{overpic}[width=0.5in]{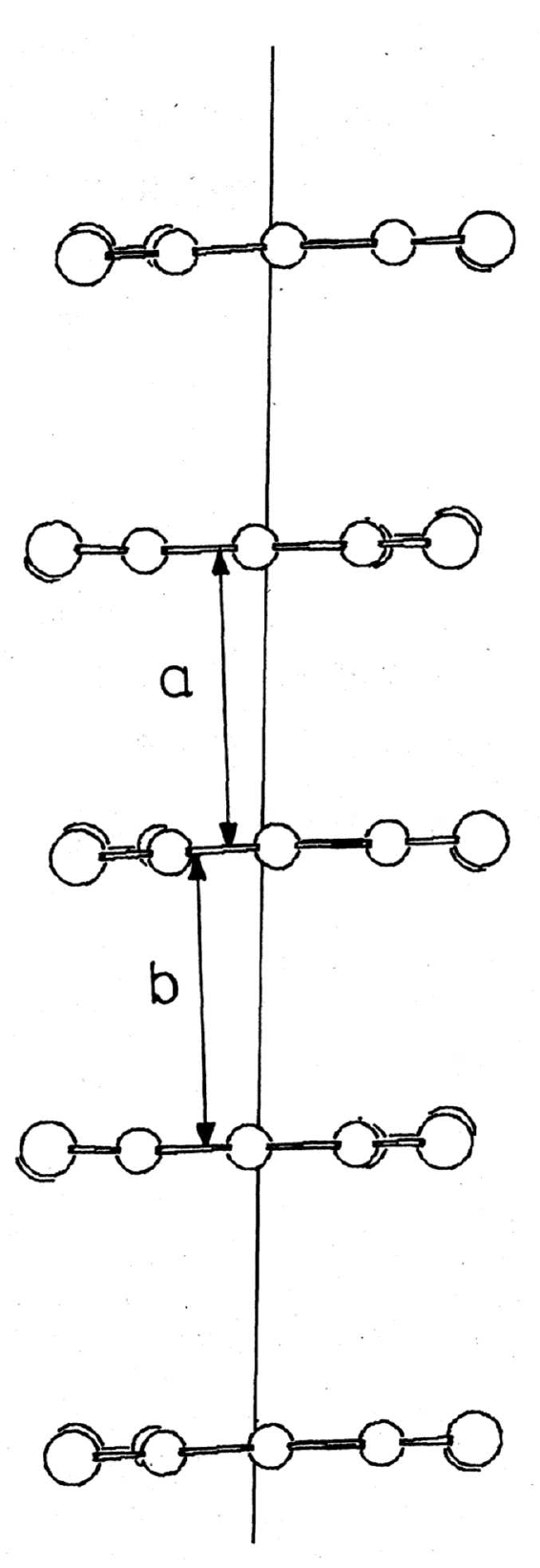}
      \put (-12,85) {\small(a)}
    \end{overpic}
    \hspace{1.0in}
    \begin{overpic}[width=0.8in]{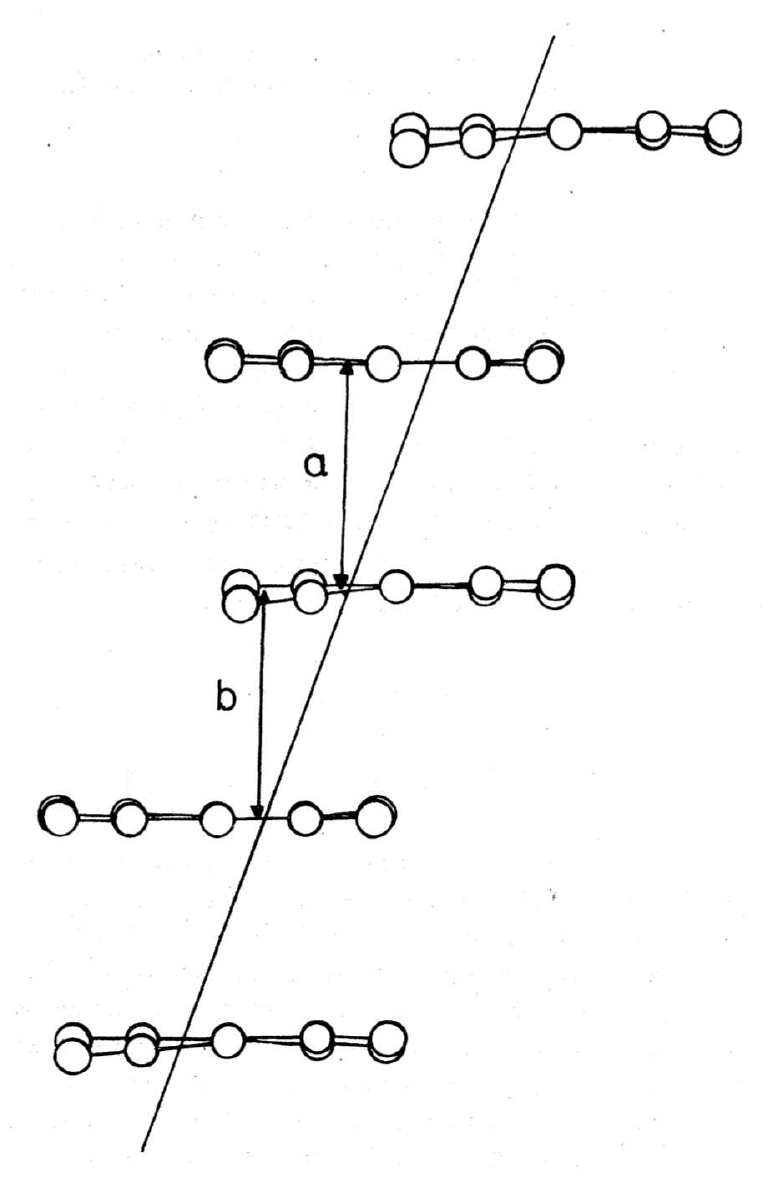}
      \put (-5,80) {\small(b)}
    \end{overpic}
  \end{center}
\caption{(a) TCNQ molecules of MEM(TCNQ)$_2$ as seen along their
  longest axis in the metallic state.  The drawn line indicates the
  chain direction. Nearest neighbor intermolecular hoppings are
  uniform in this case. (b) Same below T$_{4k_{\rm F}}$. The stack is
  now strongly dimerized.  Reprinted with permission from
  Ref.~\cite{Huizinga79a}, $\copyright$ 1979 The American Physical
  Society.}
\label{MEM}
\end{figure}

The EHM with NN intersite interaction was investigated numerically for
moderate $U$ and $V$ using quantum Monte Carlo (QMC) by Hirsch and
Scalapino \cite{Hirsch84a}, for $\rho=\frac{1}{2}$ and then for
$\rho=0.6$. The authors initially probed wavevector dependent
charge-charge correlations only, which gives the tendency to CO
\cite{Hirsch83c,Hirsch84a}, and concluded that the $\cdots 1010
\cdots$ 4k$_{\rm F}$ CO is obtained at $\rho=\frac{1}{2}$ only for
nonzero $V$ when $U$ is finite (see Fig.~\ref{HirschU}). In later work
\cite{Hirsch84a} the authors also investigated wavevector-dependent
bond order-bond order correlations, and for a limited set of $U$, $V$
parameters found signatures of both 2k$_{\rm F}$ and 4k$_{\rm F}$ BOW
instabilities (see Fig.~\ref{HirschUV}).  These results played a very
significant role in later developments of theories of spatial broken
symmetry in the correlated 1D, especially for $\rho=\frac{1}{2}$. They
showed that information about the {\it tendency to CO or BOW} can be
obtained even without explicit inclusion of electron-phonon
interactions in the Hamiltonian, as this information is built into the
purely electronic wavefunction itself. Second, the demonstration of
the BOW instability, even for nonzero $V$, indicated that CO is
obtained only for the NN interaction larger than a threshold value, a
result that would be demonstrated explicitly many years later
\cite{Mila93a,Penc94a,Clay03a}.  On the other hand, the pattern of the
2k$_{\rm F}$ BOW, or the issues of coexistence versus competition
between CO and BOW were, however, not understood at this point. One
important experimental development was the identification of the phase
below T$_{\rm 4k_{\rm F}}$ in MEM(TCNQ)$_2$ as the dimerized BOW phase
\cite{Huizinga79a}. We show in Fig.~\ref{MEM} schematics of the
structure of the TCNQ stack above and below T$_{\rm 4k_{\rm F}}$.
T-dependent magnetic susceptibility $\chi(T)$ showed Bonner-Fisher
behavior in the region T$_{\rm 2k_{\rm F}} < $T$ < $T$_{\rm 4k_{\rm
    F}}$, with an SP gap below T$_{\rm 2k_{\rm F}}$, identifying the
latter temperature as the SP temperature T$_{\rm SP}$. Neutron
diffraction studies showed that the SP transition led to dimerization
of the bond-dimerization \cite{Visser83a}.

\subsection{Intermolecular versus intramolecular electron-phonon interactions}

Two different kinds of electron-phonon (e-p) interactions have been
discussed in the literature. The so-called Su-Schrieffer-Heeger (SSH)
coupling \cite{Su79a} between electrons and stretching vibrations of
the 1D lattice modulates the intermolecular hopping.  Because of the
massive molecular natures of the sites in the CTS, displacements
orthogonal to the stacking axis
(see Fig.~\ref{MEM}(b)) as well as
librational motion can make significant contributions to the SSH
coupling. The Holstein electron-molecular vibration (EMV) coupling
\cite{Holstein59a} also plays a significant role in CO instabilities
as well as optical excitations \cite{Rice75a,Rice77a,Conwell88a}.
Strictly in $\rho=1$ these two
couplings give rise to competing BOW and site-CDW instabilities, for
$U=0$ \cite{Kivelson83a,Su83a} as well as nonzero $U$ \cite{Dixit84a}.
Coexistence is the norm at all other $\rho$ \cite{Ung93a}. The EMV
coupling dictates that specific molecular frequencies vary with the
charge density on the molecule, and at the onset of charge
disproportionation at low T there occurs a splitting of infrared
vibrational modes that are coupled strongly to the electrons
\cite{Dressel12a,Girlando11a}.

\subsection{$\rho$-dependent correlation effects}
\label{rho-dependence}

The 4k$_{\rm F}$ instability, independent of its CDW or BOW nature, is
clearly a signature of large Hubbard $U$. Yet another signature of
strong correlation effects is the static magnetic susceptibility
$\chi(T)$: $U$ enhances the susceptibility strongly over what is
expected in the noninteracting $U=0$ limit for the same hopping
integral \cite{Torrance77a}.  A mystery in this early phase was the
{\it absence} of enhanced magnetic susceptibility and the 4k$_{\rm F}$
instability in many CTS. Systems with nearly identical molecular
components and very similar crystal structures behave very
differently. Thus while the susceptibility of TTF-TCNQ at room
temperature was enhanced by a factor of 3 over the expected Pauli
value \cite{Torrance77a}, the corresponding enhancement factor for
HMTTF-TCNQ was only 1.15 \cite{Tomkiewicz77a}. The same enhancement
factor for MEM(TCNQ)$_2$ was 20! It also appeared that enhanced
(unenhanced) susceptibility and occurrence (non-occurrence) of the
4k$_{\rm F}$ instability went together. This had led to considerable
argument about the magnitude of $U$ in the CTS.

\begin{figure}[tb]
  \begin{center}
    \resizebox{3.0in}{!}{\includegraphics{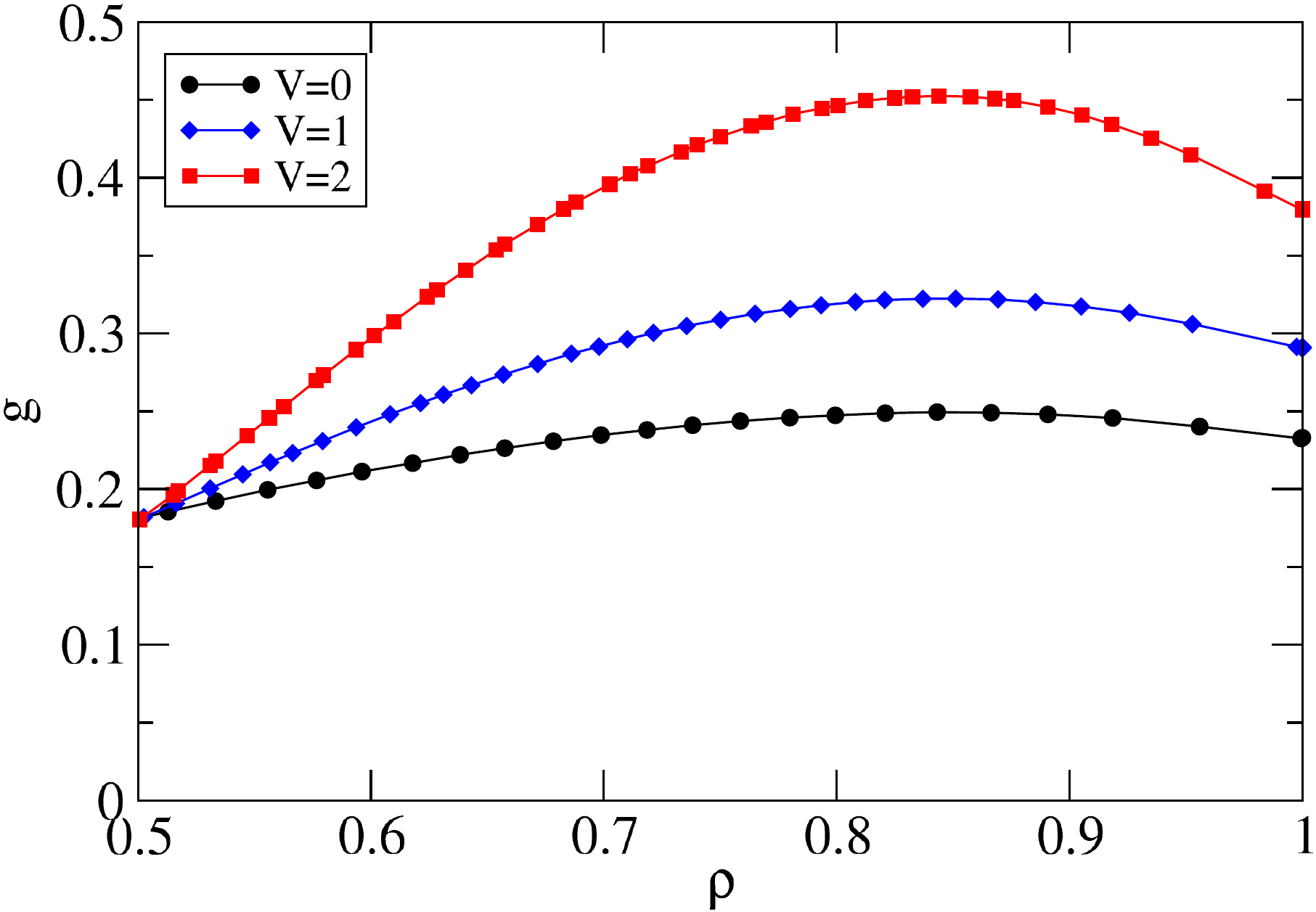}}
    \end{center}
  \caption{(Color online) Normalized double occupancy $g(\rho)$ versus
    $\rho$ for the 1D EHM of Eq~\ref{EHM} for $U=6$, and $V=0$
    (circles), 1 (diamonds) and 2 (squares), for a 64 site
    system. Lines are guides to the eye.}
\label{g-1d}
\end{figure}

The solution to this problem came from the recognition that neither the
$U>>|t|$ nor the $V=0$ approximations that are often adopted for
simplicity are justified. With realistic $U$ ($4|t| \leq U \leq 8|t|$)
and $V$ ($|t| \leq V \leq 3|t|$), correlation effects are strongly, and
{\it non-monotonically} dependent on $\rho$
\cite{Mazumdar83a,Mazumdar86a}.  At or near $\rho=\frac{1}{2}$, $V$ and
$U$ act in concert to reduce double occupancies of sites by charge
carriers; however, over the broad region $\frac{2}{3} \leq \rho \leq 0.8$
$V$ {\it increases} double occupancy. One measure of the
$\rho$-dependent correlations is the normalized probability of double
occupancy of sites by carriers $g(\rho)$, defined as,
\begin{equation}
g(\rho)=\frac{\langle n_{i,\uparrow}n_{i,\downarrow}\rangle}{\langle n_{i,\uparrow}\rangle \langle n_{i,\downarrow}\rangle}
\end{equation}
The plot in Fig.~\ref{g-1d} shows $g(\rho)$ versus $\rho$ for several
different $V$ ($V=$ 0, 1, and 2 in units of $|t|$) for $U=6|t|$
calculated using the Stochastic Series Expansion (SSE) QMC method
\cite{Sandvik92a,Syljuasen02a}.  Behavior for other $U$ is the
same. For nonzero $V$ strongly correlated behavior (small $g$) is
found for $\rho$ at or near $\frac{1}{2}$, while relatively weakly
correlated behavior (moderate to large $g$) occurs for $\frac{2}{3}
\leq \rho \leq 0.8$.  A turnaround in $g(\rho)$ is again observed as
the Mott-Hubbard region of $\rho=1$ is approached.

Table \ref{table-1d} lists the known behavior of a large number of CTS
with $\rho$ spanning from 0.5 to 1. The results agree fully with the
predictions of Fig~\ref{g-1d}. Enhanced susceptibilities and 4k$_{\rm
  F}$ scattering always appear in the range $0.5 < \rho < 0.55$ and
never appear in the range $0.63 < \rho < 0.8$.  The intervening range
$0.55 < \rho < 0.63$ forms a transition region: the symptoms appear at
$\rho = 0.59$ in TTF-TCNQ, but do not appear at $\rho = 0.57$ in
TMTSF-TCNQ with its considerably larger cation bandwidth.  Compounds in the
range $0.83 < \rho < 1.0$ are unknown, and compounds with $\rho=1$ are
universally Mott-Hubbard semiconductors.

\begin{table}
  \begin{center}
    \resizebox{2.75in}{!}{\includegraphics{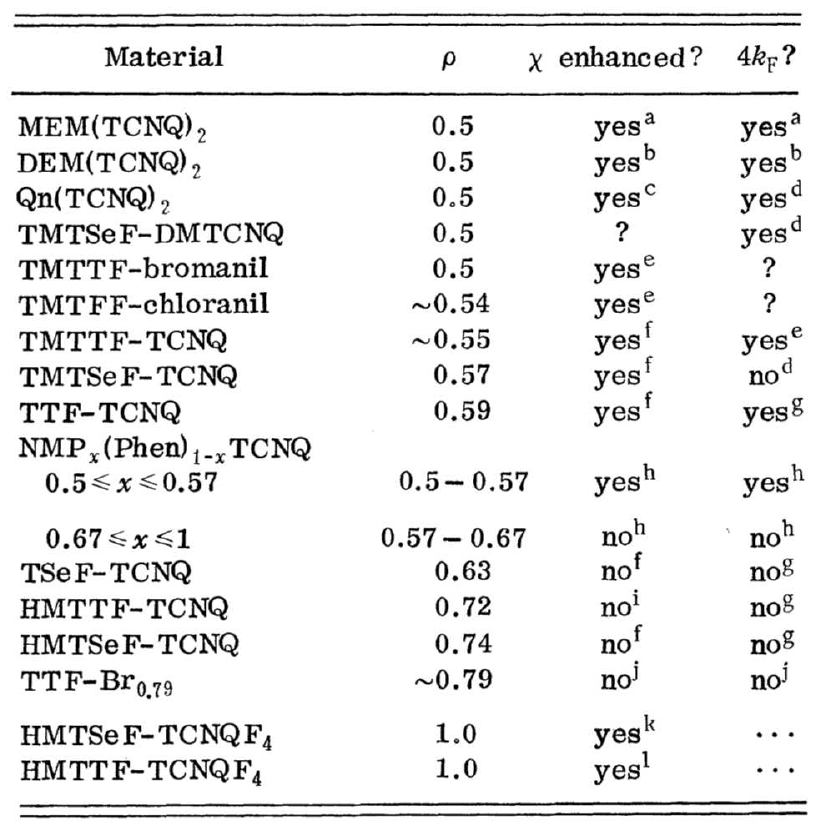}}
    \end{center}
  \caption{Summary of the properties of several quasi-1D CTS
    in the density range 0.5 $\leq$ $\rho$ $\leq$ 1 \cite{Mazumdar83a}.}
\label{table-1d}
\end{table}

The demonstration of $\rho$-dependent correlation effects is a
significant result, since SC in the CTS is restricted precisely
to the carrier concentration where correlation effects are the
strongest! This makes it unlikely that the SC mechanism can be
understood within a noninteracting electron picture. In \ref{othersc}
we will point out that many of the perplexing features that
are seen in the the $\rho=\frac{1}{2}$ CTS, including SC, are also
shared by other $\rho=\frac{1}{2}$ systems, including inorganic
materials.

\subsection{Summary} The normal state of the entire family of
  CTS, whether one-chain or two-chain, can be understood within the
  extended Hubbard model (EHM) with nonzero moderate nearest-neighbor
  Coulomb interaction $V$. The $V$ interaction enhances single
  occupancy of sites by electrons near $\rho=\frac{1}{2}$, and double
  occupancy in the range $0.67 < \rho < 0.8$, thus explaining both the
  observed ``strongly correlated'' and ``weakly correlated
  behavior''. Lattice instabilities can be driven by both
  intermolecular electron-phonon coupling and electron-molecular
  vibration (EMV) coupling.  The periodicities of the density waves
  however are determined by Coulomb interactions. Correlation effects
  are strongest at $\rho=\frac{1}{2}$, which is the carrier density in
  the superconductors (see following sections), making it unlikely
  that superconductivity in the CTS can be explained within BCS
  theory.

\section{Broken symmetries in quasi-one dimensional CTS, Theory}
\label{broken-1d}

\begin{figure}[tb]
  \center{\resizebox{4.0in}{!}{\includegraphics{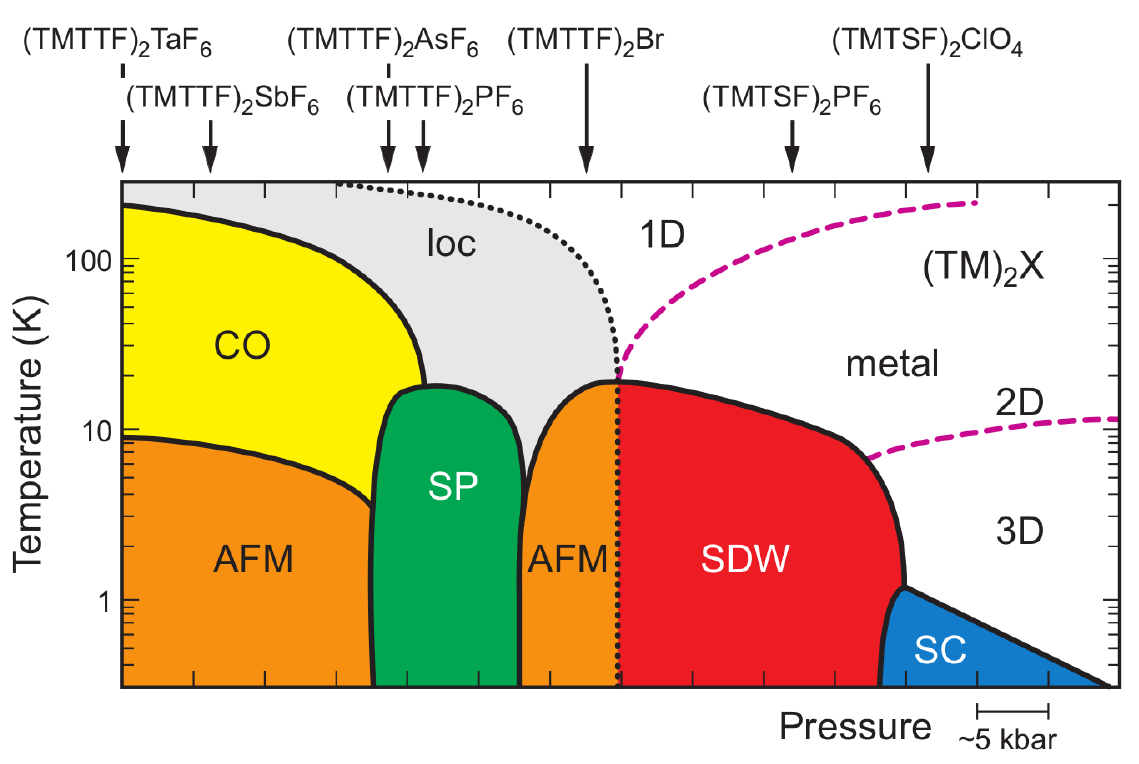}}}
  \caption{Phase diagram proposed by Jerome for the TMTTF$_2$X and
    TMTSF$_2$X.  Arrows indicate the ambient pressure position of
    several materials.  Reprinted with permission from
    Ref.~\cite{Salameh11a}, $\copyright$ 2011 The American Physical
    Society.}
\label{phasediagram}
\end{figure}

While the earliest work described in the previous section was largely
on the two-chain conductors, interest over the past two decades have
been focused on the 2:1 cationic materials (and more recently, on the
1:2 anionic materials), as these are the only ones that exhibit SC.
The Fabre and Bechgaard salts (TMTCF)$_2$X, C = S, Se, have been
intensively studied. A generic phase diagram (see
Fig.~\ref{phasediagram}) for these systems has been proposed
\cite{Jerome91a}.  The proposed phase diagram has received ample
experimental support. It is generally agreed that the pressure axis is
a measure of the interchain coupling at ambient pressure.  Different
materials shown along this axis indicate the role interchain
interactions play in specific materials. The different transition
temperatures in the phase diagram that are of interest are (i) T$_{\rm
  MI}$, the metal-insulator (MI) transition temperature, which at the
very extreme left occurs at a high temperature of 200 K or so, and is
driven by charge localization, induced perhaps by counter anion-driven
dimerization along the stacking axis; (ii) T$_{\rm CO}$, the
temperature at which charge-ordering occurs (usually $~60-100$K), and
(iii) T$_{\rm AFM}$ and T$_{\rm SP}$, temperatures at which
transitions to the antiferromagnetic and SP states occur (we refer to
the AFM/SDW phase of Fig.~\ref{phasediagram} as simply AFM-II, and the
antiferromagnetic phase at the low pressure region as AFM-I).  T$_{\rm
  AFM}$ and T$_{\rm SP}$ are nearly an order of magnitude lower than
T$_{\rm CO}$ near the low pressure region of the phase diagram.  On
the other hand, a single transition from the metallic state to the
AFM-II state occurs at the high pressure region.

One goal of the present work is to review theoretical works that have
attempted to explain the phase diagram, including the
temperature dependence and the evolution from the left to the right
along the pressure (interchain interaction) axis. As of now though the
evolution to the superconducting state still remains a mystery, and
much of the theoretical work on the superconducting phase is
independent of other works on the spatial broken symmetry states. We
believe, however, that a correct theoretical approach should be able
to explain the evolution of the SC state from the semiconducting
states.  Since the SC state is largely believed to be spin singlet, a
strong emphasis is placed in this review to understand the transitions
between the two AFM states and the singlet SP state, as the mechanisms
of these AFM--spin singlet transitions may hold the key understanding
the mechanism of the transition to SC.
A recent suggestion that like the AFM phases there also exist two different SP states 
with different charge occupancies and bond distortions that are
separated by a phase boundary, has however been shown to be incorrect \cite{Yoshimi12a,Ward14a,Seo14a}.

Recent experimental works have shown that the phase diagram in
Fig.~\ref{phasediagram} is less universal than it was originally
believed to be. There exist several quasi-1D 2:1 cationic materials
that actually {\it do not} lie in any of the regions of the phase
diagram Fig.~\ref{phasediagram}.  Summaries of experimental results on
several of these ``non-traditional'' quasi-1D systems are presented in
Section \ref{1dnontrad-section}.  In the (TMTCF)$_2$X series charge
and spin transitions are separated in high and low temperature scales,
with spin order (either AFM or SP) only occurring at low ($<$20 K)
temperature.  In the materials discussed in Section
\ref{1dnontrad-section}, there are in contrast examples of high
temperature spin gap transitions, and others where the charge and spin
gaps occur {\it simultaneously} at high temperatures.  (BCPTTF)$_2$X,
X = PF$_6$ and AsF$_6$, exhibits a magnetic transition at a
temperature (30 K) significantly higher than T$_{\rm SP}$ in the Fabre
salts, without going through a CO phase.  In (EDO-TTF)$_2$X T$_{\rm
  MI}$ coincides with the spin gap transition at approximately 270 K;
rather than a traditional second-order SP transition, the transition
in (EDO-TTF)$_2$X is strongly first order.

While some of these unusual characteristics and details of the
thermodynamics may require the inclusion of 3D interactions and anion
coupling, we argue below that much {\it can} be understood from the
underlying 1D strongly-correlated electron model.  In particular, in
Section \ref{1dlimit} we show that the $\cdots$1100$\cdots$ CO can
coexist with two {\it different} bond distortion patterns, BCDW-I and
BCDW-II.  Numerical calculations (Section \ref{1dnumerics}) show that
depending on the strength of e-e and e-p interactions, the magnitude
of the CO varies by an order of magnitude, which greatly changes the
magnitude of co-operative effects like anion coupling.  In Section
\ref{1d-tdep} we discuss the reason different thermodynamics are
expected for BCDW-I and BCDW-II.

\subsection{Theoretical model and parameters}

The theoretical model appropriate for the complete families of the
above materials incorporates single chain and interchain interactions,
as well as interaction with the counterions.
As we are not interested in material-specific behavior we ignore the 
interaction with counterions and write the Hamiltonian as
\begin{eqnarray}
H=-\sum_{\nu,\langle ij\rangle_\nu}t_\nu(1+\alpha_\nu\Delta_{ij})B_{ij} 
+\frac{1}{2}\sum_{\nu,\langle ij\rangle_\nu} K^\nu_\alpha \Delta_{ij}^2 \nonumber \\
+g \sum_i v_i n_i + \frac{K_g}{2} \sum_i v_i^2  
+ U\sum_i n_{i\uparrow}n_{i\downarrow} + 
\frac{1}{2}\sum_{\langle ij\rangle}V_{ij} n_i n_j.
 \label{ham}
\end{eqnarray}
In Eq.~\ref{ham}, $\nu$ runs over multiple lattice directions $a$, $b$
and $c$ (with the stacking axis along the $a$ direction in most of the
quasi-1D materials).
$B_{ij}=\sum_\sigma(c^\dagger_{i\sigma}c_{j\sigma}+H.c.)$,
$\alpha_\nu$ is the intersite e-p coupling, $K^\nu_\alpha$ is the
corresponding spring constant, and $\Delta_{ij}$ is the distortion of
the $i$--$j$ bond, to be determined self-consistently; $v_i$ is the
intrasite phonon coordinate and $g$ is the intrasite e-p coupling
with the corresponding spring constant $K_g$. For the Fabre salts
it is generally accepted that $t_a\simeq0.1-0.2$ eV $>> t_b >> t_{\rm c}$
\cite{Ishiguro}, while in the TMTSF $t_a$ is larger and $t_b\sim0.1t_a$.  We
will henceforth write $t_a$ and $t_b$ as $t$ and $t_{\perp}$,
respectively.  Additional interesting roles played by frustration may
be relevant \cite{Yoshimi12a}; we postpone discussions of these until
later.  Because of the strong anisotropies, we will consider only
unidirectional SSH electron-phonon coupling ($\alpha_\nu$, $K_\nu$ = 0
for $\nu \neq a$). Based on the strong $\rho$-dependent experimental
behavior in the family of CTS overall (see Section \ref{rho-dependence}) we retain only
the NN intersite Coulomb interactions, which we will write as $V$ and
$V_{\perp}$, respectively (note that longer range interactions wipe
out the $\rho$-dependence of correlation effects \cite{Hubbard78a}). The onsite Hubbard
repulsion $U$ ranges from 0.6 - 1.0 eV. This implies $U/t \sim 6-10$
in the TMTTF, with the ratio about half in TMTSF. The magnitude of $V$
is less certain, but based on the fact that $\frac{1}{2}$-filled band
materials are Mott-Hubbard semiconductors, $V<\frac{1}{2}U$
(since otherwise $\frac{1}{2}$-filled systems would have been
charge-ordered.)  Based on other experiments (see Section \ref{1d-expt}) we
believe that $\frac{U}{4} \leq V \leq \frac{U}{3}$.

The Hamiltonian in Eq.~\ref{ham}, especially the 1D version of it, has
been investigated at $\rho=\frac{1}{2}$ using a variety of approaches
including mean-field \cite{Seo97a,Seo00a,Shibata01a,Seo07a,Otsuka08a},
bosonization
\cite{Tsuchiizu01a,Sugiura04a,Sugiura05a,Yoshioka06a,Yoshioka07a,Yoshioka10a},
exact diagonalization
\cite{Mila93a,Penc94a,Sano94a,Ung94a,Riera00a,Nakamura00b,Riera01a,Clay02a,Sano04a,Li10a,Dayal11a,Yoshimi12a},
quantum Monte Carlo (QMC)
\cite{Hirsch83c,Hirsch84a,Clay01a,Clay03a,Clay05a,Mazumdar00a,Clay07a,Clay12a,Clay17a}
and the Density Matrix Renormalization Group (DMRG)
\cite{Nishimoto00a,Kuwabara03a,Ejima05a,Ejima06a,Benthien05a,Shirakawa09a}. Mean-field
approaches ignore quantum fluctuations and in general overemphasize
the AFM and the WC broken symmetries, thereby missing the subtleties
associated with the transition to spin singlet states, which is a
quantum effect.  We will not discuss the mean-field and related
approaches. Bosonization ignores the discrete nature of the lattice
and thus in most cases miss the coexistences of various broken
symmetries; it is also a weak coupling approach, while the strong e-e
interactions in the CTS demand a strong coupling approach. The two
separate requirements of retaining the lattice discreteness and the
strong coupling are met only by numerical approaches, and we will
largely limit ourselves to discussions of these.

Numerical approaches suffer from one serious disadvantage, in that
they are often for relatively small systems (especially those results
that are obtained through exact diagonalizations).  Proper finite size
scaling of numerical results is then crucial for obtaining the
extrapolated behavior of the physical system in the thermodynamic
limit. This is, however, often very difficult or even impossible, as
strong e-e interactions can preclude calculations for several
different system sizes, which is essential for finite size
scaling. QMC and DMRG have their own difficulties (for example,
self-consistent solutions to the e-p interactions can be a problem
using these approaches) and are even less useful when lattice
frustration plays a strong role. These limitations can make it
difficult to determine whether a given broken symmetry that is found
from numerical calculations is unconditional or not (an unconditional
transition is one which occurs for infinitesimally small values of the
coupling constants $\alpha$ and $g$ in the thermodynamic limit;
finite systems with discrete energy gaps between states give broken
symmetries only for nonzero coupling constants.) In general only
unconditional transitions are expected in the physical systems.  While
conditional transitions that occur only for couplings above specific
threshold values may also occur in a real system, it is necessary to
prove in such cases that the couplings in the real system do exceed
the required threshold values. This is also a difficult task within
correlated-electron models. In view of the above, prior to presenting
the numerical data we develop physical heuristic ideas of the
unconditional broken symmetries that might be expected within
Eq.~\ref{ham} for $\rho=\frac{1}{2}$ in the thermodynamic
limit. Agreement between the heuristic ideas and numerical results may
be considered a criterion for essential correctness of both.

\subsection{Spatial broken symmetries - configuration space based physical ideas}
\label{configspace}

\begin{figure}[tb]
  \centerline{\resizebox{4.0in}{!}{\includegraphics{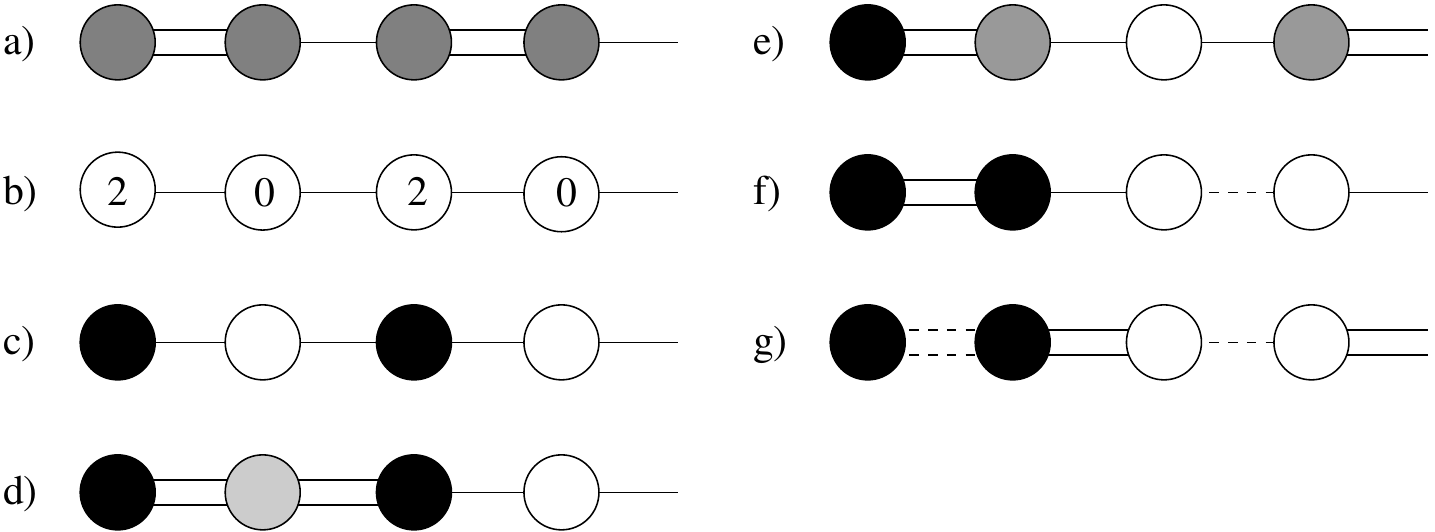}}}
  \caption{Broken symmetry states in 1D. (a) $\rho=1$ 2k$_{\rm F}$ BOW and $\rho=1$ 4k$_{\rm F}$
BOW, both with uniform charge.
    (b) $\rho=1$ 2k$_{\rm F}$ CDW. Numbers here indicate
    site charge occupancy. (c) $\rho=\frac{1}{2}$ 4k$_{\rm F}$ CDW (d)
    $\rho=\frac{1}{2}$ 4k$_{\rm F}$ CDW-SP (e) $\rho=\frac{1}{2}$ 2k$_{\rm F}$ CDW (f)
    $\rho=\frac{1}{2}$ BCDW-II (g) $\rho=\frac{1}{2}$ BCDW-I.  Lines indicate bond
    strength with double lines stronger than single and dashed bonds,
    respectively. In (c)-(g) black (white) circles indicate sites with
    charge density $0.5+\frac{1}{2}\Delta n$ ($0.5-\frac{1}{2}\Delta n$).}
  \label{1dcartoons}
\end{figure}  

There are reasons beyond justifying numerical results to
develop physical ideas concerning spatial broken symmetries in
correlated-electron materials. First, the alternative physical
explanations are usually couched in the language of Peierls
instability, with nesting as the condition for lattice or magnetic
instability. This is a strictly one-electron theory and does not apply
to the many-electron Hamiltonian of Eq.~\ref{ham}. The effect of e-e
interactions are understood best in configuration space. Second, as we
show below, coexisting, or even co-operative interactions between
broken symmetries are most easily understood within the configuration
space picture (by co-operative we refer to the situation where a given
interaction enhances two or more coexisting broken symmetries
simultaneously). Below we first discuss the 1D limit. The AFM phases
are not obtained in this limit even at T=0. Transitions to
spin-singlet states are possible, although 2D interactions are
necessary to get the same transitions at finite temperature. Our
discussions of the 1D limit will therefore be limited to charge- and
bond-order only (see Fig.\ref{1dcartoons}). Following this we will
discuss the 2D case, where the AFM phases are considered.

\subsubsection{The one dimensional limit}
\label{1dlimit}

Our configuration space arguments are based on the
observation that the ground state wavefunction of a broken symmetry
state has unequal contributions from equivalent configurations that
are related by the symmetry operator that is lost when symmetry is
broken.  For the sake of illustrating the usefulness of the arguments
we first discuss the instabilities in $\rho=1$, which is is simpler to
understand than any other $\rho$, because $\rho=1$ is an insulator for
nonzero $U$ and because there is minimal coexistence between the BOW
and the CDW. We begin with the CO in the 2-site 2-electron
molecule. The three configurations that describe the many-electron
ground state wavefunction here are written as 20, 11 and 02, where the
numbers correspond to site occupancies by electrons (for SU(2)
symmetry the 11 is the spin-singlet combination of site occupancies).
The symmetric wavefunction has equal relative weights of 20 and 02,
but in the presence of broken symmetry (either because of different
site energies or for very large EMV coupling $g$) the relative
weights will be different. This concept can be easily extended to the
4-site system, where the equivalent ``extreme'' configurations that
are now related by symmetry but make unequal contributions to the
wavefunction are 2020 and 0202. Configurations 2011 (or 1120) are
closer to 2020, from which they are obtained by a single one-electron
hop, while three such hops will be necessary to reach them from
0202. One can then construct ``paths'' between the extreme
configurations 2020 and 0202, such as:
\begin{equation}
2020 \to 2011 \to 1111 \to 0211 \to 0202  \nonumber
\end{equation}
where each arrow denotes a single one-electron hop. There exist many
such paths, as branching can occur at any intervening point:
\begin{equation}
2020 \to 2011 \to 2002 \to 1102 \to 0202 \nonumber
\end{equation}
Due to the very short path lengths between the ``extreme''
configurations, configuration mixing is very efficient in finite
molecules which do not exhibit unconditional broken symmetry.  As the
system size $N \to \infty$, however, the path lengths approach
infinity, and in spite of the very large number of paths, the system
may be ``stuck'' on one or other side of the symmetric configurations,
thus undergoing broken symmetry. In the infinite system, $g \to
0^+$ may therefore be sufficient to give the 2k$_{\rm F}$ CDW
(Fig.~\ref{1dcartoons}(b))
here in the
limit of $U=0$. The consequences of e-e interactions can now be
determined from inspection of the paths alone. $U$ diminishes double
occupancies, thus making the paths ``downhill'' to the symmetric
configurations, and therefore strongly reduces the amplitude of the
CDW.

The 2k$_{\rm F}$ BOW in $\rho=1$ (Fig.~\ref{1dcartoons}(a))
has in the past been of enormous interest
because of its applicability to polyacetylene \cite{Baeriswyl92a}. The extreme
configurations that favor one or other bond alternation phase are now
the valence bond (VB) diagrams 1--2 3--4 5--6 ....  and 2--3 4--5
6--7...., in which nearest neighbors are linked by spin-singlet
bonds. Charge-transfer between bonded sites which are spin-paired is
necessarily stronger than that between non-bonded sites, and thus the
bond-alternated wavefunction has inequivalent contributions from the
above two VB diagrams. Unlike the case of the 2k$_{\rm F}$ CDW, we now see
that to reach one extreme VB diagram from the other, creation (as
opposed to destruction) of double occupancies now becomes
necessary. This creates an energy barrier along the path connecting
the two extreme VB diagrams, and hence we expect the Hubbard $U$ to
enhance the bond alternation \cite{Dixit84a}. The predicted enhancement of
the strength of the 2k$_{\rm F}$ BOW by $U$ has been confirmed by numerical
calculations by many authors \cite{Mazumdar83b,Hirsch83b,Soos84a,Mazumdar85a}.

Other conclusions that can be drawn from the nature of these diagrams
include the absence of coexistence between the BOW and the CDW in
$\rho=1$, since each of these broken symmetries is ``favored'' by
extreme configurations that are different, and that are
non-overlapping in the thermodynamic limit.  Although we have limited
ourselves to 1D in the above, in the special case of $\rho=1$ Coulomb
effects on broken symmetries can be understood even in 2D (where AFM
becomes relevant) using similar arguments \cite{Mazumdar87a}.  The
idea that the pattern of broken symmetry and that the effects of e-e
interactions can be predicted from configuration space based physical
ideas has a firm basis.

The situation is much more complicated for $\rho\neq 1$, which can be
conductors even for nonzero $U$ in the absence of spatial broken
symmetry. Further, unlike in $\rho=1$, we will see that coexistence of
BOW and CDW is now the norm.

\paragraph{$\rho = \frac{1}{2}$, $U=0$}

There are three
classes of configurations that constitute the wavefunction of the
$\rho=\frac{1}{2}$ unit cell: 2000 (class I), 1100 (class II) and 1010
(class III). There are four each of the first two (related to one
another by C$_4$ symmetry) but only two of the last (related by
C$_2$).  At $U=0$ all individual configurations are equally preferred,
which means that classes I and II with larger multiplicities each make
larger contributions to the wavefunction than the class III
configurations. Breaking of spatial symmetry at $U=0$ then involves
either class I or II configurations. Loss of C$_4$ symmetry, which
generates 2k$_{\rm F}$ periodicity, implies that the four configurations
$\cdots2000\cdots$, $\cdots0200\cdots$, $\cdots0020\cdots$ and
$\cdots0002\cdots$ within class I make unequal contributions to the
ground state wavefunction.  The 2k$_{\rm F}$ broken symmetry can also
involve the class II configurations, and the two competing broken
symmetries are described by different phase angles (see below).

There are two inequivalent charge-poor sites labeled `0' in
$\cdots2000\cdots$ distinguished by their distances from sites labeled
`2'.  For nonzero $t$ there is significant charge-transfer across the
2-0 bonds and very little across the 0-0 bonds. Thus if 2k$_{\rm F}$
instability involving class I configurations dominates we expect strong
co-operative coexistence between the 2k$_{\rm F}$ CDW (with charge
modulations of charge-rich, intermediate charge, charge poor,
intermediate charge) and 2k$_{\rm F}$ BOW with bond strength modulations
SSWW (S =strong, W = weak), as shown in Fig.~\ref{1dcartoons}(e). The
charge-rich (and charge-poor) sites in $\cdots1100\cdots$ are
completely equivalent. The bond strengths will be modulated within
this CDW also: at $U=0$ charge-transfer across the 1--1 bond occurs in
both directions, while in the 1--0 bond this is in one direction
only. Thus type II 2k$_{\rm F}$ instability will give 2k$_{\rm F}$ CDW with charge
modulation charge-rich, charge-rich, charge-poor, charge-poor and bond
strength modulations SMWM (S =strong, M= medium, W = weak) as shown in
Fig.~\ref{1dcartoons}(f).  The two broken symmetries are mutually
exclusive, and both correspond to Peierls 2k$_{\rm F}$ CDWs, with charge
densities on sites $n$ given by $\rho_n=\bar{\rho}+\rho_0(\cos2k_{\rm F}n +
\theta)$, where $\bar{\rho}=\frac{1}{2}$ and $\theta=\pi/2$  and
$\pi/4$, respectively.  From the physical arguments alone there is no
way to predict which of the two CDWs wins at $U=0$. Actual
calculations indicate that 2000 dominates over 1100 at $U=0$, but
barely \cite{Ung93a}.

\paragraph{Weak to intermediate $U$, weak $V$}

We imagine increasing $U$ slowly from zero, in the $V=0$ limit
initially. The near degeneracy between $\cdots2000\cdots$ and
$\cdots1100\cdots$ that exists at $U=0$ is quickly destroyed by $U$
and $\cdots1100\cdots$ begins to dominate. We will refer to the
coexisting BOW and CDW -- a Bond-Charge Density wave (BCDW) -- as the
2k$_{\rm F}$ BCDW-II (Fig.~\ref{1dcartoons}(f))
\cite{Ung94a,Clay12a,Clay17a}, to distinguish it from a second BCDW that
appears at stronger correlations (see below).  As the above arguments
indicate, the appearance of the BCDW-II phase {\it does not} need
non-convex intersite Coulomb interaction ($V_1<2V_2$, where $V_1$ and
$V_2$ are NN and NNN interactions, respectively), as has sometimes
been claimed \cite{Kobayashi98a}. Non-convex Coulomb potentials
are unrealistic, as they require screening at short range to be larger
than that at long range. Our arguments above show that the BCDW-II can
be driven by a {\it quantum effect}, the tendency to form nearest
neighbor spin-singlets.

As $U$ is increased further, charge-transfer across the 1-1 bond,
which creates a double occupancy, will gradually get weaker, while the
1-0 bond gets relatively stronger at the expense of the 1-1 bond
strength. There will thus be a gradual cross-over where
at the end point is the 2k$_{\rm F}$ BCDW-I, with the bond strengths now
SW$^{\prime}$SW (S = strong, W = weak, W$^{\prime}$ = weak bond
stronger than W), where the S, W$^{\prime}$ and W bonds corresponding
to 1--0, 1--1 and 0--0 bonds, respectively (see Fig.~\ref{1dcartoons}(g)).
The longer bond length
between the charge-rich sites implies that the crossover from the
BCDW-II to the BCDW-I is facilitated by nonzero $V$ for each $U$. Note
that the BCDW-II and I are the two extreme distortion patterns, with the BCDW-II
dominating at small $U$ and BCDW-I at large $U$.

\paragraph{Large $U$, moderate $V$} This parameter region has been
investigated much more widely. As already pointed out in Section \ref{decoupling},
in the $U \to \infty$ limit the system effectively mimics a band of 1D
spinless Fermions. In the limit of $\alpha$, $g \to 0$, but
nonzero $V$, the Hamiltonian of Eq.~\ref{ham} reduces to the effective Hamiltonian,
\begin{equation}
H_{\rm eff} = V\sum_i n_in_{i+1} - t\sum_i(a^\dagger_ia_{i+1}+a^\dagger_{i+1}a_i)
\label{spinless}
\end{equation}
where as in Eq.~\ref{klein} the $a^\dagger_i$ create spinless Fermions
and $n_i=0,1$ only. For large $V$, the ground state now is the
4k$_{\rm F}$ CO $\cdots$1010$\cdots$ Jordan-Wigner transformation converts the
effective Hamiltonian Eq.~\ref{spinless} to the XXZ spin Hamiltonian
with $J_z=V$ and $J=2|t|$.  The AFM solution to the spin Hamiltonian
corresponds to the 4k$_{\rm F}$ CO in the particle language, and from
exact Bethe ansatz solution of the spin Hamiltonian it is ascertained
that the CO ground state is obtained only for $V>V_{\rm c}=2|t|$. For
$V<V_{\rm c}$ the system is a Luttinger liquid (LL)
\cite{Haldane80a,Haldane81a} (see \cite{Voit95a} for a review of
Luttinger liquids).

\begin{figure}[tb]
  \begin{center}
    \raisebox{0.2in}{
      \resizebox{1.8in}{!}{\includegraphics{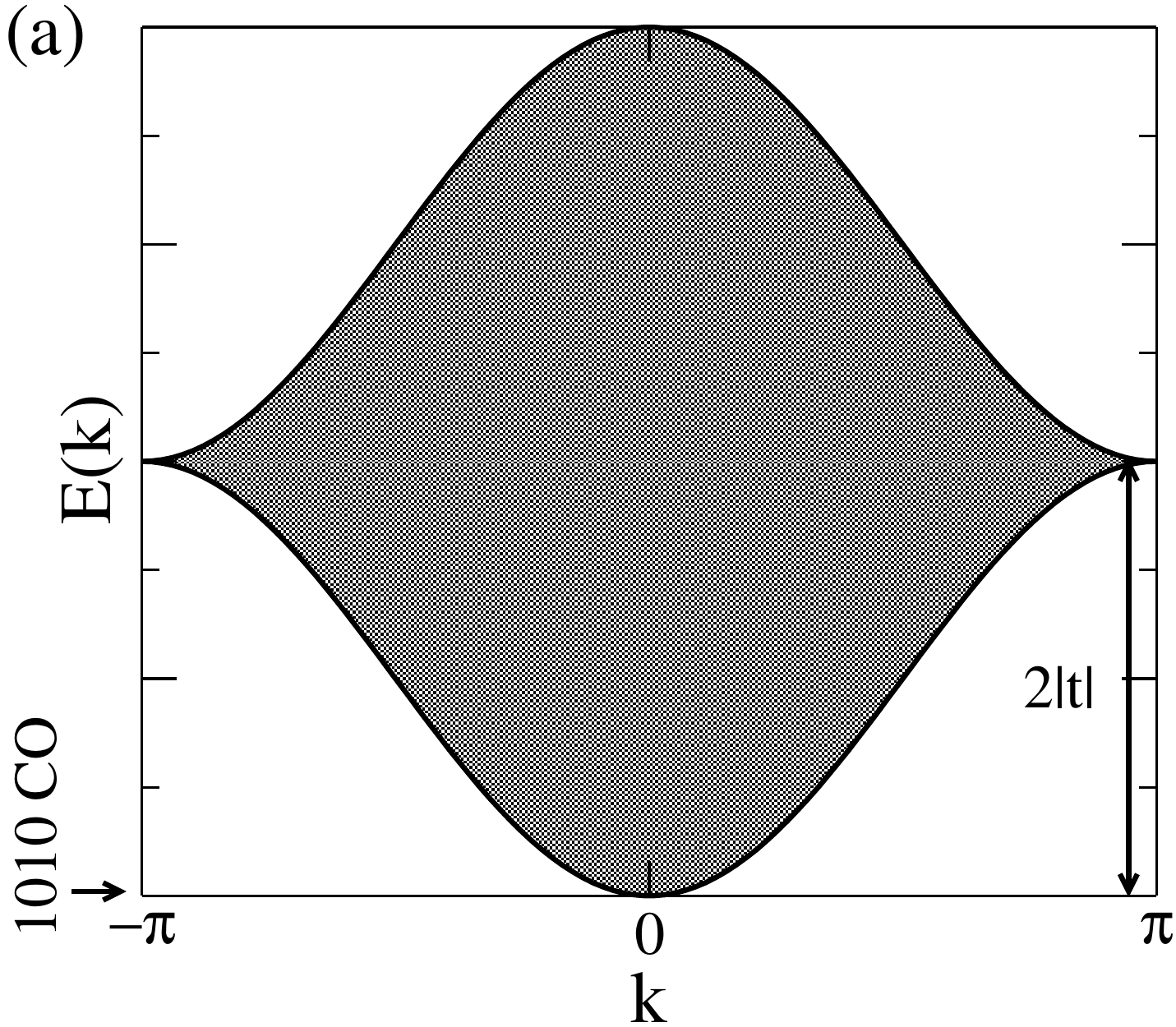}}
    }
    \hspace{0.2in}
    \resizebox{1.8in}{!}{\includegraphics{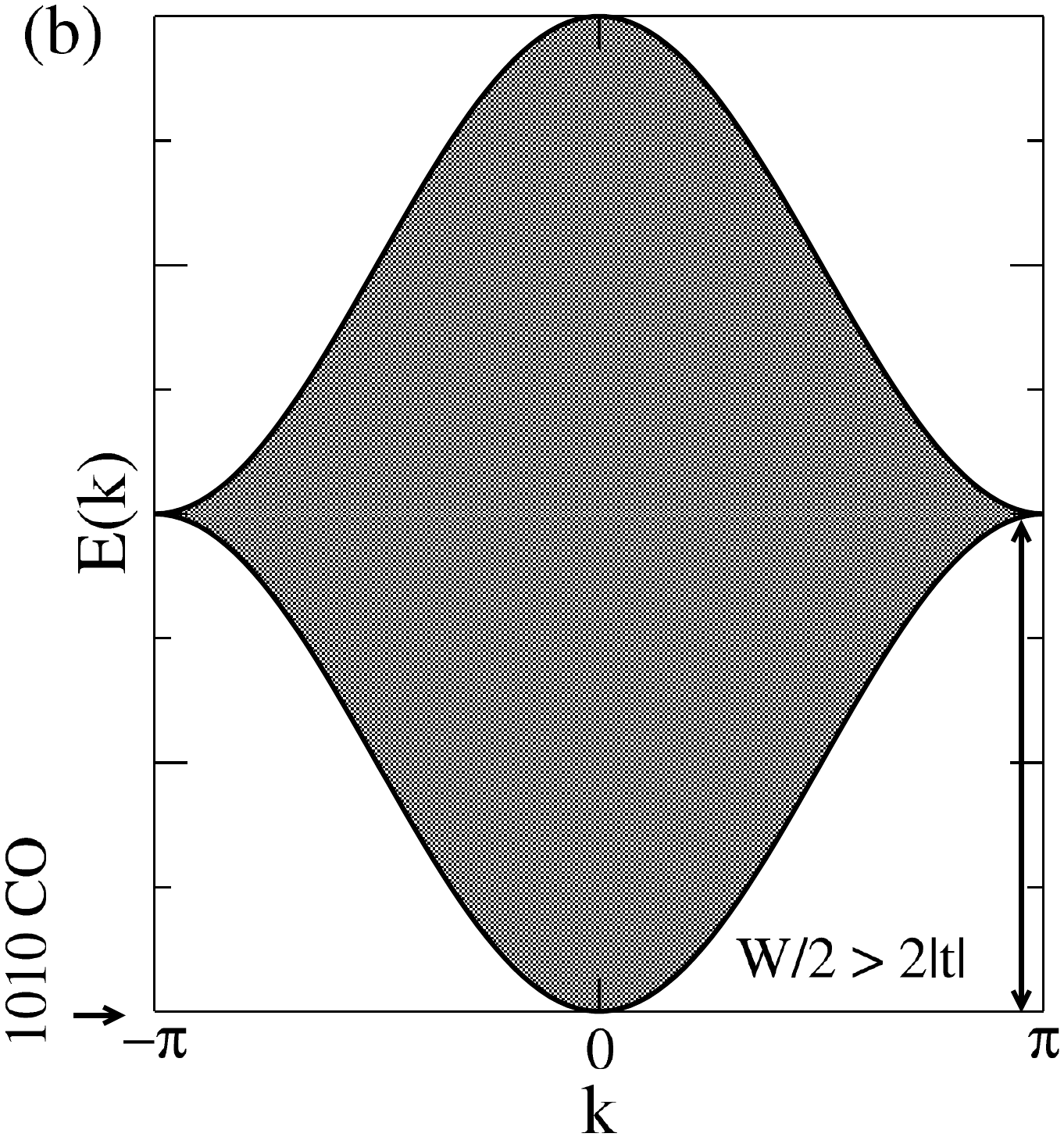}}

  \end{center}
  \caption{(a) Schematic of the energy spectrum of the spinless
    Fermion Hamiltonian at $V = 2|t|$ (Eq.\ref{spinless}).  The band
    center is at $E=V$, the energy required to create one dimer.
    Above the $\cdots$1010$\cdots$ CO, there is a continuum due to the
    band motion of the 1-1 and 0-0 dimers (see text). The width of
    this band is half that of a free electron band since each hop of
    the dimer takes it through two lattice sites.  The width of the
    energy spectrum is fixed by $|t|$ alone and hence an energy gap
    develops for $V > W/2$.  (b) Schematic of the energy spectrum for
    finite $U$. Now the CO configuration is the same as before but the
    width of the total energy dispersion is larger due to the much
    larger number of additional configurations with double occupancies
    excluded in the spinless Hamiltonian.}
  \label{bands}
\end{figure}

The exact result in the particle language is an entropy effect that
can also be understood physically, from the perspective of the extreme
configurations $\cdots1010\cdots$ and $\cdots0101\cdots$ and the paths
connecting them. The first few steps of one such path are given
below. The very first step creates a pair of nearest neighbor
occupancies ``11'' (and a pair of vacancies ``00'') and costs an
initial energy $V$, but further electronic motion is
barrierless:
  \small
\begin{equation*}
  \cdots1010101010\cdots \to \cdots1010011010\cdots \to
  \cdots1010010110\cdots \to \cdots1001010110\cdots 
\end{equation*}
\normalsize
Band-like motion of the occupied pair ``11'' and the
unoccupied pair ``00'' occurs after the first step, where each pair
moves over two lattice sites, giving an approximate energy dispersion
$E_k = V -2|t|\cos(2k_1a) - 2|t|\cos(2k_2a)$, $k_1+k_2=k$.  The schematic
energies of the CO state and of the excited band shown in Fig.~\ref{bands}(a),
where a charge gap appears only for $V \geq 2|t|$.

The point of constructing the above argument for a case where the
exact solution is available is to extend it to the {\it finite} $U$
case.  In addition to the paths between the two extreme CO
configurations that are of the type indicated above, involving only
single occupancies and vacancies, for finite $U$ there will occur
additional paths in which the intermediate configurations also possess
double occupancies. The large number of such configurations increases
the the bandwidth due to the ``excited'' configurations rapidly, and
we expect $V_{\rm c}(U)$ to increase as $U$ decreases from infinity (see Fig.~\ref{bands}(b)).

\paragraph{Spin dependence of $V_{\rm c}(U)$ and implication} 
Within the 1D EHM, $E(S) < E(S+1)$, where $E(S)$ is the lowest energy in the  
spin subspace $S$.
Fig.~\ref{bands}(b) in the above then corresponds to the
case of total spin $S=0$. One can visualize solving the Hamiltonian
separately for each different spin subspace, and determining
numerically $V_{\rm c}(U)$ for each $S$. $V_{\rm c}$ is clearly spin
dependent, as Fig.~\ref{bands}(a) continues to be valid for the
ferromagnetic subspace $S=S_{max}$ even for finite $U$, since double
occupancies are not possible in this subspace. We therefore conclude
that $V_{\rm c}(U,S_{max}) < V_{\rm c}(U,S=0)$.  Our discussion of the
entropy effect for finite $U$ shows that the entropy effect increases
with the probability of double occupancy, which decreases
monotonically with increasing $S$.  The total number of configurations
in any spin subspace decreases with increasing $S$ which implies that
the width of the continuum in Fig.~\ref{bands}(b) decreases with
increasing $S$.  It is then possible to interpolate between the two
extreme cases of $S=0$ and $S=S_{max}$ and to conclude that $V_{\rm
  c}(U,S)>V_{\rm c}(U,S+1)$.  This spin-dependence implies a {\it
  temperature dependence}, since in the temperature regime T$_{\rm
  2k_{\rm F}}<$ T $<$ T$_{\rm 4k_{\rm F}}$ the excitations of the
EHM are predominantly spin excitations, and
hence the partition function and free energy are dominated more and
more by high spin states as the temperature increases from T$_{\rm
  2k_{\rm F}}$.  The consequences for the transitions in the real
materials are then as follows.

The charge-gapped state below T$_{\rm 4k_{\rm F}}$ in
$\rho=\frac{1}{2}$ systems with moderate to large $U,~V$ is understood
within the spinless Hamiltonian of Eq.~\ref{spinless}, in the presence
of nonzero $\alpha$ and $g$. The effective $\frac{1}{2}$-filled band,
depending upon the value of $V$ in the real system, will undergo
transition to the site-diagonal CO $\cdots1010\cdots$
(Fig.~\ref{1dcartoons}(c), for $V>2|t|$) or to the state with 4k$_{\rm
  F}$ BOW (Fig.~\ref{1dcartoons}(a) with $\rho=\frac{1}{2}$, for
$V<2|t|$), with alternating short and long bonds (the 4k$_{\rm F}$ BOW
is sometimes referred to as the Dimer Mott insulator
\cite{Seo02a,Kuwabara03a}). The evolution from the charge- or
bond-alternated phase to the SP phase is, however, nontrivial.  The
straightforward scenario is that the nature of the SP phase below
T$_{\rm 2k_{\rm F}}$ is predetermined already at T$_{\rm 4k_{\rm F}}$
by the nature of the high temperature phase. Within this scenario the
4k$_{\rm F}$ CO phase upon cooling undergoes SP transition to a state
in which the overall charge distribution remains nearly the same, and
the bonds between the charge-rich sites alternate (hereafter 4k$_{\rm
  F}$ CO-SP state, Fig.~\ref{1dcartoons}(d)) \cite{Kuwabara03a}.
Similarly, the dimer units within the 4k$_{\rm F}$ BOW phase move
closer or further apart from one another below the SP transition. In
this case the larger spin exchange between NN sites on dimers that are
closer to one another leads to larger charge densities, as is shown in
Fig.~\ref{1dcartoons}(g). In addition to these straightforward
evolutions the spin dependence of $V_{\rm c}(U)$ requires that there
exists a third possibility, wherein the 4k$_{\rm F}$ CO upon cooling
evolves into the 2k$_{\rm F}$ BCDW-I instead of the 4k$_{\rm F}$ CO-SP
state.  This will occur in experimental systems in which the material
parameter $V$ is greater than 2$|t|$ but less than $V_{\rm
  c}(U,S=0)$. We will discuss all possible cases in Section
\ref{1d-expt}.

\paragraph{BCDW-II}

While the CO in BCDW-I and BCDW-II is identical, their bond patterns
are different \cite{Clay17a}.  Importantly, the location of the
singlet bond is different, being intra-dimer in BCDW-II
(Fig.~\ref{1dcartoons}(f)) and inter-dimer in BCDW-I
(Fig.~\ref{1dcartoons}(g)) \cite{Clay12a,Clay17a}.  As discussed
above, in systems where the ground state is BCDW-I, a high temperature
transition to either 4k$_{\rm F}$ BOW or 4k$_{\rm F}$ CO is expected
above the SP transition. When the high temperature state is the
4k$_{\rm F}$ BOW, the low temperature 2k$_{\rm F}$ BCDW-I can be
intuitively understood as a ``second dimerization'' of a dimer
lattice, with inter-dimer bonds alternating between W and W$^\prime$
\cite{Mazumdar00a}.  However, the bond distortion of BCDW-II {\it
  cannot} be visualized in this manner. Instead, if an effective
$\frac{1}{2}$-filled band dimer model is constructed for BCDW-II, the
dimer occupations follow the $\rho=1$ CDW pattern
$\cdots$2--0--2--0$\cdots$. Within such an effective model there would
be a single transition from uniform charge to this CO state.

Other key differences between BCDW-I and BCDW-II can be seen in
results of numerical calculations that include both e-e and e-p
interactions (see Section \ref{1dnumerics-peh}). In BCDW-I the
amplitude of the CO is always small.  However, in BCDW-II, depending
on the e-e and e-p parameters, the CO amplitude can become very
large. With the singlet forming at the strongest (intra-dimer) bond in
the chain, larger spin gaps are expected in BCDW-II.  When the CO and
BOW amplitude becomes very large, the separation of the energy scales
of 4k$_{\rm F}$ and 2k$_{\rm F}$ distortions that give low and high
temperature transitions can then possibly break down in the BCDW-II
case \cite{Clay17a}.  As we discuss in Section
\ref{1dnontrad-section}, for several 1D $\rho=\frac{1}{2}$ CTS such as
(EDO-TTF)$_2$X, a {\it single} coupled MI and spin-gap (SG) transition
is found at remarkably high temperatures. The measured bond distortion
pattern in several of these materials follows that of BCDW-II, and
large CO amplitudes are also found.  

\subsubsection{Weak two dimensional interactions}
\label{bcsdw-section1}

\begin{figure}[tb]
  \centerline{\resizebox{3.0in}{!}{\includegraphics{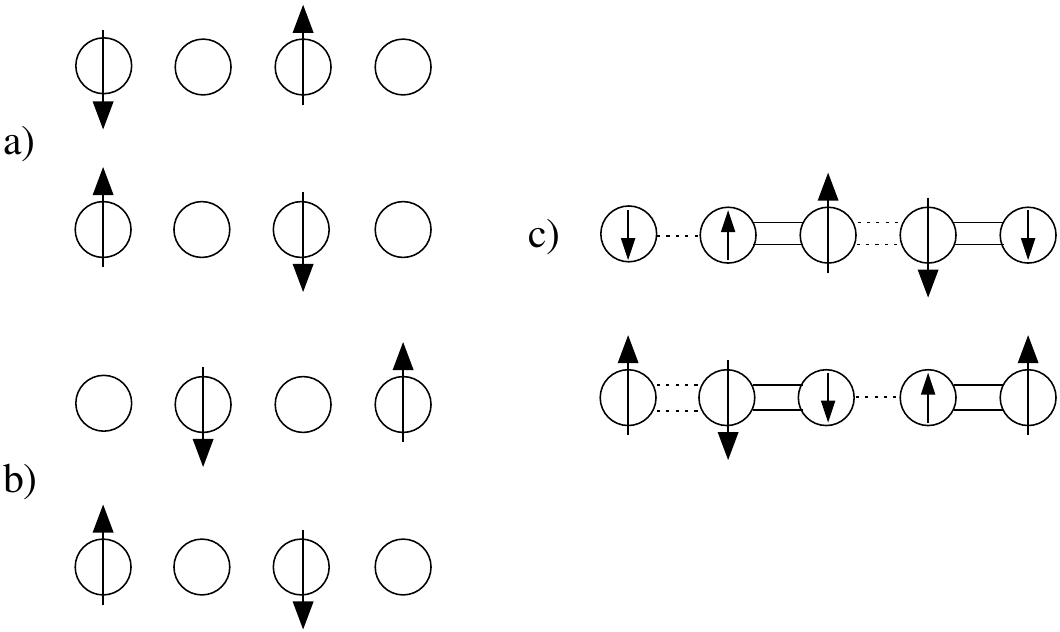}}}
  \caption{(a)-(b) AFM configurations possible between weakly coupled 1D
    chains with $\cdots1010\cdots$ CO (c) AFM coexisting with $\cdots1100\cdots$
    BCDW.}
  \label{afm}
\end{figure}

AFM becomes relevant for nonzero interchain Coulomb interactions and
carrier hopping. The crystal structure of (TMTCF)$_2$X is such that
frustration also plays a role. The AFM configurations can be derived
from both the charge occupancies $\cdots1010\cdots$ and
$\cdots1100\cdots$. Even with the single chain site charge occupancies
$\cdots1010\cdots$ there are two distinct possibilities, as shown in
Fig.~\ref{afm}(a) and (b), respectively. As we will see in the next
subsection, charge occupancy $\cdots1010\cdots$ is most relevant at
the left end of the phase diagram in Fig.~\ref{phasediagram}. The
correct 2D site occupancy for the AFM configurations in
Figs.~\ref{afm}(a) and (b) is therefore the one which evolves
naturally to the spin singlet SP state as interchain interactions are
increased. Fig.~\ref{afm}(c) shows the AFM configuration with site
charges $\cdots1100\cdots$.  In this AFM configuration the total
charge within each dimer (sites connect by double lines in
Fig.~\ref{afm}(c)) is one, which is however distributed unequally
between the two sites comprising the dimer.  Note that in this case
AFM is expected to coexist with lattice distortion, as indicated in
the Figure.  We have termed the combined BCDW and AFM state of
Fig.~\ref{afm}(c) a Bond-Charge Spin Density Wave (BCSDW)
\cite{Mazumdar00a}.
\subsection{Numerical Results}
\label{1dnumerics}

In this section we review numerical results in one dimension for $\rho=\frac{1}{2}$ and
for weakly coupled $\rho=\frac{1}{2}$ chains.

\subsubsection{One dimensional extended Hubbard model}

In the absence of e-p interactions, the zero temperature phase diagram
of Eq.~\ref{EHM} in one dimension in the range of parameters
appropriate for quasi-1D CTS is by now well characterized. The onsite
Coulomb interaction $U$/$|t|$ is expected to be in the range of 6--8
\cite{Clay07a}. Based on comparison with quasi-1D $\rho=1$ CTS, where
a CDW driven by $V$ with pattern $\cdots$2020$\cdots$
(Fig.~\ref{1dcartoons}(b)) does {\it not} occur, it is expected that
$V<U/2$ \cite{Clay07a}.

\begin{figure}[tb]
  \centerline{\resizebox{2.5in}{!}{\includegraphics{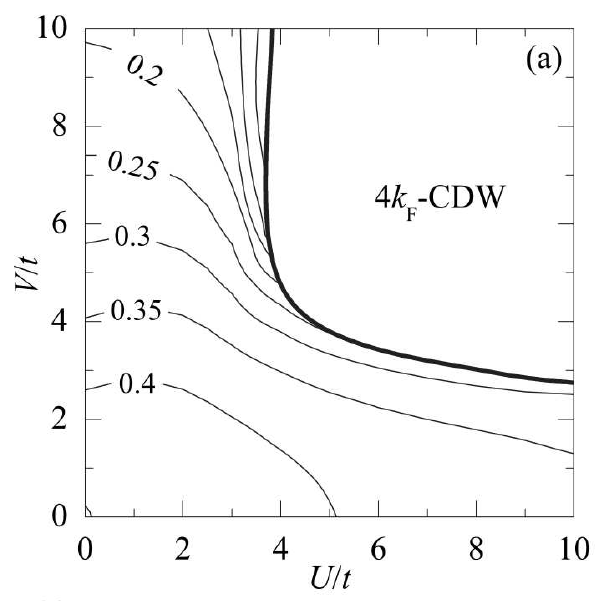}}}
  \caption{(color online) Zero temperature phase diagram of the 1D
    extended Hubbard model Eq.~\ref{EHM} in one dimension at
    $\rho=\frac{1}{2}$, as determined from DMRG calculations of up to
    60 sites. Contours show the values of the LL exponent K$_{\rho}$.
    Reprinted with permission from Ref.~\cite{Shirakawa09a},
    $\copyright$ 2009 The American Physical Society.}
  \label{tuv-phase}
\end{figure}

Except in the $\cdots$1010$\cdots$ WC phase when $V>V_{\rm c}$, the
low energy properties of Eq.~\ref{EHM} are those of a LL
\cite{Haldane80a,Haldane81a,Voit95a}. The phase diagram may then be
mapped out by calculating the LL correlation exponent K$_\rho$, which
determines the asymptotic form of the charge and spin correlations
\cite{Voit95a}. For K$_\rho<\frac{1}{3}$, 4k$_{\rm F}$ charge
correlations are more divergent than 2k$_{\rm F}$ charge correlations,
and for K$_\rho=\frac{1}{4}$ the insulating $\cdots$1010$\cdots$ phase
is reached \cite{Voit95a}.  This phase boundary has been mapped out
using exact diagonalization
\cite{Mila93a,Penc94a,Nakamura00b,Sano04a}, QMC \cite{Clay03a}, and
DMRG \cite{Ejima05a,Shirakawa09a} methods. Fig.~\ref{tuv-phase} shows
the phase diagram as obtained from finite-size scaled DMRG
calculations \cite{Shirakawa09a}.  Two important limiting cases for
the boundary to the $\cdots$1010$\cdots$ phase are the large $U$
limit, where as described above $V_{\rm c}$=2t, and in the limit of
large $V$, where $U_{\rm c}$=4t \cite{Mila93a}.

\subsubsection{One dimensional Peierls-extended Hubbard model}     
\label{1dnumerics-peh}

\begin{figure}[tb]
  \centerline{\resizebox{2.5in}{!}{\includegraphics{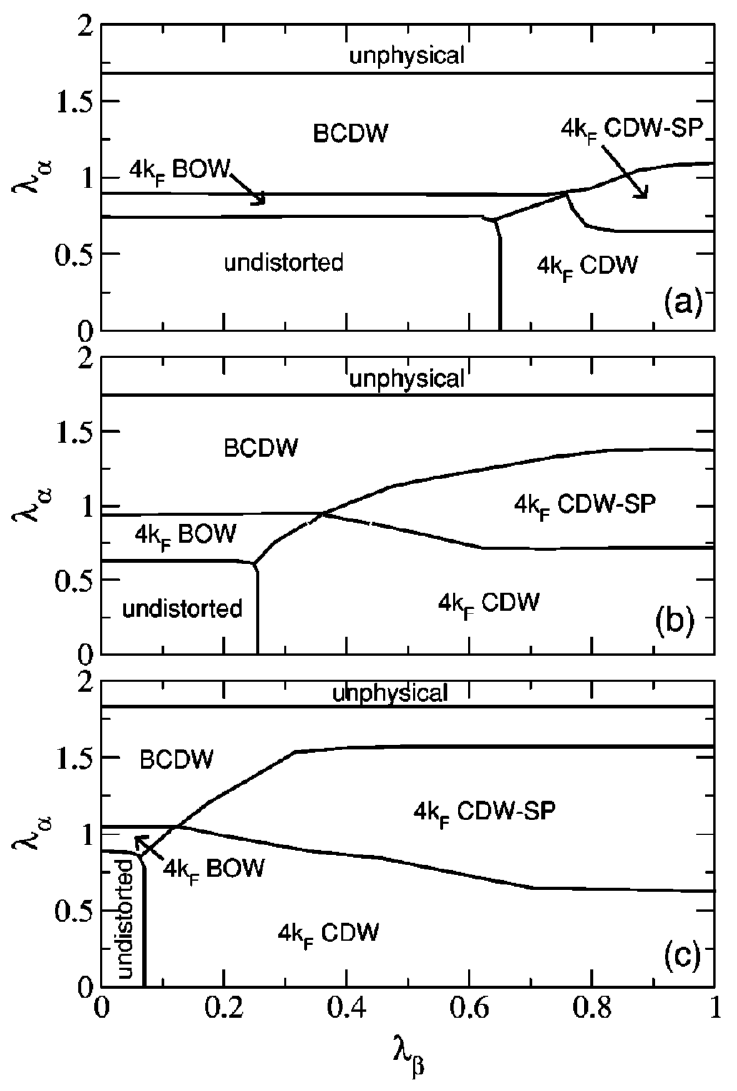}}}
  \caption{Phase diagram of Eq.~\ref{ham} in one dimension at $\rho=\frac{1}{2}$ from 16-site exact
    diagonalization calculations \cite{Clay03a}.
    Here $\lambda_\alpha=\alpha^2/(K_\alpha t)$ and $\lambda_\beta=g^2/(K_g t)$ are the
    dimensionless inter- and intra-site e-p couplings, respectively.
    Parameters are
    $U/t$=8 and (a) $V/t$=2; (b) $V/t$=3; and (c) $V/t$=4.}
  \label{16site-phase-diagram}
\end{figure}

\paragraph{4k$_{\rm F}$ phases}
The competition of the $\cdots$1010$\cdots$
CO with and without an accompanying bond distortion
(Fig.~\ref{1dcartoons}(c) and (d)) with BCDW-I and BCDW-II has also
been studied
\cite{Kuwabara03a,Clay03a}. Fig.~\ref{16site-phase-diagram} shows the
ground-state phase diagram of Eq.~\ref{ham} calculated with
self-consistent exact diagonalization for 16 sites with
$U/t=8$ \cite{Clay03a}. As $V/t$ increases, the amount of phase space
with the 4k$_{\rm F}$ CDW and 4k$_{\rm F}$ CDW-SP phases compared to
the BCDW increases (see Fig.~\ref{1dcartoons}).  For smaller values of
$V/t$ a finite intrasite e-p coupling is required for the 4k$_{\rm
  F}$ CDW-SP phase (see Fig.~\ref{16site-phase-diagram}(a) and (b)),
suggesting that the 4k$_{\rm F}$ CDW-SP may not occur unconditionally
(i.e. in the limit of 0$^+$ e-p interactions). Resolving this
issue will require finite-size scaled numerical calculations, which
to our knowledge have not been performed in this region of parameter
space.

\paragraph{Temperature-dependent WC-to-BCDW-I transition}
\label{1d-tdep}

In a 1D electron system no phase transitions are expected except at
zero temperature. However, once 2D and 3D interactions and coupling to
the anions are included, finite-temperature phase transitions must
result, as demonstrated by Fig.~\ref{phasediagram}.  To include these
additional interactions is simply not feasible without making
uncontrolled assumptions as in mean-field approximations.
However,  information as to the relative order in temperature of
different transitions can still be determined by examining the 
excited states of the purely 1D system, as well as the temperature
dependence of the charge and bond susceptibilities \cite{Clay07a}.

\begin{figure}[tb]
  \begin{center}
    \resizebox{2.0in}{!}{\includegraphics{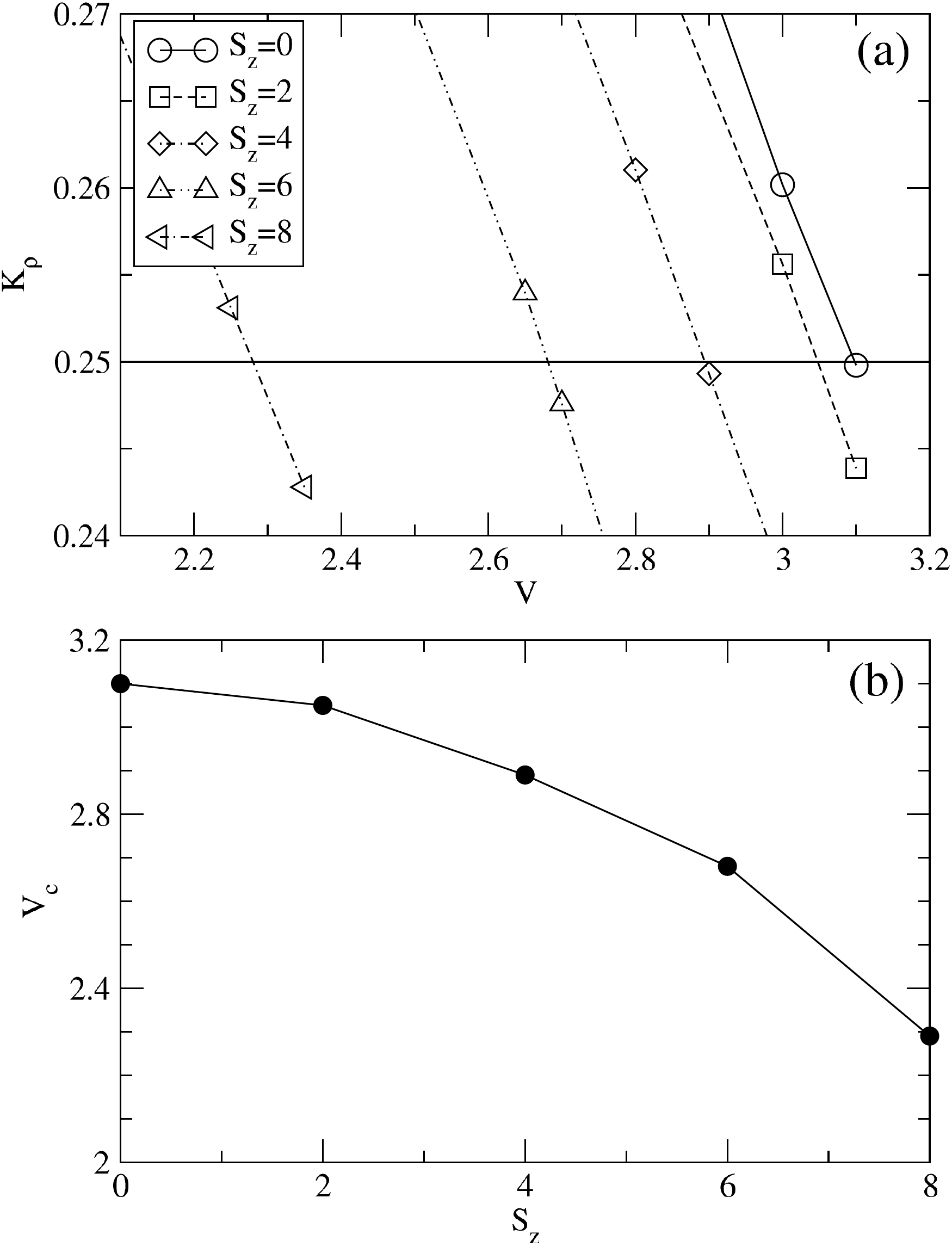}}
    \raisebox{0.3in}{
      \resizebox{2.75in}{!}{\includegraphics{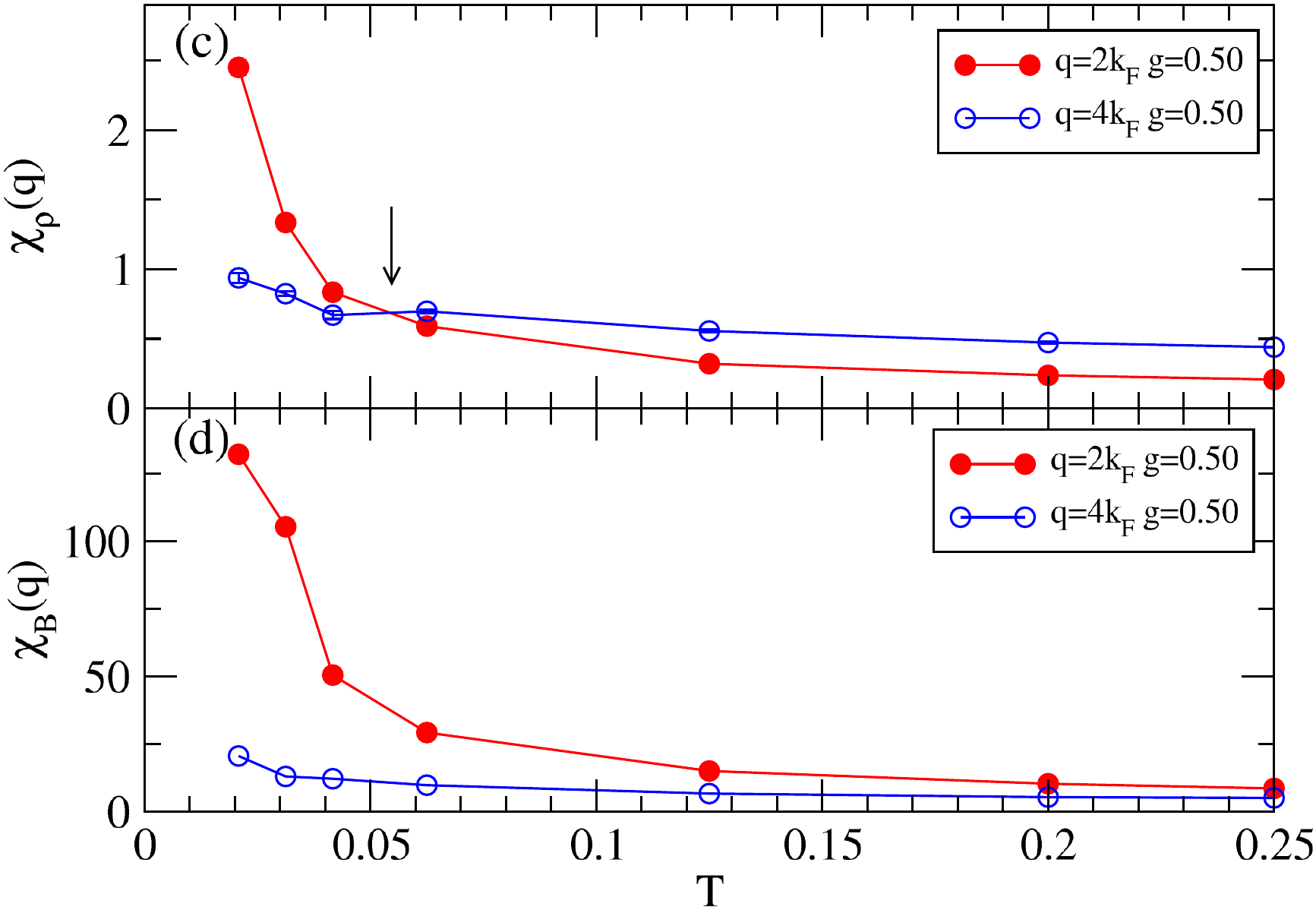}}
      }
  \end{center}
  \caption{(a) The Luttinger liquid exponent K$_\rho$ calculated
    for the EHM for $U=8t$ as a function of $V$ and S$_{\rm z}$. Results
    are for a 32-site periodic ring \cite{Clay07a}.
    (b) $V_{\rm c}$ as determined from (a) where K$_\rho=\frac{1}{4}$
    (c)-(d) Finite-temperature QMC calculation of the charge
    (c) and bond (d) susceptibilities for a 64-site periodic ring as a function
    of temperature. 
Phonons are treated dynamically, with  parameters $U=8t$, $V=2.75t$, $\alpha=0.27$,
    $\omega_{\rm S}=0.1$, $g=0.5$, and $\omega_{\rm H}=0.5$ (see text). The arrow
indicates where $\chi_\rho(2k_{\rm F})=\chi_\rho(4 k_{\rm F})$  \cite{Clay07a}.}
  \label{spinvc}
\end{figure}

The spin dependence of V$_{\rm c}$ (see Section \ref{1dlimit}) has been
verified by explicit calculations on the EHM \cite{Clay07a}.
Fig.~\ref{spinvc}(a)-(b) show  calculations of $V_{\rm c}$ for
a 32 site periodic ring for different $S_z$ values. The QMC calculation
method used here conserves $S_z$ but not total $S$. However,
the Lieb-Mattis theorem \cite{Lieb62a}, $E(S)<E(S+1)$,
where $E(S)$ is the energy of the lowest state in the spin subspace $S$,
applies to the 1D EHM. Because of the absence of a magnetic field,
all $S_z$ states for a given $S$ are degenerate, and results for
the lowest state within each $S_z$ must pertain to $S=S_z$ \cite{Clay07a}.
The apparently larger $V_{\rm c}=2.3t > 2t$
in Fig.\ref{spinvc}(b) for the fully polarized S$_{\rm z}$=8 is the result of
finite-size effects, which does not however change the ordering of
critical $V$'s with $S_z$.

In Reference \cite{Clay07a} temperature
dependent QMC calculations were also performed for Eq.~\ref{ham}
with dynamic e-p interactions within the following Hamiltonian:
:
\begin{subequations}
\begin{eqnarray}
H &=& H_{\rm SSH} + H_{\rm Hol} + H_{\rm ee} \label{ham-dynamic} \\
H_{\rm SSH} &=& t\sum_i[1+\alpha(a_i^{\dagger}+a_i)]B_{i,i+1}
+ \hbar \omega_{\rm S} \sum_i a_i^{\dagger}a_i  \\
H_{\rm Hol} &=& g \sum_i (b_i^{\dagger} + b_i)n_i + \hbar \omega_{\rm H} \sum_i b_i^{\dagger}b_i \\
H_{\rm ee} &=& U\sum_i n_{i,\uparrow}n_{i,\downarrow} + V\sum_i n_in_{i+1}. 
\end{eqnarray}
\end{subequations}
In the above, $a^\dagger_i$ and $b^\dagger_i$ create dispersionless
SSH and Holstein phonons on the $i$th bond and site, respectively,
with corresponding phonon frequencies $\omega_{\rm S}$ and
$\omega_{\rm H}$. The electron hopping and interaction terms are
otherwise identical to Eq.~\ref{ham}.  Figs.~\ref{spinvc}(c) and (d)
show the charge and bond susceptibilities at 2k$_{\rm F}$ and
4k$_{\rm F}$ as a function of temperature for $V=2.75t$. This value of
$V$ is between $V(S_z=S_z^{\rm max})=2t$ and $V_{\rm c}(U) \approx
3.0t$. At high temperature, $\chi_\rho(4k_{\rm F})>\chi_\rho(2k_{\rm
  F})$, indicating that $\cdots$1010$\cdots$ CO is favored. However,
at low temperature $\chi_\rho(2k_{\rm F})>\chi_\rho(4k_{\rm F})$, with
$\chi_\rho(2k_{\rm F})$ and $\chi_{\rm B}(2k_{\rm F})$ becoming
divergent as $T\rightarrow 0$, indicating that the
$\cdots$1100$\cdots$  BCDW is the ground state. This has important
experimental implications, see Section \ref{tmttf-sp}.

\paragraph{BCDW-I versus BCDW-II}
Within the semi-classical limit of Eq.~\ref{ham-dynamic},
the displacement of
the $i$th site from equilibrium, $\Delta_i$, can be written as a superposition of
2k$_{\rm F}$ (period 4) and 4k$_{\rm F}$ (period 2) terms \cite{Ung94a,Clay17a},
\begin{equation}
\Delta_j = \Delta_0[a_2\cos(2k_{\rm F}j-\phi_2)+a_4\cos(4k_{\rm F}j-\phi_4)].
\label{eq:r2r4}
\end{equation}
\begin{figure}[tb]
  \centerline{\raisebox{0.3in}{\resizebox{2.5in}{!}{\includegraphics{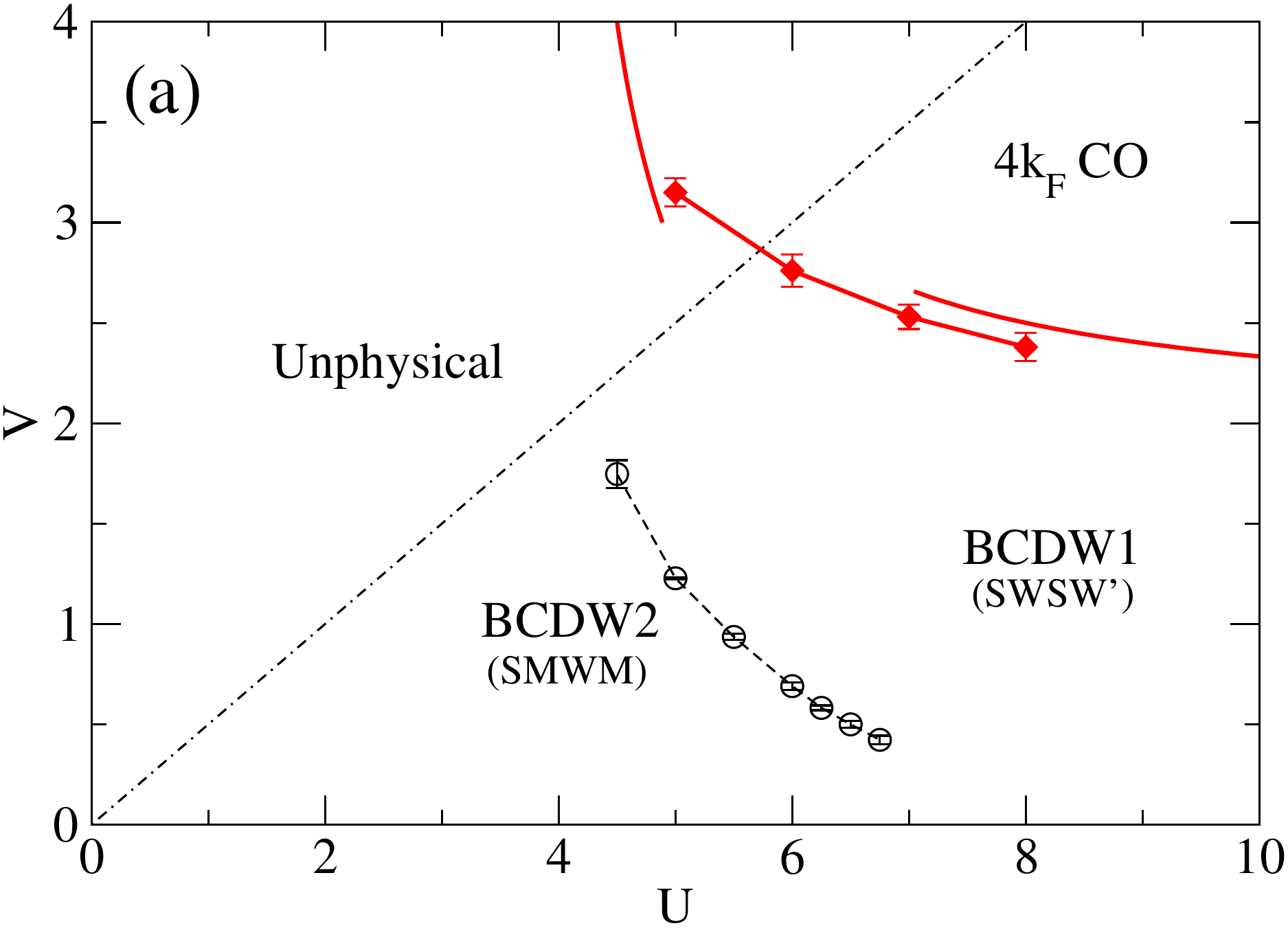}}}%
    \hspace{0.1in}%
    \resizebox{3.0in}{!}{\includegraphics{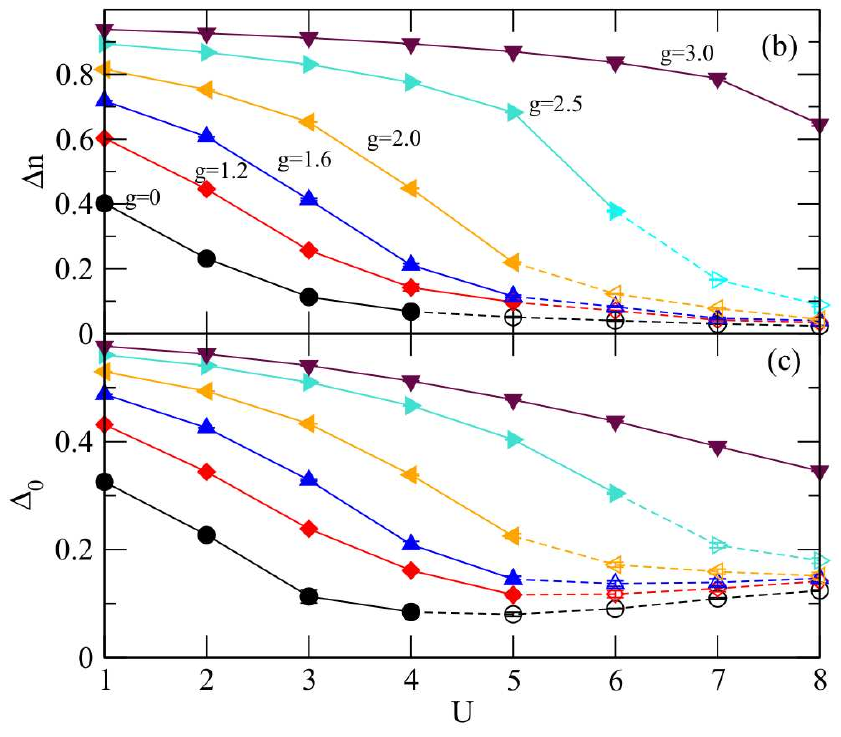}}}
  \caption{(color online) Left: Phase diagram of Eq.~\ref{ham},
    for $\rho=\frac{1}{2}$ in the limit of weak ($\alpha=g=0^{+}$) e-p
    interactions \cite{Clay17a}.  Solid lines are strong-coupling
    expansion results; solid symbols are QMC points where
    K$_\rho=\frac{1}{3}$ (see text).  Open points mark the
    crossover between BCDW-I and BCDW-II bond patterns.  Right:
    Charge order amplitude $\Delta n$ and distortion amplitude
    $\Delta_0$ (see text) for the Hamiltonian of Eq.~\ref{ham} at
    $\rho=\frac{1}{2}$ versus $U$, for $\alpha=1.2$, $K_\alpha=K_g=2.0$,
    and $V=U/4$ \cite{Clay17a}. Solid (dashed) lines denote
    parameters giving the BCDW-II (BCDW-I) bond distortion.}
  \label{bcdw1-bcdw2}
\end{figure}
In Eq.~\ref{eq:r2r4}, $\Delta_0$ is the overall amplitude and the
normalization condition $a_2+a_4=1$ is assumed. The phase angles are
$(\phi_2=\frac{\pi}{2},\phi_4=0)$ for BCDW-I and BCDW-II
\cite{Ung94a,Clay17a}.  Provided ($U$, $V$) are not close to the
$\cdots$1010$\cdots$ region of the electronic model, the ground state
of Eq.~\ref{ham} is either BCDW-II Fig.~\ref{1dcartoons}(f) or BCDW-I
Fig.~\ref{1dcartoons}(g) \cite{Ung94a,Clay17a}. In the limit of weak
e-p interactions, $a_2/a_4$ in Eq.~\ref{eq:r2r4} is controlled by $U$
and $V$, with stronger e-e interactions giving BCDW-I (see Section
\ref{1dlimit}).  In the 1D model a crossover occurs between BCDW-I and
BCDW-II, shown in Fig.~\ref{bcdw1-bcdw2} \cite{Clay17a}.

The principal difference between BCDW-I and BCDW-II is the location of
the strongest bond, which is intra-dimer in BCDW-II and inter-dimer in
BCDW-I \cite{Ung94a,Clay12a,Clay17a}. The amplitude of the CO
coexisting with BCDW-I is of small amplitude for all choices of e-e
and e-p interactions, with $\Delta$n$\sim$0.1 (we define
$\Delta$n=n$_{\rm large}$-n$_{\rm small}$ as the difference between
the large and small charge densities) \cite{Clay17a}. On the other
hand, the CO coexisting with BCDW-II can be of very large amplitude
depending on the value of the intra-site e-p coupling $g$, with
$\Delta$n$\sim$0.9 possible for large $g$.
Fig.~\ref{bcdw1-bcdw2}(b)-(c) shows finite-size scaled results for
$\Delta$n and the lattice distortion magnitude $\Delta_0$ as a
function of $U$ and the intra-site e-p coupling \cite{Clay17a}.
Strong intra-site e-p interactions in particular can greatly enlarge
the region of parameters giving BCDW-II \cite{Clay17a}.

\subsubsection{Weakly two dimensional results}
\label{bcsdw-section2}

\begin{figure}[tb]
  \begin{center}
    \raisebox{0.5in}{
      \begin{overpic}[width=2.0in]{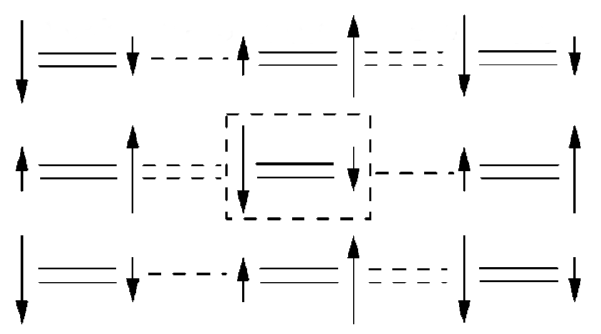}
        \put (1,60) {\small(a)}
      \end{overpic}
    }
    \hspace{0.1in}
    \begin{overpic}[width=2.5in]{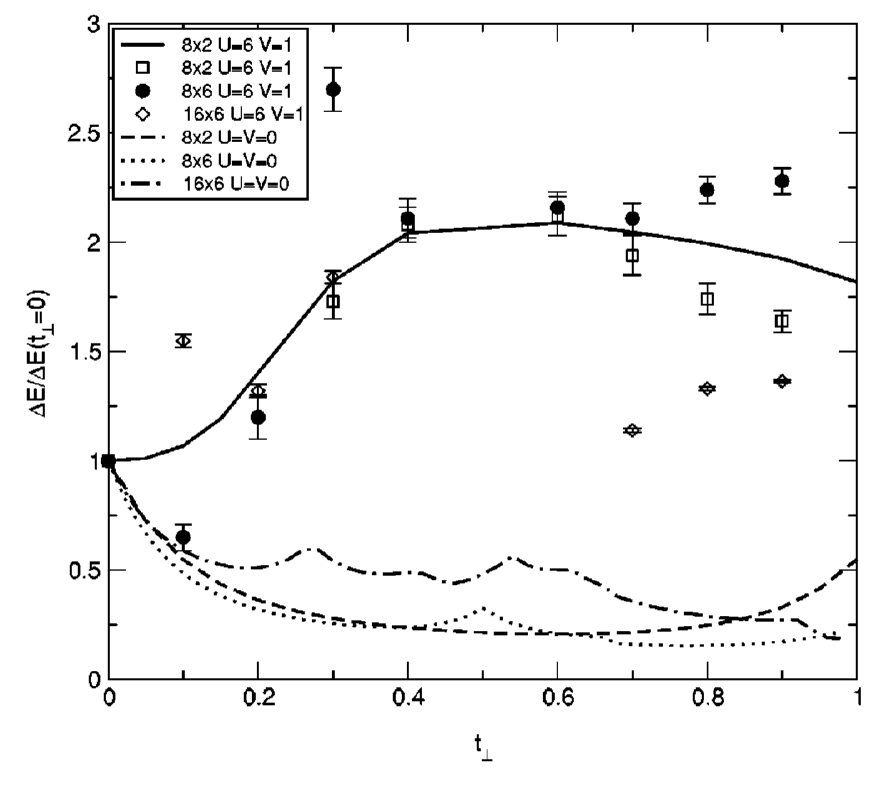}
      \put (-5,80) {\small(b)}
    \end{overpic}
  \end{center}
  \caption{(a) The Bond-Charge-Spin Density wave (BCSDW) state
    \cite{Mazumdar00a}.  (b) The per-site energy gained upon inter-dimer bond
    distortion, $\Delta E$, normalized to its value in 1D. The lattice
    is a rectangular dimer lattice with intra-dimer (inter-dimer)
    hopping of 1.2 (0.8) along the chains, and variable inter-chain
    hopping $t_\perp$. Solid lines are exact results, points are
    calculations using Constrained Path Monte Carlo (CPMC)
    \cite{Mazumdar00a}.  $\Delta E/\Delta E(t_\perp)>1$ over a wide
    range of $t_\perp$ indicates that the BCSDW state of (a) is
    stabilized by e-e interactions.}
  \label{numerics-bcsdw}
\end{figure}

The emergence of AFM in 2D depends critically on the amount of
frustration in the lattice. Numerical work shows that when
1D chains with the BCDW distortion
and e-e interactions are coupled by {\it unfrustrated} inter-chain
bonds $t_\perp$ (see Fig.~\ref{numerics-bcsdw}(a)), the result is
a 2D state combining charge distortion, bond distortion, and
AFM order, the BCSDW \cite{Mazumdar99a,Mazumdar00a}. 
The experimental relevance of this novel result is discussed in Section \ref{afm2}.

Reference \cite{Mazumdar00a} considered a series of rectangular
lattices of coupled dimers.  In Fig.~\ref{numerics-bcsdw}(b) $\Delta
E$ is defined as the energy gain upon formation of the BCDW-I state
from this dimer lattice, i.e. the difference in the energies of the
uniformly dimerized system and the system with the BCDW-I distortion
along the chains. As shown in Fig.~\ref{numerics-bcsdw}(b), $\Delta E$
per site is {\it greater} for the 2D system of coupled chains than for
a 1D chain with the same bond distortion, {\it provided} e-e
interactions are included. The ratio $\Delta E/\Delta E(t_\perp=0)$
increases rapidly above unity from $t_\perp=0$ through intermediate
$t_\perp$, suggesting that the BCSDW state is favored for weak
interchain interactions.  In the BCSDW state the spin pattern in terms
of dimers (see dotted box in Fig.~\ref{numerics-bcsdw}(b)) follows the
usual N\'eel pattern. The gain in energy over the 1D chain in the
presence of interactions comes from antiferromagnetic exchange between
the anti-parallel spins along the perpendicular bonds.

On a frustrated lattice, which is present in nearly all of the 2D CTS (see
Section \ref{2d-section}) and is also relevant for TMTSF under pressure, 
the BCSDW may no longer be the ground state. Instead, a
state composed of interchain local NN singlets can become favored.
 We return to this possibility in Section \ref{qtr-theory}.

\subsection{Summary}
(i) Within the quarter-filled one dimensional extended Hubbard model
(Eq.~\ref{EHM}) the $\cdots$1010$\cdots$ Wigner crystal (WC) of single
electrons occurs only for the nearest-neighbor Coulomb interaction $V
> V_c(U) > 2|t|$, and $V_c(U)$ increases with decreasing $U$. (ii) The
critical $V$, $V_c(U)$, is spin-dependent with $V_c(U,S) >
V_c(U,S+1)$. (iii) This implies there is a range of $V$, $2|t| < V <
V_c(U)$, where at high temperature $\cdots$1010$\cdots$ 4k$_{\rm F}$
charge order is accompanied at low temperature by a spin-Peierls phase
that is $\cdots$1100$\cdots$. (iv) For weak $t_\perp$ there exists a
bond-charge-spin density wave.

\section{Broken symmetries in quasi-one-dimensional CTS, Experiments}
\label{1d-expt}

In spite of the significant theoretical progress achieved
over the previous two decades completely consistent understanding of
the spatial broken symmetries in the quasi-1D 2:1 cationic CTS has
still not been reached. There are many reasons for this, the most
important of which, in decreasing order of importance according to us, are given
below.

(i) Not all cationic 2:1 1D materials are described by
Fig.~\ref{phasediagram}. The material (EDO-TTF)$_2$X (X=PF$_6$,
AsF$_6$), in which T$_{\rm MI}$ = T$_{\rm CO}$ = T$_{SG}$ \cite{Ota02a} is in
a class of its own. There exist other materials in which there occurs
the SP transition, but either the CO transition at a higher
temperature is absent completely, or the SP transition occurs at a
pressure where the CO has vanished. In all such cases T$_{\rm MI}$ is much
larger than T$_{\rm SP}$. The list here includes (TMTTF)$_2$I
\cite{Furukawa11a}, (BCPTTF)$_2$X (X=PF$_6$ and AsF$_6$)
\cite{Ducasse88a,Liu93a}, ($o$-DMTTF)$_2$X (X=Cl, Br,I)
\cite{Foury-Leylekian11a} and (EDT-TTF-CONMe$_2$)$_2$X (X = AsF$_6$,
Br) \cite{Auban-Senzier09a}. In spite of their differences from the
``standard'' 2:1 CTS which do belong to one or more regions of the
phase diagram in Fig.~\ref{phasediagram}, we believe these
``nonstandard'' materials should also be described within a single
``umbrella'' theoretical viewpoint. This is attempted below within the
context of the theoretical work described in the previous section,
along with explicit discussions of issues that remain to be
understood.

(ii) The charge difference $\Delta n$ between charge-rich and
charge-poor sites has been obtained mostly from measurements of
spin-lattice relaxation time T$_1$ and from infrared
spectroscopy. Optical spectroscopy in such cases gives $\Delta n$ that
is far smaller than that obtained from T$_1$ measurements
\cite{Hirose10a,Dressel12a}.  There is also disagreement between
different experimental groups over whether or not $\Delta n$ shows a
significant decrease upon entering the SP phase from the CO phase
\cite{Fujiyama06a,Nakamura07a,Iwase11a,Dressel12a,Hirose10a}.

(iv) The same disagreement exists also between theoretical groups,
which follows from a related but different disagreement over whether
or not the SP phase can have the $\cdots1010\cdots$ order
\cite{Clay07a,Yoshimi12a,Ward14a}.

In the following we discuss all of the above issues, first for the
more traditional Fabre salts, then for the Bechgaard salts, and finally for the
nonstandard materials mentioned in (i).
Fig.~\ref{our-phasediagram} shows our proposed interpretation of the
insulating phases in Fig.~\ref{phasediagram}
\cite{Clay07a,Ward14a}. We discuss our interpretation of these phases
in Sections \ref{fabre}-\ref{1dnontrad-section} below.

\subsection{Traditional Fabre salts}  
\label{fabre}

\begin{figure}[tb]
  \center{\resizebox{3.5in}{!}{\includegraphics{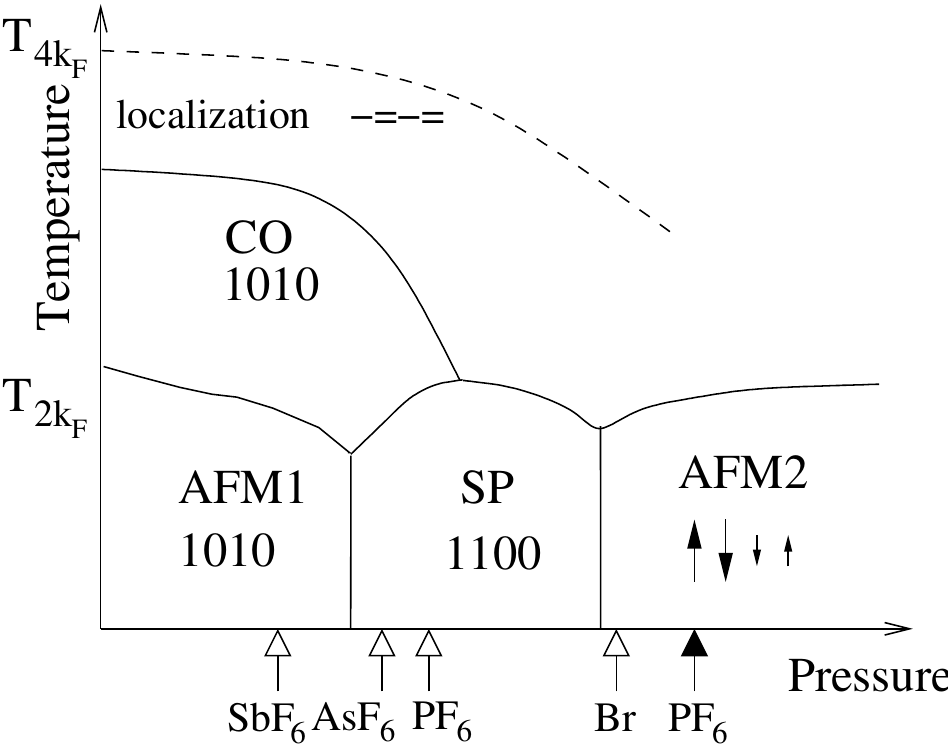}}}
  \caption{Schematic   (TMTSF)$_2$X and (TMTTF)$_2$X
    phase diagram with proposed charge occupancies.
    Open (filled) arrows indicate the ambient pressure locations of
    TMTTF (TMTSF) salts \cite{Clay07a}.}
\label{our-phasediagram}
\end{figure}

In Table~\ref{table-tmttf} we have listed the Fabre salts
which are most relevant in the context of the present review. In all
cases we have given the different temperatures at which the various
transitions are observed. T$_{\rho}$ is the temperature at which
charge-localization, as measured from transport studies, is
observed. Blanks imply that the particular state is absent in the
material under consideration (AO transitions are absent in the salts
with centrosymmetric anions; similarly, the CO transition does not
occur in X = I, see below). Before we discuss individual materials, we
point out some general trends and their implications.
\begin{table}
  \begin{center}
  \begin{tabular}{c|c|c|c|c|c}
X & T$_\rho$ & T$_{\rm AO}$ & T$_{\rm CO}$ & T$_{\rm SP}$ & T$_{\rm N}$ \\
\hline
TaF$_6$ & ? & --- & 175$^{\rm(a)}$ & --- & 9$^{\rm (a)}$ \\
SbF$_6$ & $\simeq$ 240$^{\rm (b),(c),(d)}$ & --- & 154--157$^{\rm (e),(f)}$ & --- &
8$^{\rm (c),(g)}$ \\
AsF$_6$ & 230--250$^{\rm (d),(f)}$ & --- & 102$^{\rm (h),(f)}$ & 13$^{\rm (e),(i)}$ & ---\\
PF$_6$ & 250$^{\rm(d),(f)}$ & --- & 67$^{\rm(d),(f),(h)}$ & 19$^{\rm(d),(e)}$ & --- \\
Br & $\simeq$ 200$^{\rm(f)}$ & --- & 28$^{\rm(f)}$ & --- & 13$^{\rm(j),(k)}$ \\
I & ? & ---& ---& 21$^{\rm(g)}$ & --- \\
SCN & 265$^{\rm(b)}$ & 160--170$^{\rm(e),(f)}$ & --- & --- & 7--9$^{\rm(m),(n)}$ \\
ReO$_4$ & 290$^{\rm(e),(f)}$ & 154$^{\rm(e),(f)}$ & 227.5$^{\rm(f)}$ & --- & --- \\
BF$_4$ & 220$^{\rm(e),(f)}$ & 40$^{\rm(e),(f)}$ & 83$^{\rm(f)}$ & --- & ---
  \end{tabular}
  \end{center}
  \caption{T$_{\rho}$, T$_{\rm CO}$, T$_{\rm AO}$, T$_{\rm SP}$, and T$_{\rm N}$
for different X in the series (TMTTF)$_2$X. All temperatures are in Kelvin.
(a) Reference \cite{Iwase09a}, (b) Reference \cite{Javadi88a},
(c) Reference \cite{Itoi08a}, (d) Reference \cite{Dressel12a}, (e)
Reference  \cite{Coulon85a}, (f) Reference \cite{Monceau12a},
(g) Reference \cite{Yu04a}, (h) Reference \cite{Chow00a},
(i) Reference \cite{Zamborszky02a}, (j) Reference \cite{Nakamura95a},
(k) Reference \cite{Pouget97a}, (l) Reference \cite{Furukawa11a},
(m) Reference \cite{Coulon82a}, and (n) Reference \cite{Peo84a}.}
  \label{table-tmttf}
\end{table}
First, ``CO'' in Fig.~\ref{our-phasediagram} and in
Table~\ref{table-tmttf} refers to the $\cdots1010\cdots$ 4k$_{\rm F}$
CDW. This has been confirmed from measurements of dielectric
permittivity, which show giant responses at T$_{\rm CO}$, indicating
ferroelectricity and loss of inversion symmetry
\cite{Nad06a,Monceau12a}. We shall henceforth refer to the CO phase in
Fig.~\ref{our-phasediagram} and in Table~\ref{table-tmttf} as the
ferroelectric CO (FCO), to distinguish it from the $\cdots0110\cdots$
BCDW. The latter is relevant only in the case of SP and the AFM-II
states (see below).

Second, given the charge carrier density of 0.5 per TMTTF, and T$_{\rm
  CO}$ $<$ T$_{\rho}$, there can be only one interpretation of
T$_{\rho}$, viz., it corresponds to the transition to the dimerized
4k$_{\rm F}$ BOW state which has lost its sharpness because of the
influence of the anions. Structural data from experiments such as
diffuse X-ray scattering are obscured by the intrinsic dimerization
along the TMTTF stack axis that may or may not be driven by
cation-anion interaction. The observed dimerizations even decrease
below T$_{\rho}$ in some cases, but this need not mean that the
difference in the {\it bond orders} (B$_{ij}$ in Eq.~\ref{ham}) also
decreases \cite{Pouget97a}.  In cases where metallic transport
behavior is seen at high T in spite of crystal structure-driven
dimerized structure, the bond orders of consecutive bonds must
necessarily be almost equal for the charge-gap to be absent. T$_{\rm
  CO}<T_{\rho}$ then demonstrates that $V$ in the real materials in
less than $V_{\rm c}(U,S=0)$, and that the FCO transition does not
result from a natural instability of the 1D chain, but from other
influences.

Third, among the salts with anions of octahedral symmetry (TaF$_6$,
SbF$_6$, AsF$_6$ and PF$_6$), T$_{\rm CO}$ decreases with decreasing
anion size. This can be a consequence of reduced chemical pressure
with large anions, which tend to give larger interstack separations. It
can also indicate larger electrostatic interaction between the TMTTF
molecules and larger anions, implying that the electrostatic
interaction is the driving force behind the FCO transition.  The
strong influence of the anions on the FCO is seen most clearly from
the nearly linear decrease of T$_{\rm CO}$ with increasing distance
between the ligands of the anions (O or F) and the sulfur atom in
TMTTF \cite{Kohler11a,Dressel12a}.  There is then an apparent
contradiction in the absence of the any kind of CO in X = I, and the
observation of charge-induced instability in X = Br \cite{Pouget97a},
which is smaller. As we explain below, this is due to (TMTTF)$_2$Br
belonging to the AFM-II region of the phase diagram, where the CO
pattern is different from the FCO that characterizes AFM-I.

Fourth, there appears to be no correlation between T$_{\rm CO}$ and
T$_{\rm N}$. However, for the two systems where both FCO and SP transitions
occur (X = AsF$_6$ and PF$_6$), T$_{\rm SP}$ is smaller in AsF$_6$ with
the larger T$_{\rm CO}$. T$_{\rm CO}$ is even larger in X = SbF$_6$, which
lies in the AFM-I region and exhibits the SP state only under
pressure.  In all cases therefore there is a competition between the
FCO and SP phases. In the remainder of the discussions below, we will
try to understand the pattern of the CO (if any) within each of the
phases in Fig.~\ref{our-phasediagram}.

\subsubsection{The FCO phase} Although the FCO transition was referred to as
``structureless'' in the earliest days \cite{Coulon85a}, recent
uniaxial thermal expansivity measurements in (TMTTF)$_2$AsF$_6$ and
(TMTTF)$_2$PF$_6$ have found distinct lattice effects at T$_{\rm CO}$
\cite{Souza08a}.  The lattice effects are most dramatic along the
$c^*$-direction, which is perpendicular to the plane containing the
TMTTF molecular planes, and along which direction the TMTTF planes and
the anions planes alternate.  The effects are weakest along the
stacking axis ($a$-direction) in X = PF$_6$ and could not be measured
in X = AsF$_6$ samples, which could not sustain the strain along this
direction.  A peak in $\alpha_{\rm c}^*$, the thermal expansivity along the
$c^*$-axis is seen at T slightly above T$_{\rm CO}$, with negative $d
\alpha_{\rm c}^*/dT$ at higher temperatures. The anomaly in the $c^*$-axis
thermal expansivity and the accompanying loss of inversion symmetry
are believed to be due to the uniform displacement of the anions
towards the charge-rich TMTTF sites (see Fig.~4 in reference
\cite{Souza08a}). Coupling between electrons and the anion
displacement field has been suggested as the driving force behind the
FCO transition \cite{Riera01a}. Bond alternation (4k$_{\rm F}$ BOW
instability) within the 4k$_{\rm F}$ CO phase has been suggested as an
additional mechanism for ferroelectricity, as this will remove
inversion symmetry too \cite{Monceau01a} (note however that if
T$_{\rho}$ is considered to be T$_{\rm MI}$ then actually the 4k$_{\rm F}$ CDW
is occurring here within the BOW phase and not the other way around;
the symmetry-related arguments remain valid nevertheless).

\subsubsection{The AFM-I phase} Among the existing TMTTF salts only two materials
belong in the AFM-I phase: X = TaF$_6$ and SbF$_6$. Of these, the
first was obtained as a consequence of a deliberate search, based on
the anticipation that an octahedral anion that is even larger than
SbF$_6$ should give a complex in the AFM-I region of the phase diagram
\cite{Iwase09a}. T$_{\rm CO}$ larger than in SbF$_6$ confirms the role of
the anion in the CO formation. No pressure-dependent study of
(TMTTF)$_2$TaF$_6$ has been carried out yet. We shall therefore focus
on (TMTTF)$_2$SbF$_6$, which has been taken across all the regions of
the phase diagram in Fig.~\ref{our-phasediagram} by application of
pressure (SC appears at pressure $>$ 5.4 GPa \cite{Itoi08a}, which is
the highest among Fabre salts with octahedral anions.)

Dielectric permittivity measurements have indicated appearance of the
FCO at 157 K in X = SbF$_6$ \cite{Nad06a,Monceau12a}. For a long time
it was assumed that T$_{\rm CO}$ coincides with T$_{\rho}$ here
\cite{Yu04a}, but the actual T$_{\rho}$ is higher \cite{Kohler11a}
(see also phase diagram in Fig.~9 of reference \cite{Iwase11a}.)  This
distinction is important, as it indicates that the metal-insulator
transition is likely due to bond strength alternation (as measured by
actual differences in bond orders, see above), and the FCO phase
occurs within the dimerized state, leaving open the mechanism of the
SP transition (BCDW-I versus 4k$_{\rm F}$ CDW-SP). The charge
disproportionation at ambient pressure, as detected from $^{13}$C-NMR
\cite{Yu04a,Iwase11a}, is $\Delta n \simeq 0.5$.  A significantly
smaller value, 0.25 - 0.30 is obtained from infrared spectroscopy
\cite{Dressel12a}. Pressure decreases T$_{\rm CO}$ monotonically, with
T$_{\rm CO} \sim 110$ K at P = 0.5 GPa \cite{Iwase11a} (90 K and 0.5
GPa \cite{Yu04a}).  The lowest temperature phase at ambient and low
pressures is the AFM-I phase \cite{Yu04a,Iwase11a}, and T$_{\rm N}$,
like T$_{\rm CO}$, decreases monotonically with pressure.  Above 0.8
GPa no evidence for CO could be detected from $^{13}$C-NMR
\cite{Yu04a,Iwase11a}. NMR measurements indicate that the low
temperature phase is the SP phase for pressure greater than 0.5 GPa
\cite{Yu04a,Iwase11a}.  Taken together, all of the above suggest
co-operative coexistence between the FCO and AFM-I, but competition
between FCO and the SP.  We believe that the lack of evidence for the
FCO in the region where the SP state occurs is a signature of the
$\cdots1100\cdots$ BCDW nature of the SP phase, as is the observation
that $\Delta n$ decreases severely in the SP phases of X = AsF$_6$ and
PF$_6$ (see below). Deuteration of the protons of the methyl groups in
TMTTF gives (TMTTF-d$_{12}$)$_2$SbF$_6$, in which T$_{\rm CO}$ is
enhanced by $\sim$ 20 K \cite{Nad05a,Furukawa05a}. No simple
explanation based on crystallographic structure analysis
\cite{Iwase11a} seems to suffice, and an alternate explanation for
this involves the Holstein e-p coupling. Specifically, the larger mass
of the deuterium reduces the phonon frequency taking the system closer
to the static limit.

Pressure-induced transition from the SP to the AFM-II phase in X =
SbF$_6$ at the lowest temperatures is achieved for pressure greater
than 2.0 GPa \cite{Iwase11a}.  We believe that the AFM-II phase is the
$\cdots1100\cdots$ BCSDW of Sections \ref{bcsdw-section1} and
\ref{bcsdw-section2} (see Fig.~\ref{our-phasediagram}). This is
discussed further below.

\subsubsection{The SP phase in X = PF$_6$ and AsF$_6$}
\label{tmttf-sp}
The spin-gapped states in
the Fabre salts occur in the higher chemical pressure region of the
phase diagram in Fig.~\ref{our-phasediagram}, relative to the AFM-I
phase. This suggests greater 2D character in the spin-gapped systems
than in the AFM systems, which appears to contradict the conventional
wisdom that the SP transition is associated with one-dimensionality
and antiferromagnetism with two dimensionality. This has led Nakamura
{\it et al.}  to argue that the spin-gapped states here are not true
SP states \cite{Furukawa11a,Iwase11a}. 
It is our belief that this discussion is largely semantic:
while it is true that weak 2D interactions stabilize the SP state
against the AFM-I in the Fabre salts, {\it the singlet spin-bonded
  sites lie on the same stack in these systems}, as has been confirmed
from elastic neutron-scattering studies in (TMTTF)$_2$PF$_6$
\cite{Foury-Leylekian04a}. While frustration-induced interchain
interactions, over and above the 1D e-p interactions, may indeed
contribute to the driving of the transition from the AFM to the
spin-gapped state, the occurrence of the spin-paired sites on the same
chain ensures that the dominant spin excitations are also 1D. The
overall physical behavior should therefore resemble that that of the
standard SP systems.  The intensity of the neutron scattering from
(TMTTF)$_2$PF$_6$, compared to that in (BCPTTF)$_2$PF$_6$ is, however,
very weak, indicating small amplitude of the SP stack distortion
\cite{Foury-Leylekian04a}. This may indeed be a consequence of 2D
interactions, although other possibilities exist, as discussed below.

The patterns of the CO in the SP phases X = PF$_6$ and AsF$_6$ have
been the most contentious \cite{Clay07a,Yoshimi12a}.  The high
temperature broken symmetry is certainly the FCO state
\cite{Nad06a,Monceau12a}. As with X = SbF$_6$ there is disagreement
over the amplitude of the CO. $\Delta n$, as obtained from T$_1$
studies in $^{13}$C NMR (0.30 - 0.50 in X = AsF$_6$
\cite{Zamborszky02a,Fujiyama06a} and 0.28 in PF$_6$
\cite{Nakamura07a}) are considerably larger than those obtained from
infrared spectroscopy, 0.20 - 0.25 and 0.10 - 0.15, respectively
\cite{Dressel12a}. Even smaller $\Delta n$ in X = AsF$_6$ ($\sim$ 0.11
- 0.16) has been claimed from measurements of Knight shift and
chemical shift \cite{Hirose10a}. A straightforward extrapolation,
based on the FCO above T$_{\rm SP}$ predicts the SP state to also have the
$\cdots1010\cdots$ pattern \cite{Yoshimi12a}.  This might even explain
the weak amplitude of the SP bond distortion
\cite{Foury-Leylekian04a}: the extent of the SP bond distortion is
indeed smaller in the 4k$_{\rm F}$ CDW-SP state than in the BCDW-I for the
same SSH e-p coupling (and $\Delta n$ is larger) \cite{Clay03a}.

There are, however, strong reasons to believe that the SP state is
BCDW-I, even with the FCO as the high temperature state. First, as
pointed out in the above, the charge localization below T$_{\rho}$ can
result from a crossover into a (disordered) 4k$_{\rm F}$ BOW, in which
case it is likely that the SP transition is a dimerization of the
dimerization. This could be particularly likely given that the FCO
transition is driven by interaction with anions, while the
bond-dimerization as well as the SP transition are both intrinsic
instabilities of the quasi-1D chains.  As shown in \ref{1d-tdep}, a
temperature-dependent transition from the WC to the BCDW-I is {\it
  expected} for e-e interactions with intermediate strength ($6 \leq U
\leq 10$, $2 \leq V \leq 3$).  Such a transition would explain the
experimentally noted competition between FCO and SP
\cite{Zamborszky02a,Voloshenko17a}.

\begin{figure}[tb]
  \begin{center}
    \begin{overpic}[width=3.2in]{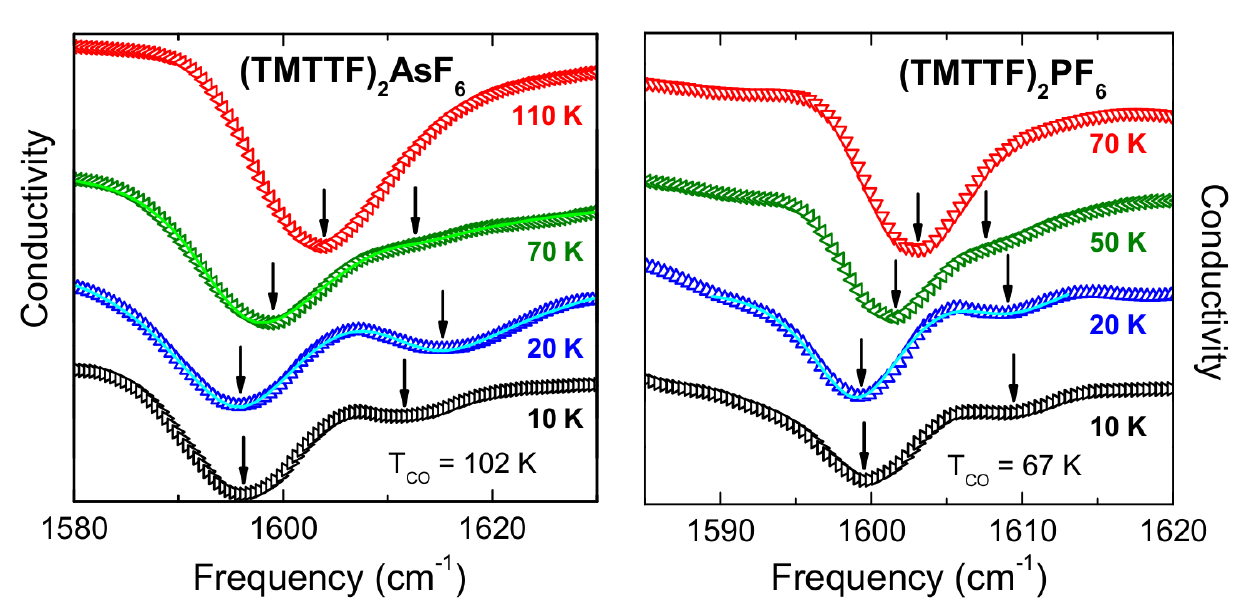}
      \put (-2,48) {\small(a)}
      \put (48,48) {\small(b)}
    \end{overpic}
    \hspace{0.1in}
    \begin{overpic}[width=2.3in]{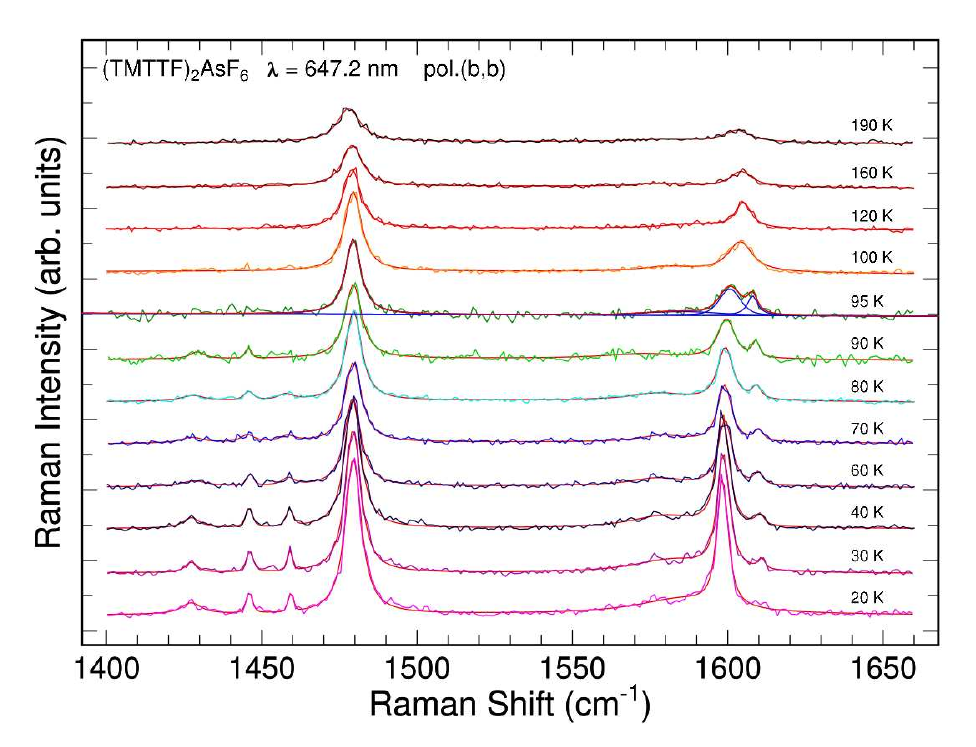}
      \put (-5,70) {\small(c)}
    \end{overpic}
    \end{center}
  \caption{(color online) Mid-infrared conductivity of (a)
    (TMTTF)$_2$AsF$_6$ and (b) (TMTTF)$_2$PF$_6$ for light polarized
    parallel to the molecular stacks. The arrows indicate the
    positions of the anti-resonance dips. T$_{\rm SP}$ = 19 K for
    PF$_6$ and 13 K for AsF$_6$. (c) Raman spectra of
    (TMTTF)$_2$AsF$_6$ as a function of temperature.  The $\nu_3$(ag)
    mode splits below T$_{\rm CO}$ = 102 K.  Reprinted with permission
    from Ref.~\cite{Dressel12a} under CC BY.}
\label{Dressel1}
\end{figure}

Theoretical work in this case predicts charge distribution leading to
much smaller $\Delta n$ in the SP state than in the FCO state
\cite{Clay03a,Clay07a}.  Charge redistribution with observable
decrease in $\Delta n$ upon entering the SP state has been claimed in
both X = AsF$_6$ \cite{Fujiyama06a} and PF$_6$ \cite{Nakamura07a} from
$^{13}$C NMR studies. Optical spectroscopy in this context appear to
give ambiguous results.  Optical conductivity obtained from reflection
studies, in the spectral region of the EMV coupled $\nu_3$ mode,
shows no charge redistribution (see Fig.~\ref{Dressel1}(a) and (b)).
Raman measurement of the same mode shows a splitting of the
vibrational frequency below T$_{\rm CO}$ but apparent vanishing of the
splitting below T$_{\rm SP}$ (see Fig.~\ref{Dressel1}(c)).  We note
however that later Raman data continues to show a splitting of the
$\nu_3({\rm ag})$ mode at 10 K, although unexplained differences were
noted with the data of Reference \cite{Dressel12a} in the low
temperature region \cite{Swietlik17a}.
Recall also the pressure dependent measurement in X = SbF$_6$ (see
above), which finds that the CO amplitude decreases strongly and is
nearly unobservable in the pressure region where the SP phase appears
\cite{Yu04a,Iwase11a}, again indicating charge redistribution in the
SP state.  Thermal expansivity measurements have found, in addition to
the FCO and SP transitions, a third likely transition at temperature
T$_{int}$ intermediate between T$_{\rm CO}$ and T$_{\rm SP}$. A
possible intermediate transition is seen also in dielectric
permittivity study of X = PF$_6$. No explanation exists in the
literature currently of this intermediate temperature transition.  It
is conceivable that the transition at T$_{int}$ is the WC-to-BCDW
transition discussed in Section \ref{1d-tdep} (theory does not require
the transition to occur {\it at} T$_{\rm SP}$ but merely requires the
charge occupancies in the SP and FCO phases to be different) and this
possibility merits further experimental investigation. Finally, there
exists a whole slew of systems that exhibit the SP transition within
the insulating phase below T$_{\rho}$ without going through the
intermediate FCO state. These materials are discussed below.

\subsection{(TMTTF)$_2$Br and (TMTSF)$_2$PF$_6$, the AFM-II region}
\label{afm2}
  
\subsubsection{(TMTTF)$_2$Br}
(TMTTF)$_2$Br lies very close to the boundary between the SP and
AFM-II phases in the phase diagram of
Fig~\ref{our-phasediagram}. T$_{\rho}$ in this case is lower than in the
other TMTTF salts ($\sim$ 100 K), although some samples exhibit
T$_{\rho}$ as high as 200 K \cite{Nad98a}. The antiferromagnetism has
been determined to be commensurate under ambient pressure from both
$^{13}$C NMR \cite{Barthel93a} and $^1$H NMR \cite{Nakamura95a}
studies, with ${{\bf q}_{\rm AFM}=(\frac{1}{2},\frac{1}{4},0)}$, and the
low T AFM amplitude of 0.14 $\mu_B$.  Coupling of the spin
fluctuations with the lattice and charge degrees of freedom was first
found from diffuse X-ray scattering \cite{Pouget97a}, which indicated
the signature of a 2k$_{\rm F}$ quasi-1D lattice instability below 70
K. This behavior is similar to the SP materials, which also show 1D SP
fluctuations at a temperature considerably higher than the the 3D
T$_{\rm SP}$. In contrast to the SP systems, the 2k$_{\rm F}$ scattering here
vanished abruptly at 17 K and long range SP order is not
observed. This, however, does not imply complete loss of structural
distortion.  Instead of 2k$_{\rm F}$ scatterings, 4k$_{\rm F}$ scatterings are
observed below T$_{\rm N}$ \cite{Pouget96a,Pouget97a}.

Tendency to coupled bond-charge-spin instability (BCSDW) in the paramagnetic
phase above T$_{\rm N}$, beginning from $\sim$ 50 K, has been confirmed from
a number of other experimental studies \cite{Nad98a,Dumm00b,Zorko01a}.
Nad {\it et al.} concluded from dielectric permittivity studies that
the growth of $\epsilon^{\prime}$ in the paramagnetic phase is due to
fluctuations of a 2k$_{\rm F}$ displacive lattice instability of ``SP
type''. These fluctuations seem to persist below T$_{\rm N}$ but vanish at
$\sim$ 10 K. Suppression of 2k$_{\rm F}$ fluctuations, as observed in the
X-ray measurements, and decrease of $\epsilon^{\prime}$ begin at the
same temperature range above T$_{\rm N}$.  Relaxation via acoustic phonons
and 1D lattice fluctuations were also seen in ESR studies in the
temperature range between approximately T$_{\rm N}$ and 50 K
\cite{Dumm00b,Zorko01a}.

The above studies confirm the short-range BCSDW nature of the broken
symmetry in (TMTTF)$_2$Br in the T$_{\rm N}$ $<$ T $<$ 50 K region. While
this apparent agreement with theory might appear to be satisfying, the
vanishing of the 2k$_{\rm F}$ instability at T$_{\rm N}$ is not predicted within
the existing theories \cite{Mazumdar99a,Kobayashi98a}. The solution to
this puzzle may however be straightforward, given that the 2k$_{\rm F}$
charge fluctuations can exist only in the presence of the 2k$_{\rm F}$ BOW
\cite{Mazumdar99a}. As the temperature is lowered and the 2D hoppings
between chains become more relevant, it is very likely that the 2k$_{\rm F}$
BOW gets weaker. While this has not been proved in the specific case
of $\rho=\frac{1}{2}$, the weakening of the 2k$_{\rm F}$ BOW in correlated
2D structures is not unanticipated \cite{Mazumdar87a}. We speculate
that the 4k$_{\rm F}$ bond dimerization, with zero phase difference between
chains, continues to persist within the EHM
of Eq.~\ref{EHM} even when weak interchain hoppings are included.  This will
be in agreement with the X-ray scattering data \cite{Pouget97a} and
will also give the observed ${\bf q_{\rm AFM}}$, with equal charge
densities on each site. This is a topic of future research.

\subsubsection{(TMTSF)$_2$PF$_6$} The compound (TMTSF)$_2$PF$_6$ has been scrutinized perhaps the most as this
was chronologically the first organic superconductor
\cite{Jerome80a,Jerome82a}. An incommensurate SDW appears here at 12.5
K under ambient pressure, with the SDW moment of 0.03 - 0.08
$\mu_B$/molecule \cite{Takahashi86a,Barthel93a}. The SDW wavevector,
though incommensurate, is close to commensurability with ${{\bf
    q}}_{\rm SDW}$=$(0.50,0.20\pm0.05, 0.06\pm0.20)$. The incommensurate SDW
has often been thought to result from nesting of the Fermi surface
\cite{Ishiguro}, as in the Peierls CDW mechanism, especially in the
context of the magnetic field induced SDW \cite{Lebed}. Of greater
relevance to the present review are however indications for the
coupling between the lattice, charge and spin degrees of freedom that
is not expected within a simple nesting mechanism for the SDW
instability, but that were indicated over the years by a variety of
experiments \cite{Pouget82a,Gruner94a,Brown92a,Odin94a,Yang99a}.
Convincing proof of this coupling was obtained from direct X-ray
measurements by Pouget and Ravy \cite{Pouget96a,Pouget97a} and
Kagoshima {\it et al.} \cite{Kagoshima99a,Kagoshima01a}, which we
briefly discuss below. Yet another subject of interest is the nature
of the very low temperature phase below 4 K
\cite{Takahashi86a,Kagoshima99a,Kagoshima01a,Valfells97a,Nomura93a,Hoshikawa00a,Lasjaunias94a},
which has only recently shown from $^{13}$C-NMR to be commensurate SDW
\cite{Nagata13a}.  {\it The pressure-induced transition to SC is
  therefore a commensurate BCSDW-SC transition, making it even more
  likely that the
  mechanisms of the superconducting transition in the
  so-called quasi-1D compounds and the strongly 2D ET and
  [Pd(dmit)$_2$]$_2$ compounds (see Section \ref{2d-section}) are related.}
X-ray scattering experiments have directly demonstrated the mixed
CDW-SDW nature of the broken symmetry below T$_{\rm SDW}$ in
(TMTSF)$_2$PF$_6$
\cite{Pouget96a,Pouget97a,Kagoshima99a,Kagoshima01a}. Pouget and Ravy
observed satellites with wave vector components $a^*/2$ (2k$_{\rm F}$)
as well as $a^*$ (4k$_{\rm F}$), thus confirming that the CDW and SDW
periodicities were identical.  The identical periodicity excludes a
simple classical explanation of the mixed CDW-SDW as a 4k$_{\rm F}$ CO
accompanied by 2k$_{\rm F}$ SDW, which merely requires the
$\cdots1010\cdots$ charge periodicity driven by the intersite Coulomb
interaction.  The common 2k$_{\rm F}$ periodicity requires that the
mixed CDW-SDW is the BCSDW of Section \ref{bcsdw-section1}.  Note that
this precludes a simple nesting mechanism for the instability, since
strong e-e interactions is a necessary requirement for the BCSDW.

The lowest temperature at which Pouget and Ravy saw the 2k$_{\rm F}$ and
4k$_{\rm F}$ scatterings was 10.7 K. The intensities of the satellite
reflections were weaker than the usual 2k$_{\rm F}$ and 4k$_{\rm F}$ reflections
by 2 to 3 orders of magnitude. Kagoshima {\it et al.} observed that
below 4 K, there was remarkable decrease in the intensities with
decreasing temperature, and the satellites were completely absent at
1.6 K \cite{Kagoshima99a,Kagoshima01a}. Kagoshima {\it et al's}
experiment gave strong evidence for the existence of a SDW subphase
below 4 K that was presumably different from the high temperature
phase, as had been conjectured already from other experiments
\cite{Takahashi86a,Valfells97a,Nomura93a,Hoshikawa00a,Lasjaunias94a}.
While the recent demonstration of the commensurate character of the
low T SDW \cite{Nagata13a} has solved the mystery of the nature of the
SDW subphase, it is has introduced two new questions, which are (i)
What is the mechanism of the T-dependent incommensurate-commensurate
transition? (ii) What does this imply for the SDW-SC transition?

We believe that the answer to the first question involves again
interchain interaction. As in (TMTTF)$_2$Br, the vanishing of the
mixed CDW-SDW below 4 K is related to increased two dimensionality,
which destroys the 2k$_{\rm F}$ BOW. Unlike (TMTTF)$_2$Br, the BCSDW
survives here below T$_{\rm SDW}$ because of the larger bandwidth of
(TMTSF)$_2$PF$_6$, which makes $U/|t|$ and $V/|t|$ smaller. Below 4 K,
the SDW then has the structure $\uparrow, \uparrow, \downarrow,
\downarrow$. As already mentioned above, this reasoning, while highly
plausible, needs confirmation from future theoretical work.

The commensurate nature of the ground state of (TMTSF)$_2$PF$_6$
strongly suggests that the mechanisms of the superconducting
transition here and in $\kappa$-(ET)$_2$Cu[N(CN)$_2$]Cl are
related. We speculate that the effect of pressure on (TMTSF)$_2$PF$_6$
is to create {\it interchain} spin singlets.  Interchain singlets are
discussed in Section \ref{zigzag} in the context of CTS ladder
materials.

\subsection{Non-traditional 2:1 cationic salts}
\label{1dnontrad-section}

\subsubsection{(TMTTF)$_2$I}
We classify (TMTTF)$_2$I among the nontraditional materials, in spite
of having the same cation as the systems discussed above, as it
exhibits the SP state without going through the FCO phase. The crystal
structure of this material is similar to the other Fabre salts
\cite{Galigne80a}.  $^{13}$C-NMR measurements indicate completely
equivalent TMTTF molecules throughout the paramamgnetic phase
\cite{Furukawa11a}.  Transition to the SP state at T$_{\rm SP}=21$ K
was detected from both T-dependent ESR susceptibility and $^{13}$C-NMR
measurements \cite{Furukawa11a}.  Line broadening at T$_{\rm SP}$, as
found in (TMTTF)$_2$PF$_6$, and ascribed to charge redistribution
\cite{Nakamura07a} was absent in this case. The temperature at which
charge localization occurs here is not known. Presumably this is
greater than 300 K.  In spite of the larger chemical pressure due to
the smaller bromide counterion, (TMTTF)$_2$Br is antiferromagnetic and
lies to the right of (TMTTF)$_2$I on the higher pressure side of the
phase diagram in Fig.~\ref{our-phasediagram}. This is a clear
indication both of two different kinds of AFM states and of
(TMTTF)$_2$Br belonging to the AFM-II region.

\subsubsection{(BCPTTF)$_2$X}
These materials exhibit semiconducting charge-gap already at room
temperature, and undergo SP transitions at T$_{\rm SP}$ = 36 K (X =
PF$_6$) and 32.5 K (A = AsF$_6$) \cite{Ducasse88a,Liu93a}.  The
molecule BCPTTF is asymmetric, but the crystal structures of
(BCPTTF)$_2$X are close to that of (TMTTF)$_2$X. The calculated
intrachain NN hopping integrals are smaller than those calculated
for TMTTF, using the same computational technique \cite{Ducasse88a}.
The FCO is nevertheless absent in (BCPTTF)$_2$X, while the effective
dimerization is larger \cite{Ducasse88a}.  The larger effective
dimerization explains both larger T$_{\rho}$ and T$_{\rm SP}$. Clearly the
charge localization, the transition to the FCO phase (or its absence,
as in the present case) and the nature of the low temperature state
(AFM versus SP) are all strongly linked and they cannot be discussed
in isolation.

\subsubsection{($o$-DMTTF)$_2$X}
\begin{figure}[tb]
  \begin{center}
    \begin{overpic}[width=1.8in]{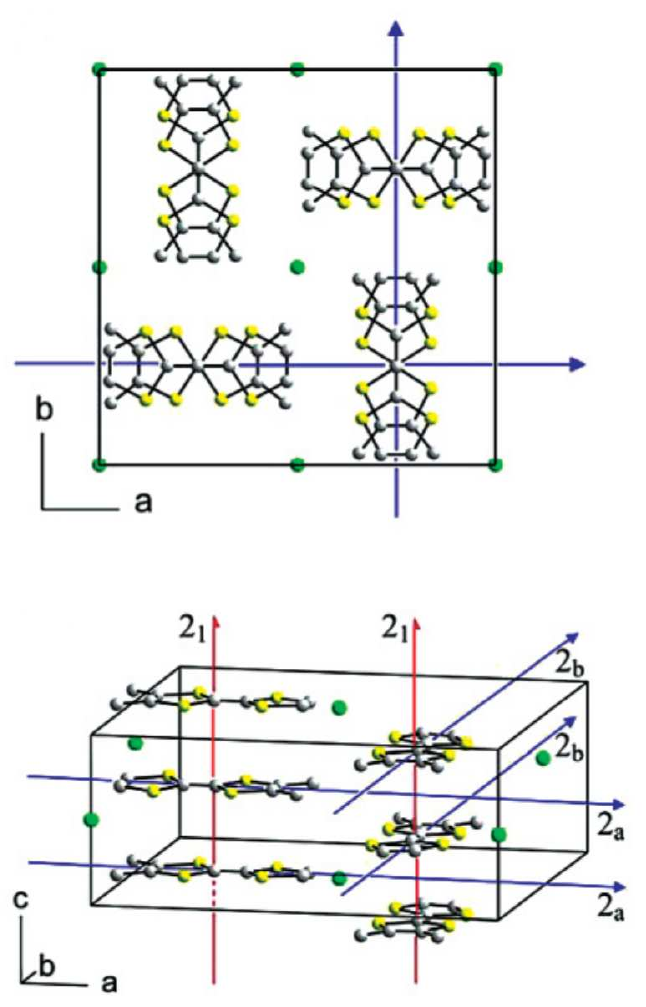}
      \put (0,90) {\small(a)}
    \end{overpic}
    \hspace{0.1in}
    \raisebox{0.4in}{
      \begin{overpic}[width=2.4in]{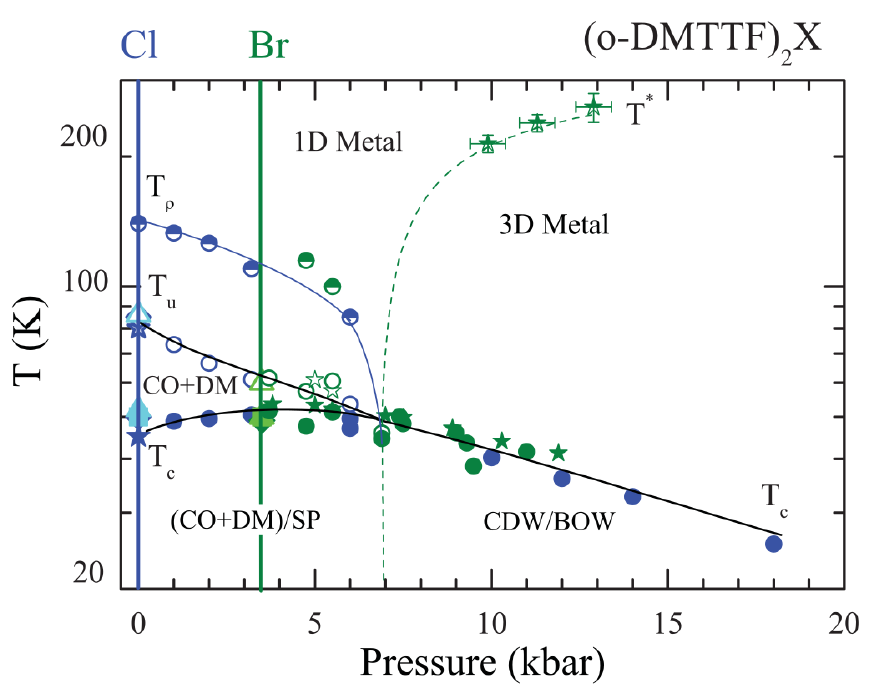}
        \put (1,70) {\small(b)}
      \end{overpic}
    }
  \end{center}
  \caption{(color online) (a) Crystal structure and (b) phase diagram
    of ($o$-DMTTF)$_2$X.  Blue (green) symbols are data for X=Cl
    (Br). The data for the Br salt is shifted by 0.35 GPa.  Circles
    (stars) are determined from the longitudinal (transverse)
    resistivity.  A ambient (zero) pressure, diamonds, triangles, and
    squares are from thermopower, ESR, and X-ray measurements,
    respectively.  Reprinted with permission from
    Ref.~\cite{Foury-Leylekian11a}, $\copyright$ 2011 The American
    Physical Society.}
\label{oDMTTF}
\end{figure}

The DMTTF molecule is a hybrid of TTF and TMTTF, with two methyl
groups attached to one side of the TTF moiety. The crystal structure
of ($o$-DMTTF)$_2$Cl is shown in Fig.~\ref{oDMTTF}(a). The space group
symmetry is $I\bar{4}2d$, and it is generally believed that the
central locations of the halogen atoms give the uniform stacking and
weak coupling between the chains.

The T vs P phase diagram for X = Cl and Br is shown in
Fig.~\ref{oDMTTF}(b), where the Br-compound is shifted in
pressure by 0.35 GPa. At the low pressure end, the system is a highly
anisotropic 1D metal, as deduced from longitudinal and transverse
resistivity.
It is believed that a smooth charge localization at T$_{\rho}$ here is
associated with 4k$_{\rm F}$ fluctuations driven by e-e interactions
\cite{Foury-Leylekian11a}. Pretransitional diffuse 4k$_{\rm F}$ lines are
seen below T$_{\rho}$ in X-ray measurements.
Additional increase in resistivity is observed as the temperature is
lowered to below T$_{u}$. This transition is considered to be the
usual 4k$_{\rm F}$ transition, although no significant structural modulation
has been observed in X-ray diffuse scattering measurements. Absence of
any change in the T-dependent magnetic susceptibility at T$_{\rho}$ or
T$_u$ indicates charge-spin decoupling. It has been argued that the
transition at T$_u$ is a combination of 4k$_{\rm F}$ FCO and BOW, with the
FCO and the BOW on separate stacks, or perhaps a combined FCO-BOW on
the same stack. Vibrational spectroscopy, however, has failed to
detect CO \cite{Jankowski11a}, indicating either pure BOW nature of
the 4k$_{\rm F}$ phase, or at the very least, small amplitude of the FCO.

The phase transition at the lowest temperature T$_{\rm c} \simeq 50$ K
occurs in both X = Cl and Br and is easily observed in magnetic
susceptibility measurements and diffuse X-ray scattering. Magnetic
measurements indicate the state below to be nonmagnetic, while X-ray
measurements find tetramerization of the stacks. Both of these suggest
that the state below T$_{\rm c}$ is the SP phase expected within the 4k$_{\rm F}$
BOW (2k$_{\rm F}$ BCDW-I). While this would agree qualitatively with the
theory presented in Sections \ref{decoupling} and \ref{4kf}, this interpretation may be
overly simplistic from a quantitative perspective. The difference
between the charge- and spin-gap in ($o$-DMTTF)$_2$X are much smaller
than what is seen in either MEM(TCNQ)$_2$ or the Fabre salts, as
evidenced from the relatively small difference between T$_u$ and T$_{\rm c}$
in Fig.~\ref{oDMTTF}(b) (particularly in the Br). The most
likely reason for the relatively strong coupling between charge- and
spin-degrees of freedom in ($o$-DMTTF)$_2$X are that $U/|t|$ and
$V/|t|$ here are smaller than in (TMTTF)$_2$X.
The actual charge and bond distortion patterns in the spin-gapped
state are currently unknown. We believe that the charge pattern here
will be found to be $\cdots0110\cdots$, but the bond distortion
pattern may be complicated and a superposition of SMWM and
SWSW${^\prime}$.

The high pressure region in Fig.~\ref{oDMTTF}(b) is very
different from the phase diagram of Fig.~\ref{phasediagram}, in that
the low temperature phase in this region is {\it not} AFM or
SDW. Furthermore, the temperatures T$_u$ and T$_{\rm c}$ have merged here.
T$^*$ here is obtained from the maxima in plots of transverse
resistivity against temperature. Whether or not the state below T$^*$
is a true metal is arguable (``bad metal'' could be one
classification). Neither is it clear whether or not there is true
boundary or a crossover from the spin-gapped phase to this
state. X-ray diffraction and magnetic measurements have not been
performed under pressure. The low T phase in this high pressure region
has been thought to be the ``usual Peierls phase''. We believe that
future experiments will find this state to be the $\cdots0110\cdots$
BCDW.

\subsubsection{$\delta$-(EDT-TTF-CONMe$_2$)$_2$X}
\begin{figure}[tb]
  \begin{center}
    \begin{overpic}[width=1.8in]{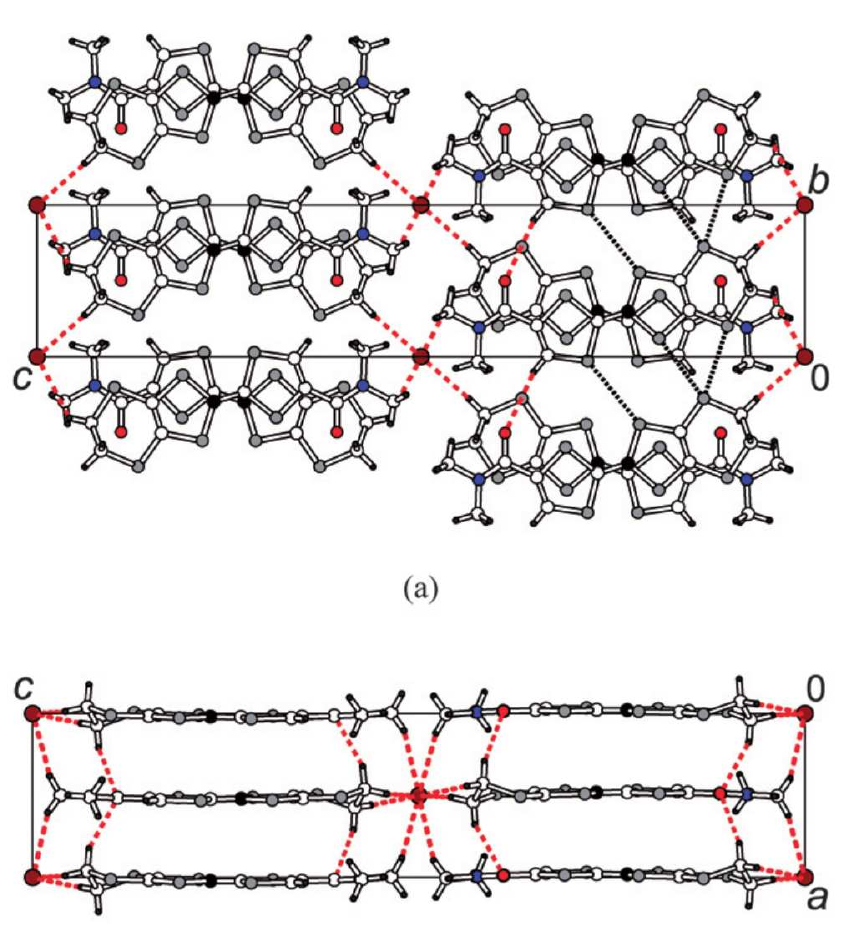}
      \put (-5,95) {\small(a)}
    \end{overpic}
    \hspace{0.1in}
    \begin{overpic}[width=3.0in]{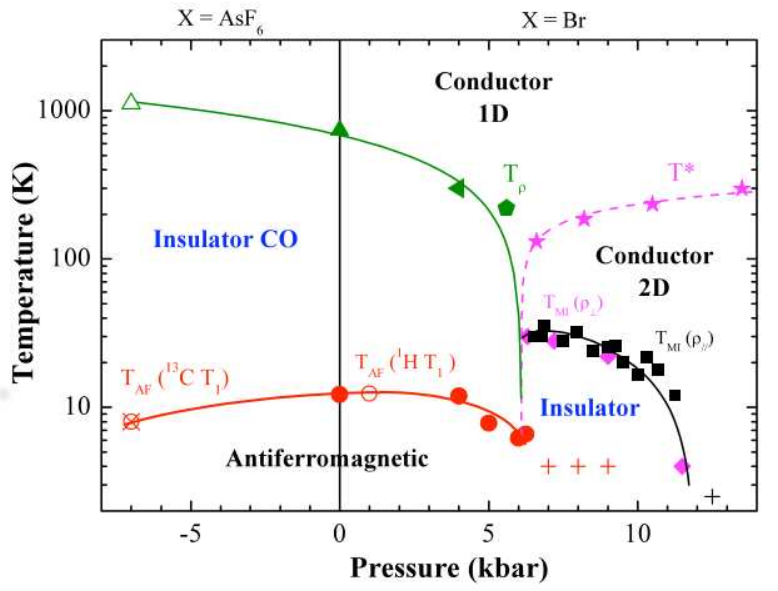}
      \put (-3,70) {\small(b)}
    \end{overpic}
    \end{center}
  \caption{(color online) (a) structure of
    $\delta$-(EDT-TTF-CONMe$_2$)$_2$X.  Reprinted with permission from
    Ref.~\cite{Zorina09a}, $\copyright$ 2009 The Royal Society of
    Chemistry.  (b) phase diagram determined from resistivity and
    $^1$H NMR measurements.  Reprinted with permission from
    Ref.~\cite{Auban-Senzier09a}, $\copyright$ 2009 The American
    Physical Society.}
\label{EDT-TTF}
\end{figure}

This material is of interest for being perhaps the only one where the
MI transition at ambient pressure is due to transition to the purely
CO state. The crystal structure of the cation is shown in
Fig.~\ref{EDT-TTF}(a).  The compounds are hereafter referred to as
EDT$_2$X (X = AsF$_6$, Br). The T vs P phase diagram, determined from
transport and NMR studies \cite{Auban-Senzier09a} is shown in
Fig.~\ref{EDT-TTF}(b).  As seen in the Fig.~\ref{EDT-TTF},
substitution of AsF$_6$ with Br is equivalent to applying hydrostatic
pressure of nearly 0.7 GPa. There is no crystal-structure driven
dimerization here and T$_{\rm MI}$ = T$_{\rm CO}$ in the low pressure
region. Furthermore, T$_{\rm CO}$ are extraordinarily high here
(compare against Fig.~\ref{phasediagram}).  The high T$_{\rm CO}$
suggests the WC as the charge-ordered phases. This is supported by
structural analysis in the insulating phase that finds uniform
intrastack intermolecular distances \cite{Zorina09a}.  The stability
of the WC here is a signature of very large $V/|t|$, with likely
$V>V_{\rm c}(U)$, as opposed to (TMTTF)$_2$X, where additional
contribution to the stabilization of the WC CO must come from
cation-anion interaction or the Holstein e-mv interaction.  This
latter conclusion is in agreement with calculations that find rather
small hopping integrals $t_{||}=71$ meV, t$_{\perp}=-25$ meV for
EDT$_2$AsF$_6$ at 300 K \cite{Heuze03a} and $t_{||}=87$ meV,
t$_{\perp}=-32$ meV for EDT$_2$Br at 150 K \cite{Zorina09a} (the
calculated hopping integrals suggest greater two dimensionality, which
reduces V$_{\rm c}$ and hence may also be responsible for stabilizing
the WC \cite{Song14a}.)

The phase diagram of Fig.~\ref{EDT-TTF} exhibits several interesting
features, which are, (i) the occurrence of CO and AFM in the same
pressure region, (ii) the steep pressure-dependence of T$_{\rm CO}$,
(iii) an apparent boundary between 1D and 2D conducting regions, and
(iv) the low T high P insulating region that is distinct from the AFM
phase, and that may be separated from the AFM phase by a real phase
boundary. Of these, (i) is to be expected from our discussions in
Section \ref{decoupling} and \ref{1d-expt}, identifying the CO phase
at low pressure as the WC and the AFM phase at low T as AFM-I.  The
rapid drop in T$_{\rm CO}$ with pressure is unique with these
compounds, suggesting a stronger role of frustration than
usual. Whether or not the curve labeled T$^*$ in the Fig. represents a
true boundary or even crossover is debatable: T$^*$ are temperatures
at which broad maxima are seen in plots of transverse resistivity.
The boundary between the high pressure conducting and insulating
phases, on the other hand, represents the true T$_{\rm MI}$ as found from
measurements of longitudinal and transverse resistivities. The
insulating phase has been found to be nonmagnetic from NMR
measurements \cite{Auban-Senzier09a}.

The absence of a SP phase {\it within} the CO phase, while it has
received relatively less attention so far, is of strong interest in
view of the theoretical discussions of Sections \ref{configspace} and
\ref{1dnumerics}. {\it This result is in apparent agreement with the
  theoretical idea that the WC CO and the SP phase are mutually
  exclusive in the real materials (Section \ref{1dnumerics-peh}).} The
high pressure nonmagnetic insulating phase has not been completely
characterized yet. We believe that this phase is a 2k$_{\rm F}$ BCDW,
as in the high pressure region of $o$-(DMTTF)$_2$X.

\subsubsection{(EDO-TTF)$_2$X and BCDW-II} 
\begin{figure}[tb]
\center{\resizebox{3.0in}{!}{\includegraphics{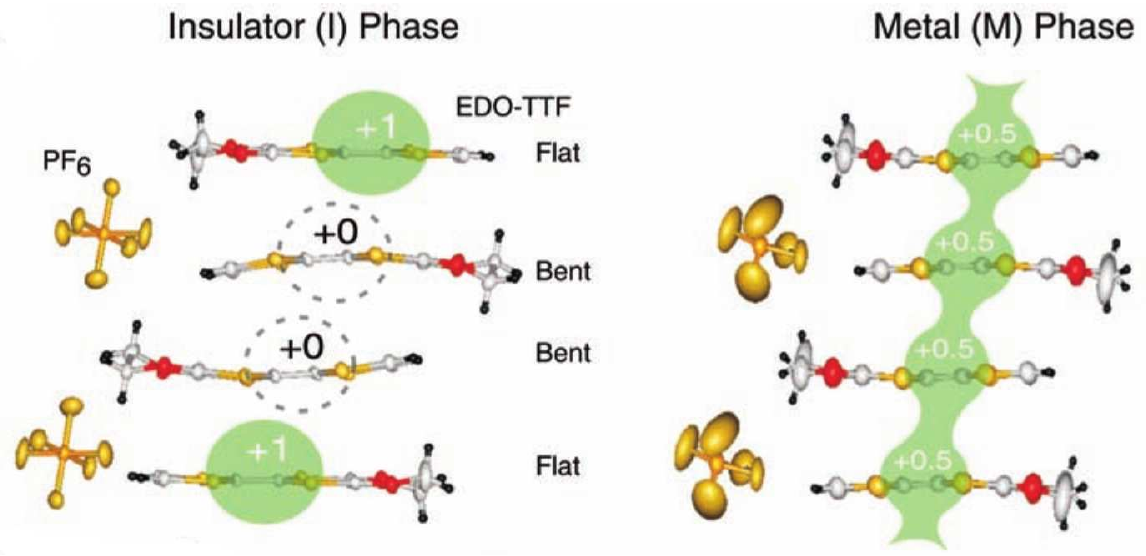}}}
\caption{(color online) Crystal structures of EDO-TTF in the low
  temperature insulating and high-temperature metallic phases.
  Reprinted with permission from Ref.~\cite{Chollet05a}, $\copyright$
  2005 AAAS.}
\label{EDO-TTF}
\end{figure}

 The crystal structure of EDO-TTF is shown in
Fig.~\ref{EDO-TTF}(a).
(EDO-TTF)$_2$X are strongly 1D with weak interstack
interaction. These compounds undergo first-order metal-insulator
transitions at a very high T$_{\rm MI}$ = 280 K and 265 K for X = PF$_6$
and AsF$_6$, respectively \cite{Ota02a,Yamochi09a}. Magnetic
susceptibility measurements indicate transition to a spin-gapped
nonmagnetic state at the {\it same} temperature T$_{\rm MI}$, a behavior
completely different from that observed in any other 2:1 cationic
material at ambient pressure (we note, however, that the high pressure
phases in $o$-(DMTTF)$_2$X and (EDT-TTF-CONMe$_2$)$_2$X do show single
transitions involving both charge and spin). The transition to the
nonmagnetic insulating state is accompanied by period 4 distortions of
both the intermolecular bonds and of the cationic molecules
themselves, as is shown in the Fig.~\ref{EDO-TTF}.

The molecules are equivalent and weakly non-planar above T$_{\rm MI}$,
where the charges on the cations are 0.5. For $T<T_{\rm MI}$ there occurs
tetramerized molecular distortions $BFFB$ ($B$ = bent and $F$ = flat),
where the bent molecules are nearly completely neutral (charge $\sim$
0), and the flat molecules are nearly completely ionic (charge $\sim$
+1.0). These assignments are obtained from vibrational spectroscopy
\cite{Drozdova04a}, which gives charges of +0.96 and +0.04 for the
charge-rich and charge-poor cations, and crystal structure analysis
\cite{Aoyagi04a}, which gives a smaller charge disproportionation with
charges of +0.8 and +0.2, respectively.  Strong role of the Holstein
e-mv coupling in driving the transition is indicated by these
experiments.  These studies have also established that the
intermolecular bond strengths are of the form SMWM, with the strongest
`S' bond between the charge-rich sites \cite{Drozdova04a}. Thus the
SSH e-p coupling also plays a strong role.

Drozdova {\it et al.} \cite{Drozdova04a} recognized that
(EDO-TTF)$_2$X can be very clearly understood as a prototype example
of 2k$_{\rm F}$ BCDW-II, with small to moderate $U$ and $V$ (see Sections \ref{1dlimit}
and \ref{1dnumerics-peh}).
 These authors measured the charge-transfer absorption
spectrum of the compound as a function of temperature, and noted that
the strength of the high energy absorption band (labeled CT$_2$ by the
authors) with peak energy at $\sim$ 1.38 eV, relative to the lower
energy band with peak energy $\sim$ 0.52 eV, was significantly higher
than in most other CTS. The authors calculated the charge-transfer
spectrum within a modified Hubbard model with modulated hopping
integrals and site energies for a tetramer consisting of only four
molecules (with charge arrangement $0110$), and justified this
boundary condition based on the weakness of the $0\cdots0$ bonds,
which would make each tetramer nearly decoupled from the chain.  In
order to obtain the CT$_1$ and CT$_2$ energies and the site charge
densities of 0.96 and 0.04 obtained from the Raman spectrum the
authors needed hopping integrals 0.15 and 0.28 eV (corresponding to
the 0--1 and 1--1 bonds, respectively), site energies $\pm$ 0.24 eV
and $U=0.93$ eV. Although the $0\cdots0$ bond was excluded entirely
from the calculations, and the site energies should have been in
principle obtained self-consistently, the values of the
Hubbard $U$ and the hopping integrals required in this calculation
were quite reasonable. The authors assigned the CT$_1$ and CT$_2$
transitions to predominantly $0110 \to 0101$ and $0110 \to 0200$
charge-transfers respectively, and ascribed the unusually large
intensity of CT$_2$ (compared to the charge-transfer absorptions in
this region in the ``standard'' 2:1 materials) to the strength of the
``isolated'' 1=1 bond.

A more recent Density Functional Theory (DFT)-based modeling of
(EDO-TTF)$_2$X treats e-e interactions at the mean-field level, and
ascribes the tetramerization at the MI transition to an ``electric
potential bias within a tetramer of EDO-TTF molecules'' as opposed to
SSH and Holstein e-p interactions \cite{Iwano08a}.  This latter work
is more difficult to understand; also, as already pointed out, if the
charge distribution of $\cdots0110\cdots$ has to originate entirely
from built in electric potential, the latter has to be non-convex,
which is unrealistic.  The authors obtained a one-electron
tight-binding Hamiltonian with modulated hopping integrals and site
energies from fitting to the DFT calculations. The largest tight
binding integral from the fitting was 0.70 eV, which is unusually
large for CTS, in which the hopping integrals range from 0.10 - 0.25
eV. This discrepancy is likely due to the neglect of e-e and e-p
interactions.

(EDO-TTF)$_2$X has also attracted recent attention as a material in
which a photoinduced transition occurs from the insulating state to a
metallic state that is very close to the high temperature state
\cite{Chollet05a}.  The transition to the metallic state takes nearly
100 picoseconds and has been thought to proceed via a metastable
charge-ordered state $\cdots1010\cdots$ \cite{Onda08a,Fukuzawa12a},
with coherent phonon interactions playing a strong role. A DFT-based
theory has attempted to explain why excitation at the CT$_2$
transition energy is more efficient in driving the transition than the
excitation at other energies \cite{Iwano13a}.  Photoinduced phase
transitions are a subject of considerably broader interest
\cite{Yonemitsu08a}, but are outside the scope of this review.

\subsection{Summary}

Wigner crystal (WC) and bond-charge density waves (BCDWs) of two
different types (BCDW-I and BCDW-II) are all observed experimentally
in quasi-one-dimensional $\rho=\frac{1}{2}$ CTS.  The WC state in
general is found in the most quasi-one dimensional of these materials,
which then exhibit an antiferromagnetic phase at still lower
temperatures.  In the (TMTCF)$_2$X series, the spin-Peierls (SP) phase
is BCDW-I and the low temperature phase adjacent to superconductivity
(SC) is BCDW-I or the bond-charge-spin density wave (BCSDW). The WC
and SP phases are mutually exclusive. In (TMTTF)$_2$PF$_6$ and AsF$_6$
there occurs a WC-to-BCDW-I transition as the temperature is
lowered. Spin gap is thus a signature of BCDW-I or II.  The
spin-density wave state found in (TMTSF)$_2$PF$_6$ under
ambient pressure is commensurate at the lowest temperatures, and the
transition to SC is a commensurate BCSDW to SC transition.  The
properties of materials exhibiting the alternate bond-distortion
pattern BCDW-II are quite different; in (EDO-TTF)$_2$X the transition
to BCDW-II coincides with a first-order metal-insulator transition at
a relatively high temperature.

\section{Broken symmetries in quasi-two dimensional CTS, Experiments}
\label{2d-section}

The 2D CTS feature a much greater variety of materials than the
quasi-1D CTS.  We will therefore first review experimental results,
and follow with a summary of theory in Section \ref{2dtheory}.  Due to
the expansive experimental literature, we must focus on a narrower
range of materials rather than all 2D molecular charge-transfer salts.
Specifically, we review the classes of 2D materials that are
superconductors or exhibit charge and spin ordered states proximate to
superconductivity.  Some topics of current interest, for example the
very large literature on magnetic field effects, are outside the scope
of this review.

\subsection{Quasi-two dimensional cationic CTS}
\label{2d-cationic}

Details of the crystal structure of 2D CTS strongly influence their
electronic properties.  Different CTS structural types are denoted by a
Greek letter prefix ($\alpha$, $\beta$, $\theta$, etc).  Below we
organize the discussion according to these different crystal
structures.

\subsubsection{$\kappa$ CTS}
\label{kappa}

\begin{figure}
  \center{
    \begin{overpic}[width=2.5in]{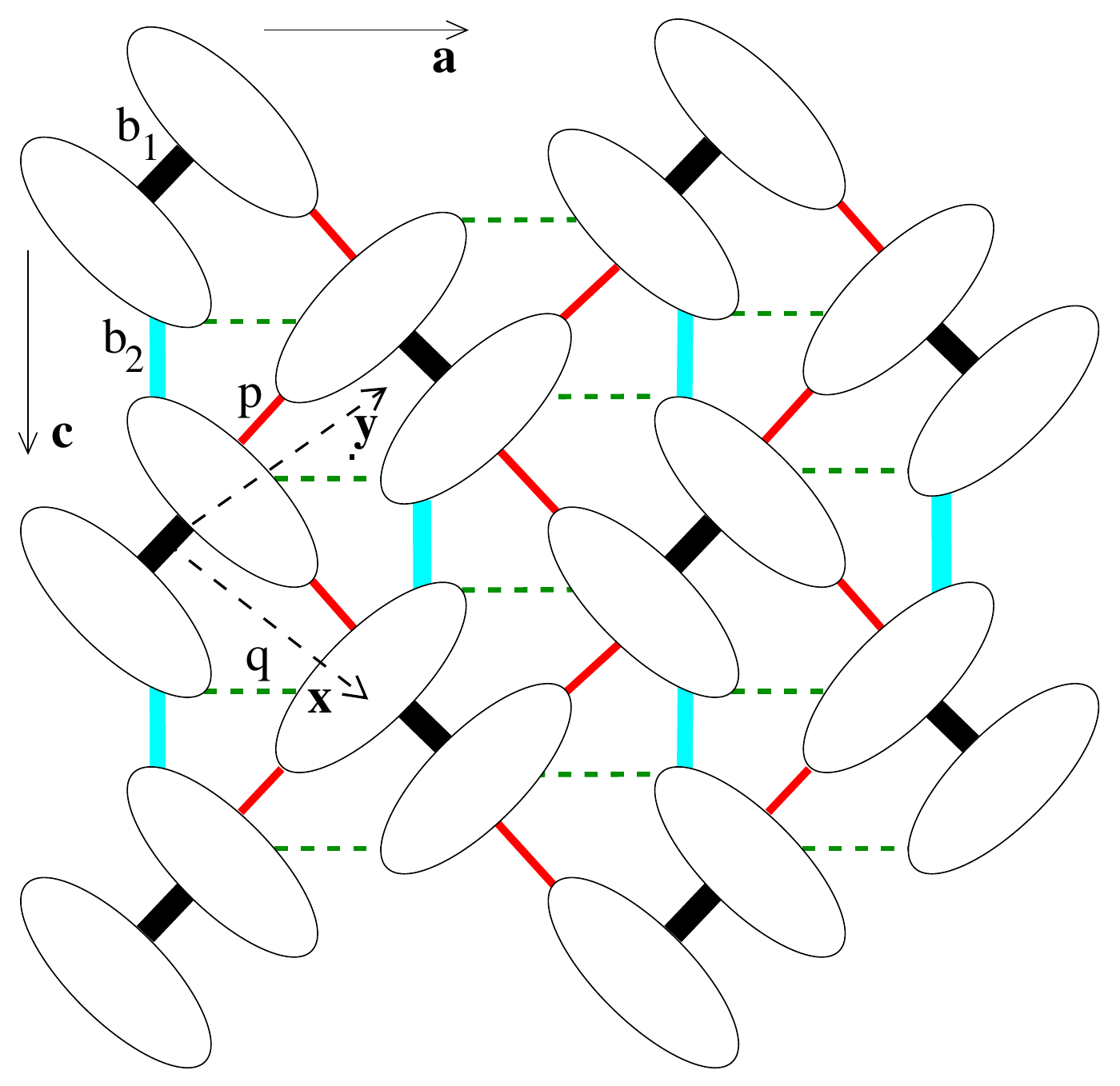}
      \put (-5,90) {\small(a)}
    \end{overpic}
    \raisebox{0.7in}{
      \begin{overpic}[width=2.0in]{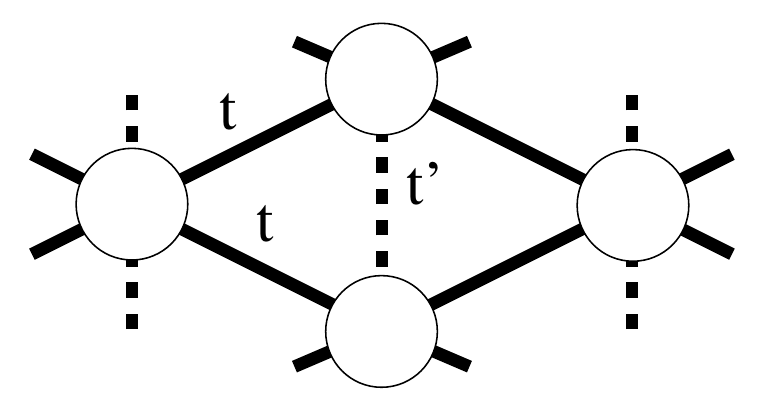}
        \put(5,50) {\small(b)}
      \end{overpic}
    }
  }
  \caption{(color online) (a) Lattice structure of
    $\kappa$-(ET)$_2$X showing individual BEDT-TTF molecules. In
    order of decreasing magnitude, $b_1$, $b_2$, $p$, and $q$ label
    the intermolecular hopping integrals. The {\bf a} and {\bf c}
    crystal axes for $\kappa$-(ET)$_2$Cu[N(CN)$_2$]X are
    indicated; for $\kappa$-(ET)$_2$Cu$_2$(CN)$_3$ the
    corresponding axes are conventionally labeled {\bf b} and {\bf c}.
    (b) effective $\frac{1}{2}$-filled dimer model. Here each site
    corresponds to a dimer in (a). $t$ is the effective hopping in the
    $\hat{x}$ and $\hat{y}$ directions and $t^\prime$ the effective
    hopping in the $-\hat{x}+\hat{y}$ direction.}
\label{kappa-lattice}
\end{figure}
The $\kappa$-(ET)$_2$X salts are perhaps the most studied family of
organic superconductors after the quasi-1D TMTSF and TMTTF
materials. The great interest in $\kappa$-(ET)$_2$ stems from both
their relatively high $T_{\rm c}$'s of up to 12.8 K (see Table
\ref{kappa-table}; a T$_{\rm c}$ of 14.2 K is however found in the
$\beta^\prime$ CTS \cite{Taniguchi03b}, see Section \ref{beta-cts}),
as well as their apparent similarity to the cuprate high-$T_{\rm c}$
materials.  As in the cuprates, the presence of AFM near to the SC
phase has led to widespread belief that Mott physics is central to
understanding the SC \cite{Kanoda11a}.  This belief has led many to
suggest that the electronic properties and SC in $\kappa$-(ET)$_2$ can
be described by an effectively $\frac{1}{2}$-filled band model.  In
this section we point out however, that many experimental observations
cannot be explained within such a model.

The schematic  crystal structure of the conducting layers in $\kappa$-(ET)$_2$X
is shown in Fig.~\ref{kappa-lattice}(a). In the $\kappa$ structure,
dimers of BEDT-TTF molecules are arranged in an anisotropic triangular
lattice. In terms of single molecules, as with other ET salts, the
conducting layer is $\frac{3}{4}$-filled with electrons.  Fig.~\ref{kappa-lattice}(b)
shows the effective dimer model commonly used to describe
$\kappa$-(ET)$_2$X; here each dimer is replaced by a single site,
leading to an effective $\frac{1}{2}$-filled band model, with an average
density of one hole per dimer. We discuss this model further in
Section \ref{2dtheory}, but introduce it here, as the anisotropy $t^\prime/t$ is
commonly used as a parameter to characterize $\kappa$-(ET)$_2$X.
Table \ref{kappa-table} lists the most commonly studied
$\kappa$-(ET)$_2$X, sorted by their effective $t^\prime/t$ as
calculated by DFT
\cite{Kandpal09a,Nakamura09a}.  A recent DFT calculation of the
hopping parameters for eight different $\kappa$-(ET)$_2$X found no
apparent correlation between $T_{\rm c}$ and any of the four $t_{ij}$,
or any of their ratios \cite{Guterding16a}. For example,
X=Ag(CN)$_2\cdot$H$_2$O and X=Cu[N(CN)$_2$]Br in Table
\ref{kappa-table} have nearly the same $t^\prime/t$, but have $T_{\rm
  c}$'s differing by more than a factor of two.

\begin{table}
  \begin{center}
    \scriptsize
    \begin{tabular}{c|c|c|c|c|c|l}
      X & $t^\prime/t$(EH) & $t^\prime/t$(DFT) & $T_{\rm c}$(K)& P$_{\rm c}$(GPa) & Notes & references \\
      \hline
      I$_3$ & 0.54 & 0.35  & 3.6& 0  & --- & \cite{Kobayashi87a,Hiramatsu15a,Guterding16a} \\
      Cu[N(CN)$_2$]Cl & 0.71 & 0.44-0.53 & 12.8& 0.03 & $T_{\rm N}$=26K& \cite{Kandpal09a,Koretsune14a,Yoshida15a}   \\
      d8-Cu[N(CN)$_2$]Br & 0.67 & 0.45-0.53 & 11& 0 &  $T_{\rm N}$=15K & \cite{Miyagawa02a,Koretsune14a,Yoshida15a} \\
      Cu[N(CN)$_2$]Br & 0.67 & 0.46 &11.6& 0 & --- & \cite{Kini90a,Hiramatsu15a,Guterding16a} \\
      Cu[N(CN)$_2$]I & --- & --- & 8.0 & 0.1 & --- & \cite{Wang91a,Kushch01a} \\      
      Ag(CN)$_2$$\cdot$H$_2$O & 0.61 & 0.47 & 5.0& 0 & --- & \cite{Mori90a,Hiramatsu15a,Guterding16a}   \\
      Cu(NCS)$_2$ & 0.81 & 0.50-0.67 & 10.4& 0 &--- & \cite{Mori99a,Kandpal09a,Nakamura09a,Koretsune14a,Yoshida15a} \\
      Cu(CN)[N(CN)$_2$] & 0.64& 0.67 & 11.2& 0 & --- &\cite{Komatsu91a,Hiramatsu15a,Guterding16a}  \\
      Hg(SCN)$_2$Cl & --- & 0.80 & ---&--- & $T_{\rm CO}$=30K&  \cite{Yasin12a,Drichko14a,Lohle17a} \\
      Ag$_2$(CN)$_3$ & 0.97 & --- & 5.2& 0.9 & --- &   \cite{Shimizu16a} \\
      Cu$_2$(CN)$_3$ & 1.07 & 0.80-0.99 & 3.9(7)& 0.6 & --- & \cite{Kandpal09a,Nakamura09a,Koretsune14a,Yoshida15a,Shimizu11a} \\
      CF$_3$SO$_3$ & 1.79 & 1.30 & 4.8& 1.1 & $T_{\rm N}=2.5$K &  \cite{Fettouhi95a,Ito16a} \\
      B(CN)$_4$ & --- & 1.80 &--- & --- & $T_{SG}$=5K &  \cite{Yoshida15a} 
    \end{tabular}
  \end{center}
  \caption{Summary of the properties of $\kappa$-(ET)$_2$X for
    different X discussed in this review. The second and third columns
    are the calculated $t^\prime/t$ using extended H\"uckel (EH) and
    density-functional theory (DFT) methods, respectively. In general,
    values calculated for low temperature  (100 K) crystal
    structures are listed; the range of values listed are accounted for
    by slightly different temperatures and theoretical methods of
    fitting the calculated bandstructure. For X=Cu$_2$(CN)$_3$, a higher T$_{\rm c}$ (7 K) has
    been found under conditions of uniaxial strain \cite{Shimizu11a}.
    $t^\prime/t$ within the EH method for  X=Cu[N(CN)$_2$]I is slightly smaller
    than  for X=Cu[N(CN)$_2$]Br \cite{Mori99a}.}
  \label{kappa-table}
\end{table}

\paragraph{AFM versus SC in $\kappa$ phase salts}

The occurrence of AFM and the proposal that it can mediate SC
has played a central role in the study of $\kappa$-(ET)$_2$X.
 The occurrence of AFM order can be simply understood
within the $\frac{1}{2}$-filled effective model of
Fig.~\ref{kappa-lattice}(b). For sufficiently large dimer $U$ and
sufficiently small frustration ($t^\prime$) between dimers, the ground
state of the $\frac{1}{2}$-filled Hubbard model on the anisotropic
triangular lattice is ${\bf Q}$=($\pi$,$\pi$) Ne\'el AFM (see Section
\ref{half-theory}).
This Ne\'el arrangement of dimer spins has been confirmed 
experimentally \cite{Miyagawa95a,Miyagawa02a}.  AFM with other
ordering wavevectors are found theoretically in the $\frac{1}{2}$-filled band Hubbard
model on the anisotropic triangular lattice for larger $t^\prime$ (for
example the 120$^\circ$ state found on the isotropic triangular
lattice \cite{Huse88a,Miyake92a,Bernu92a,Elstner93a,Capriotti99a}).
While materials in this region of the parameter space are  known
(X=CF$_3$SO$_3$) there is no information as of writing on the
ordering wavevector here.
\begin{figure}[tb]
  \center{\resizebox{2.1in}{!}{\includegraphics{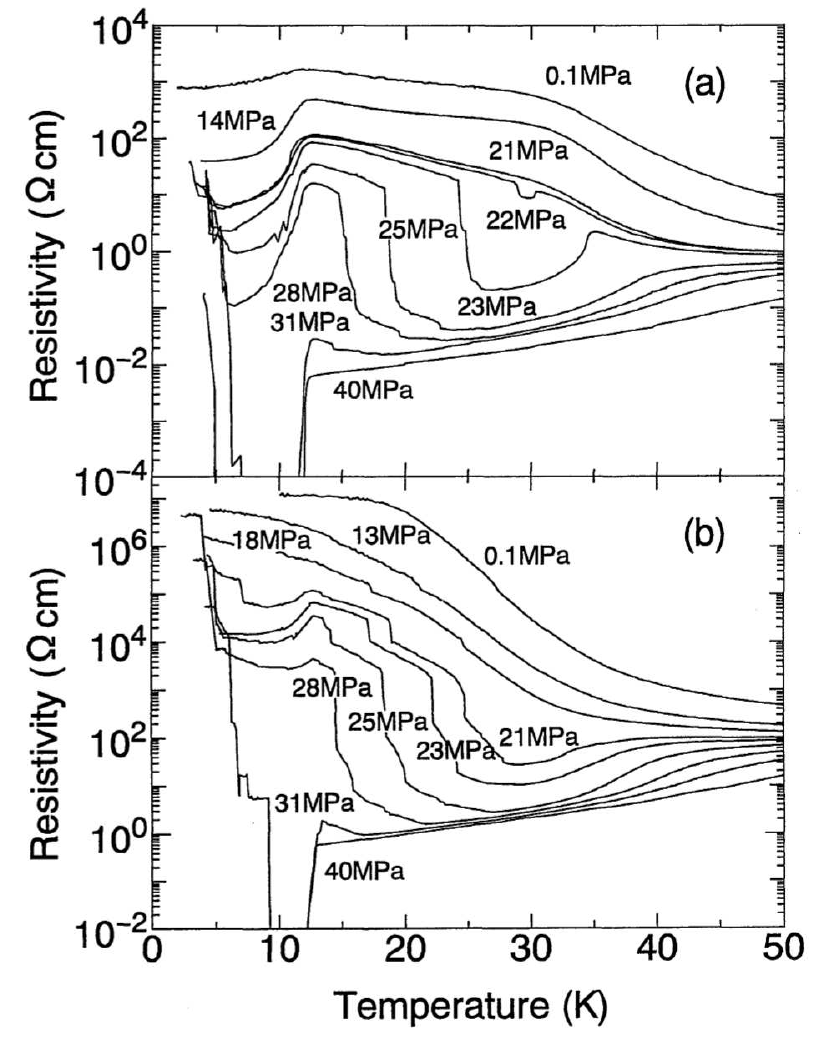}}
    \hspace{0.1in}
    \raisebox{0.25in}{
      \begin{overpic}[width=2.2in]{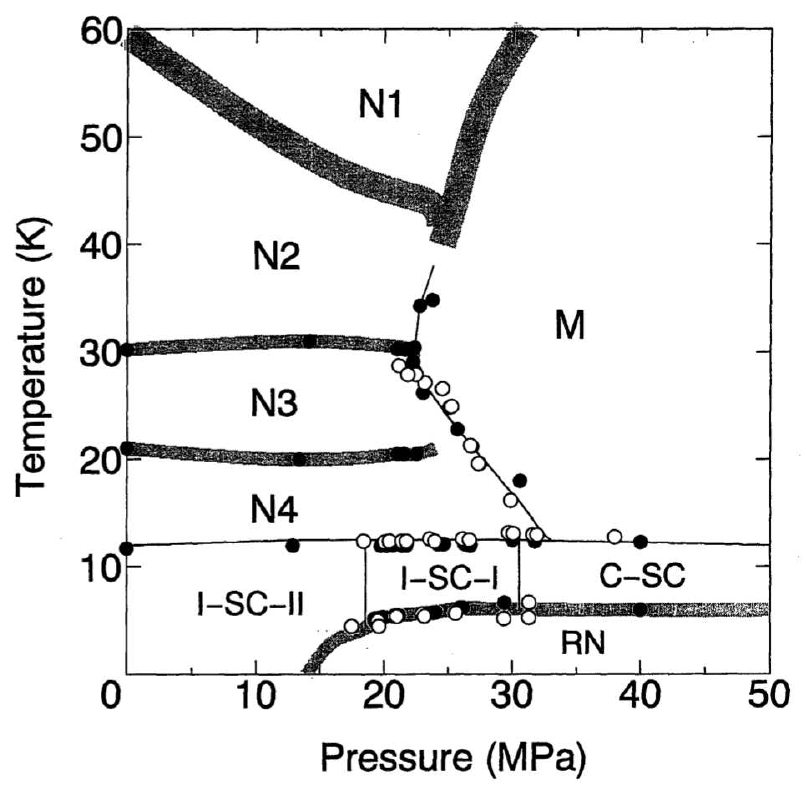}
        \put (82,82) {(c)}
      \end{overpic}
    }
  }
  \caption{Pressure dependence of (a) intralayer ($\rho_\parallel$)
    and (b) interlayer ($\rho_\perp$) resistivity of
    $\kappa$-(ET)$_2$Cu[N(CN)$_2$]Cl.  (c) Phase diagram determined
    via resistivity.  N1$\ldots$N4 refer to nonmetallic phases,
    separated by anomalies in $\rho_\parallel$ and $\rho_\perp$; M is
    the metallic phase; I-SC-II is incomplete superconductivity with
    no decrease in $\rho_{\perp}$; I-SC-I incomplete superconductivity
    with decreases in both $\rho_\parallel$ and $\rho_\perp$; C-SC
    complete superconductivity; RN recurrent nonmetallic phase.
    Reprinted with permission from Ref.~\cite{Ito96a}, $\copyright$
    1996 The Physical Society of Japan.}
  \label{kappa-cl-resistivity}
\end{figure}

\begin{figure}[tb]
  \center{
    \raisebox{0.1in}{
      \begin{overpic}[width=2.1in]{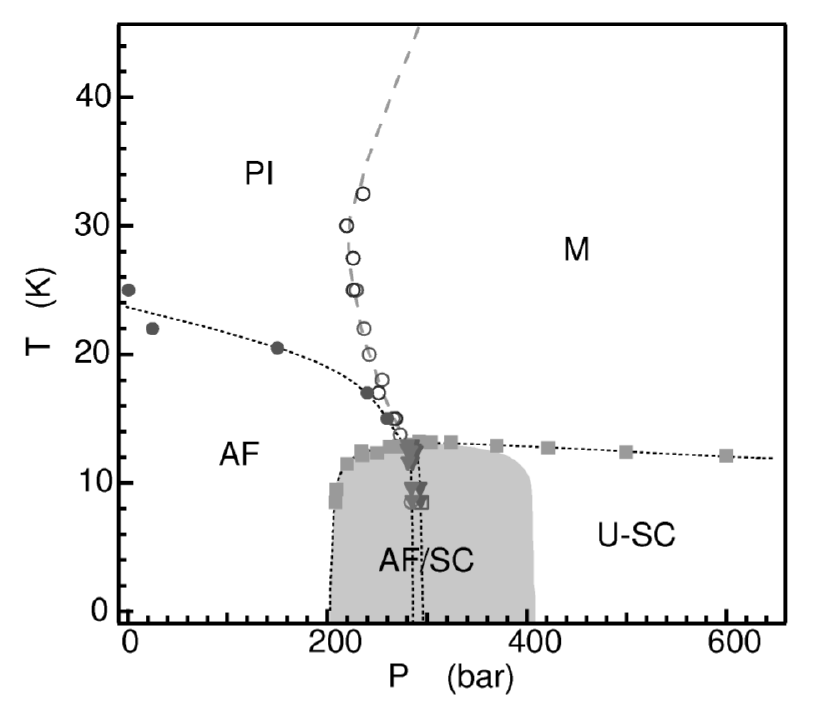}
        \put (-5,75) {\small(a)}
      \end{overpic}
    }
    \hspace{0.1in}
    \begin{overpic}[width=2.2in]{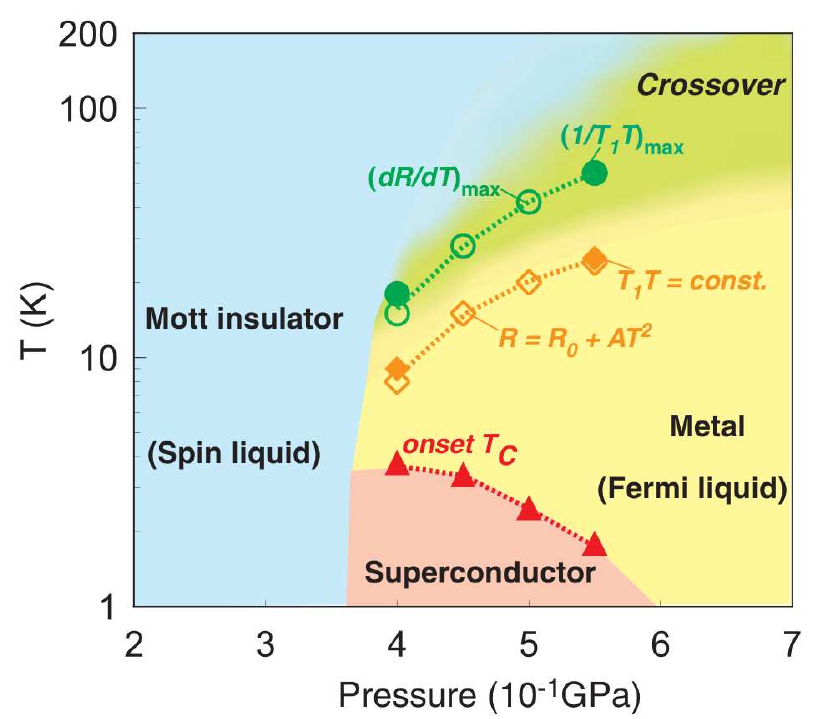}
      \put (-3,75) {\small(b)}
    \end{overpic}
  }

  \caption{(a) Temperature-pressure phase diagram of
    $\kappa$-(ET)$_2$Cu[N(CN)$_2$]Cl determined by $^1$H NMR and AC
    susceptibility measurements.  Reprinted with permission from
    Ref.~\cite{Lefebvre00a}, $\copyright$ 2000 The American Physical
    Society.  AF, PI, and M are antiferromagnetic, paramagnetic
    insulating, and metallic phases, respectively. Phase segregation
    between AF and SC takes place in the AF-SC region, while U-SC
    denotes a uniform SC state.  (b) Phase diagram for
    $\kappa$-(ET)$_2$Cu$_2$(CN)$_3$.  Reprinted with permission from
    Ref.~\cite{Kurosaki05a}, $\copyright$ 2005 The American Physical
    Society}.
  \label{kappa-phase-diagram}
\end{figure}

Despite the emphasis on AFM, experimentally antiferromagnetic ground
state is only found for a few $\kappa$ phase materials,
X=Cu[N(CN)$_2$]Cl ($\kappa$-Cl), X=Cu[N(CN)$_2$]Br ($\kappa$-Br) with
deuterated BEDT-TTF molecules (d8-$\kappa$-Br) and X=CF$_3$SO$_3$. The
most well known is X=$\kappa$-Cl with $T_{\rm N}$=26 K
\cite{Miyagawa95a,Welp92a,Williams90a}.  The AFM state is commensurate
with the lattice, with a relatively large moment of 0.4-1.0 $\mu_{\rm
  B}$ per dimer \cite{Miyagawa95a}.  Under a small pressure (even that
provided by a grease coating) of 0.03 GPa $\kappa$-Cl is
superconducting at 12.5 K \cite{Williams90a,Kobayashi91a}.  At ambient
pressure, the resistivity of $\kappa$-Cl gradually increases below 300
K, with a sharper increase below around 40 K
\cite{Williams90a,Dressel95a,Ito96a}.  The temperature and pressure
dependence of the resistivity $\kappa$-Cl, shown in
Fig.~\ref{kappa-cl-resistivity}, is complex \cite{Sushko93a,Ito96a}. A
low pressures, an ``incomplete'' SC is seen below $\sim$13 K, where
the in-plane resistivity ($\rho_{\parallel}$) decreases but does not
reach zero, and the inter-plane resistivity ($\rho_\perp$) remains
high \cite{Ito96a}.  The decrease in $\rho_\parallel$ at low
temperatures is suppressed by a magnetic field exceeding 5 T
\cite{Ito96a}. A second incomplete SC region at higher pressure has a
decrease in both $\rho_\parallel$ and $\rho_\perp$.  These incomplete
SC phases were interpreted as Cooper pair localization \cite{Ito96a}.
Several insulating phases (see Fig.~\ref{kappa-cl-resistivity}) were
identified as paramagnetic (N$_1$), fluctuating AFM (N$_2$), AFM order
(N$_3$), and AFM with weak ferromagnetic canting (N$_4$).  A reentrant
insulating phase (RN in Fig.~\ref{kappa-cl-resistivity}) was also
identified at low temperatures under pressure, which exhibited
hysteresis under thermal cycling \cite{Sushko93a,Ito96a}.  The P-T
phase diagram of $\kappa$-Cl has been confirmed with other
experimental probes, for example using $^1$H NMR and AC susceptibility
measurements as shown in Fig.~\ref{kappa-phase-diagram}(a)
\cite{Lefebvre00a}.  These results confirmed that a first order
transition occurs between the insulating AFM state and the metallic or
superconducting states \cite{Lefebvre00a}.  An S-shaped first order
transition line separates the insulating semiconducting (and AFM)
states on at low pressure from the metallic (and superconducting)
states at high pressure (see Fig.~\ref{kappa-phase-diagram}). This
line ends in at a critical point at approximately $T_{\rm c}\sim$ 40
K. At low temperature near the transition line, AFM and SC are found
to coexist (see further discussion below), giving way to pure SC under
increasing pressure \cite{Lefebvre00a}.  X=Cu[N(CN)$_2$]I ($\kappa$-I)
has a pressure-temperature phase diagram similar to $\kappa$-Cl but
shifted slightly towards the insulating phase
\cite{Kushch01a,Tanatar00a,Tanatar02a}.  Fewer experimental results
are available for this salt due to sample preparation difficulties. At
ambient pressure $\kappa$-I is insulating at low temperature; however,
to our knowledge AFM order has not yet been confirmed in this salt.

The isostructural $\kappa$-Br is superconducting at ambient pressure
with T$_{\rm c}$=11.6 K with no AFM phase \cite{Kini90a}, and this
salt is therefore placed in the U-SC region in
Fig.~\ref{kappa-phase-diagram}(a).  It is believed that the larger Br
ion, relative to Cl, induces ``chemical pressure'' on the BEDT-TTF
layer.  X=d8-$\kappa$-Br is located in the same phase diagram at lower
pressure than $\kappa$-Br and has phase segregated regions of AFM and
SC at low temperature \cite{Kawamoto97a,Miyagawa02a}.  d8-$\kappa$-Br
has a somewhat lower T$_{\rm N}$=15 K than $\kappa$-Cl and T$_{\rm
  c}$=11.5 K \cite{Kawamoto97a,Miyagawa02a}.  Below about 20K, the
d8-$\kappa$-Br NMR spectra splits into two components demonstrating
the phase segregation between AFM and metallic/SC phases
\cite{Miyagawa02a}.  This indicates that d8-$\kappa$-Br should be
placed at roughly P$\sim$250 bar in Fig.~\ref{kappa-phase-diagram}(a),
nearly exactly in between the AFM and SC phases. Partially deuterated
\cite{Taniguchi99a,Nakazawa00a} and mixed samples of the form
$\kappa$-[(h-BEDT-TTF)$_{1-x}$(d-BEDT-TTF)$_x$]Cu[N(CN)$_2$]Br
\cite{Yoneyama04a,Sasaki05a} have also been studied, which show a
range of superconducting $T_{\rm c}$'s.

An important consideration for interpreting experimental data on
$\kappa$-(ET)$_2$X is the presence of phase segregation below 35-40 K
between the AFM and metallic or superconducting phases \cite{Miyagawa02a}.  This suggests
that AFM and SC are competing phases. Phase segregation is indicated
for example by the cooling rate dependence of $T_{\rm c}$ in $\kappa$-Cl and
deuterated $\kappa$-Br \cite{Fournier07a}.  NMR results
\cite{Miyagawa02a} do not distinguish between phase segregation on a
macroscopic or nanometer length scale.  While nanometer-scale phase
segregation cannot be fully ruled out, real space optical imaging on
$\kappa$-[(h-BEDT-TTF)$_{1-x}$(d-BEDT-TTF)$_x$]Cu[N(CN)$_2$]Br found
macroscopic (micron size) AFM and metallic domains
\cite{Sasaki04a,Sasaki05a}.  Non-deuterated $\kappa$-Br is further
from the AFM phase, but the presence of stretched exponential
spin-relaxation below T$\approx$ 25 K also suggests some degree of
phase segregation \cite{Gezo13a}.  In comparison, the ambient pressure superconductor
$\kappa$-(ET)$_2$Cu(NCS)$_2$ shows a single exponential decay of
spin-relaxation, suggesting phase segregation is not present
\cite{Gezo13a}.

\paragraph{Spin liquid phase} $\kappa$-(ET)$_2$Cu$_2$(CN)$_3$ ($\kappa$-CN)
is considered as one of the best candidates for a quantum spin liquid
(QSL) \cite{Kanoda11a,Nakazawa13a}.  As shown in
Fig.~\ref{kappa-phase-diagram}(b), in the phase diagram for
$\kappa$-CN, the candidate QSL state is adjacent to SC
\cite{Kurosaki05a}.  The $^1$H NMR signal remarkably does not change
down to 32 mK (except for a small broadening below about 4 K
\cite{Shimizu05a}), which is an energy scale several orders of
magnitude smaller than the estimated inter-dimer exchange coupling, $J
\sim$ 250 K \cite{Shimizu03a}.  $\kappa$-CN is however a
superconductor under pressure with T$_{\rm c}$=3.9 K
\cite{Geiser91a,Komatsu96a}.  A higher T$_{\rm c}\sim$7 K has been found
under conditions of uniaxial strain \cite{Shimizu11a}.  As in other
$\kappa$-phase salts, the magnetic susceptibility begins to decrease
at about 50 K, although the material remains paramagnetic at low
temperature \cite{Komatsu96a,Shimizu03a}.  The disadvantage of $^1$H
NMR studies is however the small hyperfine coupling between protons at
the ends of the BEDT-TTF molecules and conduction electrons. $^{13}$C
NMR is much more sensitive because the central C atoms of the ET molecule are much more
strongly coupled to conduction electrons \cite{Miyagawa04a}. In $\kappa$-CN, the
$^{13}$C NMR signal begins to broaden below around 70 K, suggesting
the development of inhomogeneity \cite{Kawamoto04a,Shimizu06a}.  This
broadening however disappears under pressure \cite{Kawamoto06a}.  The
Knight shift, which measures the intrinsic spin susceptibility without
impurities, decreases slowly below 50 K and then sharply below 10 K
\cite{Kawamoto04a}. As in other $\kappa$ salts the NMR relaxation rate
1/(T$_1$T) increases as temperature decreases, but turns around and
decreases below 10 K \cite{Kawamoto04a}.

Many experiments have noted a transition or crossover at around 6 K,
as well as a second change at lower temperature ($\sim$ 1 K).  At 6 K,
the $^{13}$C 1/T$_1$ changes from being proportional to
T$^{\frac{1}{2}}$ to being temperature-independent
\cite{Shimizu06a}. Below about 1 K the temperature dependence of
1/T$_1$ changes again, becoming proportional to T$^{\frac{3}{2}}$
\cite{Shimizu06a}. $\mu$SR in zero or weak magnetic fields confirms
these three distinct temperature regions in $\kappa$-CN
\cite{Ohira06a,Pratt11a,Nakajima12a}. Importantly, a widened
distribution of internal fields was not observed at zero field,
meaning that the broadening and inhomogeneity observed by $^{13}$C NMR
are caused by the magnetic field. The effect of the field at low
temperature is to create a ``weak-moment antiferromagnet'' state,
which phase separates with the zero field state
\cite{Pratt11a,Nakajima12a}.  At the 6 K transition the formation of
bosonic pairs was suggested, with a second transition to a condensed
phase at lower temperature to a spin-gapped phase with $\Delta\approx
3.5$ mK \cite{Pratt11a}.

Under pressure, the resistivity of $\kappa$-CN drops suddenly below
about 10 K for $P \sim$ 0.4 GPa \cite{Kurosaki05a}. This was
interpreted as a first-order transition from Mott to a metallic state,
which under lower temperature becomes superconducting
\cite{Kurosaki05a}. Near the transition to the metallic state, two
separate components are observed in the NMR signal, at first
suggesting the presence of phase segregation into metallic and
insulating domains \cite{Kurosaki05a}. However, the absence of of
hysteresis in measurements of the resistivity and lack of two distinct
signals in $^{13}$C NMR suggest that the transition is not first order
\cite{Kawamoto06a}.  $^{13}$C NMR measurements under pressure found a
similar pressure dependence of 1/(T$_1$T) to $\kappa$-Br under
pressure, with a peak in 1/(T$_1$T) at around 10 K that moves to
higher temperatures under pressure \cite{Kawamoto06a}.

One of the unresolved puzzles concerning $\kappa$-CN is the
apparent disagreement between low temperature specific heat and
thermal conductivity measurements. The specific heat C$_p$ of $\kappa$-CN
is linear in $T$ at low temperature (0.75 to 2.5 K), characteristic
of a gapless energy spectrum \cite{Yamashita08a}. C$_p$ is also
independent of magnetic field up to 8 T \cite{Yamashita08a}.
Thermal conductivity measurements (as well as $\mu$SR) indicate
the presence of a small spin gap \cite{Yamashita09a}.

\paragraph{Pseudogap}

\begin{figure}[tb]
  \center{
    \raisebox{0.5in}{
      \begin{overpic}[width=2.1in]{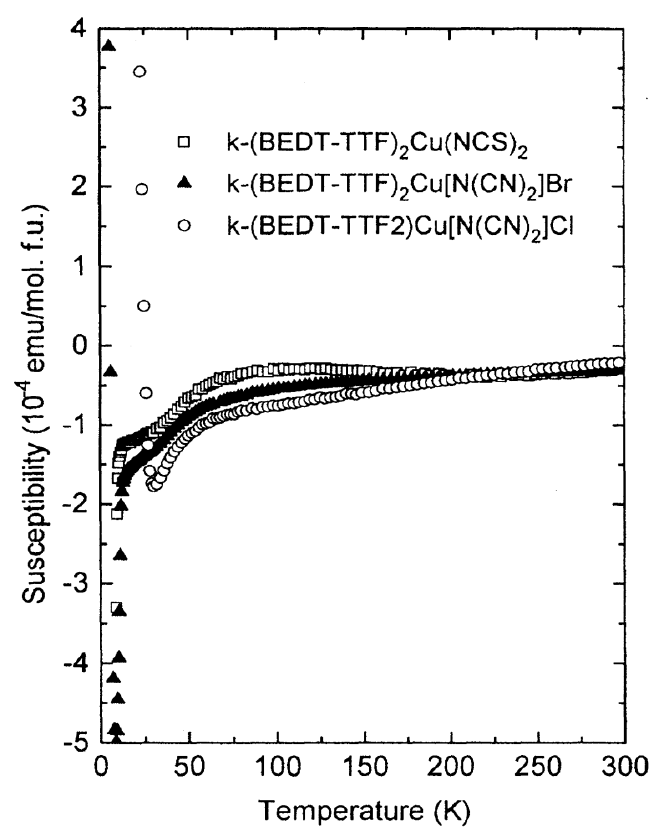}
        \put (-2,90) {(a)}
      \end{overpic}
    }
    \hspace{0.1in}
    \begin{overpic}[width=2.0in]{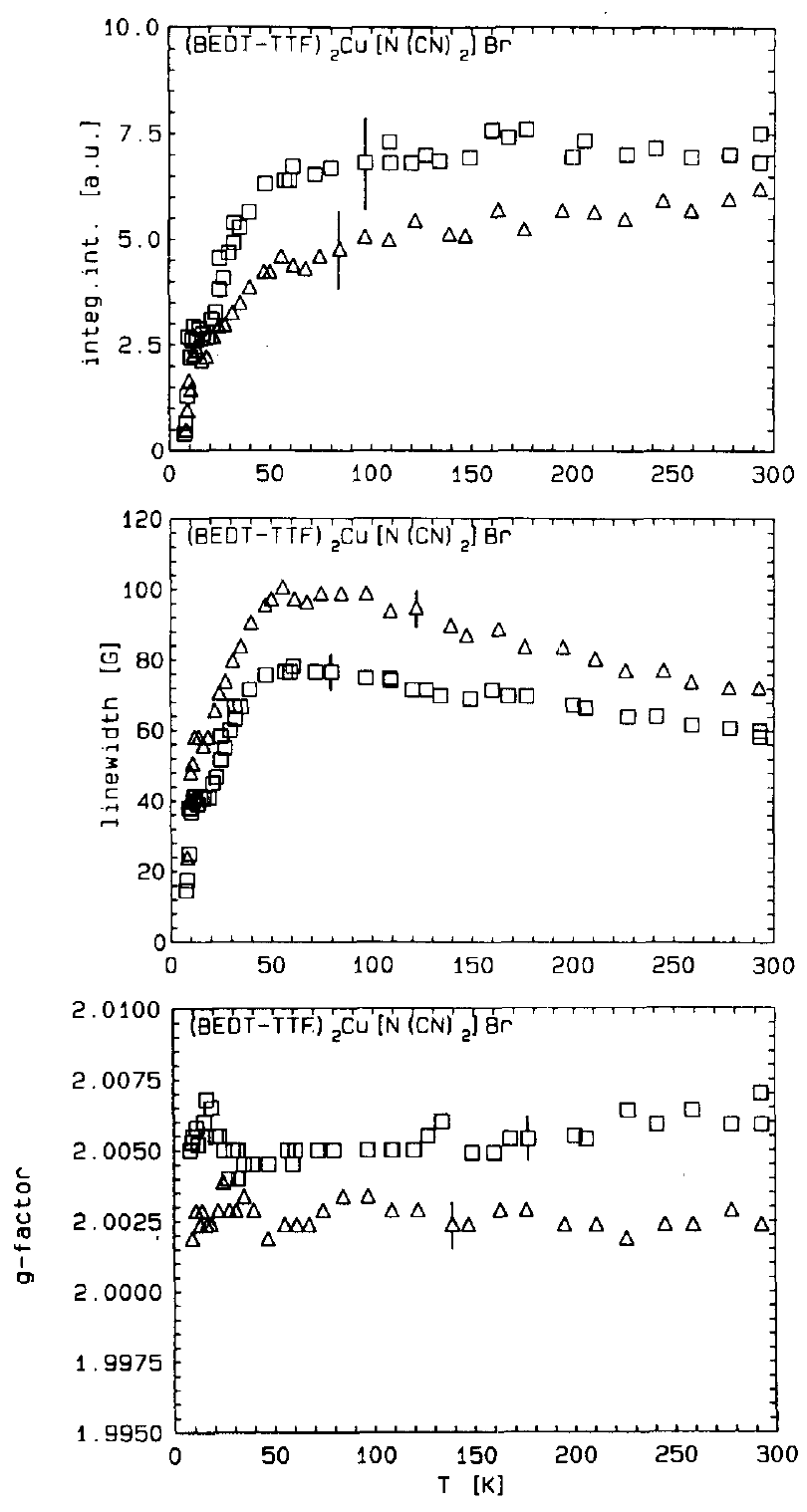}
      \put (-2,95) {\small(b)}
      \put (-2,63) {\small(c)}
      \put (-2,30) {\small(d)}
    \end{overpic}
  }
  \caption{(a) Static susceptibility of $\kappa$-Cl, $\kappa$-Br, and
    $\kappa$-NCS.  Reprinted with permission from
    Ref.~\cite{Kawamoto95b}, $\copyright$ 1995 The American Physical
    Society.  (b)-(d) ESR parameters of $\kappa$-Br versus
    temperature. Squares and triangles correspond to the field
    perpendicular and parallel to the conducting plane, respectively.
    Reprinted with permission from Ref.~\cite{Kataev92a}, $\copyright$
    1992 Elsevier.}
  \label{kappa-pg1}
\end{figure}  
\begin{figure}[tb]
  \center{
    \begin{overpic}[width=1.8in]{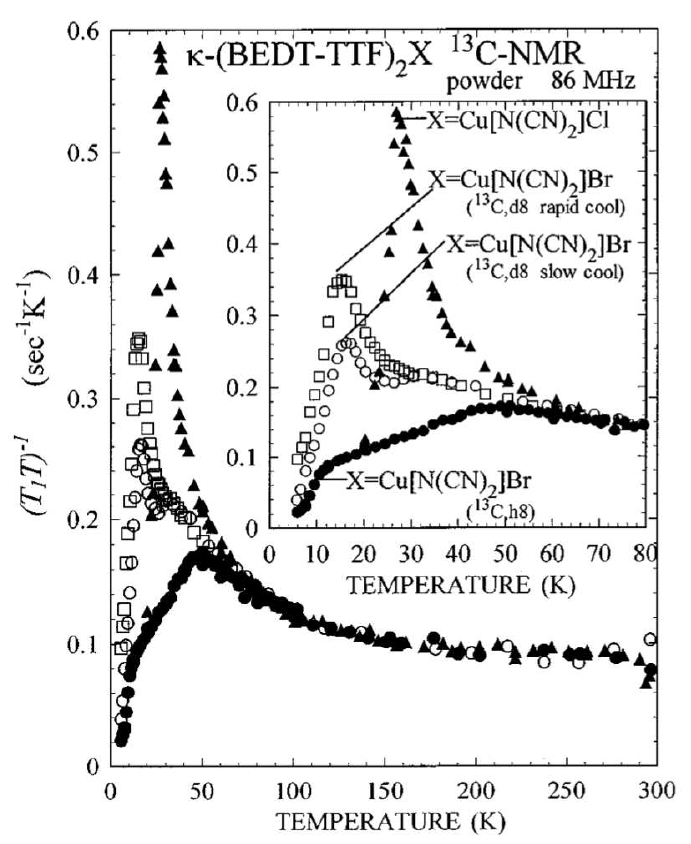}
      \put (1,90) {\scriptsize(a)}
    \end{overpic}
    \hspace{0.1in}
    \begin{overpic}[width=2.3in]{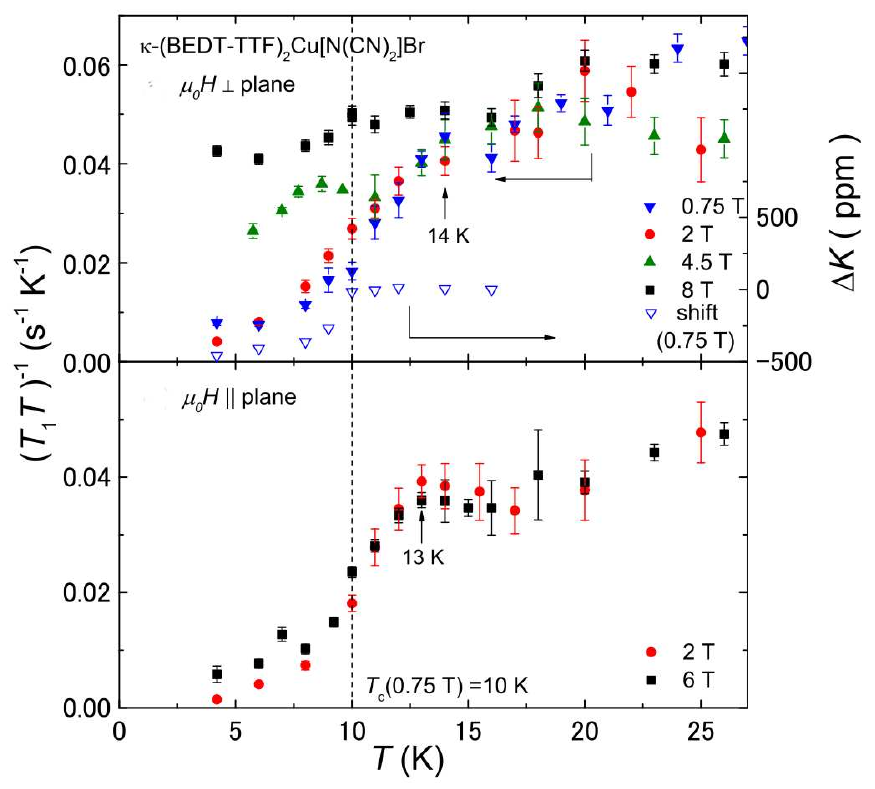}
      \put (14,76) {\scriptsize(b)}
      \put (14,44) {\scriptsize(c)}
    \end{overpic}
  }
  \caption{(a) 1/(T$_1$T) versus temperature for $\kappa$-Cl and
    $\kappa$-Br.  The inset expands the low temperature region.
    Reprinted with permission from Ref.~\cite{Kawamoto97a},
    $\copyright$ 1997 The American Physical Society.  (b)-(c)
    1/(T$_1$T) for $\kappa$-Br at low temperature. For $\kappa$-Br
    (but not $\kappa$-NCS), a depression of 1/(T$_1$T) is observed in
    the temperature range immediately above T$_{\rm c}$ Reprinted with
    permission from Ref.~\cite{Kobayashi14a}, $\copyright$ 2014 The
    American Physical Society.}
  \label{kappa-pg2}
\end{figure}
A number of experiments indicate some kind of phase transition or
crossover occurring at around 50 K within the metallic state of
$\kappa$-(ET)$_2$X. In both $\kappa$-Br and $\kappa$-NCS an anomaly in
the derivative of the resistivity is seen at around 50 K
\cite{Gartner88a,Sushko91b}.  Anomalies in the lattice expansivity
reminiscent of second-order phase transitions are seen in the same
temperature region in both of these salts \cite{Muller02a}.  In
particular, NMR experiments (see reference \cite{Miyagawa04a} for a
review) give strong evidence that this transition is to a pseudogapped
state.  As shown in Fig.~\ref{kappa-pg1}(a), the static magnetic
susceptibility begins to gradually decrease below around 100 K,
decreasing more steeply below around 50 K
\cite{Kawamoto95b}. Similarly, the ESR integrated intensity (see
Fig.~\ref{kappa-pg1}(b)) remains constant from 300 K down to about 50
K, below which it drops indicating a nonmagnetic ground state
\cite{Kataev92a}.  Similar behavior is seen in the $^{13}$C NMR
1/(T$_1$T) \cite{Kawamoto95a,Kawamoto95b}.  1/(T$_1$T) is nearly
identical for $\kappa$-Cl and $\kappa$-Br (see
Fig.~\ref{kappa-pg2}(a)) from 300 K down to 50 K ($\kappa$-NCS is
similar) \cite{Kawamoto95a,Kawamoto95b}. Below 50 K 1/(T$_1$T)
diverges as T approaches $T_{N}$ for $\kappa$-Cl, while 1/(T$_1$T)
decreases for $\kappa$-Br and $\kappa$-NCS
\cite{Kawamoto95a,Kawamoto95b}.  Closer to T$_{\rm c}$ some
differences are however observed between different superconducting
$\kappa$ salts.  In $\kappa$-NCS (T$_{\rm c}$=9.3 K), the $^{13}$C NMR
1/(T$_1$T) is nearly constant in the region approximately 5 K above
T$_{\rm c}$, with nearly the same value as the low-temperature normal
state, which was determined by suppressing SC with a magnetic field
(see Fig.~\ref{kappa-pg2}) \cite{Kawamoto97a,Kobayashi14a}.
$\kappa$-CN behaves similarly, with a nearly constant 1/T$_1$T just
above T$_{\rm c}$ \cite{Shimizu10a}. In contrast, in $\kappa$-Br
(T$_{\rm c}$=11.4 K), 1/(T$_1$T) decreases in the region approximately
5 K above T$_{\rm c}$ before reaching T$_{\rm c}$
(Figs.~\ref{kappa-pg2}(b) and (c)) \cite{Kobayashi14a}. A similar
reduction of 1/(T$_1$T) occurs just above T$_{\rm c}$ in
d8-$\kappa$-Br, where the effect appears to start at even higher
temperature ($\sim$ 30 K) \cite{Miyagawa02a}.  It has been suggested
that the apparent higher pseudogap temperature in d8-$\kappa$-Br is
due to it being closer to the AFM phase in the phase diagram
\cite{Miyagawa02a}.  However, it is also possible that the apparent
stronger temperature dependence of 1/(T$_1$T) in the 10--30 K region
is due to AFM domains in the phase segregated state
\cite{Kobayashi14a}.

The Nernst effect is another experimental probe that has been used to
detect the presence of a pseudogap in the high-T$_{\rm c}$ cuprates
above $T_{\rm c}$ \cite{Wang06a,Li10b}.  In the Nernst effect, a
transverse electric field is generated from a temperature gradient and
a perpendicular magnetic field. Normally a weak effect, the transverse
field is strongly enhanced in a superconductor due to superconducting
vortices.  In the cuprates, the large Nernst signal within the
pseudogap region but well above T$_{\rm c}$ has been interpreted by
some authors as evidence for a fluctuating superconducting
state\cite{Wang06a}. This conclusion, has however, been contradicted
by others who believe that large Nernst signal accompanies stripe
formation \cite{Cyr-Choiniere09a}.  Measurements of the Nernst effect
in $\kappa$-NCS and $\kappa$-Br both show a clear Nernst signal below
T$_{\rm c}$ \cite{Nam07a}.  For $\kappa$-Br the Nernst coefficient is
positive slightly above T$_{\rm c}$ suggesting a pseudogap, while for
$\kappa$-NCS there is no signal above T$_{\rm c}$ \cite{Nam07a}. It
has been suggested that the presence of the Nernst effect above
$T_{\rm c}$ in $\kappa$-Br but not in $\kappa$-NCS is again due to
$\kappa$-Br being closer to the AFM state in the phase diagram
\cite{Nam07a}.  Experiments on X=Cu[N(CN)$_2$]Cl$_{1-x}$Br$_x$ found a
positive Nernst signal at an even higher temperature
($\approx$5T$_{\rm c}$) for $x=0.73$, which was interpreted as an
increase in SC fluctuations due to closer proximity to the Mott
transition \cite{Nam13a}.  Another possible reason for the difference
is the different amount of disorder in these two salts \cite{Gezo13a};
in the context of cuprates it was suggested that the positive Nernst
signal tracks the disorder that accompanies stripe formation in the
La-based cuprates \cite{Cyr-Choiniere09a}.  Magnetic torque
measurements on $\kappa$-NCS did however give evidence for fluctuating
SC above T$_{\rm c}$ \cite{Tsuchiya12a}.

\paragraph{Symmetry of Superconducting order parameter}

Experiments using a wide range of probes suggest that the SC pairing
throughout the $\kappa$-(ET)$_2$X family is singlet with nodes in the
order parameter.  Experimental techniques that have been employed
include site-selective $^{13}$C NMR \cite{deSoto95a,Miyagawa04a},
specific heat measurements \cite{Taylor07a}, penetration depth
measurements \cite{Pinteric02a,Milbradt13a,Perunov12a}, as well as STM
tunneling experiments \cite{Arai01a,Ichimura08a,Oka15a}.
The temperature dependence of the NMR relaxation rate 1/T$_1$
in the superconducting state goes as T$^3$ at low temperatures,
consistent with an order parameter with nodes. This T$^3$
dependence has been verified in $\kappa$-Br \cite{deSoto95a}, 
$\kappa$-NCS \cite{Miyagawa04a}, and $\kappa$-CN \cite{Shimizu10a}.

Experiments are generally in agreement that the SC is singlet rather
than triplet, and that the order parameter has nodes.  However, there
is less agreement on the location of nodes in the conducting plane.
The order parameter symmetries $d_{x^2-y^2}$ and $d_{xy}$ have nodes
in different locations, rotated by 45$^\circ$.  An additional
complication is that the axes in the effective dimer lattice
($\hat{x}$ and $\hat{y}$ in Fig.~\ref{kappa-lattice}) are rotated at
45$^\circ$ from the crystal axes ($\hat{a}$ and $\hat{c}$ in
Fig.~\ref{kappa-lattice}).  In the experimental literature
$d_{x^2-y^2}$ symmetry is usually assumed to have nodes at 45$^\circ$
to the crystal axes, while $d_{xy}$ has nodes along the crystal
axes. In theory based on the effective half-filled model these two
symmetries are interchanged.  Magneto-optical \cite{Schrama99a} and
specific heat measurements in a magnetic field find the nodes aligned
with the crystal axes \cite{Malone10a}.  Thermal conductivity
\cite{Izawa01a} and STM measurements \cite{Arai01a,Ichimura08a}
however find the nodes between the crystal axes.  STM measurements on
partially deuterated $\kappa$-Br found two possible $d$-wave order
parameters, probably due to phase segregation \cite{Oka15a}.  Recent
STM measurements on $\kappa$-Br suggest a mixed order parameter of
$s+d_{x^2-y^2}$ form \cite{Guterding16b}. Such a symmetry is predicted
by numerical calculations within a $\frac{3}{4}$-filled model (see
Section \ref{kappa-pairpair}) as well by RPA spin-fluctuation
calculations \cite{Guterding16b,Guterding16a}.  Disagreement in the
location of the nodes is probably due to experimental factors;
determining the symmetry of the order parameter from thermodynamic
measurements is difficult because one must separate the small
electronic contribution to the specific heat from the much larger
phonon background \cite{Taylor07a,Kuhlmorgen17a}.  For example,
earlier specific heat measurements had suggested $s$-wave pairing
\cite{Elsinger00a,Muller02b}, but more recent work is consistent with
nodes in the gap function \cite{Taylor07a,Kuhlmorgen17a}.  In summary,
experiments now agree that nodes {\it are} present in the
superconducting order parameter, but more work appears to be needed to
precisely define the functional form and determine to what degree it
varies for different members of the $\kappa$-(ET)$_2$X family.  In
relation to this point, experiments under pressure that are sensitive
to the position of the nodes have yet to be performed on $\kappa$-CN.

\paragraph{Charge and lattice effects}
Within the effective $\frac{1}{2}$-filled band picture at large
Hubbard $U$ there is minimal charge fluctuation and lattice effects
are also expected to be weak. Nevertheless,
a wide range of experiments indicate that charge and lattice degrees
of freedom play important roles in the low-temperature physics
$\kappa$-(ET)$_2$X.  Most features in the phase diagram of
$\kappa$-Cl (the superconducting transition, MI transitions both as a
function of temperature and pressure, and the critical point) were
observed as anomalies in the propagation speed of ultrasound
\cite{Fournier03a}.  The AFM transition at T$_{N} \sim$ 27 K
however could not be identified \cite{Fournier03a}.
Precision lattice expansivity measurements have also been performed on
several $\kappa$-(ET)$_2$X \cite{Muller02a,Souza15a}.
As mentioned above, anomalies are seen in the lattice
  expansivity in the $\sim$50 K region where the pseudogap opens.
At lower temperatures,
in $\kappa$-CN a
large anomaly is seen in the in-plane lattice expansivity at 6 K,
along with a smaller feature at 2.8 K \cite{Manna10a}. The sharpness
of the 6 K feature suggests a real phase transition rather than a
crossover, and the shape of the feature is similar to the anomaly seen
at the SP transition in (TMTTF)$_2$X \cite{Souza08a}.  These features
were unaltered by magnetic fields up to 8 T \cite{Manna10a}.

Measurements of the frequency-dependent dielectric constant for
$\kappa$-CN found that the dielectric constant goes through a broad
maximum at temperature T$_{\rm max}$ for T$\lessapprox$ 50 K
\cite{Abdel-Jawad10a}.  The peak strengthens with decreasing
frequency, with T$_{\rm max}$ shifting to lower temperature
\cite{Abdel-Jawad10a}. The response was interpreted as being similar
to relaxor ferroelectrics, with anti-ferroelectric ordering in the
case of $\kappa$-CN.  Because of the low frequencies involved, random
domains of short-range charge order (with unequal charges on the two
BEDT-TTF molecules in a dimer) are presumed to be fluctuating rather
than the charges on individual ET molecules.  The extrapolated
temperature for antiferroelectric ordering is approximately 6 K
\cite{Abdel-Jawad10a}.  While the first measurements were performed
with the electric field perpendicular to the BEDT-TTF layers
\cite{Abdel-Jawad10a}, the effect was confirmed for fields parallel to
the layers \cite{Pinteric14a}.  A similar dielectric response is seen
in $\kappa$-Cl \cite{Lunkenheimer12a,Lang14a} (see
\cite{Lunkenheimer15a} for a review).  In the case of $\kappa$-Cl, the
ferroelectric ordering coexists with the AFM state, giving rise to the
suggestion that $\kappa$-Cl is an example of a multiferroic
\cite{Lunkenheimer12a}.  Microwave measurement of the in-plane
dielectric constant of $\kappa$-CN also confirms the activity of
charge degrees of freedom \cite{Poirier12a}. Below 6 K a peak in the
microwave dielectric function is seen, which is suppressed by a
magnetic field \cite{Poirier12a}.  Significant sample preparation,
strain, and thermal effects were noted however.

The suggestion that fluctuating charge order is present in
$\kappa$-(ET)$_2$X has led to a renewed interest in examining
signatures of CO in these materials. No signatures of static CO are
seen by NMR \cite{Miyagawa04a,Kanoda11a}. Optical spectroscopy with
light polarized perpendicular to the BEDT-TTF layers has a much
smaller background from the conduction electrons in the planes and is
therefore particularly sensitive to charge disproportionation on the
molecules.  No indications of CO to within $\Delta n = 0.01 e$ were seen in
the optical spectra down to 12 K for $\kappa$-Br, $\kappa$-Cl, and
$\kappa$-NCS \cite{Sedlmeier12a}. Fluctuating CO with a timescale
slower than 10$^{-11}$s was excluded by \cite{Sedlmeier12a}. Other
optical measurements do give evidence for a fluctuating CO
in $\kappa$-CN. Raman scattering found line broadening of the
charge-sensitive  $\nu_2$ mode below 50 K in $\kappa$-CN \cite{Yakushi15a}.
This broadening was only seen in samples cooled slowly enough,
and also disappeared under the effect of pressure once the
material entered the metallic phase \cite{Yakushi15a}.
In pump-probe experiments on $\kappa$-CN, a response at 30 cm$^{-1}$
was interpreted as evidence for fluctuating charge order
within each dimer \cite{Itoh13a}. This mode begins to increase
in amplitude at approximately the same temperature as the
relaxor dielectric response begins ($\sim$40 K), and steeply
increases and saturates at 6 K \cite{Itoh13a}. The
fast recovery time from the photoexcitation (2.6 ps) rules
out a structural transformation or formation of domains
as seen in $\alpha$-(ET)$_2$I$_3$, where the timescale
is orders of magnitude slower \cite{Ivek10a,Ivek11a}. This is
also much faster than the timescale of the measurements of
\cite{Abdel-Jawad10a}.

The possibility of static CO of very small amplitude in
$\kappa$-(ET)$_2$X also remains.  In the (TMTTF)$_2$X salts a
peak in the dielectric constant at the charge ordering
temperature is seen, which is of much larger relative magnitude than
in $\kappa$-(ET)$_2$X \cite{Nad06a}. Based on the
known magnitude of CO in (TMTTF)$_2$X,
$\Delta$n=0.26$e$ for X=AsF$_6$ and $\Delta$n=0.12$e$  X=PF$_6$
  \cite{Dressel12a},
and the smaller relative dielectric anomaly
in $\kappa$-(ET)$_2$X, a charge disproportionation smaller than
$\Delta$n$\sim$0.01$e$
is expected \cite{Lang14a}.

A second explanation for the unusual dielectric response is that it is
coming not from charge disproportionation within the
$\kappa$-(ET)$_2$X dimers, but from the domain walls between
microscopic insulating and metallic domains \cite{Tomic13a}.  This
picture is supported by the general tendency towards phase segregation
below $\sim$50 K seen in many $\kappa$-phase materials noted
above. The slower time scale seen in dielectric constant experiments
(of order kHz and lower frequencies) also supports the existence of
larger domains.  Various explanations have been given for the source
of disorder. One intrinsic source of disorder in many
$\kappa$-(ET)$_2$X is the Cu atom valence in the anion layer, which
can also lead to a mixture of Cu$_2$(CN)$_3$ and Cu(CN)[N(CN)$_2$]
anions \cite{Komatsu96a,Drozdova01a}.  However, similar relaxor
dielectric response is seen in both X=Cu$_2$(CN)$_3$ and
X=Ag$_2$(CN)$_3$ \cite{Shimizu16a,Pinteric16a}.  Another suggestion is
that the anion layer, and in particular the CN groups for
X=Cu$_2$(CN)$_3$ are a source of disorder
\cite{Pinteric14a,Pinteric16a}. Regarding the influence of the anion
layer, it is important to note that relaxor dielectric response very
similar to that found in $\kappa$-CN is also observed in other
dimerized CTS, for example $\beta^\prime$-(ET)$_2$ICl$_2$
\cite{Iguchi13a} and $\beta$-({\it meso}-DMBEDT-TTF)$_2$PF$_6$
\cite{Niizeki08a}.  Like $\kappa$-(ET)$_2$X,
$\beta^\prime$-(ET)$_2$ICl$_2$ is strongly dimerized and also has AFM
order. However, the lack of a complex anion layer in
$\beta^\prime$-(ET)$_2$ICl$_2$ shows that anion layer disorder is not
an essential component to the relaxor dielectric response seen in many
CTS.

\paragraph{Charge ordered $\kappa$ phase salts}

A family of $\kappa$-phase CTS that shows a variety of CO transitions
are the $\kappa$-(ET)$_2$Hg (SCN)$_{3-n}$X$_n$, with $n=1,2$ and X a
variety of anions (F, Cl, Br, I) \cite{Aldoshina93a}.  MI transitions
involving lattice and likely CO are found over a range of temperatures
from 30 to 140 K within this series
\cite{Aldoshina93a,Yudanova95a,Yudanova96a,Yasin12a,Drichko14a,Lohle17a}.
The lower transition temperatures suggest the possibility that the CO
pattern has unequal charges {\it within} each ET dimer.  In
$\kappa$-(ET)$_2$Hg(SCN)$_{2}$Br, the MI transition occurs at 100 K;
for a deuterated analog (D$_8$-ET)$_2$Hg(SCN)$_{2}$Br the transition
temperature is lower (60 K) \cite{Yudanova95a}.  Below the transition
the susceptibility in both materials decreases regardless of the
crystal orientation indicating a spin-gapped ground state. The
susceptibility can be fitted by the form usually assumed for a SP gap,
$\chi=C/T\exp(\Delta/T)$, with $\Delta\approx$ 496 and 340 K for the
protonated and deuterated Br salts, respectively \cite{Yudanova95a}.
We argue in Section \ref{pec2dcts} that spin gap accompanying CO is a
signature that the charge-ordering does not correspond to a simple WC.
In $\kappa$-(ET)$_2$Hg(SCN)$_{2}$I the resistivity decreases rapidly
below 50 K \cite{Yudanova96a}; the properties of this salt may be
complicated however by the presence of two crystallographically
distinct ET layers.

A number of experimental probes, as well as the effect of hydrostatic
pressure have been used to investigate the Hg(SCN)$_2$Cl salt, where
$T_{\rm MI} \sim$ 30 K \cite{Yasin12a,Drichko14a,Lohle17a}.  Optical
measurements find CO with
$\Delta$n=0.2$e$
below the transition
\cite{Drichko14a}.  No obvious change of the crystal structure is
visible in X-ray analysis, further suggesting that the MI transition
corresponds to pure charge ordering
with unequal charges within each BEDT-TTF dimer
\cite{Drichko14a}.
While
the MI transition and presence of CO is now well-confirmed in this
material, the properties at lower temperatures are less certain.
Slightly below $T_{\rm MI}$ a kink was observed in the resistivity at
27-28 K, together with a small decrease in the susceptibility which
might indicate tendency towards a spin-gapped ground state
\cite{Yasin12a}.  However, at lower temperatures an increase in the
susceptibility, as well as ESR measurements suggested that the ground
state has AFM order \cite{Yasin12a}.  While many samples have
displayed similar anomalies in $\rho(T)$ below $T_{\rm MI}$ which are
reproducible under repeated heating and cooling, details of the
properties are sample dependent \cite{Lohle17a}. This possibly
suggests that there are several competing magnetic states. Under
pressure, CO is suppressed, but to date no SC has been detected down
to 1.5 K \cite{Lohle17a}.

In some families of $\kappa$-(ET)$_2$X CO occurs at relatively high
temperatures, with the CO pattern consisting of dimers of neutral
BEDT-TTF$^0$ and dimers of BEDT-TTF$^{+1}$ \cite{Ota07a,Lapinski13a},
analogous to the CO pattern BCDW-II in 1D (see
Fig.~\ref{1dcartoons}(f)).  Such a CO state with two holes on a {\it
  single} dimer has a large charge gap and would be expected to remain
an insulator at low temperatures, as is also found for quasi-1D
BCDW-II CTS \cite{Clay12a}.  Materials with this CO state include
$\kappa$-[Et$_4$N](ET)$_4$M(CN)$_6\cdot$ 3 H$_2$O with
M=Co$^{\rm{III}}$, Fe$^{\rm{ III}}$, and Cr$^{{\rm III}}$
\cite{Magueres96a,Magueres97a,Swietlik01a,Swietlik03a,Swietlik04a},
and $\kappa$-(ET)$_4$- [M(CN)$_6$][N(C$_2$H$_5$)$_4$]$\cdot$2H$_2$O
with M=Co$^{\rm{III}}$ and Fe$^{\rm{ III}}$
\cite{Swietlik06a,Ota07a,Lapinski13a}.  To our knowledge SC has not
been observed in any of these $\kappa$-(ET)$_2$X.

\paragraph{Recent developments}
Here we highlight several $\kappa$-(ET)$_2$X CTS that have been of
interest recently, X=Ag$_2$(CN)$_3$, B(CN)$_4$, and CF$_3$SO$_3$.

The properties of $\kappa$-(ET)$_2$Ag$_2$(CN)$_3$ are quite similar to
$\kappa$-(ET)$_2$Cu$_2$(CN)$_3$: at ambient pressure Ag$_2$(CN)$_3$
displays QSL-like properties with no apparent magnetic order to low
temperatures \cite{Shimizu16a,Pinteric16a}.  Compared to
Cu$_2$(CN)$_3$, Ag$_2$(CN)$_3$ is believed to be slightly closer to an
isotropic triangular dimer lattice \cite{Shimizu16a,Pinteric16a}.  One
experimental advantage of Ag$_2$(CN)$_3$ over Cu$_2$(CN)$_3$ is the
absence of Cu valence impurity difficulties as Ag$^{2+}$ is unknown.  Under pressure
Ag$_2$(CN)$_3$ is also superconducting with T$_{\rm c}$=5.2 K, with a
slightly higher pressure required compared to Cu$_2$(CN)$_3$
\cite{Shimizu16a}.

X=B(CN)$_4$ and CF$_3$SO$_3$ are unique compared to most previously
studied $\kappa$ CTS because their effective dimer frustration $t^\prime/t$
is significantly {\it larger} that 1 \cite{Fettouhi95a,Yoshida15a,Ito16a}.
For B(CN)$_4$ $t^\prime/t \sim$ 1.44 at room temperature \cite{Yoshida15a}.
B(CN)$_4$ undergoes a second-order transition to a spin-gapped state at 5 K; the
large ratio of 2$\Delta$/T$_{\rm c}$ = 10.6 is inconsistent with a weak-coupling
mechanism for the gap \cite{Yoshida15a}. While pressure does reduce the gap,
this material remains semiconducting with no SC up to 2.5 GPa \cite{Yoshida15a}.
It is unknown whether any CO coexists with the SG at low temperatures as Raman
measurements have been performed only for T$>$T$_{\rm SG}$.

For CF$_3$SO$_3$ $t^\prime/t$ is even larger, about 1.8 at room
temperature \cite{Ito16a}. At ambient pressure AFM order is found with
$T_{\rm N}=2.5$ K \cite{Fettouhi95a,Ito16a}.  The much lower N\'eel
temperature (and $t^\prime/t>1$) may indicate this AFM state has a
different spin pattern than the dimer N\'eel state found in
$\kappa$-Cl \cite{Fettouhi95a,Ito16a}.  SC is found with $T_{\rm c} = $4.8 K
at a pressure of 2.2 GPa \cite{Ito16a}. This the only known CTS
superconductor with T$_{\rm N}<$T$_{\rm c}$.

\paragraph{Summary}
The cation layer in $\kappa$-(ET)$_2$X is characterized by strongly
dimerized anisotropic triangular lattices with one hole per
dimer. Relatively few have antiferromagnetic
($\kappa$-(ET)$_2$Cu[N(CN)$_2$]Cl,
$\kappa$-d8-(ET)$_2$Cu[N(CN)$_2$]Br, and
$\kappa$-(ET)$_2$CF$_3$SO$_3$) or spin liquid
($\kappa$-(ET)$_2$Cu$_2$(CN)$_3$ and $\kappa$-(ET)$_2$Ag$_2$(CN)$_3$)
ground states while the others exhibit ambient pressure
superconductivity (SC), charge order, or spin gap.  SC is spin-singlet
and the order parameter has nodes. There is evidence for pseudogap in
$\kappa$-(ET)$_2$Cu[N(CN)$_2$]Br and $\kappa$-(ET)$_2$Cu(NCS)$_2$ at
temperatures higher than T$_c$. In $\kappa$-(ET)$_2$Cu$_2$(CN)$_3$
there occurs a transition or crossover at 6 K whose nature is not
understood.  There is evidence for intra-dimer charge-fluctuation,
relatively strong lattice effects, and a small spin gap. A spin gap
has been observed in some charge-ordered $\kappa$-(ET)$_2$X.  The
N\'eel temperature T$_{\rm N}$ is lower than the superconducting
critical temperature T$_{\rm c}$ in $\kappa$-(ET)$_2$CF$_3$SO$_3$,
which would seem to argue against spin-fluctuation driven SC. Taken
together, the observations overall also do not support an effective
$\frac{1}{2}$-filled band model.

\subsubsection{$\beta$, $\beta^\prime$, and $\beta^{\prime\prime}$ type CTS}
\label{beta-cts}

\begin{table}
  \begin{center}
    \scriptsize
    \begin{tabular}{c|c|c|c|l}
      formula & $T_{\rm c}$(K)& P$_{\rm c}$(GPa) & Notes & references \\
      \hline
      $\beta$-(ET)$_2$I$_3$ & 1.4 (8.1) & 0 (0.1) &$\beta_{\rm L}$ ($\beta_{\rm H}$) & \cite{Yagubskii84a,Murata85a} \\
      $\beta$-(ET)$_2$IBr$_2$ & 2.7 & 0 & & \cite{Williams84a} \\
      $\beta$-(ET)$_2$I$_2$Br & --- & --- & no SC to 0.5 K & \cite{Emge85a} \\
      $\beta$-(ET)$_2$AuI$_2$ & 5 & 0 & & \cite{Wang85a} \\
      $\beta$-(ET)$_2$ReO$_4$ & 2 & 0.4 & T$_{\rm MI}$=81 K & \cite{Parkin83a}\\
      $\beta$-(BDA-TTP)$_2$SbF$_6$ & 7.5 & 0 &  & \cite{Yamada01a} \\
      $\beta$-(BDA-TTP)$_2$AsF$_6$ & 5.8 & 0 &  & \cite{Yamada01a} \\
      $\beta$-(BDA-TTP)$_2$PF$_6$ & 5.9 & 0 &  & \cite{Yamada01a} \\
      $\beta$-(BDA-TTP)$_2$FeCl$_4$ & 3.0  & 0.4 & T$_{\rm MI}$=118 K, T$_{\rm N}$=8.2 K  & \cite{Yamada01b,Choi04a} \\
      $\beta$-(BDA-TTP)$_2$GaCl$_4$ & 3.1 & 0.4  & T$_{\rm MI}$=113 K  & \cite{Yamada03a} \\
      $\beta$-(BDA-TTP)$_2$I$_3$ & $\sim$10 & 1.0 &  & \cite{Yamada06a} \\
      $\beta$-({\it meso}-DMBEDT-TTF)$_2$PF$_6$ & 4.3 & 0.1 & T$_{\rm MI}$=91 K, T$_{\rm CO}$=70 K& \cite{Kimura04a,Morinaka09a} \\
      $\beta$-({\it meso}-DMBEDT-TTF)$_2$AsF$_6$ & 4.2 & 0.4 & T$_{\rm MI}$=90 K, T$_{\rm CO}$=70 K & \cite{Shikama12a} \\
      \hline
      $\beta^\prime$-(ET)$_2$ICl$_2$ & 14.2 & 8.2 & $T_{\rm N}$=22 K & \cite{Anzai87a,Taniguchi03b} \\
      $\beta^\prime$-(ET)$_2$AuCl$_2$ & --- & --- & $T_{\rm N}$=28 K & \cite{Mori87a,Taniguchi05a} \\
      $\beta^\prime$-(ET)$_2$SF$_5$CF$_2$SO$_3$ & --- & --- & T$_{\rm SG}$=T$_{\rm MI}$=45 K & \cite{Ward00a} \\
      $\beta^\prime$-(ET)$_2$CF$_3$CF$_2$SO$_3$& ---& ---& T$_{\rm SG}$=T$_{\rm MI}$=30 K & \cite{Graja09a} \\
      \hline
      $\beta^{\prime\prime}$-(ET)$_2$AuBr$_2$ & 1.1 & 1.0$^*$ & T$_{\rm CO}\sim$ 8 K, $^*$uniaxial & \cite{Mori86a,Kondo10a} \\
      $\beta^{\prime\prime}$-(ET)$_2$SF$_5$CH$_2$CF$_2$SO$_3$ & 5.2 & 0 & T$_{\rm CO}$=200 K, fluct. CO & \cite{Geiser96a} \\
      $\beta^{\prime\prime}$-(ET)$_2$SF$_5$CHFCF$_2$SO$_3$ & --- & --- & T$_{\rm MI}$=190 K & \cite{Schlueter01a} \\
      $\beta^{\prime\prime}$-(ET)$_2$SF$_5$CH$_2$SO$_3$ & --- & --- & semiconductor 300 K & \cite{Ward00a} \\
      $\beta^{\prime\prime}$-(ET)$_2$SF$_5$CHFSO$_3$ & --- & --- & metallic, fluct. CO & \cite{Ward00a} \\
      $\beta^{\prime\prime}$-(ET)$_2$CF$_3$CH$_2$SO$_3$ & --- & --- & CO semiconductor 300K & \cite{Schlueter02a,Olejniczak09a} \\
      $\beta^{\prime\prime}$-(DODHT)$_2$AsF$_6$ & $\sim$3 & 1.5 & & \cite{Nishikawa02a} \\
      $\beta^{\prime\prime}$-(DODHT)$_2$PF$_6$ & 2.3 & 1.3 & T$_{\rm MI}$=255 K & \cite{Nishikawa02a,Nishikawa05a} \\
      $\beta^{\prime\prime}$-(DODHT)$_2$BF$_4\cdot$H$_2$O & $\sim$3 & 1.5 & & \cite{Nishikawa05a} \\
      $\beta^{\prime\prime}$-(ET)$_4$ [(H$_3$O)Ga(C$_2$O$_4$)$_3$](C$_5$H$_5$N) & $\sim$2 & 0 & incomplete SC & \cite{Akutsu02a} \\
      $\beta^{\prime\prime}$-(ET)$_4$ [(H$_3$O)Ga(C$_2$O$_4$)$_3$](C$_6$H$_5$NO$_2$) & 7.5 & 0 &  & \cite{Akutsu02a} \\
      $\beta^{\prime\prime}$-(ET)$_4$Pt(CN)$_4$H$_2$O & 2.0 & 0.65 &  & \cite{Mori91a} \\
      $\beta^{\prime\prime}$-(ET)$_4$Pd(CN)$_4$H$_2$O & 1.2 & 0.70 &  & \cite{Mori92a} \\
      $\beta^{\prime\prime}$-(ET)$_3$Cl$_2\cdot$2H$_2$O & 2-3 & 1.1 &T$_{\rm MI}\sim$130 K & \cite{Mori87b,Lubczynski96a} \\
      
    \end{tabular}
  \end{center}
  \caption{Summary of the properties of $\beta$, $\beta^\prime$, and
    $\beta^{\prime\prime}$ salts discussed in this section.}
  \label{beta-table}
\end{table}  

\begin{figure}[tb]
   \center{\resizebox{5.0in}{!}{\includegraphics{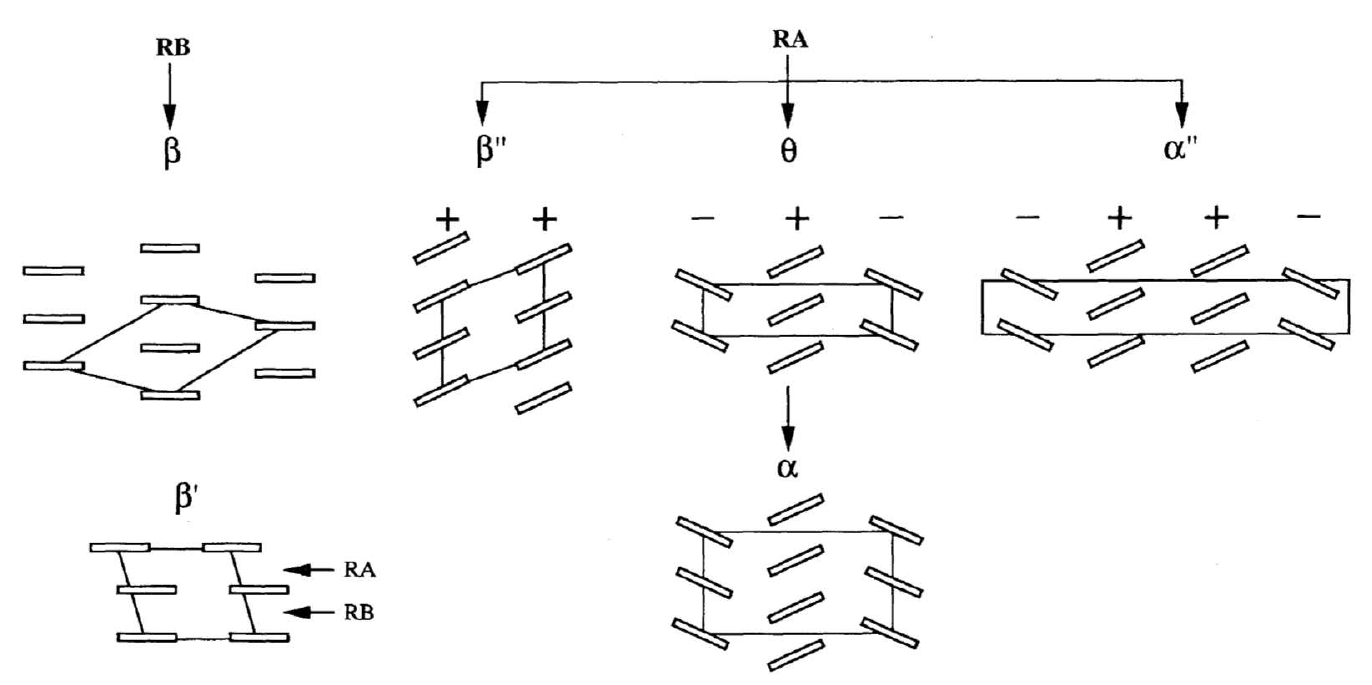}}}
  \caption{Comparison of $\alpha$, $\beta$, and $\theta$
    structures. Ring-over-atom (RA) and ring-over-bond (RB) denote two
    different kinds of inter-molecular overlaps.  Reprinted with
    permission from Ref.~\cite{Mori98c}, $\copyright$ 1998 The
    Chemical Society of Japan.}
  \label{alpha-beta-theta-structures}
\end{figure}

The $\alpha$, $\beta$, $\beta^\prime$, $\beta^{\prime\prime}$, and
$\theta$ type CTS all have structures composed of stacks of BEDT-TTF
molecules forming 2D layers (see
Fig.~\ref{alpha-beta-theta-structures}).  In the $\beta$,
$\beta^\prime$, and $\beta^{\prime\prime}$ structures, the molecules
are all parallel to each other, while in $\theta$ and $\alpha$ (see
Sections \ref{thetasection} and \ref{alpha}), molecules in adjacent
stacks are inclined at an angle to each other (see
Fig.~\ref{alpha-beta-theta-structures}) \cite{Mori98c}.  The
differences between the $\beta$, $\beta^\prime$, and
$\beta^{\prime\prime}$ structures arise from different relative
orientations and regularity of the molecules in the stacks.  The
appropriate tight-binding lattice for all of these structures is
rectangular with some amount of frustration (hopping in the diagonal
directions).  Because of the angle as well as the relative position
between two neighboring molecules strongly influences the
inter-molecular transfer integrals between them, the $t_{ij}$ for
these families can be significantly different from each other.

A large number of $\beta$-type CTS are superconducting (see Table
\ref{beta-table}).  Many undergo a MI transition at ambient pressure
to a charge- and/or bond-ordered insulating state and are
superconducting under the application of pressure. There are also many
ambient-pressure superconductors.  Antiferromagnetic ground states are
found in the $\beta^\prime$ structure, as well as in some $\beta$
structure CTS containing magnetic ions such as Fe (see below).
Besides BEDT-TTF, superconductors in these structures have been
synthesized with BEDO-TTF, DMET, {\it meso}-DMBEDT-TTF, BDA-TTP, and
other related molecules \cite{Mori06a}.

In the $\beta$ structures the molecular stacks are dimerized with
varying degrees of dimerization.  In the $\beta^\prime$ structure the
stacks are slightly staggered (see
Fig.~\ref{alpha-beta-theta-structures}), which leads to a larger
intra- than inter-stack hopping. Because of this, the
$\beta^\prime$-structure CTS are considered more one-dimensional than
the $\beta$-structure CTS \cite{Mori98c}. On the other hand, in the
$\beta^{\prime\prime}$ structure \cite{Mori86a,Ugawa86a}, the
inter-stack overlaps are larger than the intra-stack overlaps. For
this reason, $\beta^{\prime\prime}$ materials are usually considered
more 2D than $\beta$ or $\beta^\prime$.

\paragraph{$\beta$ CTS}

The $\beta$ structure CTS were the first class of organic
superconductors discovered after the TMTSF/TMTTF group of materials
\cite{Parkin83a}.  Among $\beta$-(ET)$_2$X several ambient-pressure
superconductors are found, including X=I$_3$ \cite{Yagubskii84a},
IBr$_2$ \cite{Williams84a}, and AuI$_2$ \cite{Wang85a}.  X=I$_2$Br is
not superconducting however, due to disorder in the anions
\cite{Emge85a,Schulz86a}.  X=I$_3$ is superconducting at ambient
pressure with a relatively low T$_{\rm c}$ of about 2 K
\cite{Yagubskii84a}, but a higher T$_{\rm c}$ phase (T$_{\rm c}$=8.1
K) appears under pressure \cite{Murata85a}.  X=ReO$_4$ at ambient
pressure undergoes a first-order MI transition at T$_{\rm{MI}}$ = 81 K
\cite{Parkin83a}.  The insulating state is accompanied with a period-4
lattice distortion along the stack \cite{Ravy86a}, and is non-magnetic
with a large drop in the ESR spin susceptibility accompanying the
transition \cite{Carneiro84a}. Optical studies estimated the gap at
2$\Delta\approx$ 480 cm$^{-1}$ and confirmed that the low-temperature
state is not a result of anion ordering \cite{Baker99a}.  Under a
pressure of 0.4 GPa X=ReO$_4$ becomes superconducting with $T_{\rm
  c}\sim$ 2 K \cite{Parkin83a}.

\begin{figure}
  \center{
    \begin{overpic}[width=2.0in]{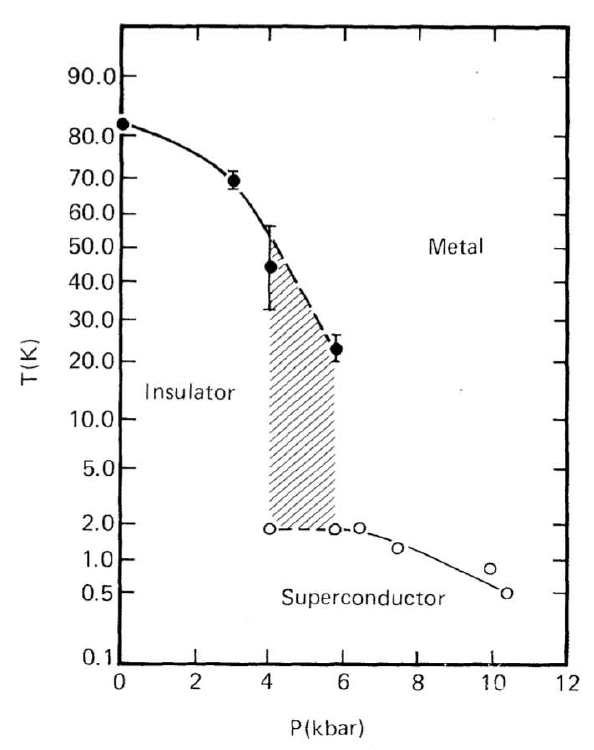}
      \put(1,95) {\small(a)}
    \end{overpic}
    \hspace{0.1in}
    \raisebox{0.2in}{
      \begin{overpic}[width=2.5in]{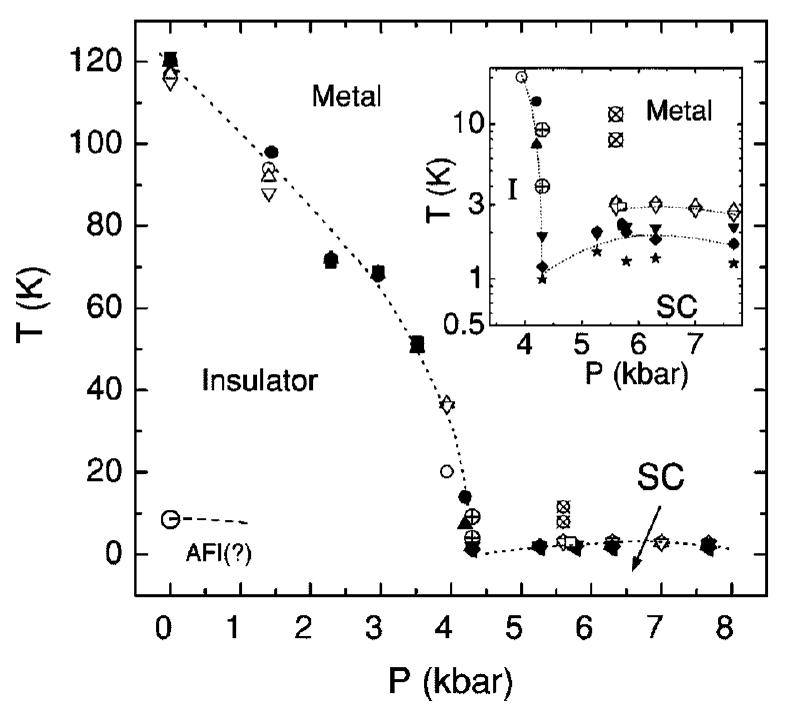}
        \put (5,90) {\small(b)}
      \end{overpic}
    }
  }
  \caption{(a) Phase diagram of $\beta$-(ET)$_2$ReO$_4$.  Thermal
    hysteresis was observed in the hatched area.  Reprinted with
    permission from Ref.~\cite{Parkin83a}, $\copyright$ 1983 The
    American Physical Society.  (b) Phase diagram of
    $\beta$-(BDA-TTP)$_2${\it M}Cl$_4$, {\it M}=Fe, Ga. Solid symbols
    are for {\it M}=Fe and open symbols for {\it M}=Ga Reprinted with
    permission from Ref.~\cite{Choi04a}, $\copyright$ 2004 The
    American Physical Society.}
  \label{beta-bda-ttp}
\end{figure}  
\begin{figure}
  \begin{center}
    \raisebox{0.5in}{
    \begin{overpic}[width=2.0in]{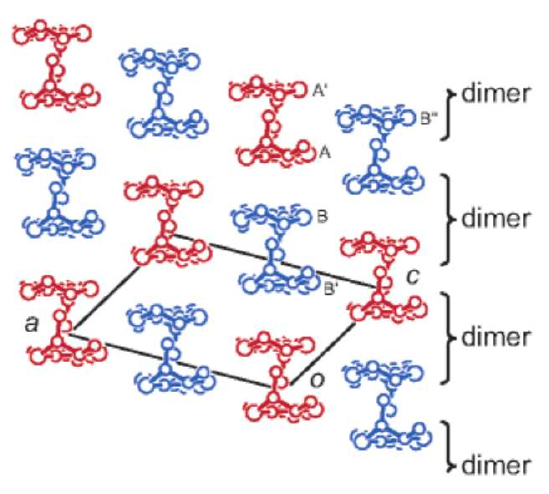}
      \put(-8,85) {\small (a)}
    \end{overpic}
    }
    \hspace{0.1in}%
    \begin{overpic}[width=2.5in]{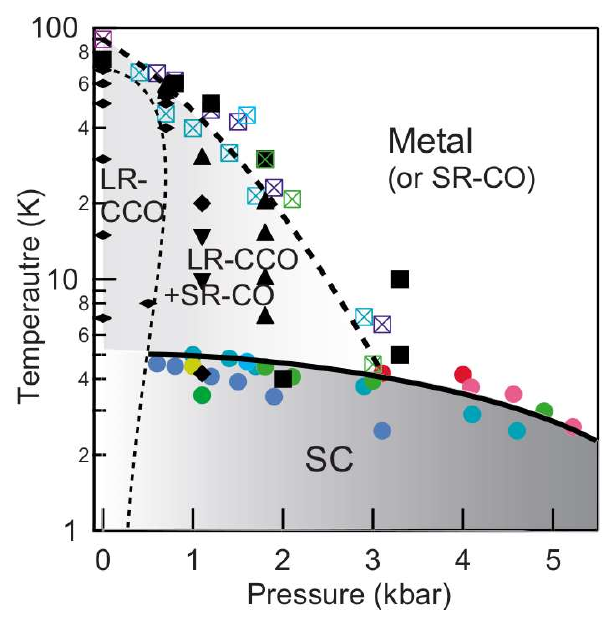}
      \put(-5,95) {\small (b)}
    \end{overpic}
  \end{center}
  \caption{(a) Charge order state found in $\beta$-({\it
      meso}-DMBEDT-TTF)$_2$PF$_6$.  Reprinted with permission from
    Ref.~\cite{Kimura06a}. $\copyright$ 2006 The American Chemical
    Society.  (b) Phase diagram.  Long-range charge order (LR-CCO)
    evolves into short range order charge order (SR-CO) and SC under
    pressure.  Reprinted with permission from
    Ref.~\cite{Morinaka09a}. $\copyright$ 2009 The American Physical
    Society.}
  \label{beta-meso}
\end{figure}  

Several superconductors as well as non-magnetic insulating states are
found in the series $\beta$-(BDA-TTP)$_2$X \cite{Yamada04a,Yamada04b}.
While these have the same $\beta$ structure as $\beta$-(ET)$_2$I$_3$,
the inter-molecular overlap is weaker, with overlap integrals roughly
half those of $\beta$-(ET)$_2$I$_3$ \cite{Yamada01a}.  Here X=SbF$_6$,
AsF$_6$, and PF$_6$ are ambient pressure superconductors with T$_{\rm
  c}$'s in the 6-7 K range \cite{Yamada01a}. X=GaCl$_4$, FeCl$_4$, and
I$_3$ are superconducting under pressure with T$_{\rm c}\approx$ 3 K,
with I$_3$ having a higher T$_{\rm c}\approx$ 8 K
\cite{Yamada01b,Yamada03a,Yamada06a}.

At ambient pressure, X=GaCl$_4$ and FeCl$_4$ undergo MI transitions at
118 K and 113 K respectively \cite{Yamada01b,Yamada03a}.  FeCl$_4$
undergoes an AFM transition at 8.5 K which is absent in the
isostructural GaCl$_4$, suggesting that the transition involves the Fe
spins \cite{Yamada01b,Yamada03a}.  In GaCl$_4$ the ambient-pressure
state at low temperatures is a nonmagnetic insulator: the ESR magnetic
susceptibility drops sharply by almost two orders of magnitude at
T$_{\rm MI}$, indicating a first-order transition
\cite{Tokumoto08a}. Raman measurements are consistent with a lowered
lattice symmetry at the transition, likely involving bond and/or
charge order \cite{Tokumoto08a}.  X-ray
structural analysis found superlattice reflections below T$_{\rm MI}$
similar to those found in $\alpha$-(ET)$_2$I$_3$
indicating a doubling of the periodicity of the BDA-TTP stacks (to
period 4) \cite{Sasamori13a,Tokumoto08a}.

$\beta$-({\it meso}-DMBEDT-TTF)$_2$X (X=AsF$_6$ and PF$_6$) is another
$\beta$-phase CTS with CO adjacent to SC (see Fig.~\ref{beta-meso})
where the CO state has been studied in detail \cite{Shikama12a}.  The
properties of AsF$_6$ and PF$_6$ are very similar \cite{Shikama12a}.
PF$_6$ undergoes a MI transition at about 90 K, and SC appears under
pressure with a T$_{\rm c}$=4.3 K \cite{Kimura04a}.  CO has been
confirmed via optical methods, with an estimated charge
disproportionation of $\Delta$n$\sim 0.5$ for X=PF$_6$ under ambient
pressure \cite{Tanaka08a}. While the pattern of CO has been described
as a ``checkerboard'', this would be true only if {\it pairs} of
charge-rich and charge-poor molecules are considered as individual
sites; considering the individual {\it molecules} as sites, the CO
follows a clear $\cdots$1100$\cdots$ pattern along the DMBEDT-TTF
stacks as shown in Fig.~\ref{beta-meso}(a) \cite{Kimura06a}.  In
infrared reflectivity and Raman measurements on PF$_6$, CO was found
in the insulating state below 70 K at ambient pressure
\cite{Tanaka08a,Morinaka09a,Okazaki13a}.  Under pressure, experiments
noted the formation of metallic regions of short-range CO (see
Fig.~\ref{beta-meso}).  The sequence of MI and CO transitions appears
similar to $\theta$-(ET)$_2$RbZn(SCN)$_4$ (see Section
\ref{thetasection}). First, in the 70-100 K region X-ray superlattice
reflections appear, along with weak diffuse scattering above 100 K
\cite{Morinaka09a}.  This state was interpreted as a dimer-Mott
insulator (without CO) \cite{Morinaka09a,Okazaki13a}.  Following the
lattice distortion and MI transition, CO appears at a lower
temperature, approximately 70 K. At the same time as CO appears the
susceptibility drops, indicating a non-magnetic state
\cite{Shikama12a}.  In spatially resolved optical measurements, phase
segregation in micron-sized domains was seen between the CO state and
the dimer-Mott state, with the volume fraction of CO increasing below
70 K as the temperature was lowered \cite{Okazaki13a}.

\paragraph{$\beta^\prime$ CTS}

\begin{figure}[tb]
  \center{\resizebox{2.75in}{!}{\includegraphics{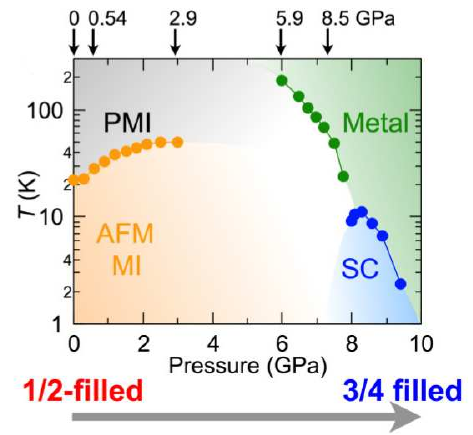}}\hspace{0.1in}}
  \caption{Phase diagram of the $\beta^{\prime}$-(ET)$_2$ICl$_2$.
    Reprinted with permission from
    Ref. \cite{Hashimoto15a}. $\copyright$ 2015 The American Physical
    Society.}
  \label{beta-prime-icl2}
\end{figure}

Among the $\beta^\prime$ CTS no ambient pressure superconductors are
known, consistent with their more quasi-1D structures.  However, under
pressure $\beta^\prime$-(ET)$_2$ICl$_2$ \cite{Anzai87a} currently has
the highest T$_{\rm c}$=14.2 K of all CTS superconductors but is only
superconducting under very high pressures (8.2 GPa) requiring a diamond anvil
pressure cell (see Fig.~\ref{beta-prime-icl2}) \cite{Taniguchi03b}.
At ambient pressure $\beta^\prime$-(ET)$_2$ICl$_2$ is a semiconductor
at room temperature. The ground state at ambient pressure is AFM with
T$_{\rm N}$=22 K \cite{Tokumoto87a}. The related salt
$\beta^\prime$-(ET)$_2$AuCl$_2$ has a slightly higher T$_{\rm N}$=28 K
with however no SC reported in pressures up to 9.9 GPa
\cite{Mori87a,Taniguchi05a}.  The AFM moment in the AFM state of
ICl$_2$ is large, approximately 1.0 $\mu_B$ per dimer \cite{Eto10a}.
Under a pressure of approximately 0.6 GPa the magnetic state however
changes \cite{Sato06a,Sato09a,Eto10a}.
both in moment (which decreases above 0.6 GPa), and N\'eel temperature
(which increases above 0.6 GPa) \cite{Sato09a,Eto10a}. At the same
time the spin susceptibility decreases with pressure
\cite{Eto10a}. These changes likely imply a change in the magnetic AFM
ordering pattern. Like the insulating $\kappa$-phase CTS,
$\beta^\prime$-(ET)$_2$ICl$_2$ displays a very similar
frequency-dependent relaxor dielectric response \cite{Iguchi13a}.
Under the application of a  static electric field at 15 K, two additional peaks
on both sides of the charge-sensitive $\nu_2$ Raman mode
were observed, indicating field-induced CO within each dimer
\cite{Hattori17a}.

Due to the very high pressure required, few experimental results are available
in the pressure region approaching superconductivity. In optical
measurements, under pressure, intra-dimer charge transfer features
shift to much lower energy \cite{Hashimoto15a}. This was interpreted
as a change of the effective electronic state of the material from an
effectively $\frac{1}{2}$-filled dimer-Mott state at ambient pressure
to a $\frac{3}{4}$-filled correlated insulator with a tendency to CO
(see Fig.~\ref{beta-prime-icl2}) \cite{Hashimoto15a}.

While $\beta^\prime$-(ET)$_2$ICl$_2$ is AFM at ambient pressure, some
$\beta^\prime$ are nonmagnetic insulators under ambient pressure.
$\beta^\prime$-(ET)$_2$SF$_5$CF$_2$SO$_3$ (not to be confused with the
$\beta^{\prime\prime}$ salts with similar anions--see below) enters a
spin-gapped state below 45 K \cite{Ward00a}. In this salt the
susceptibility increases slightly with decreasing temperature, with a
broad peak at around 150 K \cite{Ward00a}. A steep drop in the
susceptibility for all orientations of the applied magnetic field
indicates an SP-like spin-gapped state \cite{Ward00a}.  The similar
$\beta^\prime$-(ET)$_2$CF$_3$CF$_2$SO$_3$ also has a spin-gap
transition at 30 K \cite{Graja09a}.  For this salt Raman investigation
did find evidence for symmetry breaking caused by lattice distortion
in the SP phase \cite{Graja09a}. CO however was not detected in
$\beta^\prime$-(ET)$_2$CF$_3$CF$_2$SO$_3$, although CO with molecular
charges of 0.6 and 0.4 was found in a different structural variant
($\delta^\prime$ structure) of the same salt \cite{Graja09a}.

\paragraph{$\beta^{\prime\prime}$ CTS}

The first $\beta^{\prime\prime}$ salt discovered was
$\beta^{\prime\prime}$-(ET)$_2$AuBr$_2$ \cite{Mori86a}.
$\beta^{\prime\prime}$-(ET)$_2$AuBr$_2$ stays metallic until low
temperatures. SC has not been found under isotropic pressure, but a
$T_{\rm c}$ of approximately 1 K was found under conditions of uniaxial
strain \cite{Kurmoo87a,Kondo10a}.  AuBr$_2$ has some unusual properties at
low temperatures that could indicate a weak CO state: At
ambient pressure there is a kink in the resistivity at 110 K followed
by a slight upturn at around 8 K that is suppressed by pressure
\cite{Kurmoo87a}.  X-ray superlattice reflections are observed at low
temperatures, indicating bond ordering below 6.5 K \cite{Kondo10a};
optical studies are consistent with a gapped state coexisting with
free carriers \cite{Ugawa95a}.  There is also a steep decrease in the
susceptibility at low temperatures indicating a non-magnetic ground
state \cite{Kurmoo87a}.
\begin{figure}
  \center{\resizebox{5.0in}{!}{\includegraphics{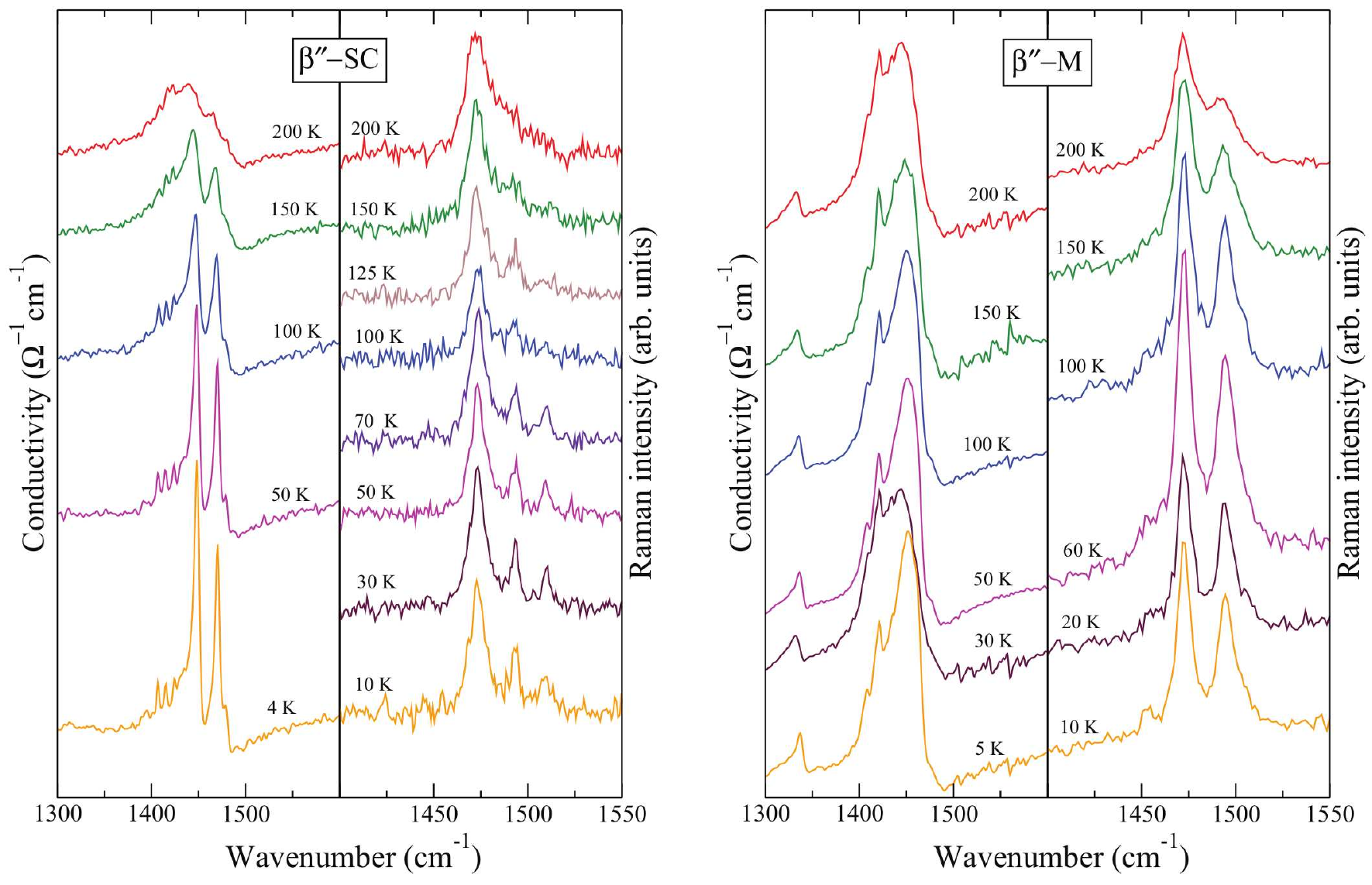}}}
  \caption{ Optical conductivity and Raman spectra as a function of
    temperature for the $\beta^{\prime\prime}$-(ET)$_2$SF$_5$R family,
    R=CH$_2$CF$_2$SO$_3$ ($\beta^{\prime\prime}$-SC) and R=CHFSO$_3$
    ($\beta^{\prime\prime}$-M).  The emergence of the doublet in the
    optical conductivity below about 150 K indicates the presence of
    charge order in $\beta^{\prime\prime}$-SC but not in
    $\beta^{\prime\prime}$-M.  Reprinted with permission from
    Ref. \cite{Girlando14a}. $\copyright$ 2014 The American Physical
    Society.}
  \label{beta-double-prime-optics}
\end{figure}

$\beta^{\prime\prime}$-(ET)$_2$SF$_5${\it R} were the first CTS
superconductors discovered with both cations and anions organic
\cite{Geiser96a,Ward00a,Schlueter01a}.  A summary of the different
known {\it R} and structures is given in reference \cite{Geiser03a} (note
that some also crystallize in  structures different from $\beta^{\prime\prime}$).
Recent investigations into this series have noted
the presence of
SC, CO, as well as CO fluctuations.

In this series of salts {\it R}=CH$_2$CF$_2$SO$_3$
($\beta^{\prime\prime}$-SC) is the only known superconductor, with
T$_{\rm c}=$ 5.2 K under ambient pressure \cite{Geiser96a}.  In this
series the cation layer is composed of two inequivalent dimerized
stacks of BEDT-TTF molecules along the $a$ axis.  Differences in the
relative distance of these two BEDT-TTF stacks to SO$_3^-$ and SF$_5$
groups in the anion layer lead to a slight charge imbalance between
the two stacks even at 300 K, with a charge density of
0.5$+\Delta n/2$ on one stack and 0.5-$\Delta n/2$ on the other
\cite{Ward00a,Schlueter01a}.  In the superconducting material the
inequivalence between the two stacks decreases at lower temperatures
\cite{Schlueter01a}.

Other members of the family include {\it R}=CHFCF$_2$SO$_3$
($\beta^{\prime\prime}$-MI), which undergoes a first-order MI
transition at 190 K, below which the stack dimerization increases
greatly \cite{Schlueter01a}.  {\it R}=CH$_2$SO$_3$
($\beta^{\prime\prime}$-I) is a charge-ordered semiconductor already
at room temperature \cite{Ward00a}, with the degree of CO estimated as
$\Delta$n $\sim$ 0.4
from optical measurements \cite{Olejniczak09a}.  The {\it
  R}=CHFSO$_3$ ($\beta^{\prime\prime}$-M) remains metallic to low
temperature \cite{Ward00a}.
$\beta^{\prime\prime}$-(ET)$_2$CF$_3$CH$_2$SO$_3$ is another salt that
is semiconducting and charge ordered at 300 K, with the CO amplitude
estimated as
$\Delta$n $\sim$ 0.6
\cite{Schlueter02a,Olejniczak09a}.

 Optical investigations have provided evidence for CO and charge
fluctuations in this series of salts \cite{Olejniczak09a,Kaiser10a,Girlando14a}.  In
$\beta^{\prime\prime}$-SC the $\nu_{27}$ IR mode splits below 200 K
without any structural change \cite{Kaiser10a}. Compared to
$\beta^{\prime\prime}$-M, in $\beta^{\prime\prime}$-SC at low
temperatures a pronounced optical conductivity in the mid-IR (300
$\sim$ 1000 cm$^{-1}$) develops \cite{Kaiser10a} (see
Fig.~\ref{beta-double-prime-optics}).  A strong signal at $\sim$40
cm$^{-1}$, related to an inter-site phonon mode, also develops at low
temperature in the superconducting salt \cite{Kaiser10a}. These
features were interpreted as the formation of a pseudogap, and
evidence for the importance of lattice coupling in the low-temperature
electronic state \cite{Kaiser10a}. Raman experiments have given more
details (see Fig.~\ref{beta-double-prime-optics})
\cite{Girlando14a}. In Raman experiments on $\beta^{\prime\prime}$-SC,
the totally-symmetric charge coupled mode splits at low temperature
\cite{Girlando14a}. This response was well fit by a two-state ``jump''
model of charge fluctuations within a dimer
\cite{Kubo69a,Girlando12b}. The authors concluded that (i) charge
fluctuations are present in both $\beta^{\prime\prime}$-SC and
$\beta^{\prime\prime}$-M which are related to the fluctuation of
charge between the two inequivalent BEDT-TTF stacks; (ii) in
$\beta^{\prime\prime}$-SC (but not $\beta^{\prime\prime}$-M), a second
static coexisting CO develops at low temperature, which is instead
characterized by an additional charge disproportionation {\it within}
each of the BEDT-TTF stacks (see Fig.~\ref{beta-double-prime-optics}).

$\beta^{\prime\prime}$-(DODHT)$_2$X has smaller intermolecular
interactions along the stacks compared to
$\beta^{\prime\prime}$-(ET)$_2$AuBr$_2$ \cite{Nishikawa02a}.  At
ambient pressure and room temperature X=AsF$_6$, PF$_6$, TaF$_6$, and
BF$_4\cdot$H$_2$O are semiconductors, and X=AsF$_6$, PF$_6$, and
BF$_4\cdot$H$_2$O become superconducting with T$_{\rm c}\approx$ 3 K
under $\sim$ 1.5 GPa pressure
\cite{Nishikawa02a,Nishikawa03a,Nishikawa05a,Nishikawa06a,Nishikawa08a}.
At ambient pressure in X=PF$_6$, the resistivity starts to increase
steeply below 255 K indicating a transition to an insulating state
\cite{Nishikawa05a}.  The low temperature state is a non-magnetic
insulator.  X-ray satellite reflections indicate structural changes at
low temperature and a charge-ordered ground state has been suggested
\cite{Nishikawa05a,Higa07a}; however to our knowledge direct
charge-sensitive measurements have not yet been performed on this
series of CTS.

\begin{figure}[tb]
  \center{
    \raisebox{0.25in}{
    \begin{overpic}[width=2.5in]{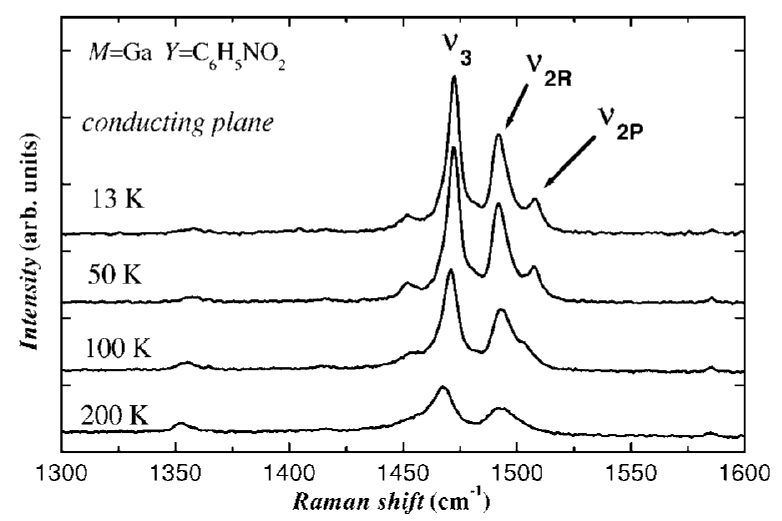}
      \put (-5,60) {\small(a)}
    \end{overpic}
    }
    \hspace{0.1in}
      \begin{overpic}[width=2.5in]{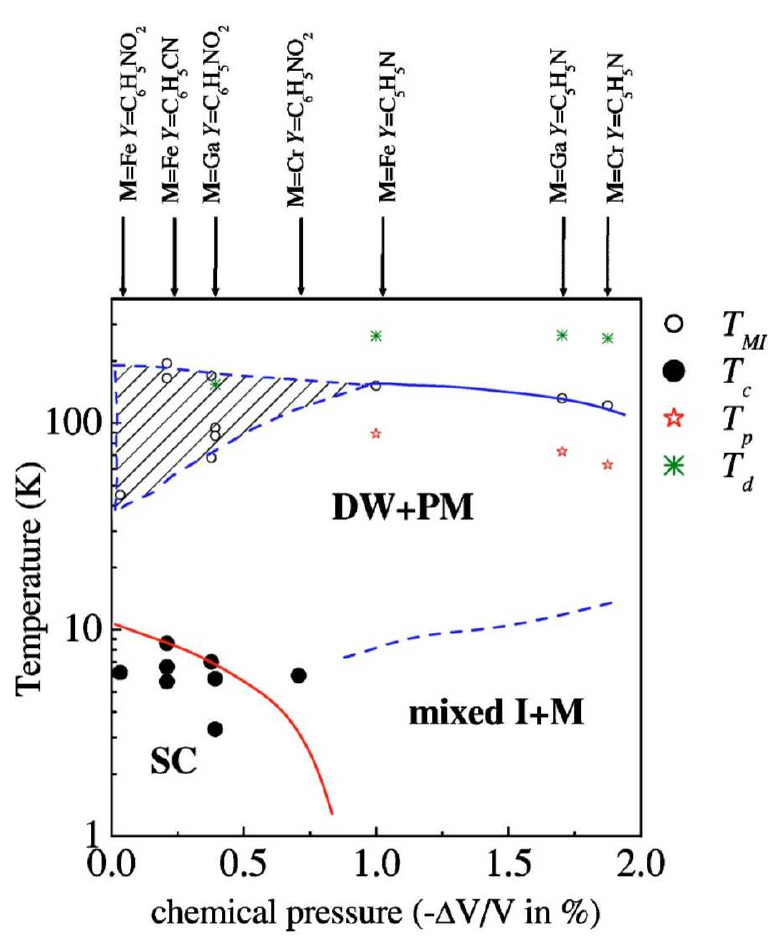}
        \put (0,65) {\small(b)}
      \end{overpic}
  }
  \caption{(a) Raman spectra of
    $\beta^{\prime\prime}$-(ET)$_4$[(H$_3$O){\it M}(C$_2$O$_4$)$_3$]
    $\cdot${\it Y}, {\it M}=Ga, {\it Y}=C$_6$H$_5$NO$_2$. The
    splitting of the $\nu_2$ mode below $\sim$100 K indicates charge
    ordering.  Reprinted with permission from
    Ref. \cite{Bangura05a}. $\copyright$ 2005 The American Physical
    Society.  (b) Phase diagram of this family of CTS. Here $\Delta
    V/V$ is the fractional change of the unit cell volume compared to
    {\it M}=Fe.  $T_p$ and $T_d$ are features observed in resistivity
    measurements. I, M, DW, and PM are insulating, metallic, density
    wave, and paramagnetic metallic phases, respectively.  Reprinted
    with permission from Ref. \cite{Coldea04a}. $\copyright$ 2004 The
    American Physical Society.}
  \label{Bangura05a-figs}
\end{figure}  

Some other superconducting $\beta^{\prime\prime}$ CTS include
$\beta^{\prime\prime}$-(ET)$_4$ [(H$_3$O){\it M}
  (C$_2$O$_4$)$_3$]{\it Y}, where {\it M}=Ga, Cr, or Fe, and {\it
  Y}=pyridine (C$_2$H$_5$N) or nitrobenzene (C$_6$H$_5$NO$_2$)
\cite{Kurmoo95a,Rashid01a,Akutsu02a}. Of these, {\it M}=Ga and Cr with
     {\it Y}=C$_6$H$_5$NO$_2$ and {\it M}=Fe with {\it Y}=C$_6$H$_5$CN
     are superconductors (see Fig.~\ref{Bangura05a-figs})
     \cite{Kurmoo95a,Akutsu02a,Coldea04a,Bangura05a}.
     In this group of CTS, the resistivity decreases from room
     temperature until around 100 K, below which it rises to a peak at
     about 10 K before SC \cite{Bangura05a}. T$_{\rm c}$'s up to about
     8 K are reported \cite{Bangura05a}.  As in $\kappa$-(ET)$_2$X,
     the low temperature state of this series appears to have a
     disordered mixture of insulating and metallic phases
     \cite{Coldea04a,Bangura05a}. A very low carrier density was also
     noted \cite{Bangura05a}.  For {\it M}=Ga and {\it
       Y}=C$_6$H$_5$NO$_2$, Raman scattering spectra split below 100 K
     indicating charge ordering (see Fig.~\ref{Bangura05a-figs}(a)); the
     splitting is more pronounced in the non-superconducting salts
     \cite{Bangura05a}.  Further Raman measurements found charge
     disproportionation of amplitude $0.5\sim \Delta$n$\sim 0.13$
     \cite{Yamamoto08a}. Based on the pattern of bonds lengths, the CO
     pattern along the stacking direction was inferred to be
     $\cdots$1100$\cdots$ \cite{Yamamoto08a}.

$\beta^{\prime\prime}$-(ET)$_3$Cl$_2\cdot$H$_2$O is an interesting CTS
superconductor, as by stoichiometry it appears to be $\frac{2}{3}$-
rather than $\frac{3}{4}$-filled
\cite{Mori87b,Mori87c}. This salt has a wide MI transition at around
100-160 K, and is superconducting with a T$_{\rm c}$ of 2-3 K at
$\sim$ 1.6 GPa \cite{Mori87b,Mori87c,Lubczynski96a,Gaultier99a}.
Optics and $^{13}$C NMR measurements indicate CO below 100 K, and a
decrease in the susceptibility shows that the low-temperature state is
a non-magnetic singlet \cite{Yamamoto08a,Nagata11a}.  The unit cell
contains three independent molecules. The charges on these three sites
in the CO state was estimated as 0.4{\it e}, 0.6{\it e}, and 1.0{\it
  e} \cite{Nagata11a}. This corresponds to a single $\rho=1$ stack of
BEDT-TTF molecules, and two stacks with $\rho=\frac{1}{2}\pm\Delta n/2$.
X-ray structural results at low temperature are consistent with one
stack having $\rho\sim$1.0{\it e} charge \cite{Gaultier99a}.  NMR
measurements under pressure indicate that the spin-singlet state
remains under pressure until the SC phase is reached \cite{Nagata11a}.

\paragraph{Summary} A
charge-order-to-superconductivity transition is common in the $\beta$,
$\beta^\prime$ and $\beta^{\prime\prime}$ materials. In all cases
where the pattern of the charge order or bond distortion is known, it
appears to be different from that of the simple Wigner crystal (WC),
in that in more than one direction the charge order pattern is
$\cdots1100\cdots$.  Spin gap (SG) often accompanies the charge order
or occurs as a separate lower temperature phase. A SG is not expected
within the WC (see Section \ref{pec2d}, in particular
Fig.~\ref{pecphasediagram}).

\subsubsection{$\theta$ CTS}
\label{thetasection}

\begin{figure}[tb]
  \center{
    \raisebox{0.6in}{
      \begin{overpic}[width=2.3in]{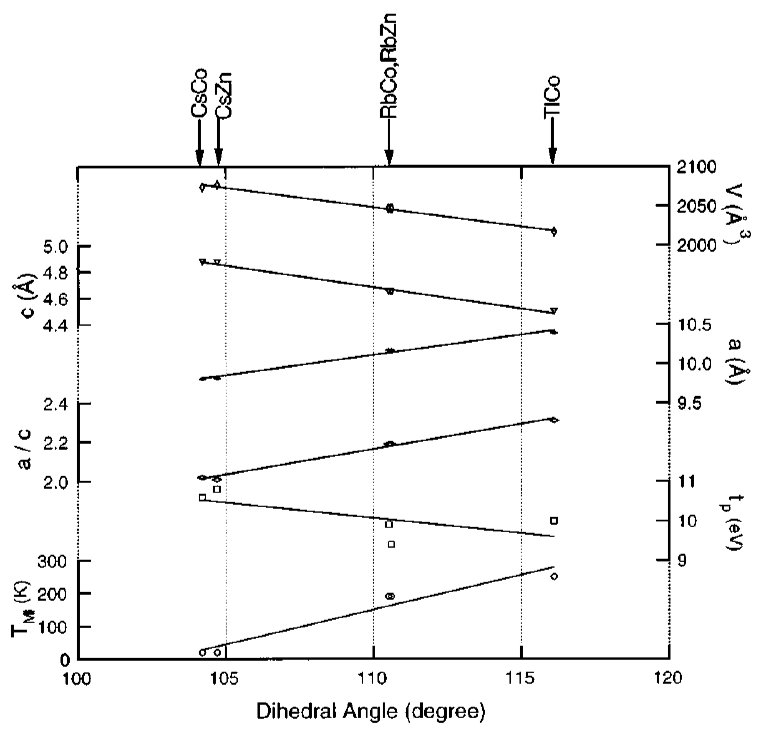}
        \put (-5,80) {\small(a)}
      \end{overpic}
    }
    \hspace{0.1in}
    \begin{overpic}[width=2.2in]{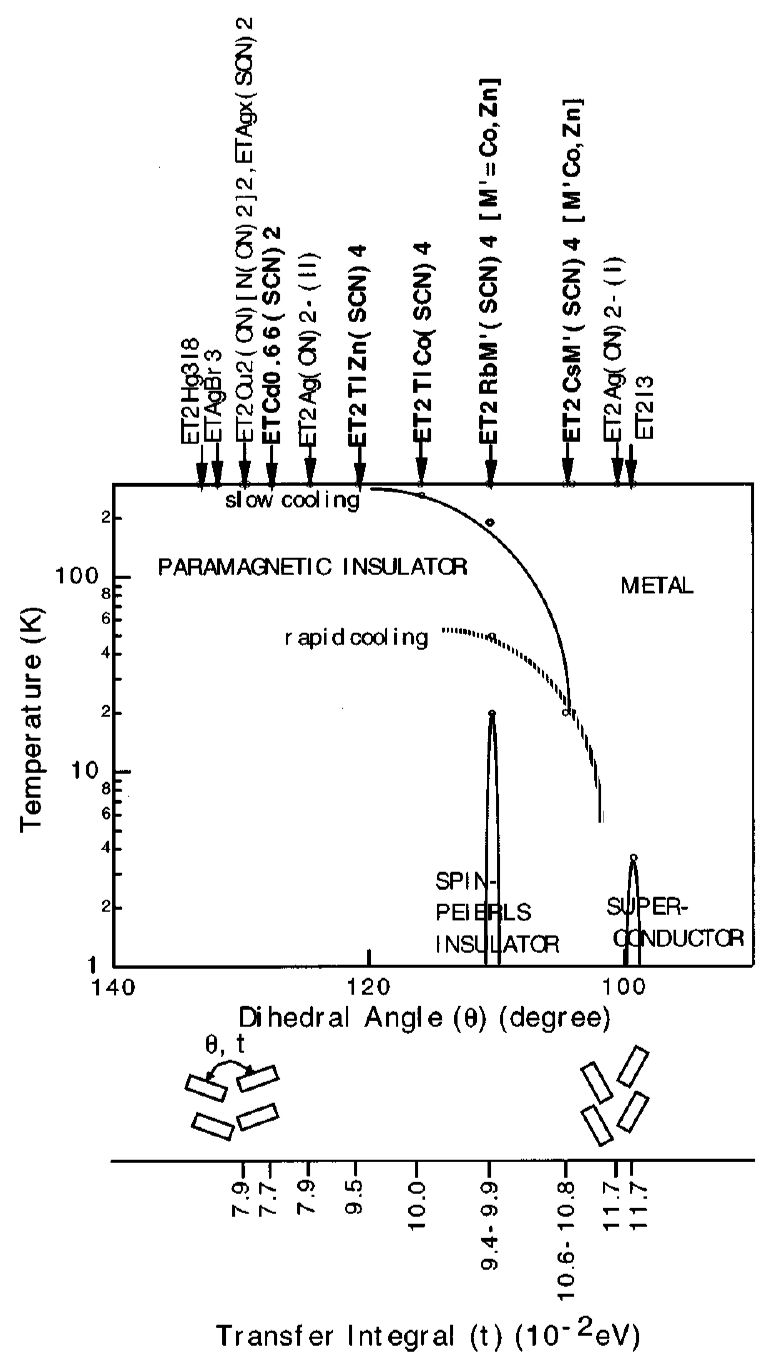}
      \put(-5,60) {\small(b)}
    \end{overpic}
  }
  \caption{ (a) Variation of structural parameters and MI transition
    temperature for several $\theta$-(ET)$_2$XMM$^\prime$(SCN)$_4$.
    (b) Phase diagram. Here the horizontal axis is the dihedral angle
    between BEDT-TTF stacks (see text).  Reprinted with permission
    from Ref. \cite{Mori98b}. $\copyright$ 1998 The American Physical
    Society.}
  \label{theta}
\end{figure}

In the $\theta$ structure (see
Fig.~\ref{alpha-beta-theta-structures}), molecules in neighboring
stacks are tilted with respect to each other by a dihedral angle of
100--140$^\circ$ (see Fig.~\ref{theta}) \cite{Kobayashi86b}. Among the
$\theta$-(ET)$_2$X, X=I$_3$ is superconducting at ambient
pressure with T$_{\rm c}$=3.6 K \cite{Kobayashi86b}. I$_3$ is metallic at
300 K. The resistivity drops with decreasing temperature, with a more
rapid decrease below around 75 K \cite{Kobayashi86b}. At low
temperatures the optical conductivity displays a Drude peak at low
frequencies \cite{Tamura88a,Takenaka05a}. Unlike conventional metals
however, the Drude response has a long tail extending to higher
frequencies \cite{Takenaka05a}. As temperature increases, the Drude
peak is replaced by a far-IR peak, which shifts to higher energy and
loses spectral weight with increasing temperature \cite{Takenaka05a}.
These features were suggested as indicating that I$_3$ is a strongly
correlated system showing ``bad metal'' characteristics
\cite{Takenaka05a}.

\begin{figure}[tb]
  \center{
    \begin{overpic}[width=2.3in]{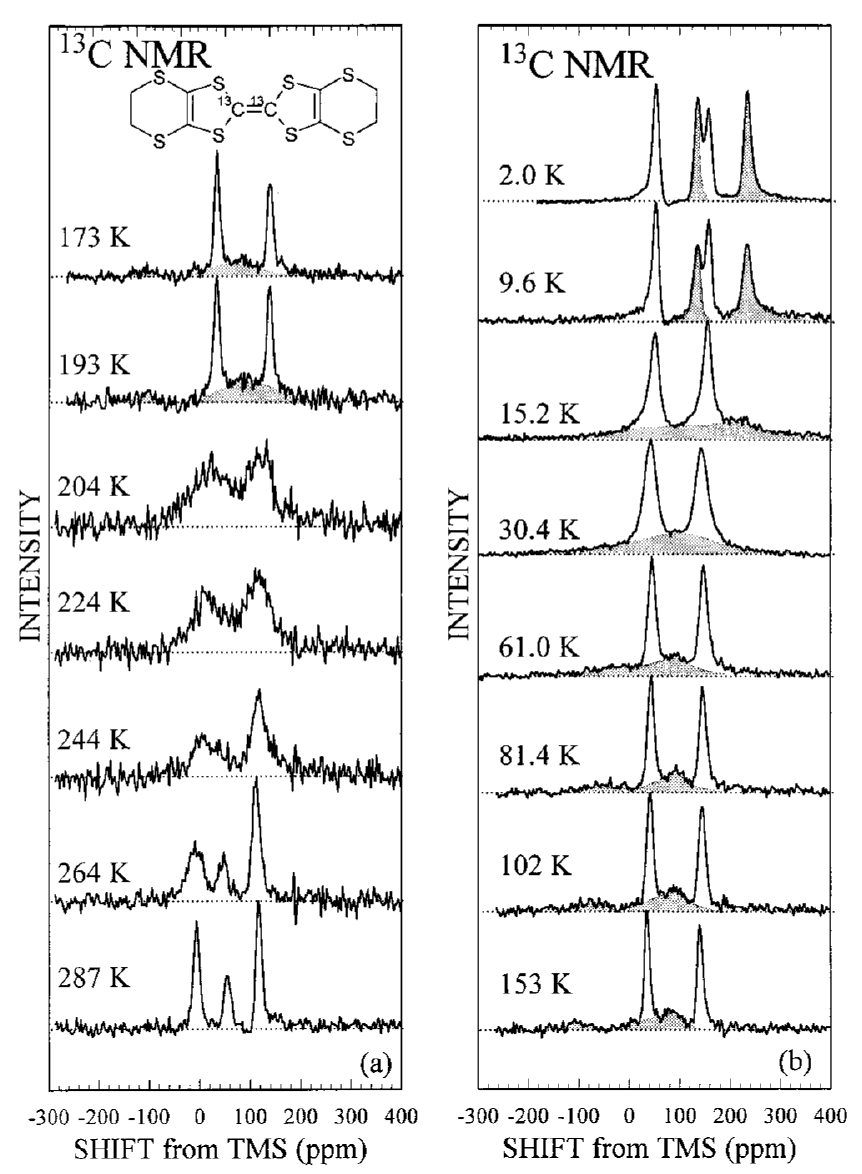}
      \put (-5,90) {\small(a)}
    \end{overpic}
    \hspace{0.1in}
    \raisebox{0.4in}{
      \begin{overpic}[width=2.2in]{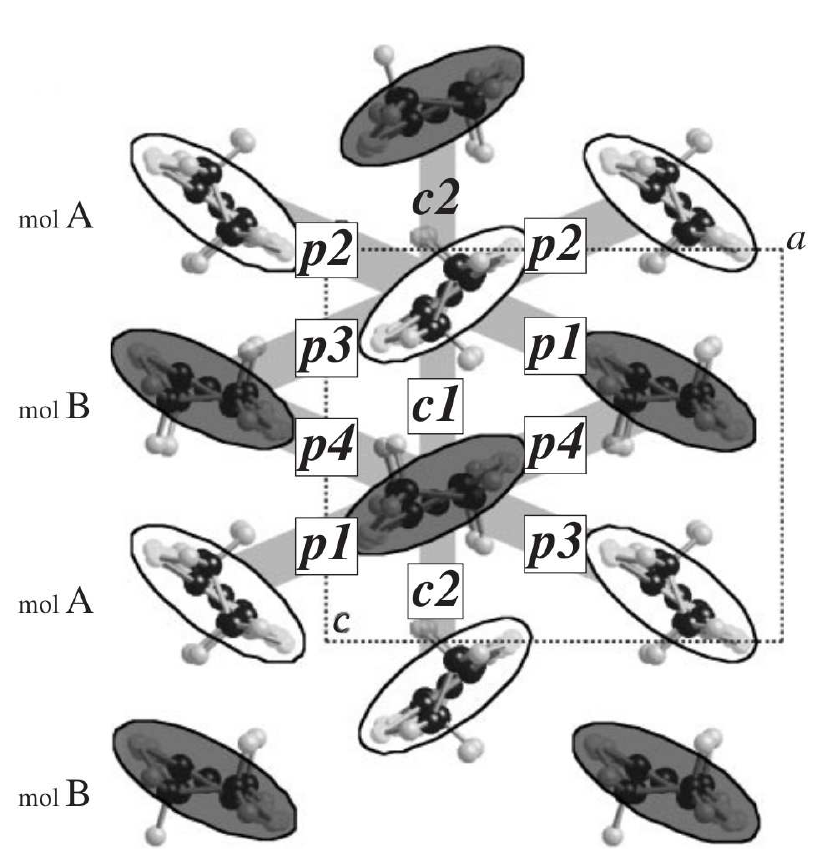}
        \put (-5,95) {\small(b)}
      \end{overpic}
    }
  }
  \caption{ (a) $^{13}$C NMR spectra of
    $\theta$-(ET)$_2$RbZn$^\prime$(SCN)$_4$ showing charge ordering at
    195 K.  Reprinted with permission from
    Ref. \cite{Miyagawa00a}. $\copyright$ 2000 The American Physical
    Society.  (b) Horizontal stripe charge order found in
    $\theta$-(ET)$_2$RbZn$^\prime$(SCN)$_4$ and other $\theta$-phase
    CTS.  Here the dark and light ellipses denote charge-rich and
    charge-poor molecules, respectively. Note the charge occupancy
    pattern $\cdots1100\cdots$ along both the $p$-directions.
    Reprinted with permission from
    Ref. \cite{Watanabe04a}. $\copyright$ 2004 The Physical Society of
    Japan.}
  \label{thetaco}
\end{figure}

Besides X=I$_3$, a variety of $\theta$ salts with X={\it M}{\it
  M}$^\prime$(SCN)$_4$ are known, which in general undergo MI
transitions between 20 and 250 K (see Fig.~\ref{theta})
\cite{Mori95a,Mori97a,Mori98b}.  The $\theta$ CTS have been grouped
into a phase diagram shown in Fig.~\ref{theta}(b). A convenient parameter
is the dihedral angle $\theta$ between neighboring molecules (see
Fig.~\ref{theta}) \cite{Mori98b}. The hopping integral $t_p$
(see Fig.~\ref{thetaco}(a)) between
stacks depends sensitively on $\theta$, with $t_p$ increasing with
decreasing $\theta$ \cite{Mori98b}. The smaller bandwidth for larger
$\theta$ was assumed to increase T$_{\rm MI}$ \cite{Mori98b}, although a
simultaneous decrease in $c$ (see Fig.~\ref{thetaco}(b)) increasing $V_{\rm c}$ also favors an
insulating state \cite{Watanabe04c}.  Unlike most other insulating
CTS, in $\theta$-(ET)$_2$X T$_{\rm MI}$ {\it increases} with
hydrostatic pressure \cite{Mori98b,Watanabe04c}.
A complication in studying the insulating phases in the $\theta$
CTS is that their properties depend strongly on cooling rate,
with slow cooling required to obtain long-range order (LRO) \cite{Mori98b}.

CO coincides with T$_{\rm MI}$ in $\theta$-(ET)$_2$X.
The CO transition has been studied most intensely for two series of $\theta$ CTS,
$\theta$-(ET)$_2$MM$^\prime$(SCN)$_4$ with either
(M,M$^\prime$)=(Rb, Zn) or (Rb, Co) and (Cs, Zn) or (Cs, Co)
\cite{Mori95a}. In the RbZn salt T$_{\rm MI}\sim$195 K, and CO was
confirmed by NMR (see Fig.~\ref{thetaco}(a)) \cite{Miyagawa00a,Chiba01a,Chiba01b}, Raman and IR
reflectance \cite{Yamamoto02a}, X-rays
\cite{Watanabe03a,Watanabe04a}, and dielectric permittivity
\cite{Nad06c}. Here the MI/CO transition is first order, with a
dimerization of the stacks occurring below T$_{\rm MI}$.  The pattern
of CO is the horizontal stripe shown in Fig.~\ref{thetaco}(b), with
charge disproportionation $\Delta n = 0.6$
\cite{Chiba01a,Yamamoto02a,Watanabe03a}. Above T$_{\rm MI}$ evidence
for fluctuating charge order is seen in NMR line broadening \cite{Miyagawa00a,Chiba04a}
(see Fig.~\ref{thetaco})
and in diffuse X-ray scattering \cite{Watanabe03c}. The periodicity of
the diffuse scattering above T$_{\rm MI}$ is however not the same as
the periodicity in the CO state, suggesting that two different charge
order patterns are competing \cite{Watanabe03c}.  In RbZn no change in
the magnetic susceptibility is seen at T$_{\rm MI}$
\cite{Mori98b}. However, in slowly cooled samples, a second 
transition occurs at T$_{\rm SG}\approx$ 20K into a non-magnetic state
\cite{Mori98b,Watanabe07a}. Below T$_{\rm SG}$ a structural symmetry
change occurs breaking the screw-axis symmetry existing along the
horizontal charge order stripe (see Fig.~\ref{thetaco})
\cite{Watanabe07a}.  Within this symmetry breaking, the bonds
$p4$ in Fig.~\ref{thetaco} become inequivalent,  consistent with a
dimerization along the horizontal stripe direction giving
a spin gap \cite{Watanabe07a}.
We discuss this further in Sections \ref{pec2d} and \ref{pec2dcts}.

In comparison to RbM$^\prime$, in the CsM$^\prime$ salts are weakly
metallic until low temperature, when a MI transition occurs at
approximately 20 K \cite{Mori98b}.  Unlike the well-ordered insulating
state in slowly cooled RbM$^\prime$ samples, experiments on
CsM$^\prime$ suggest a phase separated state at low temperature
\cite{Nad08a}.  The resistivity begins to rise slightly below $\sim$
100 K, and then increase steeply below 20 K \cite{Mori98b}. T$_{\rm
  MI}$ here also increases with pressure \cite{Mori98b}.  X-rays show
diffuse scattering at {\bf q$_1$}=(2/3, $k$, 1/3) as well as {\bf
  q$_2$}=(0, $k$, 1/2) in the 50 -- 120 K range, with {\bf q$_2$} (the
same periodicity as in RbZn) becoming stronger at low temperatures
\cite{Nogami99c}. The {\bf q$_2$} modulation however remains
short-range however in CsZn \cite{Nogami99c}.  However, a finite jump
in the specific heat is observed at 20 K in CsZn, confirming that a
phase transition takes place (as also for RbZn) \cite{Nishio99a}.  For
CsM$^\prime$ the susceptibility increases at low temperature rather
than entering a spin-gapped state \cite{Mori98b,Nakamura00c}. Sample
dependence was not noted in the low temperature susceptibility
increase, ruling out Curie-tail impurities \cite{Nakamura00c}. NMR and
EPR results suggested three different temperature regimes: first, from
300 K to 140 K a region of charge fluctuations similar to those in
RbZn above T$_{\rm MI}$; second, from 140 K to 50 K a slowing down of
the charge fluctuations; and third, below 30 K a charge reorganization
with the emergence of spin singlets {\it without} large charge
disproportionation \cite{Nakamura00c,Chiba03a,Chiba08a}.  At low
temperature some sites remain paramagnetic and cause the observed
local magnetic moments \cite{Nakamura00c,Chiba03a,Chiba08a}.  Raman
and IR investigation categorized CsZn as having short-range CO, with
properties similar to those in RbZn which is rapidly cooled; under
pressure long-range CO is stabilized \cite{Suzuki05a}.

Besides the $\theta$-(ET)$_2$MM$^\prime$(SCN)$_4$ series,
long-range horizontal stripe CO is seen in other $\theta$-(ET)$_2$X,
for example in X=Cu$_2$CN[N(CN)$_2$]$_2$ below 220 K \cite{Yamamoto04a}.

\paragraph{Summary}
A metal-insulator transition to charge order is common in
$\theta$-(ET)$_2$X and is often followed by a lower temperature
transition to a nonmagnetic state with a spin gap.  The pattern of the
charge order at the lowest temperatures in most cases is the
horizontal stripe (Fig.~\ref{thetaco}(b), which is different from a
simple Wigner crystal and has charge occupancies $\cdots1100\cdots$
along the two most strongly coupled directions.  Superconductivity
occurs in $\theta$-(ET)$_2$I$_3$.

\subsubsection{$\alpha$ CTS}
\label{alpha}

\begin{figure}[tb]
  \center{
    \begin{overpic}[width=1.75in]{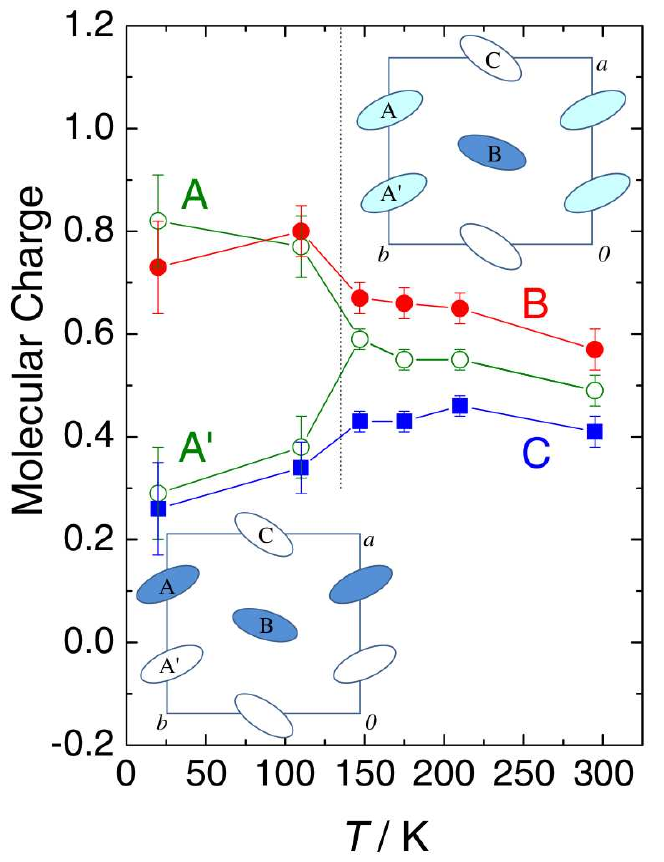}
      \put (-5,95) {\small(a)}
    \end{overpic}
    \hspace{0.1in}
    \raisebox{0.1in}{
      \begin{overpic}[width=2.75in]{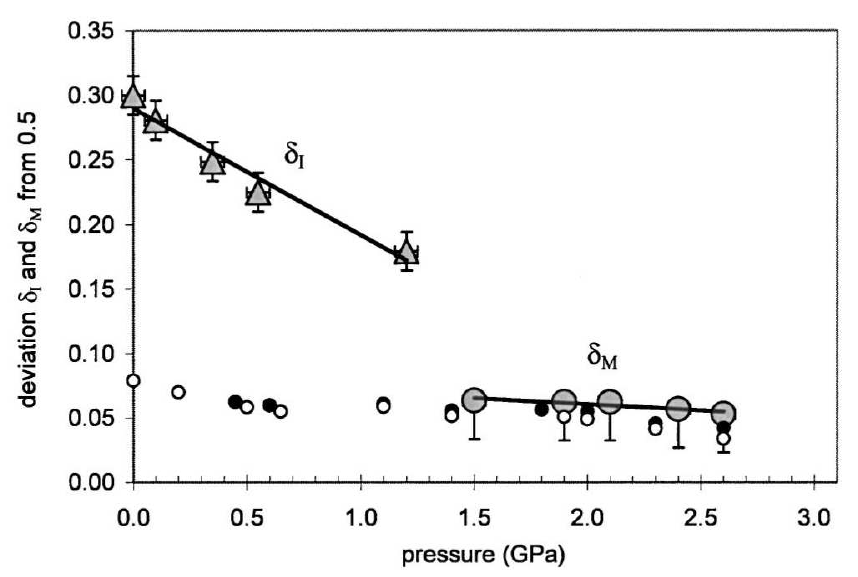}
        \put (-5,60) {\small(b)}
      \end{overpic}
    }
  }
  \caption{(a) Charge order transition in $\alpha$-(ET)$_2$I$_3$ as
    measured by synchrotron X-ray scattering.  Reprinted with
    permission from Ref. \cite{Kakiuchi07b}. $\copyright$ 2007 The
    Physical Society of Japan.  The insets show the crystal structure
    above and below the 135 K CO transition. A and A$^\prime$ are
    equivalent by inversion symmetry in the high-temperature phase.
    (b) Temperature and pressure dependence of the CO amplitude as
    determined by Raman spectroscopy.  In the high-temperature phase
    the charge densities on sites A, A$^\prime$, B, and C are given by
    $\rho_A=\rho_{A^\prime}=\rho_B=0.5 +\delta_M$ and $\rho_{\rm
      c}=0.5-3\delta_M$.  $\delta_I$ is defined in the CO phase as the
    average of the disproportionation of sites A and B,
    $\delta_I=(\delta_A+\delta_B)/2$.  The CO phase vanishes under a
    hydrostatic pressure of approximately 1.5 GPa.  Reprinted with
    permission from Ref. \cite{Wojciechowski03a}. $\copyright$ 2003
    The American Physical Society.}
  \label{alpha-i3}
\end{figure}  

The $\alpha$ structure is nearly identical to the
$\theta$ structure (see Fig.~\ref{alpha-beta-theta-structures}),
except the periodicity is doubled already in the room-temperature
structure in the stacking direction by a weak dimerization.
$\alpha$-(ET)$_2$I$_3$ was discovered as a second crystal structure in
the crystal growth of $\beta$-(ET)$_2$I$_3$ \cite{Kaminskii83a}.  This
salt has been extensively studied because of its CO transition,
unusual transport properties at low temperature, and
superconductivity.  $\alpha$-(ET)$_2$I$_3$ undergoes a sharp MI
transition at 135 K \cite{Kaminskii83a}.  The susceptibility decreases
with temperature until 135 K and then drops sharply at T$_{\rm MI}$,
indicating a nonmagnetic ground state \cite{Rothaemel86a}. The MI
transition is suppressed by hydrostatic pressure, without however the
appearance of SC. However, SC with T$_{\rm c}\sim$ 7 K is found under
conditions of $\sim$0.2 GPa uniaxial strain along the $a$ axis (the
$a$ axis is the stacking direction, see Fig.~\ref{alpha-i3})
\cite{Tajima02a}. Unlike hydrostatic pressure, $a$-axis uniaxial
strain does not change T$_{\rm MI}$ greatly, but instead greatly
reduces the resistivity in the insulating state
\cite{Tajima02a}. Strain along the other in-plane direction ($b$ axis)
reduced T$_{\rm MI}$ without the appearance of SC \cite{Tajima02a}.

Charge order in $\alpha$-(ET)$_2$I$_3$ has been confirmed by
several experimental methods including optics
\cite{Moldenhauer93a,Wojciechowski03a}, X-rays
\cite{Heidmann92a,Kakiuchi07b} and NMR \cite{Takano01a,Kawai09a}.  In
this salt CO exists in both the high and low temperature phases (see
Fig.~\ref{alpha-i3}).  Above T$_{\rm MI}$, alternate BEDT-TTF stacks
are charge ordered in the pattern $\cdots$1010$\cdots$ with a CO
amplitude of
$\Delta$n $\sim$ 0.2
\cite{Kakiuchi07b,Wojciechowski03a}.
Below T$_{\rm MI}$ all stacks become charge ordered, with the CO
pattern following a horizontal stripe as in $\theta$-(ET)$_2$X.
Unlike $\theta$ where a spin gap only opens at lower temperatures, in
$\alpha$-(ET)$_2$I$_3$ T$_{\rm SG}$ coincides with T$_{\rm MI}$.
Because of the lower symmetry of the $\alpha$ structure compared to
$\theta$, in the horizontal stripe CO in $\alpha$ the bonds in the $p$
direction (the bonds between molecules A and B in Fig.~\ref{alpha-i3})
become inequivalent at the MI transition. This is in contrast to the
$\theta$ CTS, where this inequivalence only occurs below the lower
temperature magnetic transition \cite{Watanabe07a}.

The CO amplitude $\Delta$n is approximately 0.6 (see
Fig.~\ref{alpha-i3})\cite{Kakiuchi07b,Wojciechowski03a}.  Like other
CTS, evidence for relaxor ferroelectricity is found in measurements of
the dielectric constant \cite{Lunkenheimer15b}.  Measurements of the
frequency dependence of the dielectric constant (with the electric
field perpendicular to the BEDT-TTF layers) show two general effects
\cite{Lunkenheimer15b}: first, for $T\gtrapprox$ 80 K there is a
gradual step-like decrease in the dielectric constant with decreasing
temperature. The step feature shifts to larger temperatures at higher
frequencies.  Second, for low frequencies, a peak develops in the
dielectric constant, which is located at 40--50 K for the lowest
frequencies.  Assuming the response corresponds to a relaxor
ferroelectric, the extrapolated Curie-Weiss ordering temperature is 35
K \cite{Lunkenheimer15b}. It is unknown why these effects show up at
temperatures that are much lower than T$_{\rm MI}$=135 K where the CO
first occurs.

Hydrostatic pressure suppresses the CO, which vanishes at
approximately 1.5 GPa \cite{Wojciechowski03a}.  The properties of
$\alpha$-(ET)$_2$I$_3$ under pressure are unusual. $^{13}$C NMR
under a pressure of 2.3 GPa, where the CO is completely suppressed,
show a vanishing of the spin susceptibility below approximately 50 K
\cite{Hirata15a}.  The transport properties under pressure are unusual
and have been investigated by many authors. In the high-pressure state the
resistivity becomes nearly constant in temperature
\cite{Mishima95a,Tajima00b}. This was interpreted as a semiconductor
in which as temperature decreases the mobility grows while the carrier
density simultaneously falls \cite{Tajima00b}. Recently there has been
great interest in the semi-metallic state of $\alpha$-(ET)$_2$I$_3$
under pressure, which has been proposed as a candidate realization of a
Dirac cone semi-metal \cite{Katayama06a,Kobayashi07a} (for a review
see \cite{Kajita14a}). This is an intriguing possibility but
beyond the scope of this review.

\begin{figure}[tb]
  \center{
    \begin{overpic}[width=2.3in]{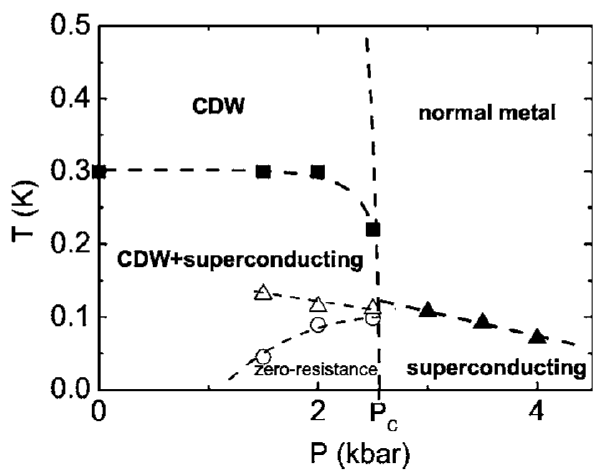}
      \put (-5,70) {\small(a)}
    \end{overpic}
    \hspace{0.1in}
    \begin{overpic}[width=2.5in]{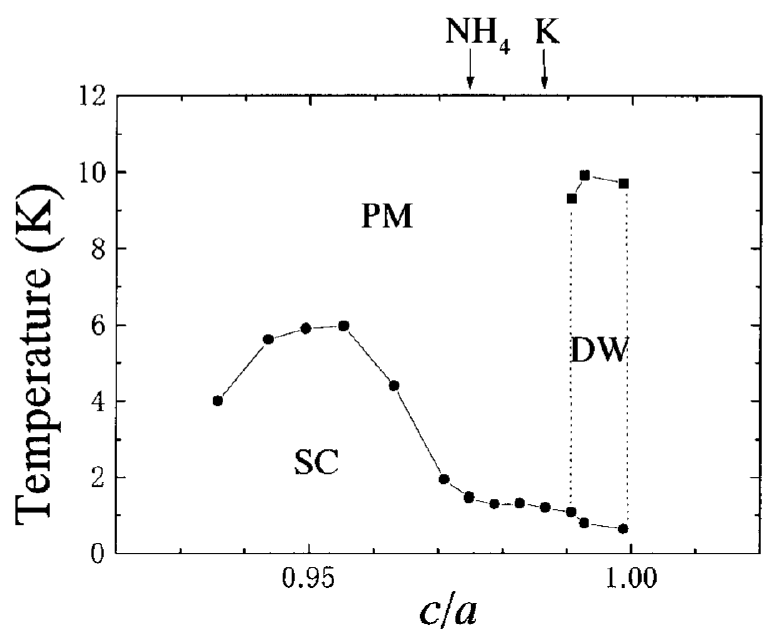}
      \put (-5,65) {\small(b)}
    \end{overpic}
    }
  \caption{(a) Low temperature-pressure phase diagram of
    $\alpha$-(ET)$_2$KHg(SCN)$_4$.  Reprinted with permission from
    Ref.~\cite{Andres05a}. $\copyright$ 2005 The American Physical
    Society.  (b) Phase diagram of $\alpha$-(ET)$_2${\it M}Hg(SCN)$_4$
    determined from uniaxial strain experiments. $c/a$ is the ratio of
    the in-plane lattice constants.  The arrows indicate the
    approximate ambient-pressure locations of {\it M}=NH$_4$ and K The
    $a$ and $c$ crystal axes in $\alpha$-(ET)$_2${\it M}Hg(SCN)$_4$
    are equivalent to the $b$ and $a$ axes, respectively, in
    $\alpha$-(ET)$_2$I$_3$ (see Fig.~\ref{alpha-i3}).  Reprinted with
    permission from Ref.~\cite{Maesato01a}. $\copyright$ 2001 The
    American Physical Society.}
  \label{alpha-mhg1}
\end{figure}  

\begin{figure}[tb]
  \center{
    \raisebox{0.3in}{
      \begin{overpic}[width=2.4in]{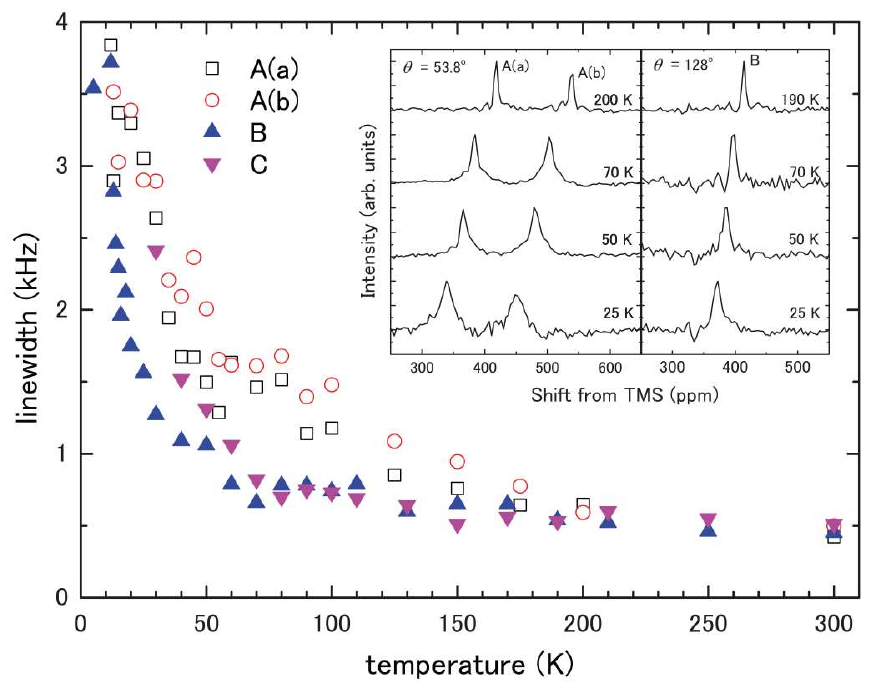}
        \put (-5,68) {\small(a)}
      \end{overpic}
      \hspace{0.1in}
      \begin{overpic}[width=1.8in]{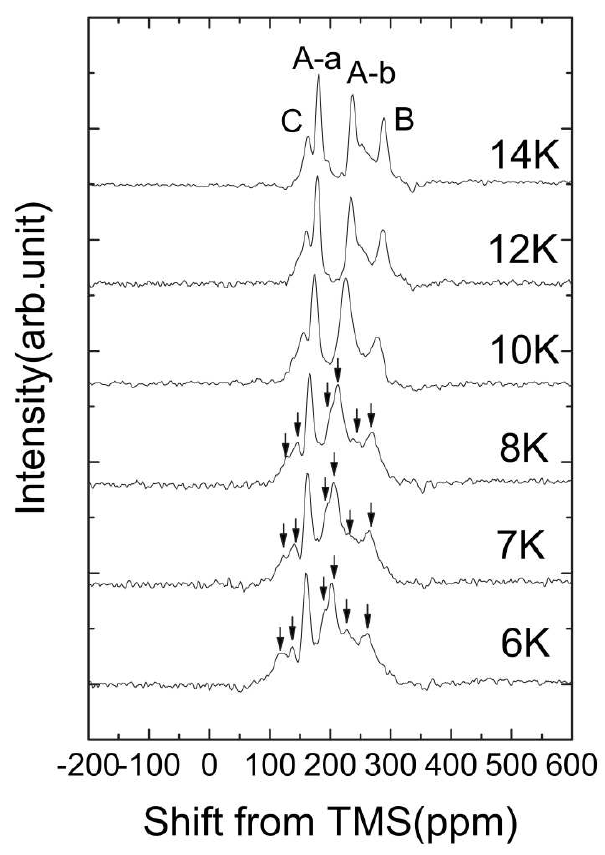}
        \put (-5,90) {\small(b)}
      \end{overpic}
    }
  }
  \caption{(a) Temperature dependence of the $^{13}$C NMR linewidth of
    $\alpha$-(ET)$_2$RbHg(SCN)$_4$ \cite{Noda13a}.  Reprinted with
    permission from Ref.~\cite{Andres05a}. $\copyright$ 2005 The
    American Physical Society.  Here the sites A(a) and A(b)
    correspond to A and A$^\prime$ in Fig.~\ref{alpha-i3}
    respectively.  (b) $^{13}$C NMR spectra of as a function of
    temperature. Peak splitting indicating CO is observed in the lines
    for the A, B, and C molecules below 10 K Reprinted with permission
    from Ref.~\cite{Kawai08a}. $\copyright$ 2008 The American Physical
    Society.}
  \label{alpha-mhg2}
\end{figure}

$\alpha$-(ET)$_2$ {\it M}Hg(SCN)$_4$ with {\it M}=K, Rb, Tl,
and NH$_4$ are another family of $\alpha$-phase conductors and
superconductors \cite{Oshima89a,Mori90b,Mori90c,Wang90b,Kushch92a}.
The {\it M}=NH$_4$ salt is an ambient pressure superconductor with
T$_{\rm c}\sim$ 1 K \cite{Wang90b,Mori90c}. For {\it M}=K, Rb, and Tl the
resistivity drops with temperature, leveling off at around 10 K; below
an anomaly about 8 K the resistance drops further
\cite{Sasaki90a,Ito95a}. The resistivity decrease in some samples at
low temperatures is suppressed by magnetic field below 100-300 mK,
suggesting the appearance of a filamentary SC state
\cite{Ito93a,Ito95a}. Hydrostatic pressure stabilizes the SC state,
leading to the phase diagram shown in Fig.~\ref{alpha-mhg1}
\cite{Andres05a,Kartsovnik07a}.  While T$_{\rm c}$ is low in this series at ambient or
under hydrostatic pressure,  the application of uniaxial stress
can increase T$_{\rm c}$ to $\sim$ 5 K \cite{Maesato01a}. Systematic studies of
uniaxial stress applied to {\it M}=K and NH$_4$ suggest that SC occurs
in this series for those salts with a smaller $c/a$ ratio (see
Fig.~\ref{alpha-mhg1}) \cite{Maesato01a}.

The transition at 8--10 K has been considered ``mysterious'' as it has
similarities with both charge- and spin-density wave systems
\cite{Sasaki95a}.  A coexistence of metallic and magnetic ordering was
also suggested \cite{Kinoshita91a}.  Satellite X-ray reflections of an
incommensurate wavevector are seen beginning at room temperature and
increase sharply below 10 K in {\it M}=K; for {\it M}=Rb diffuse
scattering at the same wavevectors is seen,
reflecting short-range order
\cite{Foury-Leylekian03a,Foury-Leylekian10a}.  In this temperature
range the resistivity decreases while the Hall resistance increases
\cite{Sasaki95a}. The spin susceptibility shows dependence on the field
direction, being constant for fields perpendicular to the BEDT-TTF
layers, and decreasing for fields within the layer
\cite{Sasaki95a}. Zero-field $\mu$SR experiments on {\it M}=K found
two step-like increases in the relaxation rate at 12 K and 8 K
\cite{Pratt95a}. These were interpreted as the onset of weak SDW
order, first within the 2D planes at 12 K and then with 3D long range
order at 8 K \cite{Pratt95a}. In the optical conductivity a
significant dip centered around 200 cm$^{-1}$ develops below 200 K and
was interpreted as a pseudogap \cite{Dressel03a}.  From infrared and
Raman studies on {\it M}=K, a breaking of the inversion symmetry
between molecules A and A$^\prime$ (see Fig.~\ref{alpha-i3}) was noted
below about 200 K \cite{Hiejima10a}. This would be consistent with
short-ranged horizontal-stripe CO as found in $\theta$-(ET)$_2$X
\cite{Hiejima10a}.

Recent $^{13}$C NMR experiments on RbHg(SCN)$_4$ have provided more
information on the density-wave state \cite{Kawai08a,Noda13a}.  Below
300 K the NMR linewidth increases with decreasing temperature, as shown
in Fig.~\ref{alpha-mhg2} \cite{Noda13a}.  This is consistent with the
development of incommensurate density waves as observed by X-rays
\cite{Noda13a,Foury-Leylekian03a,Foury-Leylekian10a}. The linewidth
increase above the 8--10 K transition is principally in the A and
A$^\prime$ molecules (see Fig.~\ref{alpha-i3}) \cite{Noda13a}.
Fig.~\ref{alpha-mhg2} shows the $^{13}$C NMR spectra for {\it M}=Rb as
a function of temperature \cite{Kawai08a}. Peak splitting is observed
below 10 K: the splitting of peak A-B indicates that the A and
A$^\prime$ molecules become crystallographically inequivalent
\cite{Kawai08a}. The splitting of peaks B and C indicates that the
inversion symmetry between molecules B and C is also broken
\cite{Kawai08a}. These features indicate that CO coexists with a
lattice-based density wave in the low temperature state of
$\alpha$-(ET)$_2${\it M}Hg(SCN)$_4$.

\paragraph{Summary}
The $\alpha$ structure has lower symmetry (inequivalent stacks) at
room temperature than the $\theta$ structure.  Charge-ordering and
spin-gap transitions occur at the same temperature in
$\alpha$-(ET)$_2$I$_3$.  In $\alpha$-(ET)$_2${\it M}Hg(SCN)$_4$ there
are likely coexisting charge- and spin-density waves.  The dominant
charge order pattern is the horizontal stripe.

\subsection{Quasi-two dimensional anionic CTS}
\label{2d-anionic}

The materials in Sections \ref{kappa}-\ref{alpha} are all based on
layers of cationic organic molecules that are $\frac{3}{4}$-filled
with electrons, or $\frac{1}{4}$-filled in terms of holes. However,
many organic conductors composed of anionic acceptor molecules are
also known (see \cite{Kato04a,Kato14a} for reviews). Here we focus on
one particular family composed of metal dithiolene molecules,
M(dmit)$_2$ (M=Ni, Pd, Pt).  Within this family a variety of quasi-2D
superconductors are known, as well as insulating states with charge
and spin order (see Table \ref{dmit-table})
\cite{Kato04a,Kanoda11a,Kato14a}. In common with the cationic BEDT-TTF
and related superconductors, SC in the M(dmit)$_2$ family also occurs
when the carrier density is 0.5 per molecule.

\begin{table}
  \begin{center}
    \scriptsize
    \begin{tabular}{c|c|c|c|c|c|l}
      Cation & {\it M} & $T_{\rm c}$(K)& P$_{\rm c}$(GPa) & T$_{\rm N}$(K) & Notes & references \\
      \hline
      Et$_2$Me$_2$N & Pd & 4 & 0.2 & --- &single layer &  \cite{Kobayashi92a} \\
      $\beta$-Me$_4$N & Ni & 5  & 0.7  & ---&solid crossing  & \cite{Kobayashi87b}  \\ 
      $\beta$-Me$_4$N & Pd & 6  & 0.7  & 12 & solid crossing  & \cite{Kobayashi91b,Nakamura01a}  \\
      $\beta^\prime$-Me$_4$P & Pd & ---  & --- & 42 & solid crossing &\cite{Kato97a,Nakamura00d}  \\
      $\beta^\prime$-Me$_4$P & Pt & ---  & --- & ---& solid crossing; T$_{\rm CO}$=218 K &\cite{Kato97a}  \\
      $\beta^\prime$-Me$_4$Sb & Pt & 3, 8$^\star$  & 1.0, 0.5$^\star$ & 18 & solid crossing  &\cite{Rouziere98a,Tajima05a}\\
      $\beta^\prime$-Me$_4$As & Pt & 4$^\star$ & 0.7$^\star$ & 35 & solid crossing  & \cite{Kobayashi90a,Kato02b} \\
      $\beta^\prime$-Et$_2$Me$_2$P & Pd & 4,6$^\star$ & 0.7,0.5$^\star$ & 18 & solid crossing & \cite{Kato98a,Nakamura00d} \\
      $\beta^\prime$-Et$_2$Me$_2$As & Pd & 5.5 & 0.8 & 18 & solid crossing & \cite{Kato04a,Nakamura01a} \\
      $\beta^\prime$-Et$_2$Me$_2$Sb & Pd & --- & --- & --- & solid crossing; T$_{\rm CO}$=70 K & \cite{Kato98a} \\
      $\beta^\prime$-EtMe$_3$As & Pd & 4  & 0.7 & 23 & solid crossing & \cite{Kato06a} \\
      $\beta^\prime$-EtMe$_3$Sb & Pd & ---  & --- & --- & solid crossing, QSL & \cite{Itou08a} \\
      {\it m}-EtMe$_3$P &Pd & 5  & 0.4 & --- &parallel column, T$_{\rm SG}$=25 K & \cite{Kato06a} \\
      {\it t}-EtMe$_3$P & Pd & ?  & ? & --- & parallel column, T$_{\rm SG}$=50 K & \cite{Yamamoto11a} \\
    \end{tabular}
  \end{center}
  \caption{Summary of the properties of the {\it M}(dmit)$_2$
    salts discussed in this section. T$_{\rm c}$'s and P$_{\rm c}$'s
    marked with $^\star$ refer to uniaxial strain rather than
    hydrostatic pressure.}
  \label{dmit-table}
\end{table}
\begin{figure}[tb]
  \center{\resizebox{2.5in}{!}{\includegraphics{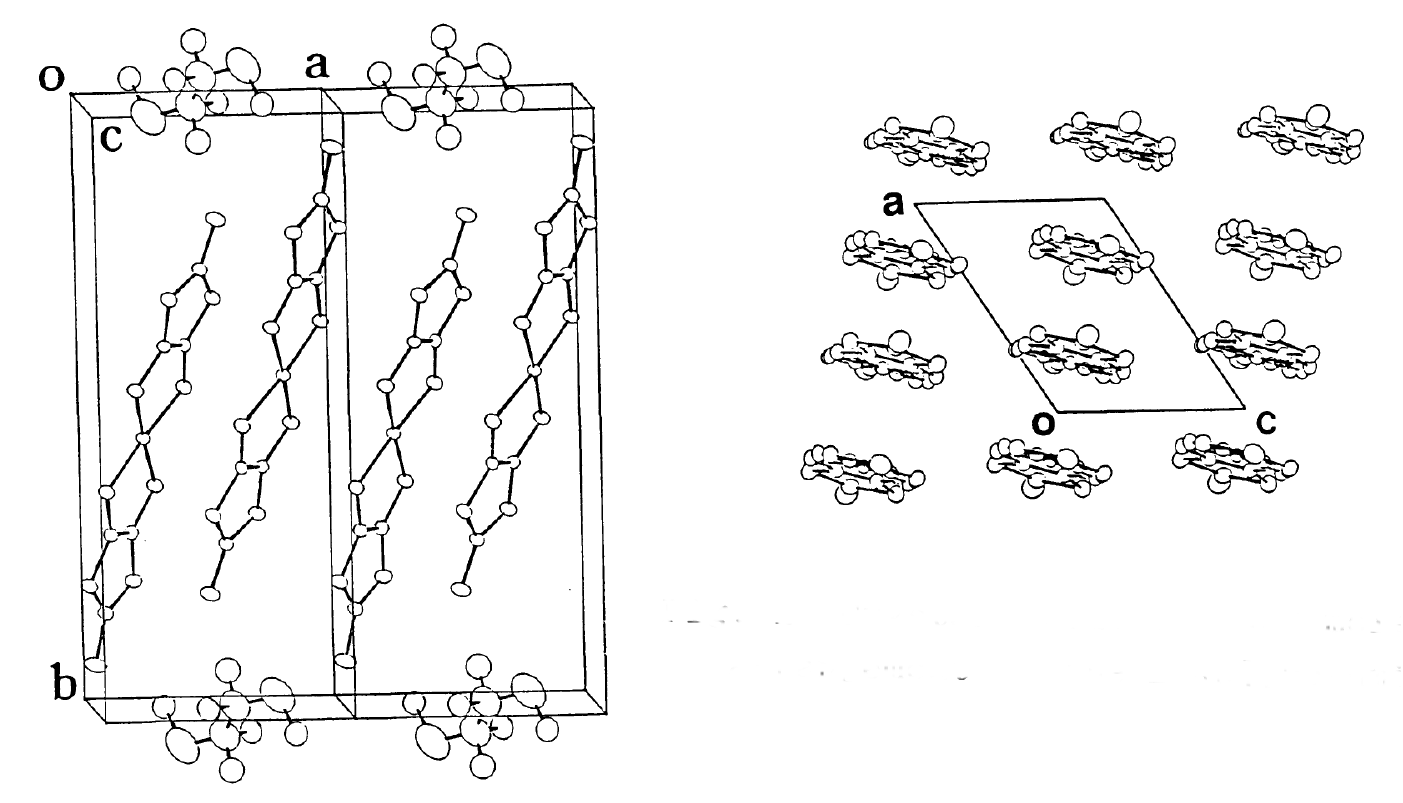}}}
  \caption{Crystal structure of Et$_2$Me$_2$N[Pd(dmit)$_2$]$_2$.
         Reprinted  with permission from Ref.~\cite{Kobayashi92a}. $\copyright$ 1992 The
Chemical Society of Japan.}
  \label{dmit-structure-1}
\end{figure}  

\begin{figure}[tb]
  \center{\resizebox{2.5in}{!}{\includegraphics{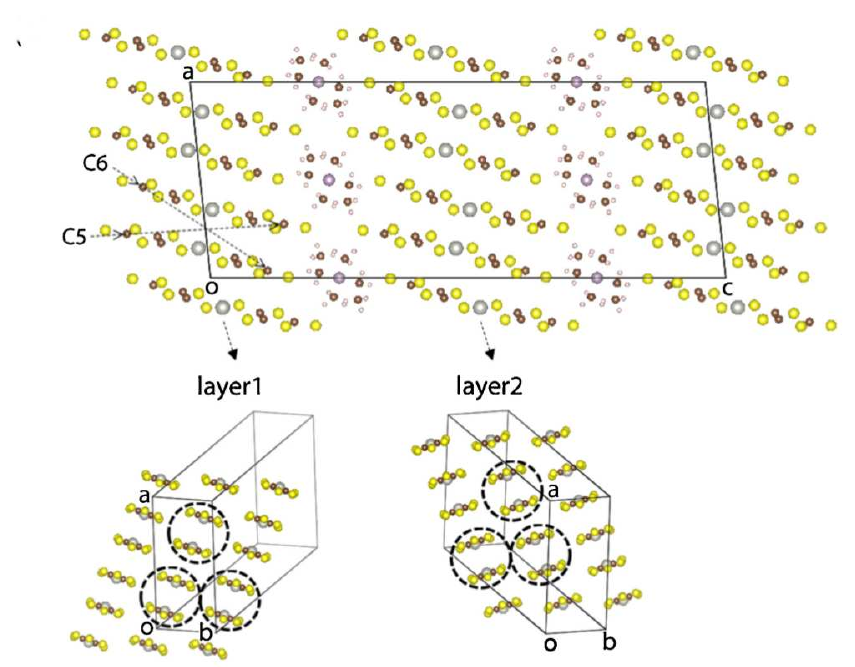}}}
  \caption{Crystal structure of
    $\beta^\prime$-Et$_2$Me$_2$P[Pd(dmit)$_2$]$_2$ with
    ``solid-crossing'' stacking in adjacent layers.  Reprinted with
    permission from Ref.~\cite{Otsuka14a}. $\copyright$ 2014 The
    Physical Society of Japan.  \cite{Otsuka14a}.}
  \label{dmit-structure-2}
\end{figure}  

\begin{figure}[tb]
  \center{\resizebox{2.5in}{!}{\includegraphics{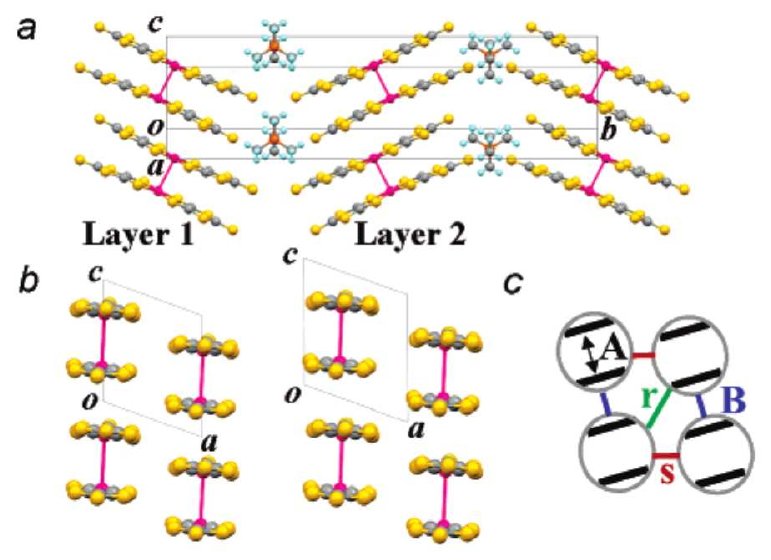}}}
  \caption{Crystal structure of EtMe$_3$P[Pd(dmit)$_2$]$_2$ with
    parallel stacking in adjacent layers.  Reprinted with permission
    from \cite{Kato06a}. $\copyright$ 2006 The American Chemical
    Society.}
  \label{dmit-structure-3}
\end{figure}  

M(dmit)$_2$ salts have several important similarities and differences
compared to BEDT-TTF salts. Compared to BEDT-TTF the HOMO-LUMO gap in
the M(dmit)$_2$ molecule is smaller by a factor of about 3 \cite{Kanoda11a}. In
general the M(dmit)$_2$ salts for M=Pd are strongly dimerized, with
more tightly coupled dimers than in
$\kappa$-(ET)$_2$X. Structurally, within the M(dmit)$_2$ planes,
the molecular packing is similar for all the materials in Table
\ref{dmit-table}, with columnar packing of M(dmit)$_2$ molecules
similar to the $\beta$ and $\beta^\prime$ structures
(Fig.~\ref{alpha-beta-theta-structures}). However, there are several
different arrangements of the planes in the 3D crystal structures as
shown in Figs.~\ref{dmit-structure-1}-\ref{dmit-structure-3} (see
\cite{Kanoda11a}).  These include single-layer structures
(Fig.~\ref{dmit-structure-1}), the so-called ``solid-crossing'' layer
structure, where molecules stack in different directions in alternate
planes (Fig.~\ref{dmit-structure-2}), and a parallel column structure
with two crystallographically distinct planes with stacking in the
same direction (Fig.~\ref{dmit-structure-3}).  Within the group of
solid-crossing structures there are several types, including very
similar $\beta$ and $\beta^\prime$ structures \cite{Kanoda11a}.  An
$\alpha$-type solid-crossing structure is also known (unrelated to the
$\alpha$-(ET)$_2$X structure) \cite{Kanoda11a}.

\subsubsection{AFM, CO, and SC in M(dmit)$_2$}

SC was first observed in this class of CTS in
Me$_4$N[Ni(dmit)$_2$]$_2$ with T$_{\rm c}=$ 5 K
\cite{Kobayashi87b,Kobayashi88a}. The resistivity undergoes a jump
(with thermal hysteresis) at around 100 K, followed by a sharp rise in
resistivity at 20 K. SC with T$_{\rm c}$ up to 5.0 K was observed,
with however large sample dependence \cite{Kobayashi87b,Kobayashi88a}.

Compared to Ni(dmit)$_2$ salts, in Pd(dmit)$_2$ salts the dimerization
is typically very strong, with the ratio of the extended Huck\"el
overlaps for intra- to inter-bonds within one stack of Pd(dmit)$_2$
molecules being of order 10:1 \cite{Nakamura01a}. The strong
dimerization suggests that an effectively $\frac{1}{2}$-filled band
picture is appropriate, and AFM should be observed in
many Pd(dmit)$_2$ salts provided the dimer lattice is not frustrated
\cite{Kato04a}. This is indeed seen, with many of the Pd(dmit)$_2$
materials having AFM transitions in the 10--40 K range (see Table
\ref{dmit-table}).

\begin{figure}[tb]
  \center{
    \begin{overpic}[width=2.5in]{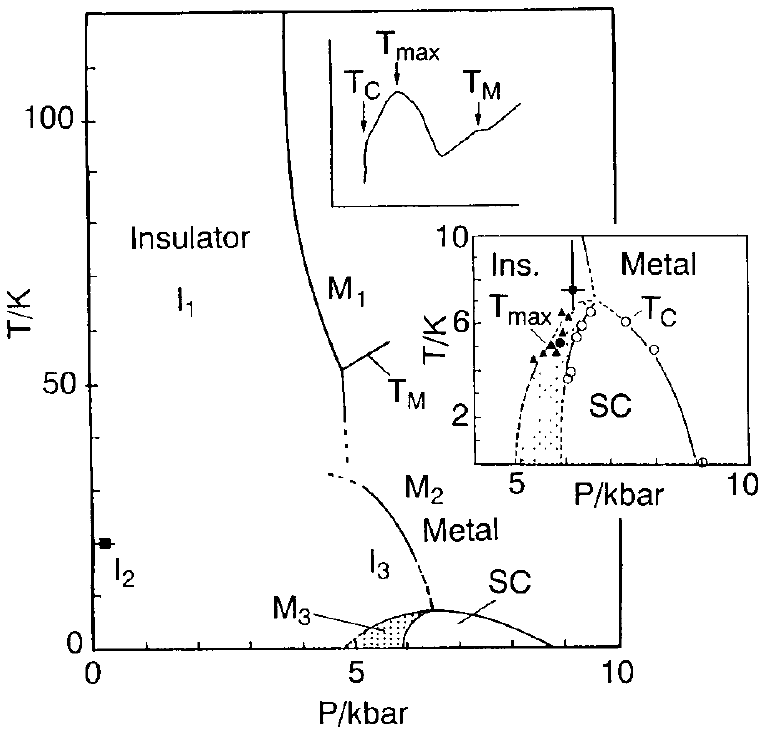}
      \put (-5,90) {\small(a)}
    \end{overpic}
    \hspace{0.1in}%
    \raisebox{0.3in}{
      \begin{overpic}[width=2.1in]{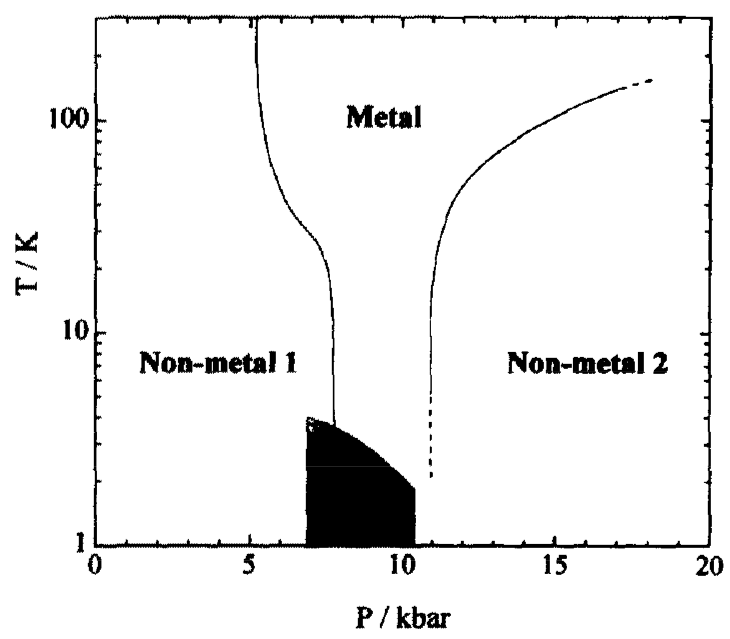}
        \put (-5,78) {\small(b)}
      \end{overpic}
    }
  }
  \caption{(a) Phase diagram of $\beta$-Me$_4$N[Pd(dmit)$_2$]$_2$.
    Reprinted with permission from Ref.~\cite{Kobayashi93a},
    $\copyright$ 1993 Elsevier.  I$_{\rm x}$ and M$_{\rm x}$ indicate
    different insulating and metallic phases based on resistivity
    measurements. The AFM transition is indicated by I$_{\rm 2}$ at
    ambient pressure. The shaded region indicates a drop in the
    resistivity without full SC.  (b) Phase diagram of
    $\beta^\prime$-Et$_2$Me$_2$P[Pd(dmit)$_2$]$_2$ based on
    resistivity measurements.  Reprinted with permission from
    Ref.~\cite{Kato98a}, $\copyright$ 1998 Elsevier.}
    \label{dmit-phase-diag-1}
\end{figure}  

SC is found under pressure in several of the Pd(dmit)$_2$ salts
showing AFM under ambient pressure. It is, however, not clear that the
transition to SC occurs from the AFM phase, or if there exist other
intermediate phases. Fig.~\ref{dmit-phase-diag-1} shows the
pressure-temperature phase diagrams determined by resistivity
measurements for Me$_4$N.  Under pressure, the AFM insulating phase in
Me$_4$N is suppressed at about 0.4 GPa (see
Fig.\ref{dmit-phase-diag-1}) \cite{Kobayashi93a}.  After an anomaly in
the resistivity at about 60 K a second metallic phase M$_2$ is
reached, followed by a different insulating phase I$_{\rm 3}$ below 30
K \cite{Kobayashi93a}. It was noted that near onset the resistivity
was very noisy, possibly due to a phase segregation of insulating and
superconducting regions \cite{Kobayashi93a}.

The phase diagram for Et$_2$Me$_2$P[Pd(dmit)$_2$]$_2$ is shown in
Fig.~\ref{dmit-phase-diag-1}(b) \cite{Kato98a}.  At ambient pressure the
resistivity increases with decreasing temperature below 300 K
\cite{Kato98a}. Under pressure, metallic behavior is observed above 0.5
GPa \cite{Kato98a}. At 0.69 GPa the resistivity reaches a minimum at
about 30 K and then increases at lower temperature before the
superconducting transition, becoming larger at low temperature than at
room temperature \cite{Kato98a}. An unusual feature of this salt is
that at larger pressures, a second insulating phase is reached (see
Fig.~\ref{dmit-phase-diag-1}(b)). The salt Et$_2$Me$_2$N[Pd(dmit)$_2$]$_2$ also has a
similar phase diagram with SC surrounded by two insulating phases,
although it is not isostructural to Et$_2$Me$_2$P as it has a
single-layer structure \cite{Kobayashi93a}. Et$_2$Me$_2$As[Pd(dmit)$_2$]$_2$ is
isostructural with Et$_2$Me$_2$P and is also superconducting under
pressure \cite{Kato04a}.

Several of the Pd(dmit)$_2$ salts are superconducting only under the
application of uniaxial strain, or reach higher T$_{\rm c}$'s under the
application of uniaxial strain (see Table \ref{dmit-table})
\cite{Kato02b,Tajima05a}. The highest T$_{\rm c}$ reached is 8.4 K for
Me$_4$Sb \cite{Tajima05a}. In all cases, the enhancement of SC occurs
when the salts are compressed along the $b$ axis, which is the
inter-stack direction (see Fig.~\ref{dmit-structure-2})
\cite{Kato02b,Tajima05a}. Application of pressure along the
stacking ($a$) direction resulted in enhanced non-metallic behavior
in the low-pressure region \cite{Kato02b}.

As in the dimerized BEDT-TTF materials \cite{Lunkenheimer15a}, the
frequency dependence of the dielectric constant displays relaxor
characteristics in many [Pd(dmit)$_2$]$_2$ CTS \cite{Abdel-Jawad13a}.
Extrapolated Curie-Weiss ordering temperatures are found in the range
10 $\sim$ 30 K; in general a larger T$_{\rm N}$ appears to correlate
with a smaller T$_{\rm CW}$ \cite{Abdel-Jawad13a}.  Assuming the
relaxor behavior is due to fluctuating CO (with the same caveats as
discussed in Section \ref{kappa}), a
$\Delta$n $\sim$0.1$e$
was estimated, relatively independent of
the cation \cite{Abdel-Jawad13a}. Static CO has been confirmed for
EtMe$_3$P as discussed in the following section.

\subsubsection{EtMe$_3$P[Pd(dmit)$_2$]$_2$}
\label{etme3p}

\begin{figure}[tb]
  \center{
    \begin{overpic}[width=2.0in]{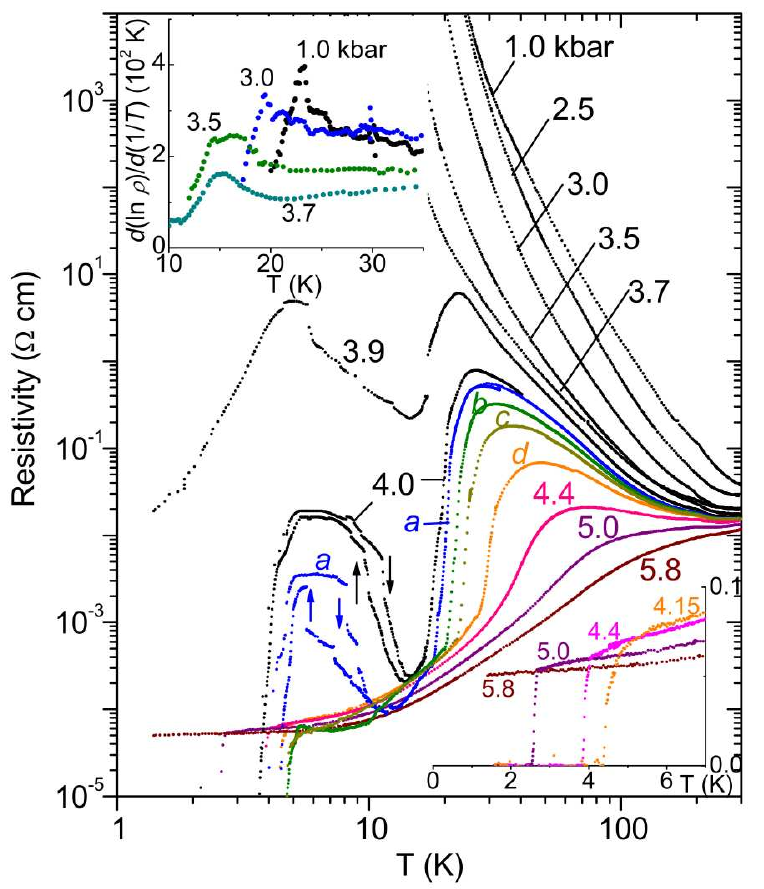}
      \put (-5,95) {\small(a)}
    \end{overpic}
      \hspace{0.1in}
      \raisebox{0.2in}{
        \begin{overpic}[width=2.5in]{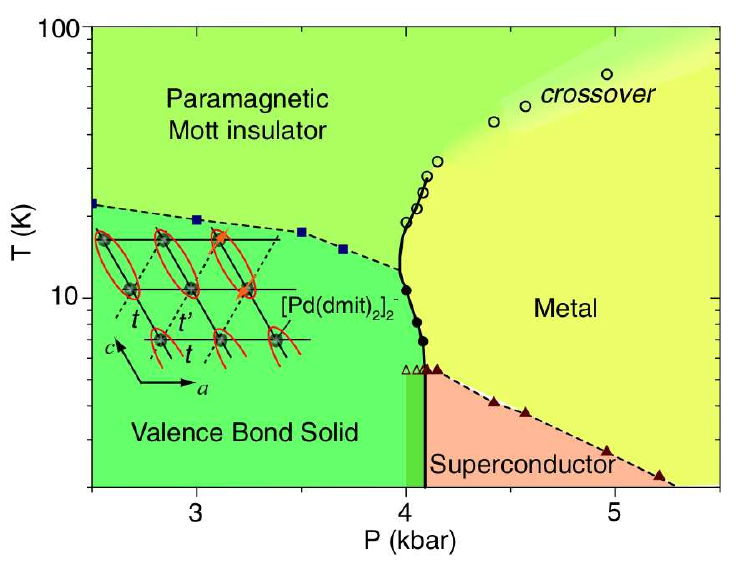}
          \put (-5,65) {\small(b)}
        \end{overpic}
      }
  }
  \caption{(a) Resistivity of EtMe$_3$P[Pd(dmit)$_2$]$_2$ as a
    function of temperature and pressure.  (b) Phase diagram. Solid
    and dashed lines denote first and second order transitions,
    respectively.  Reprinted with permission from
    Ref.~\cite{Shimizu07a}, $\copyright$ 2007 The American Physical
    Society.}
    \label{etme3p-phase}
\end{figure}  

\begin{figure}[tb]
  \center{
    \begin{overpic}[width=2.0in]{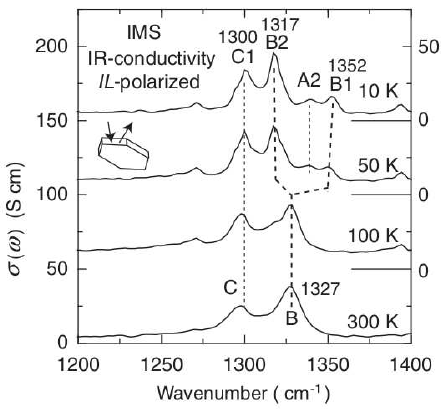}
      \put (-5,85) {\small(a)}
    \end{overpic}
    \hspace{0.1in}
    \raisebox{0.4in}{
      \begin{overpic}[width=2.5in]{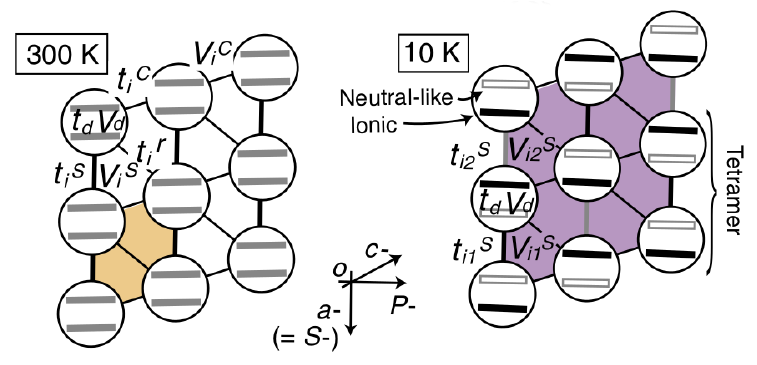}
        \put (1,50) {\small(b)}
      \end{overpic}
    }
    }
  \caption{(a) Temperature dependence of the optical conductivity of
    {\it t}-EtMe$_3$P[Pd(dmit)$_2$]$_2$. The peak splitting below 50 K
    indicates CO. (b) The pattern of CO is $\cdots1100\cdots$ along
    the stack.  Reprinted with permission from
    Ref.~\cite{Yamamoto11a}, $\copyright$ 2011 The Physical Society of
    Japan.}
  \label{etme3p-co}
\end{figure}

EtMe$_3$P[Pd(dmit)$_2$]$_2$ is unique among this class of materials
\cite{Kato06a}. Instead of the solid crossing structure found in the
$\beta^\prime$ [M(dmit)$_2$] CTS, stacks of [Pd(dmit)$_2$] molecules
in EtMe$_3$P[Pd(dmit)$_2$]$_2$ are parallel to those in other layers
as shown in Fig.~\ref{dmit-structure-3}.  Two different crystal
structures of EtMe$_3$P are known, one with monoclinic structure ({\it
  m}-EtMe$_3$P) \cite{Kato06a} and one with triclinic structure({\it
  t}-EtMe$_3$P) \cite{Yamamoto11a}.  Under ambient pressure both
variants are semiconducting at room temperature 
\cite{Kato06a,Yamamoto11a}.  Under pressure {\it m}-EtMe$_3$P is
superconducting with T$_{\rm c}$=5 K at 0.4 GPa
\cite{Kato06a,Shimizu07a}.  The resistivity under pressure is shown in
Fig.~\ref{etme3p-phase}.  At about 0.4 GPa, the resistivity drops by
three orders of magnitude at around 20 K, but then increases further
at lower temperatures.  For pressures near the appearance of SC
thermal hysteresis is observed, suggesting that the transition between
the insulating state and the metallic state appears to be first order
\cite{Shimizu07a}.  To our knowledge, experiments under pressure for
({\it t}-EtMe$_3$P) have not been reported yet.

Under ambient pressure, EtMe$_3$P[Pd(dmit)$_2$]$_2$ undergoes a
second-order phase transition to a non-magnetic ground state that
breaks translational symmetry at 25 K (50 K) in the monoclinic
(triclinic) forms
\cite{Kato06a,Tamura06a,Ishii07a,Yamamoto11a,Yamamoto14a}.  The
isotropic field dependence of the susceptibility rules out magnetic
long-range order (i.e. AFM) for the ground state \cite{Tamura06a}. The
spin gap for the monoclinic salt was estimated as $\Delta \sim$40 K
\cite{Tamura06a}.  This non-magnetic state was originally claimed to
be a valence-bond solid (VBS) state, with formation of alternating
spin-singlet inter-dimer bonds and with electron occupancy of 1 per
dimer \cite{Tamura06a}.  However, optical studies (see
Fig.~\ref{etme3p-co}) have found that CO coexists with the bond
ordering in the non-magnetic state \cite{Yamamoto11a,Yamamoto14a}.
Again, considering the individual molecules and not dimers as units,
the charge densities on the molecules have the pattern
$\cdots$0110$\cdots$ along the stacks. The individual molecules within
each dimer are different, as shown in Fig.~\ref{etme3p-co}, indicating
that the description as VBS is overly simplistic.  Very recent careful
optical studies \cite{Yamamoto17a} have clearly indicated the
$\frac{1}{4}$-filled nature of this material.  The same pattern is
also found in the CO state of $\beta$-({\it meso}-DMBEDT-TTF)$_2$X
\cite{Shikama12a} (see Fig.~\ref{beta-meso}(a)).  The amplitude of the
CO is estimated as $\delta\rho\sim 0.4 - 0.7$ in {\it t}-EtMe$_3$P and
$\sim$ 0.1 in {\it m}-EtMe$_3$P \cite{Yamamoto11a,Yamamoto14a}.

\subsubsection{$\beta^\prime$-EtMe$_3$Sb[Pd(dmit)$_2$]$_2$}

\begin{figure}[tb]
  \center{
    \begin{overpic}[width=2.3in]{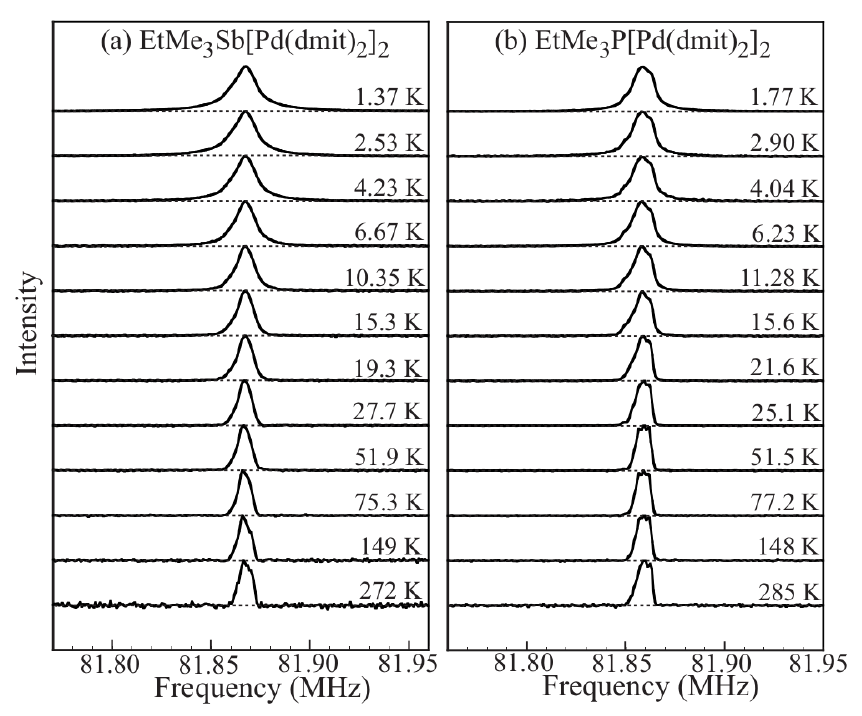}
      \put (-5,79) {\small(a)}
    \end{overpic}
    \hspace{0.1in}%
    \begin{overpic}[width=2.0in]{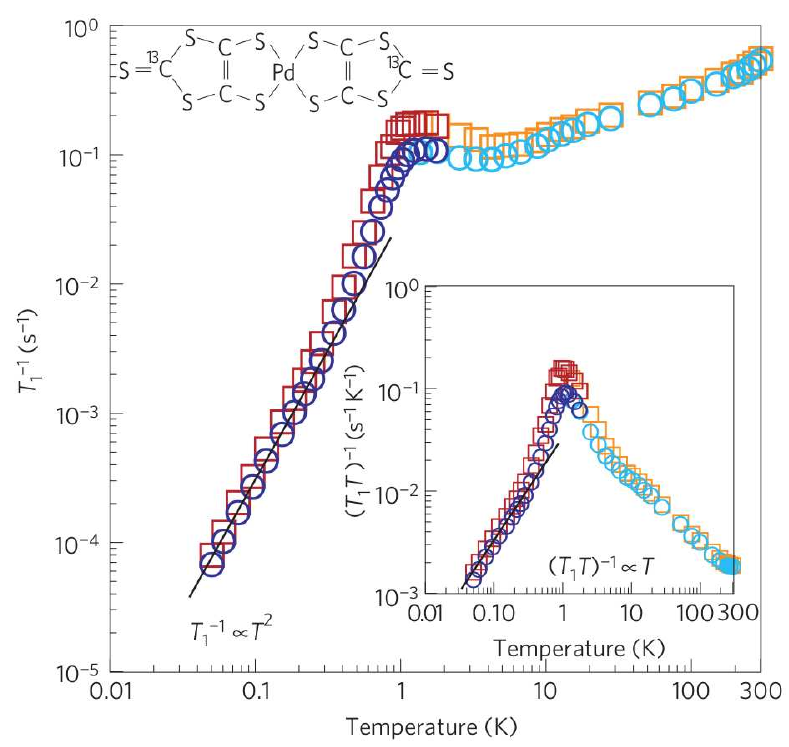}
      \put(-5,90) {\small(b)}
    \end{overpic}
    }
  \caption{(a) $^{13}$C NMR spectra of EtMe$_3$Sb and {\it
      m}-EtMe$_3$P.  Reprinted with permission from
    Ref.~\cite{Itou08a}, $\copyright$ 2008 The American Physical
    Society.  (b) relaxation time T$_1^{-1}$ versus temperature.
    Reprinted with permission from Ref.~\cite{Itou10a}, $\copyright$
    2010 MacMillan Publishers Ltd.}
  \label{etme3sb-1}
\end{figure}

EtMe$_3$Sb[Pd(dmit)$_2$]$_2$ has been the most studied CTS of this
group because of the possibility that it is an experimental
realization of a QSL \cite{Kanoda11a}. As in
$\kappa$-(ET)$_2$Cu$_2$(CN)$_3$, in EtMe$_3$Sb the $^{13}$C NMR
spectra hardly changes (see Fig.~\ref{etme3sb-1}) down to $\sim$1 K,
despite an estimated inter-dimer $J$ of 220--250 \cite{Itou08a}.  As
for Cu$_2$(CN)$_3$, the NMR line in EtMe$_3$Sb also broadens at low
temperature (broadening is also seen for {\it m}-EtMe$_3$P in the
$\delta\rho\sim$ 0.1 CO state below 25 K, see Fig.~\ref{etme3sb-1}).
A kink is found in the  $^{13}$C NMR relaxation time at $\sim$1 K
indicating some kind of continuous phase transition \cite{Itou10a}.
While (T$_1$)$^{-1}$ is consistent with gapless behavior above 1K,
the rapid decrease below 1K indicates gap formation, although the
power-law rather than exponential decrease suggests that the
gap is nodal rather than complete \cite{Itou10a}.
$^{13}$C NMR experiments using the inner instead of outer carbon atoms
of the Pd(dmit) molecule reached the same conclusion, suggesting that
the results are not affected by the cation molecules \cite{Itou11a}.

\begin{figure}[tb]
  \center{
    \begin{overpic}[width=2.1in]{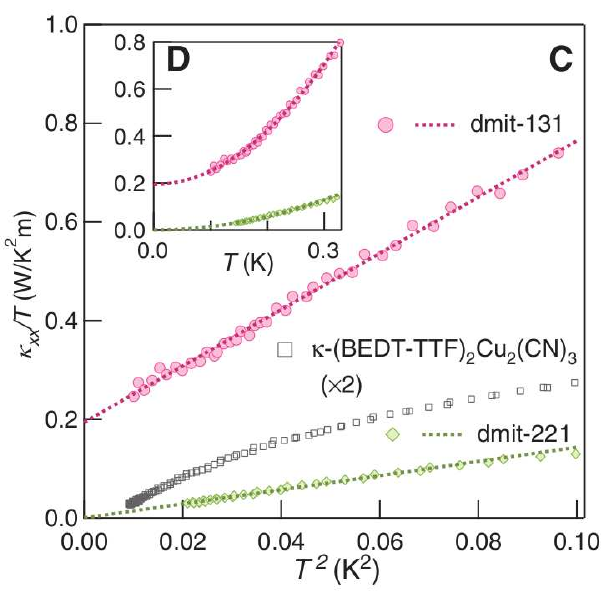}
      \put (-5,90) {\small(a)}
    \end{overpic}
    \hspace{0.1in}%
    \begin{overpic}[width=2.5in]{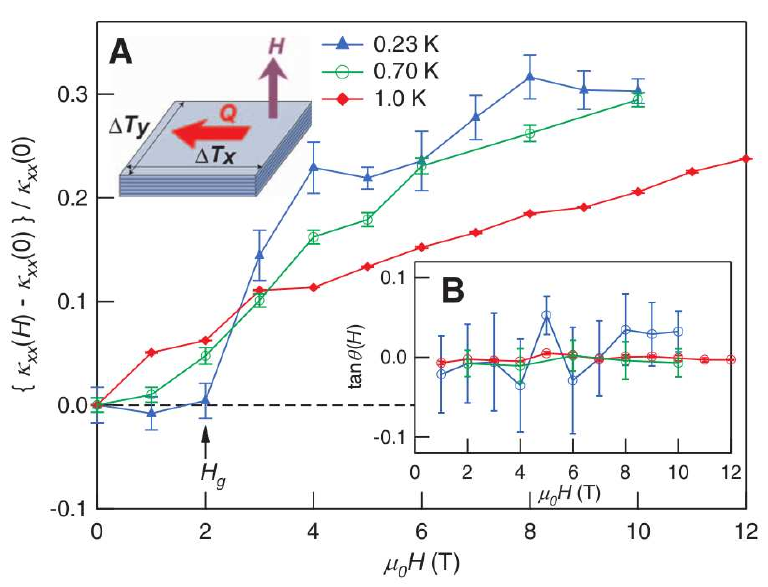}
      \put (-5,70) {\small(b)}
    \end{overpic}
  }
  \caption{(a) Temperature dependence of the thermal conductivity of
    EtMe$_3$Sb (dmit-131), Et$_2$Me$_2$Sb (dmit-221), and
    $\kappa$-(ET)$_2$Cu$_2$(CN)$_3$ (b) Magnetic field dependence of
    the thermal conductivity, normalized to its zero-field value.
    Reprinted with permission from Ref.~\cite{Yamashita10a},
    $\copyright$ 2010 AAAS.}
  \label{etme3sb-2}
\end{figure}

Thermodynamic measurements are consistent with a mix of gapped and
gapless behavior at low temperature
\cite{Yamashita10a,Yamashita11a,Yamashita12a,Watanabe12a,Nakazawa13a}.  The magnetic
susceptibility measured using a torque magnetometer, which cancels
impurity Curie contributions, remains paramagnetic to very low
temperatures (30 mK) \cite{Watanabe12a}.  The specific heat of
EtMe$_3$Sb has a finite linear in temperature term as T$\rightarrow$0,
consistent with gapless excitations \cite{Yamashita11a}.  The thermal
conductivity $\kappa$ similarly extrapolates to a finite value for
T$\rightarrow$0 (see Fig.~\ref{etme3sb-2})
\cite{Yamashita10a}. However, under a magnetic field gap-like behavior
is seen, with a sudden increase in $\kappa$ once the field exceeds a
critical value (Fig.~\ref{etme3sb-2}) \cite{Yamashita10a}.

\paragraph{Summary} A variety of crystal structures and normal state behavior
(antiferromagnetism and charge ordering) are seen in the anionic CTS,
in spite of much stronger dimerization than in the $\kappa$-(ET)$_2$X.
A charge-order to superconductivity transition is seen most clearly in
{\it m}-EtMe$_3$P [Pd(dmit)$_2$]$_2$, where the pattern of the charge
order is not that of a simple valence bond solid, which requires equal
charge densities on all the organic molecules. The charge pattern on
the molecules is very clearly $\cdots$0110$\cdots$, as expected in the
$\frac{1}{4}$-filled band. Recent experiments in the spin liquid
compound $\beta^\prime$-EtMe$_3$Sb indicate intradimer charge
fluctuations and $\frac{1}{4}$-filled behavior \cite{Yamamoto17a}.

\subsection{Summary}

Only in $\kappa$-(ET)$_2$X is superconductivity (SC) found adjacent to
antiferromagnetism.  Antiferromagnetic order is found in very few
$\kappa$-(ET)$_2$X, specifically X= Cu[N(CN)$_2$]Cl, Cu[N(CN)$_2$]Br
(deuterated), and X=CF$_3$SO$_3$. For X=CF$_3$SO$_3$, T$_{\rm N}$ is
lower than T$_{\rm c}$. There is evidence for intradimer charge
fluctuations and pseudogap at temperatures greater than T$_c$ in
$\kappa$-(ET)$_2$X.  In a greater number of 2D CTS, SC is reached upon
applying pressure to a charge-ordered or density-wave state.
Experimental phase diagrams of the anionic [M(dmit)$_2$]$_2$
superconductors are similar in many ways to the ET and other cationic
CTS. Here charge order is again found adjacent to SC, for
example in EtMe$_3$P[Pd(dmit)$_2$]$_2$.  In all cases where a
charge-order-to-SC transition occurs, and the charge order pattern is
known, it does not correspond to a Wigner crystal; rather the charge
order pattern $\cdots1100\cdots$ is common among these materials, and
there occur spin gaps at low temperature.  Finally, in general T$_{\rm
  c}$ decreases with increasing pressure, inconsistent with the
prediction of BCS theory.  We return to the theoretical interpretation
of these observations in the next Section.

\section{Broken symmetries in quasi-two dimensional CTS, Theory}
\label{2dtheory}

There is by now strong evidence that SC in the organic superconductors
cannot be described by BCS theory and that any successful theory must
incorporate the effect of e-e interactions from the beginning (see
Reference \cite{Powell06a} for a review of failures of the BCS model
in the context of the CTS superconductors). In this Section we first
review the most popular theories proposed to explain SC in the CTS.
We review numerical results for 2D systems at $\rho=1$ and argue that
understanding SC in the CTS requires going beyond concepts derived
from the RVB theory or theories of spin-fluctuation mediated SC.  We
further present a VB theory of SC that is unique to
$\rho=\frac{1}{2}$, that is inspired by the RVB theory but that is
also substantially different (even as it has conceptual overlaps with
VBS-to-SC theories).  We present results of $\rho$-dependent numerical
calculations of superconducting pair-pair correlations that strongly
support our VB theory.

\paragraph{Strongly correlated theories of SC in the CTS}

Correlated-electron theories of SC were largely born out of efforts to
explain SC in the high T$_{\rm c}$ cuprates.  Broadly speaking, one
can distinguish between two different classes of theory.  The first
class emphasizes site-diagonal density waves, AFM/SDW, and CDW, as the
mechanism for pairing. Within these approaches, SC pairing is mediated
by the coupling of charge carriers with {\it fluctuations} or
excitations out of the density wave. Within the second class of
theories, SC emerges out of a proximate {\it spin-singlet}
semiconductor. In the context of the cuprates this second theory has
been known as the RVB theory. Our theoretical approach belongs to this
class and is based on VB theory but is yet substantially different
from the RVB theory.

\paragraph{Class I theories}

In the case of theories of AFM-driven SC the excitations exchanged 
between charge carriers are spin fluctuations or magnons. This
approach has been widely used in the context of the cuprates, usually
within the model of a weakly doped Mott-Hubbard semiconductor. In its
application to the CTS, a pressure-induced increase in lattice
frustration drives  SC within the effectively $\frac{1}{2}$-filled band 
$\kappa$-phase CTS.  The chief conceptual difficulty with
spin-fluctuation mediated pairing is that unlike in BCS theory, where
the phonons that lead to an attractive interaction between electrons
are physically distinct particles from the electrons, the AFM
fluctuations assumed to bind electrons into Cooper pairs are
themselves generated by the same electrons. Most theoretical
calculations in this class of theories are based on mean-field theory.
It is however questionable as to whether the problem can be reduced to
weak coupling by the separation of magnetic and charge degrees of
freedom.
From the perspective of experiments, it is not clear 
whether the pseudogap in either the cuprates or the organics (see
Section \ref{kappa}) can be  explained within this approach, although
it is claimed that dynamical cluster approximation (DCA) results prove
that there is a pseudogap within the Hubbard model \cite{Maier00a,Gull13a,Merino14a}.
Recent experimental work on the cuprates (see Section \ref{cuprates}),
in particular the finding of a CO phase within the pseudogap phase,
are extremely difficult to explain within a weakly doped Mott-Hubbard
semiconductor picture.

In the context of the 2D CTS, there are far more materials that
exhibit a CO to SC transition than an AFM to SC transition (see
Sections \ref{2d-cationic} and \ref{2d-anionic}). In spite of this,
the bulk of theories have emphasized the $\frac{1}{2}$-filled Hubbard
model and the AFM state found there.  It has also been proposed that
charge rather than AFM fluctuations mediate SC in certain CTS
superconductors, and that SC may be found adjacent to WC CO driven by
the NN Coulomb interaction $V$ in a $\frac{1}{4}$-filled model
\cite{Merino01a}.  Here it is assumed that charge carriers couple to
``charge fluctuations'' of the WC \cite{Merino01a}, although unlike
magnons there is no equivalent theory or experimental signature of the
collective excitations out of a WC. In this theory, the application of
a slave-boson transformation to Eq.~\ref{EHM} in the limit of large
$U$ replaces the electron creation operator $c^\dagger_{i,\sigma}$
with the product of two operators $f^\dagger_{i,\sigma}b_i$, where
$f^\dagger_{i,\sigma}$ represents a fermion and $b_i$ a boson
associated with the electron charge \cite{Merino01a}. The bosons
introduced in this manner are coupled to the fermions, which is argued
to produce SC.  We will point out however in Section \ref{qtr-theory}
that in the CTS that display a transition from CO to SC, the CO phase
is in fact {\it not} a WC.

\paragraph{Class II theories}

In this class of theories, SC is assumed to result from the
destabilization of a spin-gapped or spin-liquid insulating state.
Anderson proposed that a strongly correlated system could best be
understood in a valence bond basis of spin-paired carriers
\cite{Anderson73a,Fazekas74a}.  At large $U$ and $\rho=1$ the Hubbard
model reduces to the $S=\frac{1}{2}$ Heisenberg spin Hamiltonian with
exchange constant $J\propto-t^2/U$.  Here the energy of a pair of
spins in a singlet state is $-\frac{3}{4}$J while the classical energy
of two anti-parallel spins is $-\frac{1}{4}J$.  Thus, in a 1D $\rho=1$
system where each site has two nearest neighbors, the spin-paired SP
state is lower in energy than an antiferromagnetically ordered
state. A given spin can only be in a singlet bond with one neighbor,
which limits the energy gain per site in the singlet valence bond
state. Once the number of neighbors is large enough (for example the
square lattice with four nearest neighbors), the AFM state becomes
lower in energy than the valence bond singlet state.  Anderson and
Fazekas proposed a RVB wavefunction consisting of a superposition of
valence bonds as a variational wavefunction for the Heisenberg model
on the triangular lattice \cite{Anderson73a,Fazekas74a}.  The actual
ground state however, turned out to not be a singlet, but rather the
120$^\circ$ AFM state
\cite{Huse88a,Miyake92a,Bernu92a,Elstner93a,Capriotti99a}. The idea
remained however, that the $\rho=1$ system with large $U$ must be
close to a spin-gapped RVB state.

Doping the $\frac{1}{2}$-filled band square lattice destroys the
AFM. Anderson further proposed that in this case the RVB state would
become the ground state, and that this would be a superconductor
\cite{Anderson87a}.  In theories of RVB SC applied to
$\kappa$-(ET)$_2$X, the assumption is made that the same or somewhat
modified theory applies to a $\frac{1}{2}$-filled anisotropic
triangular lattice of BEDT-TTF dimers, where the controlling parameter
responsible for destroying AFM is lattice frustration rather than
doping (see below).  Such a theory in principle naturally explains
preformed pairs in the pseudogap phase of the cuprates
\cite{Chatterjee11a,Mishra14a} and probably also in the CTS.  There
are several problems however: first, no singlet state has been seen in
calculations on the $\frac{1}{2}$-filled model.  Second, we show below
that numerical studies on frustrated lattices in general find no SC
either in the $\frac{1}{2}$-filled band.  The most difficult
experimental observation to explain is how to extend the $\rho=1$ RVB
theory to account for those CTS where a CO to SC transition is seen.
This same problem has become relevant in the context of the cuprates
since the discovery of CO phases \cite{Keimer15a}.  Some authors have
proposed the concept of a WC of Cooper pairs
\cite{Anderson04b,Franz04a,Tesanovic04a,Tesanovic04a,Chen04a,Vojta08a,Hamidian16a,Cai16a,Mesaros16a}.
The attractive feature of this idea is that it can explain both the
presence of preformed pairs (giving the pseudogap) and an apparently
competing phase (CO). The key problem however is that the precise
nature of such an ordered state remains unknown.

The idea of SC emerging from a proximate spin-singlet has nevertheless
remained popular. For instance, it has been proclaimed by multiple
authors that SC emerges from a 2D SP state
\cite{Hirsch87a,Imada91a}. The problem is that once again no such SP
state in 2D is known. The only case where interactions give a $\rho=1$
$S=0$ singlet state outside 1D are even-leg ladders.  For doped
two-leg ladders numerical studies using DMRG do find dominant
superconducting correlations \cite{Noack97a,Dolfi15a}.  Theories of
ladders however cannot be extended to 2D. To the best of our knowledge
there is no satisfactory theory of $S=0$ in 2D except ours that we
present below.

\paragraph{Our VB theory of SC}

The basis of our theory is the observation that a PEC is particularly
stable in frustrated $\rho=\frac{1}{2}$ systems. The frustration
destroys the WC of single electrons, as well as the AFM that may occur
in strongly dimerized $\rho=\frac{1}{2}$. The occurrence of the PEC in
2D uniquely at $\rho=\frac{1}{2}$ can be understood physically.  Only
at the carrier density of $\rho=\frac{1}{2}$ such a paired CO, in
which singlet pairs are separated by pairs of vacancies, is
commensurate, conferring it the exceptional stability that is
necessary to dominate over both the metallic state as well as the WC
of single electrons driven by the NN Coulomb interaction $V$.  In
other words, at this density there exists a range of parameters where
there is a balance between the exchange energy gained from
spin-singlet formation and the repulsive interaction between the
electrons.  Alternatively, this state is the equivalent of a 2D SP
state, with nearly decoupled 1D insulating stripes separated by
stripes of vacancies (see Fig.~\ref{thetaco}(b) and below). We have
called this paired WC the Paired Electron Crystal (PEC). In analogy
with theories of VBS-to-SC transitions, we believe that the
spin-singlet pairs of the PEC can behave as hard-core bosons, and the
``melting'' of the PEC, driven by further frustration or very weak
doping, then gives SC (the key difference between our work and
proposed VBS-to-SC transition theories is that we will demonstrate the
transition to the PEC numerically). Note that the gain in kinetic
energy from the crystal-to-fluid transition in such a lattice of hard
core bosons is largest when the number of bosons is exactly half the
number of effective sites. This is also $\rho=\frac{1}{2}$ in terms of
the electrons that constitute the spin-singlets.  The chief advantage
of our theory is that it has the potential of explaining SC in all CTS
within a unified approach, irrespective of whether the broken symmetry
in the insulating state is proximate to SC is AFM or CO.

In the next sections we present the results of precise numerical
calculations to substantiate the above. We will begin with
calculations for the $\rho=1$ Hubbard model, within which we show the
absence of SC for any correlations or frustration. Following this, we
present numerical results demonstrating the PEC and correlation-driven
enhancement of superconducting pair-pair correlations uniquely for
$\rho=\frac{1}{2}$.

\subsection{Half-filled band}
\label{half-theory}

The simplest $\rho=1$ model in the context of the 2D CTS
superconductors is the Hubbard model on the anisotropic triangular
lattice with hopping integrals $t$ and $t^\prime$ (see the inset,
Fig.~\ref{rho1-anisotropic}(a)).  This lattice becomes equivalent to
the simple square lattice when $t^\prime=0$.  We have performed exact
calculations for the 4$\times$4 periodic lattice, and Path Integral
renormalization group (PIRG) calculations for periodic 6$\times$6 and
8$\times$8 lattices, for $t^\prime$=0.2, 0.5, and 0.8
\cite{Clay08a,Dayal12a}, whose results are discussed below.  The PIRG
method is an unbiased method free from the fermion sign problem which
approximates the ground state wavefunction as a sum of Slater
determinants \cite{Kashima01b}.  Within each basis size PIRG is
strictly variational and variance extrapolation is used to remove the
basis size bias \cite{Kashima01b}.

\subsubsection{Magnetic Phases}
\begin{figure}[tb]
  \begin{center}
    \begin{overpic}[width=2.75in]{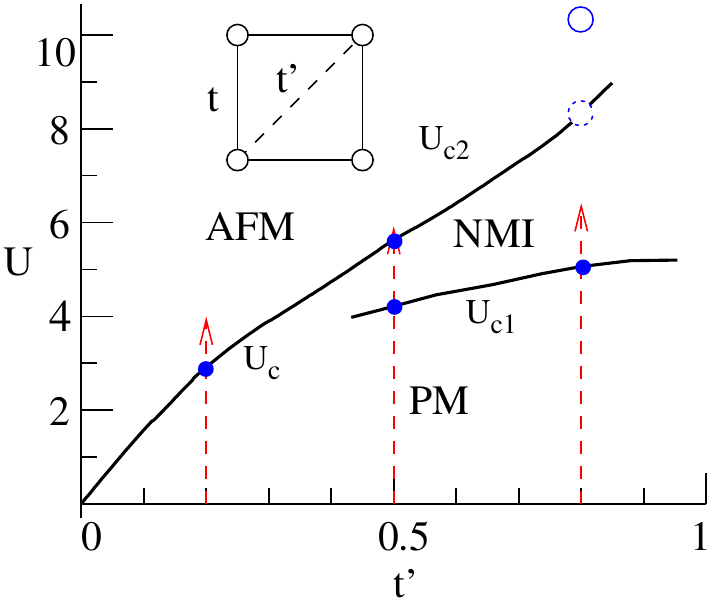}
      \put(-5,80)  {\small(a)}
    \end{overpic}
    \hspace{0.2in}%
    \raisebox{0.3in}{
      \begin{overpic}[width=1.7in]{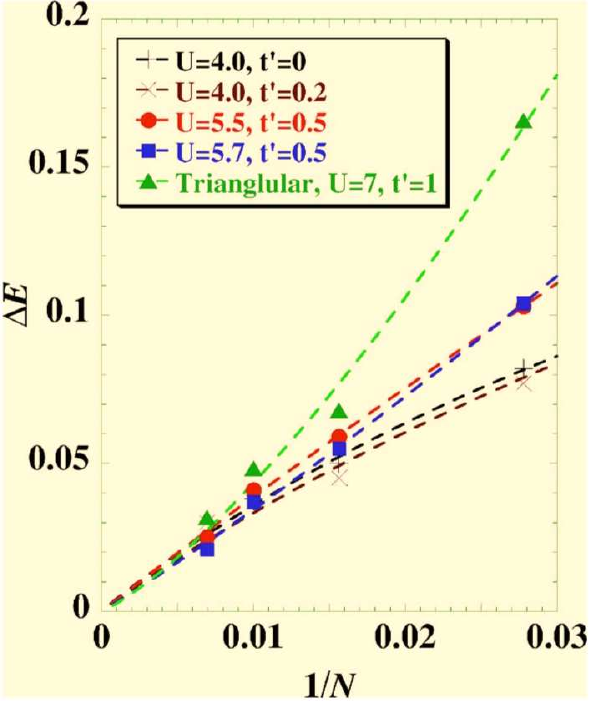}
        \put(-7,95) {\small(b)}
      \end{overpic}
      }
  \end{center}
  \caption{(color online) (a) Phase diagram of the $\rho=1$ Hubbard
    model on the anisotropic triangular lattice \cite{Dayal12a}.
    Filled circles are finite-size scaled values for the phase
    boundaries. At $t^\prime=0.8$ the solid circle is an exact upper
    bound for the phase boundary, and the dashed circle shows the
    expected finite-size scaled value. The lines linking the points
    are schematic guides to the eye. The dashed lines with arrows show
    the parameters chosen for pair-pair correlation calculations (see
    Fig.~\ref{pirg-half-filled}).  A 120$^\circ$ AFM phase not shown
    here exists for large $U$ at $t^\prime=1$ (see text).  (b)
    Finite-size scaling of the singlet-triplet gap for $N$ site
    lattices using the PIRG method \cite{Mizusaki06a}.  Circles,
    squares, and triangles are for parameters within the NMI phase.
    Triangles are data for the the lattice with a single diagonal
    $t^\prime$ bond as in panel (a) and others are for the lattice
    with two $t^\prime$ bonds. The dashed lines are fits to the data.
    Reprinted with permission from Ref.~\cite{Mizusaki06a},
    $\copyright$ 2006 The American Physical Society.}
 \label{rho1-anisotropic}
\end{figure}

In the $t^\prime=0$
limit AFM order is found in the ground state for all $U>0$
as shown in the phase diagram in Fig.~\ref{rho1-anisotropic}(a)
(for a comparison of different calculations of the phase diagrams
see Fig.~16 of Reference  \cite{Hotta12a}).
As $t^\prime/t$ increases frustration reduces the 
region of the phase diagram over which AFM dominates,
which occurs for $U>U_{\rm c}$ with finite $U_{\rm c}$ for
$t^\prime/t>0$. At $t^\prime/t\approx0.5$ a third phase enters,
referred to as either a non-magnetic insulator (NMI) or QSL
\cite{Morita02a}.  Compared to what is seen on finite size lattices,
in the thermodynamic limit the NMI phase extends to a smaller $U$
\cite{Morita02a,Dayal12a}.  The PIRG method has been used by several
authors to study the phase diagram of this model
\cite{Kashima01b,Morita02a,Watanabe03b,Mizusaki06a,Dayal12a}.  These
calculations find that the NMI phase extends for $U$ as small as 4.0
at $t^\prime$=0.5 \cite{Dayal12a}.  PIRG calculations have further
shown that in the NMI phase there is no dimerization, plaquette
singlet state, staggered flux state, or charge order
\cite{Watanabe03b}.  As shown in Fig.~\ref{rho1-anisotropic}(b), the
NMI phase also appears to be gapless. For these reasons the NMI phase
was proposed to explain the QSL behavior seen in
$\kappa$-(ET)$_2$Cu$_2$(CN)$_3$ \cite{Morita02a}.

In the isotropic limit ($t^\prime/t=1$) three successive phases are
found as $U$ increases \cite{Yoshioka09a}. For $U<U_{\rm c1}\sim$
7.4$t$ the ground state is PM; for $U_{\rm c1}<U<U_{\rm c2}\sim 9.2t$
the ground state is the NMI; and for $U>U_{\rm c2}$ the ground state
is a 120$^\circ$ N\'eel ordered state \cite{Yoshioka09a}.
Recent variational Monte Carlo (VMC) calculations find a similar value
$U_{\rm c2}\sim 8$ \cite{Tocchio14a}.  As
discussed in Section \ref{kappa}, 120$^\circ$ AFM ordering has not
been observed experimentally in any CTS, despite
the existence of a wide range of $t^\prime/t$ for different X (see
Table \ref{kappa-table} in Section \ref{kappa}). The range of $t^\prime/t$ experimentally investigated
is further expanded by experiments under pressure.  Assuming that the
effective dimer model continues to be valid for
$\kappa$-(ET)$_2$X for large $U$ and $t^\prime/t\approx 1$, this
implies that the effective dimer $U$ must be less than about 9.
 The region of the phase diagram
for $U\sim U_{\rm c2}$ close to $t^\prime/t=1$ is difficult for numerical
methods to access. Different methods give somewhat different ground
states in this region, involving 120$^\circ$ AFM, QSL, and AFM states
\cite{Tocchio14a,Laubach15a,Acheche16a,Goto16a}. Given that
the effective $U$ here is likely larger than that 
in $\kappa$-(ET)$_2$X, it is unlikely that this region of
parameters is relevant for any 2D CTS.

\subsubsection{Superconductivity: suppression of superconducting pair-pair correlations}

Numerous mean-field, cluster DMFT, and variational calculations have
suggested that $d$-wave SC occurs within the $\rho=1$ Hubbard model on
the anisotropic triangular lattice (or on the related lattice with two
frustrating diagonals) usually near the boundary of the metallic and
insulating phases
\cite{Kino98a,Schmalian98a,Kondo98a,Vojta99a,Baskaran03a,Liu05a,Kyung06a,Yokoyama06a,Watanabe06a,Sahebsara06a,Nevidomskyy08a,Sentef11a,Hebert15a}.
In these works the mechanism of SC is assumed to be AFM spin
fluctuations or RVB.  A similar SC phase is also suggested for the
related Hubbard-Heisenberg model (see below).

Superconducting pair-pair correlations calculated numerically
from unbiased methods provide a way to investigate  SC. We define
singlet  pair creation  operators
\begin{equation}
  \Delta^\dagger_i=\frac{1}{N_\nu}\sum_\nu g(\nu)\frac{1}{\sqrt{2}}(c^\dagger_{i,\uparrow}c^\dagger_{i+\vec{r}_\nu,\downarrow}
  -c^\dagger_{i,\downarrow}c^\dagger_{i+\vec{r}_\nu,\uparrow}),
  \label{ppdelta}
\end{equation}
where $g(\nu)$ determines the pairing symmetry.  For $d_{x^2-y^2}$
symmetry, $g(\nu)=\{1,-1,1,-1\}$ for
$\vec{r}_\nu={\hat{x},\hat{y},-\hat{x},-\hat{y}}$, respectively.  For
$d_{xy}$ symmetry, $g(\nu)=\{1,-1,1,-1\}$ for
$\vec{r}_\nu={\hat{x}+\hat{y},-\hat{x}+\hat{y},-\hat{x}-\hat{y},\hat{x}-\hat{y}}$,
respectively.  $N_\nu$ is a normalization factor, with $N_\nu=2$ for a
superposition of four singlets.  The equal-time pair-pair correlation
function is then defined as
\begin{equation}
  P(r)\equiv P(|\vec{r}_i-\vec{r}_j|)=\langle \Delta^\dagger_i \Delta_j\rangle
  \label{pr}
\end{equation}
Two criteria must be satisfied in the ground state for SC to exist:
First, the pair-pair correlations must extrapolate to a finite value
at long range ($r\rightarrow\infty$).  Using currently available
unbiased methods it is unfortunately not possible to perform a full
finite-size extrapolation of pair-pair correlations.  Second, if the
SC pairing is to be driven by e-e interactions, the pair-pair
correlations must be enhanced over their $U=0$ value. This second
criterion can be tested with precise calculations of pair-pair
correlations. We show below the results of 4$\times$4 exact and up to
8$\times$8 PIRG calculations.

Exact diagonalization calculations on a 4$\times$4 lattice at $\rho=1$ show
that the second criteria is not met for any value of $U$ and $t^\prime$,
excluding very short-range correlations:
$P(r)$ decreases {\it monotonically} with $U$ in all cases
\cite{Clay08a}.
\begin{figure}[tb]
  \centerline{\resizebox{3.0in}{!}{\includegraphics{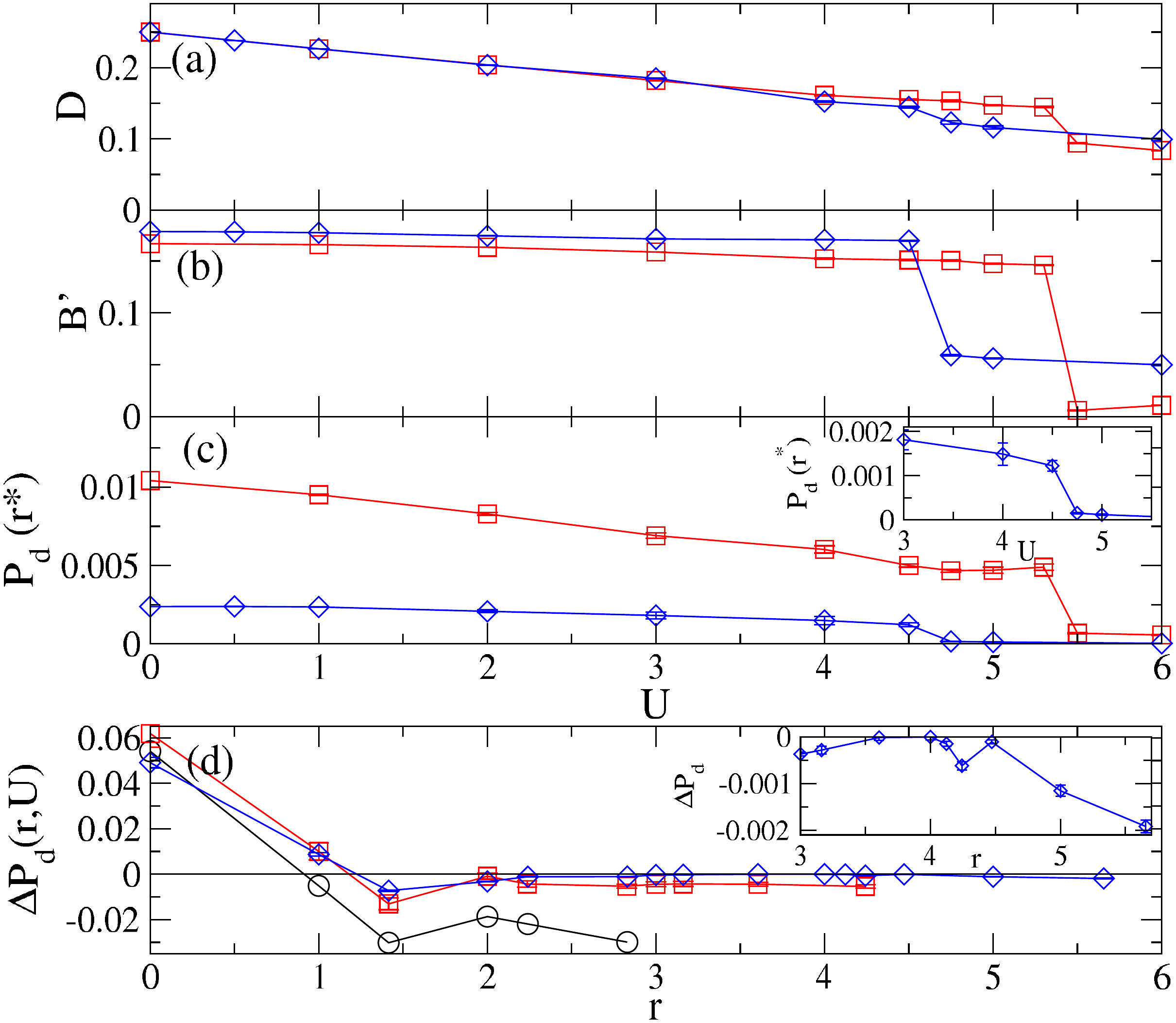}}}
  \caption{(color online) QP-PIRG results for the $\rho=1$ Hubbard
    model on the anisotropic triangular lattice with $t^\prime$=0.5
    \cite{Dayal12a}.  Circles, squares, and diamonds are for
    4$\times$4, 6$\times$6, and 8$\times$8 lattices, respectively. (a)
    Double occupancy $D$ (b) $t^\prime$ bond order $B^\prime$ (c) $d_{x^2-y^2}$
    pair-pair correlation $P_{\rm d}(r^\star)$ for the next-to-furthest distance $r^\star$
    on each lattice (d) enhancement of pair-pair correlations over the non-interacting system as
    a function of distance $r$ for $U=4.5$.}
  \label{pirg-half-filled}
\end{figure}  
Early PIRG calculations suggested that at least for $U<4$ pair-pair
correlations did not show long-range order \cite{Morita02a}.
Reference \cite{Dayal12a} calculated pair-pair correlations using an
improved method, QP-PIRG, that includes space and spin symmetries that
significantly improve the accuracy \cite{Mizusaki04a}.
Fig.~\ref{pirg-half-filled} shows results for $t^\prime=0.5$ for up to
8$\times$8 lattices \cite{Dayal12a}.
Fig.~\ref{pirg-half-filled}(a)-(b) show a discontinuous drop in the
double occupancy and bond order along the diagonal bond direction,
which indicates the transition from the PM to AFM state as $U$
increases. Fig.~\ref{pirg-half-filled}(c) shows the long-range
$d_{x^2-y^2}$ pair-pair correlation versus $U$, which decreases
monotonically with $U$ with a further drop at the MI transition.
Fig.~\ref{pirg-half-filled}(d) plots $\Delta P_d(r,U)=P_u(r,U)-P_u(r,U=0)$.
Except for the very short range region, the pair-pair correlation
function is always smaller for nonzero $U$ than at $U=0$. Results for
other values of $t^\prime$ are similar \cite{Dayal12a}.  It is
important to note that the pair-pair correlations are trivially
enhanced by $U$ for the $r=0$ point, and for some parameters for $r=1$
(see Fig.~\ref{pirg-half-filled}(d)). At such short distances however,
the pairs physically overlap. In particular, at $r=0$ the
$d_{x^2-y^2}$ pair-pair correlation function reduces to a linear
combination of spin and charge correlations \cite{Aimi07a}.  The
increase of the anti-parallel spin correlation in the AFM state causes
the $r=0$ $d_{x^2-y^2}$ correlation to increase
\cite{Aimi07a,Dayal12a}. This short-range increase with $U$ does not
imply true SC but is perhaps the reason why mean-field or cluster DMFT
methods find SC.

VMC calculations provide one way to investigate larger lattices.
However, in VMC a specific form of the wavefunction must be assumed,
and there is no way to assess the accuracy of the assumed functional
form besides the comparison of the variational energies.  Even with
small differences in the variational energy there can be large
differences in correlation functions.  Some VMC
calculations have claimed SC within the $\rho=1$ model
\cite{Yokoyama06a,Watanabe06a}, but several more recent calculations have not
found SC \cite{Tocchio09a,Tocchio14a}. It is of particular note that
the authors of Reference \cite{Watanabe06a} after improving their
trial wavefunction found that SC was no longer favored
\cite{Watanabe08a}.  Similarly, other groups using VMC also found that
the tendency to SC went {\it down} after improving their wavefunctions
\cite{Tocchio09a,Tocchio14a}.

Due to lack of space we have not discussed the lightly doped
half-filled model ($\rho\lessapprox 1$), where many of the same issues
regarding SC exist.  Recently several authors using the VMC
\cite{Yokoyama13a,Yanagisawa16a,Tocchio16a} and DCA \cite{Gull13a}
methods have suggested that SC is present in the Hubbard model for
$\rho\sim 0.8$ only when $U>U_c$, with $U_c \sim $ 4--6 $t$. While our
PIRG calculations are limited to $U\leq6t$, we remark that in exact
diagonalization calculations for the $\rho=1$ Hubbard model on the
anisotropic triangular lattice \cite{Clay08a}, there was no sign of
any enhancement of SC at large $U$; pairing correlations continued to
decrease with $U$ up to the largest values of $U$ considered,
$U\sim20t$.  Furthermore, we note that recent DMRG calculations for
the Hubbard Hamiltonian with $\rho=0.875$ on cylinders of width 4 and
6 sites found that pairing correlations decay exponentially with
distance even for $U$ as large as 8$t$ \cite{Ehlers17a}.

\subsubsection{$\rho=1$ $U$-$t$-$J$ models}

It has been claimed that the mean-field treatment of the simple
Hubbard model does not correctly describe AFM (spin-spin)
correlations, but a mean-field treatment of Hubbard-Heisenberg
($U$-$t$-$J$) model does
\cite{Gan05a,Powell05a,Gan06a,Powell07a,Guertler09a}.  The $U$-$t$-$J$
model includes an additional interaction between electron spins of the
form $J_{ij}\sum_{\langle ij\rangle} \vec{S}_i\cdot\vec{S}_j$, with
$J_{ij}$ an independent interaction from the Hubbard $U$.  Proponents
of this model find SC within a wavefunction obtained by a Gutzwiller
projection on a BCS wavefunction, which is a wavefunction of the RVB
form \cite{Powell05a,Gan05a}.  Other methods used to treat this model
include slave-rotor theory, which finds a region of VBS order in
addition to metal, QSL, and superconducting phases \cite{Rau11a}.  The
authors proposed that the calculated VBS order explained the observed
CO in EtMe$_3$P[Pd(dmit)$_2$]$_2$ (see Section \ref{etme3p}).

Exact calculations within the $U$-$t$-$J$ model again show no sign of
enhancement of pair-pair correlations over the noninteracting results
\cite{Gomes13a}. No tendency towards VBS order was found either
\cite{Gomes13a}.  The AFM region of the phase diagram of
Fig.~\ref{rho1-anisotropic}(a) gets broader for $J\neq0$ but no SC
emerges.  If VBS order is present in the $U$-$t$-$J$ model, it would
also be expected to be found {\it with greater tendencies} in the pure
Heisenberg model on frustrated lattices. The most recent large-scale
numerical results for the $J_1$-$J_2$ Heisenberg model do not find any
evidence for a VBS phase \cite{Jiang12a,Hu13a}.  This strongly
suggests that the ``valence-bond'' insulator phase in
EtMe$_3$P[Pd(dmit)$_2$]$_2$ (see Section \ref{etme3p}) requires a
quarter-filled description in order be understood.  This last
observation is in agreement with the latest work of Yamamoto {\it et
  al.}, who have concluded that this material should be considered a
$\frac{1}{4}$-filled band system \cite{Yamamoto17a}.  We consider the
properties of $\frac{1}{4}$-filled band systems in the next Section.

\subsection{Quarter-filled band}
\label{qtr-theory}

Since SC is found in the CTS at fixed carrier concentration $\rho$,
and since in the previous subsection we have already seen that the
effective $\frac{1}{2}$-filled model does not give enhanced
superconducting correlations, the only choice is then to investigate
the interacting $\frac{1}{4}$-filled band.  We have already argued
against the idea of charge fluctuation-induced SC \cite{Merino01a}, as
such fluctuations out of the WC are simply not defined.  Another
approach taken by several authors is to argue that the basic principle
of spin-fluctuation (Class I theories above) mediated SC in the
$\frac{1}{2}$-filled band is correct, but that incorporating the
actual bandfilling and crystal structure of the 2D CTS is necessary to
reproduce details of the superconducting state, such as the pairing
symmetry or location of nodes of the order parameter
\cite{Kondo01a,Kuroki06a,Sekine13a,Guterding16a,Guterding16b}.  We do
not see, given the failures of the spin-fluctuation approach we have
pointed out above in the $\frac{1}{2}$-filled model, how any
modification of this approach can lead to a successful theory of
correlated electron SC.  Going beyond the spin-fluctuation model
leaves the idea of SC emerging from spin singlet VB states as the only
option. In the rest of the section we show how such a state emerges in
the interacting $\frac{1}{4}$-filled band.

\subsubsection{Coupled spin-singlet and charge order at $\rho=\frac{1}{2}$}
\label{zigzag}

\begin{figure}[tb]
  \begin{center}
    \resizebox{3.0in}{!}{\includegraphics{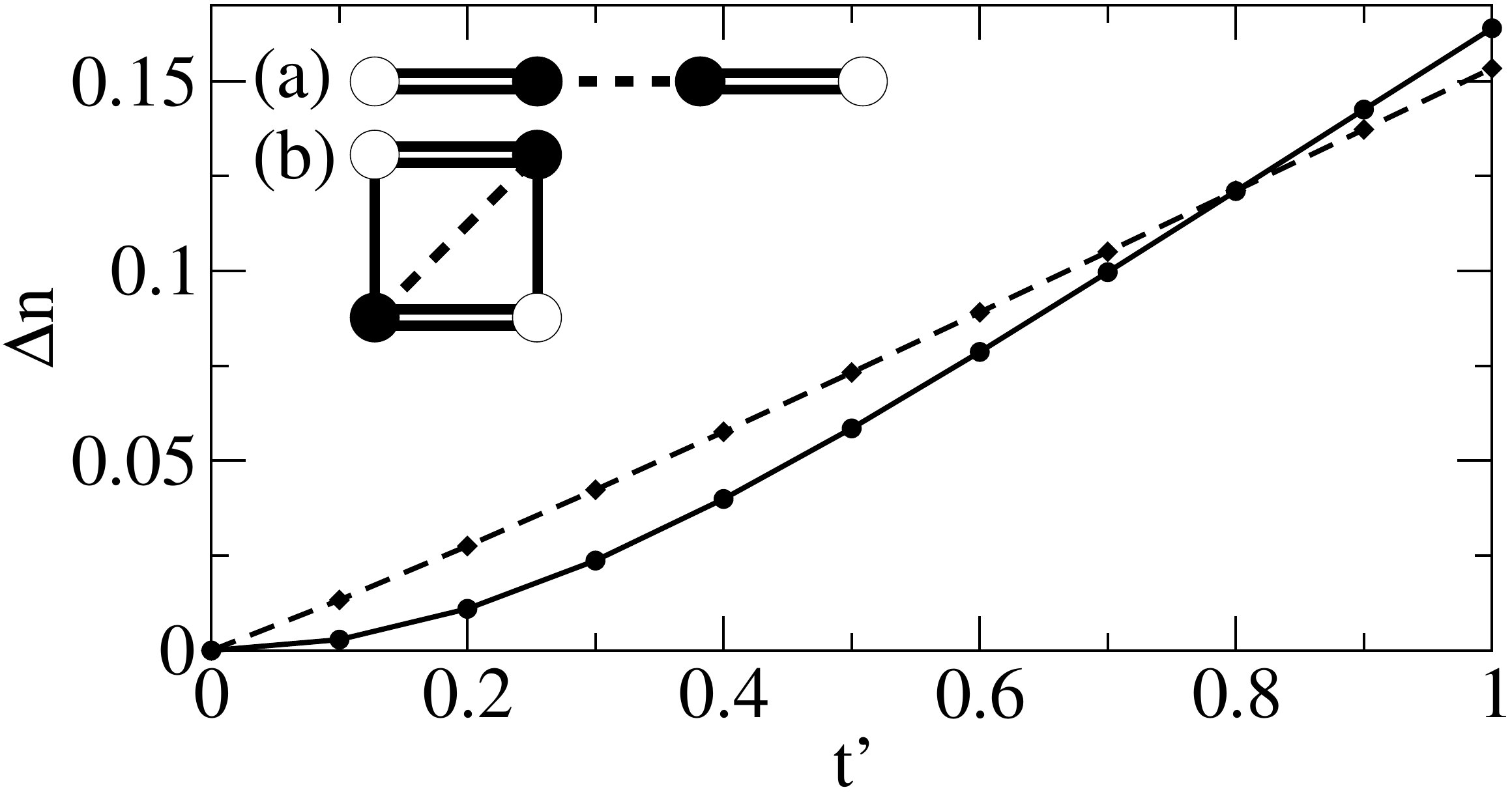}}
    \end{center}
  \caption{Charge difference $\Delta$n as a function of the hopping
    $t^\prime$ for 4-site clusters. The intradimer (double line) bonds
    have strength $t_1=1.5$ and the single bonds in (b) are $t_2$=0.5.
    The dashed line has strength $t^\prime$.  Interaction parameters
    in \ref{EHM} are $U$=4 and $V$=0.  Filled and empty circles are
    sites with charge densities 0.5+$\frac{\Delta\rm{n}}{2}$ and
    0.5-$\frac{\Delta\rm{n}}{2}$, respectively.  Solid (dashed) curves
    are for clusters (a) ((b)) \cite{Dayal11a}.}
  \label{4sites}
\end{figure}

In Section \ref{broken-1d} we showed the occurrence of the stable
BCDW-I and II states in the interacting 1D $\frac{1}{4}$-filled
band. These states are the natural 2k$_{\rm F}$ states in 1D, but can
also be thought to emerge because of the strong tendency to form spin
singlets in the $\frac{1}{4}$-filled band, where the periodic
arrangement of the spin singlets is stabilized because of the high
order commensurability of BCDW. There is a natural tendency to have
co-operative charge migration and spin-singlet formation in the
$\frac{1}{4}$-filled band. In Fig.~\ref{4sites} we consider two
coupled dimers with two electrons. The wavefunction for large $U$ can
be written as $\frac{1}{2}|1010 + 1001 + 0110 + 0101\rangle$.  As the
bond coupling the dimers (dashed line in Fig.~\ref{4sites}(a)) is
turned on, the energy of the configuration 0110 is lowered relative to
the other three, via antiferromagnetic superexchange. We assume here
that the relevant Hamiltonian is the EHM (Eq.~\ref{EHM}) with $V<V_c$
so that the Wigner crystal WC CO is of higher energy (for the
triangular lattice with isotropic $V$, the classical energies of the
PEC and WC of single electrons are equal, and upon taking into account
the quantum effects due to electron hopping this statement is
trivially true \cite{Clay02a}).  The stronger contribution by 0110 to
the overall wavefunction, leads to charge disproportionation between
the outer and inner sites in Fig.~\ref{4sites}(a).  The same effect
also takes place between the coupled dimers in the lattice of
Fig.~\ref{4sites}(b).  Fig.~\ref{4sites} shows the resulting $\Delta$n
calculated for both 4-site systems for $U=4$ and $V=0$ as a function
of $t^\prime$, the bond coupling the two dimers. As $t^\prime$ becomes
stronger $\Delta$n increases.

\begin{figure}[b!]
  \begin{center}
    \raisebox{0.5in}{
      \resizebox{1.5in}{!}{\includegraphics{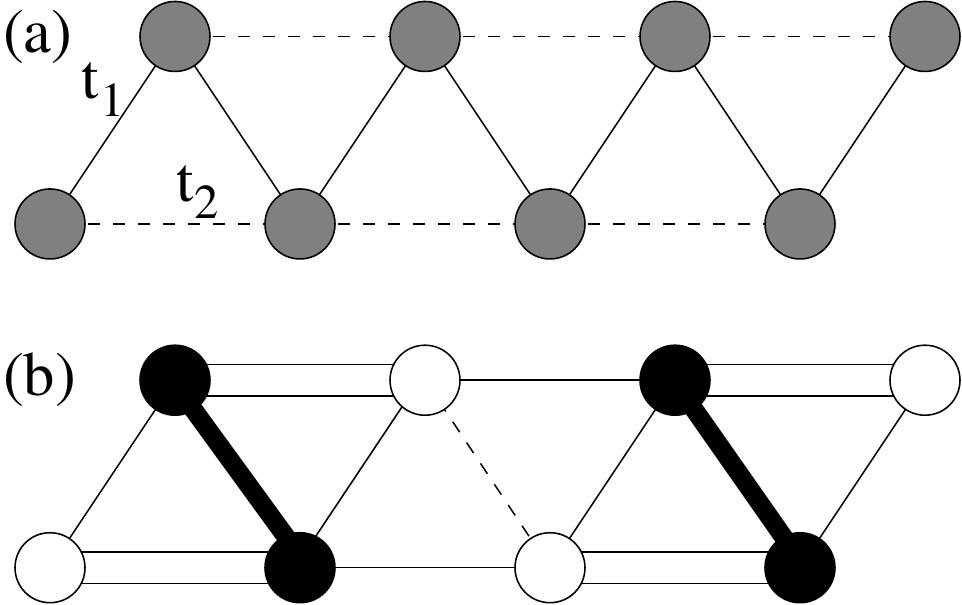}}\hspace{0.1in}
    }
    \resizebox{2.5in}{!}{\includegraphics{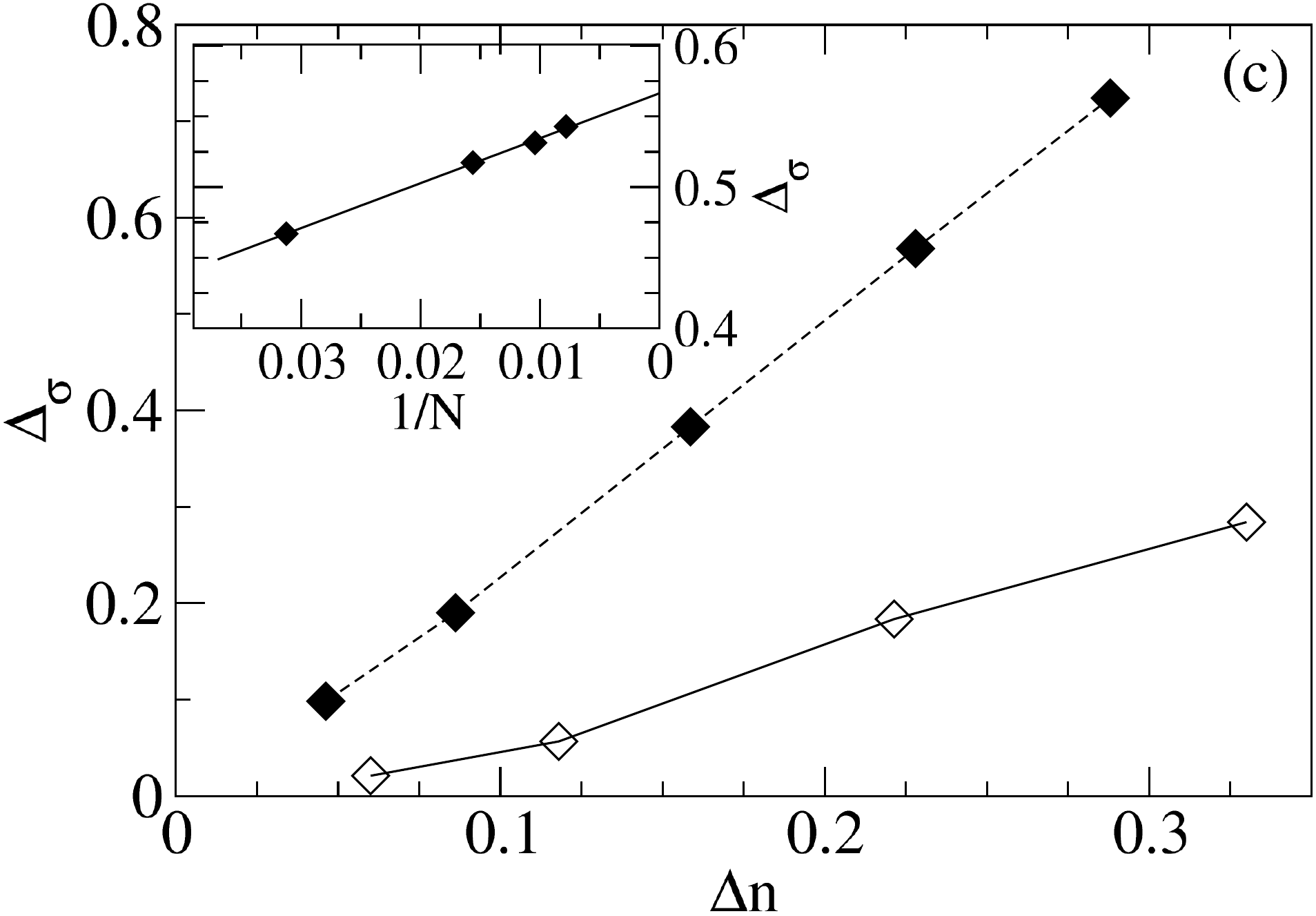}}
  \end{center}
  \caption{Zigzag ladder lattice with $\rho=\frac{1}{2}$ (a) The
    ground state with $t_2/t_1>1.707$ has uniform charges and bonds
    (b) For $t_2/t_1<1.707$ the ground state is a PEC with bond order
    of the BCDW-II type along the zigzag direction. (c) Finite-size
    scaled spin gap $\Delta_\sigma$ for $t_2/t_1=1.43$, $U=6$, and
    $V=1$ versus $\Delta$n for the zigzag ladder (filled symbols) and
    in a 1D chain with the same $U$ and $V$ (open symbols).  The inset
    shows the finite-size scaling for the zigzag ladder with
    $|\epsilon|$=0.3.  The unit of energy is taken as the average
    hopping integral along the zigzag direction (see text)
    \cite{Clay05a}.}
  \label{zzladder}
\end{figure}

An extended system that shows a broken symmetry very similar to the
4-site square cluster of Fig.~\ref{4sites}(b) is the zigzag ladder of
Fig.~\ref{zzladder} \cite{Clay05a,Clay12a}.  This lattice has two
different bonds with $t_1$ along the rung (zigzag) direction of the
lattice and $t_2$ along the rail direction.  The zigzag ladder is also
of interest at $\rho$=1, since for $t_2 \geq$ 0.25 $t_1$ the ground
state is a valence bond solid \cite{Chitra95a,White96a}. We have shown
that a coexisting CO-singlet occurs at $\rho=\frac{1}{2}$ for $t_2 < 1.707
t_1$.  (see Fig.~\ref{zzladder}) \cite{Clay05a,Clay12a}.  The energy
dispersion in the noninteracting limit is
\begin{equation}
E(k) = -2t_1\cos(q)-2t_2\cos(2q),
\end{equation}  
where $q$ refers to a wavenumber defined along the $t_1$ direction.
The topology of the bandstructure for $\rho=\frac{1}{2}$ changes when
$t_2/t_1=(t_2/t_1)_c=(2+\sqrt{2})/2=1.707$.  For $t_2/t_1<1.707\ldots$ there are two
points on the Fermi surface, while for $t_2/t_1>1.707$ there are
four. In the presence of e-p coupling a Peierls distorted state
therefore occurs when $t_2/t_1<1.707$, and as in the pure 1D case this
distorted state has $\cdots$1100$\cdots$ CO.

The distorted state in the zigzag ladder is shown in
Fig.~\ref{zzladder}(b).  The bond pattern along the zigzag direction
is the same as that of BCDW-II in the purely 1D case, with the
strongest bond between the two large charges \cite{Clay12a}. This
distorted state continues to exist in the presence of e-e
interactions, with the same bond and charge distortion pattern
\cite{Clay05a,Clay12a}. As in the 1D BCDW-II, the amplitude of charge
order can be quite large, accompanied by a large spin
gap. Fig.~\ref{zzladder}(c) shows the finite-size scaled spin gap
$\Delta_\sigma$ as a function of the charge disproportionation
$\Delta$n for the zigzag ladder, compared to a 1D chain with the same
$U$ and $V$.  In the calculation of Fig.~\ref{zzladder}(c) the unit of
energy is taken as the average hopping integral along the zigzag
direction, $\frac{t_1+t_2}{2}$.  In these calculations charge order
was externally imposed by adding a term $\sum_i \epsilon_i
(n_{i,\uparrow} +n_{i,\downarrow})$ to Eq.~\ref{EHM}, with the sign of
$\epsilon_i$ chosen to produce the CO pattern shown in
Fig.~\ref{zzladder}(b).  In the comparison 1D system the CO is
$\cdots$1100$\cdots$ and the bond pattern was BCDW-I. The resulting
spin gap $\Delta_\sigma$ is plotted as a function of the resulting
$\Delta n$. $\Delta_\sigma$ is found to be considerably larger in the
ladder than in 1D for the {\it same} CO amplitude.

Importantly, the bond distortion in the ladder remains that of BCDW-II
even for strong e-e interactions ($U$=6 and $V$=1 in
Fig.\ref{zzladder}(c)), as compared to 1D where the pattern is BCDW-I
(see Section \ref{1dnumerics-peh}).  This is the reason the spin gap
is large in the zigzag ladder.  Large spin gaps can be expected in
$\rho=\frac{1}{2}$ CTS with zigzag ladder crystal structures.  Similar
to quasi-1D $\rho=\frac{1}{2}$ CTS that show the BCDW-II distortion
\cite{Clay17a}, a single coupled charge-bond-spin-gap transition would
be expected in the zigzag ladder \cite{Clay12a}.  While not many CTS
are known experimentally that have the zigzag structure, one example
are the (DTTF)$_2M$(mnt)$_2$ series \cite{Rovira00a}, where large spin
gaps are found experimentally: T$_{\rm SG}\sim$ 90 K for $M$=Cu and
$\sim$ 70 K for $M$=Au, respectively \cite{Ribas05a}.

\subsubsection{Paired Electron Crystal in two dimensions}
\label{pec2d}

The absence of a spin-singlet phase in the phase diagram of
Fig.~\ref{rho1-anisotropic}, and the suppression of superconducting
pair-pair correlations in the frustrated $\frac{1}{2}$-filled band
Hubbard model are related. There is simply no evidence for either the
proposed RVB state, or a VBS state that is driven by lattice
frustration at $\rho=1$ (Given the tendency to form the 120$^\circ$
AFM at large t$^\prime$, it may be argued that there is no physical
reason why such a singlet state should form at all.). Yet the concept
of mobile nearest neighbor singlets being the equivalents of Cooper
pairs in configuration space remains attractive. Theoretical work on
both rectangular \cite{Dagotto99a} and zigzag
\cite{Chitra95a,White96a} $\rho=1$ ladders suggests that singlet
formation, instead of being driven by lattice frustration, can be due
to ``effective one-dimensionalization'' within a 2D lattice. The only
way to have this in an extended lattice is to have $\rho\neq 1$, where
the one-dimensionalization is due to formation of segregated stripes
of sites occupied by charge carriers. This is a form of
charge-ordering and will not occur at arbitrary $\rho$: for
$\rho\approx 1$ it is impossible for an ``occupied'' site to not have
more than two neighbors, while for $\rho$ very small the kinetic
energy of individual carriers will overwhelm the gain in exchange
energy that drives local singlet formation. The optimum concentration
is exactly $\rho=\frac{1}{2}$, where one can visualize occupied and
unoccupied stripes perfectly alternating, thereby conferring
commensurability driven stability to the structure. Furthermore, we
will show in this section that in the presence of geometric lattice
frustration such a striped structure is simultaneously a PEC, in which
pairs of singlet-bonded sites are separated by pairs of vacant
sites. In the presence of e-p coupling the insulating striped
structure undergoes lattice distortion that also resembles a 2D SP
state, which we have already pointed out has in the past been thought
to be behind correlated-electron SC \cite{Hirsch87a,Imada91a}.  We
have termed this charge-ordered state as the Paired Electron Crystal
(PEC).  It was shown that in the context of the interacting continuous
electron gas, a PEC state has a variational energy that is {\it lower}
than that of the WC \cite{Moulopoulos92a,Moulopoulos93a}.  The PEC we
describe below gains additional energy over the continuum case due to
the lattice commensurability at $\rho=\frac{1}{2}$.
\begin{figure}[tb!]
  \begin{center}
    \resizebox{3.0in}{!}{\includegraphics{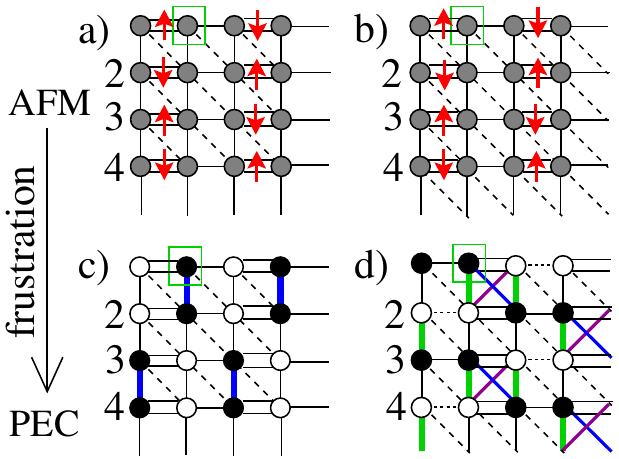}}
  \end{center}
  \caption{(color online) PEC formation in a 4$\times$4 lattice \cite{Li10a}.
    (a) OBC and (b) PBC lattices for $t^\prime<t^\prime_c$. Double
    bonds and thick lines indicate strong bonds; thin lines are
    weak bonds. Dashed lines indicate diagonal bonds whose strength
    is varied. Small, large, and uniform charge densities are indicated
    by white, black, and gray circles, respectively. (c) and (d) PEC
    state for $t^\prime>t^\prime_c$. Numbers correspond to the chain
    indices in Fig.~\ref{4x4pecdata}. Boxes mark the site 2 of
    chain 1; spin-spin correlations between this site and other chains
    are shown in Fig.~\ref{4x4pecdata} \cite{Li10a}.}
  \label{4x4pec}
\end{figure}

\begin{figure}[htb!]
  \begin{center}
    \resizebox{2.7in}{!}{\includegraphics{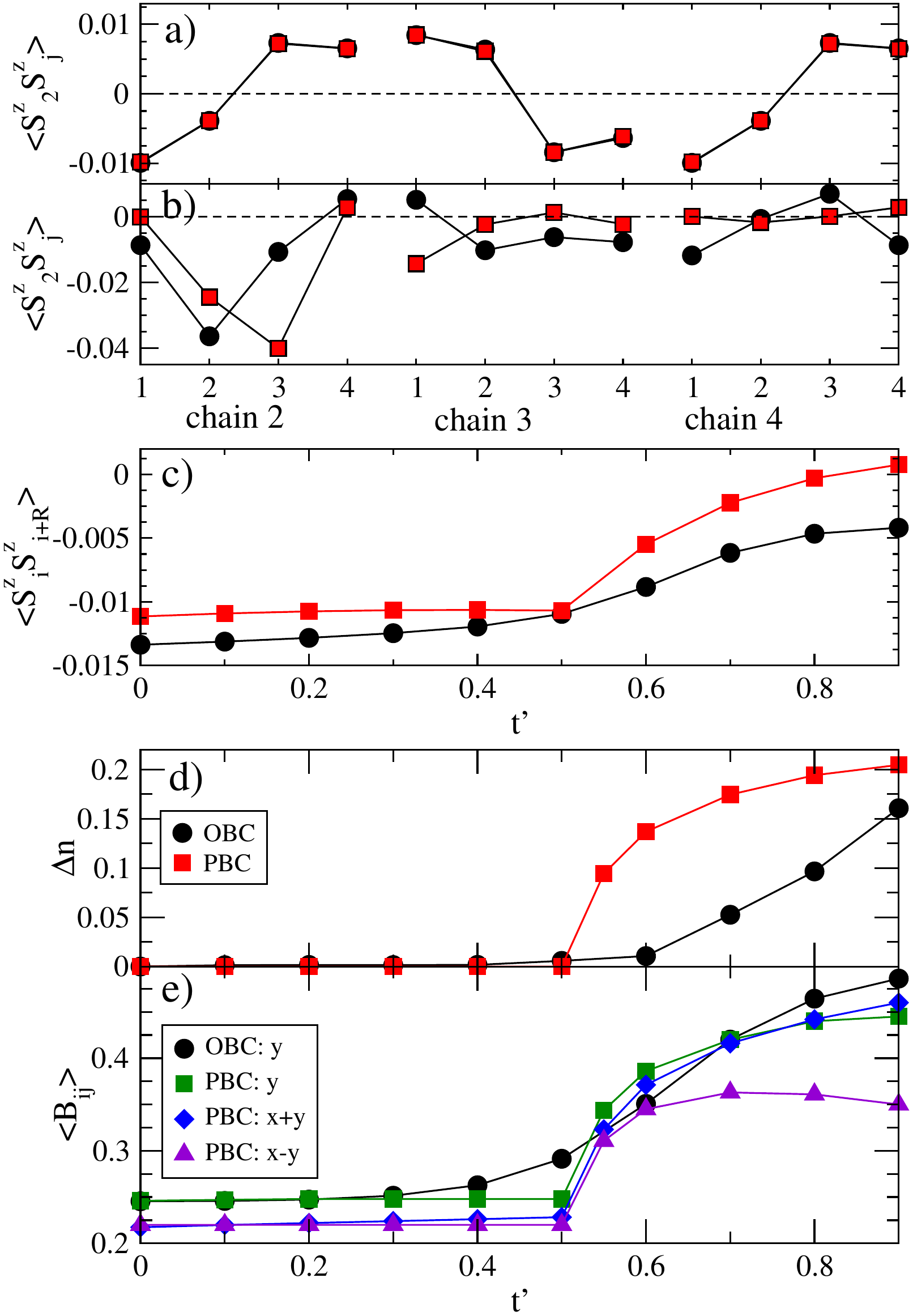}}
  \end{center}
  \caption{(color online) Correlation functions from exact 4$\times$4
    lattice calculations corresponding to Fig.~\ref{4x4pec}
    \cite{Li10a}.  Chain indexing is shown in Fig.~\ref{4x4pec}.  (a)
    Z-Z spin correlations between site 2 on chain 1 (marked with a box
    on Fig.~\ref{4x4pec}) and sites on chains 2, 3, and 4, for
    $t^\prime=0$ (b) same as (a) except $t^\prime=0.7$.  (c) Spin-spin
    correlations between sites that are members of the most distant
    dimers (d) difference in charge density between charge-rich and
    charge-poor sites. (e) bond orders between nearest neighbor sites
    in the indicated directions.  In all panels, circles and squares
    correspond to OBC and PBC calculations, respectively. In (e) bond
    orders for other directions in the PBC calculations are shown with
    additional symbols \cite{Li10a}.}
    \label{4x4pecdata}
\end{figure}

We first summarize the results of our calculations, after which we
present further details.  The
Hamiltonian is
\begin{eqnarray}
H=-\sum_{\nu,\langle ij\rangle_\nu}t_\nu(1+\alpha_\nu\Delta_{ij})B_{ij} 
+\frac{1}{2}\sum_{\nu,\langle ij\rangle_\nu} K^\nu_\alpha \Delta_{ij}^2 \nonumber \\
+g \sum_i v_i n_i + \frac{K_g}{2} \sum_i v_i^2  
+ U\sum_i n_{i\uparrow}n_{i\downarrow} + 
\frac{1}{2}\sum_{\langle ij\rangle}V_{ij} n_i n_j.
 \label{hamagain}
\end{eqnarray}
In Eq.~\ref{hamagain} the sums $\nu$ run over three lattice directions in two
dimensions, $\hat{x}$, $\hat{y}$, and $\hat{x}-\hat{y}$.
$B_{ij}=\sum_\sigma(c^\dagger_{i\sigma}c_{j\sigma}+H.c.)$,
$\alpha_x$, $\alpha_y$, and  $\alpha_{x-y}$ are the intersite e-p couplings.
Unless denoted otherwise we choose
$\alpha_x=\alpha_y\equiv\alpha$ and $\alpha_{x-y}=0$.
$K_\alpha^x=K_\alpha^y=K_\alpha^{x-y}\equiv K_\alpha$ is the
corresponding spring constant, and $\Delta_{ij}$ is the distortion of
the $i$--$j$ bond, to be determined self-consistently; $v_i$ is the
intrasite phonon coordinate and $g$ is the intrasite e-p coupling
with the corresponding spring constant $K_g$.
We assume the simplest form of frustration, with a bond strength $t^\prime$ in
the $\hat{x}-\hat{y}$ direction on a square lattice.  However, we
begin here with the simple, unfrustrated, square lattice ($t^\prime=0$
and $V_{x-y}=0$) limit of Eq.~\ref{hamagain}.  In this case two different
semiconducting states are possible: in-phase dimerized AFM with all
site charge densities equal (see Fig.~\ref{4x4pec}(a) and (b)) and for
sufficiently large $V_x$ and $V_y$, the WC with checkerboard site
occupancies.  These two insulating phases are mutually exclusive.  We
choose parameters that give AFM in the weakly frustrated limit, as in
2D CTS with experimentally observed AFM lattices are universally
dimerized.  Fig.~\ref{4x4pec} shows schematically the results of exact
ground state calculations on 4$\times$4 lattices with two different
boundary conditions, open (OBC) and periodic (PBC).  Both are periodic
along $x$ and $y$ directions, but the OBC (PBC) is open (periodic)
along $x-y$ with 12 (16) $t^\prime$ bonds.  For both boundary
conditions we find a frustration-driven AFM-PEC transition as we
increase the strength of the frustrating hopping $t^\prime$.  For the
OBC we take $\alpha_\nu=g=0$ in Eq.~\ref{hamagain}; in order to have AFM
order at $t^\prime=0$ here we incorporate intrinsic dimerization
$t_x=t\pm \delta_t$ as indicated in Fig.~\ref{4x4pec}(a). For the PBC
lattice, lattice distortion is not assumed but results from the e-p
interactions.  Spin-spin correlations, charge densities and bond
orders for the PBC are obtained self-consistently \cite{Li10a}.
\begin{figure}[tb]
  \centerline{\resizebox{4.5in}{!}{\includegraphics{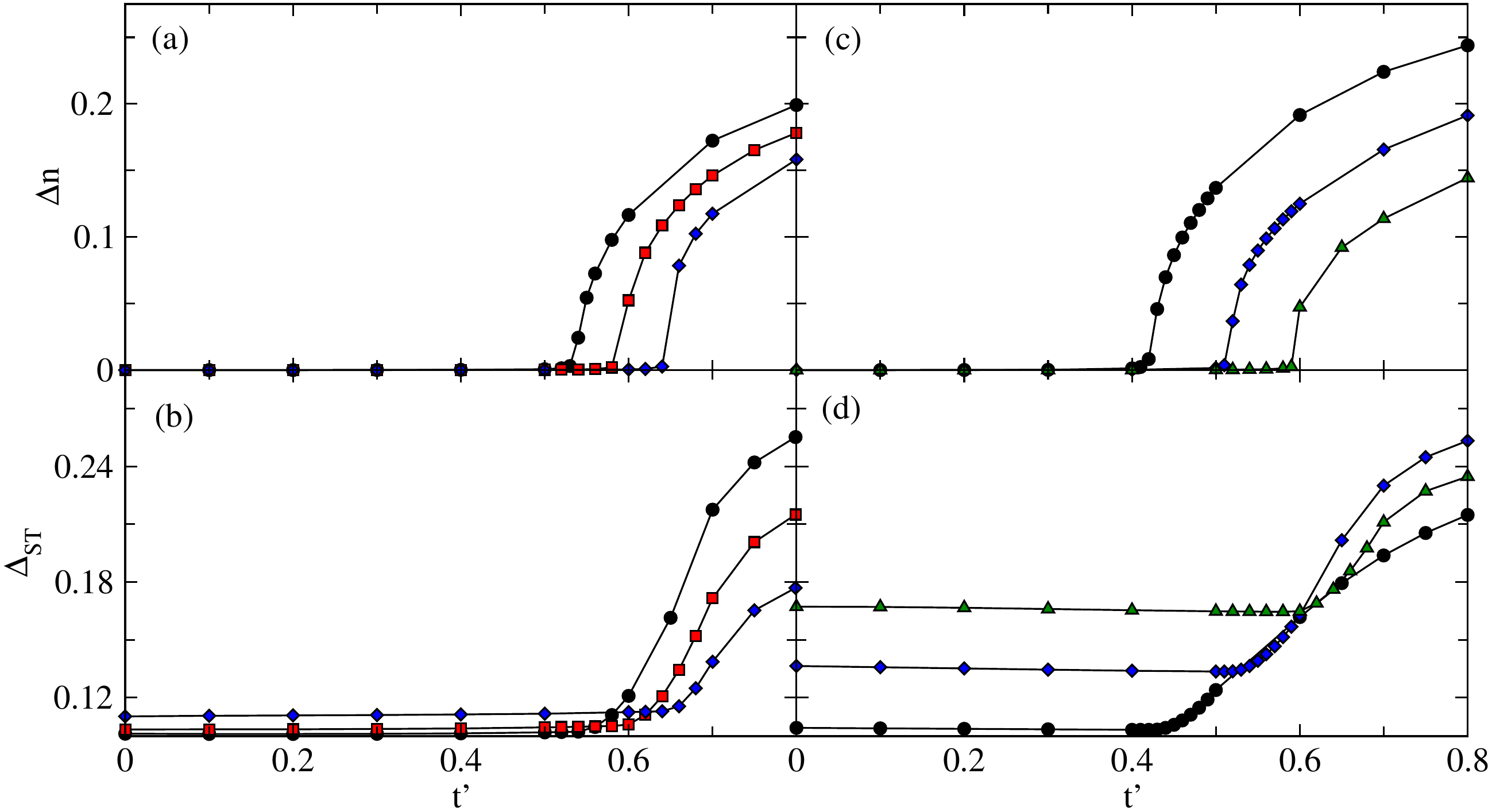}}}
  \caption{(color online) Charge order amplitude $\Delta$n ((a) and
    (c)) and singlet-triplet gap $\Delta_{\rm ST}$ ((b) and (d)) for
    the 4$\times$4 lattice as a function of $t^\prime$. Circles,
    squares, diamonds, and triangles are for $U$=2, 3, 4, and 6,
    respectively. For all
    $\alpha=1.1$, $g$=0.1, and
    K$_\alpha$=K$_g$=2. In panels (a) and (b), $V_x=V_y=V^\prime$=0,
    and in panels (c) and (d), $V_x=V_y=1$ and $V^\prime$=0.  The
    average (undistorted) hopping integral in the $x$ direction is
    taken as the unit of energy \cite{Dayal11a}.}
  \label{pecspingap}
\end{figure}

In Fig.~\ref{4x4pecdata} we show numerical results that prove the
AFM--PEC transition indicated in Fig.~\ref{4x4pec}.  These results are
for $U=4$, $V_x=V_y=1, V_{x-y}=0$ within Eq.~\ref{hamagain}, where the
undistorted hopping integral in the $x$ direction ($t_\nu$ of
Eq.~\ref{hamagain} with $\nu=x$) is taken as the unit of energy.  For the
OBC lattice $\delta_t=0.2$ and $\alpha_\nu=g=0$ as described above,
and for the PBC $\alpha_x=1.3$, $\alpha_y=1.0$, $\alpha_{x-y}=0$,
$K^x_\alpha=K^y_\alpha=2$, $g=0.1$, and $K_g=2$.  In
Fig.~\ref{4x4pecdata}(a) we plot the z-z spin-spin correlation
functions for $t^\prime=0$ between a fixed site (marked with box on
each lattice in Fig.~\ref{4x4pec}) and sites $j$, labeled sequentially
1, 2, 3, 4 from the left, on neighboring chains labeled 2, 3, 4 in
Fig.~\ref{4x4pec}.  In Fig.~\ref{4x4pecdata}(a) only, the average
spin-spin correlation with each chain has been shifted to zero (note
dotted line for $\langle S^z_2S^z_j\rangle=0$ in
Fig.~\ref{4x4pecdata}(a)-(b)) in order to clearly show the AFM
pattern, which is clearly $\cdots- - + + \cdots$ and $\cdots+ + - -
\cdots$, indicating N\'eel ordering of the dimer spin moments in both
lattices.  The loss of this pattern in Fig.~\ref{4x4pecdata}(b) for
large $t^\prime=0.7$ indicates loss of AFM order.  In Fig.~
\ref{4x4pecdata}(c) we plot the spin-spin correlation between
maximally separated dimers, which measures the strength of the AFM
moment.  This correlation is nearly constant until $t^\prime_c\sim
0.5$, beyond which the AFM order is destroyed.

Fig.~\ref{4x4pecdata}(d) shows $\Delta$n as a function of $t^\prime$.
There is a rapid increase in $\Delta n$, starting from zero, for
$t^\prime>t^\prime_c$ with both lattices.  Simultaneously with CO,
there is a jump in the bond orders $\langle B_{ij} \rangle$ between
the sites that form the localized spin-singlets. This is shown in
Fig.~\ref{4x4pecdata}(e).  These bond orders are by far the strongest
in both lattices for $t^\prime > t^\prime_c$.  The spin-spin
correlation between the same pairs of sites becomes strongly negative
at the same $t^{\prime}$, even as all other spin-spin correlations
approach zero (Fig.~\ref{4x4pecdata}(b)), indicating spin-singlet
character of the strongest bonds.  The {\it simultaneous} jumps in
$\langle S^z_iS^z_{i+R}\rangle$, $\Delta n$, and $B_{ij}$ {\it at this
  same $t^\prime$} give conclusive evidence for the AFM-PEC transition
shown in Fig.~\ref{4x4pec}.

Figs.~\ref{pecspingap}(b) and (d) show the singlet-triplet gap
$\Delta_{\rm ST}$ compared with $\Delta n$, for the 4$\times$4 lattice
with periodic boundaries and $V=0$ ((a) and (b)) and $V\neq0$ ((c) and
(d)). Although $\Delta_{\rm ST}$ is nonzero for all values of
$t^\prime$ (as expected in a finite-size lattice), the spin gap
increases sharply in the PEC state for $t^\prime>t^\prime_{\rm c}$.
The increase of $\Delta_{\rm ST}$ between the AFM and PEC states is
also proportional to the amplitude of the distortion $\Delta n$ as
expected.

In the PEC the CO pattern is $\cdots$1100$\cdots$ in two of three
lattice directions, $\hat{x}$ and $(-\hat{x}+\hat{y})$ in
Fig.~\ref{4x4pec}(d), and $\hat{y}$ and $(\hat{x}+\hat{y})$ in
Fig.~\ref{4x4pec}(c). In the third direction the CO pattern is
$\cdots$1010$\cdots$.  The singlets that coexist with the CO are of
the interdimer type as is found in BCDW-I in 1D, and the bond order
pattern found here is the same SWSW$^\prime$ pattern (note that in
Fig.~\ref{4x4pec} distortions of the dashed $t^\prime$ bonds are not
indicated). The results here are for the ground state
only. As in 1D (Sections \ref{1dlimit} and \ref{1dnumerics}) it is
possible that the charge and bond distortion may occur at a higher
temperature, with the SG only opening at a lower temperature. We
discuss this further in relation to 2D CTS below.

\begin{figure}[tb]
  \begin{center}
    \raisebox{0.3in}{
      \begin{overpic}[width=2.5in]{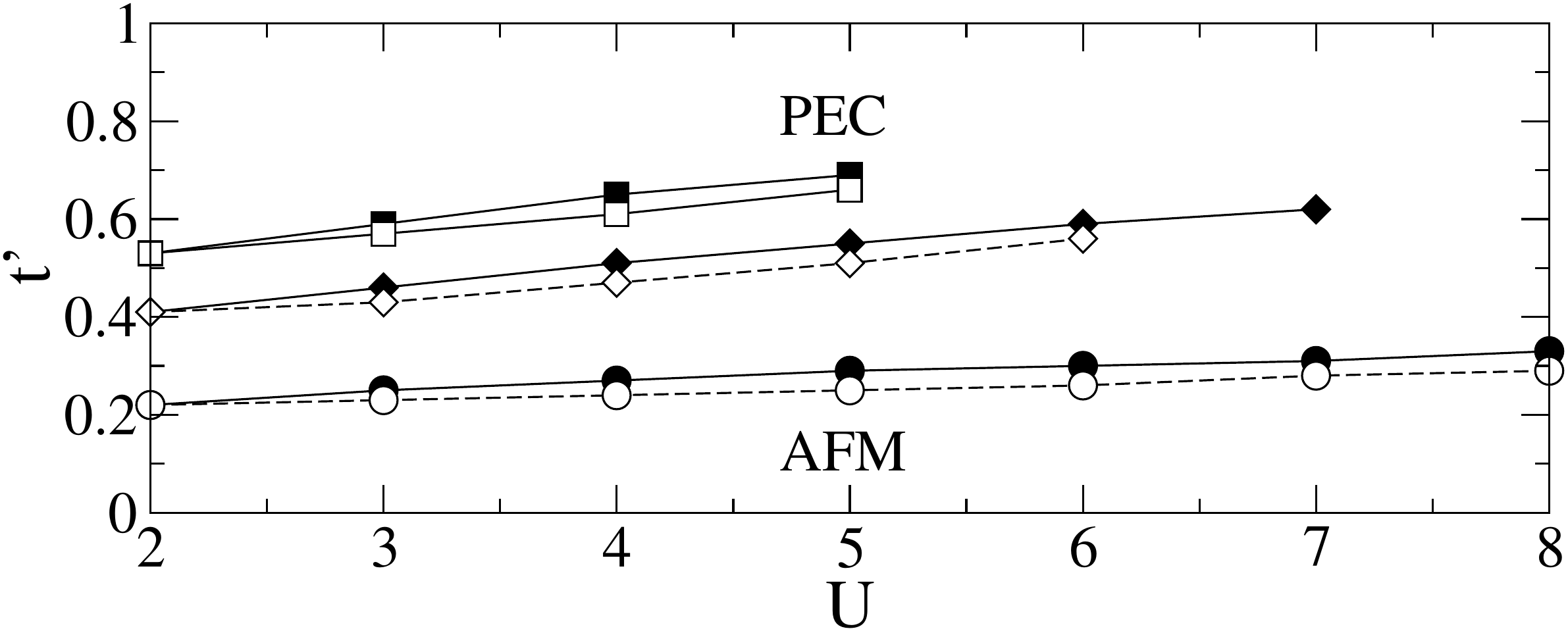}
        \put(-5,38) {\small(a)}
      \end{overpic}
    }
    \hspace{0.1in}
    \begin{overpic}[width=2.5in]{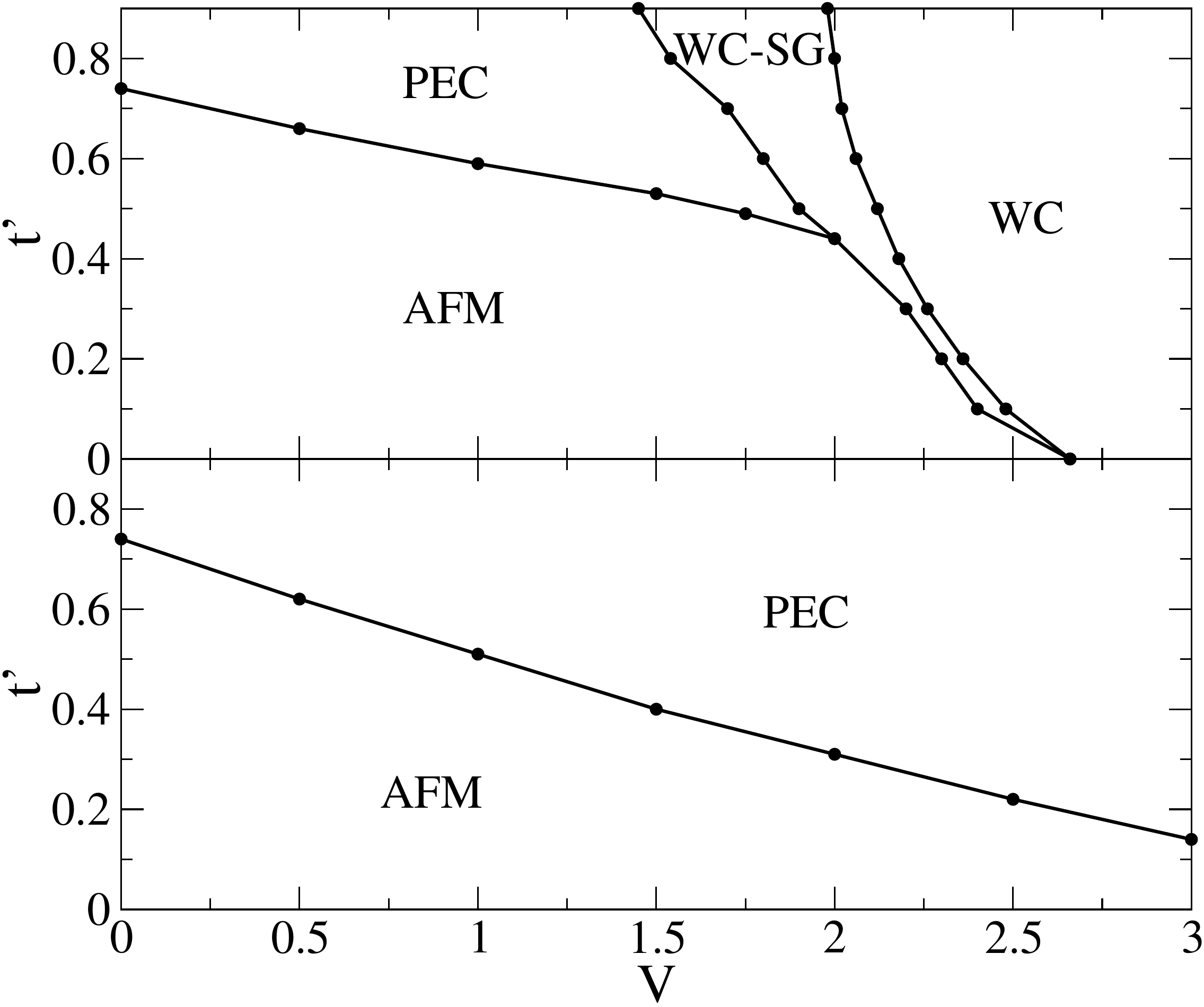}
      \put(-5,80) {\small(b)}
      \put(-5,41) {\small(c)}
    \end{overpic}
  \end{center}
  \caption{Ground state phase diagrams from self-consistent
    calculations of Eq.~\ref{hamagain} on periodic 4$\times$4 lattice,
    showing AFM and PEC phases \cite{Dayal11a} For all plots $g$=0.1
    and $K_\alpha=K_g=2$.  (a) Phase diagram as a function of
    $t^\prime$ and $U$.  Squares are the AFM-PEC phase boundary for
    $\alpha=1.1$ and $V_x=V_y=V^\prime=0$. Diamonds are for
    $\alpha=1.1$, $V_x=V_y=1$, and $V^\prime=0$. Circles are for
    $\alpha=1.2$, $V_x=V_y=1$, and $V^\prime=0$. Filled (open) points
    correspond to positive (negative) $t^\prime$.  For $U$ values
    where no points are plotted, the lattice was undistorted.  (b)
    Phase diagram for $V_x=V_y=V$ and $V^\prime=0$ with $U=6$ and
    $\alpha=1.1$.  WC and WC-SG are Wigner crystal of single electrons
    and the coexisting single electron Wigner crystal-spin-gap phases,
    respectively.  (c) Same as (b), except $V=V_x=V_y=V^\prime$. }
  \label{pecphasediagram}
\end{figure}

Fig.~\ref{pecphasediagram} shows how e-e interactions $U$ and $V$
affect the PEC \cite{Dayal11a}. As shown in
Fig.~\ref{pecphasediagram}(a), increasing $U$ moderately increases
$t^\prime_{\rm c}$. The points not plotted for larger $U$ indicate
that a finite e-p coupling is required for the PEC state, which is
expected due to the finite-size gap in this small
lattice. Fig.~\ref{pecphasediagram} shows the effect of the NN Coulomb
interaction $V$. Here the phase diagram depends critically on the form
assumed for $V_{ij}$. With $V_x=V_y=V$ but $V^\prime=0$, the WC is
found for sufficiently strong $V$, along with a narrow region where it
coexists with a SG (Fig.~\ref{pecphasediagram}(b)).
The very narrow width of the WC-SG phase in the figure indicates the
small likelihood of the WC coexisting with SG in real
materials. Equally importantly, most CTS lattices are partially
triangular, and the assumption that $V^\prime=0$ is not realistic.
For $V_x=V_y=V^\prime$ the WC and the PEC have the same classical
energies; as shown in Fig.~\ref{pecphasediagram}(c) in this case the
WC is completely replaced by the PEC.
    
\subsubsection{Paired Electron Crystal in quasi-two dimensional CTS}
\label{pec2dcts}

Evidence for the PEC with CO, bond order, and spin gap, has been found
experimentally in many 2D CTS.  CO by itself does not necessarily
imply that the PEC is present.  As discussed in the previous section,
the pattern of charge order in the PEC is {\it distinct} from that of
the WC; in the PEC pattern there are NN charge-rich sites which are
not present in the WC.  The frustrated crystal lattice in all of the
2D CTS have significant inter-molecular hopping in three different
lattice directions.  Using the principle that in two lattice
directions the CO pattern is $\cdots$1100$\cdots$ while
$\cdots$1010$\cdots$ in one direction, the predicted pattern for
different CTS lattices can be derived.  However, even if the exact
pattern of CO is not known, a period four lattice distortion occurring
in a 2D crystal indicates PEC formation.  This is because the
competing CO state, the WC, does not coexist with a bond distortion.
Similarly, a spin gap is a convincing signature of the PEC, even when
accompanying CO has not been found. In contrast, only AFM can coexist
with the WC \cite{McKenzie01a} (the AFM pattern here is however
different from that in Fig.~\ref{4x4pec}).

\paragraph{$\beta$, $\beta^\prime$, and  $\beta^{\prime\prime}$ CTS}
The largest group of CTS where the PEC has been found are the $\beta$,
$\beta^\prime$, and $\beta^{\prime\prime}$ CTS.  As discussed in
Section \ref{beta-cts}, charge- and bond-ordered insulating states and
spin gaps are frequently found adjacent to SC.  In cases where the CO
pattern has been determined, it matches the pattern of
Fig.~\ref{4x4pec} exactly.  For example, $\beta$-({\it
  meso}-DMBEDT-TTF)$_2$X is one well characterized example where the
experimental CO and bond pattern is identical (compare
Fig.~\ref{beta-meso}(a) and Fig.~\ref{4x4pec}). Here the bond pattern
along the stacking direction goes $\cdots$SWSW$^\prime\cdots$, exactly
as in BCDW-I \cite{Shikama12a}.  The ground state of $\beta$-({\it
  meso}-DMBEDT-TTF)$_2$X in the ambient pressure insulating phase is
also spin gapped.

Although it is $\frac{1}{4}$-filled in terms of electrons rather than
holes, EtMe$_3$P [Pd(dmit)$_2$]$_2$ is another example of the PEC with
exactly the same CO and bond order pattern (see Section \ref{etme3p}
and Fig.~\ref{etme3p-co}(b)).  Again, under pressure,
EtMe$_3$P [Pd(dmit)$_2$]$_2$ becomes superconducting.
Kato et al.  in their original work had referred to the CO
in this system as a valence bond solid, as would be expected in
a $\frac{1}{2}$-filled band \cite{Tamura09a}. In their recent work
\cite{Yamamoto17a} the authors find that a $\frac{1}{4}$-filled description is
more appropriate.

In
$\beta$-(ET)$_2$ReO$_4$ (Fig.~\ref{beta-bda-ttp}(a)) and
$\beta$-(BDA-TTP)$_2$X (Fig.~\ref{beta-bda-ttp}(b)) the PEC is 
found adjacent to SC.  In both of these CTS a period four lattice
distortion and spin gap is found in the insulating state.  Similarly,
in the $\beta^{\prime\prime}$ materials, strong evidence for a PEC
ground state exists: here a $\cdots$1100$\cdots$ CO pattern along the
stacking direction has also been determined from optical experiments
\cite{Yamamoto08a}.  CO coexisting with spin gap has also recently been found
in $\beta^{\prime\prime}$-(ET)$_2$Hg(SCN)$_2$Cl \cite{Li17a}.

Less is known about $\beta^\prime$-(ET)$_2$ICl$_2$ in its phase
adjacent to SC due to the high pressure required to access this region
of the phase diagram. It is clear however that the AFM state that
exists at ambient pressure disappears under high pressure before the
superconducting state, and a different form of insulating state enters
above around 6 GPa (see Fig.~\ref{beta-prime-icl2}).  Based on
similarities with the $\beta$-phase CTS we predict that PEC-type CO
and a spin gap will be found in the insulating state adjacent to SC
here as well.  The phase diagram of $\beta$-Me$_4$NPd(dmit)$_2$]$_2$
  (Fig.~\ref{dmit-phase-diag-1}) appears somewhat similar, as here
  again the ambient pressure AFM is suppressed by about 0.4 GPa,
  before the SC region is reached.  Here and also in
  $\beta^\prime$-Et$_2$Me$_2$P[Pd(dmit)$_2$]$_2$ (see also
  Fig.~\ref{dmit-phase-diag-1}) we suggest that the phase adjacent to
  SC is the PEC.

\begin{figure}[tb]
  \begin{center}
  \begin{overpic}[width=3.0in]{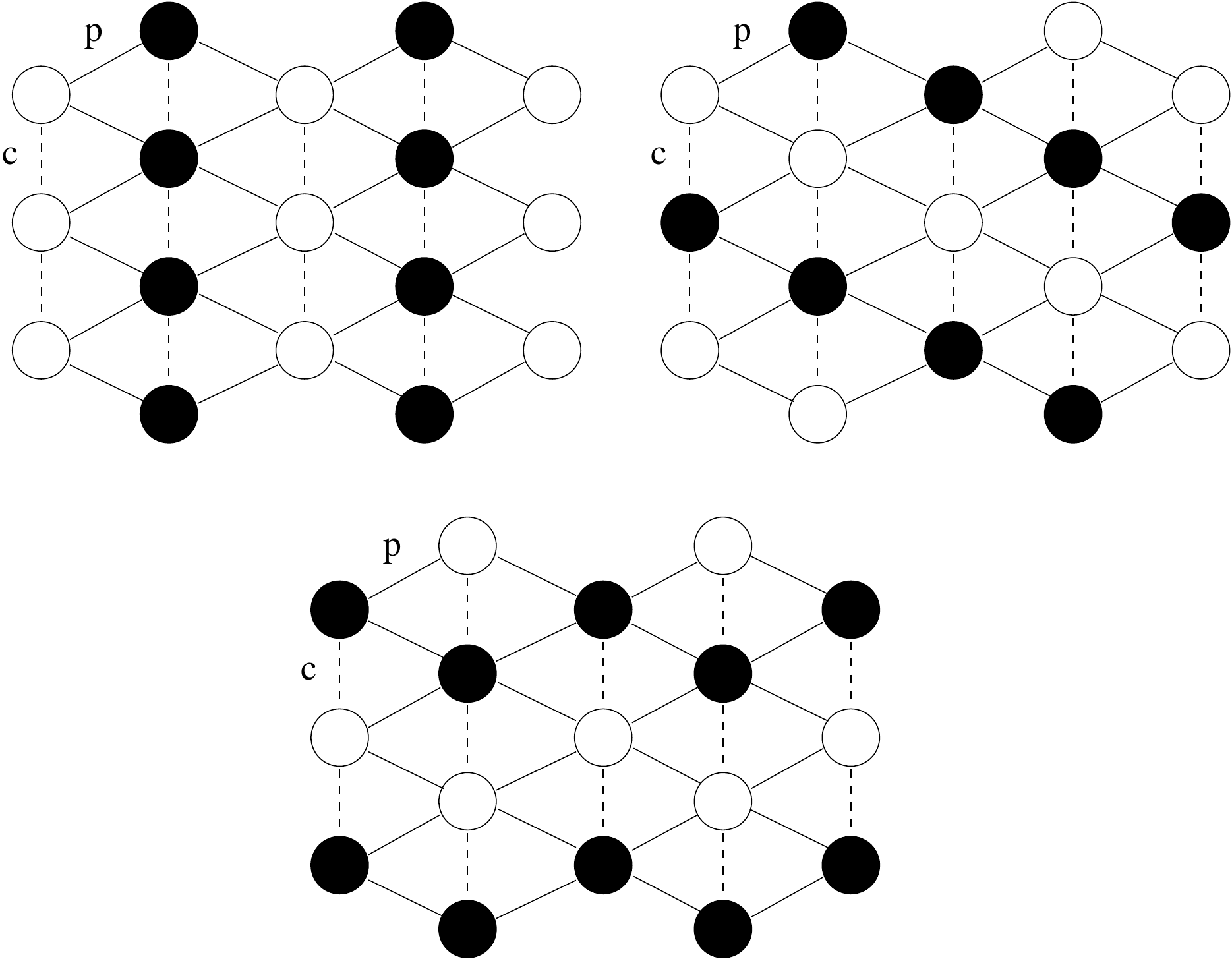}
    \put(-5,75) {\small (a)}
    \put(48,75) {\small (b)}
    \put(15,30) {\small (c)}
  \end{overpic}
  \hspace{0.4in}
  \raisebox{0.2in}{
    \begin{overpic}[width=2.0in,angle=90]{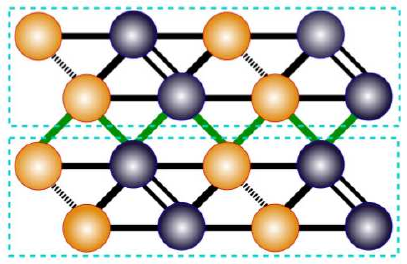}
      \put(-10,90) {\small (d)}
    \end{overpic}
  }
  \end{center}
  \caption{CO patterns and PEC in $\theta$-(ET)$_2$X.  (a) vertical
    stripe (b) diagonal stripe (c) horizontal stripe. The horizontal
    stripe has charge occupancy $\cdots1100\cdots$ along the two
    $p$-directions with strong intermolecular coupling. (d) PEC within
    the horizontal stripe CO, which can be visualized as repeated
    zigzag ladders \cite{Gomes16a}. Note that the bond dimerization
    along the vertical bonds (crystal {\bf c} direction) is not shown
    here.}
    \label{thetapec}
\end{figure}

\paragraph{$\theta$ and $\alpha$ CTS}

In the $\theta$ (Section \ref{thetasection}) and $\alpha$ (Section
\ref{alpha}) CTS the most stable CO pattern is the horizontal stripe
(see Fig.~\ref{thetapec}).  In $\theta$-(ET)$_2$ MM$^\prime$(SCN)$_4$
with MM$^\prime$ = RbZn, the horizontal stripe CO transition
temperature is 195 K, but the PEC with spin gap only appears at $\sim$
20 K within the horizontal stripe CO.  The appearance of CO at a
relatively high temperature without a spin gap suggests that NN
Coulomb interactions ($V_{\rm ij}$) play an important role in
determining the pattern of CO \cite{Seo00a,Clay02a}.  All of the CO
patterns shown in Fig.~\ref{thetapec}(a)-(c) have identical energies
within the classical limit (assuming $V_{ij}$ is identical in all
three lattice directions). There is a key difference between the
vertical and diagonal stripes (Fig.~\ref{thetapec}(a) and (b)) and the
horizontal stripe (Fig.~\ref{thetapec}(c)) however: in both the
vertical and diagonal stripes, the CO pattern is $\cdots$1010$\cdots$
in {\it two} lattice directions, as is true for the checkerboard WC on
the square lattice \cite{Clay02a}. If mapped to a square lattice, both
the vertical and diagonal CO patterns are identical to the usual WC.
Because of this, mean field (Hartree or Hartree-Fock) calculations
tend to predict the vertical or diagonal stripe patterns
\cite{Clay02a,Seo00a}. In contrast, the horizontal stripe has CO
patterns $\cdots$1010$\cdots$ in the direction of weakest
intermolecular hopping and $\cdots$1100$\cdots$ along the two
directions with strong intermolecular hopping.  The horizontal stripe
pattern thus gains additional stabilization energy because of the
stronger tendency to spin-singlet formation.  The horizontal stripe CO
is only found as a solution in mean-field calculations provided the
dimerization along the {\bf c} axis (which only appears for $T<T_{\rm
  CO}$) is explicitly included \cite{Seo00a}. This dimerization is a
{\it consequence} rather than a cause of the CO \cite{Clay02a}.

In order for the horizontal stripe CO to have a spin gap, a further
symmetry breaking accompanying  NN singlet formation must take place
along the direction of the stripe, as shown in Fig.~\ref{thetapec}(d).
Within the horizontal stripe there are two possible locations for
NN singlets; in Fig.~\ref{thetaco} the NN singlet pair could be
located at {\it either} of the two bonds labeled ``p4'' in the
figure. This second symmetry breaking and the spin gap state occurs at
$\sim$ 20 K in RbZn.  The horizontal stripe PEC in this case can also
be visualized as being composed of coupled zigzag ladders (see Fig.~\ref{thetapec}(d))
\cite{Gomes16a}.

Compared to RbM$^\prime$, the CsM$^\prime$ $\theta$ salts remain
metallic to much lower temperature and have much more confusing
low-temperature behavior with short-range rather than long-range
charge order (see Section \ref{thetasection}).  Diffuse X-ray
scattering is also found at several wavevectors in the CsM$^\prime$
salts indicating a competition between several CO patterns.  This
competition may be particularly important to understanding the unusual
transport properties \cite{Inagaki04a}.  Several works by Hotta and
collaborators have explored the competition between different CO
states as a function of $V_{ij}$
\cite{Hotta06b,Nishimoto09a,Yoshida14a}. A partially ordered ``pinball
liquid'' phase was suggested to describe the unusual semi-metallic
properties of $\theta$-(ET)$_2$CsZn(SCN)$_4$
\cite{Hotta06a,Nishimoto09a}.  Further work focused on the effect of
thermal fluctuations \cite{Yoshida14a}.

The PEC is also found in $\alpha$-(ET)I$_3$
(Fig.~\ref{alpha-i3}). Here the CO pattern is again the horizontal
stripe with a coexisting SG.  Unlike $\theta$-(ET)$_2$X, the SG
transition is at the same temperature as the CO (135 K). The presence
of a single combined CO/SG PEC transition in $\alpha$ as compared to
$\theta$ may be due to the lower symmetry of the $\alpha$ crystal
structure and the presence of partial CO already above 135 K (see
Figs.~\ref{alpha-beta-theta-structures} and \ref{alpha-i3}).

\paragraph{$\kappa$ CTS} The stabilization energy the PEC gains from
\begin{figure}[tb]
  \begin{center}
    \begin{overpic}[width=2.25in]{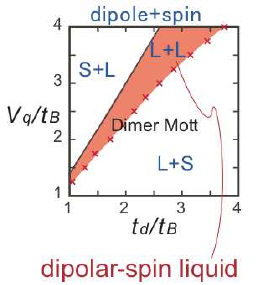}
      \put(5,90) {\small (a)}
    \end{overpic}
    \hspace{0.3in}
    \raisebox{0.25in}{
      \begin{overpic}[width=2.5in]{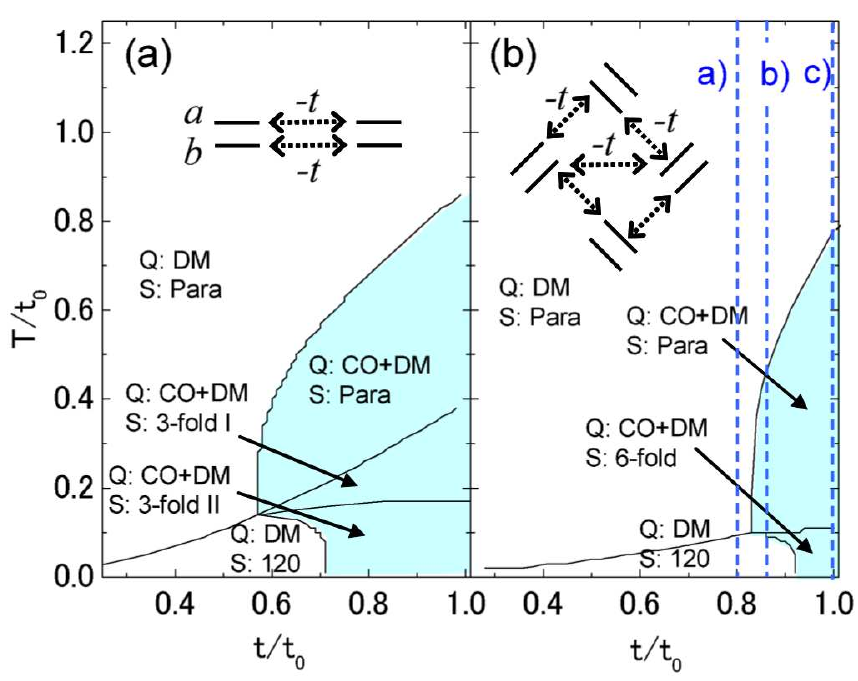}
        \put(0,70) {\small (b)}
      \end{overpic}
    }
  \end{center}
  \caption{(a) Exact diagonalization ground state phase diagram for
    the effective strong-coupling Hamiltonian for $\kappa$-(ET)$_2$X
    of Reference \cite{Hotta10a}.  $t_B$, $t_d$, and $V_q$ are the
    inter-dimer hopping integral $t_{b_2}$, intra-dimer hopping
    integral, and n.n. Coulomb interaction, respectively (see
    Fig.~\ref{kappa-lattice}).  S+L denotes a charge ordered spin
    liquid phase, L+S a spin ordered charge liquid phase, and L+L a
    charge-spin liquid phase.  Reprinted with permission from
    Ref.~\cite{Hotta10a}, $\copyright$ 2010 The American Physical
    Society.  (b) Finite-temperature phase diagram calculated within
    mean-field theory \cite{Naka10a,Ishihara14a}.  $t/t_0$ is the
    ratio of the inter- to intra-dimer hopping integral.  Reprinted
    with permission from Ref.~\cite{Naka10a}, $\copyright$ 2010 The
    Physical Society of Japan.}
  \label{hotta-ishihara}
\end{figure}  
NN  singlet formation will be less on the $\kappa$ lattice due to the
many nearly degenerate ways of connecting NN molecules with singlet
bonds \cite{Li10a,Dayal11a}.  Because dimers of BEDT-TTF molecules are
nearly perpendicular to each other, any static insulating
PEC state would be expected to
be very weak \cite{Li10a}.
Nevertheless, some experimental evidence for an insulating PEC
state exists for two $\kappa$ phase CTS.
Recent experiments have shown that
$\kappa$-(ET)$_2$B(CN)$_4$, which is more quasi-1D than
most $\kappa$ CTS, is spin-gapped below 5 K strongly
suggesting a PEC ground state \cite{Yoshida15a}. Raman scattering
measurements were  performed to 10 K, which is above the spin-gap
transition \cite{Yoshida15a}, leaving open the possibility of CO
below 5 K.  CO is explicitly found in
$\kappa$-(ET)$_2$Hg(SCN)$_2$Cl (see Section \ref{kappa}),
although the pattern of CO has not been determined yet.  The fairly
low T$_{\rm CO}$=30 K here suggests that the CO state has 1--0 charges
within each dimer, rather than 1--1 and 0--0 dimer CO.  So far a spin
gap at low temperature has not been found however.

A PEC-like CO pattern was suggested to explain the relaxor-like
dielectric response found in $\kappa$-(ET)$_2$Cu$_2$(CN)$_3$, where CO
within each dimer is the source of the dielectric polarization (see
Fig.~5 of Reference \cite{Abdel-Jawad10a}).  Several theoretical works
have proposed models based on a $\frac{3}{4}$-filled description with
the goal of capturing the coupling between dimer charge dipoles and
inter-dimer spin interactions \cite{Naka10a,Hotta10a,Ishihara14a}. In
these models a strong-coupling limit is assumed, with only those
states that have exactly one electron per dimer retained. While both
groups found fluctuating charge ordered phases approximating the PEC
CO pattern predicted in Reference \cite{Li10a}, the possibility of
spin-singlet formation was not considered.
Fig.~\ref{hotta-ishihara}(a) shows the ground state phase diagram of
an effective model where the effects of $t_{ij}$ and $V_{ij}$ between
dimers are treated perturbatively \cite{Hotta10a}.  The phase diagram
is governed by a competition between $V$ and the inter-dimer $t_{ij}$
\cite{Hotta10a}.  The ``S+L'' phase has ordered charge dipoles on each
dimer, and has the same CO pattern as would be expected for a PEC
state on the $\kappa$ lattice (compare Fig.~1(d) of reference
\cite{Hotta10a} with Fig.~4(c) of reference \cite{Li10a}). The ``L+S''
phase on the other hand has AFM order but no CO; this is the
conventional dimer-Mott AFM state.  A state with disordered charge and
spin degrees of freedom is found at the boundary of these phases; this
was proposed to apply to $\kappa$-(ET)$_2$Cu$_2$(CN)$_3$.

Naka and Ishihara \cite{Naka10a} developed a similar effective model
for a dimerized $\frac{1}{4}$-filled lattice, with four states per
dimer and interactions between dimers treated perturbatively.  The
resulting model was studied under the mean-field approximation and
using classical Monte Carlo.  Fig.~\ref{hotta-ishihara}(b) shows the
mean-field phase diagram for this model on a lattice appropriate for
the $\kappa$ CTS. As in Fig.~\ref{hotta-ishihara}(a), an increase of
the ratio of inter- to intra- dimer hopping favors CO.  Two different
possible scenarios were proposed to describe
$\kappa$-(ET)$_2$Cu$_2$(CN)$_3$ \cite{Naka10a}. In the first, at high
temperature the system has no CO (Q:DM phase in
Fig.~\ref{hotta-ishihara}(b)).  CO occurs at low temperature, followed
by a spin ordered CO state at the lowest temperature. In the other
suggested scenario, CO does not occur at low temperatures, with the
system staying in the unshaded phases of Fig.~\ref{hotta-ishihara}(b)
as $T$ decreases.  However, due to proximity to the CO phases some
charge fluctuation features would be observed experimentally.  The
timescale of the dielectric constant experiments is much longer than
the expected fluctuation time of a single BEDT-TTF dimer; this
suggests that in any case rather than long-range PEC order, short
ranged domains might exist in $\kappa$-(ET)$_2$Cu$_2$(CN)$_3$, and it
is these domains that fluctuate rather than single molecules.  The
question of whether CO exists within a single dimer, or if CO exists
but is smaller than the limits ($\Delta n = 0.01$) set by optical
measurements still remains \cite{Sedlmeier12a}.

As discussed in Section \ref{kappa}, in $\kappa$-(ET)$_2$X AFM and
metallic/SC regions are phase segregated at low temperature.  In
Sections ~\ref{enhancedpp} and ~\ref{kappa-pairpair} below we will
show that on the $\frac{3}{4}$-filled $\kappa$ lattice superconducting
pair-pair correlations are enhanced by Coulomb interactions ($U$) in
this metallic state. There we will propose that in the majority of
$\kappa$ phase CTS, the ground state when not AFM is a Paired Electron
Liquid (PEL) of singlets, which is either a superconductor or a
metallic state that is a precursor to SC.

\subsubsection{Selective enhancement of superconducting pair-pair correlations at $\rho=\frac{1}{2}$}
\label{enhancedpp}

The above results show that the PEC is a distinct phase in the
correlated $\rho=\frac{1}{2}$ systems and is actually observed in 2D
CTS with CO.  In analogy with the conjectures built around the concept
of a density wave of Cooper pairs
\cite{Anderson04b,Franz04a,Tesanovic04a,Tesanovic04a,Chen04a,Vojta08a,Hamidian16a,Cai16a,Mesaros16a},
we have further proposed that SC will occur upon the destabilization
of the PEC state
\cite{Li10a,Dayal11a,Mazumdar12a,Mazumdar14a,Gomes16a}.  Note that
this idea also has overlaps with the idea of SC emerging from a 2D SP
state (which we have emphasized does not exist at $\rho=1$), as the
dimerized horizontal stripes of Figs.~\ref{thetapec}(c) and (d) can be
thought of as weakly coupled SP stripes. In contrast to these other
theories, we believe that the proposed density wave of Cooper pairs is
unique to $\rho=\frac{1}{2}$ in 1D and frustrated 2D systems.  With
further increase in frustration, there occurs a transition from the
PEC to a Paired Electron Liquid, with mobile singlet pairs
constituting the equivalent of Cooper pairs.  We have therefore
studied in detail the superconducting pair-pair correlations on
multiple frustrated lattices, as a function of $\rho$, $0<\rho<1$.  In
these calculations we do not explicitly include e-p interactions, with
the result that the insulating PEC of Figs.~\ref{4x4pec} does not
occur.

\begin{figure}[tb]
  \begin{center}
    \resizebox{4.5in}{!}{\includegraphics{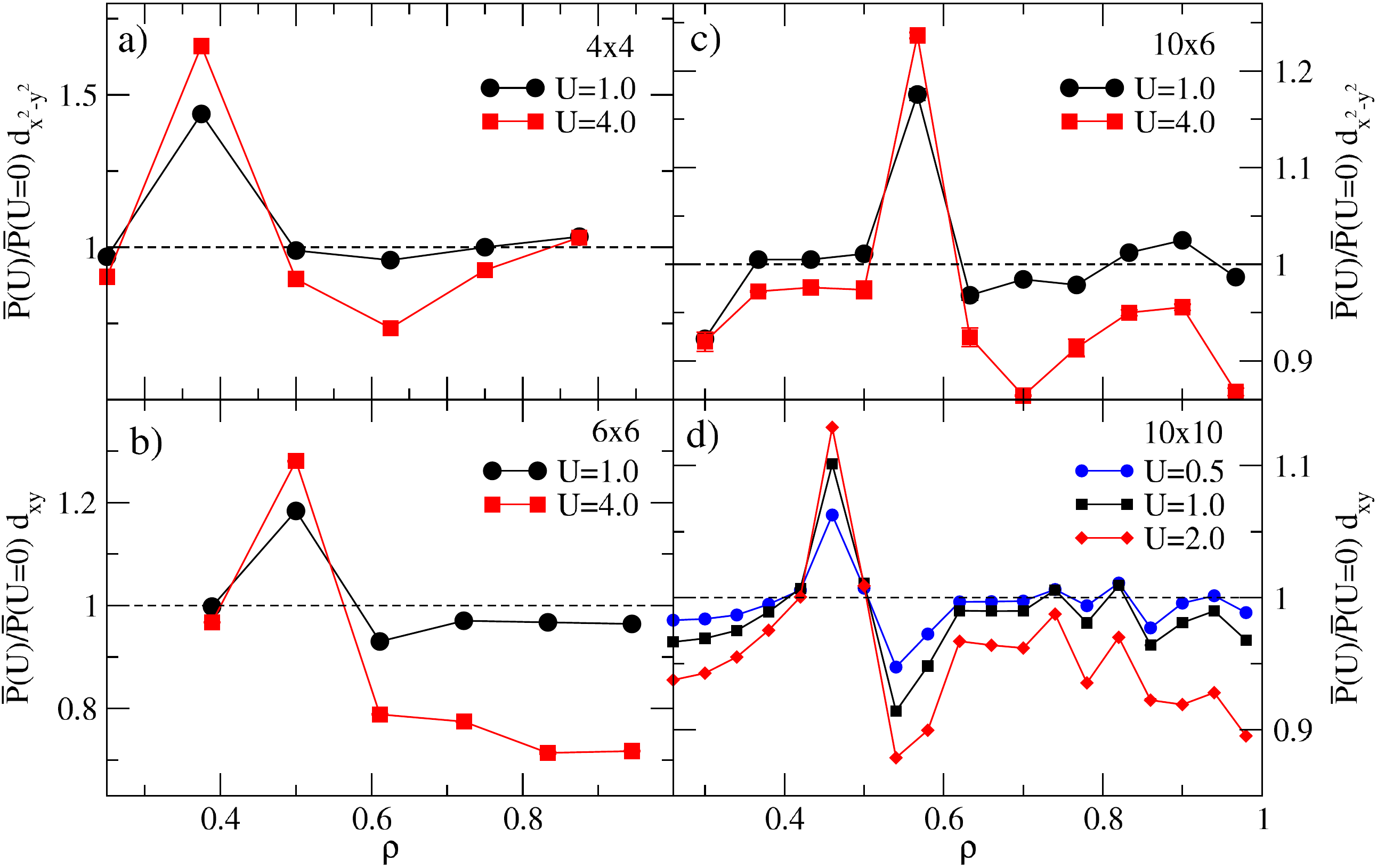}}
  \end{center}
  \caption{(color online) Average long-range pair-pair correlation
    $\bar{P}(U)$ normalized by its noninteracting value versus $\rho$
    on the anisotropic triangular lattice with $t_{\rm x}=1$, $t_{\rm
      y}=0.9$, and $t^\prime$=0.8. The lattices and methods used are
    (a) 4$\times$4 (exact), (b) 6$\times$6 (QP-PIRG), (c) 10$\times$6
    (QP-PIRG), and (d) 10$\times$10 (CPMC).}
  \label{pp-fig1}
\end{figure}

\begin{figure}[tb]
  \begin{center}
    \resizebox{4.5in}{!}{\includegraphics{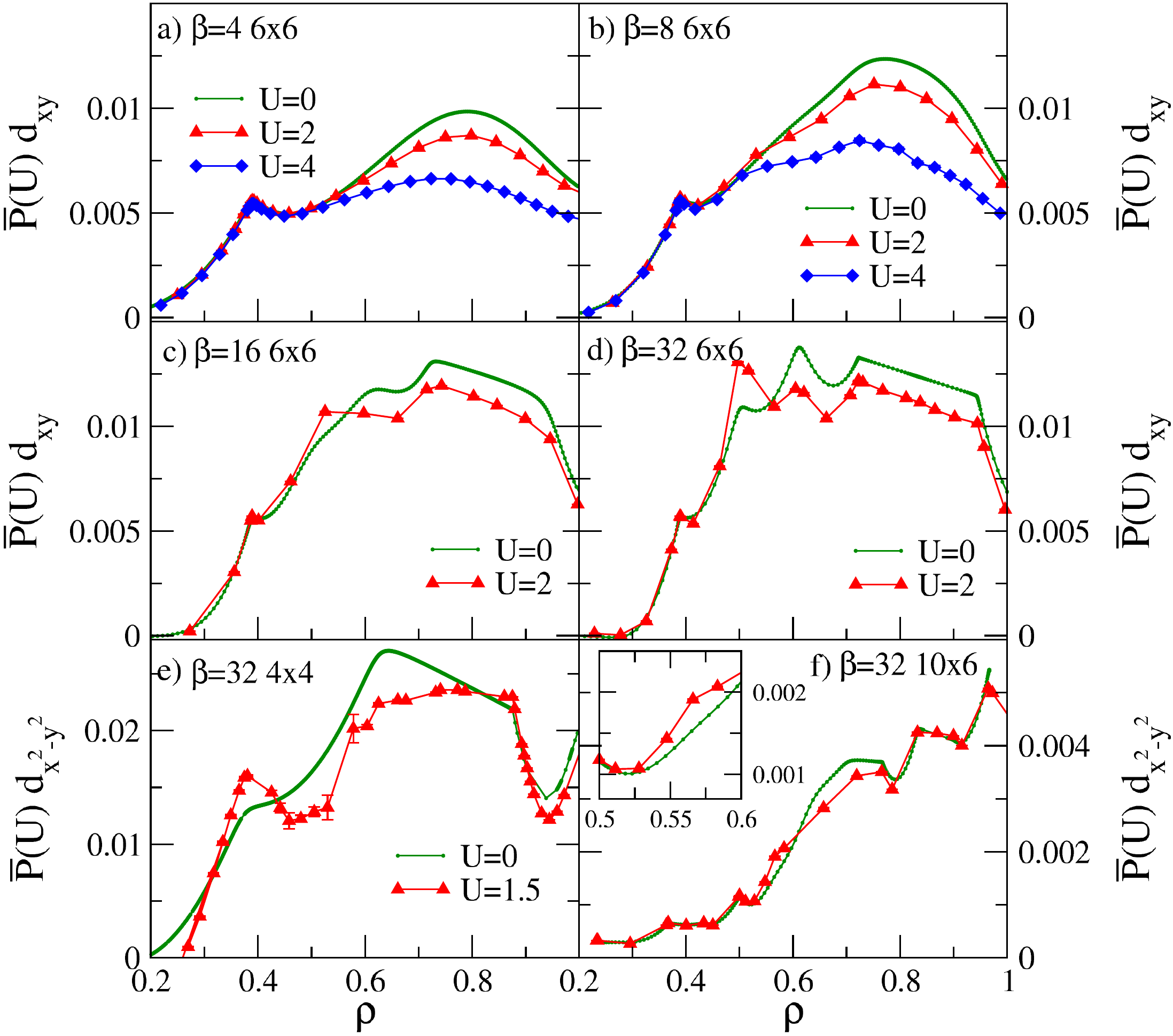}}
  \end{center}
  \caption{(color online) (a)-(d) $\bar{P}(U)$ for $d_{xy}$ pairing
    on the 6$\times$6 lattice as a function of $\rho$
    and inverse temperature $\beta$, calculated using DQMC.
    $\bar{P}(U)$ for $\rho\approx0.5$ is enhanced by $U$ at low
    temperatures, while it is suppressed at other $\rho$.  (e)
    $\bar{P}(U)$ for $d_{x^2-y^2}$ pairing on the 4$\times$4 lattice,
    and (f) $\bar{P}(U)$ for $d_{x^2-y^2}$ pairing for the 10$\times$6
    lattice \cite{Gomes16a}. The inset enlarges the density range near
    $\rho\sim0.5$.}
  \label{pp-fig2}
\end{figure}

While many previous calculations have examined superconducting
pair-pair correlations for $\rho\sim 1$, comparatively few
calculations exist for $\rho=\frac{1}{2}$ on strongly frustrated lattices.  We
consider again the anisotropic triangular lattice of
Fig.~\ref{4x4pec}, with $t^\prime\sim$0.8 \cite{Gomes16a}. We
performed calculations on up to 10$\times$10 lattices periodic in all
three directions with four different numerical methods (exact
diagonalization, QP-PIRG \cite{Mizusaki04a}, Determinantal Quantum
Monte Carlo (DQMC) \cite{Blankenbecler81a}, and Constrained Path Monte
Carlo (CPMC) \cite{Zhang97a} \cite{Gomes16a}).  These particular
lattices were chosen using several criteria, including the requirement
that the noninteracting system be nondegenerate at $\rho=\frac{1}{2}$
\cite{Gomes16a}.  Results from these calculations are shown in
Fig.~\ref{pp-fig1} and Fig.~\ref{pp-fig2}.

In order to compare different lattices and mitigate finite-size
effects, we average the pair-pair correlations $P(r)$
(see Eq.~\ref{pr}) over all pair separations with distances
$r>2$ lattice spacings,
\begin{equation}
  \bar{P}=N^{-1}_P\sum_{|\vec{r}|>2}P(r),
  \label{pbar}
\end{equation}
where $N_P$ is the number of terms in the sum. A similar averaged
pair-pair correlation $\bar{P}$ has been used in previous studies of
pair-pair correlations near $\rho=1$ \cite{Huang01a,Yokoyama06a}.  In
these calculations we found either $d_{x^2-y^2}$ or $d_{xy}$ pairing
symmetries to dominate over $s$-wave pairing.  Fig.~\ref{pp-fig1}
summarizes these calculations.  Here $\bar{P}$ is normalized by its
$U=0$ value; $\bar{P}(U)/\bar{P}(U=0)>1$ indicates enhancement. As
seen in Fig.~\ref{pp-fig1}, for every lattice we find that pairing is
enhanced {\it only} for $\rho=\frac{1}{2}$ or one of two immediately
adjacent available densities on that particular lattice. At all other
densities, pair-pair correlations weaken continuously with increasing
$U$. The suppression of pairing in the range $0.75 < \rho < 1$ is
consistent with
previous
PIRG
\cite{Dayal12a} results, and consistent with many calculations
  using DQMC \cite{White89b}, CPMC \cite{Zhang97b,Guerrero99a}, and
  DMRG \cite{Ehlers17a} that find suppression of pairing by $U$ in
  the unfrustrated model for the same density range.
Ignoring the 6$\times$6 lattice where
$\bar{P}$ is enhanced at exactly $\rho=\frac{1}{2}$, for all other lattices
the $\rho$ where enhancement occurs becomes closer to $\rho=\frac{1}{2}$ with
increasing lattice size.  On different lattices either $d_{x^2-y^2}$
or $d_{xy}$ pairing is enhanced by $U$; it would be expected in this
highly frustrated lattice ($t^\prime=0.8$ in Fig.~\ref{pp-fig1} and
\ref{pp-fig2}) that the optimum pair symmetry is a superposition of
$d_{x^2-y^2}$ and $d_{xy}$.  The effect of nonzero $V_{ij}$ was also
investigated for the 4$\times$4 lattice \cite{Gomes16a}. With
$V_x=V_y=V^\prime=V$ the magnitude of the pair-pair correlations
decreases somewhat, but the unique enhancement at $\rho\sim0.5$
remains \cite{Gomes16a}.

Fig.~\ref{pp-fig2} shows the results of finite-temperature DQMC
calculations. Here the sign problem becomes severe at large $U$, but
low temperatures (an inverse temperature of $\beta=32$) can be reached
for $U=2$. As seen in Fig.~\ref{pp-fig2}(a)-(d), with decreasing
temperature there is a progressive enhancement of $\bar{P}(U)$ for
$\rho\sim \frac{1}{2}$ in the 6$\times$6 lattice. Fig.~\ref{pp-fig2}(e) and (f)
show similar results for the 4$\times$4 and 10$\times$6
lattices. These results are in complete agreement with the PIRG
results in Fig.~\ref{pp-fig1}: the $\rho$ at which the enhancement
occurs is identical.

To the best of our knowledge, the above results are the only
demonstrations to date of enhancement of $\bar{P(U)}$ by $U$ for any
lattice.  The question remains as to whether the first criterion for
SC given above is satisfied, e.g. do the pair-pair correlations show
long-range order at $\rho=\frac{1}{2}$? Unfortunately, the results of
Figs.~\ref{pp-fig1} and \ref{pp-fig2} are not sufficient to perform a
finite-size scaling of $P(r)$: the enhancement occurs at slightly
different densities from $\rho=\frac{1}{2}$ on different lattices, and
in different pairing symmetries. It is possible that larger lattices
and an optimized superposition of $d_{xy}$ and $d_{x^2-y^2}$ pair
symmetries would resolve these difficulties.  If superconducting
long-range order (LRO) is present at finite $U$, $\bar{P}(U)$ would
converge to a constant value as the system size increases while
$\bar{P}(U=0)$ would continue to decrease. In this case
$\bar{P}(U)/\bar{P}(U=0)$ would increase with increasing system size.
Our finding that $\bar{P}(U)/\bar{P}(U=0)$ at its peak value instead
decreases with increasing system size (compare the $y$-axis scales on
Fig.~\ref{pp-fig1}(a)-(d)) on the other hand seems to suggest that
long-range order is not present. The issue of the long range order is,
however, complicated.  We discuss this further in Section
\ref{conclusions}.

\subsubsection{Superconducting pair-pair correlations in the  $\kappa$-(ET)$_2$X lattice}
\label{kappa-pairpair}

\begin{figure}[tb]
  \begin{center}
    \resizebox{1.5in}{!}{\includegraphics{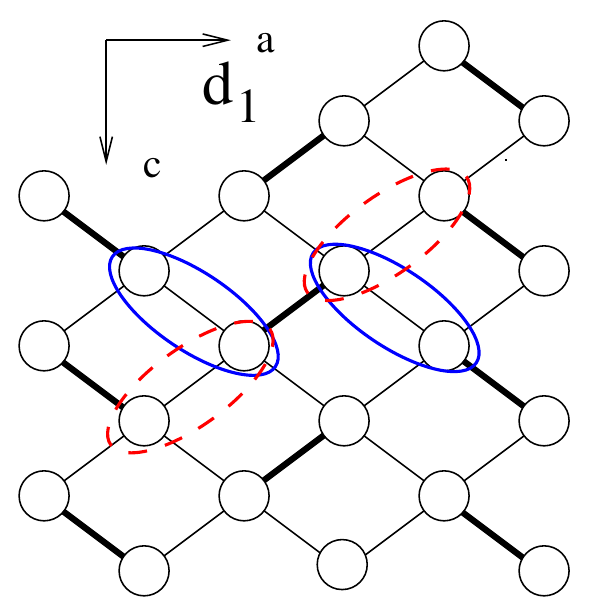}}\hspace{0.1in}%
    \resizebox{1.5in}{!}{\includegraphics{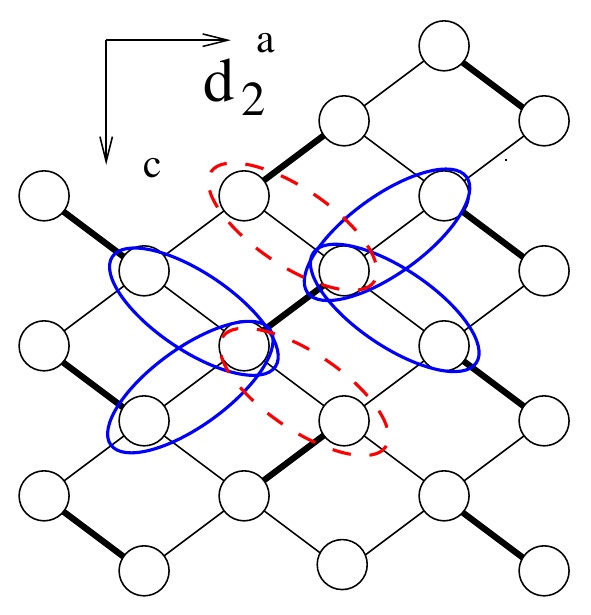}}
  \end{center}
  \begin{center}
    \resizebox{1.5in}{!}{\includegraphics{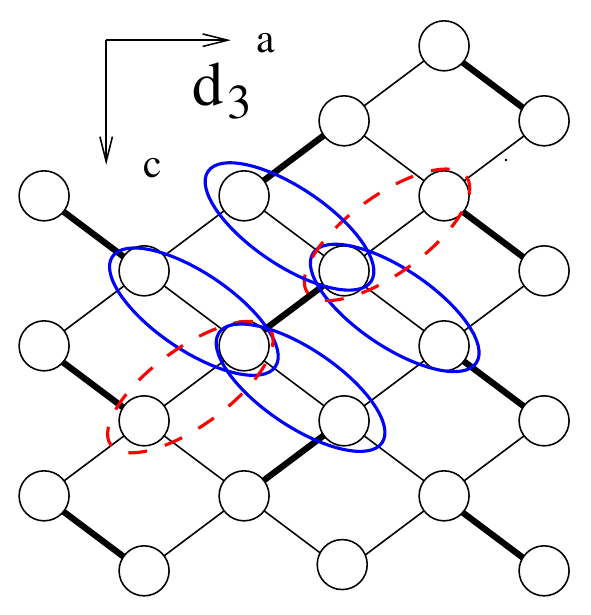}}\hspace{0.1in}%
    \resizebox{1.5in}{!}{\includegraphics{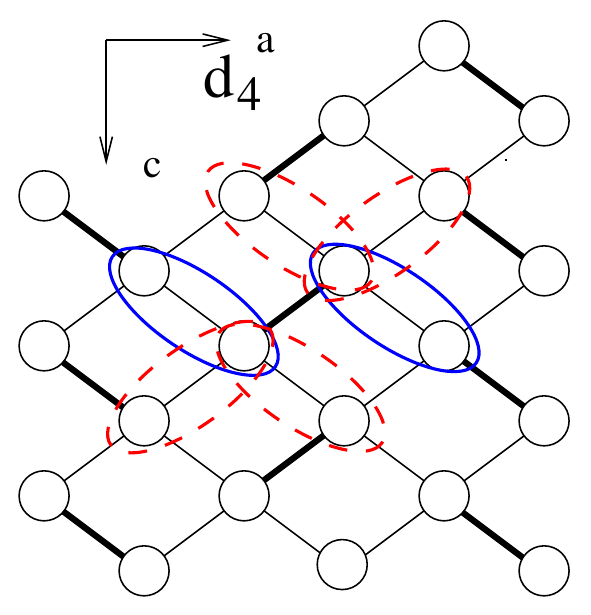}}
  \end{center}
  \caption{(color online) Pairing symmetries considered in
    Fig.~\ref{pp-kappa}. Here the thick lines correspond to BEDT-TTF
    dimers. Each ellipse surrounding two sites indicates
    a singlet in the superposition of the pair creation operator
    $\Delta^\dagger_i$ centered at dimer $i$. Blue, solid (red,
    dashed) singlets have opposite signs \cite{DeSilva16a}.}
  \label{kappa-symmetries}
\end{figure}

\begin{figure}[tb]
  \begin{center}
    \resizebox{2.6in}{!}{\includegraphics{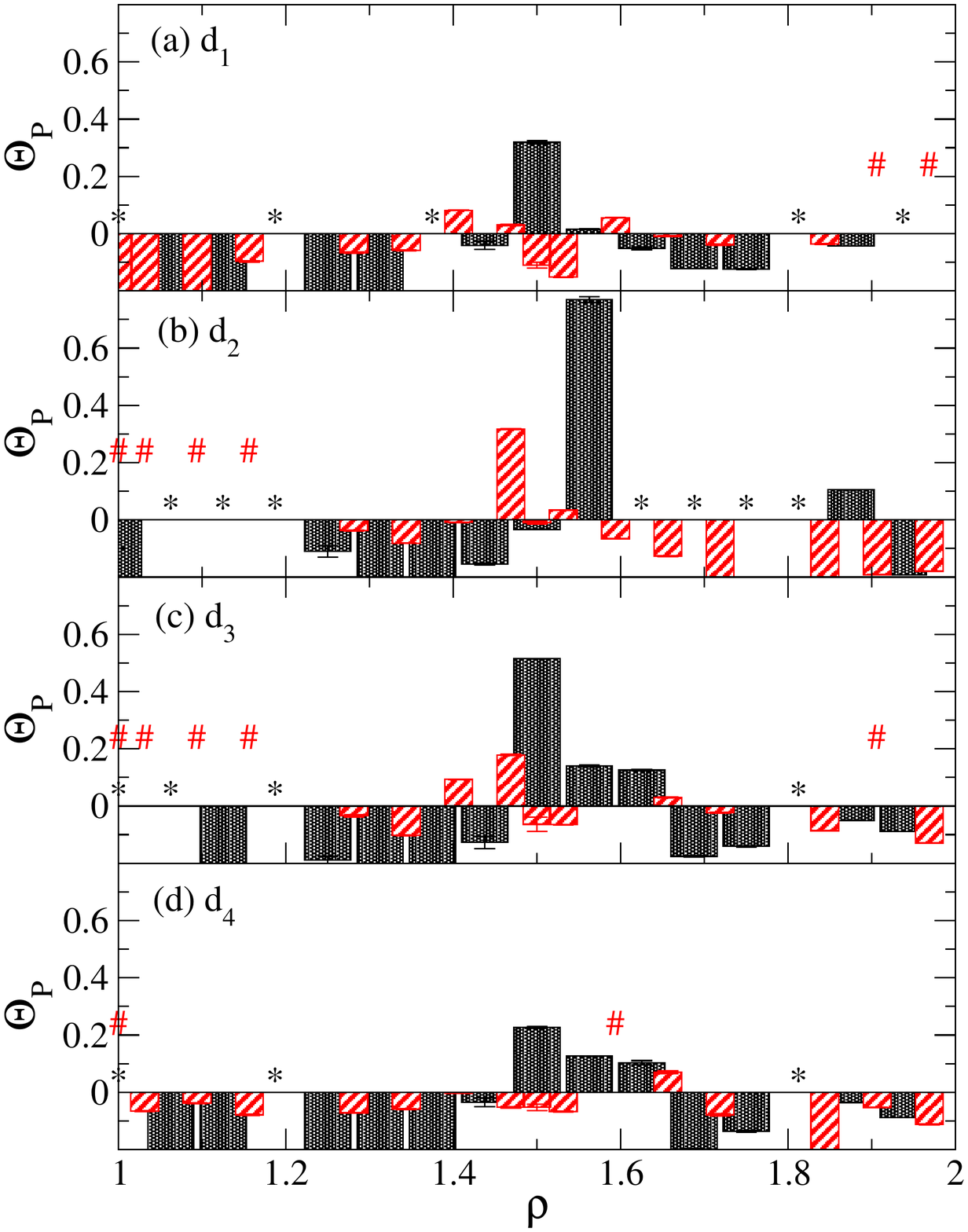}}\hspace{0.1in}%
    \resizebox{2.6in}{!}{\includegraphics{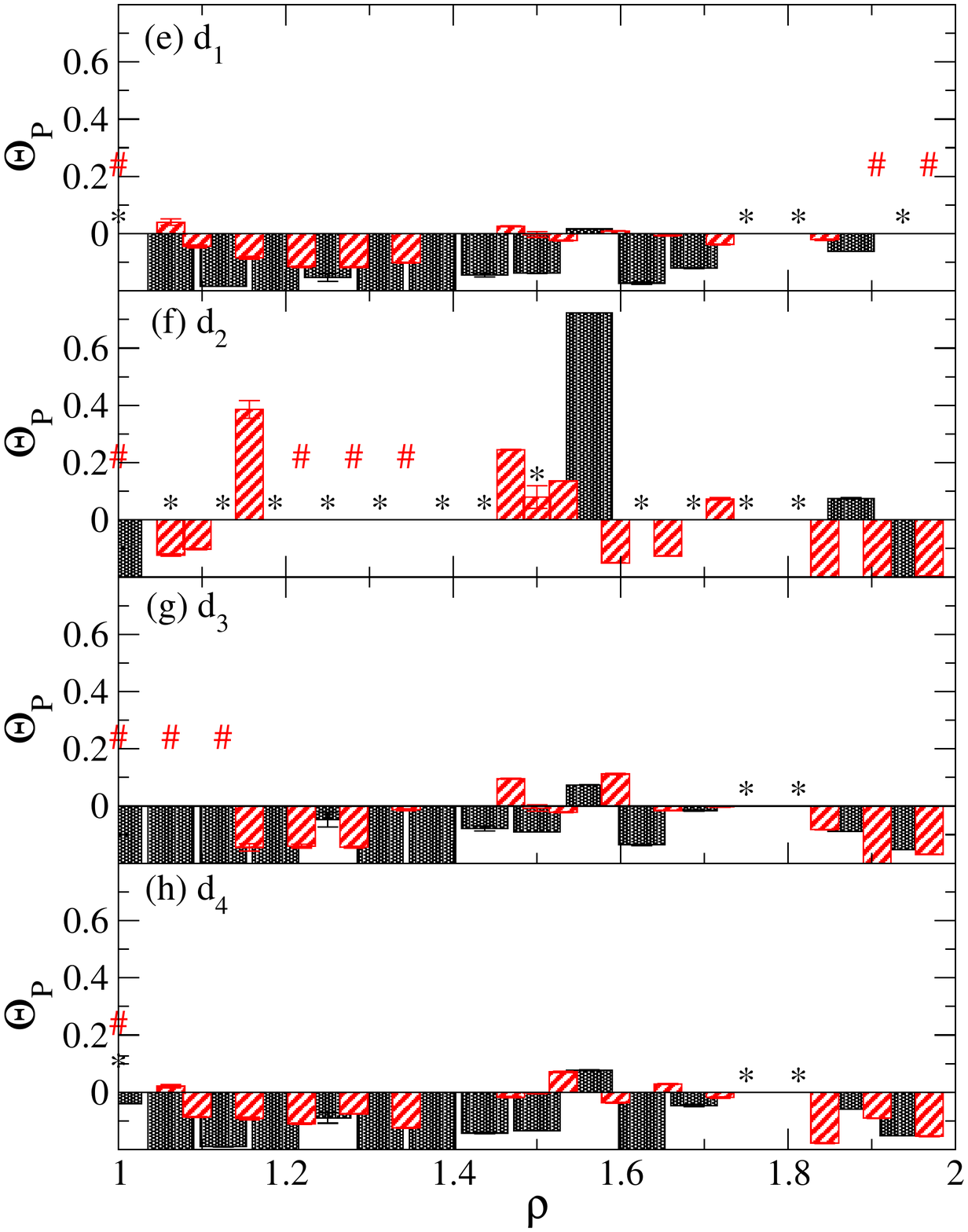}}
  \end{center}
  \caption{(color online) Enhancement factor $\Theta_{\rm P}$ (see
    text) versus electron density $\rho$ for the long-range component
    of pair-pair correlations for a $\frac{3}{4}$-filled monomer model
    of $\kappa$-(ET)$_2$X for four different pairing symmetries (see
    Fig.~\ref{kappa-symmetries}) \cite{DeSilva16a}.  Shaded (striped)
    bars are for 32 (64) site lattices using QP-PIRG for 32 sites and
    CPMC for 64 sites; $U=0.5$ eV for all results.  Symbols `*' and
    `\#' indicate densities not shown, for 32 and 64 sites,
    respectively, due to strong finite-size effects at these $\rho$
    that result in $\bar{P}(U=0)\approx0$.  Panels (a)-(d) are for
    X=Cu[N(CN)$_2$]Cl and (e)-(h) are for X=Cu$_2$(CN)$_3$
    \cite{DeSilva16a}.  We are ignoring the positive $\Theta_{\rm P}$
    in panel (f) at $\rho=1.156$.  At this density, a discontinuous
    transition occurs for small $U$, suggesting that the apparent
    enhancement is a finite-size effect connected to the band
    structure \cite{DeSilva16a}.}
  \label{pp-kappa}
\end{figure}

The above results strongly suggest that SC is driven by e-e
interactions uniquely at $\rho=\frac{1}{2}$. It is therefore important
to see if they persist on other lattices, and in particular for
$\kappa$-(ET)$_2$X, whose monomer lattice (see
Fig.~\ref{kappa-lattice}(a)) is significantly different from the
simple anisotropic triangular lattice of Fig.~\ref{pp-fig1}. The first
question is what is an appropriate form for the pair wavefunction. In
SC emerging from a charge ordered PEC state with strong e-e
interactions, $U$ for holes on a single {\it molecule} will be strong.
The dominant configurations in the wavefunction will still have one
hole per dimer, but while the charge densities are homogeneous in the
configurations that give AFM, in the superconducting state, which is
spin singlet, the intradimer charge densities in dominant wavefunction
configurations are necessarily inhomogeneous, with singlet bonds
between charge-rich sites on neighboring molecules (see
Fig.~\ref{4x4pec}).  Fig.~\ref{kappa-symmetries} shows possible
$d$-wave (i.e. with four nodes) symmetries constructed using this
constraint. Each ellipse in Fig.~\ref{kappa-symmetries} corresponds to
one term in the construction of $\Delta^\dagger_i$ (Eq.~\ref{ppdelta})
that creates a superconducting pair centered at dimer $i$
\cite{DeSilva16a}. The $d_1$ symmetry consists of a superposition of
four NN singlets, while the other symmetries ($d_2$, $d_3$, and $d_4$)
are a superposition of 6 singlets. The nodes of these order parameters
occur in different locations. For $d_1$ they are aligned with the
crystal axes, but for the other order parameters one or both of the
nodal lines are at an angle to the crystal axes.  The $d_1$ symmetry
is equivalent to the ``$d_{x^2-y^2}$'' symmetry in the effective
$\frac{1}{2}$-filled band model; $d_2\ldots d_4$ correspond to mixed
symmetries of $s+d_{x^2-y^2}$ or $s+d_{xy}$ form.  We also
investigated $s$-wave pair symmetries, but found no enhancement of
$s$-wave pairing.

We performed calculations for the monomer $\kappa$-(ET)$_2$X lattice
as shown in Fig.~\ref{kappa-lattice}(a) for two lattice sizes, 32 and
64 monomer sites.  Calculations were done in the electron representation.
As with the triangular lattice calculations (Fig.~\ref{pp-fig1}) we performed
the $\kappa$-lattice calculations for a wide range of $\rho$, even as only 
$\rho=1.5$ corresponds to the carrier density in the CTS.
Both lattices used periodic
boundary conditions, chosen so that the corresponding effective dimer
model would support N\'eel AFM order without frustration
\cite{DeSilva16a}.  The 32 and 64 site lattices were arranged as
4$\times$4 and 8$\times$4 dimer lattices, respectively.  The hopping
parameters we used were taken from recent DFT calculations of
Reference \cite{Koretsune14a}. We investigated two different sets of
parameters, for the crystal structures of $\kappa$-(ET)$_2$X with
X=Cu[N(CN)$_2$]Cl and X=Cu$_2$(CN)$_3$, in order to assess the
importance of AFM order to SC in $\kappa$-(ET)$_2$X.  We calculated
ground-state correlation functions with two different methods, QP-PIRG
and CPMC. The CPMC was restricted to only closed-shell configurations
where it is more accurate \cite{Zhang97a}.  As in the results of
Section \ref{enhancedpp} we calculated the average of long-range
pair-pair correlations, $\bar{P}$.  Because $\Delta^\dagger_i$ creates
a pair centered at dimer $i$, we define the pair-pair correlation
function $P(r)$ with the distance $r$ measured in units of dimer-dimer
spacing.  The minimum distance cutoff in Eq.~\ref{pbar} is again two
in units of the dimer-dimer distance.

Fig.~\ref{pp-kappa} summarizes the results of these calculations.
Instead of $\bar{P}$ we plot $\Theta_P=[\bar{P}(U)/\bar{P}(U=0)]-1$,
which is positive only if e-e interactions enhance the pairing.  The
strongest enhancement for both lattices ($\kappa$-Cl and $\kappa$-CN)
is found for $\rho\approx 1.5$.  The results in Fig.~\ref{pp-kappa}
show that as in the anisotropic triangular lattice, pairing is
enhanced by e-e interactions uniquely at or near quarter filling by
holes, here at $\rho=1.5$ \cite{DeSilva16a}.  The enhancement is
strongest for the $d_2$ symmetry, which notably is the {\it only} pair
symmetry that shows any significant enhancement for $\kappa$-CN.  We
note that a similar $s+d_{x^2-y^2}$ order parameter was found in RPA
spin-fluctuation calculations in a $\frac{3}{4}$-filled model
\cite{Guterding16b,Guterding16a,Zantout18a} and in a subsequent VMC
calculation \cite{Watanabe17a}.  The nodes of the order parameter in
the $d_2$ symmetry (Fig.~\ref{kappa-symmetries}) are located at an
angle to the crystal axes \cite{DeSilva16a}.
\begin{figure}[tb]
  \begin{center}
    \resizebox{3.5in}{!}{\includegraphics{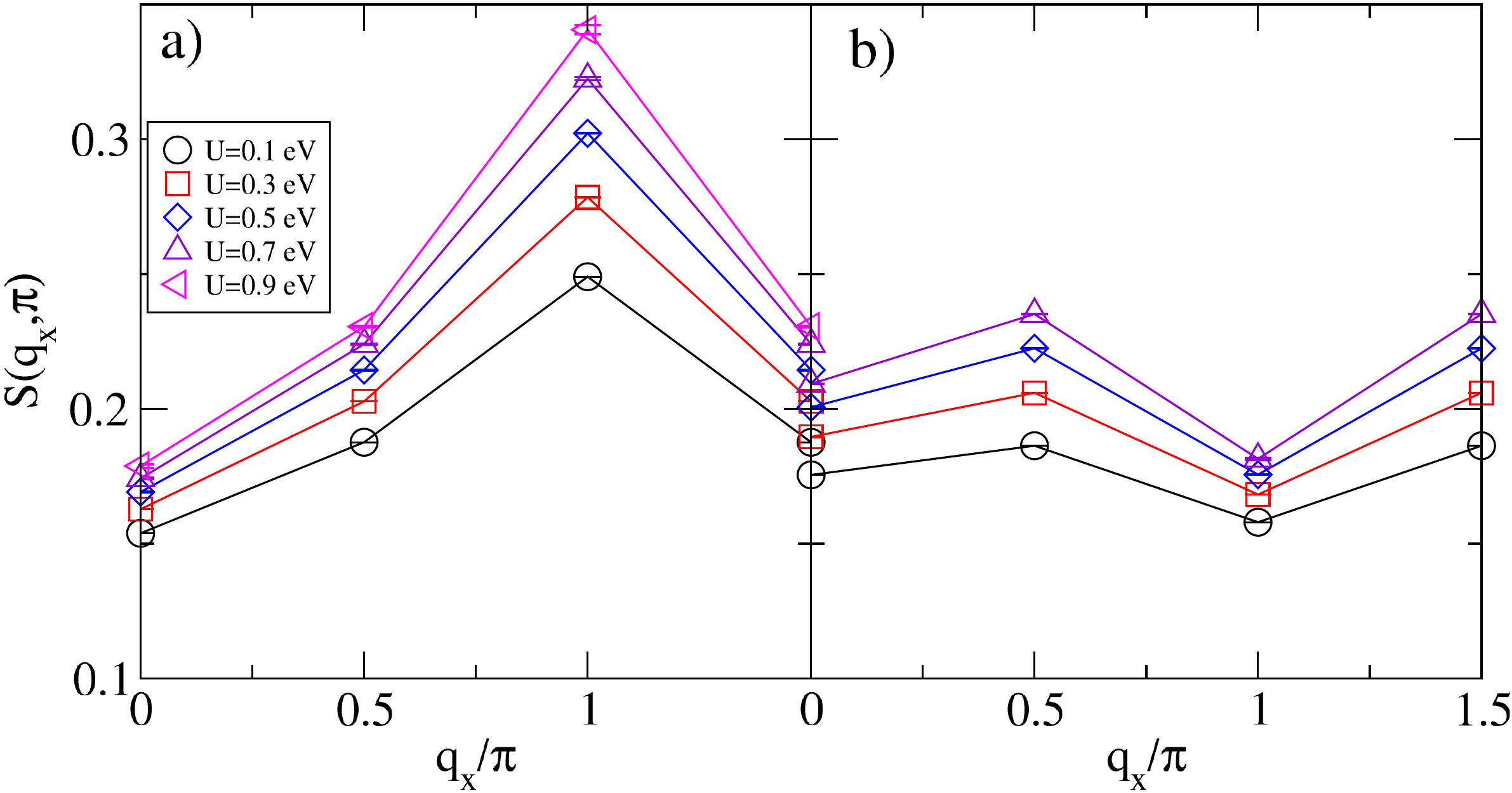}}
  \end{center}
  \caption{(color online) Dimer spin structure factor $S({\bf q})$
    calculated by QP-PIRG for a 32 site monomer lattice of (a)
    $\kappa$-Cl and (b) $\kappa$-CN. Wavevectors are defined in terms
    of the effective dimer lattice (see text). }
  \label{kappa-sfac}
\end{figure}

Dimer-dimer AFM correlations can be measured by the dimer spin
structure factor,
$$
S({\bf q})=\frac{1}{N_d}\sum_{j,l}e^{i{\bf q}\cdot({\bf r}_j-{\bf r}_l)}\langle S^z_jS^z_l\rangle,
$$ where the sums run over all dimers in the system and $N_d$ is the
total number of dimers. $S^z_j=n_{j_1,\uparrow}+n_{j_2,\uparrow}
-n_{j_1,\downarrow}-n_{j_2,\downarrow}$ is the operator for the $z$
component of the total spin on dimer $j$ composed of molecules $j_1$
and $j_2$ and {\bf q} the wavevector defined in terms of dimer
coordinates (the $\hat{x}$ and $\hat{y}$ axes of
Fig.~\ref{kappa-lattice}(a)) . Results for $S({\bf q})$ are shown in
Fig.~\ref{kappa-sfac}. For $\kappa$-Cl (Fig.~\ref{kappa-sfac}(a)), a
clear peak at {\bf q}=$(\pi,\pi)$ is seen indicating the tendency to N\'eel AFM
order. For $\kappa$-CN (Fig.~\ref{kappa-sfac}(b)) this peak is absent
as expected due to the strong frustration in the effective dimer
lattice. The lack of peaks in $S({\bf q})$ for $\kappa$-CN is
consistent with the QSL behavior found in this salt.
However, we find strong pairing enhancement in the $d_2$ symmetry
 for both $\kappa$-Cl and $\kappa$-CN, which would not occur in
a $\rho=1$ model of SC  mediated by dimer-dimer AFM fluctuations.

Finally, we present preliminary results for calculations of SC
pair-pair correlations on the lattice of $\kappa$-(ET)$_2$CF$_3$SO$_3$
($\kappa$-SO$_3$) \cite{Gomes17a}.  $\kappa$-SO$_3$ is unique among
$\kappa$-(ET)$_2$X superconductors for several reasons (see Section
\ref{kappa}). First, this material undergoes a structural transition
at $\sim$ 230 K, below which it has a bilayer structure. Second, the
superconducting critical temperature T$_{\rm c}$ under pressure
(T$_{\rm c}$ =4.8 K) is {\it higher} than the ambient pressure N\'eel
temperature T$_{\rm N}$ (T$_{\rm N}$ = 2.5 K). This makes the
applicability of any effective $\frac{1}{2}$-filled band theory, or
spin-fluctuation mediated SC questionable.  The lattice structure of
$\kappa$-SO$_3$ is strongly anisotropic, with $t^\prime/t$ within the
effective dimer lattice (see inset of Fig.~\ref{rho1-anisotropic})
greater than one; it is the largest $t^\prime/t$ of all known
$\kappa$-(ET)$_2$X \cite{Ito16a}).  In the $\rho=1$ Hubbard model on
the anisotropic triangular lattice, $t^\prime/t>1$ gives AFM ordering
with wavevector {\bf Q}=($\pi$,0) rather than {\bf Q}=($\pi$,$\pi$)
\cite{Clay08a,Acheche16a}.  Within $\rho=1$ models of spin-fluctuation
induced SC on the anisotropic triangular lattice, large $t^\prime$
strongly reduces the T$_{\rm c}$ of $d_{x^2-y^2}$ SC
\cite{Kino98a}. Other theoretical works suggest that the symmetry of
the SC order parameter changes for $t^\prime/t>1$, from $d_{x^2-y^2}$
to $d_{xy}-s$ \cite{Powell07a} or a $d+id$ state \cite{Gan06a}. It is
however hard to reconcile the $T_{\rm c}>T_{\rm N}$ within any theory
of SC mediated by AFM fluctuations.

\begin{figure}[tb]
  \begin{center}
    \resizebox{2.6in}{!}{\includegraphics{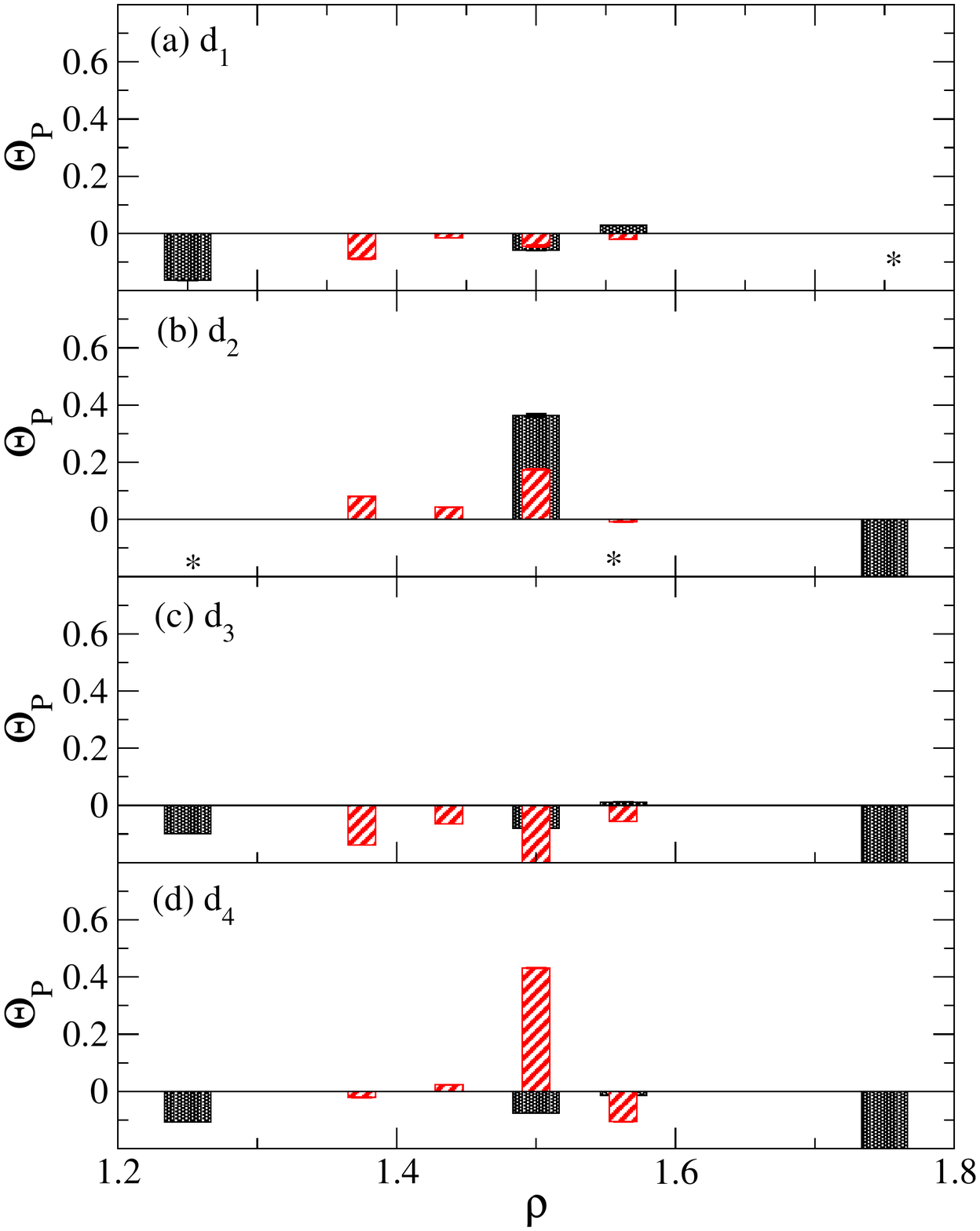}}\hspace{0.1in}%
    \resizebox{2.55in}{!}{\includegraphics{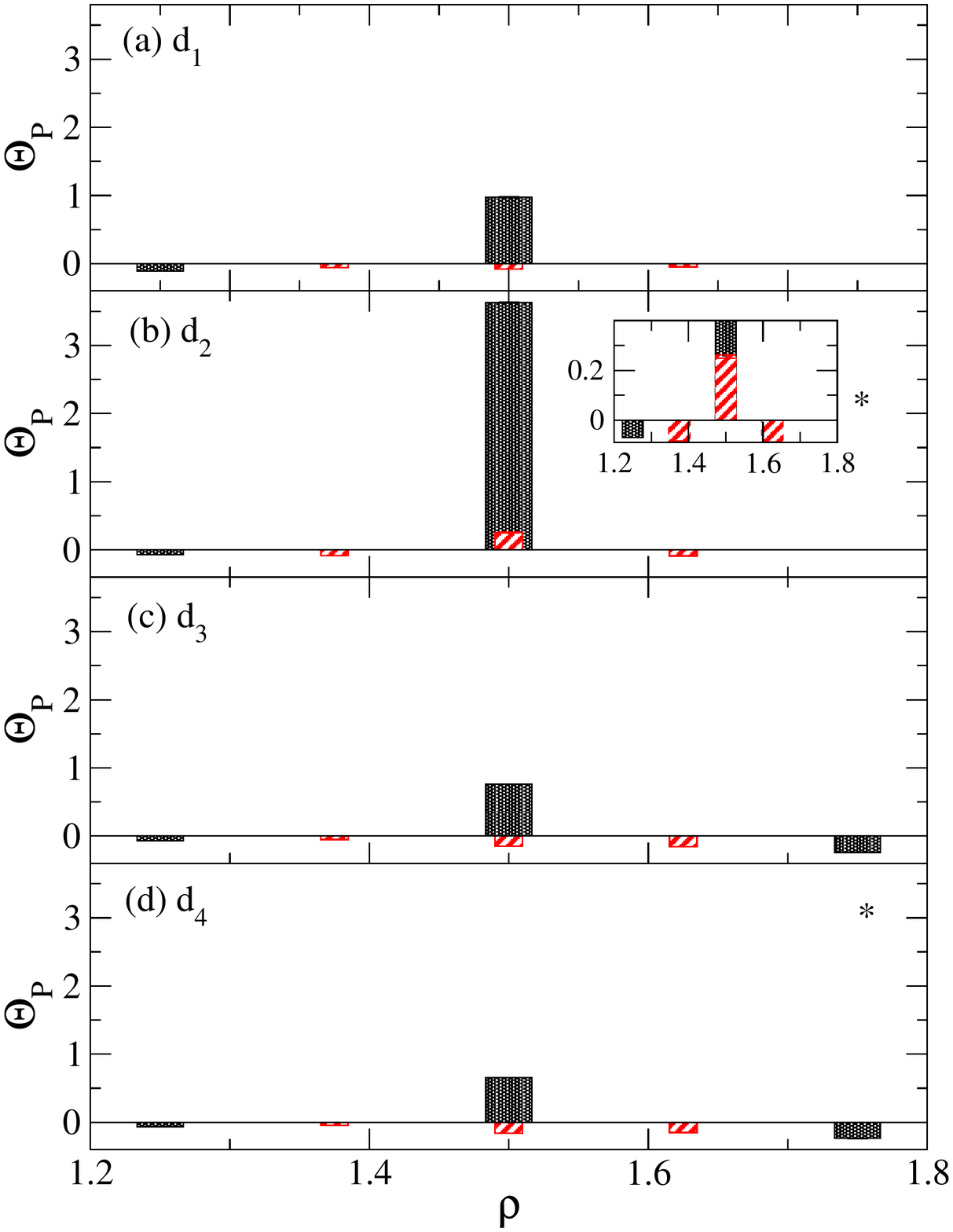}}
  \end{center}
  \caption{(color online) Preliminary results for the enhancement
    factor $\Theta_{\rm P}$ versus electron density $\rho$ for the bilayer
    $\kappa$-(ET)$_2$CF$_3$SO$_3$ lattice. Panels (a)-(d) are for layer A and
    panels (e)-(h) for layer B (see text).  Shaded (striped) bars are
    for 32 (64) site lattices using QP-PIRG for 32 sites and CPMC for
    64 sites; $U=0.4$ eV for all results.
    64 site results are restricted to the range $1.35 < \rho < 1.65$.
    As in Fig.~\ref{pp-kappa}, the
    symbols `*' indicate 32 site points removed due to finite-size effects.}
  \label{pp-cf3so3}
\end{figure}

Because of the unusual properties of $\kappa$-SO$_3$ calculations
within the full monomer lattice are desirable. We performed QP-PIRG
and CPMC calculations on 32 and 64 site systems with the same
methodology as detailed above for $\kappa$-Cl and $\kappa$-CN.  We
took hopping parameters from Reference \cite{Ito16a}.  The two
distinct layers, labeled A and B here as in Reference \cite{Ito16a},
have slightly different hopping integrals. $t^\prime/t$ within the
effective dimer model is 1.50 and 1.77 for layers A and B respectively
\cite{Ito16a}. Fig.~\ref{pp-cf3so3} shows the results of preliminary
calculations of ground-state superconducting pair-pair correlations
both layers.  As with the calculations for $\kappa$-Cl and
$\kappa$-CN, the most striking feature of these results is that
enhancement of SC by $U$ is restricted to $\rho=1.5$.  For
$\kappa$-SO$_3$ the greatest enhancement is again found for the $d_2$
pair symmetry.  Comparing the layers A and B, the enhancement is
significantly larger for layer B: for monomer $U=0.4$ eV, the $d_2$
enhancement factor $\Theta_P$ is greater than 3 for layer B, while it
is only $\sim$ 0.4 for layer A.  A direct comparison of the amount of
enhancement between $\kappa$-Cl, $\kappa$-CN, and $\kappa$-SO$_3$ is
difficult with the lattice sizes we have access to, as the $U=0$ Fermi
level degeneracies are different which leads to slightly different
finite-size effects in each case. For $\kappa$-SO$_3$ both A and B
layers {\it do} have similar Fermi level degeneracies however, so we
believe that the finding of stronger pairing enhancement in layer B is
significant.

\subsection{Summary}

Direct many-body numerical calculations do not find superconductivity
(SC) in $\rho=1$ models. The effective $\rho=1$ model for dimerized
CTS also fails to correctly describe some of their insulating phases:
while $\rho=1$ effective model can explain the presence of
antiferromagnetic and metallic states in the CTS, a ``valence-bond
solid'' (VBS) phase is not found in the effective $\rho=1$ model.  The
correct theoretical description of the VBS phase found in
EtMe$_3$P[Pd(dmit)$_2$]$_2$ {\it requires} $\rho=\frac{1}{2}$
\cite{Yamamoto17a}. The charge-ordered and spin-gapped state found
here is the Paired Electron crystal (PEC).  Our many-body calculations
with monomer molecules as units reproduce the strong tendency to
antiferromagnetism in $\kappa$-(ET)$_2$Cu[N(CN)$_2$]Cl and the absence
of the same in $\kappa$-(ET)$_2$Cu$_2$(CN)$_3$.  SC in CTS is a
consequence of the spin-singlets in the PEC becoming mobile to give a
paired-electron liquid, driven by further increase in frustration.
This feature is a unique characteristic of the correlated
$\frac{1}{4}$-filled band (see also \ref{othersc}), although it may be
perhaps observed at other fillings depending on lattice geometry.

\subsubsection{Application of theory to CO-to-SC transitions.} 

As discussed extensively in Section~\ref{pec2dcts} the pattern of the
CO in CTS that exhibit CO-to-SC transition ($\beta$, $\beta^\prime$,
$\beta^{\prime\prime}$, $\theta$ and $\alpha$ BEDT compounds, as well
as in EtMe$_3$P[Pd(dmit)$_2$]$_2$) in every case corresponds to the
PEC. This cannot be a coincidence.  Note also that numerical
calculations find that the PEC is favored over the WC of single
electrons in the frustrated lattice (Fig.~\ref{pecphasediagram}(c)).
The VB theory of SC, and the calculations reported in Sections
\ref{pec2d} and \ref{enhancedpp} apply most directly to these systems.

\subsubsection{Where is the PEC in the $\kappa$-(ET)$_2$X?}

Considering $\kappa$-(ET)$_2$X, the results shown in
Figs.~\ref{pp-kappa} and \ref{pp-cf3so3} show that the resulting
picture of SC is quite different from that in the effective dimer
model.  Within the actual $\rho=1.5$ system in order to understand SC
one must go beyond the effective dimer model, explicitly consider
charge imbalance between the molecules within dimers, and the the
formation of interdimer spin-singlet pairs between the charge-rich
sites of nearest-neighbor dimers. We emphasize that ours is a
strong-correlation model very different from theories based on
mean-field and random phase approximation.  Such a model however
raises the question: where is the PEC in these materials?  The answer
lies in the recognition that the frustration in the superconducting
$\kappa$-(ET)$_2$X, even when considering monomers and not dimers as
the units, is significantly larger than in the other ET compounds,
such that there are many equivalent and degenerate PEC configurations
here \cite{Li10a}.  Instead of crystallization to a single such
configuration, there is thus transition to what we believe is a
superposition of PEC-like configurations, which is however a
spin-singlet state. It can also be thought of as a transition from an
effective $\frac{1}{2}$-filled band to an effective
$\frac{3}{4}$-filled band, as visualized in the schematic phase
diagram of Fig.~\ref{beta-prime-icl2}. This is perhaps the only
scenario that can give consistent explanations of (i) the pseudogap
in some compounds, whose spin-singlet nature should be obvious from
Figs.~\ref{kappa-pg1}-\ref{kappa-pg2}, (ii) strong lattice effects
including anomalies in the lattice expansivity in the same temperature
region \cite{Muller02a,Souza15a}; and (iii) resistivity anomalies.
Note also that the natures of the various broken symmetries below 6 K
in $\kappa$-CN remain unresolved. It is not inconceivable that in this
highly frustrated system the same transition occurs at a much lower
temperature. Finally, the occurrence of CO in X = Hg(SCN)$_2$Cl and
spin gap in X = B(CN)$_4$ give indirect evidence for the proximity of
the PEC to SC.

\subsubsection{Application to the quasi-1D superconductors.}
Although we have not directly calculated superconducting pair-pair correlations
for the quasi-1D (TMTSF)$_2$X, we believe that our demonstration of
the BCDW in (TMTTF)$_2$X and the BCSDW in (TMTSF)$_2$X (Section
\ref{1d-expt}), taken together with our calculations for the
$t^\prime/t > 1$ ET compound $\kappa$-SO$_3$ (Fig.~\ref{pp-cf3so3}) indicate very
strongly that the mechanism of SC in the quasi-1D and quasi-2D CTS is
the same. The only difference is that the superconducting wavefunction
in the quasi-1D compounds must consist of predominantly intrastack
spin-singlets with significantly smaller contribution from interdimer
singlets. This will obviously change the gap symmetry.

\subsubsection{Similarities with high-T$_{\rm c}$ cuprates}
\label{cuprates}

The obvious similarities between SC in the cuprates and in the organic
CTS are the layered natures of both families and the proximity of SC
to AFM and QSL states in the $\kappa$-(ET)$_2$X. The occurrence of a
possible pseudogap phase in the $\kappa$-Br and $\kappa$-(NCS)$_2$
that has been ascribed to fluctuating SC \cite{Kawamoto95a,Kataev92a}
is also reminiscent of the preformed pair-based explanations of the
pseudogap behavior in the cuprates
\cite{Wang06a,Li10b,Chatterjee11a,Mishra14a}.  These similarities have
led to the effective $\frac{1}{2}$-filled band theories of SC in the
CTS that we discussed in Section \ref{half-theory}.  Interestingly,
the experimental literature on the cuprates itself has meanwhile seen
some very dramatic turns, with the discovery of an ubiquitous CO phase
{\it in between the AFM and superconducting phase} in the hole-doped
cuprates
\cite{Chang12a,Ghiringelli12a,Wu11a,Wu13a,Wu15a,Blanco-Canosa13a,Blanco-Canosa14a,Hucker14a,Comin14a,SilvaNeto14a,SilvaNeto15a,Comin15a,Hanaguri04a,Shen05a}.
Discovery of the CO phase in the electron-doped cuprates
\cite{SilvaNeto15a,SilvaNeto16a} further indicates that the idea of
the superconducting state evolving directly from the AFM state may be
simplistic.  Very interestingly, several groups have found that the CO
in cuprates is {\it commensurate}, and has a periodicity of four
Cu-O-Cu lattice spacings {\it independent of doping}
\cite{Hanaguri04a,Shen05a,Cai16a,Mesaros16a}, which, as we have
indicated throughout this review, is the CO periodicity in the
quasi-1D and quasi-2D CTS at the lowest temperatures. It has been
suggested that the CO and SC in the cuprates are isoenergetic, and
perhaps there exists even a CO-SC duality in the cuprates
\cite{Wu13a}.  Given the preponderance of CTS that show CO $\to$ SC as
well as AFM $\to$ SC transitions it is not inconceivable that the same
mechanism of SC, or at least some variant of the same mechanism is
also shared by the CTS.  One of us has recently proposed a Cu$^{2+}
\to$ Cu$^{1+}$ valence transition in the layered cuprates at the
pseudogap phase transition, following which they the Cu-band becomes
electronically inactive, and the CO and SC both emerge from the
$\frac{1}{4}$-filled O-band \cite{Mazumdar18a}. This is discussed
further in \ref{appendix-cuprates}.

\section{Conclusions}
\label{conclusions}

In this review we have focused on the large family of superconducting
organic CTS. These unconventional superconductors feature a large
diversity of crystal structures including quasi-1D as well as quasi-2D
lattices, and both hole and electron charge carriers. We have argued
that the most important unifying element of the organic CTS that
exhibit SC is their unique carrier density, $\rho=\frac{1}{2}$ per
molecule (which includes both electron density of 1.5 per molecule in
the hole carriers and 0.5 per molecule in the electron
carriers). Deviations from this carrier density are both rare and tiny
in real materials.  From the perspective of the search for a theory of
SC in strongly correlated-electron systems, CTS are attractive because
of their relative simplicity: because no doping is required to achieve
SC, a superconducting phase should be expected to emerge simply from
changing the parameters of the same model Hamiltonian that describes
the proximate insulating phases.  The complex lattice structures and
in particular presence of lattice frustration however make these
challenging systems to study.  The review therefore focused first on
understanding the insulating phases found in those CTS that exhibit
SC.

Examination of the normal state of the entire family of CTS, including
the ``old'' TCNQ-based materials, as well as the $\rho=\frac{1}{2}$ 2D
cationic and anionic CTS, indicates that there does exist a single
theoretical model that explains the global behavior of the entire
family. This is the EHM with significant Hubbard $U$ and moderate
Hubbard $V$. In addition to strong electronic correlations, e-p
coupling is essential to understand the non-superconducting broken
symmetry states. Coupling of the charge carriers with both
intramolecular and intermolecular vibrations are relevant.  At
$\rho=\frac{1}{2}$ these two e-p couplings act co-operatively, each
enhancing the effect of the other. Equally importantly, although the
CO broken symmetry is frequently taken to be a ``classical'' Coulomb
effect driven by the Hubbard $V$, specifically at $\rho=\frac{1}{2}$
there occurs also a CO {\it different from the classical WC} and that
is driven by a {\it quantum effect}, the tendency of electrons to form
spin-singlet pairs. We have presented numerical evidence for this
Paired Electron Crystal (PEC), in 1D as well as in frustrated 2D
lattices.  The PEC competes with both AFM and WC. In 2D CTS, the
manifestation of the PEC is the occurrence of charge periodicity
$\cdots1100\cdots$ in more than one direction, which in many cases
leads to the formation of (dimerized) insulating stripes with spin
gaps. We have pointed out (Section \ref{pec2dcts}) that the pattern of
the CO in all CTS that exhibit CO-to-SC transition corresponds to the
PEC. It is unlikely that this is a coincidence.

The identification of the CO in superconducting CTS as the PEC implies
that there exists a single unified theory of SC for the entire CTS
family, in which SC emerges upon destabilization (``melting'') of the
PEC, the spin-singlet pairs of which behave as hard core
bosons. Existing theories of SC in the CTS have overwhelmingly
(actually almost entirely) focused on the $\kappa$-(ET)$_2$X, in which
the superconducting state has been proclaimed to emerge from the AFM
state. This in spite of the fact that few of the $\kappa$-(ET)$_2$X
are actually antiferromagnetic, at least two
($\kappa$-(ET)$_2$Hg(SCN)$_2$Cl and $\kappa$-(ET)$_2$B(CN)$_4$) are
likely PECs, and the critical temperature T$_c$ in one
($\kappa$-(ET)$_2$CF$_3$SO$_3$) is larger than the N\'eel temperature
T$_N$. Precise numerical calculations show that SC is absent in
theoretical models of $\kappa$-(ET)$_2$X that treat them within
effective $\rho=1$ models. We have presented limited, but numerically
precise calculations that show that within the standard Hubbard
Hamiltonian, in both the triangular lattice and the $\kappa$-lattice,
superconducting pair-pair correlations are enhanced by the Hubbard $U$
only at $\rho \simeq \frac{1}{2}$ and are suppressed at all other
$\rho$, which supports our basic premise.

We reemphasize that our fundamental premise, viz., that
correlated-electron SC emerges from a density wave of Cooper pairs has
existed in the cuprate literature for some time now
\cite{Anderson04b,Franz04a,Tesanovic04a,Chen04a,Vojta08a}, and has
also been suggested from very recent experimental observations
\cite{Hamidian16a,Cai16a,Mesaros16a}. The list of investigators who
have expressed the opinion that SC emerges from the melting of a VBS
(which in the context of cuprates is different from a PEC) is equally
long or even longer. The reason for this is that this hypothesis
resolves a key controversy concerning the pseudogap phase in the
cuprates, viz., whether the pseudogap is due to pre-existing pairs or
competing broken symmetry; the density wave of Cooper pairs satisfies
both requirements simultaneously!  Our work demonstrates explicitly
that the PEC emerges naturally in 1D or frustrated 2D from the EHM at
$\rho=\frac{1}{2}$.  Beyond the conceptual overlap between the RVB
theory and our VB theory of SC, there are also similarities between
our picture and the original bipolaron theories of metal-insulator
transitions and SC
\cite{Chakraverty78a,Chakraverty79a,Chakraverty80b}. The latter had
hypothesized the formation of nearest neighbor spin-singlets, driven
by e-p interactions that overscreened the nearest neighbor e-e
repulsion. Our proposed mechanism of SC thus has overlaps with two
very disparate theories, even as the actual mechanism behind the
formation of the PEC is quite different from what had been proposed
before. The driving force behind PEC formation in CTS and other
$\rho=\frac{1}{2}$ systems is a quantum effect - the tendency to form
spin-singlets - that is strongly enhanced at this carrier
concentration due to the commensurate nature of the PEC. Similarly,
the kinetic energy gained from the motion of the effective hard core
bosons is also largest when their effective density is $\frac{1}{2}$.
It is interesting to note in this context that the original
experimental materials that were considered to exhibit prototype
bipolaron behavior were uniformly $\rho=\frac{1}{2}$ (see
~\ref{othersc}), even though this feature was not emphasized by the
authors.

Our 2D numerical calculations of pairing correlations show an
enhancement by e-e interactions selectively at $\rho\approx
\frac{1}{2}$ for multiple different lattice geometries and lattice
sizes. To the best of our knowledge, there exists no other unbiased
calculation that has shown correlation-induced enhancement of
superconducting pairing correlations. ~\ref{othersc} further indicates
that the calculated enhancement is not a coincidence.  Unfortunately,
however, currently we are unable to present evidence for LRO in the
calculated superconducting correlation functions, and it is even
conceivable that true LRO is absent in our calculations. At the same
time, we cannot rule LRO out completely within the present results,
because of the approximation introduced by the free-electron trial
wave function we used with the CPMC method in the 10$\times$10 lattice
results of Fig.~\ref{pp-fig1}; in general we found that CPMC with the
free-electron trial wavefunction slightly underestimated $\bar{P}$
\cite{Gomes16a}.  Recent improvements in the CPMC method may help to
improve the accuracy of pairing calculations \cite{Shi14a}.  An
alternate possibility is that the theory of the ground state as it
exists now is incomplete, which itself can have two
implications. First, even as the unique enhancement of SC pair
correlations at $\rho\approx \frac{1}{2}$ is meaningful, true SC may
require additional interactions (e.g., e-p coupling) ignored in the
purely electronic Hamiltonian. Because e-p coupling is required to
realize the bond distorted PEC state, some role of e-p coupling in a
PEC-to-SC transition might be expected. Second, even as the current
calculations indicate the likelihood of pair formation, the question
LRO has to be settled by calculations of a correlation function that
is slightly different, because of the unconventional nature of the
superconducting state.  Note that the pair motion in the PEL has to be
{\it correlated} (to minimize the repulsion between pairs). Thus the
long range component of the order parameter for the correlated SC may
very well be oscillating \cite{Berg10a}.  Elsewhere we have attempted
to simulate the PEC-to-SC transition within an effective $U<0, V>0$
model where the singlet bonds of $\rho\approx \frac{1}{2}$ are
replaced with double occupancies and the pairs of vacancies with
single vacant sites \cite{Mazumdar08a}. Transition from a WC of double
occupancies to a $s$-wave superconductor occurs as the frustrating
hopping integral is slowly increased. Analysis of the exact
wavefunctions shows however that the superconducting state at $V>0$ is
substantially different from the well understood case of $V=0$; only a
subset of the many-electron configurations that describe the
superconducting state at $V=0$ dominate the $V>0$ wavefunction. This
is because for $V>0$ pair motion has to occurs within a background
PEC.  This last result gives partial support to the conjecture that
the $\rho=\frac{1}{2}$ PEL might be a true superconducting state, but
more elaborate pairing correlations will be necessary to prove
this. These and related topics are being investigated currently.

\addcontentsline{toc}{section}{\bf Acknowledgments}
\noindent{\bf Acknowledgments}
\vskip 1pc

Much of the early research on BCDW and BCSDW done in collaboration
with D. K. Campbell. The authors also acknowledge collaborations with
S.  Dayal, W. Wasanthi De Silva, T. Dutta, N. Gomes, R. P. Hardikar,
H. Li, S.  Ramasesha, J.-P. Song, K.-C. Ung (deceased), and
A. B. Ward. Over the many years that the research on organic
charge-transfer solids was conducted the authors have benefited from
numerous fruitful discussions with M. Abdel-Jawad, S.  E. Brown,
E. M. Conwell (deceased), N. Drichko, A. J. Epstein, M. Dressel,
H. Fukuyama, A. Girlando, K. Kanoda, H. Mori, J. P. Pouget, R. Kato,
S.  Tomic, and T. Yamamoto. Parts of this work received financial
support from U.S. National Science Foundation and Department of
Energy.

\appendix
\newpage
%RTC1114 note that Elsevier's table of contents code is broken for appendices
% to fix: include an alternate title for the TOC with a horizontal space 
\section[\hspace{0.6in}Molecule abbreviations]{Molecule abbreviations}
\label{molecules}

\begin{center}
  \begin{tabular}{ll}
    abbreviation & molecule \\
    \hline
    BCPTTF & benzocyclopentyltetrathiafulvalene \\
    BDA-TTP & 2,5-Bis(1,3-dithian-2-ylidene)-1,3,4,6-tetrathiapentalene \\
    BEDO-TTF  & bis(ethylenedioxy)tetrathiafulvalene \\
    BEDT-TSF & bis(ethylenedithio)tetraselenafulvalene \\
    BEDT-TTF & bis(ethylenedithio)tetrathiafulvalene \\
    DMEDO-TTF & dimethyl(ethylenedioxy)tetrathiafulvalene \\
    DMET & dimethyl(ethylenedithio)diselenadithiafulvalene \\
    M(dmit)$_2$ & bis(4,5-dimercapto-1,3-dithiole-2-thione)-M \\
    DMTTF & 3,4-dimethyl-tetrathiafulvalene \\
    DODHT & (1,4-dioxan-2,3-diyldithio)dihydrotetrathiafulvalene \\
    EDO-TTF & ethylenedioxytetrathiafulvalene \\
    EDT-TTF & ethylenedithiotetrathiafulvalene \\
    HMTSF & hexamethylene-tetraselenafulvalene \\
    HMTTF & hexamethylene-tetrathiafulvalene\\
    MEM & N-methyl-N-ethyl-morpholinium \\
    {\it meso}-DMBEDT-TTF & 2-(5,6-dihydro-1,3-dithiolo[4,5-b][1,4]dithiin-2-ylidene)\\
    & -5,6-dihydro-5,6-dimethyl-1,3-dithiolo[4,5-b][1,4]dithiin \\
    NMP & N-methylphenazinum \\
    TCNQ & tetracyanoquinodimethane \\
    TMTSF & tetramethyltetraselenafulvalene \\
    TMTTF &  tetramethyltetrathiafulvalene \\
    TSF & tetraselenafulvalene \\
    TTF & tetrathiafulvalene 
\end{tabular}
\end{center}
\newpage

\section[\hspace{0.6in}Other $\rho=\frac{1}{2}$ superconductors]{Other $\rho=\frac{1}{2}$ superconductors}
\label{othersc}
 The principal thesis of the present work is that among strongly
 correlated materials the tendency to form spin-singlets, leading
 either to the PEC or the PEL, is strongest at $\rho=\frac{1}{2}$ (or
 1.5) in both 1D and 2D. It follows that there should be other
 $\rho=\frac{1}{2}$ systems or families of materials where also
 correlated-electron SC is observed. We describe in this Appendix
 precisely such systems.  Each material or family of materials has
 been of strong interest to the condensed matter physics community
 individually, but to date the common link between them, the same
 carrier density, has not been noted. In every case we discuss below,
 the following are true, (i) in structurally related materials with
 variable $\rho$, SC is limited to very narrow carrier concentration
 range about $\rho=\frac{1}{2}$, and (ii) in nearly all cases SC is
 proximate to a correlated CO state, and in those cases where the
 nature of the CO is understood there are reasons to believe that the
 CO is a PEC (either the periodicity is $\cdots0110\cdots$, or the CO
 coexists with a spin gap). The absence of any other carrier density
 where there exist so many systems with these common features gives
 credence to our VB theory of SC.

\subsection[\hspace{0.5in}Superconducting layered cobalt oxide hydrate]{Superconducting layered cobalt oxide hydrate}
\label{cobaltates}

\begin{figure}
  \begin{center}
    \raisebox{0.2in}{
      \begin{overpic}[width=2.0in]{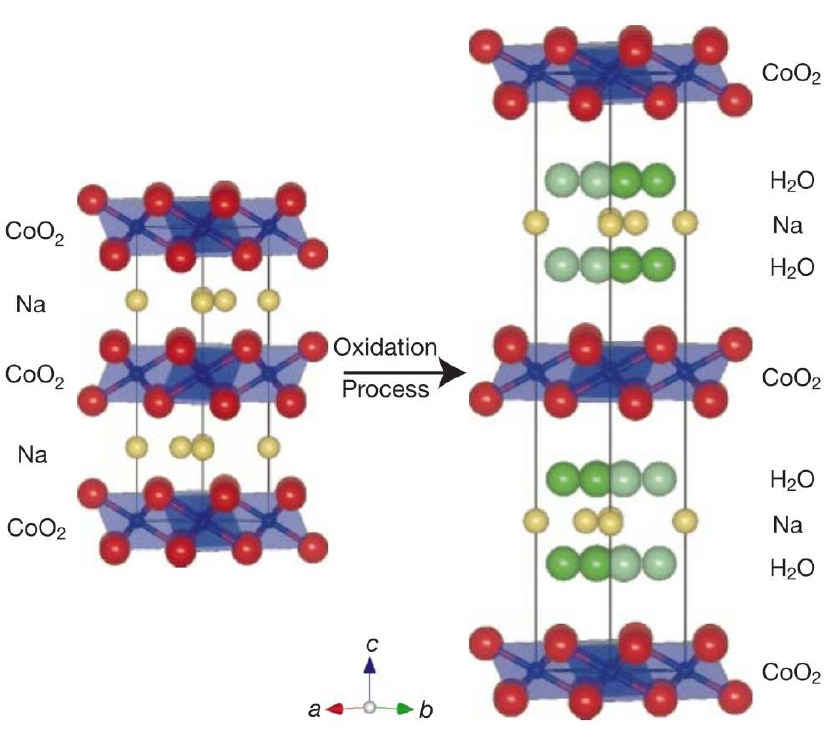}
        \put(5,80) {\small(a)}
      \end{overpic}
    }
    \raisebox{0.1in}{
    \begin{overpic}[width=2.0in]{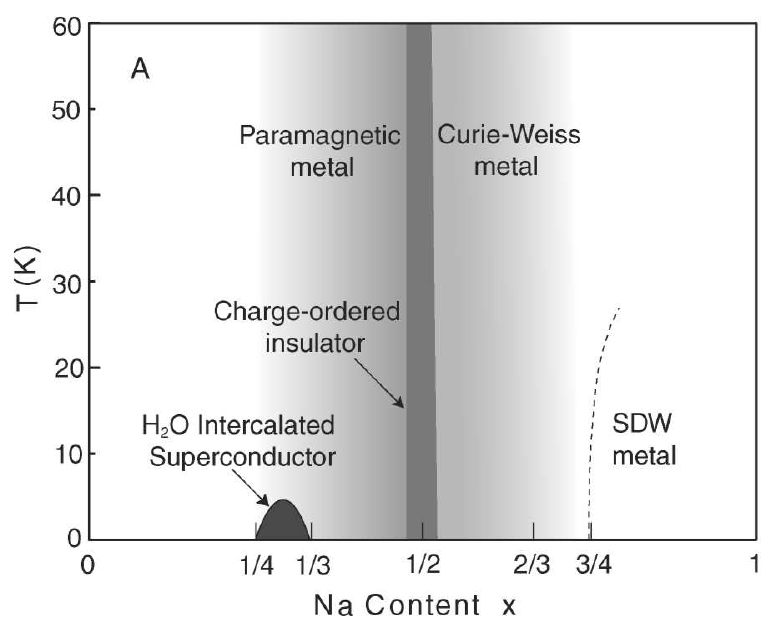}
      \put(-2,75) {\small(b)}
    \end{overpic}
    }
    \hspace{0.1in}
    \begin{overpic}[width=2.0in]{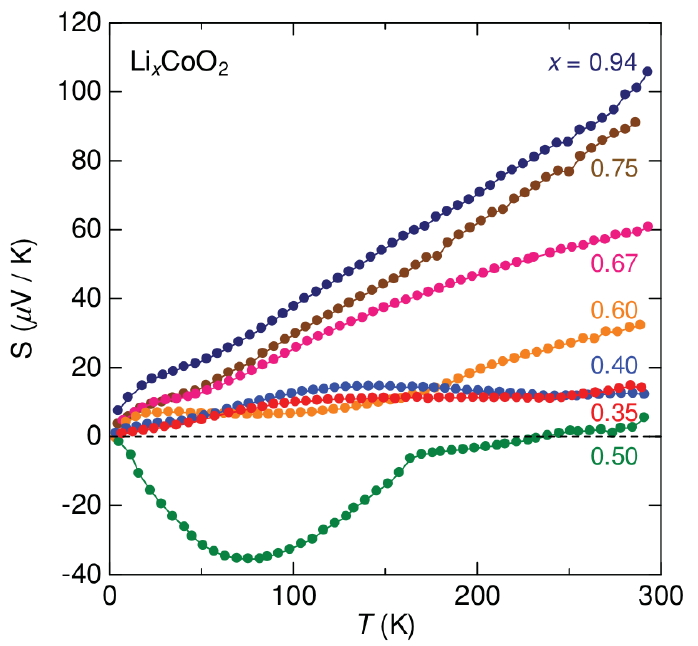}
      \put(0,85) {\small(c)}
    \end{overpic}
  \end{center}    
  \caption{(a) Structure of Na$_x$CoO$_2$ and the hydrated
    superconductor Reprinted with permission from
    Ref.~\cite{Takada03a}, $\copyright$ 2003 MacMillan Publishers Ltd.
    (b) phase diagram.  Reprinted with permission from
    Ref.~\cite{Foo04a}, $\copyright$ 2004 The American Physical
    Society.  (c) Temperature dependence of the Seebeck coefficient of
    Li$_x$CoO$_2$.  Reprinted with permission from
    Ref.~\cite{Motohashi11a}, $\copyright$ 2011 The American Physical
    Society.}
  \label{cobaltate-seebeck}
\end{figure}  

Layered cobaltates M$_x$CoO$_2$ (M = Li, Na, K) have attracted strong
interest as strongly correlated-electron materials in which the
carrier concentration can be varied over a wide range by varying the
Na-concentration $x$ \cite{Foo04a,Sakurai15a,Li11a}. Na$_x$CoO$_2$ is
of particular interest as in the hydrated state it is an
unconventional superconductor with T$_c=4.7$ K \cite{Takada03a}.  In
the anhydrous materials the Co-ions in the CoO$_2$ layers separated by
the Na$^+$-ions occupy an isotropic triangular lattice as shown in
Fig.~\ref{cobaltate-seebeck}(a) and have charge ranging from 3+ (at
$x=1$) to 4+ (at $x=0$). The corresponding electron configurations are
$t_{2g}^6$ with spin $S=0$, and $t_{2g}^5$ with $S=\frac{1}{2}$,
respectively. Trigonal distortion splits the $t_{2g}$ orbitals into
degenerate low-lying $e_g^\prime$ levels and a higher $a_{1g}$ level.
Charge carriers are holes on the Co$^{4+}$ sites \cite{Hasan04a}, with
hole density $\rho=1-x$. ARPES studies indicate that the $e_g^\prime$
levels are completely filled and therefore should be electronically
inactive \cite{Hasan04a,Shimojima06a}, although this is somewhat
controversial \cite{Laverock07a}. Band calculations find a larger
Fermi surface due to $a_{1g}$ levels, and a smaller Fermi surface due
to the $e_g^\prime$ levels \cite{Singh00a}. The physical behavior of
anhydrous Li$_x$CoO$_2$ and Na$_x$CoO$_2$ are strongly $x$-dependent
(see Fig.~\ref{cobaltate-seebeck}(b)) \cite{Foo04a}.  The density
$\rho=\frac{1}{2}$ has unique properties here, as seen for example in
its completely different thermoelectric properties as a function of
temperature (see Fig.~\ref{cobaltate-seebeck}(c)) \cite{Motohashi11a}.
The global feature of the $x$-dependence, with paramagnetic weakly
correlated behavior at small $x$ (large $\rho$) and Curie-Weiss
strongly correlated behavior at large $x$ (small $\rho$) was
considered mysterious for a while. Since $\rho=1$ is a Mott-Hubbard
semiconductor densities close to but not equal to $\rho=1$ would have
been expected to be strongly correlated. Similarly, since $\rho=0$ is
a band semiconductor, small $\rho$ was expected to be be weakly
correlated. The observed behavior is exactly the opposite, see Fig.~\ref{cobaltate-seebeck}(b)).
The $\rho$-dependence was successfully explained by Li
{\it et al.}, who showed that the observed behavior was exactly what
would be expected within both one-band and two-band extended Hubbard
models with realistic nearest neighbor Coulomb repulsion $V$ over and
above the onsite Hubbard interaction $U$ \cite{Li11a}. Identical
$\rho$-dependent physical behavior are seen in the family of
conducting organic charge-transfer solids \cite{Mazumdar83a}.  Exactly
as in the CTS (see discussions in Section \ref{rho-dependence}) for
fixed Hubbard $U$, Hubbard $V$ enhances double occupancies at large
$\rho$ away from 1, and suppresses double occupancies at small $\rho$
in the triangular Co-lattice (see Figs.~3 and 4 in reference
\cite{Li11a}).  Importantly, the fraction of holes occupying the
$e_g^\prime$ levels is found to be $\rho$-dependent and is smaller for
smaller $\rho$ \cite{Lee05b,Li11a}. The $e_g^\prime$-occupancy is less
than 10\% at $\rho \sim \frac{1}{2}$.

SC in the hydrated Na$_x$CoO$_2$ was found by Takada {\it et al.}
\cite{Takada03a}. The chemical composition was initially thought to be
Na$_x$CoO$_2 \cdot$ $y$H$_2$O ($x \sim 0.35$, $y \sim 1.3$), with the
H$_2$O entering in between the CoO$_2$ layers. Assuming that $x$ alone
determines the Co-ion valency as in the anhydrous material this gives
$\rho=0.65$. RVB theories of SC were proposed initially
\cite{Baskaran03c,Kumar03a,Ogata03a,Wang04c} even as this value of
$\rho$ is far from $\rho=1$ where the RVB state is supposed to occur
\cite{Anderson73a}. It is now known that a significant proportion of
the water in the hydrated material enters between the CoO$_2$ layers
as H$_3$O$^+$ and the true superconducting composition
\cite{Sakurai15a} is Na$_x$(H$_3$O)$_z$CoO$_2 \cdot$$y$H$_2$O, which
would make Co-ion valency of 3.65+ an absolute lower limit ({i.e.},
$\rho> 0.35$).  Significant effort has gone into finding the true
Co-ion valency in the bilayered hydrate that is
superconducting. Several chemical studies
\cite{Sakurai06a,Banobre-Lopez09a,Sakurai15a} have found the valency
to be very close to 3.5+. One ARPES study puts the Co valence at 3.56
$\pm$ 0.05, making $\rho$ extremely close to 0.5.  Spin fluctuation
theories of SC based on RPA or FLEX calculations predict spin-triplet
$p$- or $f$-wave SC \cite{Kuroki04a,Ogata07a}. There is however no
evidence experimentally for triplet pairing in these materials
\cite{Sato10a,Sakurai15a}. These theories also require the presence of
$e_g^\prime$ orbitals at the Fermi surface \cite{Ogata07a}.  The
observed limitation of SC to Co-ion valency $\sim$ 3.5+ and the
evidence for strong e-e interaction \cite{Li11a} once again suggest
that the mechanism of SC here too is related to our proposed
mechanism, viz., correlated Co$^{4+}$--Co$^{4+}$ spin-singlet motion
at $\rho=\frac{1}{2}$.

\subsection[\hspace{0.5in}Spinel superconductors]{Spinel superconductors}
\label{spinelsc}

\begin{figure}
  \begin{center}
    \resizebox{3.5in}{!}{\includegraphics{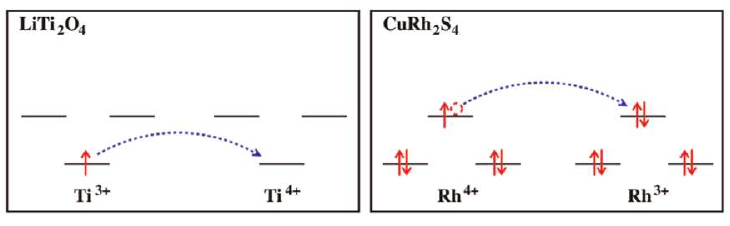}}
  \end{center}
  \caption{Effective $\frac{1}{4}$ ($\frac{3}{4}$)-filled bands of the
     spinel superconductors. The $t_{2g}$ orbitals are split by Jahn-Teller distortion in (a) 
and by Jahn-Teller distortion or SOC in (b).}
  \label{spinelorbitals}
\end{figure}

Spinels are inorganic ternary compounds AB$_2$X$_4$, with the
B-cations as the active sites. The B sublattice in the spinels forms
corner-sharing tetrahedra, giving rise to a geometrically frustrated
pyrochlore lattice.  Out of several hundred spinel compounds with
transition metals as the B-cations only three undoped compounds are
confirmed superconductors, LiTi$_2$O$_4$ with T$_c \simeq 12$ K
\cite{Johnston73a}, CuRh$_2$S$_4$ (T$_c = 4.8$ K \cite{Hagino95a} and
up to 6.4 K under pressure \cite{Ito03a}), and CuRh$_2$Se$_4$ (T$_c =
3.5$ K \cite{Hagino95a}).  While early reports on powder samples
indicated SC at $\sim$ 4 K in CuV$_2$S$_4$, later measurements on
single crystals found no SC \cite{Seki92a,Hagino94a}.  The Cu-ion in
the latter compounds are known to be monovalent \cite{Hart00a}.  Band
calculations have shown that the transition metal's partially occupied
t$_{2g}$ $d$-bands are well separated from the empty e$_g$ bands by
$\sim$ 3 eV, and also from the $p$-bands due to O and S by about the
same amount \cite{Satpathy87a,Massidda88a,Hart00a}.  The extreme
selectivity of SC cannot be a coincidence, especially considering that
(i) while LiTi$_2$O$_4$ superconducts at 12 K, structurally related
LiV$_2$O$_4$ does not show SC down to 10 mK \cite{Kondo97a}, (ii) SC
in Li$_{1+x}$Ti$_{2-x}$O$_4$ disappears for $x$ as small as $\sim 0.1$
\cite{McCallum76a}, and (ii) Ti$^{3.5+}$ in LiTi$_2$O$_4$ has one
$d$-electron per two Ti's, and Rh$^{3.5+}$ in CuRh$_2$S$_4$ and
CuRh$_2$Se$_4$ is in the low-spin state with one hole per two t$_{2g}$
orbitals (see Fig.~\ref{spinelorbitals}).  We give detailed
discussions below why we believe that these materials should be
considered correlated-electron $\frac{1}{4}$ ($\frac{3}{4}$)-filled
band superconductors just like the CTS.

As with the CTS, a hint to the mechanism of SC in the spinels is
obtained by examining the metal-insulator transitions that occur in
structurally related compounds. In Ti$_4$O$_7$, which also has
Ti-valency of 3.5+, and which is structurally related to
LiTi$_2$O$_4$, there occurs charge disproportionation into Ti$^{3+}$
(d$^1$) and Ti$^{4+}$ (d$^0$) and the formation of Ti$^{3+}$-Ti$^{3+}$
{\it spin singlets} \cite{Lakkis76a,Chakraverty80a}. The most likely
explanation of this is orbitally-induced band Jahn-Teller distortion
\cite{Khomskii05a,Khomskii05b,Radaelli05a,Croft07a} that lifts the
degeneracy of the t$_{2g}$ $d$-band, giving a $\frac{1}{4}$-filled
nondegenerate $d_{xy}$-band (see Fig.~\ref{spinelorbitals})
\cite{Mazumdar12a,Mazumdar14a}. Importantly, the band Jahn-Teller
distortion at this filling requires strong e-e interaction, which is
the driver of singlet formation in $\frac{1}{4}$-filled band
\cite{Clay10a}.  As shown in Fig.~\ref{spinelorbitals} the occupancies
of the t$_{2g}$ orbitals on individual Rh$^{3.5+}$ ions in
CuRh$_2$S$_4$ and CuRh$_2$Se$_4$ fluctuate between t$_{2g}^6$ with
closed shell and t$_{2g}^5$ with one hole.  Metal-insulator
transitions in isoelectronic CuIr$_2$S$_4$
\cite{Radaelli02a,Khomskii05a} and LiRh$_2$O$_4$
\cite{Okamoto08a,Knox13a} are accompanied by cubic-to-tetragonal
structural transition {\it as well as PEC formation,} as evidenced by
$M^{4+}-M^{4+}-M^{3+}-M^{3+}$ ($M=$ Rh, Ir) charge and bond
tetramerization and $M^{4+}-M^{4+}$ spin-bonded dimers along specific
directions. The charge ordering in CuIr$_2$S$_4$ was originally also
ascribed to orbitally-induced lifting of t$_{2g}$ degeneracy that left
filled degenerate d$_{xz}$ and d$_{yz}$ bands lowered in energy by
band Jahn-Teller distortion, and a nondegenerate $\rho=\frac{1}{2}$ d$_{xy}$
band, separated by an energy gap
\cite{Khomskii05a,Khomskii05b,Radaelli05a,Croft07a,Okamoto08a}. More
recent theoretical work has ascribed the degeneracy lifting in
Ir$^{4+}$ to spin-orbit coupling (SOC), with the splitting between
$J=\frac{3}{2}$ and $J=\frac{1}{2}$ states
\cite{Rau16a,Kim08b,Witczak-Krempa14a}.  For our purpose, {\it it is
  irrelevant whether the lifting of the degeneracy is due to band
  Jahn-Teller distortion or SOC}. More relevant is the effective
$\rho=\frac{1}{2}$ hole occupancy of the active orbital, and the implied
proximity of SC to PEC.

We believe that the extreme selectivity of SC in the spinels, taken
together with the PEC formation in isostructural isoelectronic
compounds CuIr$_2$S$_4$ and LiRh$_2$O$_4$, are strong indicators that
the mechanism of SC in these materials is the same in these materials
as in the CTS.  Particularly in the context of LiTi$_2$O$_4$, there is
ample evidence for strong e-e interactions among the 3$d$-electrons
\cite{Wu94a,Mazumdar89a,Fazileh06a}, and that e-p interactions may not
be the driver of SC (see Reference \cite{Moshopoulou99a} for a
review).  A very recent experiment in LiTi$_2$O$_4$ has found a
transition from isotropic negative magnetoresistance to anisotropic
positive magnetoresistance at 50 K \cite{Jin15a}, which is also the
temperature where a Raman study has shown a strong anomaly in the
T$_{2g}$ e-p coupling \cite{Chen17a}.  The authors of both studies
have suggested an orbital ordering transition at this temperature,
which would give strong though indirect support to our proposed
orbital ordering in Fig.~\ref{spinelorbitals}. We propose similar
magnetoresistance and Raman measurements in CuRh$_2$S$_4$ and
CuRh$_2$Se$_4$.  Within our theory SC is due to the correlated motion
of n.n. singlets Ti$^{3+}$--Ti$^{3+}$ and Rh$^{4+}$--Rh$^{4+}$.

\subsection[\hspace{0.5in}Superconducting vanadium bronzes]{Superconducting vanadium bronzes}
\label{vanadates}

\begin{figure}
  \begin{center}
    \raisebox{0.5in}{
      \begin{overpic}[width=2.25in]{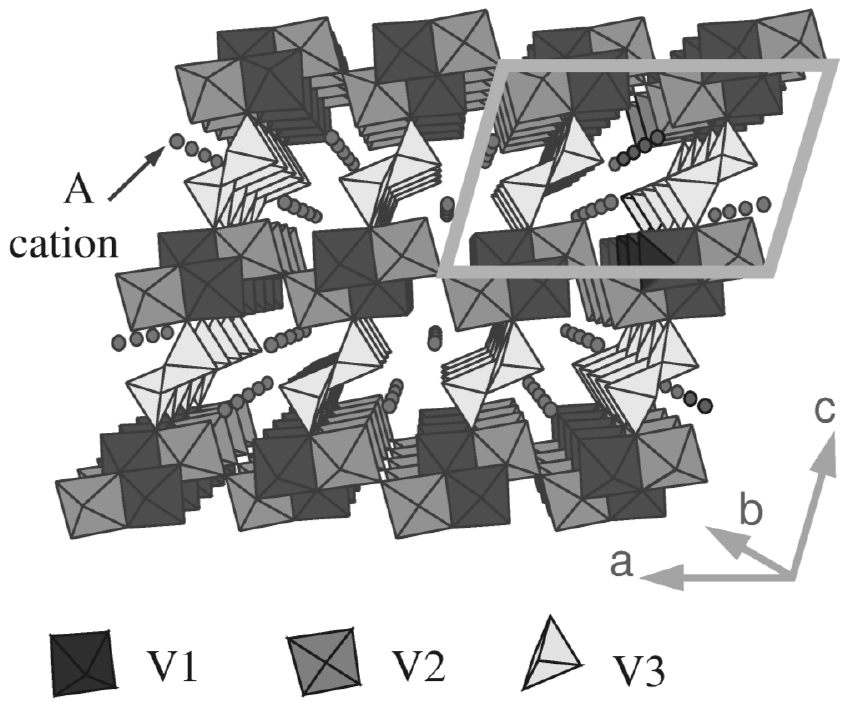}
        \put(5,75) {\small(a)}
      \end{overpic}
    }
    \hspace{0.1in}
    \begin{overpic}[width=1.5in]{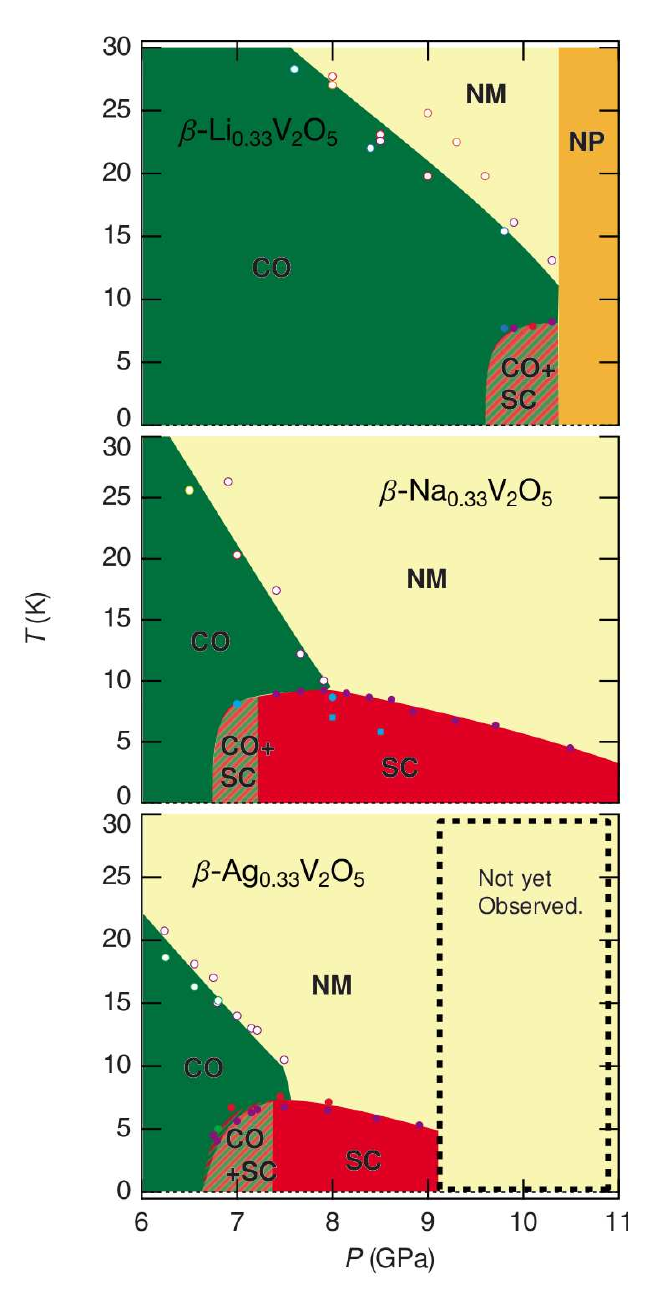}
      \put(0,95) {\small(b)}
      \put(0,65) {\small(c)}
      \put(0,35) {\small(d)}
    \end{overpic}
  \end{center}
  \caption{(a) Crystal structure of $\beta$-A$_{0.33}$V$_2$O$_5$.
    Reprinted with permission from Ref.~\cite{Yamauchi02a},
    $\copyright$ 2002 The American Physical Society.  (b)-(d)
    Temperature-pressure phase diagrams for (b) A=Li (c) A=Na and (d)
    A=Ag. Here NM is a normal metal phase and NP is a high-pressure
    non-superconducting phase.  Reprinted with permission from
    Ref.~\cite{Yamauchi08a}, $\copyright$ 2008 The American Physical
    Society.}
  \label{vanadate-fig}
\end{figure}  
Yet another family of materials that are $\frac{1}{4}$-filled and
exhibits pressure-induced CO-to-SC transition are the series of
vanadium bronzes $\beta$-A$_{0.33}$V$_2$O$_5$, A = Li, Na, Ag
\cite{Yamauchi02a,Yamauchi08a}.  Structurally, the materials consist
of the V$_2$O$_5$ framework with three distinct V sites (referred to
as V1, V2 and V3, respectively as shown in
Fig.~\ref{vanadate-fig}(a)), with the V-ions forming quasi-1D chains
along the b-axis of the crystal. The electronic configuration of the
V$^{5+}$-ions in pure V$_2$O$_5$ is 3d$^0$.  The metal ions occupy
straight ``tunnels'' along the b-axis. Given that there are three
different kinds of V-ions the compositions
$\beta$-A$_{0.33}$V$_2$O$_5$ are stoichiometric, even as other
compounds with compositions that are slightly different are known.

The most interesting observations with this compounds are as
follows. First, all vanadium bronzes show quasi-1D conduction at
ambient pressure. It has, however, been proposed that there occurs a
dimensional crossover to quasi-2D behavior under pressure
\cite{Yamauchi08a}.  Second, angle-resolved photoemission study of
$\beta$-Na$_{0.33}$V$_2$O$_5$ indicates that the perfectly
stoichiometric compound consists of $\frac{1}{4}$-filled band 1D
chains, with exactly equal populations of V$^{5+}$ (3d$^0$) and
V$^{4+}$ (3d$^1$) \cite{Okazaki04a}. Which of the three chains becomes
$\frac{1}{4}$-filled on ``doping'' remains however unresolved. Third,
both the CO phase and the SC phases are extremely sensitive to
A-cation off stoichiometry, with smallest non-stoichiometry destroying
SC, though not necessarily CO. Finally, even as the CO-to-SC
transition is pressure-induced, T$_{\rm SC}$ decreases with pressure
(see Figs.~\ref{vanadate-fig}(b)-(d)), in
contrast to what is expected from a BCS-superconductor.  All of these
observations, in particular SC limited to commensurate
$\frac{1}{4}$-filling and being destabilized by pressure, are
reminiscent of the behavior of the CTS family and the other compounds
listed above.  It is relevant in this context to now point out that
vanadium bronzes were among the first compounds in which nearest
neighbor spin-paired bipolarons (V$^{4+}$-V$^{4+}$) were hypothesized
\cite{Chakraverty78a}. Similar nearest neighbor spin-pairings between
Ti$^{3}$-Ti$^{3+}$ had been proposed for Ti$_4$O$_7$, which is
structurally related to LiTi$_2$O$_4$ and also contains a
$\frac{1}{4}$-filled Ti-band \cite{Lakkis76a}. It was suggested that
overscreening of e-e repulsion by e-p interactions drove the formation
of the bipolarons \cite{Chakraverty80a,Chakraverty80b}, but as has
been shown in our work here, $\frac{1}{4}$-filling and moderate e-e
interactions together can drive this transition to the PEC, which the
earlier group of authors had referred to as the bipolaronic insulator.
We also note that Hague {\it et al} \cite{Hague07a} showed that 
{\it intersite} spin-singlets can have high mobility and small effective mass.

\subsection[\hspace{0.5in}Superconducting fullerides]{Superconducting fullerides}
\label{c60}

\begin{figure}
  \begin{center}
    \begin{overpic}[width=2.5in]{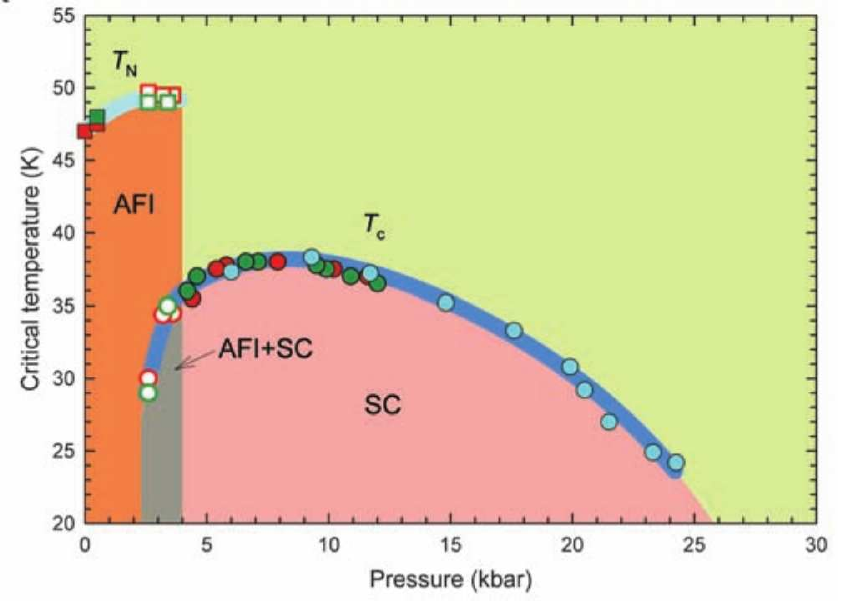}
      \put (-5,70) {\small(a)}
    \end{overpic}
    \hspace{0.1in}
    \raisebox{0.5in}{
      \resizebox{2.5in}{!}{\includegraphics{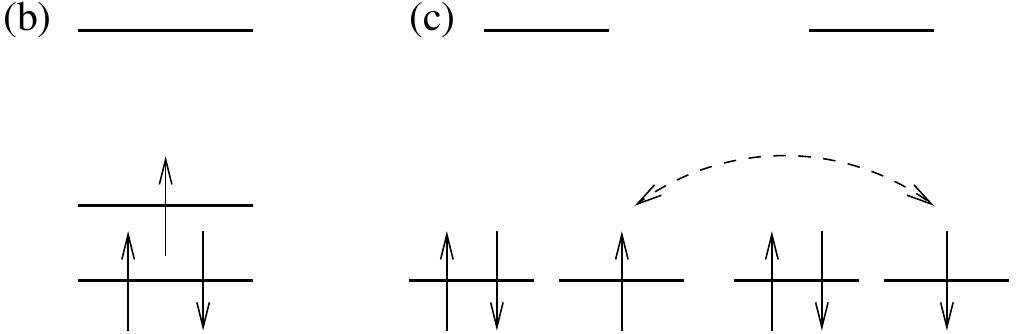}}
      }
  \end{center}
  \caption{(color online) (a) Phase diagram of A15 Cs$_3$C$_{60}$.
    Reprinted with permission from Ref.~\cite{Takabayashi09a},
    $\copyright$ 2009 AAAS.  In the Mott insulator state (AFI region
    in (a)), two electrons occupy the lowest orbital, and one the
    second-lowest orbital on each C$_{ 60}$ molecule.  (b) The orbital
    occupation in the same Mott insulator state, but with finite $U$,
    resulting in a reduced gap between the lowest two orbitals. (c)
    The inclusion of off-diagonal terms between MOs equalizes
    populations between the lowest two MOs, resulting in a
    $\frac{3}{4}$-filled band (see text).}
    \label{c60orbitals}
\end{figure}  
One peculiarity of superconducting fullerides is that SC is limited to
trivalent C$_{60}^{3-}$ \cite{Gunnarsson97a}.  Since there is nothing
unique to this valence within purely e-p coupled models this should
already be an indicator that e-e interactions may be relevant
here. This has been further confirmed by the observation of AFM in
Cs$_3$C$_{60}$ with T$_{\rm N}$ $\sim$ 46 K \cite{Takabayashi09a}, as
well as of pressure-induced AFM-to-SC transition with T$_{\rm C} = 38$
K \cite{Takabayashi09a,Ganin08a,Ganin10a}. Careful experimental
measurements have indicated that the AFM Cs$_3$C$_{60}$ has a single
unpaired electron per C$_{60}^{3-}$ \cite{Ganin10a,Klupp12a}. This has a unique
explanation, viz., the three-fold degeneracy of the antibonding
$t_{1u}$ MOs of C$_{60}$ is lifted by Jahn- Teller instability in the
trianion, with occupancies of 2, 1 and 0 electrons in the three
nondegenerate MOs, in increasing order of energies (See
Fig.~\ref{c60orbitals}(a)) \cite{Klupp12a}.  AFM is now due to
spin-coupling between the unpaired electrons on neighboring
ions. While it is true that given the low-spin nature of
Cs$_3$C$_{60}$ there is no alternative to the characterization of the
insulating AFM state as a magnetic Mott-Jahn-Teller insulator
\cite{Klupp12a}, the nature of the so-called Jahn-Teller-metal that is
reached under pressure remains not understood.  Existing theories of
SC
\cite{Chakravarty91a,Varma91a,Schluter92a,White92b,Gunnarsson97a,Capone09a}
explicitly or implicitly assume that the Jahn-Teller-metal has
regained the degeneracy of the undoped state, and the SC is due to
``on-ball pairing'', driven largely by the Jahn-Teller phonons, with
the Mott-Hubbard $U$ playing either a competing or co-operative
role. Such a quasi-BCS mechanism would seem to be in disagreement with
the observed decrease of T$_c$ with pressure \cite{Ganin08a,Ganin10a}.
Even more importantly, we believe that this picture fails to explain
why T$_c$ should be so sharply peaked at anionic charge of -3.  The
strong sensitivity of SC to commensurability is a characteristic
feature of all the materials discussed in this review, and it should
be clear by now that these cannot be merely shared coincidences.

We agree with the viewpoint that the orbital degeneracy of the
so-called Jahn-Teller-metal is different from the Mott-Jahn-Teller
insulator. We believe, however, that that the perfect nondegeneracy of
the former is replaced with partial degeneracy in the latter, exactly
as in the case of the spinels within a mechanism that is an analogue
of the band Jahn-Teller instability. Recall first that the Hubbard $U$
suppresses Jahn-Teller instability when the orbital occupancy is even,
but not when the occupancy is odd \cite{Dixit84a}. Thus, unlike in
Fig.~\ref{c60orbitals}(a), where the gaps between the antibonding MOs
of C$_{60}$ are nearly identical (as they would be in the strictly
$U=0$ case, where the energy gained upon distortion due to adding two
electrons is exactly twice that gained when a single electron is
added), for $U \neq 0$ the gap between the lowest doubly occupied
level and the singly occupied level is much smaller than that between
the latter and the unoccupied level (see Fig.~\ref{c60orbitals}(b)).
Now as one includes intermolecular hopping of electrons, there will
also be off-diagonal terms in the MO representation, hopping between
MOs with different symmetry. Given the small intrinsic gap between the
two lowest levels, this would equalize the populations of the two
lowest MOs which are now degenerate, giving a population of 1.5
electrons per MO, with the unoccupied MO at a higher level.  This
equalization of populations of the two lower levels is further
enhanced by e-e interactions \cite{Dutta14a}.  Comparing
Figs.~\ref{c60orbitals}(c) and \ref{spinelorbitals}, the similarity
between superconducting spinels and a lattice of C$_{60}^{3-}$ is
readily observed.  Note that within our proposed mechanism, pairing is
``inter-ball''. The mechanism that we have proposed for the AFM-to-SC
transition here is related to that proposed for $\kappa$-(ET)$_2$X: in
both cases the transition from the AFM insulator is driven by and
accompanied with orbital (re)ordering that makes the ``metallic''
state effectively $\frac{1}{4}$ ($\frac{3}{4}$)-filled.

\subsection[\hspace{0.5in}Li$_{0.9}$Mo$_6$O$_{17}$]{Li$_{0.9}$Mo$_6$O$_{17}$}
\label{LMO}

This is yet another low-dimensional superconductor where SC is
proximate to an insulating state, with the active electrons belonging
to the $d$-orbitals of a transition metal, and the bandfilling nearly
$\frac{1}{4}$.  The crystal structure is monoclinic
\cite{Onoda87a,daLuz11a}. The conduction electrons are located mostly
on two Mo-ions that occupy octahedral sites and form what has been
described in the literature as double-zigzag chains or two-leg zigzag
ladders along the b-axis, giving the system its strong one-dimensional
behavior. The peculiarities of the system include a low temperature
metal-insulator transition, as observed from resistivity studies at
T$_{\rm MI} \sim 25$ K
\cite{Greenblatt88a,Schlenker85a,Sato87a,Choi04b} that has remained
poorly understood, and a superconducting transition below 2 K
\cite{Schlenker85a}. There is no signature of structural anomaly at
T$_{\rm MI}$ from X-ray scattering and thermal expansion studies
\cite{dosSantos07a} or neutron scattering \cite{daLuz11a}. The high
temperature behavior (above T$_{\rm MI}$) can be understood within a
Luttinger liquid picture \cite{dosSantos08a}. The absence of the
structural anomaly at T$_{\rm MI}$ has led to suggestions that the
transition is ``purely electronic'' or a dimensional crossover.  The
upper critical field for magnetic fields parallel to the conducting
chains is five times larger than the Pauli paramagnetic para-magnetic
limit \cite{Mercure12a}, indicating that the SC is
unconventional. Triplet pairing has been suggested \cite{Lebed13a},
although this has not been confirmed experimentally yet.
Note that large upper critical fields are also characteristic of CTS, 
where it is now known that SC is not due to triplet pairing
\cite{Zuo00a,Shinagawa07a}.

The strong 1D behavior at high temperatures can be understood from the
crystal structure (see Fig.~2 in \cite{daLuz11a}). The unit cell
contains six distinct Mo ions, of which two (labeled Mo3 and Mo6 in
the literature) are on tetrahedral sites. Of the remaining four
Mo-ions in octahedral sites Mo1 and Mo4 form highly 1D structures that
have been described as double zigzag chains \cite{daLuz11a} or two-leg
zigzag ladders \cite{Merino12a}. Based on DFT calculations
\cite{Popovic06a}, Merino and McKenzie concluded that the chemical
formula of LiMo$_6$O$_{17}$ can be written as
Li$^{1+}$(Mo$^{4.5+}$)$_2$Mo$^{\prime 6+}_4$(O$^{2-}$)$_{17}$ where Mo
but not Mo$^{\prime}$ constitute the coupled zigzag chains. Note that
Mo-ion valence of 4.5+ implies electron configurations of 4d$^0$ and
4d$^1$. Band structure calculations indicate that only the d$_{xy}$
orbitals are the active bands, which are then exactly
$\frac{1}{4}$-filled in LiMo$_6$O$_{17}$ and slightly away from this
in Li$_{0.9}$Mo$_6$O$_{17}$.  We speculate that the transition at
T$_{\rm MI}$ is {\it simultaneously} purely electronic as well as a
dimensional crossover. As pointed out by Merino and McKenzie, the 3D
electron-electron interactions and hoppings between the zigzag chains
on different planes increase the frustration. Within our theory
increased frustration leads to charge disproportionation in
$\rho=\frac{1}{2}$, with spin-singlet formation between charge-rich
sites that are NN on {\it different} chain pairs or ladders. The
resultant electronic state that is no longer 1D can be either a PEC or
PEL. This viewpoint would be in agreement with the observation that
the superconducting transition in Li$_{0.9}$Mo$_6$O$_{17}$ is
insulator-superconductor \cite{dosSantos08a}. It is also tempting to
assign the giant Nernst effect in Li$_{0.9}$Mo$_6$O$_{17}$
\cite{Cohn12a} to preformed pairs \cite{Wang06a}.  Clearly further
theoretical and experimental research are warranted here; what is
however true is that Li$_{0.9}$Mo$_6$O$_{17}$ exhibits the ``generic''
behavior of $\rho=\frac{1}{2}$ superconductors.

\subsection[\hspace{0.5in}Intercalated and doped IrTe$_2$]{Intercalated 
and doped IrTe$_2$}
\label{IrTe2}

One other material of interest is IrTe$_2$, consisting of edge-sharing
IrTe$_6$ octahedra, with Ir layers sandwiched between Te layers
\cite{Ko15a}. The interlayer Te-Te distance is smaller than the
intralayer distance.  While IrTe$_2$ itself does not exhibit SC, it
undergoes a CDW-like transition at $\sim$ 260 K which remains poorly
understood. There is broad agreement \cite{Oh13a,Fang13a} that at high
temperatures the valence state of Ir is Ir$^{3+}$, which would make
its electron configuration closed shell t$_{2g}^6$.  Conversely, the
ionic state of Te is Te$^{1.5-}$, which suggests equal populations of
Te$^{2-}$ and Te$^{1-}$ and a hole concentration $\rho=\frac{1}{2}$ on the Te
sites.  The transition at 260 K is accompanied by strong diamagnetism
and structural anomaly \cite{Oh13a,Fang13a}, and one way to understand
the diamagnetism would be to assume spin-singlet formations between
the open shell Te$^{1-}$, which would be in accordance with our theory
of singlets in $\rho=\frac{1}{2}$, provided a DW with periodicity
Te$^{2-}$-Te$^{2-}$-Te$^{1-}$-Te$^{1-}$ was also found.  There is,
however, controversy as to whether the 260 K transition involves only
the Te ions \cite{Fang13a}, or both Ir and Te; the latter would make
the Ir ions mixed valence \cite{Ko15a}.  Pd intercalation into IrTe$_2$,
giving Pd$_x$IrTe$_2$, or weak Ir-site doping, giving
Ir$_{1-y}$Pd$_y$Te$_2$, give SC at $\sim$ 3K for $x$ and $y$ as small
as $\sim 0.02$, while beyond $x=0.1$ the SC vanishes \cite{Yang12a}.
Similar behavior is also seen with very weak Pt doping. Clearly
whether or not the structural anomaly involves only Te, or both Ir and
Te, is a very important question when it comes to ascertaining the
mechanism of SC here. It is nevertheless interesting that the parent
semiconductor is again $\frac{1}{4}$-filled
at high temperature.

\subsection[\hspace{0.5in}Superconducting hole- and electron-doped cuprates]{Superconducting hole- and electron-doped cuprates}
\label{appendix-cuprates}

Even after three decades of their discovery \cite{Bednorz86a}, high
T$_c$ superconductivity in the cuprates remains a formidable
problem. There is little understanding of the pseudogap phase in the
hole-doped materials, which is accompanied by a partial loss of
density of states as seen in some experiments. The origin of the
pseudogap in early studies had been assigned to preformed pairs and
fluctuating SC, as would occur within RVB theories
\cite{Wang06a,Li10b,Chatterjee11a,Mishra14a}.  More recent experiments
have found a ubiquitous CO in all cuprates within the pseudogap phase
\cite{Chang12a,Ghiringelli12a,Wu11a,Wu13a,Wu15a,Blanco-Canosa13a,Blanco-Canosa14a,Hucker14a,Comin14a,SilvaNeto14a,SilvaNeto15a,Comin15a,Hanaguri04a,Shen05a},
that coexists with broken rotational symmetry \cite{Kohsaka12a}, which
has been ascribed to inequivalent oxygens within the 2D unit CuO$_2$
cell. The bulk of the theoretical work on the hole-doped cuprates has
been within the weakly doped 2D Hubbard model. There is no
satisfactory explanation of the various coexisting spatial broken
symmetry phases within the pseudogap phase within these theories. To
obviate the preformed pairs versus competing order conundrum some
investigators have proposed that the CO is a density wave of Cooper
pairs
\cite{Anderson04b,Franz04a,Tesanovic04a,Tesanovic04a,Chen04a,Vojta08a,Hamidian16a,Cai16a,Mesaros16a}.
This is reminiscent of our concept of the PEC.

The most difficult issues with the electron-doped cuprates involve,
(i) robust AFM over a very broad doping range (up to dopant
concentration $\sim 0.14$ instead of the $\sim 0.03$ in the hole-doped
cuprates), and SC over a very narrow carrier concentration ($\sim
0.15-0.17$) in the Ce-doped bulk Nd$_2$CuO$_4$ and Pr$_2$CuO$_4$
\cite{Armitage10a}, and yet (ii) SC in the {\it undoped} specially
prepared thin films \cite{Naito16a}, and (iii) holelike charge
transport at optimal doping \cite{Dagan07a}. Understanding the
similarities as well as differences between the hole- and
electron-doped materials has not been possible within either the
one-band or the so-called three-band weakly doped Mott-Hubbard models.

\begin{figure}
  \begin{center}
    \includegraphics[width=3.2in]{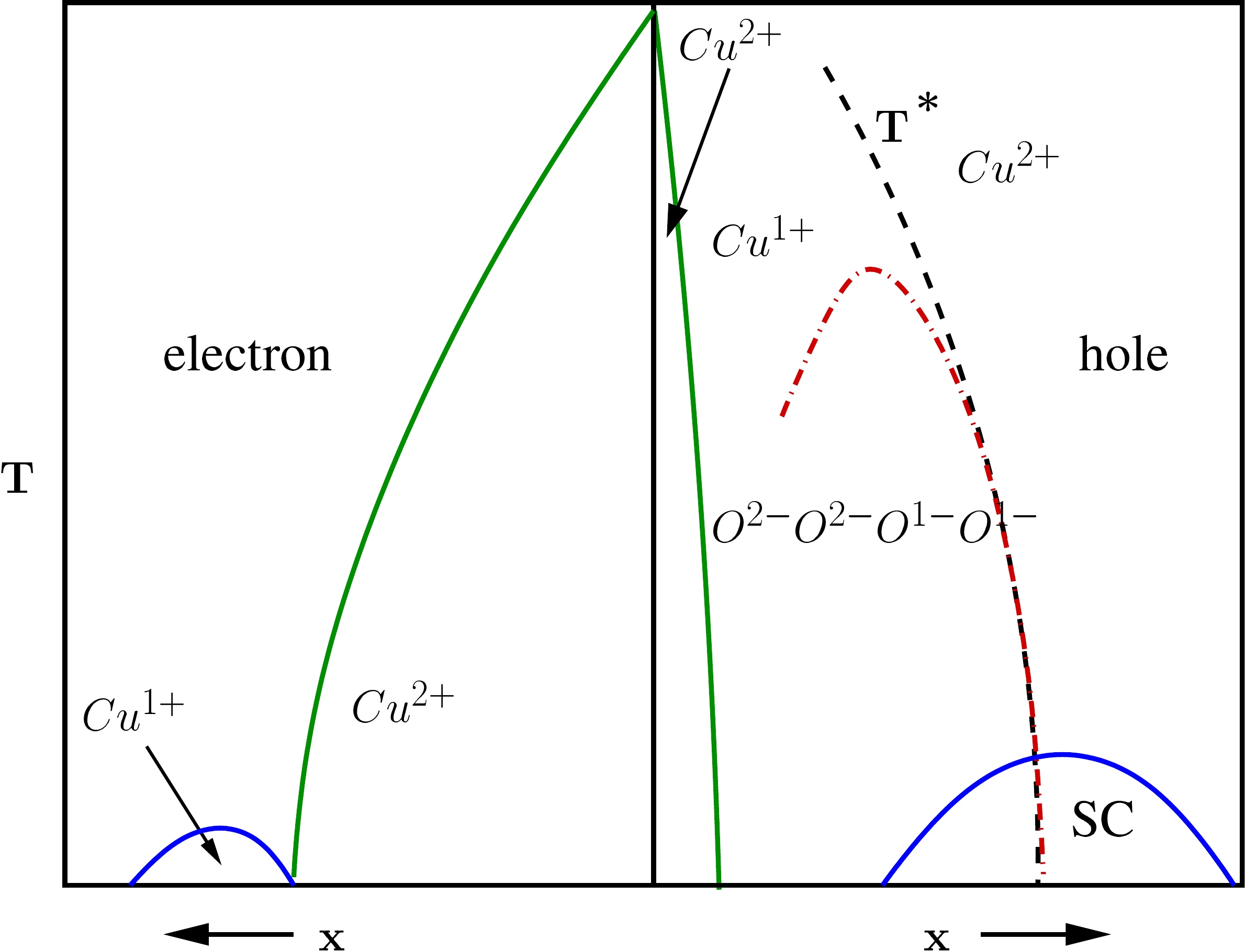}
  \end{center}
\caption{Schematic phase diagram of the electron- and hole-doped
  superconducting cuprates within the valence transition model.  The
  CO region has charge modulation O$^{2-}$-O$^{2-}$-O$^{1-}$-O$^{1-}$
  along any Cu-O bond direction. The precise quantum critical point at
  which T$^*$ intersects the dopant axis in the hole-doped systems
  cannot be evaluated without detailed calculations, and is not
  relevant.}
\label{phasediagram-cuprates}
\end{figure}

One of us has recently presented a valence transition model that
attempts a unified explanation of all the difficult experiments in
both hole- and electron-doped cuprates \cite{Mazumdar18a}. The theory
is based on the concept of transition from positive to negative
charge-transfer gap in strongly correlated charge-transfer insulators
in which cation-anion hopping is much smaller than electron
correlations.  Such a transition has been documented over three
decades in 1D mixed-stack donor-acceptor CTS, where it is known as the
neutral-ionic transition \cite{Masino17a}. The valence transition
theory begins with the observation that the second ionization energy
of the cuprate monoanion Cu$^{1+}$ is unusually high because of its
closed shell nature (higher than the monocations of Ni and Zn by
$\sim$ 2 eV). Thus for small Cu-O $d$-$p$ hopping it is only because of
the large gain in Madelung energy in the undoped parent cuprates that
the ionic charges Cu$^{2+}$ and O$^{2-}$ are obtained (the second
electron affinity of oxygen, O$^{1-} \to$ O$^{2-}$ is actually
positive because of the repulsion between the excess electrons). With
doping, there is reduction in the Madelung energy and at a crystal
structure-dependent (T vs T$^\prime$) doping concentration there
occurs a discrete jump in ionicity from Cu$^{2+}$ to Cu$^{1+}$ in both
the hole- and electron-doped compounds.  The hole in the erstwhile
Cu$^{2+}$ transfers to the oxygens in the CuO$_2$ layer.
In analogy with the neutral-ionic transition, the model envisages two
distinct phases, even though the true Cu-charges are different from
precisely 2+ and 1+. The essential point, however, is that in the
undoped parent semiconductor the O$^{2-}$ anions are closed shell and
electronically inactive, and the Cu-band can be thought of as an
effective $\frac{1}{2}$-filled band Mott-Hubbard semiconductor, a
theoretical picture that we are all familiar with. The new conceptual
development is that in the second phase the Cu$^{1+}$ ions with
3d$^{10}$ electron configuration are similarly electronically
inactive, and the effective model now corresponds to a
$\frac{1}{4}$-filled O-band, with nearly half the O-ions as
O$^{1-}$. The pseudogap transition in the hole-doped cuprates and the
sudden transition from AFM semiconductor to superconductor in the
electron-doped materials are both ascribed to this valence
transition. Fig.~\ref{phasediagram-cuprates} shows a schematic of the phase
diagram of both hole- and electron-doped cuprates within the model.
The O-sublattice is a frustrated checkerboard lattice and therefore is
susceptible to the transition to the PEC at $\frac{1}{4}$-filling. As
discussed in the original reference \cite{Mazumdar18a}, the
experimentally observed period 4 CO
\cite{Hanaguri04a,Comin14a,Shen05a,Mesaros16a} is due to the nominally
O$^{2-}$-O$^{2-}$-O$^{1-}$-O$^{1-}$ charge distribution along both the
Cu-O bonds, which also simultaneously destroys C$_4$ symmetry
\cite{Kohsaka12a}. Similarly, Hall coefficient measurements that
indicate hole like transport in the optimally doped electron-doped
cuprates \cite{Dagan07a} is expected since at this carrier
concentration the charge carriers are the same holes on the O$^{1-}$
as in the hole-doped materials. SC within the model is a consequence
of the destabilization of the PEC on the $\frac{1}{4}$-filled
O-sublattice, with O$^{1-}$-O$^{1-}$ singlets being the equivalents of
the Cooper pairs in configuration space.  The theoretical work makes a
large number of experimental predictions that can be found in the
original paper \cite{Mazumdar18a}.


\newpage

\addcontentsline{toc}{section}{\bf References}
\noindent{\bf References}

\begin{thebibliography}{100}
\expandafter\ifx\csname url\endcsname\relax
  \def\url#1{\texttt{#1}}\fi
\expandafter\ifx\csname urlprefix\endcsname\relax\def\urlprefix{URL }\fi
\expandafter\ifx\csname href\endcsname\relax
  \def\href#1#2{#2} \def\path#1{#1}\fi

\bibitem{Bednorz86a}
J.~G. Bednorz, K.~A. \protect{M\"uller}, Possible high \protect{T$_c$}
  superconductivity in the \protect{Ba-La-Cu-O} system, Z. Phys. B 64 (1986)
  189--193.

\bibitem{Timusk99a}
T.~Timusk, B.~Statt, The pseudogap in high-temperature superconductors: an
  experimental survey, Rep. Prog. Phys. 62 (1999) 61--122.

\bibitem{Norman05a}
M.~R. Norman, D.~Pines, C.~Kallin, The pseudogap: friend or foe of high
  \protect{T$_c$}?, Adv. Phys. 54 (2005) 714--733.

\bibitem{Keimer15a}
B.~Keimer, S.~A. Kivelson, M.~R. Norman, S.~Uchida, J.~Zaanen, From quantum
  matter to high-temperature superconductivity in copper oxides, Nature 518
  (2015) 179--186.

\bibitem{Stewart11a}
G.~R. Stewart, Superconductivity in iron compounds, Rev.\ Mod.\ Phys. 83 (2011)
  1589.

\bibitem{Si16a}
Q.~Si, R.~Yu, E.~Abrahams, High-temperature superconductivity in iron pnictides
  and chalcogenides, Nature Rev. Mater. 1 (2016) 16017.

\bibitem{Kobayashi17a}
T.~Kobayashi, Study of Electronic Properties of 122 Iron Pnictide Through
  Structural, Carrier-Doping, and Impurity-Scattering Effects, Springer, New
  York, 2017.

\bibitem{Dagotto05a}
E.~Dagotto, Open questions in \protect{CMR} manganites, relevance of clustered
  states and analogies with other compounds including the cuprates, New J.
  Phys. 7 (2005) 67.

\bibitem{Mazumdar89a}
S.~Mazumdar, A unified theoretical approach to superconductors with strong
  {C}oulomb correlations: the organics, \protect{LiTi$_2$O$_4$}, electron- and
  hole-doped copper oxides and doped \protect{BaBiO$_3$}, in: D.~Baeriswyl,
  D.~K. Campbell (Eds.), Interacting Electrons in Reduced Dimensions.
  Proceedings of a \protect{NATO} {A}dvanced {R}esearch {W}orkshop, Plenum, New
  York, 1989, pp. 315--329.

\bibitem{Uemura91a}
Y.~J. Uemura, L.~P. Le, G.~M. Luke, B.~J. Sternlieb, W.~D. Wu, J.~H. Brewer,
  T.~M. Riseman, C.~L. Seaman, M.~B. Maple, M.~Ishikawa, D.~G. Hinks, J.~D.
  Jorgensen, G.~Saito, H.~Yamochi, Basic similarities among cuprate,
  bismuthate, organic, {C}hevrel-phase, and heavy-fermion superconductors shown
  by penetration-depth measurements, Phys.\ Rev.\ Lett. 66 (1991) 2665--2668.

\bibitem{Fukuyama87a}
H.~Fukuyama, Y.~Hasegawa, Superconductivity in organics and oxides: similarity
  and dissimilarity, Physica B \& C 148B+C (1987) 204--211.

\bibitem{Fukuyama90a}
H.~Fukuyama, On some organic conductors in the light of oxide superconductors,
  in: G.~Saito, S.~Kagoshima (Eds.), Physics and Chemistry of Organic
  Superconductors. Proceedings of the ISSP International Symposium,
  Springer-Verlag, Berlin, 1990, pp. 15--20.

\bibitem{McKenzie97a}
R.~H. McKenzie, Similarities between organic and cuprate superconductors,
  Science 278 (1997) 820.

\bibitem{Iwasa03a}
Y.~Iwasa, T.~Takenobu, Superconductivity, {M}ott-{H}ubbard states, and
  molecular orbital order in intercalated fullerides, J. Phys.: Condens. Matter
  15 (2003) R495.

\bibitem{Capone09a}
M.~Capone, M.~Fabrizio, C.~Castellani, E.~Tosatti, Colloquium: Modeling the
  unconventional superconducting properties of expanded {A}$_3${C}$_{60}$
  fullerides, Rev.\ Mod.\ Phys. 81 (2009) 943.

\bibitem{Mazumdar12a}
S.~Mazumdar, R.~T. Clay, Is there a common theme behind the correlated-electron
  superconductivity in organic charge-transfer solids, cobaltates, spinels and
  fullerides?, Phys. Stat. Solidi 249 (2012) 995--998.

\bibitem{Baskaran16a}
G.~Baskaran, \protect{RVB} states in doped band insulators from {C}oulomb
  forces: theory and a case study of superconductivity in \protect{BiS$_2$}
  layers, Supercond. Sci. Technol. 29 (2016) 124002.

\bibitem{Jerome80a}
D.~J\'erome, A.~Mazaud, M.~Ribault, K.~Bechgaard, Superconductivity in a
  synthetic organic conductor \protect{(TMTSF)$_2$PF$_6$}, J. Phys. (Paris)
  Lett. 41 (1980) L95--L98.

\bibitem{Jerome80b}
M.~Ribault, J.~P. Pouget, D.~Jerome, K.~Bechgaard, Superconductivity and
  absence of a {K}ohn anomaly in the quasi-one-dimensional organic conductor -
  \protect{(TMTSF)$_2$AsF$_6$}, J. Phys. (Paris) Lett. 41 (1980) L607--610.

\bibitem{Parkin83a}
S.~S.~P. Parkin, E.~M. Engler, R.~R. Schumaker, R.~Lagier, V.~Y. Lee, J.~C.
  Scott, R.~L. Greene, Superconductivity in a new family of organic conductors,
  Phys.\ Rev.\ Lett. 50 (1983) 270–273.

\bibitem{Kini90a}
A.~M. Kini, U.~Geiser, H.~H. Wang, K.~D. Carlson, J.~M. Williams, W.~K. Kwok,
  K.~G. Vandervoort, J.~E. Thompson, D.~L. Stupka, New ambient-pressure organic
  superconductor, \protect{$\kappa$-(ET)$_2$Cu[N(CN)$_2$]Br}, with the highest
  transition temperature yet observed (inductive onset \protect{T$_c$} = 11.6
  {K}, resistive onset = 12.5 {K}), Inorg. Chem. 29 (1990) 2555--2557.

\bibitem{Ishiguro}
T.~Ishiguro, K.~Yamaji, G.~Saito, Organic Superconductors, Springer-Verlag, New
  York, 1998.

\bibitem{Williams}
J.~M. Williams, J.~R. Ferraro, R.~J. Thron, K.~D. Carlson, U.~Geiser, H.~H.
  Wang, A.~M. Kini, M.~H. Whangbo, Organic Superconductors (Including
  {F}ullerenes): Synthesis, Structure, Properties, and Theory, Prentice-Hall,
  Englewood Cliffs, NJ, 1992.

\bibitem{Saito07a}
G.~Saito, Y.~Yoshida, Development of conductive organic molecular assemblies:
  organic metals, superconductors, and exotic functional materials, Bull. Chem.
  Soc. Jpn. 80 (2007) 1–137.

\bibitem{Lebed}
A.~G. Lebed (Ed.), The Physics of Organic Superconductors and Conductors,
  Springer, Berlin, 2008.

\bibitem{Kato04a}
R.~Kato, Conducting metal dithiolene complexes: Structural and electronic
  properties, Chem. Rev. 104 (2004) 5319--5346.

\bibitem{Lyubovskaya87a}
R.~N. Lyubovskaya, E.~A. Zhilyaeva, A.~V. Zvarykina, V.~N. Laukhin, R.~B.
  Lyubovskii, S.~I. Pesotskii, Is the organic metal {(ET)$_4$Hg$_3$Br$_8$} a
  quasi-{2D} superconductor?, JETP Lett. 45 (1987) 530.

\bibitem{Naito05a}
A.~Naito, Y.~Nakazawa, K.~Saito, H.~Taniguchi, K.~Kanoda, M.~Sorai, Anomalous
  enhancement of electronic heat capacity in the organic conductors
  \protect{$\kappa$-(BEDT-TTF)$_4$Hg$_{3−\delta}$X$_8$ (X=Br,Cl)}, Phys.\
  Rev.\ B 71 (2005) 054514.

\bibitem{Taniguchi07a}
H.~Taniguchi, T.~Okuhata, T.~Nagai, K.~Satoh, N.~Mori, Y.~Shimizu, M.~Hedo,
  Y.~Uwatoko, Anomalous pressure dependence of superconductivity in layered
  organic conductor, \protect{$\kappa$-(BEDT-TTF)$_4$Hg$_{2:89}$Br$_8$}, J.\
  Phys.\ Soc.\ Jpn. 76 (2007) 11370.

\bibitem{Torrance77a}
J.~B. Torrance, Y.~Tomkiewicz, B.~D. Silverman, Enhancement of the magnetic
  susceptibility of {TTF-TCNQ} (tetrathiafulvalene-tetracyanoquinodimethane) by
  {C}oulomb correlations, Phys.\ Rev.\ B 15 (1977) 4738--4749.

\bibitem{Torrance78a}
J.~B. Torrance, Spin waves, scattering at \protect{4$k_F$}, and spin-{P}eierls
  fluctuations in an organic metal: tetrathiafulvalene-tetracyanoquinodimethane
  {(TTF-TCNQ)}, Phys.\ Rev.\ B 17 (1978) 3099--3103.

\bibitem{Mazumdar83a}
S.~Mazumdar, A.~N. Bloch, Systematic trends in short-range {C}oulomb effects
  among nearly one-dimensional organic conductors, Phys.\ Rev.\ Lett. 50 (1983)
  207--210.

\bibitem{Armitage10a}
N.~P. Armitage, P.~Fournier, R.~L. Greene, Progress and perspectives on
  electron-doped cuprates, Rev.\ Mod.\ Phys. 82 (2010) 2421--2487, and
  references therein.

\bibitem{Clay12a}
R.~T. Clay, J.~P. Song, S.~Dayal, S.~Mazumdar, Ground state and finite
  temperature behavior of 1/4-filled band zigzag ladders, J.\ Phys.\ Soc.\ Jpn.
  81 (2012) 074707.

\bibitem{Bourbonnais08a}
C.~Bourbonnais, D.~J\'erome, Interacting electrons in quasi-one-dimensional
  organic superconductors, in: A.~G. Lebed (Ed.), The Physics of Organic
  Superconductors and Conductors, Springer, Berlin, 2008, pp. 357--412.

\bibitem{Bourbonnais11a}
C.~Bourbonnais, A.~Sedeki, Superconductivity and antiferromagnetism as
  interfering orders in organic conductors, Comptex Rendus Physique 12 (2011)
  532--541.

\bibitem{Clay08a}
R.~T. Clay, H.~Li, S.~Mazumdar, Absence of superconductivity in the half-filled
  band {H}ubbard model on the anisotropic triangular lattice, Phys.\ Rev.\
  Lett. 101 (2008) 166403.

\bibitem{Dayal12a}
S.~Dayal, R.~T. Clay, S.~Mazumdar, Absence of long-range superconducting
  correlations in the frustrated \protect{$\frac{1}{2}$}-filled band {H}ubbard
  model, Phys.\ Rev.\ B 85 (2012) 165141.

\bibitem{Anderson73a}
P.~W. Anderson, Resonating valence bonds: a new kind of insulator?, Mater. Res.
  Bull. 8 (1973) 153--160.

\bibitem{Anderson87a}
P.~W. Anderson, The resonating valence bond state in \protect{La$_2$CuO$_4$}
  and superconductivity, Science 235 (1987) 1196.

\bibitem{Alexandrov94a}
A.~S. Alexandrov, N.~F. Mott, High Temperature Superconductors And Other
  Superfluids, Taylor and Francis, London, 1994.

\bibitem{Fradkin15a}
E.~Fradkin, S.~A. Kivelson, J.~M. Tranquada, Colloquium: Theory of intertwined
  orders in high temperature superconductors, Rev.\ Mod.\ Phys. 87 (2015)
  457--482.

\bibitem{Cai16a}
P.~Cai, W.~Ruan, Y.~Peng, C.~Ye, X.~Li, Z.~Hao, X.~Zhou, D.-H. Lee, Y.~Wang,
  Visualizing the evolution from the \protect{M}ott insulator to a
  charge-ordered insulator in lightly doped cuprates, Proc. Natl. Acad. Sci. 12
  (2016) 1047--1052.

\bibitem{Mesaros16a}
A.~Mesaros, K.~Fujita, S.~D. Edkins, M.~H. Hamidian, H.~Eisaki, S.~Uchida,
  J.~C.~S. Davis, M.~J. Lawler, E.-A. Kim, Commensurate {4$a_0$} period charge
  density modulations throughout the {Bi$_2$Sr$_2$CaCu$_2$O$_{8+x}$} pseudogap
  regime, Proc. Natl. Acad. Sci. 113 (2016) 12661–12666.

\bibitem{Soos74a}
Z.~G. Soos, Theory of $\pi$-molecular charge-transfer crystals, Ann. Rev. Phys.
  Chem. 25 (1974) 121--153.

\bibitem{Soos75a}
Z.~G. Soos, D.~J. Klein, Charge-transfer in solid state complexes, in:
  R.~Foster (Ed.), Molecular Association, Vol.~1, Academic, New York, 1975,
  p.~1.

\bibitem{Torrance79a}
J.~B. Torrance, The difference between metallic and insulating salts of
  tetracyanoquinodimethane \protect{(TCNQ)}: How to design an organic metal,
  Acc. Chem. Res. 12 (1979) 79–86.

\bibitem{Torrance81a}
J.~B. Torrance, J.~E. V. J.~J. Mayerle, V.~Y. Lee, Discovery of a
  neutral-to-ionic phase transition in organic materials, Phys.\ Rev.\ Lett. 46
  (1981) 253–257.

\bibitem{Torrance81b}
J.~B. Torrance, A.~Girlando, J.~J. Mayerle, J.~I. Crowley, V.~Y. Lee,
  P.~Batail, Anomalous nature of neutral-to-ionic phase transition in
  tetrathiafulvalene-chloranil, Phys.\ Rev.\ Lett. 47 (1981) 1747--1750.

\bibitem{Koshihara90a}
S.~Koshihara, Y.~Tokura, T.~Mitani, G.~Saito, T.~Koda, Photoinduced valence
  instability in the organic molecular compound tetrathiafulvalene-p-chloranil
  {(TTF-CA)}, Phys.\ Rev.\ B 42 (1990) 6853–6856.

\bibitem{Torrance75a}
J.~B. Torrance, B.~A. Scott, F.~B. Kaufman, Optical properties of
  charge-transfer salts of tetracyanoquinodimethane {(TCNQ)}, Solid St. Commun.
  17 (1975) 1369--1373.

\bibitem{Kumar11a}
M.~Kumar, B.~J. Topham, R.~H. Yu, B.~D.~H. Quoc, Z.~G. Soos, Magnetic
  susceptibility of alkali-tetracyanoquinodimethane salts and extended
  {H}ubbard models with bond order and charge density wave phases, J. Chem.
  Phys. 134 (2011) 234304.

\bibitem{Epstein71a}
A.~Epstein, S.~Etemad, A.~Garito, A.~Heeger, Metal-insulator transition in an
  organic solid: Experimental realization of the one-dimensional {H}ubbard
  model, Solid St.\ Comm. 9 (1971) 1803--1808.

\bibitem{Epstein81a}
A.~Epstein, J.~S. Miller, J.~P. Pouget, R.~Comes, X-ray observation of
  crossover of \protect{2k$_{\rm F}$} to \protect{4k$_{\rm F}$} scattering in
  \protect{(N-Methylphenazinium)$_x$}\protect{(Phenazine)$_{1-x}$}
  \protect{(Tetracyanoquinodimethane)}\protect{[(NMP)$_x$(Phen)$_{1-x}$(TCNQ)]},
  \protect{$0.5 \leq x \leq 1.0$}, Phys.\ Rev.\ Lett. 47 (1981) 741--744.

\bibitem{Ferraris73a}
J.~P. Ferraris, D.~O. Cowan, J.~V.~Walatka, J.~H. Perlstein, Electron transfer
  in a new highly conducting donor-acceptor complex, J. Am. Chem. Soc. 95
  (1973) 948.

\bibitem{Coleman73a}
L.~B. Coleman, M.~J. Cohen, D.~J. Sandman, F.~G. Yamagishi, A.~F. Garito, A.~J.
  Heeger, Superconducting fluctuations and the {P}eierls instability in an
  organic solid, Solid St.\ Comm. 12 (1973) 1125.

\bibitem{Keller74a}
H.~J. Keller (Ed.), Low-dimensional cooperative phenomena, Springer, New York,
  1975.

\bibitem{Torrance80a}
J.~B. Torrance, J.~J. Mayerle, K.~Bechgaard, B.~D. Silverman, Y.~Tomkiewicz,
  Comparison of two isostructural organic compounds, one metallic and the other
  insulating, Phys.\ Rev.\ B 22 (1980) 4960--4965.

\bibitem{Hawley78a}
M.~E. Hawley, T.~O. Poehler, T.~F. Carruthers, A.~N. Bloch, D.~O. Cowan, T.~J.
  Kistenmacher, Magnetic and electrical behavior of a new organic
  charge-transfer salt, Bull. Am. Phys. Soc. 23 (1978) 424.

\bibitem{Keller77a}
H.~J. Keller (Ed.), Chemistry and Physics of One Dimensional Metals, Plenum,
  New York, 1977.

\bibitem{Devreese79a}
J.~T. Devreese, R.~P. Evrard, V.~E. van Doren (Eds.), Highly conducting
  one-dimensional solids, Plenum, New York, 1979.

\bibitem{Bernasconi75a}
J.~Bernasconi, M.~J. Rice, W.~R. Schneider, S.~Strassler, \protect{P}eierls
  transition in the strong-coupling \protect{Hubbard} chain, Phys.\ Rev.\ B 12
  (1975) 1090--1092.

\bibitem{Klein74a}
D.~J. Klein, W.~A. Seitz, Partially filled linear \protect{Hubbard} model near
  the atomic limit, Phys.\ Rev.\ B 10 (1974) 3217--3227.

\bibitem{Bray75a}
J.~W. Bray, H.~R.~H. Jr., L.~V. Interrante, I.~S. Jacobs, J.~S. Kasper, G.~D.
  Watkins, S.~H. Wee, J.~C. Bonner, Observation of a spin-{P}eierls transition
  in a {H}eisenberg antiferromagnetic linear-chain system, Phys.\ Rev.\ Lett.
  35 (1975) 744--747.

\bibitem{Bray83a}
J.~W. Bray, L.~V. Interrante, I.~S. Jacobs, J.~C. Bonner, The spin-{P}eierls
  transition, in: J.~S. Miller (Ed.), Extended linear chain compounds, Vol.~3,
  Plenum, New York, 1983, Ch.~7, pp. 353--415.

\bibitem{Cross79a}
M.~C. Cross, D.~S. Fisher, A new theory of the spin-{P}eierls transition with
  special relevance to the experiments on \protect{TTFCuBDT}, Phys.\ Rev.\ B 19
  (1979) 402--419.

\bibitem{Pouget76a}
J.~P. Pouget, S.~K. Khanna, F.~Denoyer, R.~Com\'es, X ray observation of
  {2k$_F$} and {4$k_F$} scatterings in
  tetrathiafulvalene-tetracyanoquinodimethane {(TTF-TCNQ)}, Phys.\ Rev.\ Lett.
  37 (1976) 437--440.

\bibitem{Comes77a}
R.~Com\'es, X-ray and neutron scattering investigation of the charge density
  waves in {TTF-TCNQ}, in: H.~J. Keller (Ed.), Chemistry and Physics of
  One-Dimensional Conductors, Plenum, New York, 1977, pp. 315--339.

\bibitem{Kagoshima76a}
S.~Kagoshima, T.~Ishiguro, H.~Anzai, X-ray scattering study of phonon anomalies
  and superstructures in {TTF-TCNQ}, J.\ Phys.\ Soc.\ Jpn. 41 (1976) 2061.

\bibitem{Mook76a}
H.~A. Mook, J.~C.~R.~Watson, Neutron inelastic scattering study of
  tetrathiafulvalene tetracyanoquinodimethane \protect{(TTF-TCNQ)}, Phys.\
  Rev.\ Lett. 36 (1976) 801–803.

\bibitem{Shirane76a}
G.~Shirane, S.~M. Shapiro, R.~Com\'es, A.~F. Garito, A.~J. Heeger, Phonon
  dispersion and \protect{K}ohn anomaly in
  tetrathiafulvalene-tetracyanoquinodimethane {(TTF-TCNQ)}, Phys.\ Rev.\ B 14
  (1976) 2325--2334.

\bibitem{Bosch77a}
A.~Bosch, B.~v.~Bodegom, The crystal structure of the {1:2} complex of
  {N-methyl-N-ethylmorpholinium} and {7,7,8,8-tetracyanoquinodimethane},
  {MEM(TCNQ)$_2$}, at {-160$^\circ$} {C}, Acta Crystallogr. Sect. B 33 (1977)
  3013.

\bibitem{Huizinga79a}
S.~Huizinga, J.~Kommandeur, G.~A. Sawatzky, B.~T. Thole, K.~Kopinga, W.~J.~M.
  de~Jonge, J.~Roos, Spin-{P}eierls transition in
  {N-methyl-N-ethyl-morpholinium-ditetracyanoquinodimethanide}
  {[MEM-(TCNQ)$_2$]}, Phys. Rev. B 19~(9) (1979) 4723--4732.

\bibitem{Huizinga82a}
S.~Huizinga, J.~Kommandeur, H.~T. Jonkman, C.~Haas, Magnetic ({2$k_F$}) and
  electronic ({4$k_F$}) {P}eierls transitions from a {H}ubbard {H}amiltonian
  extended with intersite-dependent transfer, Phys.\ Rev.\ B 25 (1982) 1717.

\bibitem{Epstein82a}
A.~Epstein, J.~W. Kaufer, H.~Rommelmann, I.~A. Howard, E.~M. Conwell, J.~S.
  Miller, J.~P. Pouget, R.~Comes, Evidence for solitons in conducting organic
  charge-transfer crystals, Phys.\ Rev.\ Lett. 49 (1982) 1037--1041.

\bibitem{Conwell83a}
E.~M. Conwell, I.~A. Howard, Solitons in quasi one-dimensional crystals, J.
  Phys. (Colloque) C3 (1983) 1487--1492.

\bibitem{Rice82a}
M.~J. Rice, E.~J. Mele, Possibility of solitons with charge \protect{$\pm e/2$}
  in highly correlated 1:2 salts of tetracyanoquinodimethane \protect{(TCNQ)},
  Phys.\ Rev.\ B 25 (1982) 1339--1343.

\bibitem{Hirsch84a}
J.~E. Hirsch, D.~J. Scalapino, \protect{2p$_F$} and \protect{4p$_F$}
  instabilities in the one-dimensional \protect{Hubbard} model, Phys.\ Rev.\ B
  29 (1984) 5554--5561.

\bibitem{Bloch83a}
A.~N. Bloch, S.~Mazumdar, Short-range {C}oulomb effects and the chemical
  systematics of conducting organic charge-transfer salts, J. Physique C-3 44
  (1983) 1273--1279.

\bibitem{Hirsch83c}
J.~E. Hirsch, D.~J. Scalapino, \protect{2p$_{\rm F}$} and \protect{4p$_{\rm
  F}$} instabilities in a one-quarter-filled-band {H}ubbard model, Phys.\ Rev.\
  B 27 (1983) 7169.

\bibitem{Mila93a}
F.~Mila, X.~Zotos, Phase diagram of the one-dimensional extended {H}ubbard
  model at quarter-filling, Europhys.\ Lett. 24 (1993) 133.

\bibitem{Penc94a}
K.~Penc, F.~Mila, Phase diagram of the one-dimensional extended
  \protect{Hubbard} model with attractive and/or repulsive interactions at
  quarter filling, Phys.\ Rev.\ B 49 (1994) 9670--9678.

\bibitem{Clay03a}
R.~T. Clay, S.~Mazumdar, D.~K. Campbell, The pattern of charge ordering in
  quasi-one dimensional organic charge-transfer solids, Phys.\ Rev.\ B 67
  (2003) 115121.

\bibitem{Visser83a}
R.~J.~J. Visser, S.~Oostra, C.~Vettier, J.~Voiron, Determination of the
  \protect{spin--Peierls} distortion in \protect{N-methyl-N-ethyl-morpholinium}
  ditetracyanoquinodimethanide \protect{[MEM(TCNQ)$_2$]}: Neutron diffraction
  study at 6 {K}, Phys.\ Rev.\ B 28 (1983) 2074--2077.

\bibitem{Su79a}
W.~P. Su, J.~R. Schrieffer, A.~J. Heeger, Solitons in polyacetylene, Phys.\
  Rev.\ Lett. 42 (1979) 1698.

\bibitem{Holstein59a}
T.~Holstein, Studies of polaron motion: Part {I}. the molecular-crystal model,
  Ann. Phys.(N.Y.) 8 (1959) 325.

\bibitem{Rice75a}
M.~J. Rice, C.~B. Duke, N.~O. Lipari, Intramolecular vibrational stabilization
  of the charge density wave state in organic metals, Solid St.\ Comm. 17
  (1975) 1089--1093.

\bibitem{Rice77a}
M.~J. Rice, N.~O. Lipari, Electron—molecular-vibration coupling in
  tetrathiafulvalene-tetracyanoquinodimethane \protect{(TTF-TCNQ)}, Phys.\
  Rev.\ Lett. 38 (1977) 437--439.

\bibitem{Conwell88a}
E.~M. Conwell, Introduction to highly conducting quasi-one-dimensional organic
  crystals, Semiconductors and Semimetals 27 (1988) 1--27.

\bibitem{Kivelson83a}
S.~Kivelson, Solitons with adjustable charge in a commensurate {P}eierls
  insulator, Phys.\ Rev.\ B 28 (1983) 2653.

\bibitem{Su83a}
W.-P. Su, Soliton excitations in a commensurability 2 mixed {P}eierls system,
  Solid St.\ Comm. 48 (1983) 479.

\bibitem{Dixit84a}
S.~N. Dixit, S.~Mazumdar, Electron-electron interaction effects on {P}eierls
  dimerization in a half-filled band, Phys.\ Rev.\ B 29 (1984) 1824--1839.

\bibitem{Ung93a}
K.-C. Ung, S.~Mazumdar, D.~K. Campbell, Coexisting \protect{CDW} and
  \protect{BOW} in organic conductors with non-half-filled bands, Solid St.\
  Comm. 85 (1993) 917--920.

\bibitem{Dressel12a}
M.~Dressel, M.~Dumm, T.~Knoblauch, M.~Masino, Comprehensive optical
  investigations of charge order in organic chain compounds {(TMTTF)$_2$X},
  Crystals 2 (2012) 528.

\bibitem{Girlando11a}
A.~Girlando, Charge sensitive vibrations and electron-molecular vibration
  coupling in {B}is(ethylenedithio)-tetrathiafulvalene {(BEDT-TTF)}, J. Phys.
  Chem. C 115 (2011) 19371--19378.

\bibitem{Tomkiewicz77a}
Y.~Tomkiewicz, B.~Welber, P.~E. Seiden, R.~Schumaker, Electron-electron
  interactions in the fulvalene family of organic metals, Solid St. Commun. 23
  (1977) 471.

\bibitem{Mazumdar86a}
S.~Mazumdar, S.~N. Dixit, Unified theory of segregated-stack organic
  charge-transfer solids: magnetic properties, Phys.\ Rev.\ B 34 (1986)
  3683--3699.

\bibitem{Sandvik92a}
A.~W. Sandvik, A generalization of {H}andscomb's quantum {M}onte {C}arlo
  scheme-application to the 1{D} {H}ubbard model, J. Phys. A 25 (1992)
  3667--3682.

\bibitem{Syljuasen02a}
O.~F. Syljuasen, A.~W. Sandvik, Quantum {M}onte {C}arlo with directed loops,
  Phys.\ Rev.\ E 66 (2002) 046701.

\bibitem{Jerome91a}
D.~\protect{J\'erome}, The physics of organic superconductors, Science 252
  (1991) 1509--1514.

\bibitem{Salameh11a}
B.~Salameh, S.~Yasin, M.~Dumm, G.~Untereiner, L.~Montgomery, M.~Dressel, Spin
  dynamics of the organic linear chain compounds \protect{(TMTTF)$_2$X}
  (\protect{X=SbF$_6$}, \protect{AsF$_6$}, \protect{BF$_4$}, \protect{ReO$_4$},
  and \protect{SCN}), Phys.\ Rev.\ B 83 (2011) 205126.

\bibitem{Yoshimi12a}
K.~Yoshimi, H.~Seo, S.~Ishibashi, S.~E. Brown, Tuning the magnetic
  dimensionality by charge ordering in the molecular {T}{M}{T}{T}{F} salts,
  Phys.\ Rev.\ Lett. 108 (2012) 096402.

\bibitem{Ward14a}
A.~B. Ward, R.~Clay, S.~Mazumdar, Comment on ``{T}uning the magnetic
  dimensionality by charge ordering in the molecular {TMTTF} salts'', Phys.\
  Rev.\ Lett. 113 (2014) 029701.

\bibitem{Seo14a}
H.~Seo, S.~Ishibashi, S.~E. Brown, Seo \protect{\it et al.} reply:, Phys.\
  Rev.\ Lett. 113 (2014) 029702.

\bibitem{Hubbard78a}
J.~Hubbard, Generalized \protect{Wigner} lattices in one dimension and some
  applications to tetracyanoquinodimethane \protect{(TCNQ)} salts, Phys.\ Rev.\
  B 17 (1978) 494--505.

\bibitem{Seo97a}
H.~Seo, H.~Fukuyama, Antiferromagnetic phases of one-dimensional quarter-filled
  organic conductors, J.\ Phys.\ Soc.\ Jpn. 66 (1997) 1249--1252.

\bibitem{Seo00a}
H.~Seo, Charge ordering in organic \protect{ET} compounds, J.\ Phys.\ Soc.\
  Jpn. 69 (2000) 805--820.

\bibitem{Shibata01a}
Y.~Shibata, S.~Nishimoto, Y.~Ohta, Charge ordering in the one-dimensional
  extended \protect{Hubbard} model: Implication to the \protect{TMTTF} family
  of organic conductors, Phys.\ Rev.\ B 64 (2001) 235107.

\bibitem{Seo07a}
H.~Seo, Y.~Motome, T.~Kato, Finite-temperature phase transitions in
  quasi-one-dimensional molecular conductors, J.\ Phys.\ Soc.\ Jpn. 76 (2007)
  013707.

\bibitem{Otsuka08a}
Y.~Otsuka, H.~Seo, Y.~Motome, T.~Kato, Finite-temperature phase diagram of
  quasi-one-dimensional molecular conductors: {Q}uantum {M}onte {C}arlo study,
  J.\ Phys.\ Soc.\ Jpn. 77 (2008) 113705.

\bibitem{Tsuchiizu01a}
M.~Tsuchiizu, H.~Yoshioka, Y.~Suzumura, Crossover from quarter-filling to
  half-filling in a one-dimensional electron system with a dimerized and
  quarter-filled band, J.\ Phys.\ Soc.\ Jpn. 70 (2001) 1460--1463.

\bibitem{Sugiura04a}
M.~Sugiura, M.~Tsuchiizu, Y.~Suzumura, Interplay of lattice dimerization and
  charge ordering in one-dimensional quarter-filled electron systems, J.\
  Phys.\ Soc.\ Jpn. 73 (2004) 2487--2493.

\bibitem{Sugiura05a}
M.~Sugiura, M.~Tsuchiizu, Y.~Suzumura, Spin-{P}eierls transition temperature in
  quarter-filled organic conductors, J.\ Phys.\ Soc.\ Jpn. 74 (2005) 983--987.

\bibitem{Yoshioka06a}
H.~Yoshioka, M.~Tsuchiizu, H.~Seo, Finite-temperature charge-ordering
  transition and fluctuation effects in quasi one-dimensional electron systems
  at quarter filling, J.\ Phys.\ Soc.\ Jpn. 75 (2006) 063706.

\bibitem{Yoshioka07a}
H.~Yoshioka, M.~Tsuchiizu, H.~Seo, Charge-ordered state versus dimer-{M}ott
  insulator at finite temperatures, J.\ Phys.\ Soc.\ Jpn. 76 (2007) 103701.

\bibitem{Yoshioka10a}
H.~Yoshioka, M.~Tsuchiizu, Y.~Otsuka, H.~Seo, Finite-temperature properties
  across the charge ordering transition ---combined bosonization,
  renormalization group, and numerical methods, J.\ Phys.\ Soc.\ Jpn. 79 (2014)
  094714.

\bibitem{Sano94a}
K.~Sano, Y.~Ono, Electronic structure of one-dimensional extended
  \protect{Hubbard} model, J.\ Phys.\ Soc.\ Jpn. 63 (1994) 1250--1253.

\bibitem{Ung94a}
K.~C. Ung, S.~Mazumdar, D.~Toussaint, Metal-insulator and insulator-insulator
  transitions in the quarter-filled band organic conductors, Phys.\ Rev.\ Lett.
  73 (1994) 2603--2606.

\bibitem{Riera00a}
J.~Riera, D.~Poilblanc, Coexistence of charge-density waves, bond-order waves,
  and spin-density waves in quasi-one-dimensional charge-transfer salts, Phys.\
  Rev.\ B 62 (2000) R16243--R16246.

\bibitem{Nakamura00b}
M.~Nakamura, Tricritical behavior in the extended {H}ubbard chains, Phys.\
  Rev.\ B 61 (2000) 16377--16392.

\bibitem{Riera01a}
J.~Riera, D.~Poilblanc, Influence of the anion potential on the charge ordering
  in quasi-one-dimensional charge-transfer salts, Phys.\ Rev.\ B 63 (2001)
  241102.

\bibitem{Clay02a}
R.~T. Clay, S.~Mazumdar, D.~K. Campbell, Charge ordering in
  \protect{$\theta$-(BEDT-TTF)$_2$X} materials, J.\ Phys.\ Soc.\ Jpn. 71 (2002)
  1816--1819.

\bibitem{Sano04a}
K.~Sano, Y.~Ono, Critical behavior near the metal-insulator transition in the
  one-dimensional extended {H}ubbard model at quarter filling, Phys.\ Rev.\ B
  70 (2004) 155102.

\bibitem{Li10a}
H.~Li, R.~T. Clay, S.Mazumdar, The paired-electron crystal in the
  two-dimensional frustrated quarter-filled band, J. Phys.: Condens. Matter 22
  (2010) 272201.

\bibitem{Dayal11a}
S.~Dayal, R.~T. Clay, H.~Li, S.~Mazumdar, Paired electron crystal: Order from
  frustration in the quarter-filled band, Phys.\ Rev.\ B 83 (2011) 245106.

\bibitem{Clay01a}
R.~T. Clay, S.~Mazumdar, D.~K. Campbell, Re-integerization of fractional
  charges in the correlated quarter-filled band, Phys.\ Rev.\ Lett. 86 (2001)
  4084--4087.

\bibitem{Clay05a}
R.~T. Clay, S.~Mazumdar, Co-operative bond-charge density wave and giant spin
  gap in the quarter-filled zigzag electron ladder, Phys.\ Rev.\ Lett. 94
  (2005) 207206.

\bibitem{Mazumdar00a}
S.~Mazumdar, R.~T. Clay, D.~K. Campbell, Bond-order and charge-density waves in
  the isotropic interacting two-dimensional quarter-filled band and the
  insulating state proximate to organic superconductivity, Phys.\ Rev.\ B 62
  (2000) 13400--13425.

\bibitem{Clay07a}
R.~T. Clay, R.~P. Hardikar, S.~Mazumdar, Temperature-driven transition from the
  {W}igner crystal to the bond-charge-density wave in the quasi-one-dimensional
  quarter-filled band, Phys.\ Rev.\ B 76 (2007) 205118.

\bibitem{Clay17a}
R.~T. Clay, A.~B. Ward, N.~Gomes, S.~Mazumdar, Bond patterns and charge order
  amplitude in $\frac{1}{4}$-filled charge-transfer solids, Phys.\ Rev.\ B 95
  (2017) 125114.

\bibitem{Nishimoto00a}
S.~Nishimoto, M.~Takahashi, Y.~Ohta, Charge gap of the quasi-one-dimensional
  organic conductors: a density-matrix renormalization-group study, J.\ Phys.\
  Soc.\ Jpn. 69 (2000) 1594--1597.

\bibitem{Kuwabara03a}
M.~Kuwabara, H.~Seo, M.~Ogata, Coexistence of charge order and spin-{P}eierls
  lattice distortion in one-dimensional organic conductors, J.\ Phys.\ Soc.\
  Jpn. 72 (2003) 225.

\bibitem{Ejima05a}
S.~Ejima, F.~Gebhard, S.~Nishimoto, Tomonaga-{L}uttinger parameters for doped
  {M}ott insulators, Eur.\ Phys.\ Lett. 70 (2005) 492--498.

\bibitem{Ejima06a}
S.~Ejima, F.~Gebhard, S.~Nishimoto, Tomonaga-{L}uttinger parameters and spin
  excitations in the dimerized extended {H}ubbard model, Phys.\ Rev.\ B 74
  (2006) 245110.

\bibitem{Benthien05a}
H.~Benthien, E.~Jeckelmann, Optical conductivity of the one-dimensional
  dimerized {H}ubbard model at quarter filling, Eur. Phys. J. B 44 (2005)
  287--297.

\bibitem{Shirakawa09a}
T.~Shirakawa, E.~Jeckelmann, Charge and spin {D}rude weight of the
  one-dimensional extended {H}ubbard model at quarter filling, Phys.\ Rev.\ B
  79 (2009) 195121.

\bibitem{Baeriswyl92a}
D.~Baeriswyl, D.~K. Campbell, S.~Mazumdar, Conjugated Conducting Polymers,
  Kiess, H., Ed.; Springer: Berlin, 1992.

\bibitem{Mazumdar83b}
S.~Mazumdar, S.~N. Dixit, Coulomb effects on one-dimensional \protect{P}eierls
  instability: the \protect{P}eierls-\protect{H}ubbard model, Phys.\ Rev.\
  Lett. 51 (1983) 292--295.

\bibitem{Hirsch83b}
J.~E. Hirsch, Effect of {C}oulomb interactions on the {P}eierls instability,
  Phys.\ Rev.\ Lett. 51 (1983) 296--299.

\bibitem{Soos84a}
Z.~G. Soos, S.~Ramasesha, Valence-bond theory of linear {H}ubbard and
  {P}ariser-{P}arr-{P}ople models, Phys.\ Rev.\ B 29 (1984) 5410--5422.

\bibitem{Mazumdar85a}
S.~Mazumdar, D.~K. Campbell, Broken symmetries in a one-dimensional half-filled
  band with arbitrarily long-range {C}oulomb interactions, Phys.\ Rev.\ Lett.
  55 (1985) 2067--2070.

\bibitem{Mazumdar87a}
S.~Mazumdar, Valence-bond approach to two-dimensional broken symmetries:
  Application to {${\mathrm{La}}_{2}$Cu${\mathrm{O}}_{4}$}, Phys. Rev. B 36
  (1987) 7190--7193.

\bibitem{Kobayashi98a}
N.~Kobayashi, M.~Ogata, K.~Yonemitsu, Coexistence of \protect{SDW} and
  purely-electronic \protect{CDW} in quarter-filled organic conductors, J.\
  Phys.\ Soc.\ Jpn. 67 (1998) 1098--1101.

\bibitem{Haldane80a}
F.~D.~M. Haldane, General relation of correlation exponents and spectral
  properties of one-dimensional {F}ermi systems: application to the anisotropic
  \protect{$S=\frac{1}{2}$} {H}eisenberg chain, Phys.\ Rev.\ Lett. 45 (1980)
  1358--1362.

\bibitem{Haldane81a}
F.~D.~M. Haldane, '{L}uttinger liquid theory' of one-dimensional quantum
  fluids: {I}. properties of the {L}uttinger model and their extension to the
  general 1{D} interacting spinless {F}ermi gas, J. Phys. C 14 (1981)
  2585--2609.

\bibitem{Voit95a}
J.~Voit, One-dimensional {F}ermi liquids, Rep. Prog. Phys. 58 (1995) 977.

\bibitem{Seo02a}
H.~Seo, M.~Kuwabara, M.~Ogata, Co-existence of charge order and spin {P}eierls
  lattice distortion in one-dimensional organic compounds, J. Phys. IV France
  12 (2002) Pr9--205.

\bibitem{Lieb62a}
E.~H. Lieb, D.~Mattis, Ordering energy levels of interacting spin systems, J.
  Math. Phys. (NY) 3 (1962) 749.

\bibitem{Mazumdar99a}
S.~Mazumdar, S.~Ramasesha, R.~T. Clay, D.~K. Campbell, Theory of coexisting
  charge- and spin-density waves in \protect{(TMTTF)$_2$Br},
  \protect{(TMTSF)$_2$PF$_6$}, and
  \protect{$\alpha$-(BEDT-TTF)$_2$MHg(SCN)$_4$}, Phys.\ Rev.\ Lett. 82 (1999)
  1522--1525.

\bibitem{Ota02a}
A.~Ota, H.~Yamochi, G.~Saito, A novel metal-insulator phase transition observed
  in {(EDO-TTF)$_{2}$PF$_{6}$}, J. Mater. Chem. 12 (2002) 2600--2602.

\bibitem{Furukawa11a}
K.~Furukawa, K.~Sugiura, F.~Iwase, T.~Nakamura, Structural investigation of the
  spin-singlet phase in \protect{(TMTTF)$_2$I}, Phys.\ Rev.\ B 83 (2011)
  184419.

\bibitem{Ducasse88a}
L.~Ducasse, C.~Coulon, D.~Chasseau, R.~Yagbasan, J.~M. Fabre, A.~K. Gousamia,
  Structural, electronic and physical properties of \protect{(BCPTTF)$_2$MF$_6$
  (M = P, As)}, occurrence of a spin-{P}eierls transition
  \protect{(T$_{\rm{SP}} \simeq 30$K)}, Synth.\ Metals 27 (1988) B543--B548.

\bibitem{Liu93a}
Q.~Liu, S.~Ravy, J.~P. Pouget, C.~Coulon, Structural fluctuations and
  spin-{P}eierls transitions revisited, Synth.\ Metals 55-57 (1993) 1840--1845.

\bibitem{Foury-Leylekian11a}
P.~Foury-Leylekian, P.~Auban-Senzier, C.~Coulon, O.~Jeannin, M.~Fourmigu\'e,
  C.~Pasquier, J.-P. Pouget, Phase diagram of the correlated
  quarter-filled-band organic salt series {(o-DMTTF)$_2$X} ({X} = {Cl}, {Br},
  {I}), Phys.\ Rev.\ B 84 (2011) 195134.

\bibitem{Auban-Senzier09a}
P.~Auban-Senzier, C.~R. Pasquier, D.~J\'erome, S.~Suh, S.~E. Brown, C.~Meziere,
  J.~P. Pouget, Phase diagram of quarter-filled band organic salts
  {[EDT-TTF-CONMe$_2$]$_2$X}, $x=$ {AsF$_6$} and {B}r, Phys.\ Rev.\ Lett. 102
  (2009) 257001.

\bibitem{Hirose10a}
S.~Hirose, A.~Kawamoto, N.~Matsunaga, K.~Nomura, K.~Yamamoto, K.~Yakushi,
  Reexamination of {$^{13}$C}-{NMR} in {(TMTTF)$_2$AsF$_6$}: Comparison with
  infrared spectroscopy, Phys.\ Rev.\ B 81 (2010) 205107.

\bibitem{Fujiyama06a}
S.~Fujiyama, T.~Nakamura, Redistribution of electronic charges in
  spin-{P}eierls state in \protect{(TMTTF)$_2$AsF$_6$} observed by
  \protect{$^{13}$C} {N}{M}{R}, J.\ Phys.\ Soc.\ Jpn. 75 (2006) 014705.

\bibitem{Nakamura07a}
T.~Nakamura, K.~Furukawa, T.~Hara, Redistribution of charge in the proximity of
  the spin-{P}eierls transition: \protect{$^{13}$C} \protect{NMR} investigation
  of \protect{(TMTTF)$_2$PF$_6$}, J.\ Phys.\ Soc.\ Jpn. 76 (2007) 064715.

\bibitem{Iwase11a}
F.~Iwase, K.~Sugiura, K.~Furukawa, T.~Nakamura, {$^{13}$C} {NMR} study of the
  magnetic properties of the quasi-one-dimensional conductor
  {(TMTTF)$_2$SbF$_6$}, Phys.\ Rev.\ B 84 (2011) 115140.

\bibitem{Iwase09a}
F.~Iwase, K.~Sugiura, K.~Furukawa, T.~Nakamura, Electronic properties of a
  {TMTTF}-family salt, {(TMTTF)$_2$TaF$_6$}: New member located on the modified
  generalized phase-diagram, J.\ Phys.\ Soc.\ Jpn. 78~(10) (2009) 104717.

\bibitem{Javadi88a}
H.~H.~S. Javadi, R.~Laversanne, A.~J. Epstein, Microwave conductivity and
  dielectric constant of tetramethyltetrathiafulvalene salts
  [{(TMTTF${)}_{2}$X}, {X}={SCN}, {ReO$_4$}, {SbF$_6$}], Phys. Rev. B 37 (1988)
  4280--4283.

\bibitem{Itoi08a}
M.~Itoi, C.~Araki, M.~Hedo, Y.~Uwatoko, T.~Nakamura, Anomalously wide
  superconducting phase of one-dimensional organic conductor
  {(TMTTF)$_2$SbF$_6$}, J.\ Phys.\ Soc.\ Jpn. 77 (2008) 023701.

\bibitem{Coulon85a}
C.~Coulon, S.~S.~P. Parkin, R.~Laversanne, Structureless transition and strong
  localization effects in bis-tetramethyltetrathiafulvalenium salts
  {[(TMTTF)$_2$X]}, Phys.\ Rev.\ B 31 (1985) 3583.

\bibitem{Monceau12a}
P.~Monceau, Electronic crystals: an experimental overview, Adv. Phys. 61 (2012)
  325.

\bibitem{Yu04a}
W.~Yu, F.~Zhang, F.~Zamborszky, B.~Alavi, A.~Baur, C.~A. Merlic, S.~E. Brown,
  Electron-lattice coupling and broken symmetries of the molecular salt
  \protect{(TMTTF)$_2$SbF$_6$}, Phys.\ Rev.\ B 70 (2004) 121101.

\bibitem{Chow00a}
D.~S. Chow, F.~Zamborszky, B.~Alavi, D.~J. Tantillo, A.~Baur, C.~A. Merlic,
  S.~E. Brown, Charge ordering in the \protect{TMTTF} family of molecular
  conductors, Phys.\ Rev.\ Lett. 85 (2000) 1698--1701.

\bibitem{Zamborszky02a}
F.~Zamborszky, W.~Yu, W.~Raas, S.~E. Brown, B.~Alavi, C.~A. Merlic, A.~Baur,
  Competition and coexistence of bond and charge orders in
  \protect{(TMTTF)$_2$AsF$_6$}, Phys.\ Rev.\ B 66 (2002) 081103.

\bibitem{Nakamura95a}
T.~Nakamura, T.~Nobutoki, Y.~Kobayashi, T.~Takahashi, G.~Saito, {$^1$H}-{NMR}
  investigation of the {SDW} wave-number in {(TMTTF)$_2$Br}, Synth. Metals 70
  (1995) 1293--1294.

\bibitem{Pouget97a}
J.~P. Pouget, S.~Ravy, X-ray evidence of charge density wave modulations in the
  magnetic phases of \protect{(TMTSF)$_2$PF$_6$} and \protect{(TMTTF)$_2$Br},
  Synth.\ Metals 85 (1997) 1523.

\bibitem{Coulon82a}
C.~Coulon, A.~Maaroufi, J.~Amiell, E.~Dupart, S.~Flandrois, P.~Delhaes,
  R.~Moret, J.~P. Pouget, J.~P. Morand, Antiferromagnetic and structural
  instabilities in tetramethyltetrathiafulvalene thiocyanate
  {[(TMTTF)$_{2}\mathrm{S}$CN]}, Phys. Rev. B 26 (1982) 6322--6325.

\bibitem{Peo84a}
M.~Peo, J.~C. Scott, E.~M. Engler, Proton {NMR} linewidth and relaxation-rate
  study of an organic conductor with an antiferromagnetic ground state, Phys.
  Rev. B 30 (1984) 3639--3643.

\bibitem{Nad06a}
F.~Nad, P.~Monceau, Dielectric response of the charge ordered state in
  quasi-one-dimensional organic conductors, J.\ Phys.\ Soc.\ Jpn. 75 (2006)
  051005.

\bibitem{Kohler11a}
B.~\protect{K\"ohler}, E.~Rose, M.~Dumm, G.~Untereiner, M.~Dressel,
  Comprehensive transport study of anisotropy and ordering phenomena in
  quasi-one-dimensional {(TMTTF)$_2$X} salts ({X=PF$_6$},
  {AsF$_6$},{SbF$_6$},{BF$_4$},{ClO$_4$},{ReO$_4$}), Phys.\ Rev.\ B 84 (2011)
  035124.

\bibitem{Souza08a}
M.~de~Souza, P.~Foury-Leylekian, A.~Moradpour, J.-P. Pouget, M.~Lang, Evidence
  for lattice effects at the charge-ordering transition in
  \protect{(TMTTF)$_2$X}, Phys.\ Rev.\ Lett. 101 (2008) 216403.

\bibitem{Monceau01a}
P.~Monceau, F.~Y. Nad, S.~Brazovskii, Ferroelectric {M}ott-{H}ubbard phase of
  organic {(TMTTF)$_2${X}} conductors, Phys.\ Rev.\ Lett. 86 (2001) 4080.

\bibitem{Nad05a}
F.~Nad, P.~Monceau, T.~Nakamura, K.~Furukawa, The effect of deuteration on the
  transition into a charge ordered state of \protect{(TMTTF)$_2$X} salts, J.
  Phys.: Condens. Matter 17 (2005) L399--L406.

\bibitem{Furukawa05a}
K.~Furukawa, T.~Hara, T.~Nakamura, Deuteration effect and possible origin of
  the charge-ordering transition of \protect{(TMTTF)$_2$X}, J.\ Phys.\ Soc.\
  Jpn. 74 (2005) 3288--3294.

\bibitem{Foury-Leylekian04a}
P.~Foury-Leylekian, D.~L. Bolloc'h, B.~Hennion, S.~Ravy, A.~Moradpour, J.-P.
  Pouget, Neutron-scattering evidence for a spin-{P}eierls ground state in
  \protect{(TMTTF)$_2$PF$_6$}, Phys.\ Rev.\ B 70~(18) (2004) 180405.

\bibitem{Voloshenko17a}
I.~Voloshenko, M.~Herter, R.~Breyer, A.~Pustogow, M.~Dressel,
  Pressure-dependent optical investigations of {F}abre salts in the
  charge-ordered state, J. Phys.: Condens. Matter 29 (2017) 115601.

\bibitem{Swietlik17a}
R.~\protect{\'Swietlik}, B.~\protect{Barszcz}, A.~Pustogow, M.~Dressel, Raman
  spectroscopy evidence of domain walls in the organic ferroelectrics
  \protect{(TMTTF)$_2$X} ({X}=\protect{SbF$_6$}, \protect{AsF$_6$},
  \protect{PF$_6$}), Phys.\ Rev.\ B 95 (2017) 085205.

\bibitem{Nad98a}
F.~Nad, P.~Monceau, J.~M. Fabre, Low frequency dielectric permittivity of
  quasi-one-dimensional conductor {(TMTTF)$_2$Br}, Eur. Phys. J. B 3 (1998)
  301--306.

\bibitem{Barthel93a}
E.~Barthel, G.~Quirion, P.~Wzietek, D.~{J\'erome}, J.~B. Christensen,
  M.~Jorgensen, K.~Bechgaard, {NMR} in commensurate and incommensurate spin
  density waves, Eur.\ Phys.\ Lett. 21 (1993) 87.

\bibitem{Pouget96a}
J.~P. Pouget, S.~Ravy, Structural aspects of the {B}echgaard salts and related
  compounds, J. Physique I 6 (1996) 1501.

\bibitem{Dumm00b}
M.~Dumm, A.~Loidl, B.~W. Fravel, K.~P. Starkey, L.~K. Montgomery, M.~Dressel,
  Electron spin resonance studies on the organic linear-chain compounds
  \protect{(TMTCF)$_2$X} \protect{(C=S, Se; X=PF$_6$, AsF$_6$, ClO$_4$, Br)},
  Phys.\ Rev.\ B 61 (2000) 511.

\bibitem{Zorko01a}
A.~Zorko, D.~Ar\ifmmode~\check{c}\else \v{c}\fi{}on,
  K.~Biljakovi\ifmmode~\acute{c}\else \'{c}\fi{}, C.~Carcel, J.~M. Fabre,
  J.~Dolin\ifmmode~\check{s}\else \v{s}\fi{}ek, Spin-{P}eierls fluctuations in
  {${(\mathrm{TMTTF})}_{2}\mathrm{Br}$} studied by pulsed electron spin
  resonance spin-lattice relaxation, Phys. Rev. B 64 (2001) 172404.

\bibitem{Jerome82a}
D.~J\'erome, H.~J. Schulz, Organic conductors and superconductors, Adv. Physics
  31 (1982) 299.

\bibitem{Takahashi86a}
T.~Takahashi, Y.~Maniwa, H.~Kawamura, G.~Saito, Determination of {SDW}
  characteristics in {(TMTSF)$_2$PF$_6$} by {$^1$H}--{NMR} analysis, J.\ Phys.\
  Soc.\ Jpn. 55 (1986) 1364--1373.

\bibitem{Pouget82a}
J.~Pouget, R.~Moret, R.~\protect{Com\'es}, K.~Bechgaard, J.~Fabre, L.~Giral,
  X-ray diffuse scattering study of some \protect{(TMTSF)$_2$X} and
  \protect{(TMTTF)$_2$X} salts, Mol. Cryst. Liq. Cryst. 79 (1982) 485--499.

\bibitem{Gruner94a}
G.~Gr\"uner, The dynamics of spin-density waves, Rev. Mod. Phys. 66 (1994)
  1--24.

\bibitem{Brown92a}
S.~E. Brown, B.~Alavi, G.~Gruner, K.~Bartholomew, Softening of {Y}oung's
  modulus and collective spin-density-wave transport in {(TMTSF)$_2$PF$_6$}
  (where {TMTSF} is tetramethyltetraselenafulvalene), Phys. Rev. B 46 (1992)
  10483--10486.

\bibitem{Odin94a}
J.~Odin, J.~C. Lasjaunias, K.~Biljakovic, P.~Monceau, K.~Bechgaard, Low
  temperature specific heat of the spin-density-wave compound
  {(TMTSF)$_2$PF$_6$}, Solid St.\ Comm. 91 (1994) 523--527.

\bibitem{Yang99a}
H.~Yang, J.~C. Lasjaunias, P.~Monceau, Specific heat measurements of the
  lattice contribution and spin-density-wave transition in {(TMTSF)$_2$X} (x =
  {PF$_6$} and {AsF$_6$}) and {(TMTTF)$_2$Br} salts, J. Phys.:Condens. Matter
  11 (1999) 5083.

\bibitem{Kagoshima99a}
S.~Kagoshima, Y.~Saso, M.~Maesato, R.~Kondo, Low-temperature diffuse
  \protect{X-ray} studies of charge-density waves coexisting with spin-density
  waves in the organic conductors \protect{(TMTSF)$_2$PF$_6$} and
  \protect{(TMTSF)$_2$AsF$_6$}, Solid St.\ Comm. 110 (1999) 479--483.

\bibitem{Kagoshima01a}
S.~Kagoshima, Y.~Saso, M.~Maesato, R.~Kondo, T.~Yamaguchi, V.~A. Bondarenko,
  T.~Hasegawa, Electronic states in the spin-density wave phase of organic
  conductors: roles of the coexisting 2{$k_F$} and 4{$k_F$} charge-density
  waves, Synth. Metals 117 (2001) 39--43.

\bibitem{Valfells97a}
S.~Valfells, P.~Kuhns, A.~Kleinhammes, J.~S. Brooks, W.~Moulton, S.~Takasaki,
  J.~Yamada, H.~Anzai, Spin-density-wave state in
  {${(\mathrm{TMTSF})}_{2}{\mathrm{PF}}_{6}$}: {A} {${}^{77}$Se} {NMR} study at
  high magnetic fields, Phys. Rev. B 56 (1997) 2585--2593.

\bibitem{Nomura93a}
K.~Nomura, N.~Keitoku, T.~Shimizu, T.~Sanbongi, M.~Tokumoto, N.~Kinoshita,
  H.~Anzai, {NMR} in spin-density wave phase of {(TMTSF)$_2$ClO$_4$}, J. de
  Phys. IV 3 (1993) C2--C21.

\bibitem{Hoshikawa00a}
A.~Hoshikawa, K.~Nomura, S.~Takasaki, J.~Yamada, S.~Nakatsuji, H.~Anzai,
  M.~Tokumoto, N.~Kinoshita, Collective mode of spin-density-wave in
  {(TMTSF)$_2$ClO$_4$}, J.\ Phys.\ Soc.\ Jpn. 69 (2000) 1457--1461.

\bibitem{Lasjaunias94a}
J.~C. Lasjaunias, K.~{Biljakovi\ifmmode \acute{c}\else \'{c}\fi{}}, F.~{Nad'},
  P.~Monceau, K.~Bechgaard, Glass transition in the spin-density wave phase of
  {(TMTSF)$_2$PF$_6$}, Phys. Rev. Lett. 72 (1994) 1283--1286.

\bibitem{Nagata13a}
S.~Nagata, M.~Misawa, Y.~Ihara, A.~Kawamoto, Commensurability of the
  spin-density-wave state of {(TMTSF)$_2$PF$_6$} observed by
  {$^{13}\mathbf{C}$}-{NMR}, Phys. Rev. Lett. 110 (2013) 167001.

\bibitem{Galigne80a}
J.~L. Galigne, S.~Peytavin, B.~Litautard, G.~Brun,
  Tetramethyltetrathiofulvalene iodide, \protect{(C$_{10}$H$_{12}$S$_4$)$_2$I},
  Cryst. Struct. Commun. 9 (1980) 61--63.

\bibitem{Jankowski11a}
D.~Jankowski, R.~Swietlik, E.~W. Reinheimer, M.~Fourmigu\'e, Involvement of
  weak {C---H$\cdots$X} hydrogen bonds in metal-to-semiconductor regime change
  in one-dimensional organic conductors {(o-DMTTF)$_2$X} ({X}= {Cl}, {Br}, and
  {I}): combined {IR} and {R}aman studies, J. Raman Spectros. 42 (2011) 1518.

\bibitem{Zorina09a}
L.~Zorina, S.~Simonov, C.~Meziere, E.~Canadell, S.~Suh, S.~E. Brown,
  P.~Foury-Leylekian, P.~Fertey, J.-P. Pouget, P.~Batail, Charge ordering,
  symmetry and electronic structure issues and {W}igner crystal structure of
  the quarter-filled band {M}ott insulators and high pressure metals
  {$\delta$-(EDT-TTF-CONMe$_2$)$_2$X}, {X} = {Br} and {AsF$_6$}, J. Mater.
  Chem. 19 (2009) 6980.

\bibitem{Heuze03a}
K.~\protect{Heuz\'e}, M.~Fourmigue, P.~Batail, C.~Coulon, R.~Cl'erac,
  E.~Canadell, P.~Auban-Senzier, S.~Ravy, D.~J'erome, A genuine quarter-filled
  band {M}ott insulator, {(EDT-TTF-CONMe$_2$)$_2$AsF$_6$}: {W}here the
  chemistry and physics of weak intermolecular interactions act in unision,
  Adv. Mater. 15 (2003) 1251.

\bibitem{Song14a}
J.~P. Song, R.~Clay, Monte {C}arlo simulations of two-dimensional fermion
  systems with string-bond states, Phys.\ Rev.\ B 89 (2014) 075101.

\bibitem{Yamochi09a}
H.~Yamochi, S.~Koshihara, Organic metal \protect{(EDO-TTF)$_2$PF$_6$} with
  multi-instability, Sci. Technol. Adv. Mater 10 (2009) 024305.

\bibitem{Chollet05a}
M.~Chollet, L.~Guerin, N.~Uchida, S.~Fukaya, H.~Shimoda, T.~Ishikawa,
  K.~Matsuda, T.~Hasegawa, A.~Ota, H.~Yamochi, G.~Saito, R.~Tazaki, S.~Adachi,
  S.~Koshihara, Gigantic photoresponse in {$\frac{1}{4}$}-filled-band organic
  salt {(EDO-TTF)$_2$PF$_6$}, Science 307 (2005) 86--89.

\bibitem{Drozdova04a}
O.~Drozdova, K.~Yakushi, K.~Yamamoto, A.~Ota, H.~Hideki, Y.~Yamochi, G.~Saito,
  H.~Tashiro, D.~B. Tanner, Optical characterization of \protect{2k$_F$}
  bond-charge-density wave in quasi-one-dimensional
  \protect{$\frac{3}{4}$}-filled \protect{(EDO-TTF)$_2$X} ({X} =
  \protect{PF$_6$} and \protect{AsF$_6$}), Phys.\ Rev.\ B 70 (2004) 075107.

\bibitem{Aoyagi04a}
S.~Aoyagi, K.~Kato, A.~Ota, H.~Yamochi, G.~saito, H.~Suematsu, M.~Sakata,
  M.~Takata, Direct observation of bonding and charge ordering in
  {(EDO-TTF)$_2$PF$_6$}, Angew. Chem. Int. Ed. 116 (2004) 3756.

\bibitem{Iwano08a}
K.~Iwano, Y.~Shimoi, Large electric-potential bias in an {EDO-TTF} tetramer as
  a major mechanism of charge ordering observed in its {PF$_6$} salt: A density
  functional theory study, Phys.\ Rev.\ B 77 (2008) 075120.

\bibitem{Onda08a}
K.~Onda, S.~Ogihara, K.~Yonemitsu, N.~Maeshima, T.~Ishikawa, Y.~Okimoto,
  X.~Shao, Y.~Nakano, H.~Yamochi, G.~Saito, S.~Koshihara, Photoinduced change
  in the charge order pattern in the quarter-filled organic conductor
  {(EDO−TTF)$_2$PF$_6$} with a strong electron-phonon interaction, Phys.\
  Rev.\ Lett. 101 (2008) 067403.

\bibitem{Fukuzawa12a}
N.~Fukazawa, M.~Shimizu, T.~Ishikawa, Y.~Okimoto, S.~Koshihara, T.~Hiramatsu,
  Y.~Nakano, H.~Yamochi, G.~Saito, K.~Onda, Charge and structural dynamics in
  photoinduced phase transition of {(EDO-TTF)$_2$PF$_6$} examined by picosecond
  time-resolved vibrational spectroscopy, J. Phys. Chem. C 116 (2012)
  5892−5899.

\bibitem{Iwano13a}
K.~Iwano, Y.~Shimoi, Revealing the photorelaxation mechanism in a molecular
  solid using density-functional theory, Phys.\ Rev.\ Lett. 110 (2013) 116401.

\bibitem{Yonemitsu08a}
K.~Yonemitsu, K.~Nasu, Theory of photoinduced phase transitions in itinerant
  electron systems, Phys. Reports 465 (2008) 1--60.

\bibitem{Taniguchi03b}
H.~Taniguchi, M.~Miyashita, K.~Uchiyama, K.~Satoh, N.~Mori, H.~Okamoto,
  K.~Miyagawa, K.~Kanoda, M.~Hedo, Y.~Uwatoko, Superconductivity at 14.2 {K} in
  layered organics under extreme pressure, J.\ Phys.\ Soc.\ Jpn. 72 (2003)
  469--471.

\bibitem{Kanoda11a}
K.~Kanoda, R.~Kato, Mott physics in organic conductors with triangular
  lattices, Annu. Rev. Condens. Matter Phys. 2 (2011) 167--188.

\bibitem{Kandpal09a}
H.~C. Kandpal, I.~Opahle, Y.-Z. Zhang, H.~O. Jeschke, R.~Valent\'\i, Revision
  of model parameters for $\kappa$-type charge-transfer solids: An ab initio
  study, Phys.\ Rev.\ Lett. 103 (2009) 067004.

\bibitem{Nakamura09a}
K.~Nakamura, Y.~Yoshimoto, T.~Kosugi, R.~Arita, M.~Imada, \textit{Ab initio}
  derivation of low-energy model for $\kappa$-{ET} type organic conductors, J.\
  Phys.\ Soc.\ Jpn. 78 (2009) 083710.

\bibitem{Guterding16a}
D.~Guterding, M.~Altmeyer, H.~O. Jeschke, R.~\protect{Valent\'i},
  Near-degeneracy of extended \protect{$s$+$d_{x^2-y^2}$} and
  \protect{$d_{xy}$} order parameters in quasi-two-dimensional organic
  superconductors, Phys.\ Rev.\ B 94 (2016) 024515.

\bibitem{Kobayashi87a}
A.~Kobayashi, R.~Kato, H.~Kobayashi, S.~Moriyama, Y.~Nishio, K.~Kajita,
  W.~Sasaki, Crystal and electronic structures of a new molecular
  superconductor, \protect{$\kappa$-(BEDT-TTF)$_2$I$_3$}, Chem. Lett. 16 (1987)
  459--462.

\bibitem{Hiramatsu15a}
T.~Hiramatsu, Y.~Yoshida, G.~Saito, A.~Otsuka, H.~Yamochi, M.~Maesato,
  Y.~Shimizu, H.~Ito, H.~Kishida, Quantum spin liquid: design of a quantum spin
  liquid next to a superconducting state based on a dimer-type {E}{T} {M}ott
  insulator, J. Mater. Chem. C 2015 (2015) 1378.

\bibitem{Koretsune14a}
T.~Koretsune, C.~Hotta, Evaluating model parameters of the \protect{$\kappa$}
  and \protect{$\beta^\prime$}-type {M}ott insulating organic solids, Phys.\
  Rev.\ B 89 (2014) 045102.

\bibitem{Yoshida15a}
Y.~Yoshida, H.~Ito, M.~Maesato, Y.~Shimizu, H.~Hayama, T.~Hiramatsu,
  Y.~Nakamura, H.~Kishida, T.~Koretsune, C.~Hotta, G.~Saito, Spin-disordered
  quantum phases in a quasi-one-dimensional triangular lattice, Nat. Phys. 11
  (2015) 679--684.

\bibitem{Miyagawa02a}
K.~Miyagawa, A.~Kawamoto, K.~Kanoda, Proximity of pseudogapped superconductor
  and commensurate antiferromagnet in a quasi-two-dimensional organic system,
  Phys.\ Rev.\ Lett. 89 (2002) 017003.

\bibitem{Wang91a}
H.~H. Wang, K.~D. Carlson, U.~Geiser, A.~M. Kini, A.~J. Schultz, J.~M.
  Williams, L.~K. Montgomery, W.~K. Kwok, U.~Welp, K.~G. Vandervoort, S.~J.
  Boryschuk, A.~V.~S. Crouch, J.~M. Kommers, D.~M. Watkins, New $\kappa$-phase
  materials, \protect{$\kappa$-(ET)$_2$Cu[N(CN)$_2$]X}, \protect{X=Cl, Br, and
  I.} {T}he synthesis, structure and superconductivity above 11 {K} in the {C}l
  \protect{(T$_c$=12.8 K, 0.3 kbar)} and {B}r \protect{(T$_c$=11.6 {K})} salts,
  Synth.\ Metals 41--43 (1991) 1983--1990.

\bibitem{Kushch01a}
N.~D. Kushch, M.~A. Tanatar, E.~B. Yagubskii, T.~Ishiguro, Superconductivity of
  \protect{$\kappa$-(BEDT-TTF)$_2$Cu[N(CN)$_2$]I} under pressure, JETP Lett. 73
  (2001) 429.

\bibitem{Mori90a}
H.~Mori, I.~Hirabayashi, S.~Tanaka, T.~Mori, H.~Inokuchi, A new
  ambient-pressure organic superconductor,
  \protect{$\kappa$-(BEDT-TTF)$_2$Ag(CN)$_2$H$_2$O} \protect{(T$_c$=5.0K)},
  Solid St.\ Comm. 76 (1990) 35--37.

\bibitem{Mori99a}
T.~Mori, H.~Mori, S.~Tanaka, Structural genealogy of \protect{BEDT-TTF}-based
  organic conductors {I}{I}. {I}nclined: {$\theta$}, {$\alpha$}, {$\kappa$}
  phases, Bull. Chem. Soc. Jpn. 72 (1999) 179--197.

\bibitem{Komatsu91a}
T.~Komatsu, T.~Nakamura, N.~Matsukawa, H.~Yamochi, G.~Saito, H.~Ito,
  T.~Ishiguro, M.~Kusunoki, K.~Sakaguchi, New ambient-pressure organic
  superconductors based on \protect{BEDT-TTF}, {C}u, \protect{N(CN)$_2$}, and
  \protect{CN} with \protect{T$_c$}=10.7{K} and 3.8{K}, Solid St.\ Comm. 80
  (1991) 843.

\bibitem{Yasin12a}
S.~Yasin, E.~Rose, M.~Dumm, N.~Drichko, M.~Dressel, J.~A. Schlueter, E.~I.
  Zhilyaeva, S.~Torunova, R.~N. Lyubovskaya, Electronic and magnetic studies of
  \protect{$\kappa$-(BEDT-TTF)$_2$Hg(SCN)$_2$Cl}, Physica B 407 (2012) 1689.

\bibitem{Drichko14a}
N.~Drichko, R.~Beyer, E.~Rose, M.~Dressel, J.~A. Schlueter, S.~A. Turunova,
  E.~I. Zhilyaeva, R.~N. Lyubovskaya, Metallic state and charge-order
  metal-insulator transition in the quasi-two-dimensional conductor
  {$\kappa$}-{(BEDT-TTF)$_2$Hg(SCN)$_2$Cl}, Phys.\ Rev.\ B 89 (2014) 075133.

\bibitem{Lohle17a}
A.~\protect{L\"ohle}, E.~Rose, S.~Singh, R.~Beyer, E.~Tafra, E.~I. Zhilyaeva,
  R.~N. Lyubovskaya, M.~Dressel, Pressure dependence of the metal-insulator
  transition in \protect{$\kappa$-(BEDT-TTF)$_2$Hg(SCN)$_2$Cl}: optical and
  transport studies, J. Phys.: Condens. Matter 29 (2017) 055601.

\bibitem{Shimizu16a}
Y.~Shimizu, T.~Hiramatsu, M.~Maesato, A.~Otsuka, H.~Yamochi, A.~Ono, M.~Itoh,
  M.~Yoshida, M.~Takigawa, Y.~Yoshida, G.~Saito, Pressure-tuned exchange
  coupling of a quantum spin liquid in the molecular triangular lattice
  \protect{$\kappa$-(ET)$_2$Ag$_2$(CN)$_3$}, Phys. Rev. Lett. 117 (2016)
  107203.

\bibitem{Shimizu11a}
Y.~Shimizu, M.~Maesato, G.~Saito, Uniaxial strain effects on {M}ott and
  superconducting transitions in \protect{$\kappa$-(ET)$_2$Cu$_2$(CN)$_3$}, J.\
  Phys.\ Soc.\ Jpn. 80 (2011) 074702.

\bibitem{Fettouhi95a}
M.~Fettouhi, L.~Ouahab, C.~Gomez-Garcia, L.~Ducasse, P.~Delhaes, Structural and
  physical properties of \protect{$\kappa$-(BEDT-TTF)$_2$(CF$_3$SO$_3$)},
  Synth.\ Metals 70 (1995) 1131--1132.

\bibitem{Ito16a}
H.~Ito, T.~Asai, Y.~Shimizu, H.~Hayama, Y.~Yoshida, G.~Saito, Pressure-induced
  superconductivity in the antiferromagnet
  \protect{$\kappa$-(ET)$_2$CF$_2$(SO)$_3$}, Phys. Rev. B 94 (2016) 020503.

\bibitem{Miyagawa95a}
K.~Miyagawa, A.~Kawamoto, Y.~Nakazawa, K.~Kanoda, Antiferromagnetic ordering
  and spin structure in the organic conductor,
  \protect{$\kappa$-(BEDT-TTF)$_2$Cu[N(CN)$_2$]Cl}, Phys.\ Rev.\ Lett. 75
  (1995) 1174--1177.

\bibitem{Huse88a}
D.~A. Huse, V.~Elser, Simple variational wave functions for two-dimensional
  {H}eisenberg $\frac{1}{2}$ antiferromagnets, Phys. Rev. Lett. 60~(24) (1988)
  2531--2534.

\bibitem{Miyake92a}
S.~J. Miyake, Spin-wave results for the staggered magnetization of triangular
  {H}eisenberg antiferromagnet, J.\ Phys.\ Soc.\ Jpn. 61 (1992) 983--988.

\bibitem{Bernu92a}
B.~Bernu, C.~Lhuillier, L.~Pierre, Signature of \protect{N\'eel} order in exact
  spectra of quantum antiferromagnets on finite lattices, Phys.\ Rev.\ Lett. 69
  (1992) 2590.

\bibitem{Elstner93a}
N.~Elstner, R.~R.~P. Singh, A.~P. Young, Finite temperature properties of the
  \protect{spin-$\frac{1}{2}$} \protect{Heisenberg} antiferromagnet on the
  triangular lattice, Phys.\ Rev.\ Lett. 71 (1993) 1629--1623.

\bibitem{Capriotti99a}
L.~Capriotti, A.~E. Trumper, A.~Sorella, Long-range \protect{N\'eel} order in
  the triangular {H}eisenberg model, Phys.\ Rev.\ Lett. 82 (1999) 3899.

\bibitem{Ito96a}
H.~Ito, T.~Ishiguro, M.~Kobota, G.~Saito, Metal-nonmetal transition and
  superconductivity localization in the two-dimensional conductor
  \protect{$\kappa$-(BEDT-TTF)$_2$Cu[N(CN)$_2$]Cl} under pressure, J.\ Phys.\
  Soc.\ Jpn. 65 (1996) 2987--2993.

\bibitem{Welp92a}
U.~Welp, S.~Fleshler, W.~Kwok, G.~W. Crabtree, K.~D. Carlson, H.~H. Wang,
  U.~Geiser, J.~M. Williams, V.~M. Hitsman, Weak ferromagnetism in
  \protect{$\kappa$-(ET)$_2$Cu[N(CN)$_2$]Cl}, where \protect{(ET)} is
  bis(ethylenedithio)tetrathiafulvalene, Phys.\ Rev.\ Lett. 69 (1992) 840--843.

\bibitem{Williams90a}
J.~M. Williams, A.~M. Kini, H.~H. Wang, K.~D. Carlson, U.~Geiser, L.~K.
  Montgomery, G.~J. Pyrka, D.~M. Watkins, J.~M. Kommers, From
  semiconductor-semiconductor transition (42 {K}) to the
  highest-\protect{T$_c$} organic superconductor,
  \protect{$\kappa$-(ET)$_2$Cu[N(CN)$_2$]Cl} (\protect{T$_c$} = 12.5 {K}),
  Inorg. Chem. 29 (1990) 3272--3274.

\bibitem{Kobayashi91a}
H.~Kobayashi, K.~Bun, A.~Miyamoto, T.~Naito, R.~Kato, A.~Kobayashi, J.~M.
  Williams, Superconducting transition of a grease-coated crystal of
  \protect{$\kappa$-(BEDT-TTF)$_2$Cu[N(CN)$_2$]Cl}, Chem. Lett. 20 (1991)
  1997--2000.

\bibitem{Dressel95a}
M.~Dressel, G.~Gruner, K.~D. Carlson, H.~H. Wang, Studies of the microwave
  resistivity of \protect{$\kappa$-(BEDT-TTF)$_2$Cu[N(CN)$_2$]Cl}, Synth.\
  Metals 70 (1995) 927--928.

\bibitem{Sushko93a}
Y.~V. Sushko, H.~Ito, T.~Ishiguro, S.~Horiuchi, G.~Saito, Reentrant
  superconductivity in \protect{$\kappa$-(BEDT-TTF)$_2$Cu[N(CN)$_2$]Cl} and its
  pressure phase diagram, Solid St.\ Comm. 87 (1993) 997--1000.

\bibitem{Lefebvre00a}
S.~Lefebvre, P.~Wzietek, S.~Brown, C.~Bourbonnais, D.~\protect{J\'{e}rome},
  C.~\protect{M\`{e}zi\'{e}re}, M.~\protect{Fourmigu\'{e}}, P.~Batail, Mott
  transition, antiferromagnetism, and unconventional superconductivity in
  layered organic superconductors, Phys.\ Rev.\ Lett. 85 (2000) 5420--5423.

\bibitem{Tanatar00a}
M.~A. Tanatar, S.~Kagoshima, T.~Ishiguro, H.~Ito, V.~S. Yefanov, V.~A.
  Bondarenko, N.~D. Kushch, E.~B. Yagubskii, Electronic transport properties
  and structural transformations of
  \protect{$\kappa$-(BEDT-TTF)$_2$Cu[N(CN)$_2$]I}, Phys.\ Rev.\ B 62 (2000)
  15561--15568.

\bibitem{Tanatar02a}
M.~A. Tanatar, T.~Ishiguro, S.~Kagoshima, N.~D. Kushch, E.~B. Yagubskii,
  Pressure-temperature phase diagram of the organic superconductor
  \protect{$\kappa$-(BEDT-TTF)$_2$Cu[N(CN)$_2$]I}, Phys.\ Rev.\ B 65 (2002)
  064516.

\bibitem{Kawamoto97a}
A.~Kawamoto, K.~Miyagawa, K.~Kanoda, Deuterated
  \protect{$\kappa$-(BEDT-TTF)$_2$Cu[N(CN)$_2$]Br}: a system on the border of
  the superconductor-magnetic-insulator transition, Phys.\ Rev.\ B 55 (1997)
  14140--14143.

\bibitem{Taniguchi99a}
H.~Taniguchi, A.~Kawamoto, K.~Kanoda, Superconductor-insulator phase
  transformation of partially deuterated
  \protect{$\kappa$-(BEDT-TTF)$_2$Cu[N(CN)$_2$]Br} by control of the cooling
  rate, Phys.\ Rev.\ B 59 (1999) 8424--8427.

\bibitem{Nakazawa00a}
Y.~Nakazawa, H.~Taniguchi, A.~Kawamoto, K.~Kanoda, Electronic specific heat at
  the boundary region of the metal-insulator transition in the two-dimensional
  electronic system of \protect{$\kappa$-(BEDT-TTF)$_2$Cu[N(CN)$_2$]Br}, Phys.\
  Rev.\ B 61 (2000) R16295.

\bibitem{Yoneyama04a}
N.~Yoneyama, T.~Sasaki, N.~Kobayashi, Substitution effect by deuterated donors
  on superconductivity in \protect{$\kappa$-(BEDT-TTF)$_2$Cu[N(CM)$_2$]Br}, J.\
  Phys.\ Soc.\ Jpn. 73 (2004) 1434--1437.

\bibitem{Sasaki05a}
T.~Sasaki, N.~Yoneyama, A.~Suzuki, N.~Kobayashi, Y.~Ikemoto, H.~Kimura, Real
  space imaging of the metal-insulator phase separation in the band width
  controlled organic {M}ott system
  \protect{$\kappa$-(BEDT-TTF)$_2$Cu[N(CN)$_2$]Br}, J.\ Phys.\ Soc.\ Jpn. 74
  (2005) 2351--2360.

\bibitem{Kurosaki05a}
Y.~Kurosaki, Y.~Shimizu, K.~Miyagawa, K.~Kanoda, G.~Saito, {M}ott transition
  from a spin liquid to a {F}ermi liquid in the spin-frustrated organic
  conductor $\kappa$-({ET})$_2${C}u$_2$({CN})$_3$, Phys. Rev. Lett. 95 (2005)
  177001.

\bibitem{Fournier07a}
D.~Fournier, M.~Poirier, K.~D. Truong, Competition between magnetism and
  superconductivity in the organic metal
  \protect{$\kappa$-[BEDT-TTF]$_2$Cu[N(CN)$_2$]Br}, Phys.\ Rev.\ B 76 (2007)
  054509.

\bibitem{Sasaki04a}
T.~Sasaki, N.~Yoneyama, A.~Suzuki, N.~Kobayashi, Y.~Ikemoto, H.~Kimura, Imaging
  phase separation near the {M}ott boundary of the correlated organic
  superconductors \protect{$\kappa$-(BEDT-TTF)$_2$X}, Phys.\ Rev.\ Lett. 92
  (2004) 227001.

\bibitem{Gezo13a}
J.~Gezo, T.-K. Lui, B.~Wolin, C.~P. Slichter, R.~Giannetta, Stretched
  exponential spin relaxation in organic superconductors, Phys.\ Rev.\ B 88
  (2013) 140504(R).

\bibitem{Nakazawa13a}
Y.~Nakazawa, S.~Yamashita, Thermodynamics of a liquid-like spin state in
  molecule-based magnets with goemetric frustrations, Chem. Lett. 42 (2013)
  1446--1454.

\bibitem{Shimizu05a}
Y.~Shimizu, Y.~Kurosaki, K.~Miyagawa, K.~Kanoda, M.~Maesato, G.~Saito,
  \protect{NMR} study of the spin-liquid state and {M}ott transition in the
  spin-frustrated organic system \protect{$\kappa$-(ET)$_2$Cu$_2$(CN)$_3$},
  Synth.\ Metals 152 (2005) 393--396.

\bibitem{Shimizu03a}
Y.~Shimizu, K.~Miyagawa, K.~Kanoda, M.~Maesato, G.~Saito, Spin liquid state in
  an organic {M}ott insulator with a triangular lattice, Phys.\ Rev.\ Lett. 91
  (2003) 107001.

\bibitem{Geiser91a}
U.~Geiser, H.~H. Wang, K.~D. Carlson, J.~M. Williams, H.~A. Charlier, J.~E.
  Heindl, G.~A. Yaconi, B.~J. Love, M.~W. Lathrop, Superconductivity at 2.8 {K}
  and 1.5 kbar in \protect{$\kappa$-(BEDT-TTF)$_2$Cu$_2$(CN)$_3$}: the first
  organic superconductor containing a polymeric copper cyanide anion, Inorg.
  Chem. 30 (1991) 2586--2588.

\bibitem{Komatsu96a}
T.~Komatsu, N.~Matsukawa, T.~Inoue, G.~Saito, Realization of superconductivity
  at ambient pressure by band-filling control in
  \protect{$\kappa$-(BEDT-TTF)$_2$Cu$_2$(CN)$_3$}, J.\ Phys.\ Soc.\ Jpn. 65
  (1996) 1340--1354.

\bibitem{Miyagawa04a}
K.~Miyagawa, K.~Kanoda, A.~Kawamoto, \protect{NMR} studies on two-dimensional
  molecular conductors and superconductors: {M}ott transition in
  \protect{$\kappa$-(BEDT-TTF)$_2$X}, Chem. Rev. 104 (2004) 5635--5654.

\bibitem{Kawamoto04a}
A.~Kawamoto, Y.~Honma, K.~Kumagai, Electron localization in the strongly
  correlated organic system \protect{$\kappa$-(BEDT-TTF)$_2$X} probed with
  nuclear magnetic resonance $^{13}${C}-{N}{M}{R}, Phys.\ Rev.\ B 70 (2004)
  060510R.

\bibitem{Shimizu06a}
Y.~Shimizu, K.~Miyagawa, K.~Kanoda, M.~Maesato, G.~Saito, Emergence of
  inhomogeneous moments from spin liquid in the triangular-lattice {M}ott
  insulator \protect{$\kappa$-(ET)$_2$Cu$_2$(CN)$_3$}, Phys.\ Rev.\ B 73 (2006)
  140407.

\bibitem{Kawamoto06a}
A.~Kawamoto, Y.~Honma, K.~Kumagai, N.~Matsunaga, K.~Nomura, Suppression of
  inhomogeneous electron localization in
  \protect{$\kappa$-(BEDT-TTF)$_2$Cu$_2$(CN)$_3$} under pressure, Phys.\ Rev.\
  B 74 (2006) 212508.

\bibitem{Ohira06a}
S.~Ohira, Y.~Shimizu, K.~Kanoda, G.~Saito, Spin liquid state in
  \protect{$\kappa$-(BEDT-TTF)$_2$Cu$_2$(CN)$_3$} studied by muon spin
  relaxation method, J. Low Temp. Phys. 142 (2006) 153.

\bibitem{Pratt11a}
F.~L. Pratt, P.~J. Baker, S.~J. Blundell, T.~Lancaster, S.~Ohira-Kawamura,
  C.~Baines, Y.~Shimizu, K.~Kanoda, I.~Watanabe, G.~Saito, Magnetic and
  non-magnetic phases of a quantum spin liquid, Nature 471 (2011) 612.

\bibitem{Nakajima12a}
S.~Nakajima, T.~Suzuki, K.~Ohishi, I.~Watanabe, T.~Goto, A.~Oosawa,
  N.~Yoneyama, N.~Kobayashi, F.~L. Pratt, T.~Sasaki, Microscopic phase
  separation in triangular-lattice quantum spin magnet
  \protect{$\kappa$-(BEDT-TTF)$_2$Cu$_2$(CN)$_3$} probed by muon spin
  relaxation, J.\ Phys.\ Soc.\ Jpn. 81 (2012) 063706.

\bibitem{Yamashita08a}
S.~Yamashita, Y.~Nakazawa, M.~Oguni, Y.~Oshima, H.~Nojiri, Y.~Shimizu,
  K.~Miyagawa, K.~Kanoda, Thermodynamic properties of a spin-1/2 spin-liquid
  state in a \protect{$\kappa$}-type organic salt, Nature Phys. 4 (2008)
  459--462.

\bibitem{Yamashita09a}
M.~Yamashita, N.~Nakata, Y.~Kasahara, T.~Sasaki, N.~Yoneyama, N.~Kobayashi,
  S.~Fujimoto, T.~Shibauchi, Y.~Matsuda, Thermal-transport measurements in a
  quantum spin-liquid state of the frustrated triangular magnet
  \protect{$\kappa$-(BEDT-TTF)$_2$Cu$_2$(CN)$_3$}, Nature Phys. 5 (2009)
  44--47.

\bibitem{Kawamoto95b}
A.~Kawamoto, K.~Miyagawa, Y.~Nakazawa, K.~Kanoda, Electron correlation in the
  $\kappa$-phase family of \protect{BEDT-TTF} compounds studied by
  \protect{$^{13}$C} \protect{NMR} where \protect{BEDT-TTF} is
  bis(ethylenedithio)tetrafulvalene, Phys.\ Rev.\ B 52 (1995) 15522.

\bibitem{Kataev92a}
V.~Kataev, G.~Winkel, D.~Khomskii, S.~Wohlleben, W.~Crump, K.~F. Tebbe,
  J.~Hahn, \protect{ESR} of single crystals of
  \protect{$\kappa$-(BEDT-TTF)$_2$Cu[N(CN)$_2$]X} ({X}={B}r and {I}), Solid
  St.\ Comm. 83 (1992) 435--439.

\bibitem{Kobayashi14a}
T.~Kobayashi, Y.~Ihara, Y.~Saito, A.~Kawamoto, Microscopic observation of
  superconducting fluctuations in
  \protect{$\kappa$-(BEDT-TTF)$_2$Cu[N(CN)$_2$]Br} by \protect{$^{13}$C}
  \protect{NMR} spectroscopy, Phys.\ Rev.\ B 89 (2014) 165141.

\bibitem{Gartner88a}
S.~\protect{G\"artner}, E.~Gogu, I.~Heinen, H.~J. Keller, T.~Klutz,
  D.~Schweitzer, Superconductivity at 10 {K} and ambient pressure in the
  organic metal \protect{(BEDT-TTF)$_2$Cu(SCN)$_2$}, Solid St.\ Comm. 65 (1988)
  1531--1534.

\bibitem{Sushko91b}
Y.~V. Sushko, V.~A. Bondarenko, R.~A. Petrosov, N.~D. Kushch, E.~Yagubskii,
  Hydrostatic pressure effect on \protect{T$_c$} and resistance anomaly in
  normal state of \protect{$\kappa$-(BEDT-TTF)$_2$CuN(CN)$_2$Br}, J. de
  Physique I 1 (1991) 1375--1379.

\bibitem{Muller02a}
J.~\protect{M\"uller}, M.~Lang, F.~Steglich, J.~A. Schlueter, A.~M. Kini,
  T.~Sasaki, Evidence for structural and electronic instabilities at
  intermediate temperatures in \protect{$\kappa$-(BEDT-TTF)$_2$X} for
  \protect{X = Cu[N(CN)$_2$]Cl, Cu[N(CN)$_2$]Br} and \protect{Cu(NCS)$_2$}:
  Implications for the phase diagram of these quasi-two-dimensional organic
  superconductors, Phys.\ Rev.\ B 65 (2002) 144521.

\bibitem{Kawamoto95a}
A.~Kawamoto, K.~Miyagawa, Y.~Nakazawa, K.~Kanoda, \protect{$^{13}$C}
  \protect{NMR} study of layered organic superconductors based on
  \protect{BEDT-TTF} molecules, Phys.\ Rev.\ Lett. 74 (1995) 3455.

\bibitem{Shimizu10a}
Y.~Shimizu, H.~Kasahara, T.~Furuta, K.~Miyagawa, K.~Kanoda, M.~Maesato,
  G.~Saito, Pressure-induced superconductivity and {M}ott transition in
  spin-liquid \protect{$\kappa$-(ET)$_{2}$Cu$_{2}$(CN)$_{3}$} probed by
  \protect{$^{13}$C} {NMR}, Phys. Rev. B 81~(22) (2010) 224508.

\bibitem{Wang06a}
Y.~Wang, L.~Li, N.~P. Ong, Nernst effect in high-{T$_c$} superconductors,
  Phys.\ Rev.\ B 73 (2006) 024510.

\bibitem{Li10b}
L.~Li, Y.~Wang, S.~Komiya, Y.~A. abd G.~D.~Gu, N.~P. Ong, Diamagnetism and
  {C}ooper pairing above \protect{T$_c$} in cuprates, Phys.\ Rev.\ B 81 (2010)
  054510.

\bibitem{Cyr-Choiniere09a}
O.~Cyr-Choiniere, R.~Daou, F.~\protect{Lalibert\'e}, D.~LeBoeuf,
  N.~Doiron-Leyraud, J.~Chang, J.-Q. Yan, J.-G. Cheng, J.-S. Zhou, J.~B.
  Goodenough, S.~Pyon, T.~Takayama, H.~Takagi, Y.~Tanaka, L.~Taillefer,
  Enhancement of the {N}ernst effect by stripe order in a high-\protect{T$_c$}
  superconductor, Nature 458 (2009) 743--745.

\bibitem{Nam07a}
M.-S. Nam, A.~Ardavan, S.~J. Blundell, J.~A. Schlueter, Fluctuating
  superconductivity in organic molecular metals close to the {M}ott transition,
  Nature 449 (2007) 584--587.

\bibitem{Nam13a}
M.-S. Nam, C.~\protect{M\'ezi\`ere}, P.~Batail, L.~Zorina, S.~Simonov,
  A.~Ardavan, Superconducting fluctuations in organic molecular metals enhanced
  by {M}ott criticality, Scientific Reports 3 (2013) 3390.

\bibitem{Tsuchiya12a}
S.~Tsuchiya, J.~Yamada, S.~Tanda, K.~Ichimura, T.~Terashima, N.~Kurita,
  K.~Kodama, S.~Uji, Fluctuating superconductivity in the strongly correlated
  two-dimensional organic superconductor
  \protect{$\kappa$-(BEDT-TTF)$_2$Cu(NCS)$_2$} in an in-plane magnetic field,
  Phys.\ Rev.\ B 85 (2012) 220506(R).

\bibitem{deSoto95a}
S.~M.~D. Soto, C.~P. Slichter, A.~M. Kini, H.~H. Wang, U.~Geiser, J.~M.
  Williams, \protect{$^{13}$C} \protect{NMR} studies of the normal and
  superconducting states of the organic superconductor
  \protect{$\kappa$-(BEDT-TTF)$_2$Cu[N(CN)$_2$]Br}, Phys.\ Rev.\ B 52 (1995)
  10364.

\bibitem{Taylor07a}
O.~J. Taylor, A.~Carrington, J.~A. Schlueter, Specific-heat measurements of the
  gap structure of the organic superconductors
  \protect{$\kappa$-(ET)$_2$Cu[N(CN)$_2$]Br} and
  \protect{$\kappa$-(ET)$_2$Cu(NCS)$_2$}, Phys.\ Rev.\ Lett. 99 (2007) 057001.

\bibitem{Pinteric02a}
M.~\protect{Pinteri\'c}, S.~\protect{Tomi\'c}, M.~Prester, D.~Drobac, K.~Maki,
  Influence of internal disorder on the superconducting state in the organic
  layered superconductor \protect{$\kappa$−(BEDT−TTF)$_2$Cu[N(CN)$_2$]Br},
  Phys.\ Rev.\ B 66 (2002) 174521.

\bibitem{Milbradt13a}
S.~Milbradt, A.~A. Bardin, C.~J.~S. Truncik, W.~A. Huttema, A.~C. Jacko, P.~L.
  Burn, S.-C. Lo, B.~J. Powell, D.~M. Broun, In-plane superfluid density and
  microwave conductivity of the organic superconductor
  \protect{$\kappa$-(BEDT-TTF)$_2$Cu[N(CN)$_2$]Br}: {E}vidence for d-wave
  pairing and resilient quasiparticles, Phys. Rev. B 88 (2013) 064501.

\bibitem{Perunov12a}
N.~V. Perunov, A.~F. Shevchun, N.~D. Kushch, M.~R. Trunin, Surface impedance of
  \protect{$\kappa$-(BEDT-TTF)$_2$Cu[N(CN)$_2$]Br} crystals, JETP Letters 96
  (2012) 184.

\bibitem{Arai01a}
T.~Arai, K.~Ichimura, K.~Nomura, S.~Takasaki, J.~Yamada, S.~Nakatsuji,
  H.~Anzai, Tunneling spectroscopy on the organic superconductor
  \protect{$\kappa$-(BEDT-TTF)$_2$Cu(NCS)$_2$} using {S}{T}{M}, Phys. Rev. B 63
  (2001) 104518.

\bibitem{Ichimura08a}
K.~Ichimura, M.~Takami, K.~Nomura, Direct observation of d-wave superconducting
  gap in \protect{$\kappa$-(BEDT-TTF)$_2$Cu[N(CN)$_2$]Br} with scanning
  tunneling microscopy, Journal of the Physical Society of Japan 77~(11) (2008)
  114707.

\bibitem{Oka15a}
Y.~Oka, H.~Nobukane, N.~Matsunaga, K.~Nomura, K.~Katono, K.~Ichimura,
  A.~Kawamoto, Tunneling spectroscopy in organic superconductor
  \protect{$\kappa$-(BEDT-TTF-d[3,3])$_2$Cu[N(CN)$_2$]Br}, J.\ Phys.\ Soc.\
  Jpn. 84 (2015) 064713.

\bibitem{Schrama99a}
J.~M. Schrama, E.~Rzepniewski, R.~S. Edwards, J.~Singleton, A.~Ardavan,
  M.~Kurmoo, P.~Day, Millimeter-wave magneto-optical determination of the
  anisotropy of the superconducting order parameter in the molecular
  superconductor \protect{$\kappa$-(BEDT-TTF)$_2$Cu(NCS)$_2$}, Phys. Rev. Lett.
  83 (1999) 3041--3044.

\bibitem{Malone10a}
L.~Malone, O.~J. Taylor, J.~A. Schlueter, A.~Carrington, Location of gap nodes
  in the organic superconductors \protect{$\kappa$-(ET)$_2$Cu(NCS)$_2$} and
  \protect{$\kappa$-(ET)$_2$Cu[N(CN)$_2$]Br} determined by magnetocalorimetry,
  Phys. Rev. B 82 (2010) 014522.

\bibitem{Izawa01a}
K.~Izawa, H.~Yamaguchi, T.~Sasaki, Y.~Matsuda, Superconducting gap structure of
  \protect{$\kappa$-(BEDT-TTF)$_2$Cu(NCS)$_2$} probed by thermal conductivity
  tensor, Phys.\ Rev.\ Lett. 88 (2001) 027002.

\bibitem{Guterding16b}
D.~Guterding, S.~Diehl, M.~Altmeyer, T.~Methfessel, U.~Tutsch, H.~Schubert,
  M.~Lang, J.~\protect{M\"uller}, M.~Huth, H.~O. Jeschke,
  R.~\protect{Valent\'i}, M.~Jourdan, H.-J. Elmers, Evidence for eight-node
  mixed-symmetry superconductivity in a correlated organic metal, Phys.\ Rev.\
  Lett. 116 (2016) 237001.

\bibitem{Kuhlmorgen17a}
S.~\protect{K\"uhlmorgen}, R.~\protect{Sch\"onemann}, E.~L. Greene,
  J.~\protect{M\"uller}, J.~Wosnitza, Investigation of the superconducting gap
  structure in \protect{$\kappa$-(BEDT-TTF)$_2$Cu(NCS)$_2$} and
  \protect{$\kappa$-(BEDT-TTF)$_2$Cu[N(CN)$_2$]Br} by means of
  thermal-conductivity measurements, J. Phys.: Condens. Matter 29 (2017)
  405604.

\bibitem{Elsinger00a}
H.~Elsinger, J.~Wosnitza, S.~Wanka, J.~Hagel, D.~Schweitzer, W.~Strunz,
  \protect{$\kappa$-(BEDT-TTF)$_2$Cu[N(CN)$_2$]Br}: A fully gapped
  strong-coupling superconductor, Phys. Rev. Lett. 84 (2000) 6098--6101.

\bibitem{Muller02b}
J.~\protect{M\"uller}, M.~Lang, R.~Helfrich, F.~Steglich, T.~Sasaki,
  High-resolution ac-calorimetry studies of the quasi-two-dimensional organic
  superconductor \protect{$\kappa$-(BEDT-TTF)$_2$Cu(NCS)$_2$}, Phys.\ Rev.\ B
  65 (2002) 140509.

\bibitem{Fournier03a}
D.~Fournier, M.~Poirier, M.~Castonguay, K.~D. Truang, Mott transition,
  compressibility divergence, and the {P}-{T} phase diagram of layered organic
  superconductors: an ultrasonic investigation, Phys.\ Rev.\ Lett. 90 (2003)
  127002.

\bibitem{Souza15a}
M.~de~Souza, L.~Bartosch, Probing the {M}ott physics in
  \protect{$\kappa$-(BEDT-TTF)$_2$X} salts via thermal expansion, J. Phys.:
  Condens. Matter 27 (2015) 053203.

\bibitem{Manna10a}
R.~S. Manna, M.~de~Souza, A.~Bruhl, J.~A. Schlueter, M.~Lang, Lattice effects
  and entropy release at the low-temperature phase transition in the
  spin-liquid candidate \protect{$\kappa$-(BEDT-TTF)$_2$Cu$_2$(CN)$_3$}, Phys.\
  Rev.\ Lett. 104 (2010) 016403.

\bibitem{Abdel-Jawad10a}
M.~Abdel-Jawad, I.~Terasaki, T.~Sasaki, N.~Yoneyama, N.~Kobayashi, Y.~Uesu,
  C.~Hotta, Anomalous dielectric response in the dimer {M}ott insulator
  $\kappa$-({BEDT-TTF})$_2${C}u$_2$({CN})$_3$, Phys.\ Rev.\ B 82 (2010) 125119.

\bibitem{Pinteric14a}
M.~\protect{Pinteri\'{c}}, M.~\protect{\u{C}ulo}, M.~Basleti\'{c},
  B.~\protect{Korin-Hamzi\'{c}}, E.~Tafra, A.~\protect{Hamzi\'{c}}, T.~Ivek,
  T.~Peterseim, K.~Miyagawa, K.~Kanoda, J.~A. Schlueter, M.~Dressel,
  S.~\protect{Tomic\'{c}}, Anisotropic charge dynamics in the quantum
  spin-liquid candidate \protect{$\kappa$-(BEDT-TTF)$_2$Cu$_2$(CN)$_3$}, Phys.\
  Rev.\ B 90 (2014) 195139.

\bibitem{Lunkenheimer12a}
P.~Lunkenheimer, J.~M{\"u}ller, S.~Krohns, F.~Schrettle, A.~Loidl, B.~Hartmann,
  R.~Rommel, M.~de~Souza, C.~Hotta, J.~Schlueter, M.~Lang, Multiferroicity in
  an organic charge-transfer salt: Electric-dipole-driven magnetism, Nat.
  Mater. 11 (2012) 755--758.

\bibitem{Lang14a}
M.~Lang, P.~Lunkenheimer, J.~\protect{M\"uller}, A.~Loidl, B.~Hartmann, N.~H.
  Hoang, E.~Gati, H.~Schubert, J.~A. Schlueter, Multiferroicity in the {M}ott
  insulating charge-transfer salt
  \protect{$\kappa$-(BEDT-TTF)$_2$Cu[N(CN)$_2$]Cl}, IEEE Trans. Magnetics 50
  (2014) 2700107.

\bibitem{Lunkenheimer15a}
P.~Lunkenheimer, A.~Loidl, Dielectric spectroscopy on organic charge-transfer
  salts, J. Phys.: Condens. Matter 27 (2015) 373001(21 pages).

\bibitem{Poirier12a}
M.~Poirier, S.~Parent, A.~Cote, K.~Miyagawa, K.~Kanoda, Y.~Shimizu,
  Magnetodielectric effects and spin-charge coupling in the spin-liquid
  candidate \protect{$\kappa$-(BEDT-TTF)$_2$Cu$_2$(CN)$_3$}, Phys.\ Rev.\ B 85
  (2012) 134444.

\bibitem{Sedlmeier12a}
K.~Sedlmeier, S.~Els{\"a}sser, D.~Neubauer, R.~Beyer, D.~Wu, T.~Ivek, S.~Tomic,
  J.~A. Schlueter, M.~Dressel, Absence of charge order in the dimerized
  {$\kappa$}-phase \protect{BEDT-TTF} salts, Phys.\ Rev.\ B 86 (2012) 245103.

\bibitem{Yakushi15a}
K.~Yakushi, K.~Yamamoto, T.~Yamamoto, Y.~Saito, A.~Kawamoto, Raman spectroscopy
  study of charge fluctuation in the spin-liquid
  \protect{$\kappa$-(BEDT-TTF)$_2$Cu$_2$(CN)$_3$}, J.\ Phys.\ Soc.\ Jpn. 84
  (2015) 084711.

\bibitem{Itoh13a}
K.~Itoh, H.~Itoh, M.~Naka, S.~Saito, I.~Hosako, N.~Yoneyama, S.~Ishihara,
  T.~Sasaki, S.~Iwai, Collective excitation of and electric dipole on a
  molecular dimer in an organic dimer-{M}ott insulator, Phys.\ Rev.\ Lett. 110
  (2013) 106401.

\bibitem{Ivek10a}
T.~Ivek, B.~Korin-Hamzic, O.~Milat, S.~Tomic, C.~Clauss, N.~Drichko,
  D.~Schweitzer, M.~Dressel, Collective excitations in the charge-ordered phase
  of \protect{$\alpha$-(BEDT-TTF)$_2$I$_3$}, Phys.\ Rev.\ Lett. 104 (2010)
  206406.

\bibitem{Ivek11a}
T.~Ivek, B.~Korin-Hamzic, O.~Milat, S.~Tomic, C.~Clauss, N.~Drichko,
  D.~Schweitzer, M.~Dressel, Electrodynamic response of the charge ordering
  phase: Dielectric and optical studies of
  \protect{$\alpha$-(BEDT-TTF)$_2$I$_3$}, Phys.\ Rev.\ B 83 (2011) 165128.

\bibitem{Tomic13a}
S.~\protect{Tomic\'{c}}, M.~\protect{Pinteri\'{c}}, T.~Ivek, K.~Sedlmeier,
  R.~Beyer, D.~Wu, J.~A. Schlueter, D.~Schweitzer, M.~Dressel, Magnetic
  ordering and charge dynamics in
  \protect{$\kappa$-(BEDT-TTF)$_2$Cu[N(CN)$_2$]Cl}, J. Phys.: Condens. Matter
  25 (2013) 436004(7 pages).

\bibitem{Drozdova01a}
O.~Drozdova, G.~Saito, H.~Yamochi, K.~Ookubo, K.~Yakushi, M.~Uruichi,
  L.~Ouahab, Composition and structure of the anion layer in the organic
  superconductor \protect{$\kappa^\prime$-(ET)$_2$Cu$_2$(CN)$_3$}: optical
  study, Inorg. Chem. 40 (2001) 3265--3266.

\bibitem{Pinteric16a}
M.~Pinteri\ifmmode~\acute{c}\else \'{c}\fi{}, P.~Lazi\ifmmode~\acute{c}\else
  \'{c}\fi{}, A.~Pustogow, T.~Ivek, M.~Kuve\ifmmode \check{z}\else
  \v{z}\fi{}di\ifmmode~\acute{c}\else \'{c}\fi{}, O.~Milat, B.~Gumhalter,
  M.~Basleti\ifmmode~\acute{c}\else \'{c}\fi{}, M.~\ifmmode~\check{C}\else
  \v{C}\fi{}ulo, B.~Korin-Hamzi\ifmmode~\acute{c}\else \'{c}\fi{}, A.~L\"ohle,
  R.~H\"ubner, M.~Sanz~Alonso, T.~Hiramatsu, Y.~Yoshida, G.~Saito, M.~Dressel,
  S.~Tomi\ifmmode~\acute{c}\else \'{c}\fi{}, Anion effects on electronic
  structure and electrodynamic properties of the {M}ott insulator
  \protect{$\kappa$-(BEDT-TTF)$_2$Ag$_2$(CN)$_3$}, Phys. Rev. B 94 (2016)
  161105.

\bibitem{Iguchi13a}
S.~Iguchi, S.~Sasaki, N.~Yoneyama, H.~Taniguchi, T.~Nishizaki, T.~Sasaki,
  Relaxor ferroelectricity induced by electron correlations in a molecular
  dimer {M}ott insulator, Phys.\ Rev.\ B 87 (2013) 075107.

\bibitem{Niizeki08a}
S.~Niizeki, F.~Yoshikane, K.~Kohno, K.~Takahashi, H.~Mori, Y.~Bando,
  T.~Kawamoto, T.~Mori, Dielectric response and electric-field-induced
  metastable state in an organic conductor \protect{$\beta$-({\it
  meso}-DMBEDT-TTF)$_2$PF$_6$}, J.\ Phys.\ Soc.\ Jpn. 77 (2008) 073710.

\bibitem{Aldoshina93a}
M.~Z. Aldoshina, R.~N. Lyubovskaya, S.~V. Konovalikhin, O.~A. Dyachenko, G.~V.
  Shilov, M.~K. Makova, R.~B. Lyubovskii, A new series of {E}{T}-based organic
  metals: synthesis, crystal structure and properties, Synth.\ Metals 55-57
  (1993) 1905--1909.

\bibitem{Yudanova95a}
E.~I. Yudanova, S.~K. Hoffmann, A.~Graja, S.~V. Konavalikhin, O.~A. Dyachenko,
  R.~B. Lyubovskii, R.~N. Lyubovskaya, Crystal structure, {E}{S}{R} and
  conductivity studies of bis(ethylenedithio)tetrafulvalene ({E}{T}) organic
  conductor \protect{(H$_8$ET$_2$)$_2$[Hg(SCN)$_2$Br]} and its deuterated
  analogue \protect{(D$_8$ET)$_2$[Hg(SCN)$_2$Br]}, Synth.\ Metals 73 (1995)
  227.

\bibitem{Yudanova96a}
E.~I. Yudanova, L.~M. Makarova, S.~V. Konavalikhin, O.~A. Dyachenko, R.~B.
  Lyubovskii, R.~N. Lyubovskaya, The new salt
  \protect{$\kappa$-ET$_2$[Hg(SCN)$_2$I]}: crystal structure and physical
  properties, Synth.\ Metals 79 (1996) 201.

\bibitem{Ota07a}
A.~Ota, L.~Ouahab, S.~Golhen, Y.~Yoshida, M.~Maesato, G.~Saito, R.~Swietlik,
  Phase transition from {M}ott insulating phase into the charge ordering phase
  with molecular deformation in charge-transfer salts
  \protect{$\kappa$-(ET)$_4$[M(CN)$_6$][N(C$_2$H$_5$)$_4$]$\cdot$ 2H$_2$O} ({M}
  = \protect{Co$^{\rm{III}}$} and \protect{Fe$^{\rm{III}}$}), Chemistry of
  Materials 19 (2007) 2455--2462.

\bibitem{Lapinski13a}
A.~Lapinski, R.~Swietlik, L.~Ouahab, S.~Golhen, Spectroscopic studies of the
  phase transition from the {M}ott insulator state to the charge-ordering state
  of \protect{$\kappa$-(ET)$_4$[M(CN)$_6$][N(C$_2$H$_5$)$_4$]$\cdot$ 2H$_2$O}
  ({M} = \protect{Co$^{\rm{III}}$} and \protect{Fe$^{\rm{III}}$}) salts, J.
  Phys. Chem. A 117 (2013) 5241--5250.

\bibitem{Magueres96a}
P.~L. Maguer{\`{e}}s, L.~Ouhab, N.~Corian, C.~J. Gomez-Garcia, P.~Delhaes,
  Phase transitions in new \protect{BEDT-TTF} {$\kappa$}--phase salts with
  hexacyanometalate anions \protect{[M(CN)$_6^{3-}$ M=\protect{Co$^{\rm{III}}$}
  and \protect{Fe$^{\rm{III}}$}]}, Solid St.\ Comm. 97 (1996) 27--32.

\bibitem{Magueres97a}
P.~L. Maguer{\`{e}}s, L.~Ouhab, P.~Briard, J.~Even, M.~Bertault, L.~Touper, ,
  J.~R{\'{a}}mos, C.~J. Gomez-Garcia, P.~Delhaes, T.~Mallah, Phase transitions
  \protect{[$\kappa$-(ET$_4$N)ET$_4$M(CN)$_6$ 3 H$_2$O; M=Fe$^{\rm{III}}$,
  Co$^{\rm{III}}$, Cr$^{\rm{III}}$]}, Synth.\ Metals 86 (1997) 1859--1860.

\bibitem{Swietlik01a}
R.~Swietlik, M.~Polomska, L.~Ouahab, J.~Guillevic, Infrared and {R}aman studies
  of $\kappa$-phase charge-transfer salts formed by \protect{BEDT-TTF} and
  magnetic anions \protect{M(CN)$_6^{3-}$} (where \protect{M=Co$^{{\rm III}}$,
  Fe$^{{\rm III}}$, Cr$^{{\rm III}}$}), J. Mater. Chem. 11 (2001) 1313--1318.

\bibitem{Swietlik03a}
R.~Swietlik, A.~Lapinski, M.~Polomska, L.~Ouahab, J.~Guillevic, Spectroscopic
  evidence of the charge ordering in
  \protect{$\kappa$-[Et$_4$N](BEDT-TTF)$_4$M(CN)$_6$$\cdot$ 3H$_2$O} ({M} =
  \protect{Co$^{\rm{III}}$}, \protect{Fe$^{\rm{III}}$},
  \protect{Cr$^{\rm{III}}$}), Synth. Metals 133 (2003) 273--275.

\bibitem{Swietlik04a}
R.~Swietlik, L.~Ouahab, J.~Guillevic, K.~Yakushi, Infrared and {R}aman studies
  of the charge ordering in the organic semiconductor
  \protect{$\kappa$-[Et$_4$N][Co(CN)$_6$(ET)$_4$] $\cdot$ 3H$_2$O}, Macromol.
  Symp. 212 (2004) 219--224.

\bibitem{Swietlik06a}
R.~Swietlik, A.~Lapinski, M.~Polomska, L.~Ouahab, A.~Ota, Infrared and {R}aman
  investigations of the charge ordering phenomena in the monoclinic salts
  \protect{$\kappa$-(ET)$_4$[M(CN)$_6$][N(C$_2$H$_5$)$_4$]$\cdot$ 2H$_2$O} ({M}
  = \protect{Co$^{\rm{III}}$} and \protect{Fe$^{\rm{III}}$}), J. Low Temp.
  Phys. 142 (2006) 653--656.

\bibitem{Yagubskii84a}
E.~B. Yagubskii, I.~F. Shchegolev, V.~N. Laukhin, P.~A. Konovich, M.~V.
  Karatsovnik, A.~V. Zvarykina, L.~I. Buravov, Normal-pressure
  superconductivity in an organic metal \protect{(BEDT-TTF)$_2$I$_3$} [bis
  (ethylene dithiolo) tetrathiofulvalene triiodide], JETP Lett. 39 (1984)
  12--15.

\bibitem{Murata85a}
K.~Murata, M.~Tokumoto, H.~Anzai, H.~Bando, G.~Saito, K.~Kajimura, T.~Ishiguro,
  Superconductivity with the onset at 8 {K} in the organic conductor
  \protect{$\beta$-(BEDT-TTF)$_2$I$_3$} under pressure, J.\ Phys.\ Soc.\ Jpn.
  54 (1985) 1236.

\bibitem{Williams84a}
J.~M. Williams, H.~H. Wang, M.~A. Beno, T.~J. Emge, L.~M. Sowa, P.~T. Copps,
  F.~Behroozi, L.~N. Hall, K.~D. Carlson, G.~W. Crabtree, Ambient-pressure
  superconductivity at 2.7 {K} and higher temperatures in derivatives of
  \protect{(BEDT-TTF)$_2$IBr$_2$}: synthesis, structure, and detection of
  superconductivity, Inorg. Chem. 23 (1983) 3839--3841.

\bibitem{Emge85a}
T.~J. Emge, H.~H. Wang, M.~A. Beno, P.~C.~W. Leung, M.~A. Firestone, H.~C.
  Jenkins, J.~D. Cook, J.~M. Williams, E.~L. Venturini, A test of
  superconductivity vs. molecular disorder in \protect{(BEDT-TTF)$_2$X}
  synthetic metals: synthesis, structure (298, 120 {K}), and
  microwave/{E}{S}{R} conductivity of \protect{(BEDT-TTF)$_2$2I$_2$Br}, Inorg.
  Chem. 24 (1985) 1736--1738.

\bibitem{Wang85a}
H.~H. Wang, M.~A. Beno, U.~Geiser, M.~A. Firestone, K.~S. Webb, L.~Nunez, G.~W.
  Crabtree, K.~D. Carlson, J.~M. Williams, Ambient-pressure superconductivity
  at the highest temperature (5 {K}) observed in an organic system:
  \protect{$\beta$-(BEDT-TTF)$_2$AuI$_2$}, Inorg. Chem. 24 (1985) 2465--2466.

\bibitem{Yamada01a}
J.~Yamada, M.~Watanabe, H.~Akutsu, S.~Nakatsuji, H.~Nishikawa, I.~Ikemoto,
  K.~Kikuchi, New organic superconductors \protect{$\beta$-(BDA-TTP)$_2$X}
  \protect{[BDA-TTP= 2,5-Bis(1,3-dithian-2-ylidene)-1,3,4,6-tetrathiapentalene;
  X$^-$=SbF$_6^-$, AsF$_6^-$, and PF$_6^-$]}, J. Am. Chem. Soc. 123 (2001)
  4174--4180.

\bibitem{Yamada01b}
J.~Yamada, T.~Toita, H.~Akutsu, S.~Nakatsuji, H.~Nishikawa, I.~Ikemoto,
  K.~Kikuchi, The crystal structure and physical properties of
  \protect{$\beta$-(BDA-TTP)$_2$FeCl$_4$} [\protect{BDA-TTP} =
  2,5-bis(1,3-dithian-2-ylidene)-1,3,4,6-tetrathiapentalene], Chem. Commun.
  2001 (2001) 2538--2539.

\bibitem{Choi04a}
E.~S. Choi, D.~Graf, J.~S. Brooks, J.~Yamada, H.~Akutsu, K.~Kikuchi,
  M.~Tokumoto, Pressure-dependent ground states and fermiology in
  \protect{$\beta$-(BDA-TTP)$_2${\it M}Cl$_4$} ({{\it M}}={F}e,{G}a), Phys.\
  Rev.\ B 70 (2004) 024517.

\bibitem{Yamada03a}
J.~Yamada, T.~Toita, H.~Akutsu, S.~Nakatsuji, H.~Nishikawa, I.~Ikemoto,
  K.~Kikuchi, E.~S. Choi, D.~Graf, J.~S. Brooks, A new organic superconductor,
  \protect{$\beta$-(BDA-TTP)$_2$GaCl$_4$} [\protect{BDA-TTP} =
  2,5-(1,3-dithian-2-ylidene)-1,3,4,6-tetrathiapentalene], Chem. Commun. 2003
  (2003) 2230--2231.

\bibitem{Yamada06a}
J.~Yamada, K.~Fujimoto, H.~Akutsu, S.~Nakatsuji, A.~Miyazaki, M.~Aimatsu,
  S.~Kudo, T.~Enoki, K.~Kikuchi, Pressure effect on the electrical conductivity
  and superconductivity of \protect{$\beta$-(BDA-TTP)$_2$I$_3$}, Chem. Commun.
  2006 (2006) 1331--1333.

\bibitem{Kimura04a}
S.~Kimura, T.~Maejima, H.~Suzuki, R.~Chiba, H.~Mori, T.~Kawamoto, T.~Mori,
  H.~Moriyama, Y.~Nishio, K.~Kajita, A new organic superconductor
  \protect{$\beta$-({\it meso}-DMBEDT-TTF)$_2$PF$_6$}, Chem. Commun. 2004
  (2004) 2454.

\bibitem{Morinaka09a}
N.~Morinaka, K.~Takahashi, R.~Chiba, F.~Yoshikane, S.~Niizeki, M.~Tanaka,
  K.~Yakushi, M.~Koeda, M.~Hedo, T.~Fujiwara, Y.~Uwatoko, Y.~Nishio, K.~Kajita,
  H.~Mori, Superconductivity competitive with checkerboard-type charge ordering
  in the organic conductor \protect{$\beta$-({\it meso}-DMBEDT-TTF)$_2$PF$_6$},
  Phys.\ Rev.\ B 80 (2009) 092508.

\bibitem{Shikama12a}
T.~Shikama, T.~Shimokawa, S.~Lee, T.~Isono, A.~Ueda, K.~Takahashi, A.~Nakao,
  R.~Kumai, H.~Nakao, K.~Kobayashi, Y.~Murakami, M.~Kimata, H.~Tajima,
  K.~Matsubayashi, Y.~Uwatoko, Y.~Nishio, K.~Kajita, H.~Mori, Magnetism and
  pressure-induced superconductivity of checkerboard-type charge-ordered
  molecular conductor {$\beta$}-(meso-{DMBEDT-TTF}){$_2$}{X} ({X} = {PF}{$_6$}
  and {AsF}{$_6$}), Crystals 2 (2012) 1502--1513.

\bibitem{Anzai87a}
H.~Anzai, M.~Tokumoto, T.~Ishiguro, G.~Saito, H.~Kobayashi, R.~Kato,
  A.~Kobayashi, Crystal growth of \protect{BEDT-TTF} trihalides, Synth.\ Metals
  19 (1987) 611--616.

\bibitem{Mori87a}
T.~Mori, H.~Inokuchi, Crystal structures of \protect{AuCl$_2$} salts of
  bis(ethylenedithio)-tetrathiafulvalene \protect{BEDT-TTF}. {E}xistence of
  divalent gold, \protect{Au(II)}, Solid St.\ Comm. 62 (1987) 525--529.

\bibitem{Taniguchi05a}
H.~Taniguchi, M.~Miyashita, K.~Uchiyama, R.~Sato, Y.~Ishi, K.~Satoh, N.~Mori,
  M.~Hedo, Y.~Uwatoko, High-pressure study up to 9.9 {G}{P}a of organic {M}ott
  insulator, \protect{$\beta^\prime$-(BEDT-TTF)$_2$AuCl$_2$}, J.\ Phys.\ Soc.\
  Jpn. 74 (2005) 1370--1373.

\bibitem{Ward00a}
B.~H. Ward, J.~A. Schlueter, U.~Geiser, H.~Wang, E.~Morales, J.~P. Parakka,
  S.~Thomas, J.~Williams, P.~Nixon, R.~W. Winter, G.~L. Gard, H.~J. Koo, M.~H.
  Whangbo, Comparison of the crystal and electronic structures of three 2:1
  salts of the organic donor molecule \protect{BEDT-TTF} with
  pentafluorothiomethylsulfonate anions \protect{SF$_5$CH$_2$SO$_3^-$},
  \protect{SF$_5$CHFSO$_3^-1$}, and \protect{SF$_5$CF$_2$SO$_3^-$}, Chem.
  Mater. 2000 (2000) 343--351.

\bibitem{Graja09a}
A.~Graja, I.~Olejniczak, B.~Barszcz, J.~A. Schlueter, Vibrational spectra of
  two \protect{BEDT-TTF-based} organic conductors: charge order, Cent. Eur. J.
  Phys. 7 (2009) 663--667.

\bibitem{Mori86a}
T.~Mori, F.~Sakai, G.~Saito, H.~Inokuchi, Crystal and band structures of an
  organic conductor \protect{$\beta^{\prime\prime}$-(BEDT-TTF)$_2$AuBr$_2$},
  Chem. Lett. 15 (1986) 1037--1040.

\bibitem{Kondo10a}
R.~Kondo, M.~Higa, S.~Kagoshima, N.~Hanasaki, Y.~Nogami, H.~Nishikawa,
  Interplay of charge-density waves and superconductivity in the organic
  conductor \protect{$\beta^{\prime\prime}$-(BEDT-TTF)$_2$AuBr$_2$}, Phys.\
  Rev.\ B 81 (2010) 024519.

\bibitem{Geiser96a}
U.~Geiser, J.~A. Schlueter, H.~H. Wang, A.~M. Kini, J.~M. Williams, P.~P. Sche,
  H.~I. Zakowicz, M.~L. VanZile, J.~D. Dudek, P.~G. Nixon, R.~W. Winter, G.~L.
  Gard, J.~Ren, M.~H. Whangbo, Superconductivity at 5.2 {K} in an electron
  donor radical salt of bis(ethylenedithio)tetrathiafulvalene
  \protect{(BEDT-TTF)} with the novel polyfluorinated organic anion
  \protect{SF$_5$CH$_2$CF$_2$SO$_3$$^-$}, J.\ Am.\ Chem.\ Soc. 118 (1996)
  9996--9997.

\bibitem{Schlueter01a}
J.~A. Schlueter, B.~H. Ward, U.~Geiser, H.~H. Wang, A.~M. Kini, J.~Parakka,
  E.~Morales, H.-J. Koo, M.-H. Whangbo, R.~W. Winter, J.~Mohtasham, G.~L. Gard,
  Crystal structure, physical properties and electronic structure of a new
  organic conductor \protect{(BEDT-TTF)$_2$SF$_5$CHFCF$_2$SO$_3$}, J. Chem.
  Mater. 11 (2001) 2008--2013.

\bibitem{Schlueter02a}
J.~A. Schlueter, B.~H. Ward, U.~Geiser, A.~M. Kini, H.~H. Wang, A.~N. Hata,
  J.~Mohtasham, R.~W. Winter, G.~L. Gard, Chemical modification of the
  superconducting
  \protect{$\beta^{\prime\prime}$-(ET)$_2$SF$_5$CH$_2$CF$_2$SO$_3$} structure
  through use of \protect{CF$_3$C RR$^\prime$SO$_3^-$} anions, Mol. Cryst. Liq.
  Cryst. 380 (2002) 129.

\bibitem{Olejniczak09a}
I.~Olejniczak, B.~Barszcz, A.~Szutarska, A.~Graja, R.~Wojciechowski, J.~A.
  Schlueter, A.~N. Hata, B.~H. Ward, I{R} and {R}aman spectra of
  \protect{$\beta^{\prime\prime}$-(BEDT-TTF)$_2$RCH$_2$SO$_3$} \protect{(R =
  SF$_5$, CF$_3$)}: dimerization related to hydrogen bonding, Phys. Chem. Chem.
  Phys. 11 (2009) 3910--3920.

\bibitem{Nishikawa02a}
H.~Nishikawa, T.~Morimoto, T.~Kodama, I.~Ikemoto, K.~Kikuchi, J.~Yamada,
  H.~Yoshino, K.~Murata, New organic superconductors consisting of an
  unprecedented $\pi$-electron donor, J. Am. Chem. Soc. 124~(5) (2002) 730.

\bibitem{Nishikawa05a}
H.~Nishikawa, Y.~Sato, K.~Kikuchi, T.~Kodama, I.~Ikemoto, J.~Yamada, H.~Oshio,
  R.~Kondo, S.~Kagoshima, Charge ordering and pressure-induced
  superconductivity in \protect{$\beta^{\prime\prime}$-(DODHT)$_2$PF$_6$},
  Phys.\ Rev.\ B 72 (2005) 052510.

\bibitem{Akutsu02a}
H.~Akutsu, A.~Akutsu-Sato, S.~S. Turner, D.~L. Pevelen, P.~Day, A.-K. Klehe,
  J.~Singleton, D.~A. Tocher, M.~R. Probert, J.~A.~K. Howard, Effect of
  included guest molecules on the normal state conductivity and
  superconductivity of
  \protect{$\beta^{\prime\prime}$-(ET)$_4$[(H$_3$O)Ga(C$_2$O$_4$)$_3$]$\cdot$G}
  \protect{(G=Pyridine, Nitrobenzene)}, J.\ Am.\ Chem.\ Soc. 124 (2002)
  12430--12431.

\bibitem{Mori91a}
H.~Mori, I.~Hirabayashi, S.~Tanaka, T.~Mori, Y.~Maruyama, H.~Inokuchi,
  Superconductivity in \protect{(BEDT-TTF)$_4$Pt(CN)$_4$H$_2$O}, Solid St.\
  Comm. 80 (1991) 411--415.

\bibitem{Mori92a}
T.~Mori, K.~Karo, Y.~Maruyama, H.~Inokuchi, H.~Mori, I.~Hirabayashi, S.~Tanaka,
  Strucural and physical properties of a new organic superconductor,
  \protect{(BEDT-TTF)$_4$Pd(CN)$_4$H$_2$O}, Solid St.\ Comm. 82 (1992) 177.

\bibitem{Mori87b}
T.~Mori, M.~Inokuchi, Superconductivity in
  \protect{(BEDT-TTF)$_3$Cl$_2$2H$_2$O}, Solid St.\ Comm. 64 (1987) 335--337.

\bibitem{Lubczynski96a}
W.~Lubczynski, S.~V. Demishev, J.~Singleton, J.~M. Caulfield, L.~d.~C.~de
  Jongh, C.~J. Kepert, S.~J. Blundell, W.~Hayes, M.~Kurmoo, P.~Day, A study of
  the magnetoresistance of the charge-transfer salt
  \protect{(BEDT-TTF)$_3$Cl$_2\cdot$2H$_2$O} at hydrostatic pressures of up to
  20 kbar: evidence for a charge-density-wave ground state and the observation
  of pressure-induced superconductivity, J. Phys.: Condens. Matter 8 (1996)
  6005.

\bibitem{Mori98c}
T.~Mori, Structural genealogy of \protect{BEDT-TTF-based} organic conductors
  {I}. {P}arallel molecules: $\beta$ and $\beta^{\prime\prime}$ phases, Bull.
  Chem. Soc. Jpn. 71 (1998) 2509--2526.

\bibitem{Mori06a}
H.~Mori, Materials viewpoint of organic superconductors, J.\ Phys.\ Soc.\ Jpn.
  75 (2006) 051003.

\bibitem{Ugawa86a}
A.~Ugawa, K.~Yakushi, H.~Kuroda, A.~Kawamoto, J.~Tanaka, Crystal structure and
  reflectance spectrum of \protect{$\beta^{\prime\prime}$-(BEDT-TTF)$_2$IAuBr},
  Chem. Lett. 15 (1986) 1875--1878.

\bibitem{Schulz86a}
A.~J. Schulz, T.~J. Emge, P.~C.~W. Leung, M.~A. Beno, H.~H. Wang, J.~M.
  Williams, A neutron diffraction study of the extent of disorder in the low
  temperature (20 {K}) structure of \protect{$\beta$-(BEDT-TTF)$_2$I$_2$Br},
  Physica B+C 143 (1986) 351--353.

\bibitem{Ravy86a}
S.~Ravy, R.~Moret, J.~P. Pouget, R.~Comes, S.~S.~P. Parkin, Competition between
  organic superconductivity and a displacive structural modulation in
  bis(ethylenedithio) tetrathiafulvalene perrhenate,
  \protect{(BEDT-TTF)$_2$ReO$_4$}, Phys.\ Rev.\ B 33 (1986) 2049.

\bibitem{Carneiro84a}
K.~Carneiro, J.~C. Scott, E.~M. Engler, Comparative \protect{ESR} study of
  three \protect{(BEDT-TTF):ReO$_4$} salts: an organic superconductor, a
  {P}eierls metal and a semiconductor, Solid St.\ Comm. 50 (1984) 477--481.

\bibitem{Baker99a}
S.~M. Baker, J.~Dong, G.~Li, Z.~Zhu, J.~L. Musfeldt, J.~A. Schlueter, M.~E.
  Kelly, R.~G. Daugherty, J.~M. Williams, Infrared studies of low-temperature
  symmetry breaking in the perrhenate family of {E}{T}-based organic molecular
  conductors, Phys.\ Rev.\ B 60 (1999) 931.

\bibitem{Kimura06a}
S.~Kimura, H.~Suzuki, T.~Maejima, H.~Mori, J.~Yamaura, T.~Kakiuchi, H.~Sawa,
  H.~Moriyama, Checkerboard-type charge-ordered state of a pressure-induced
  superconductor, \protect{$\beta$-(meso-DMBEDT-TTF)$_2$PF$_6$}, J. Am. Chem.
  Soc. 128 (2006) 1456.

\bibitem{Yamada04a}
J.~Yamada, New approach to achievement of organic superconductivity, J. Mater.
  Chem. 14 (2004) 2951--2953.

\bibitem{Yamada04b}
J.~Yamada, H.~Akutsu, H.~Nishikawa, K.~Kikuchi, New trends in the synthesis of
  $\pi$-eletron donors for molecular conductors and superconductors, Chem. Rev.
  104 (2004) 5057--5083.

\bibitem{Tokumoto08a}
T.~Tokumoto, J.~S. Brooks, Y.~Oshima, E.~S. Choi, L.~C. Brunel, H.~Akutsu,
  T.~Kaihatsu, J.~Yamada, J.~can Tol, Antiferromagnetic {\it d}-electron
  exchange via a spin-singlet $\pi$-electron ground state in an organic
  conductor, Phys.\ Rev.\ Lett. 100 (2008) 147602.

\bibitem{Sasamori13a}
K.~Sasamori, K.~Takahashi, T.~Kodama, W.~Fujita, K.~Kikuchi, J.~Yamada,
  Structural variations in \protect{$\beta$-(BDA-TTP)$_2$FeCl$_4$} at low
  temperature and under pressure: charge-ordered state with a two-fold crystal
  structure, J.\ Phys.\ Soc.\ Jpn. 82 (2013) 054705.

\bibitem{Tanaka08a}
M.~Tanaka, K.~Yamamoto, M.~Uruichi, T.~Yamamoto, K.~Yakushi, S.~Kimura,
  H.~Mori, Infrared and {R}aman study of the charge-ordered state in the
  vicinity of the superconducting state in the organic conductor
  \protect{$\beta$-(meso-DMBEDT-TTF)$_2$PF$_6$}, J.\ Phys.\ Soc.\ Jpn. 77
  (2008) 024714.

\bibitem{Okazaki13a}
R.~Okazaki, Y.~Ikemoto, T.~Moriwaki, T.~Shikama, K.~Takahashi, H.~Mori,
  H.~Nakaya, T.~Sasaki, Y.~Yasui, I.~Terasaki, Optical conductivity measurement
  of a dimer {M}ott-insulator to charge-order phase transition in a
  two-dimensional quarter-filled organic salt compound, Phys.\ Rev.\ Lett. 111
  (2013) 217801.

\bibitem{Tokumoto87a}
M.~Tokumoto, H.~Anzai, T.~Ishiguro, G.~Saito, H.~Kobayashi, R.~Kato,
  A.~Kobayashi, Electrical and magnetic properties of organic semiconductors,
  \protect{(BEDT-TTF)$_2$X} (\protect{X=IBr$_2$}, \protect{IBrCl}, and
  \protect{ICl$_2$}), Synth.\ Metals 19 (1987) 215--220.

\bibitem{Eto10a}
Y.~Eto, A.~Kawamoto, Antiferromagnetic phase in
  \protect{$\beta^\prime$-(BEDT-TTF)$_2$ICl$_2$} under pressure as seen via
  \protect{$^{13}$C} {N}{M}{R}, Phys.\ Rev.\ B 81 (2010) 020512(R).

\bibitem{Sato06a}
K.~Satoh, H.~Taniguchi, A.~Kawamoto, W.~Higemoto, \protect{$\mu$SR} study of an
  antiferromagnetic {M}ott insulator
  \protect{$\beta^\prime$-(BEDT-TTF)$_2$ICl$_2$}, Physica B 374--375 (2006)
  99--101.

\bibitem{Sato09a}
K.~Satoh, K.~Sato, T.~Yoshida, H.~Taniguchi, T.~Goko, T.~U. Ito, K.~Ohishi,
  W.~Higemoto, \protect{$\mu$SR} study of organic antiferromagnet
  \protect{$\beta^\prime$-(BEDT-TTF)$_2$ICl$_2$} under high pressure, Physica B
  404 (2009) 600--602.

\bibitem{Hashimoto15a}
K.~Hashimoto, R.~Kobayashi, H.~Okamura, H.~Taniguchi, Y.~Ikemoto, T.~Moriwaki,
  S.~Iguchi, M.~Naka, S.~Ishihara, T.~Sasaki, Emergence of charge degrees of
  freedom under high pressure in the organic dimer-{M}ott insulator
  \protect{$\beta^{\prime}$-(BEDT-TTF)$_2$ICl$_2$}, Phys.\ Rev.\ B 92 (2015)
  085149.

\bibitem{Hattori17a}
Y.~Hattori, S.~Iguchi, T.~Sasaki, S.~Iwai, H.~Taniguchi, H.~Kishida,
  Electric-field-induced intradimer charge disproportionation in the
  dimer-{M}ott insulator \protect{$\beta^\prime$-(BEDT-TTF)$_2$ICl$_2$}, Phys.\
  Rev.\ B 95 (2017) 085149.

\bibitem{Kurmoo87a}
M.~Kurmoo, D.~R. Talhamn, P.~Day, I.~D. Parker, R.~H. Friend, A.~M. Stringer,
  J.~Howard, Structure and properties of a new conducting organic
  charge-transfer salt \protect{$\beta$-(BEDT-TTF)$_2$AuBr$_2$}, Solid St.\
  Comm. 61 (1987) 459--464.

\bibitem{Ugawa95a}
A.~Ugawa, D.~B. Tanner, K.~Yakushi, Far-infrared reflectance of
  \protect{$\beta^{\prime\prime}$-(BEDT-TTF)$_2$AuBr$_2$}: co-existence of free
  carriers and a single-particle gap at \protect{2$\Delta$=130cm$^{-1}$},
  Synth.\ Metals 70 (1995) 979--980.

\bibitem{Girlando14a}
A.~Girlando, M.~Masino, J.~A. Schlueter, N.~Drichko, S.~Kaiser, M.~Dressel,
  Charge-order fluctuations and superconductivity in two-dimensional organic
  metals, Phys.\ Rev.\ B 89 (2014) 174503.

\bibitem{Geiser03a}
U.~Geiser, J.~A. Schlueter, A.~M. Kini, H.~H. Wang, B.~H. Ward, M.~A. Whited,
  J.~Mohtasham, G.~L. Gard, \protect{BEDT-TTF} salts with flourinated sulfonate
  anions, Synth.\ Metals 133-134 (2003) 401--403.

\bibitem{Kaiser10a}
S.~Kaiser, M.~Dressel, Y.~Sun, A.~Greco, J.~A. Schlueter, G.~L. Gard,
  N.~Drichko, Bandwidth tuning triggers interplay of charge order and
  superconductivity in two-dimensional organic materials, Phys.\ Rev.\ Lett.
  105 (2010) 206402.

\bibitem{Kubo69a}
R.~Kubo, A stochastic theory of line shape, in: K.~E. Shuler (Ed.), Stochastic
  processes in chemical physics, Vol.~15 of Advances in Chemical Physics,
  Wiley, New York, 1969, pp. 101--128.

\bibitem{Girlando12b}
A.~Girlando, M.~Masino, S.~Kaiser, Y.~Sun, N.~Drichko, M.~Dressel, H.~Mori,
  Spectroscopic characterization of charge order fluctuations in
  \protect{BEDT-TTF} metals and superconductors, Phys. Status Solidi B 249
  (2012) 953--956.

\bibitem{Nishikawa03a}
H.~Nishikawa, A.~Machida, T.~Morimoto, K.~Kikuchi, T.~Kodama, I.~Ikemoto,
  J.~Yamada, H.~Yoshino, K.~Murata, A new organic superconductor,
  \protect{(DODHT)$_2$BF$_4\cdot$H$_2$O}, Chem. Commun. 2003 (2003) 494--495.

\bibitem{Nishikawa06a}
H.~Nishikawa, Y.~Sato, T.~Kodama, K.~Kikuchi, I.~Ikemoto, J.~Yamada, H.~Oshio,
  R.~Kondo, S.~Kagoshima, Charge ordered insulating state in \protect{DODHT}
  salts, J. Low Temp. Phys. 142 (2006) 633--636.

\bibitem{Nishikawa08a}
H.~Nishikawa, H.~Oshio, M.~Higa, R.~Kondo, S.~Kagoshima, A.~Nakao, H.~Sawa,
  S.~Yasuzuka, K.~Murata, Charge ordered insulating phases of \protect{DODHT}
  salts with octahedral anions and a new radical salt,
  \protect{$\beta^{\prime\prime}$-(DODHT)$_2$TaF$_6$}, Journal of Physics:
  Conf. Series 132 (2008) --12--23.

\bibitem{Higa07a}
M.~Higa, R.~Kondo, S.~Kagoshima, H.~Nishikawa, Structural studies of
  pressure-induced organic superconductor
  \protect{$\beta^{\prime\prime}$-(DODHT)$_2$PF$_6$} having charge-ordered
  phase at ambient pressure, J.\ Phys.\ Soc.\ Jpn. 76 (2007) 034709.

\bibitem{Bangura05a}
A.~F. Bangura, A.~I. Coldea, J.~Singleton, A.~Ardavan, A.~Akutsu-Sato,
  H.~Akutsu, S.~S. Turner, P.~Day, T.~Yamamoto, K.~Yakushi, Robust
  superconducting state in the low-quasiparticle-density organic metals
  \protect{$\beta^{\prime\prime}$-(BEDT-TTF)$_4$[(H$_3$O)M(C$_2$O$_4$)$_3$]-Y}
  : {S}uperconductivity due to proximity to a charge-ordered state, Phys.\
  Rev.\ B 72 (2005) 014543.

\bibitem{Coldea04a}
A.~I. Coldea, A.~F. Bangura, J.~Singleton, A.~Ardavan, A.~Akutsu-Sato,
  H.~Akutsu, S.~S. Turner, P.~Day, Fermi-surface topology and the effects of
  intrinsic disorder in a class of charge-transfer salts containing magnetic
  ion: \protect{$\beta^{\prime\prime}$-(BEDT-TTF)$_4$[(H$_3$O){\it
  M}(C$_2$O$_4$)$_3$]{\it Y}} \protect{({\it M}=Ga, Cr, Fe; {\it
  Y}=C$_5$H$_5$N)}, Phys.\ Rev.\ B 69 (2004) 085112.

\bibitem{Kurmoo95a}
M.~Kurmoo, A.~W. Graham, P.~Day, S.~J. Coles, M.~B. Hursthouse, J.~L.
  Caulfield, J.~Singleton, F.~L. Pratt, W.~Hayes, Superconducting and
  semiconducting magnetic charge transfer salts:
  \protect{(BEDT-TTF)$_4$AFe(C$_2$O$_4$)$_3\cdot$C$_6$H$_5$CN} ({A} =
  \protect{H$_2$O}, {K}, \protect{NH$_4$}), J.\ Am.\ Chem.\ Soc. 117 (1995)
  12209--12217.

\bibitem{Rashid01a}
S.~Rashid, S.~S. Turner, P.~Day, J.~A.~K. Howard, P.~Guionneau, E.~J.~L.
  McInnes, F.~E. Mabbs, J.~H. Clark, S.~Firth, T.~Biggs, New superconducting
  charge-transfer salts
  \protect{(BEDT-TTF)$_4$[A$\cdot$M(C$_2$O$_4$)$_3$]$\cdot$C$_6$H$_5$NO$_2$}
  \protect{(A=H$_3$O or NH$_4$, M=Cr or Fe,
  BEDT-TTF=bis(ethylenedithio)tetrathiafulvalene)}, J. Mater. Chem. 2001 (2001)
  2095--2101.

\bibitem{Yamamoto08a}
T.~Yamamoto, H.~M. Yamamoto, R.~Kato, M.~Uruichi, K.~Yakushi, H.~Akutsu,
  A.~Sato-Akutsu, A.~Kawamoto, S.~S. Turner, P.~Day, Inhomogeneous site charges
  at the bounary between insulating, superconducting, and metallic phases of
  \protect{$\beta^{\prime\prime}$-type} bis-ethylenedithio-tetrathiafulvalene
  molecular charge-transfer salts, Phys.\ Rev.\ B 77 (2008) 205120.

\bibitem{Mori87c}
T.~Mori, H.~Inokuchi, Structural and electrical properties of
  \protect{(BEDT-TTF)$_3$Cl$_2$(H$_2$O)$_2$}, Chem. Lett. 16 (1987) 1657.

\bibitem{Gaultier99a}
J.~Gaultier, S.~\protect{H\'ebrard-Bracchetti}, P.~Guionneau, C.~J. Kepert,
  D.~Chasseau, L.~Ducasse, Y.~Barrans, M.~Kurmoo, P.~Day, Structural properties
  of the superconducting salt \protect{(BEDT-TTF)$_3$Cl$_3\cdot$(H$_2$O)$_2$}
  at low temperatures, J. Solid State Chem. 145 (1999) 496--502.

\bibitem{Nagata11a}
S.~Nagata, T.~Ogura, A.~Kawamoto, H.~Taniguchi, \protect{$^{13}$C-NMR} studies
  of the paramagnetic and charge-ordered states of the organic superconductor
  \protect{$\beta^{\prime\prime}$-(BEDT-TTF)$_3$Cl$_2\cdot$2H$_2$O} under
  pressure, Phys.\ Rev.\ B 84 (2011) 035105.

\bibitem{Kobayashi86b}
H.~Kobayashi, R.~Kato, A.~Kobayashi, Y.~Nishio, K.~Kajita, W.~Sasaki, A new
  molecular superconductor,
  \protect{(BEDT-TTF)$_2$(I$_3$)$_{1-x}$(AuI$_2$)$_x$} \protect{($x<0.02$)},
  Chem. Lett. 15~(5) (1986) 789--792.

\bibitem{Tamura88a}
M.~Tamura, K.~Yakushi, H.~Kuroda, A.~Kobayashi, R.~Kato, H.~Kobayashi,
  Temperature dependence of the polarized reflectance spectra of the
  $\theta$-type of bis(ethylenedithio)tetrathiafulvalenium triiodide
  \protect{$\theta$-(BEDT-TTF)$_2$I$_3$}: estimation of band parameters, J.\
  Phys.\ Soc.\ Jpn. 57 (1988) 3239--3247.

\bibitem{Takenaka05a}
K.~Takenaka, M.~Tamura, N.~Tajima, H.~Takagi, Collapse of coherent
  quasiparticle states in \protect{$\theta$-(BEDT-TTF)$_2$I$_3$} observed by
  optical spectroscopy, Phys.\ Rev.\ Lett. 95 (2005) 227801.

\bibitem{Mori95a}
H.~Mori, I.~Hirabayashi, S.~Tanaka, T.~Mori, Y.~Maruyama, Crystal structures
  and electrical resistivities of three-component organic conductors:
  \protect{(BEDT-TTF)$_2$MM$^\prime$(SCN)$_4$} [{M} = {K}, {R}b, {C}s;
  {M$^\prime$} = {C}o, {Z}n, {C}d], Bull. Chem. Soc. Jpn. 68 (1995) 1136.

\bibitem{Mori97a}
H.~Mori, S.~Tanaka, T.~Mori, A.~Fuse, Crystal structure and physical properties
  of \protect{$\alpha$-(BEDT-TTF)$_2$ MM$^\prime$(SCN)$_4$} [{M}={R}b, {C}s,
  {M}$^\prime$={C}o, {Z}n], Synth.\ Metals 86 (1997) 1789.

\bibitem{Mori98b}
H.~Mori, S.~Tanaka, T.~Mori, Systematic study of the electronic state in
  \protect{$\theta$}---type \protect{BEDT-TTF} organic conductors by changing
  the electronic correlation, Phys.\ Rev.\ B 57 (1998) 12023--12029.

\bibitem{Watanabe04c}
M.~Watanabe, Y.~Noda, Y.~Nogami, H.~Mori, Structural phase transition of
  \protect{$\theta$-(BEDT-TTF)$_2$RbZn(SCN)$_4$} under high pressure, J.\
  Phys.\ Soc.\ Jpn. 73 (2004) 921--925.

\bibitem{Miyagawa00a}
K.~Miyagawa, A.~Kawamoto, K.~Kanoda, Charge ordering in a quasi-two-dimensional
  organic conductor, Phys.\ Rev.\ B 62 (2000) R7679--R7682.

\bibitem{Watanabe04a}
M.~Watanabe, Y.~Noda, Y.~Nogami, H.~Mori, Transfer integrals and the spatial
  pattern of charge ordering in \protect{$\theta$-(BEDT-TTF)$_2$RbZn(SCN)$_4$}
  at 90 {K}, J.\ Phys.\ Soc.\ Jpn. 73 (2004) 116--122.

\bibitem{Chiba01a}
R.~Chiba, H.~M. Yamamoto, K.~Hiraki, T.~Nakamura, Charge ordering in
  \protect{$\theta$-(BEDT-TTF)$_2$RbZn(SCN)$_4$}, Synth.\ Metals 120 (2001)
  919--920.

\bibitem{Chiba01b}
R.~Chiba, H.~Yamamoto, K.~Hiraki, T.~Takahashi, Charge disproportionation in
  \protect{(BEDT-TTF)$_2$RbZn(SCN)$_4$}, J. Phys. Chem. Solids 62 (2001)
  389--391.

\bibitem{Yamamoto02a}
K.~Yamamoto, K.~Yakushi, K.~Miyagawa, K.~Kanoda, A.~Kawamoto, Charge ordering
  in \protect{$\theta$-(BEDT-TTF)$_2$RbZn(SCN)$_4$} studied by vibrational
  spectroscopy, Phys.\ Rev.\ B 65 (2002) 085110.

\bibitem{Watanabe03a}
M.~Watanabe, Y.~Noda, Y.~Nogami, H.~Mori, S.~Tanaka, Structural phase
  transition in \protect{$\theta$-(BEDT-TTF)$_2$RbM$^\prime$(SCN)$_4$}
  \protect{(M$^\prime$=Zn,Co)}, Synth.\ Metals 133--134 (2003) 283--285.

\bibitem{Nad06c}
F.~Nad, P.~Monceau, H.~M. Yamamoto, Dielectric response in the charge-ordered
  \protect{$\theta$-(BEDT-TTF)$_2$RbZn(SCN)$_4$} organic compound, J. Phys.:
  Condens. Matter 18 (2006) L509--L514.

\bibitem{Chiba04a}
R.~Chiba, K.~Hiraki, T.~Takahashi, H.~M. Yamamoto, T.~Nakamura, Extremely slow
  charge fluctuations in the metallic state of the two-dimensional molecular
  conductor \protect{$\theta$-(BEDT-TTF)$_2$RbZn(SCN)$_4$}, Phys.\ Rev.\ Lett.
  93 (2004) 216405.

\bibitem{Watanabe03c}
M.~Watanabe, Y.~Noda, Y.~Nogami, H.~Mori, Investigation of {X}-ray diffuse
  scattering in \protect{$\theta$-(BEDT-TTF)$_2$RbM$^\prime$(SCN)$_4$}, Synth.\
  Metals 135--136 (2003) 665--666.

\bibitem{Watanabe07a}
M.~Watanabe, Y.~Noda, Y.~Nogami, H.~Mori, Spin-{P}eierls transition in the
  charge-ordered organic conductor
  \protect{$\theta$-BEDT-TTF$_{2}$RbZn(SCN)$_{4}$}, J.\ Phys.\ Soc.\ Jpn. 76
  (2007) 124602.

\bibitem{Nad08a}
F.~Nad, P.~Monceau, H.~M. Yamamoto, A possible glass-like state in
  \protect{$\theta$-(BEDT-TTF)$_2$CsZn(SCN)$_4$} at low temperature, J. Phys.:
  Condens. Matter 20 (2008) 485211.

\bibitem{Nogami99c}
Y.~Nogami, J.~P. Pouget, M.~Watanabe, K.~Oshima, Structural modulation in
  \protect{$\theta$-(BEDT-TTF)$_2$CsM'(SCN)$_4$[M' = Co, Zn]}, Synth.\ Metals
  103 (1999) 1911.

\bibitem{Nishio99a}
Y.~Nishio, Y.~Nihei, M.~Tamura, K.~Kajita, T.~Nakamura, T.~Takahashi, Specific
  heat and metal-insulator transition of \protect{(BEDT-TTF)$_2$MZn(SCN)$_4$},
  Synth.\ Metals 103 (1999) 1907--1908.

\bibitem{Nakamura00c}
T.~Nakamura, W.~Minagawa, R.~Kinami, T.~Takahashi, Possible charge
  disproportionation and new type charge localization in
  \protect{$\theta$-(BEDT-TTF)$_2$CsZn(SCN)$_4$}, J.\ Phys.\ Soc.\ Jpn. 69
  (2000) 504--509.

\bibitem{Chiba03a}
R.~Chiba, K.~Hiraki, T.~Takanashi, H.~M. Yamamoto, T.~Nakamura, Charge ordering
  in \protect{$\theta$-(BEDT-TTF)$_2$MZn(SCN)$_4$} [{M}={R}b, {C}s], Synth.\
  Metals 133--134 (2003) 305--306.

\bibitem{Chiba08a}
R.~Chiba, K.~Hiraki, T.~Takahashi, H.~M. Yamamoto, T.~Nakamura, Charge
  disproportionation and dynamics in
  \protect{$\theta$-(BEDT-TTF)$_2$CsZn(SCN)$_4$}, Phys.\ Rev.\ B 77 (2008)
  115113.

\bibitem{Suzuki05a}
K.~Suzuki, K.~Yamamoto, K.~Yakushi, A.~Kawamoto, Infrared and {R}aman studies
  of \protect{$\theta$-(BEDT-TTF)$_2$CsZn(SCN)$_4$}: comparison with the frozen
  state of \protect{$\theta$-(BEDT-TTF)$_2$RbZn(SCN)$_4$}, J.\ Phys.\ Soc.\
  Jpn. 74 (2005) 2631--2639.

\bibitem{Yamamoto04a}
K.~Yamamoto, K.~Yakushi, Y.~Shimizu, G.~Saito, Infrared and {R}aman study of
  the charge-ordered state of
  \protect{$\theta$-(ET)$_2$Cu$_2$CN[N(CN)$_2$]$_2$}, J.\ Phys.\ Soc.\ Jpn. 73
  (2004) 2326--2332.

\bibitem{Kakiuchi07b}
T.~Kakiuchi, Y.~Wakabayashi, H.~Sawa, T.~Takahashi, T.~Nakamura, Charge
  ordering in \protect{$\alpha$-(BEDT-TTF)$_{2}$I$_{3}$} by synchrotron {X}-ray
  diffraction, J.\ Phys.\ Soc.\ Jpn. 76 (2007) 113702.

\bibitem{Wojciechowski03a}
R.~Wojciechowski, K.~Yamamoto, K.~Yakushi, M.~Inokuchi, A.~Kawamoto,
  High-pressure {R}aman study of the charge ordering in
  \protect{$\alpha$-(BEDT-TTF)$_2$I$_3$}, Phys.\ Rev.\ B 67 (2003) 224105.

\bibitem{Kaminskii83a}
V.~F. Kaminskii, T.~G. Prokhorova, R.~P. Shibaeva, E.~B. Yagubskii, Crystal
  structure of the organic superconductor \protect{(BEDT-TTF)$_2$I$_3$}, JETP
  Lett. 39 (1983) 17--20.

\bibitem{Rothaemel86a}
B.~Rothaemel, L.~Forr\'o, J.~Cooper, J.~Schilling, M.~Weger, P.~Bele,
  H.~Brunner, D.~Schweitzer, H.~Keller, Magnetic susceptibility of $\alpha{}$
  and $\beta{}$ phases of di[bis(ethylenediothiolo)tetrathiafulvalene]
  tri-iodide \protect{[(BEDT-TTF)$_2$I$_3$]} under pressure, Phys. Rev. B
  34~(2) (1986) 704--712.

\bibitem{Tajima02a}
N.~Tajima, A.~Ebina-Tajima, M.~Tamura, Y.~Nishio, K.~Kajita, Effects of
  uniaxial strain on transport properties of organic conductor
  \protect{$\alpha$-(BEDT-TTF)$_2$I$_3$} and discovery of superconductivity,
  J.\ Phys.\ Soc.\ Jpn. 71 (2002) 1832--1835.

\bibitem{Moldenhauer93a}
J.~Moldenhauer, C.~Horn, K.~I. Pokodnia, D.~Schweitzer, I.~Heinen, H.~J.
  Keller, \protect{FT-IR} absorption spectroscopy of \protect{BEDT-TTF} radical
  salts: charge transfer and donor-anion interaction, Synth.\ Metals 60 (1993)
  31--38.

\bibitem{Heidmann92a}
C.-P. Heidmann, A.~Barnsteiner, F.~\protect{Gro{\ss}-Alltag}, B.~S.
  Chandrasekhar, E.~Hess, Anisotropic thermal expansion of the organic
  conductor \protect{$\alpha$-(BEDT-TTF)$_2$I$_3$}: new aspects of the
  metal-insulator transition, Solid St.\ Comm. 84 (1992) 711--716.

\bibitem{Takano01a}
Y.~Takano, K.~Hiraki, H.~M. Yamamoto, T.~Nakamura, T.~Takahashi, Charge
  disproportionation in the organic conductor,
  \protect{$\alpha$-(BEDT-TTF)$_2$I$_3$}, J. Phys. Chem. Solids 62 (2001) 393.

\bibitem{Kawai09a}
T.~Kawai, A.~Kawamoto, \protect{$^{13}$C-NMR} study of charge ordering state in
  the organic conductor, \protect{$\alpha$-(BEDT-TTF)$_2$I$_3$}, J.\ Phys.\
  Soc.\ Jpn. 78 (2009) 074711.

\bibitem{Lunkenheimer15b}
P.~Lunkenheimer, B.~Hartmann, M.~Lang, J.~\protect{M\"uller}, D.~Schweitzer,
  S.~Krohns, A.~Loidl, Ferroelectric properties of charge-ordered
  \protect{$\alpha$-(BEDT-TTF)$_2$I$_3$}:, Phys.\ Rev.\ B 91 (2015) 245132.

\bibitem{Hirata15a}
M.~Hirata, K.~Ishikawa, K.~Miyagawa, M.~Tamura, C.~Berthier, D.~Basko,
  A.~Kobayashi, G.~Matsuno, K.~Kanoda, Observation of an anisotropic {D}irac
  cone reshaping and ferrimagnetic spin polarization in an organic conductor,
  Nature Commun. 7 (2015) 12666.

\bibitem{Mishima95a}
T.~Mishima, T.~Ojiro, Y.~Nishio, Y.~Iye, Temperature independent conductivity
  of \protect{$\alpha$-(BEDT-TTF)$_2$I$_3$} -compensation of the change in the
  carrier density and the mobility-, Synth.\ Metals 70 (1995) 771--774.

\bibitem{Tajima00b}
N.~Tajima, M.~Tamura, Y.~Nishio, K.~Kajita, Y.~Iye, Transport property of an
  organic conductor \protect{$\alpha$-(BEDT-TTF)$_2$I$_3$} under high pressure
  --{D}iscovery of a novel type of conductor--, J.\ Phys.\ Soc.\ Jpn. 69 (2000)
  543--551.

\bibitem{Katayama06a}
S.~Katayama, A.~Kobayashi, Y.~Suzumura, Pressure-induced zero-gap
  semiconducting state in organic conductor
  \protect{$\alpha$-(BEDT-TTF)$_2$I$_3$} salt, J.\ Phys.\ Soc.\ Jpn. 75 (2006)
  054705.

\bibitem{Kobayashi07a}
A.~Kobayashi, S.~Katayama, Y.~Suzumura, H.~Fukuyama, Massless {F}ermions in
  organic conductor, J.\ Phys.\ Soc.\ Jpn. 76 (2007) 034711.

\bibitem{Kajita14a}
K.~Kajita, Y.~Nishio, N.~Tajima, Y.~Suzumura, A.~Kobayashi, Molecular {D}irac
  {F}ermion systems -- theoretical and experimental approaches --, J.\ Phys.\
  Soc.\ Jpn. 83 (2014) 072002.

\bibitem{Andres05a}
D.~Andres, M.~V. Kartsovnik, W.~Biberacher, K.~Neumaier, E.~Schuberth,
  H.~M\protect{\"u}ller, Superconductivity in the charge-density-wave state of
  the organic metal \protect{$\alpha$-(BEDT-TTF)$_2$KHg(SCN)$_4$}, Phys.\ Rev.\
  B 72 (2005) 174513.

\bibitem{Maesato01a}
M.~Maesato, Y.~Kaga, R.~Kondo, S.~Kagoshima, Control of electronic properties
  of \protect{$\alpha$-(BEDT-TTF)$_2${\it M}Hg(SCN)$_4$} \protect{({\it M}=K,
  NH$_4$)} by the uniaxial strain method, Phys.\ Rev.\ B 64 (2001) 155104.

\bibitem{Noda13a}
K.~Noda, Y.~Ihara, A.~Kawamoto, Charge disproportionation with lattice
  distortion of \protect{$\alpha$-(BEDT-TTF)$_2$RbHg(SCN)$_4$} observed by
  \protect{$^{13}$C-NMR}, Phys.\ Rev.\ B 87 (2013) 085105.

\bibitem{Kawai08a}
T.~Kawai, A.~Kawamoto, Charge disproportionation and inversion symmetry
  breaking in organic conductor \protect{$\alpha$-(BEDT-TTF)$_2$RbHg(SCN)$_4$},
  Phys.\ Rev.\ B 78 (2008) 165119.

\bibitem{Oshima89a}
M.~Oshima, H.~Mori, G.~Saito, K.~Oshima, Crystal structures and electrical
  properties of \protect{BEDT-TTF} salts of mercury({I}{I}) and thiocyanate
  with and without {K} ion, Chem. Lett. 18 (1989) 1159--1162.

\bibitem{Mori90b}
H.~Mori, S.~Tanaka, M.~Oshima, G.~Saito, T.~Mori, Y.~Maruyama, H.~Inokuchi,
  Crystal and electronic structures of \protect{(BEDT-TTF)$_2$[MHg(SCN)$_4$]}
  ({M}={K} and \protect{NH$_4$}), Bull. Chem. Soc. Jpn. 63 (1990) 2183--2190.

\bibitem{Mori90c}
H.~Mori, S.~Tanaka, K.~Oshima, M.~Oshima, G.~Saito, T.~Mori, Y.~Maruyama,
  H.~Inokuchi, Electrical properties and crystal structures of mercury({I}{I})
  and thiocyantate salts based upon \protect{BEDT-TTF} with \protect{Li$^+$},
  \protect{K$^+$}, \protect{NH$_4^+$}, \protect{Rb$^+$}, and \protect{Cs$^+$},
  Solid St.\ Comm. 74 (1990) 1261--1264.

\bibitem{Wang90b}
H.~H. Wang, K.~D. Carlson, U.~Geiser, W.~K. Kwok, M.~D. Vashon, J.~E. Thompson,
  N.~F. Larsen, G.~D. \protect{McCabe}, R.~S. Hulscher, J.~M. Williams, A new
  ambient-pressure organic superconductor:
  \protect{(BEDT-TTF)$_2$(NH)$_4$Hg(SCN)$_4$}, Physica C 166 (1990) 57--61.

\bibitem{Kushch92a}
N.~D. Kushch, L.~I. Buravov, M.~V. Kartsovnik, V.~N. Laukhin, S.~I. Pesotskii,
  R.~P. Shibaeva, L.~P. Rozenberg, E.~B. Yagubskii, A.~V. Zvarikina, Resistance
  and magnetoresistance anomaly in a new stable organic metal
  \protect{(ET)$_2$TlHg(SCN)$_4$}, Synth.\ Metals 46 (1992) 271--276.

\bibitem{Sasaki90a}
T.~Sasaki, N.~Toyota, M.~Tokumoto, N.~Kinoshita, H.~Anzai, Transport properties
  of organic conductor \protect{(BEDT-TTF)$_2$KHg(SCN)$_4$}: {I}. {R}esistance
  and magnetoresistance anomaly, Solid St.\ Comm. 75 (1990) 93--96.

\bibitem{Ito95a}
H.~Ito, M.~V. Kartosvnik, H.~Ishimoto, K.~Kono, H.~Mori, N.~D. Kushch,
  G.~Saito, T.~Ishiguro, S.~Tanaka, Superconductivity in
  \protect{$\alpha$-(BEDT-TTF)$_2$MHg(SCN)$_4$} ({M}={K}, {R}b, {T}l,
  {N}h$_4$), Synth.\ Metals 70 (1995) 899--902.

\bibitem{Ito93a}
H.~Ito, H.~Kaneko, T.~Ishiguro, H.~Ishimoto, K.~Kono, S.~Horiuchi, T.~Komatsu,
  G.~Saito, On superconductivity of the organic conductor
  \protect{$\alpha$-(BEDT-TTF)$_2$KHg(SCN)$_4$}, Solid St.\ Comm. 85 (1993)
  1005--1009.

\bibitem{Kartsovnik07a}
M.~V. Kartsovnik, D.~Andres, W.~Biberacher, Hierarchy of the density-wave
  states and superconductivity in the organic conductor
  \protect{$\alpha$-(BEDT-TTF)$_2$KHg(SCN)$_4$}, C. R. Chimie 10 (2007) 6--14.

\bibitem{Sasaki95a}
T.~Sasaki, N.~Toyota, Mysterious ground states in the organic conductor
  \protect{$\alpha$-(BEDT-TTF)$_2$KHg(SCN)$_4$}: {M}ixed \protect{SDW} and
  \protect{CDW}?, Synth.\ Metals 70 (1995) 849--852.

\bibitem{Kinoshita91a}
N.~Kinoshita, M.~Tokumoto, H.~Anzai, Electron spin resonance and electric
  resistance anomaly of \protect{(BEDT-TTF)$_2$RbHg(SCN)$_4$}, J.\ Phys.\ Soc.\
  Jpn. 60 (1991) 2131--2134.

\bibitem{Foury-Leylekian03a}
P.~Foury-Leylekian, S.~Ravy, J.-P. Pouget, H.~\protect{M\"uller}, X-ray study
  of the density wave instability of
  \protect{$\alpha$-(BEDT-TTF)$_2$MHg(SCN)$_4$} with {M}={K} and {R}b, Synth.\
  Metals 137 (2003) 1271--1272.

\bibitem{Foury-Leylekian10a}
P.~Foury-Leylekian, J.-P. Pouget, Y.-J. Lee, R.~M. Nieminen,
  P.~\protect{Ordej\'on}, E.~Canadell, Density-wave instability in
  \protect{$\alpha$-(BEDT-TTF)$_2$KHg(SCN)$_4$} studied by x-ray diffuse
  scattering and by first-principles calculations, Phys.\ Rev.\ B 82 (2010)
  134116.

\bibitem{Pratt95a}
F.~L. Pratt, T.~Sasaki, N.~Toyota, K.~Nagamine, Zero field muon spin relaxation
  study of the low temperature state in
  \protect{$\alpha$-(BEDT-TTF)$_2$KHg(SCN)$_4$}, Phys.\ Rev.\ Lett. 74 (1995)
  3892.

\bibitem{Dressel03a}
M.~Dressel, N.~Drichko, J.~Schlueter, J.~Merino, Proximity of the layered
  organic conductors \protect{$\alpha$-(BEDT-TTF)$_2${\it M}Hg(SCN)$_4$}
  (\protect{{\it M}}={K}, protect{NH$_4$}) to a charge-ordering transition,
  Phys.\ Rev.\ Lett. 90.

\bibitem{Hiejima10a}
T.~Hiejima, S.~Yamada, M.~Uruichi, K.~Yakushi, Infrared and {R}aman studies of
  \protect{$\alpha$-(BEDT-TTF)$_2$MHg(SCN)$_4$} with \protect{M=NH$_4$} and {K}
  at low temperature: breaking of inversion symmetry due to charge-ordering
  fluctuation, Physica B 405 (2010) S153--S156.

\bibitem{Kato14a}
R.~Kato, Development of $\pi$-electron systems based on \protect{[M(dmit)$_2$]}
  ({M}={N}i and {P}d; dmit: 1,3-dithiole-2-thione-4,5-dithiolate) anion
  radicals, Bull. Chem. Soc. Jpn. 87 (2014) 335--374.

\bibitem{Kobayashi92a}
H.~Kobayashi, K.~Bun, T.~Naito, R.~Kato, A.~Kobayashi, New molecular
  superconductor, \protect{[Me$_2$Et$_2$N][Pd(dmit)$_2$]$_2$}, Chem. Lett. 21
  (1992) 1909--1912.

\bibitem{Kobayashi87b}
A.~Kobayashi, H.~Kim, Y.~Sasaki, R.~Kato, H.~Kobayashi, S.~Moriyama, Y.~Nishio,
  K.~Kajita, W.~Sasaki, The first molecular superconductor based on
  $\pi$-acceptor molecules and closed-shell cations,
  \protect{[(CH$_3$)$_4$N][Ni(dmit)$_2$]$_2$}, low-temperature {X}-ray studies
  and superconducting transition, Chem. Lett. 16 (1987) 1819--1822.

\bibitem{Kobayashi91b}
A.~Kobayashi, H.~Kobayashi, A.~Miyamoto, R.~Kato, R.~A. Clark, A.~E. Underhill,
  New molecular superconductor,
  \protect{$\beta$-[(CH$_3$)$_4$N][Pd(dmit)$_2$]$_2$}, Chem. Lett. 20 (1991)
  2163--2166.

\bibitem{Nakamura01a}
T.~Nakamura, T.~Takahashi, S.~Aonuma, R.~Kato, \protect{EPR} investigation of
  the electronic states in $\beta^\prime$-type \protect{[Pd(dmit)$_2$]$_2$}
  compounds (where dmit is 2-thioxo-1,3-dithiole-4,5-dithiolate), J. Mater.
  Chem. 11 (2001) 2159--2162.

\bibitem{Kato97a}
R.~Kato, Y.-L. Liu, Y.~Hosokoshi, S.~Aonuma, H.~Sawa, Se-substitution and
  cation effects on the high-pressure molecular superconductor,
  \protect{$\beta$-Me$_4$N[Pd(dmit)$_2$]$_2$} -- {A} unique two-band system,
  Molec. Cryst. and Liq. Cryst. Science and Technology, Sect. A 296 (1997)
  217--244.

\bibitem{Nakamura00d}
T.~Nakamura, H.~Tsukada, T.~Takahashi, S.~Aonuma, R.~Kato, Low temperature
  electronic states of $\beta^\prime$-type {P}d(dmit)$_2$ compounds, Molec.
  Cryst. Liq. Cryst 343 (2006) 187--192.

\bibitem{Rouziere98a}
S.~\protect{Rouzi\`ere}, J.-I. Yamaura, R.~Kato, Low-temperature structural
  studies of \protect{$\beta$-Pd(dmit)$_2$} conductors, Phys.\ Rev.\ B 60
  (1999) 3113.

\bibitem{Tajima05a}
A.~Tajima, A.~Nakao, R.~Kato, Uniaxial strain effects in the conducting
  \protect{Pd(dmit)$_2$} system (dmit = 1,2-dithiol-2-thione-4,5-dithiolate),
  J.\ Phys.\ Soc.\ Jpn. 74 (2005) 412--416.

\bibitem{Kobayashi90a}
A.~Kobayashi, H.~Kim, Y.~Sasaki, K.~Murata, R.~Kato, H.~Kobayashi, Crystal and
  electronic structures of new molecular conductors tetramethylammonium and
  tetramethylarsonium complexes of \protect{Pd(dmit)$_2$}, J. Chem. Soc.,
  Faraday Trans. 86 (1990) 361.

\bibitem{Kato02b}
R.~Kato, N.~Tajima, M.~Tamura, J.-I. Yamaura, Uniaxial strain effect in a
  strongly correlated two-dimensional system
  \protect{$\beta^\prime$-(CH$_3$)$_4$As[P(dmit)$_2$]$_2$}, Phys.\ Rev.\ B 66
  (2002) 020508(R).

\bibitem{Kato98a}
R.~Kato, Y.~Kashimura, S.~Aonuma, N.~Hanasaki, H.~Tajima, A new molecular
  superconductor \protect{$\beta^\prime$-Et$_2$Me$_2$P[Pd(dmit)$_2$]$_2$}
  (dmit=2-thioxo-1,3-dithiole-4,5-dithiolate), Solid St.\ Comm. 105 (1998)
  561--565.

\bibitem{Kato06a}
R.~Kato, A.~Tajima, A.~Nakao, M.~Tamura, Two pressure-induced superconducting
  anion radical salts exhibiting different spin states at ambient pressure, J.
  Am. Chem. Soc. 128 (2006) 10016--10017.

\bibitem{Itou08a}
T.~Itou, A.~Oyamada, S.~Maegawa, M.~Tamura, R.~Kato, Quantum spin liquid in the
  spin-1/2 triangular antiferromagnet \protect{EtMe$_3$Sb[Pd(dmit)$_2$]$_2$},
  Phys.\ Rev.\ B 77 (2008) 104413.

\bibitem{Yamamoto11a}
T.~Yamamoto, Y.~Nakazawa, M.~Tamura, A.~Nakao, Y.~Ikemoto, T.~Moriwaki,
  A.~Fukaya, R.~Kato, K.~Yakushi, Intradimer charge disproportionation in
  \textit{Triclinic}-{E}t{M}e$_{3}${P}[{P}d(dmit)$_{2}$]$_{2}$ (dmit:
  1,3-dithiole-2-thione-4,5-dithiolate), J.\ Phys.\ Soc.\ Jpn. 80 (2011)
  123709.

\bibitem{Otsuka14a}
K.~Otsuka, H.~Iikubo, T.~Kogure, Y.~Takano, K.~Hiraki, T.~Takahashi, H.~Cui,
  R.~Kato, Antiferromagnetic ordering in quasi-triangular localized spin system
  \protect{$\beta^\prime$-Et$_2$Me$_2$P[Pd(dmit)$_2$]$_2$}, studied by
  $^{13}${C} \protect{NMR}, J.\ Phys.\ Soc.\ Jpn. 83 (2014) 054712.

\bibitem{Kobayashi88a}
A.~Kobayashi, H.~Kim, Y.~Sasaki, S.~Moriyama, Y.~Nishio, K.~Kajita, W.~Sasaki,
  R.~Kato, H.~Kobayashi, The first molecular superconductor bases on the
  $\pi$-acceptor molecule and the closed shell cation,
  \protect{[(CH$_3$)$_4$N][Ni(dmit)$_2$]$_2$}, Synth.\ Metals 27 (1988) 339.

\bibitem{Kobayashi93a}
A.~Kobayashi, R.~Kato, R.~A. Clark, A.~E. Underhill, A.~Miyamoto, K.~Bun,
  T.~Naito, H.~Kobayashi, New molecular superconductors,
  \protect{$\beta$-[(CH$_3$)$_4$N][Pd(dmit)$_2$]$_2$} and
  \protect{[(CH$_3$)$_2$(C$_2$H$_5$)$_2$N][Pd(dmit)$_2$]$_2$}, Synth.\ Metals
  55-57 (1993) 2927--2932.

\bibitem{Abdel-Jawad13a}
M.~Abdel-Jawad, N.~Tajima, R.~Kato, I.~Terasaki, Disordered conduction in
  single-crystalline dimer {M}ott compounds, Phys.\ Rev.\ B 88 (2013) 075139.

\bibitem{Shimizu07a}
Y.~Shimizu, H.~Akimoto, H.~Tsujii, A.~Tajima, R.~Kato, Mott transition in a
  valence-bond solid insulator with a triangular lattice, Phys.\ Rev.\ Lett. 99
  (2007) 256403.

\bibitem{Tamura06a}
M.~Tamura, A.~Nakao, R.~Kato, Frustration-induced valence-bond ordering in a
  new quantum triangular antiferromagnet based on \protect{[Pd(dmit)$_2$]}, J.\
  Phys.\ Soc.\ Jpn. 75 (2006) 093701.

\bibitem{Ishii07a}
Y.~Ishii, M.~Tamura, R.~Kato, Magnetic study of pressure-induced
  superconductivity in the \protect{[Pd(dmit)$_2$]} salt with spin-gapped
  ground state, J.\ Phys.\ Soc.\ Jpn. 76 (2007) 033704.

\bibitem{Yamamoto14a}
T.~Yamamoto, Y.~Nakazawa, M.~Tamura, A.~Nakao, A.~Fukaya, R.~Kato, K.~Yakushi,
  Property of the valence-bond ordering in molecular superconductor with a
  quasi-triangular lattice, J.\ Phys.\ Soc.\ Jpn. 83 (2014) 053703.

\bibitem{Yamamoto17a}
T.~Yamamoto, T.~Fujimoto, T.~Naito, Y.~Nakazawa, M.~Tamura, K.~Yakushi,
  Y.~Ikemoto, T.~Moriwaki, R.~Kato, Charge and lattice fluctuations in
  molecule-based spin liquids, Scientic Reports 7 (2017) 12930.

\bibitem{Itou10a}
T.~Itou, A.~Oyamada, S.~Maegawa, R.~Kato, Instability in a quantum spin liquid
  in an organic triangular-lattice antiferromagnet, Nat. Phys. 6 (2010)
  673--676.

\bibitem{Itou11a}
T.~Itou, K.~Yamashita, M.~Nishiyama, A.~Oyamada, S.~Maegawa, K.~Kubo, R.~Kato,
  Nuclear magnetic resonance of the inequivalent carbon atoms in the organic
  spin-liquid material \protect{EtMe$_3$Sb[Pd(dmit)$_2$]$_2$}, Phys.\ Rev.\ B
  84 (2011) 094405.

\bibitem{Yamashita10a}
M.~Yamashita, N.~Nakata, Y.~Senshu, M.~Nagata, H.~M. Yamamoto, R.~Kato,
  T.~Shibauchi, Y.~Matsuda, Highly mobile gapless excitations in a
  two-dimensional candidate quantum spin liquid, Science 328 (2010) 1246--1248.

\bibitem{Yamashita11a}
S.~Yamashita, T.~Yamamoto, Y.~Nakazawa, M.~Tamura, R.~Kato, Gapless spin liquid
  of an organic triangular compound evidenced by thermodynamic measurements,
  Nature Commun. 2 (2011) 275.

\bibitem{Yamashita12a}
M.~Yamashita, T.~Shibauchi, Y.~Matsuda, Thermal-transport studies on
  two-dimensional quantum spin liquids, Chem. Phys. Chem. 13 (2012) 74--78.

\bibitem{Watanabe12a}
D.~Watanabe, M.~Yamashita, S.~Tonegawa, Y.~Oshima, H.~M. Yamamoto, R.~Kato,
  I.~Sheikin, K.~Behnia, T.~Terashima, S.~Uji, T.~Shibauchi, Y.~Matsuda, Novel
  {P}auli-paramagnetic quantum phase in a {M}ott insulator, Nature Commun. 3
  (2012) 1090.

\bibitem{Powell06a}
B.~J. Powell, R.~H. McKenzie, Strong electronic correlations in superconducting
  organic charge transfer salts, J. Phys: Condens. Matter 18 (2006) R827--R866.

\bibitem{Maier00a}
T.~Maier, M.~Jarrell, T.~Pruschke, J.~Keller, d-wave superconductivity in the
  {H}ubbard model, Phys.\ Rev.\ Lett. 85 (2000) 1524--1527.

\bibitem{Gull13a}
E.~Gull, O.~Parcollet, A.~J. Millis, Superconductivity and the pseudogap in the
  two-dimensional {H}ubbard model, Phys.\ Rev.\ Lett. 110 (2013) 216405.

\bibitem{Merino14a}
J.~Merino, O.~Gunnarsson, Pseudogap and singlet formation in organic and
  cuprate superconductors, Phys.\ Rev.\ B 89 (2014) 245130.

\bibitem{Merino01a}
J.~Merino, R.~H. McKenzie, Superconductivity mediated by charge fluctuations in
  layered molecular crystals, Phys.\ Rev.\ Lett. 87 (2001) 237002.

\bibitem{Fazekas74a}
P.~Fazekas, P.~W. Anderson, Ground state properties of anisotropic triangular
  antiferromagnet, Phil. Mag. 30 (1974) 423--440.

\bibitem{Chatterjee11a}
U.~Chatterjee, D.~Ai, J.~Zhao, S.~Rosenkranz, A.~Kaminski, H.~Raffy, Z.~Li,
  K.~Kadowaki, M.~Randeria, M.~R. Norman, J.~C. Compuzano, Electronic phase
  diagram of high temperature copper oxide superconductors, Proc. Natl. Acad.
  Sci. 108 (2011) 9346--9349.

\bibitem{Mishra14a}
V.~Mishra, U.~Chatterjee, J.~C. Campuzano, M.~R. Norman, Effect of the
  pseudogap on the transition temperature in the cuprates and the implications
  for its origin, Nature Physics 10 (2014) 357--360.

\bibitem{Anderson04b}
P.~W. Anderson, A suggested 4$\times$4 structure in underdoped cuprate
  superconductors: a {W}igner supersolid, arXiv:cond-mat/0406038 (2004).

\bibitem{Franz04a}
M.~Franz, Crystalline electron pairs, Science 305 (2004) 1410--1411.

\bibitem{Tesanovic04a}
A.~Tesanovic, Charge modulation, spin response, and dual {H}ofstader butterfly
  in \protect{high-$T_c$} cuprates, Phys.\ Rev.\ Lett. 93 (2004) 217004.

\bibitem{Chen04a}
H.-D. Chen, O.~Vafek, A.~Yazdani, S.-C. Zhang, Pair density wave in the
  pseudogap state of high temperature superconductors, Phys.\ Rev.\ Lett. 93
  (2004) 187002.

\bibitem{Vojta08a}
M.~Vojta, O.~\protect{R\"{o}sch}, Superconducting d-wave stripes in cuprates:
  {V}alence bond order coexisting with nodal quasiparticles, Phys.\ Rev.\ B 77
  (2008) 094504.

\bibitem{Hamidian16a}
M.~H. Hamidian, S.~D. Edkins, S.~H. Joo, A.~Kostin, H.~Eisaki, S.~Uchida, M.~J.
  Lawler, E.-A. Kim, A.~P. Mackenzie, K.~Fujita, J.~Lee, J.~C.
  \protect{S\'eamus} Davis, Detection of a {C}ooper-pair density wave in
  \protect{Bi$_2$Sr$_2$CaCu$_2$O$_{8+x}$}, Nature 532 (2016) 343--347.

\bibitem{Hirsch87a}
J.~E. Hirsch, Antiferromagnetic singlet pairs, high-frequency phonons, and
  superconductivity, Phys.\ Rev.\ B 35 (1987) 8726--8729.

\bibitem{Imada91a}
M.~Imada, Spin dimer state model and spin-phonon coupling model of
  superconductivity, J.\ Phys.\ Soc.\ Jpn. 60 (1991) 1877--1880.

\bibitem{Noack97a}
R.~M. Noack, N.~Bulut, D.~J. Scalapino, M.~G. Zacher, Enhanced
  \protect{$d_{x^2-y^2}$} pairing correlations in the two-leg {H}ubbard ladder,
  Phys.\ Rev.\ B 56 (1997) 7162.

\bibitem{Dolfi15a}
M.~Dolfi, B.~Bauer, S.~Keller, M.~Troyer, Pair correlations in doped {H}ubbard
  ladders, Phys.\ Rev.\ B 92 (2015) 195139.

\bibitem{Kashima01b}
T.~Kashima, M.~Imada, Path-integral renormalization group method for numerical
  study on ground states of strongly correlated electronic systems, J.\ Phys.\
  Soc.\ Jpn. 70 (2001) 2287--2299.

\bibitem{Hotta12a}
C.~Hotta, Theories on frustrated electrons in two-dimensional organic solids,
  Crystals 2012 (2012) 1155--1200.

\bibitem{Mizusaki06a}
T.~Mizusaki, M.~Imada, Gapless quantum spin liquid, stripe, and
  antiferromagnetic phases in frustrated {H}ubbard models in two dimensions,
  Phys.\ Rev.\ B 74 (2006) 014421.

\bibitem{Morita02a}
H.~Morita, S.~Watanabe, M.~Imada, Nonmagnetic insulating states near the {M}ott
  transitions on lattices with geometrical frustration and implications for
  \protect{$\kappa$-(ET)$_2$Cu$_2$(CN)$_3$}, J.\ Phys.\ Soc.\ Jpn. 71 (2002)
  2109--2112.

\bibitem{Watanabe03b}
S.~Watanabe, Absence of unit-cell doubling in nonmagnetic insulator phase on
  two-dimensional lattice with geometrical frustration, J.\ Phys.\ Soc.\ Jpn.
  72 (2003) 2042--2045.

\bibitem{Yoshioka09a}
T.~Yoshioka, A.~Koga, N.~Kawakami, Quantum phase transitions in the {H}ubbard
  model on a triangular lattice, Phys.\ Rev.\ Lett. 103 (2009) 036401.

\bibitem{Tocchio14a}
L.~F. Tocchio, C.~Gros, R.~\protect{Valent\'i}, F.~Becca, One-dimensional spin
  liquid, collinear, and spiral phases from uncoupled chains to the triangular
  lattice, Phys.\ Rev.\ B 89 (2014) 235107.

\bibitem{Laubach15a}
M.~Laubach, R.~Thomale, C.~Platt, W.~Hanke, G.~Li, Phase diagram of the
  {H}ubbard model on the anisotropic triangular lattice, Phys.\ Rev.\ B 91
  (2015) 245125.

\bibitem{Acheche16a}
S.~Acheche, A.~Reymbaut, M.~Charlebois, D.~\protect{S\'en\'echal}, A.-M.~S.
  Tremblay, Mott transition and magnetism on the anisotropic triangular
  lattice, Phys.\ Rev.\ B 94 (2016) 245133.

\bibitem{Goto16a}
S.~Goto, S.~Kurihara, D.~Yamamoto, Incommensurate spiral magnetic order on
  anisotropic triangular lattice: Dynamical mean-field study in a spin-rotating
  frame, Phys.\ Rev.\ B 94 (2016) 245145.

\bibitem{Kino98a}
H.~Kino, H.~Kontani, Phase diagram of superconductivity on the anisotropic
  triangular lattice {H}ubbard model: An effective model of
  \protect{$\kappa$-(BEDT-TTF)} salts, J.\ Phys.\ Soc.\ Jpn. 67 (1998) 3691.

\bibitem{Schmalian98a}
J.~Schmalian, Pairing due to spin fluctuations in layered organic
  superconductors, Phys.\ Rev.\ Lett. 81 (1998) 4232--4235.

\bibitem{Kondo98a}
H.~Kondo, T.~Moriya, Spin fluctuation-induced superconductivity in organic
  compounds, J.\ Phys.\ Soc.\ Jpn. 67 (1998) 3695--3698.

\bibitem{Vojta99a}
M.~Vojta, E.~Dagotto, Indications of unconventional superconductivity in doped
  and undoped triangular antiferromagnets, Phys.\ Rev.\ B 59 (1999) R713--716.

\bibitem{Baskaran03a}
G.~Baskaran, {M}ott insulator to high \protect{$T_c$} superconductor via
  pressure: Resonating valence bond theory and prediction of new systems,
  Phys.\ Rev.\ Lett. 90 (2003) 197007.

\bibitem{Liu05a}
J.~Liu, J.~Schmalian, N.~Trivedi, Pairing and superconductivity driven by
  strong quasiparticle renormalization in two-dimensional organic charge
  transfer salts, Phys.\ Rev.\ Lett. 94 (2005) 127003.

\bibitem{Kyung06a}
B.~Kyung, A.~M.~S. Tremblay, {M}ott transition, antiferromagnetism, and d-wave
  superconductivity in two-dimensional organic conductors, Phys.\ Rev.\ Lett.
  97 (2006) 046402.

\bibitem{Yokoyama06a}
H.~Yokoyama, M.~Ogata, Y.~Tanaka, Mott transitions and $d$-wave
  superconductivity in half-filled-band {H}ubbard model on square lattice with
  geometric frustration, J.\ Phys.\ Soc.\ Jpn. 75~(11) (2006) 114706.

\bibitem{Watanabe06a}
T.~Watanabe, H.~Yokoyama, Y.~Tanaka, J.~Inoue, Superconductivity and a {M}ott
  transition in a {H}ubbard model on an anisotropic triangular lattice, J.\
  Phys.\ Soc.\ Jpn. 75 (2006) 074707.

\bibitem{Sahebsara06a}
P.~Sahebsara, D.~\protect{S\'en\'echal}, Antiferromagnetism and
  superconductivity in layered organic conductors: Variational cluster
  approach, Phys.\ Rev.\ Lett. 97 (2006) 257004.

\bibitem{Nevidomskyy08a}
A.~H. Nevidomskyy, C.~Scheiber, D.~S\'en\'echal, A.-M.~S. Tremblay, Magnetism
  and d-wave superconductivity on the half-filled square lattice with
  frustration, Phys.\ Rev.\ B 77 (2008) 064427.

\bibitem{Sentef11a}
M.~Sentef, P.~Werner, E.~Gull, A.~Kampf, Superconducting phase and pairing
  fluctuations in the half-filled two-dimensional {H}ubbard model, Phys.\ Rev.\
  Lett. 107 (2001) 126401.

\bibitem{Hebert15a}
C.~D. Hebert, P.~Semon, A.~M.~S. Tremblay, Superconducting dome in doped
  quasi-two-dimensional organic {M}ott insulators: A paradigm for strongly
  correlated superconductivity, Phys.\ Rev.\ B 92 (2015) 195112.

\bibitem{Mizusaki04a}
T.~Mizusaki, M.~Imada, Quantum-number projection in the path-integral
  renormalization group method, Phys.\ Rev.\ B 69 (2004) 125110.

\bibitem{Aimi07a}
T.~Aimi, M.~Imada, Does simple two-dimensional {H}ubbard model account for
  high-\protect{T$_c$} superconductivity in copper oxides?, J.\ Phys.\ Soc.\
  Jpn. 76 (2007) 113708.

\bibitem{Tocchio09a}
L.~F. Tocchio, A.~Parola, C.~Gros, F.~Becca, Spin-liquid and magnetic phases in
  the anisotropic triangular lattice: The case of $\kappa$-({ET})$_2${X},
  Phys.\ Rev.\ B 80 (2009) 064419.

\bibitem{Watanabe08a}
T.~Watanabe, H.~Yokoyama, Y.~Tanaka, J.~Inoue, Predominant magnetic states in
  the {H}ubbard model on anisotropic triangular lattices, Phys.\ Rev.\ B 77
  (2008) 214505.

\bibitem{Yokoyama13a}
H.~Yokoyama, M.~Ogata, Y.~Tanaka, K.~Kobayashi, H.~Tsuchiura, Crossover between
  {B}{C}{S} superconductor and doped {M}ott insulator of $d$-wave pairing state
  in two-dimensional {H}ubbard model, J.\ Phys.\ Soc.\ Jpn. 82 (2013) 014707.

\bibitem{Yanagisawa16a}
T.~Yanagisawa, Crossover from weakly to strongly correlated regions in the
  two-dimensional {H}ubbard model---off-diagonal wavefunction {M}onte {C}arlo
  studies of {H}ubbard model {I}{I}---, J.\ Phys.\ Soc.\ Jpn. 85 (2016) 114707.

\bibitem{Tocchio16a}
L.~F. Tocchio, F.~Becca, S.~Sorella, Hidden {M}ott transition and
  \protect{large-$U$} superconductivity in the two-dimensional {H}ubbard model,
  Phys.\ Rev.\ B 94 (2016) 195126.

\bibitem{Ehlers17a}
G.~Ehlers, S.~R. White, R.~M. Noack, Hybrid-space density matrix
  renormalization group study of the doped two-dimensional {H}ubbard model,
  Phys.\ Rev.\ B 95 (2017) 125125.

\bibitem{Gan05a}
J.~Y. Gan, Y.~Chen, Z.~B. Su, F.~C. Zhang, Gossamer superconductivity near
  antiferromagnetic {M}ott insulator in layered organic conductors, Phys.\
  Rev.\ Lett. 94 (2005) 067005.

\bibitem{Powell05a}
B.~J. Powell, R.~H. McKenzie, Half-filled layered organic superconductors and
  the resonating-valence-bond theory of the {H}ubbard-{H}eisenberg model,
  Phys.\ Rev.\ Lett. 94 (2005) 047004.

\bibitem{Gan06a}
J.~Y. Gan, Y.~Chen, F.~C. Zhang, Superconducting pairing symmetries in
  anisotropic triangular quantum antiferromagnets, Phys.\ Rev.\ B 74 (2006)
  094515.

\bibitem{Powell07a}
B.~J. Powell, R.~H. McKenzie, Symmetry of the superconducting order parameter
  in frustrated systems determined by the spatial anisotropy of spin
  correlations, Phys.\ Rev.\ Lett. 98 (2007) 027005.

\bibitem{Guertler09a}
S.~Guertler, Q.~H. Wang, F.~C. Zhang, Variational {M}onte {C}arlo studies of
  gossamer superconductivity, Phys.\ Rev.\ B 79 (2009) 144526.

\bibitem{Rau11a}
J.~G. Rau, H.-Y. Kee, Emergence of superconductivity, valence bond order, and
  {M}ott insulators in $\mathrm{Pd}[(\mathrm{dmit}{)}_{2}]$ based organic
  salts, Phys.\ Rev.\ Lett. 106 (2011) 056405.

\bibitem{Gomes13a}
N.~Gomes, R.~T. Clay, S.~Mazumdar, Absence of superconductivity and valence
  bond order in the {H}ubbard-{H}eisenberg model for organic charge-transfer
  solids, J. Phys. Condens. Matter 25 (2013) 385603.

\bibitem{Jiang12a}
H.-C. Jiang, H.~Yao, L.~Balents, Spin liquid ground state of the
  spin-$\frac{1}{2}$ square \protect{$J_1$-$J_2$} {H}eisenberg model, Phys.\
  Rev.\ B 86 (2012) 024424.

\bibitem{Hu13a}
W.-J. Hu, F.~Becca, A.~Parola, S.~Sorella, Direct evidence for a gapless
  \protect{$Z_2$} spin liquid by frustrating \protect{N\'eel}
  antiferromagnetism, Phys.\ Rev.\ B 88 (2013) 060402(R).

\bibitem{Kondo01a}
H.~Kondo, T.~Moriya, Essential importance of dimerization for the
  superconductivity in organic compounds \protect{(ET)$_2$X}, J.\ Phys.\ Soc.\
  Jpn. 70 (2001) 2800--2801.

\bibitem{Kuroki06a}
K.~Kuroki, Pairing symmetry competition in organic superconductors, J.\ Phys.\
  Soc.\ Jpn. 75 (2006) 051013.

\bibitem{Sekine13a}
A.~Sekine, J.~Nasu, S.~Ishihara, Polar charge fluctuation and superconductivity
  in organic conductors, Phys.\ Rev.\ B 87 (2013) 085133.

\bibitem{Chitra95a}
R.~Chitra, S.~Pati, H.~R. Krishnamurthy, D.~Sen, S.~Ramasesha, Density-matrix
  renormalization-group studies of the spin-1/2 {H}eisenberg system with
  dimerization and frustration, Phys.\ Rev.\ B 52 (1995) 6581--6587.

\bibitem{White96a}
S.~R. White, I.~Affleck, Dimerization and incommensurate spiral spin
  correlations in the zigzag spin chain: analogies to the {K}ondo lattice,
  Phys.\ Rev.\ B 54 (1996) 9862--9869.

\bibitem{Rovira00a}
C.~Rovira, Molecular spin ladders, Chem. Eur. J. 6 (2000) 1723--1729.

\bibitem{Ribas05a}
X.~Ribas, M.~Mas-Torrent, A.~Prez-Bentez, J.~C. Dias, H.~Alves, E.~B. Lopes,
  R.~T. Henriques, E.~Molins, I.~C. Santos, K.~Wurst, P.~Foury-Leylekian,
  M.~Almeida, J.~Veciana, C.~Rovira, Organic spin ladders from
  tetrathiafulvalene \protect{(TTF)} derivatives, Adv. Functional Materials
  15~(6) (2005) 1023--1035.

\bibitem{Dagotto99a}
E.~Dagotto, Experiments on ladders reveal a complex interplay between a
  spin-gapped normal state and superconductivity, Rep. Prog. Phys. 62 (1999)
  1525.

\bibitem{Moulopoulos92a}
K.~Moulopoulos, N.~W. Ashcroft, New low density phase of interacting electrons:
  The paired electron crystal, Phys.\ Rev.\ Lett. 69 (1992) 2555.

\bibitem{Moulopoulos93a}
K.~Moulopoulos, N.~W. Ashcroft, Many-body theory of paired electron crystals,
  Phys.\ Rev.\ B 48 (1993) 11646--11665.

\bibitem{McKenzie01a}
R.~H. McKenzie, J.~Merino, J.~B. Marston, O.~P. Sushkov, Charge ordering and
  antiferromagnetic exchange in layered molecular crystals of the theta type,
  Phys.\ Rev.\ B 64 (2001) 085109.

\bibitem{Tamura09a}
M.~Tamura, R.~Kato, Variety of valence bond states formed of frustrated spins
  on triangular lattices based on a two-level system \protect{Pd(dmit)$_2$},
  Sci. Technol. Adv. Mater. 10 (2009) 024304.

\bibitem{Li17a}
W.~Li, E.~Rose, M.~V. Tran, R.~\protect{H\"ubner}, A.~Lapinski, R.~Swietlik,
  S.~A. Torunova, E.~I. Zhilyaeva, R.~N. Lyubovskaya, M.~Dressel, The
  metal-insulator transition in the organic conductor
  \protect{$\beta^{\prime\prime}$-(BEDT-TTF)$_2$Hg(SCN)$_2$Cl}, J. Chem. Phys.
  147 (2017) 064503.

\bibitem{Gomes16a}
N.~Gomes, W.~W. \protect{De Silva}, T.~Dutta, R.~T. Clay, S.~Mazumdar, Coulomb
  enhanced superconducting pair correlations in the frustrated quarter-filled
  band, Phys.\ Rev.\ B 93 (2016) 165110.

\bibitem{Inagaki04a}
K.~Inagaki, I.~Terasaki, H.~Mori, T.~Mori, Large dielectric constant and giant
  nonlinear conduction in the organic conductor
  \protect{$\theta$-(BEDT-TTF)$_2$CsZn(SCN)$_4$}, J.\ Phys.\ Soc.\ Jpn. 73
  (2004) 3364--3369.

\bibitem{Hotta06b}
C.~Hotta, N.~Furukawa, Strong coupling theory of the spinless charges on
  triangular lattice: possible formation of a gapless charge-ordered liquid,
  Phys.\ Rev.\ B 74 (2006) 193107.

\bibitem{Nishimoto09a}
S.~Nishimoto, C.~Hotta, Density-matrix renormalization study of frustrated
  fermions on a triangular lattice, Phys.\ Rev.\ B 79 (2009) 195124.

\bibitem{Yoshida14a}
T.~Yoshida, C.~Hotta, Frustrated electrons on a spatially anisotropic
  triangular lattice: {E}mergent competition of charge orders and exotic
  disorders due to thermal fluctuations, Phys.\ Rev.\ B 90 (2014) 245115.

\bibitem{Hotta06a}
T.~Hotta, Orbital ordering phenomena in d- and f-electron systems, Rep. Prog.
  Phys. 69 (2006) 2061--2155.

\bibitem{Hotta10a}
C.~Hotta, Quantum electric dipoles in spin-liquid dimer {M}ott insulator
  \protect{$\kappa$-(ET)$_2$Cu$_2$(CN)$_3$}, Phys. Rev. B 82 (2010) 241104.

\bibitem{Naka10a}
M.~Naka, S.~Ishihara, Electronic ferroelectricity in a dimer {M}ott insulator,
  J.\ Phys.\ Soc.\ Jpn. 79 (2010) 063707.

\bibitem{Ishihara14a}
S.~Ishihara, Electronic ferroelectricity in molecular organic crystals, J.
  Phys.: Condens. Matter 26 (2014) 493201.

\bibitem{Mazumdar14a}
S.~Mazumdar, R.~T. Clay, The chemical physics of unconventional
  superconductivity, Int. J. Quant. Chem. 2014 (2014) 1053.

\bibitem{Blankenbecler81a}
R.~Blankenbecler, D.~J. Scalapino, R.~L. Sugar, Monte {C}arlo calculations of
  coupled boson-fermion systems. {I}, Phys. Rev. D 24 (1981) 2278.

\bibitem{Zhang97a}
S.~Zhang, J.~Carlson, J.~E. Gubernatis, Constrained path \protect{Monte}
  \protect{Carlo} method for fermion ground states, Phys.\ Rev.\ B 55 (1997)
  7464--7477.

\bibitem{Huang01a}
Z.~B. Huang, H.~Q. Lin, J.~E. Gubernatis, Quantum {M}onte {C}arlo study of
  spin, charge, and pairing correlations in the {t}-{t$^\prime$}-{$U$}
  {H}ubbard model, Phys.\ Rev.\ B 64 (2001) 205101.

\bibitem{White89b}
S.~R. White, D.~J. Scalapino, R.~L. Sugar, N.~E. Bickers, R.~T. Scalettar,
  Attractive and repulsive pairing interaction vertices for the two-dimensional
  {H}ubbard model, Phys.\ Rev.\ B 39 (1989) R839--R842.

\bibitem{Zhang97b}
S.~Zhang, J.~Carlson, J.~E. Gubernatis, Pairing correlations in the
  two-dimensional {H}ubbard model, Phys.\ Rev.\ Lett. 78 (1997) 4486.

\bibitem{Guerrero99a}
M.~Guerrero, G.~Ortiz., J.~E. Gubernatis, Correlated wave functions and the
  absence of long-range order in numerical studies of the {H}ubbard model,
  Phys.\ Rev.\ B 59~(3) (1999) 1706--1711.

\bibitem{DeSilva16a}
W.~W. \protect{De Silva}, N.~Gomes, S.~Mazumdar, R.~T. Clay, Coulomb
  enhancement of superconducting pair-pair correlations in a
  $\frac{3}{4}$-filled model for \protect{$\kappa$-(BEDT-TTF)$_2$X}, Phys.\
  Rev.\ B 93 (2016) 205111.

\bibitem{Zantout18a}
K.~Zantout, M.~Altmeyer, S.~Backes, R.~\protect{Valent\'i}, Superconductivity
  in correlated \protect{BEDT-TTF} molecular conductors: critical temperatures
  and gap symmetries, Phys.\ Rev.\ B 97 (2018) 014530.

\bibitem{Watanabe17a}
H.~Watanabe, H.~Seo, S.~Yunoki, Phase competition and superconductivity in
  \protect{$\kappa$-(BEDT-TTF)$_2$X}: importance of intermolecular {C}oulomb
  interactions, J.\ Phys.\ Soc.\ Jpn. 86 (2017) 033703.

\bibitem{Gomes17a}
N.~Gomes, Superconductivity in strongly correlated quarter filled systems,
  Ph.D. thesis, University of Arizona (2017).

\bibitem{Chang12a}
J.~Chang, E.~Blackburn, A.~T. Holmes, N.~B. Christensen, J.~Larsen, J.~Mesot,
  R.~Liang, D.~A. Bonn, W.~N. Hardy, A.~Watenphul, M.~v.~Zimmermann, E.~M.
  Forgan, S.~M. Hayden, Direct observation of competition between
  superconductivity and charge density wave order in {YBa$_2$Cu$_3$O$_{6.67}$},
  Nature Physics 8 (2012) 871--876.

\bibitem{Ghiringelli12a}
G.~Ghiringhelli, M.~L. Tacon, M.~Minola, S.~Blanco-Canosa, C.~Mazzoli, N.~B.
  Brookes, G.~M.~D. Luca, A.~Frano, D.~G. Hawthorn, F.~He, T.~Loew, M.~M. Sala,
  D.~C. Peets, M.~Salluzzo, E.~Schierle, R.~Sutarto, G.~A. Sawatzky,
  E.~Weschke, B.~Keimer, L.~Braicovich, Long-range incommensurate charge
  fluctuations in {(Y,Nd)Ba$_2$Cu$_3$O$_{6+x}$}, Science 337 (2012) 821--825.

\bibitem{Wu11a}
T.~Wu, H.~Mayaffre, S.~\protect{Kr\"amer}, M.~\protect{Horvati\'c},
  C.~Berthier, W.~N. Hardy, R.~Liang, D.~A. Bonn, M.-H. Julien,
  Magnetic-field-induced charge-stripe order in the high-temperature
  superconductor {YBa$_2$Cu$_3$O$_y$}, Nature 477 (2011) 191.

\bibitem{Wu13a}
T.~Wu, H.~Mayaffre, S.~\protect{Kr\"amer}, M.~\protect{Horvati\'c},
  C.~Berthier, P.~L. Kuhns, A.~P. Reyes, R.~Liang, W.~N. Hardy, D.~A. Bonn,
  M.~H. Julien, Emergence of charge order from the vortex state of a
  high-temperature superconductor, Nature Communications 4 (2013) 2113.

\bibitem{Wu15a}
T.~Wu, H.~Mayaffre, S.~\protect{Kr\"amer}, M.~\protect{Horvati\'c},
  C.~Berthier, W.~N. Hardy, R.~Liang, D.~Bonn, M.-H. Julien, Incipient charge
  order observed by \protect{NMR} in the normal state of
  \protect{YBa$_2$Cu$_3$O$_y$}, Nature Communications 6 (2015) 6438.

\bibitem{Blanco-Canosa13a}
S.~Blanco-Canosa, A.~Frano, E.~Schierle, J.~Porras, T.~Loew, M.~Minola,
  M.~Bluschke, E.~Weschke, B.~Keimer, M.~Le~Tacon, Momentum-dependent charge
  correlations in \protect{YBa$_2$Cu$_3$O$_{6+\delta}$} superconductors probed
  by resonant x-ray scattering: Evidence for three competing phases, Phys.\
  Rev.\ Lett. 110 (2013) 187001.

\bibitem{Blanco-Canosa14a}
S.~Blanco-Canosa, A.~Frano, E.~Schierle, J.~Porras, T.~Loew, M.~Minola,
  M.~Bluschke, E.~Weschke, B.~Keimer, M.~L. Tacon, Resonant x-ray scattering of
  charge-density wave correlations in \protect{YBa$_2$Cu$_3$O$_{6+x}$}, Phys.\
  Rev.\ B 90 (2014) 054513.

\bibitem{Hucker14a}
M.~H\protect{\"u}cker, N.~B. Christensen, A.~T. Holmes, E.~Blackburn, E.~M.
  Forgan, R.~Liang, D.~A. Bonn, W.~N. Hardy, O.~Gutowski, M.~v.~Zimmermann,
  S.~M. Hayden, J.~Chang, Competing charge, spin, and superconducting orders in
  underdoped \protect{YBa$_2$Cu$_3$O$_y$}, Phys.\ Rev.\ B 90 (2014) 054514.

\bibitem{Comin14a}
R.~Comin, A.~Frano, M.~M. Yee, Y.~Yoshida, H.~Eisaki, E.~Schierle, E.~Weschke,
  R.~Sutarto, F.~He, A.~Soumyanarayanan, Y.~He, M.~L. Tacon, I.~S. Elfimov,
  J.~E. Hoffman, G.~A. Sawatzky, B.~Kemier, A.~Damascelli, Charge order driven
  by {F}ermi-arc instability in {Bi$_2$Sr$_{2-x}$La$_x$CuO$_{6+\delta}$},
  Science 343 (2014) 390--392.

\bibitem{SilvaNeto14a}
E.~H. D.~S. Neto, P.~Aynajian, A.~Frano, R.~Comin, E.~Schierle, E.~Weschke,
  A.~Gyenis, J.~Wen, J.~Schneeloch, Z.~Xu, S.~Ono, G.~Gu, M.~L. Tacon,
  A.~Yazdani, Ubiquitous interplay between charge ordering and high-temperature
  superconductivity in cuprates, Science 343 (2014) 393--396.

\bibitem{SilvaNeto15a}
E.~H. D.~S. Neto, , R.~Comin, F.~He, R.~Sutaro, Y.~Jiang, R.~L. Greene, G.~A.
  Sawatzky, A.~Damascelli, Charge ordering in the electron-doped
  {Nd$_{2-x}$Ce$_x$CuO$_4$}, Science 347 (2015) 282--285.

\bibitem{Comin15a}
R.~Comin, R.~Sutarto, F.~He, E.~H. da~Silva~Neto, L.~Chauviere, A.~Frano,
  R.~Liang, W.~N. Hardy, D.~A. Bonn, Y.~Yoshida, H.~Eisaki, A.~J. Achkar, D.~G.
  Hawthorn, B.~Keimer, G.~A. Sawatzky, A.~Damascelli, Symmetry of charge order
  in cuprates, Nature Materials 14 (2015) 796--800.

\bibitem{Hanaguri04a}
T.~Hanaguri, C.~Lupien, Y.~Kohsaka, D.-H. Lee, M.~Azuma, M.~Takano, H.~Takagi,
  J.~C. Davis, A `checkerboard' electronic crystal state in lightly hole-doped
  \protect{Ca$_{2-x}$Na$_x$CuO$_2$Cl$_2$}, Nature 430 (2004) 1001--1005.

\bibitem{Shen05a}
K.~M. Shen, F.~Ronning, D.~H. Lu, F.~Baumberger, N.~J.~C. Ingle, W.~S. Lee,
  W.~Meevasana, Y.~Kohsaka, M.~Azuma, M.~Takano, H.~Takagi, Z.-X. Shen, Nodal
  quasiparticles and antinodal charge ordering in
  \protect{Ca$_{2-x}$Na$_x$CuO$_2$Cl$_2$}, Science 307 (2005) 901--904.

\bibitem{SilvaNeto16a}
E.~H. D.~S. Neto, B.~Yu, M.~Minola, R.~Sutarto, E.~Schierle, F.~Boschini,
  M.~Zonno, M.~Bluschke, J.~Higgins, Y.~Li, G.~Yu, E.~Weschke, F.~He, M.~L.
  Tacon, R.~L. Greene, M.~Greven, G.~A. Sawatzky, B.~Keimer, A.~Damascelli,
  Doping-dependent charge order correlations in electron-doped cuprates, Sci.
  Adv 2 (2016) e1600782.

\bibitem{Mazumdar18a}
The valence transition model of pseudogap, charge-order and superconductivity
  in electron- and hole-doped copper oxides, preprint, arXiv:1807.00872 (2018).

\bibitem{Chakraverty78a}
B.~K. Chakraverty, M.~J. Sienko, J.~Bonnerot, Low-temperature specific heat and
  magnetic susceptibility of nonmetallic vanadium bronzes, Phys.\ Rev.\ B 17
  (1978) 3781.

\bibitem{Chakraverty79a}
B.~K. Chakraverty, Possibility of insulator to superconductor phase transition,
  J. Phys. (Paris) 40 (1979) L--99.

\bibitem{Chakraverty80b}
B.~K. Chakraverty, Insulating ground states and nonmetal-metal transitions,
  Nature 287 (1980) 393--396.

\bibitem{Shi14a}
H.~Shi, C.~A. Jimenez-Hoyos, R.~Rodriguez-Guzman, G.~E. Scuseria, S.~Zhang,
  Symmetry-projected wave functions in quantum {M}onte {C}arlo calculations,
  Phys.\ Rev.\ B 89 (21014) 125129.

\bibitem{Berg10a}
E.~Berg, E.~Fradkin, S.~A. Kivelson, Pair-density-wave correlations in the
  {K}ondo-{H}eisenberg model, Phys.\ Rev.\ Lett. 105 (2010) 146403.

\bibitem{Mazumdar08a}
S.~Mazumdar, R.~T. Clay, Quantum critical transition from charge-ordered to
  superconducting state in the negative-{U} extended {H}ubbard model on a
  triangular lattice, Phys.\ Rev.\ B 77 (2008) 180515(R).

\bibitem{Takada03a}
K.~Takada, H.~Sakurai, E.~Takayama-Muromachi, F.~Izumi, R.~A. Dilanian,
  T.~Sasaki, Superconductivity in two-dimensional \protect{CoO$_2$} layers,
  Nature 422 (2003) 53--55.

\bibitem{Foo04a}
M.~L. Foo, Y.~Wang, S.~Watauchi, H.~W. Zandbergen, T.~He, R.~J. Cava, N.~P.
  Ong, Charge ordering, commensurability, and metallicity in the phase diagram
  of the layered \protect{Na$_x$CoO$_2$}, Phys.\ Rev.\ Lett. 92 (2004) 247001.

\bibitem{Motohashi11a}
T.~Motohashi, Y.~Sugimoto, Y.~Masubuchi, T.~Sasagawa, W.~Koshibae, T.~Tohyama,
  H.~Yamauchi, S.~Kikkawa, Impact of lithium composition on the thermoelectric
  properties of the layered cobalt oxide system {L}i$_x${C}o{O}$_2$, Phys.\
  Rev.\ B 83 (2011) 195128.

\bibitem{Sakurai15a}
H.~Sakurai, Y.~Ihara, K.~Takada, Superconductivity of cobalt hydrate,
  \protect{Na$_x$(H$_3$O)$_z$CoO$_2\cdot$yH$_2$O}, Physica C 514 (2015)
  378--387.

\bibitem{Li11a}
H.~Li, R.~T. Clay, S.~Mazumdar, Theory of carrier concentration-dependent
  electronic behavior in layered cobaltates, Phys.\ Rev.\ Lett. 106 (2011)
  216401.

\bibitem{Hasan04a}
M.~Z. Hasan, Y.-D. Chuang, D.~Qian, Y.~W. Li, Y.~Kong, A.~Kuprin, A.~V.
  Fedorov, R.~Kimmerling, E.~Rotenberg, K.~Kossnagel, Z.~Hussain, H.~Koh, N.~S.
  Rogado, M.~L. Foo, R.~J. Cava, Fermi surface and quasiparticle dynamics of
  {N}a$_{0.7}${C}o{O}$_2$ investigated by angle-resolved photoemission
  spectroscopy, Phys.\ Rev.\ Lett. 92 (2004) 246402.

\bibitem{Shimojima06a}
T.~Shimojima, K.~Ishizaka, S.~Tsuda, T.~Kiss, T.~Yokoya, A.~Chainani, S.~Shin,
  P.~Badica, K.~Yamada, K.~Togano, Angle-resolved photoemission study of the
  cobalt oxide superconductor {N}a$_x${C}o{O}$_2\cdot y${H}$_2${O}: Observation
  of the {F}ermi surface, Phys. Rev. Lett. 97 (2006) 267003.

\bibitem{Laverock07a}
J.~Laverock, S.~B. Dugdale, J.~A. Duffy, J.~Wooldridge, G.~Balakrishnan, M.~R.
  Lees, G.~q.~Zheng, D.~Chen, C.~T. Lin, A.~Andrejczuk, M.~Itou, Y.~Sakurai,
  Elliptical hole pockets in the {F}ermi surfaces of unhydrated and hydrated
  sodium cobalt oxides, Phys.\ Rev.\ B 76 (2007) 052509.

\bibitem{Singh00a}
D.~J. Singh, Electronic structure of {N}a{C}o$_2${O}$_4$, Phys. Rev. B 61
  (2000) 13397--13402.

\bibitem{Lee05b}
K.-W. Lee, W.~E. Pickett, \protect{Na$_x$CoO$_2$} in the $x\rightarrow{}0$
  regime: Coupling of structure and correlation effects, Phys. Rev. B 72~(11)
  (2005) 115110.

\bibitem{Baskaran03c}
G.~Baskaran, Electronic model for {C}o{O}$_2$ layer based systems: Chiral
  resonating valence bond metal and superconductivity, Phys.\ Rev.\ Lett. 91
  (2003) 097003.

\bibitem{Kumar03a}
B.~Kumar, B.~S. Shastry, Superconductivity in \protect{CoO$_2$} layers and the
  resonating valence bond mean-field theory of the triangular lattice t-{J}
  model, Phys.\ Rev.\ B 68 (2003) 104508.

\bibitem{Ogata03a}
M.~Ogata, Superconducting states in frustrating \protect{$t-J$} model: a model
  connecting high-{T$_c$} cuprates, organic conductors and
  \protect{Na$_x$CoO$_2$}, J.\ Phys.\ Soc.\ Jpn. 72 (2003) 1839--1842.

\bibitem{Wang04c}
Q.-H. Wang, D.-H. Lee, P.~A. Lee, Doped \protect{$t-J$} model on a triangular
  lattice: possible application to \protect{Na$_x$CoO$_2\cdot$H$_2$O} and
  \protect{Na$_{1-x}$TiO$_2$}, Phys.\ Rev.\ B 69 (2004) 092504.

\bibitem{Sakurai06a}
H.~Sakurai, N.~Tsujii, O.~Suzuki, H.~Kitazawa, G.~Kido, K.~Takada, T.~Sasaki,
  E.~Takayama-Muromachi, Valence and {N}a content dependences of
  superconductivity in {N}a$_x${C}o{O}$_2\cdot$y{H}$_2${O}, Phys. Rev. B 74
  (2006) 092502.

\bibitem{Banobre-Lopez09a}
M.~Banobre-Lopez, F.~Rivadulla, M.~A. \protect{L\'opez-Quintela}, J.~Rivas,
  Competing magnetism and superconductivity in {N}a$_x${C}o{O}$_2$ at half
  doping, J. Am. Chem. Soc. 131 (2009) 9632.

\bibitem{Kuroki04a}
K.~Kuroki, Y.~Tanaka, R.~Arita, Possible spin-triplet $f$-wave pairing due to
  disconnected {F}ermi surfaces in \protect{Na$_x$CoO$_2\cdot$H$_2$O}, Phys.\
  Rev.\ Lett. 93 (2004) 077001.

\bibitem{Ogata07a}
M.~Ogata, A new triangular system: \protect{Na$_x$CoO$_2$}, J. Phys. Condens.
  Matt. 19 (2007) 145282.

\bibitem{Sato10a}
M.~Sato, Y.~Kobayashi, T.~Moyoshi, Studies on the superconducting state of
  \protect{Na$_x$CoO$_2\cdot$yH$_2$O} -- overview, Physica C 470 (2010)
  S673--S677.

\bibitem{Johnston73a}
D.~C. Johnston, H.~Prakash, W.~H. Zachariahsen, R.~Viswanathan, High
  temperature superconductivity in the \protect{Li-Ti-O} ternary system, Mater.
  Res. Bull. 8 (1973) 77.

\bibitem{Hagino95a}
T.~Hagino, Y.~Seki, N.~Wada, S.~Tsuji, T.~Shirane, K.~Kumagai, S.~Nagata,
  Superconductivity in spinel-type compounds \protect{CuRh$_2$S$_4$} and
  \protect{CuRh$_2$Se$_4$}, Phys.\ Rev.\ B 51 (1995) 12673--12684.

\bibitem{Ito03a}
M.~Ito, J.~Hori, H.~Kurisaki, H.~Okada, A.~J.~P. Kuroki, N.~Ogita, M.~Udagawa,
  H.~Fujii, F.~Nakamura, T.~Fujita, T.~Suzuki, Pressure-induced
  superconductor-insulator transition in the spinel compound
  \protect{CuRh$_2$S$_4$}, Phys.\ Rev.\ Lett. 91 (2003) 077001.

\bibitem{Seki92a}
Y.~Seki, T.~Hagino, S.~Takayanagi, S.~Nagata, Electrical and thermal properties
  in thiospinel \protect{CuV$_2$S$_4$}, J.\ Phys.\ Soc.\ Jpn. 61 (1992)
  2597--2598.

\bibitem{Hagino94a}
T.~Hagino, Y.~Seki, S.~Takayanagi, N.~Wada, S.~Nagata, Electrical-resistivity
  and low-temperature specific-heat measurements of single crystals of
  thiospinel \protect{CuV$_2$S$_4$}, Phys.\ Rev.\ B 49 (1994) 6822--6829.

\bibitem{Hart00a}
G.~L.~W. Hart, W.~E. Pickett, E.~Z. Kurmaev, D.~Hartmann, N.~Neumann,
  A.~Moewes, D.~L. Ederer, R.~Endoh, K.~Taniguchi, S.~Nagata, Electronic
  structure of \protect{Cu$_{1-x}$Ni$_x$Rh$_2$S$_4$} and
  \protect{CuRh$_2$Se$_4$}: band-structure calculations, x-ray photoemission,
  and fluorescence measurements, Phys.\ Rev.\ B 61 (2000) 4230.

\bibitem{Satpathy87a}
S.~Satpathy, R.~M. Martin, Electronic structure of the superconducting oxide
  spinel \protect{LiTi$_2$O$_4$}, Phys.\ Rev.\ B 36 (1987) R7269.

\bibitem{Massidda88a}
S.~Massidda, J.~Yu, A.~J. Freeman, Electronic structure and properties of
  superconducting \protect{LiTi$_2$O$_4$}, Phys.\ Rev.\ B 38 (1988)
  11352--11357.

\bibitem{Kondo97a}
S.~Kondo, D.~C. Johnston, C.~A. Swenson, F.~Borsa, A.~V. Mahajan, L.~L. Miller,
  T.~Gu, A.~I. Goldman, M.~B. Maple, D.~A. Gajeski, E.~J. Freeman, N.~R.
  Dilley, R.~P. Dickey, J.~Merrin, K.~Kojima, G.~M. Luke, Y.~J. Uemura,
  O.~Chmaissem, J.~D. Jorgensen, \protect{LiV$_2$O$_4$}: A heavy {F}ermion
  transition metal oxide, Phys.\ Rev.\ Lett. 78 (1997) 3729--3732.

\bibitem{McCallum76a}
R.~W. McCallum, D.~C. Johnston, C.~A. Luengo, M.~B. Maple, Superconducting and
  normal state properties of \protect{Li$_{1+x}$Ti$_{2−x}$O$_4$} spinel
  compounds. {I}{I}. {L}ow-temperature heat capacity, J. Low Temp. Phys. 25
  (1976) 177--193.

\bibitem{Lakkis76a}
S.~Lakkis, C.~Schlenker, B.~K. Chakraverty, R.~Buder, M.~Marezio,
  Metal-insulator transitions in \protect{Ti$_4$O$_7$} single crystals: Crystal
  characterization, specific heat, and electron paramagnetic resonance, Phys.\
  Rev.\ B 14 (1976) 1429.

\bibitem{Chakraverty80a}
B.~K. Chakraverty, Charge ordering in \protect{Fe$_3$O$_4$},
  \protect{Ti$_4$O$_7$} and bipolarons, Phil. Mag. B 42 (1980) 473--478.

\bibitem{Khomskii05a}
D.~I. Khomskii, T.~Mizokawa, Orbitally induced {P}eierls state in spinels,
  Phys.\ Rev.\ Lett. 94 (2005) 156402.

\bibitem{Khomskii05b}
D.~I. Khomskii, Role of orbitals in the physics of correlated electron systems,
  Physica Scripta 72 (2005) CC8--14.

\bibitem{Radaelli05a}
P.~G. Radaelli, Orbital ordering in transition-metal spinels, New J. Phys. 7
  (2005) 53.

\bibitem{Croft07a}
M.~Croft, V.~Kiryukhin, Y.~Horibe, S.-W. Cheong, Universality in
  one-dimensional orbital wave ordering in spinel and related compounds: an
  experimental perspective, New. J. Phys. 9 (2007) 86.

\bibitem{Clay10a}
R.~T. Clay, H.~Li, S.~Sarkar, S.~Mazumdar, T.~Saha-Dasgupta, Cooperative
  orbital ordering and {P}eierls instability in the checkerboard lattice with
  doubly degenerate orbitals, Phys.\ Rev.\ B 82 (2010) 035108.

\bibitem{Radaelli02a}
P.~G. Radaelli, Y.~Horibe, M.~J. Gutmann, H.~Ishibashi, C.~H. Chen, R.~M.
  Ibberson, Y.~Koyama, Y.-S. Hor, V.~Kiryukhin, S.-W. Cheong, Formation of
  isomorphic \protect{Ir$^{3+}$} and \protect{Ir$^{4+}$} octamers and spin
  dimerization in the spinel \protect{CuIr$_2$S$_4$}, Nature 416 (2002)
  155--158.

\bibitem{Okamoto08a}
Y.~Okamoto, S.~Niitaka, M.~Uchida, T.~Waki, M.~Takigawa, Y.~Nakatsu,
  A.~Sekiyama, S.~Suga, R.~Arita, H.~Takagi, Band {J}ahn-{T}eller instability
  and formation of valence bond solid in a mixed-valent spinel oxide
  \protect{LiRh$_2$O$_4$}, Phys.\ Rev.\ Lett. 101 (2008) 086404.

\bibitem{Knox13a}
K.~R. Knox, A.~M.~M. Abeykoon, H.~Zheng, W.-G. Yin, A.~M. Tsvelik, J.~F.
  Mitchell, S.~J.~L. Billinge, E.~S. Bozin, Local structural evidence for
  strong electronic correlations in spinel \protect{LiRh$_2$O$_4$}, Phys.\
  Rev.\ B 88 (2013) 174114.

\bibitem{Rau16a}
J.~G. Rau, E.~K.-H. Lee, H.-Y. Kee, Spin-orbit physics giving rise to novel
  phases in correlated systems: iridates and related materials, Ann. Rev. Cond.
  Matter Phys. 7 (2016) 195--221.

\bibitem{Kim08b}
B.~J. Kim, H.~Jin, S.~J. Moon, J.-Y. Kim, B.-G. Park, C.~S. Leem, J.~Yu, T.~W.
  Noh, C.~Kim, S.-J. Oh, J.~H. Park, V.~Durairaj, G.~Cao, E.~Rotenberg, Novel
  \protect{$J_{\rm eff}$=1/2} {M}ott state induced by relativistic spin-orbit
  coupling in \protect{Sr$_2$IrO$_4$}, Phys.\ Rev.\ Lett. 101 (2008) 076402.

\bibitem{Witczak-Krempa14a}
W.~Witczak-Krempa, G.~Chen, Y.~B. Kim, L.~Balents, Correlated quantum phenomena
  in the strong spin-orbit regime, Ann. Rev. Cond. Matter Phys. 5 (2014)
  57--82.

\bibitem{Wu94a}
W.~D. Wu, A.~Keren, L.~P. Le, G.~M. Luke, B.~J. Sternlieb, Y.~J. Uemura,
  Magnetic penetration depth in \protect{V$_3$Si} and \protect{LiTi$_2$O$_4$}
  measured by \protect{$\mu$SR}, Hyperfine Interactions 86 (1994) 615--621.

\bibitem{Fazileh06a}
F.~Fazileh, R.~J. Gooding, W.~A. Atkinson, D.~C. Johnston, Role of strong
  electronic correlations in the metal-to-insulator transition in disordered
  \protect{LiAl$_y$Ti$_{2-y}$O$_4$}, Phys.\ Rev.\ Lett. 96 (2006) 046410.

\bibitem{Moshopoulou99a}
E.~G. Moshopoulou, Superconductivity in the spinel compound
  \protect{LiTi$_2$O$_4$}, J. Am. Ceram. Soc. 82 (1999) 3317--3320.

\bibitem{Jin15a}
K.~Jin, G.~He, X.~Zhang, S.~Maruyama, S.~Yasui, R.~Suchoski, J.~Shin, Y.~Jiang,
  H.~S. Yu, J.~Yuan, L.~Shan, F.~V. Kusmartsev, R.~L. Greene, I.~Takeuchi,
  Anomalous magnetoresistance in the spinel superconductor
  \protect{LiTi$_2$O$_4$}, Nat. Commun. 6 (2015) 7183--7191.

\bibitem{Chen17a}
D.~Chen, Y.-L. Jia, T.-T. Zhang, Z.~Fang, K.~Jin, P.~Richard, H.~Ding, Raman
  study of electron-phonon coupling in thin films of the spinel oxide
  superconductor \protect{LiTi$_2$O$_4$}, Phys.\ Rev.\ B 96 (2017) 094501.

\bibitem{Yamauchi02a}
T.~Yamauchi, Y.~Ueda, N.~Mori, Pressure-induced superconductivity in
  \protect{$\beta$-Na$_{0.33}$V$_2$O$_5$} beyond charge ordering, Phys.\ Rev.\
  Lett. 89 (2002) 057002.

\bibitem{Yamauchi08a}
T.~Yamauchi, Y.~Ueda, Superconducting $\beta(\beta^\prime)$-vanadium bronzes
  under pressure, Phys.\ Rev.\ B 77 (2008) 104529.

\bibitem{Okazaki04a}
K.~Okazaki, A.~Fujimori, T.~Yamauchi, Y.~Ueda, Angle-resolved photoemission
  study of the quasi-one-dimensional superconductor
  \protect{$\beta$-Na$_{0.33}$V$_2$O$_5$}, Phys. Rev. B 69 (2004) 140506.

\bibitem{Hague07a}
J.~P. Hague, P.~E. Kornilovitch, J.~H. Samson, A.~S. Alexandrov, Superlight
  small bipolarons in the presence of a strong {C}oulomb repulsion, Phys.\
  Rev.\ Lett. 98 (2007) 037002.

\bibitem{Takabayashi09a}
Y.~Takabayashi, A.~Y. Ganin, P.~\protect{Jegli\u{c}}, D.~\protect{Ar\u{c}on},
  T.~Takano, Y.~Iwasa, Y.~Ohishi, M.~Takata, N.~Takeshita, K.~Prassides, M.~J.
  Rosseinsky, The disorder-free non-{BCS} superconductor {C}s$_3${C}$_{60}$
  emerges from an antiferromagnetic insulator parent state, Science 323 (2009)
  1585--1590.

\bibitem{Gunnarsson97a}
O.~Gunnarsson, Superconductivity in fullerides, Rev.\ Mod.\ Phys. 69 (1997)
  575--606.

\bibitem{Ganin08a}
A.~Y. Ganin, Y.~Takabayashi, Y.~Z. Khimyak, S.~Margadonna, A.~Tamai, M.~J.
  Rosseinsky, K.~Prassides, Bulk superconductivity at 38 {K} in a molecular
  system, Nat.\ Mater. 7~(5) (2008) 367--371.

\bibitem{Ganin10a}
A.~Y. Ganin, Y.~Takabayashi, P.~\protect{Jegli\u{c}}, D.~\protect{Ar\u{c}on},
  A.~\protect{Poto\u{c}nik}, P.~J. Baker, Y.~Ohishi, M.~T. McDonald, M.~D.
  Tzirakis, A.~McLennan, G.~R. Darling, M.~Takata, M.~J. Rosseinsky,
  K.~Prassides, Polymorphism control of superconductivity and magnetism in
  {C}s$_3${C}$_{60}$ close to the {M}ott transition, Nature 466 (2010)
  221--225.

\bibitem{Klupp12a}
G.~Klupp, P.~Matus, K.~\protect{Kama\'ras}, A.~Y. Ganin, A.~\protect{McLennan},
  M.~J. Rosseinsky, Y.~Takabayashi, M.~T. \protect{McDonald}, K.~Prassides,
  Dynamic {J}ahn-{T}eller effect in the parent insulating state of the
  molecular superconductor \protect{Cs$_3$C$_{60}$}, Nature Commun. 3 (2012)
  912.

\bibitem{Chakravarty91a}
S.~Chakravarty, M.~Gelfand, S.~Kivelson, Electronic correlation effects and
  superconductivity in doped fullerenes, Science 254 (1991) 970--974.

\bibitem{Varma91a}
C.~M. Varma, J.~Zaanen, K.~Raghavachari, Superconductivity in the fullerenes,
  Science 254 (1991) 989--992.

\bibitem{Schluter92a}
M.~Schluter, M.~Lannoo, M.~Needels, G.~A. Baraff, D.~\protect{Tom\'anek},
  Electron-phonon coupling and superconductivity in alkalai-intercalated
  \protect{C$_{60}$} solid, Phys.\ Rev.\ Lett. 68 (1992) 526.

\bibitem{White92b}
S.~R. White, S.~Chakravarty, M.~P. Gelfand, S.~A. Kivelson, Pair binding in
  small {H}ubbard-model molecules, Phys.\ Rev.\ B 45 (1992) 5062--5065.

\bibitem{Dutta14a}
T.~Dutta, S.~Mazumdar, Theory of metal-intercalated phenacenes: Why molecular
  valence 3 is special, Phys.\ Rev.\ B 89 (2014) 245129.

\bibitem{Onoda87a}
M.~Onoda, K.~Toriumi, Y.Matsuda, M.~Sato, Crystal-structure of lithium
  molybdenum purple bronze \protect{Li$_{0.9}$Mo$_6$O$_{17}$}, J. Solid St.
  Chem. 66 (1987) 163--170.

\bibitem{daLuz11a}
M.~S. da~Luz, J.~J. Neumeier, C.~A.~M. dos Santos, B.~D. White, H.~J.~I. Filho,
  J.~B. Leao, Q.~Huang, Neutron diffraction study of quasi-one-dimensional
  lithium purple bronze: Possible mechanism for dimensional crossover, Phys.\
  Rev.\ B 84 (2011) 014108.

\bibitem{Greenblatt88a}
M.~Greenblatt, Molybdenum oxide bronzes with quasi-low-dimensional properties,
  Chem. Rev. 88 (1988) 31--53.

\bibitem{Schlenker85a}
C.Schlenker, H.Schwenk, C.Escribe-Filippini, J.Marcus, Superconducting
  properties of the low dimensional purple bronze
  \protect{Li$_{0.9}$Mo$_6$O$_{17}$}, Physica B+C 135 (1985) 511.

\bibitem{Sato87a}
M.~Sato, Y.~Matsuda, H.~Fukuyama, Localization and superconductivity in
  \protect{Li$_{0.9}$Mo$_6$O$_{17}$}, J. Phys. C 20 (1987) L137--L142.

\bibitem{Choi04b}
J.~Choi, J.~L. Musfeldt, J.~He, R.~Jin, J.~R. Thompson, D.~Mandrus, X.~N. Lin,
  V.~A. Bondarenko, J.~W. Brill, Probing localization effects in
  \protect{Li$_{0.9}$Mo$_6$O$_{17}$} purple bronze: An optical-properties
  investigation, Phys.\ Rev.\ B 69 (2004) 085120.

\bibitem{dosSantos07a}
C.~A.~M. dos Santos, B.~D. White, Y.-K. Yu, J.~J. Neumeier, J.~A. Souza,
  Dimensional crossover in the purple bronze
  \protect{Li$_{0.9}$Mo$_6$O$_{17}$}, Phys.\ Rev.\ Lett. 98 (2007) 266405.

\bibitem{dosSantos08a}
C.~A.~M. dos Santos, M.~S. da~Luz, Y.-K. Yu, J.~J. Neumeier, J.~Moreno, B.~D.
  White, Electrical transport in single-crystalline
  \protect{Li$_{0.9}$Mo$_6$O$_{17}$}: A two-band {L}uttinger liquid exhibiting
  {B}ose metal behavior, Phys.\ Rev.\ B 77 (2008) 193106.

\bibitem{Mercure12a}
J.~F. Mercure, A.~F. Bangura, X.~F. Xu, N.~Wakeham, A.~Carrington, P.~Walmsley,
  M.~Greenblatt, N.~E. Hussey, Upper critical magnetic field far above the
  paramagnetic pair-breaking limit of superconducting one-dimensional
  \protect{Li$_{0.9}$Mo$_6$O$_{17}$} single crystals, Phys.\ Rev.\ Lett. 108
  (2012) 187003.

\bibitem{Lebed13a}
A.~G. Lebed, O.~Sepper, Possible triplet superconductivity in the
  quasi-one-dimensional conductor \protect{Li$_{0.9}$Mo$_6$O$_{17}$}, Phys.\
  Rev.\ B 87 (2013) 100511(R).

\bibitem{Zuo00a}
F.~Zuo, J.~S. Brooks, R.~H. McKenzie, J.~A. Schlueter, J.~M. Williams,
  Paramagnetic limiting of the upper critical field of the layered organic
  superconductor \protect{$\kappa$-(BEDT-TTF)$_2$Cu(SCN)$_2$}, Phys.\ Rev.\ B
  61 (2000) 750--755.

\bibitem{Shinagawa07a}
J.~Shinagawa, Y.~Kurosaki, F.~Zhang, C.~Parker, S.~E. Brown,
  D.~\protect{J\'{e}rome}, J.~B. Christensen, K.~Bechgaard, Superconducting
  state of the organic conductor \protect{(TMTSF)$_2$ClO$_4$}, Phys.\ Rev.\
  Lett. 98 (2007) 147002.

\bibitem{Merino12a}
H.~Merino, R.~H. McKenzie, Effective {H}amiltonian for the electronic
  properties of the quasi-one-dimensional material
  \protect{Li$_{0.9}$Mo$_6$O$_{17}$}, Phys.\ Rev.\ B 85 (2012) 235128.

\bibitem{Popovic06a}
Z.~S. Popovic, S.~Satpathy, Density-functional study of the {L}uttinger liquid
  behavior of the lithium molybdenum purple \protect{Li$_{0.9}$Mo$_6$O$_{17}$},
  Phys.\ Rev.\ B 74 (2006) 045117.

\bibitem{Cohn12a}
J.~Cohn, B.~D. White, C.~A.~M. dos Santos, J.~J. Neumeier, Giant {N}ernst
  effect and bipolarity in the quasi-one-dimensional metal
  \protect{Li$_{0.9}$Mo$_6$O$_{17}$}, Phys.\ Rev.\ Lett. 108 (2012) 056604.

\bibitem{Ko15a}
K.-T. Ko, H.-H. Lee, D.-H. Kim, J.-J. Yang, S.-W. Cheong, M.~Eom, J.~Kim,
  R.~Gammag, K.-S. Kim, H.-S. Kim, T.-H. Kim, H.-W. Yeom, T.-Y. Koo8, H.-D.
  Kim, J.-H. Park, Charge-ordering cascade with spin–orbit {M}ott dimer
  states in metallic iridium ditelluride, Nat. Commun. 6 (2015) 7342.

\bibitem{Oh13a}
Y.~S. Oh, J.~J. Yang, Y.~Horibe, S.-W. Cheong, Anionic depolymerization
  transition in \protect{IrTe$_2$}, Phys.\ Rev.\ Lett. 110 (2013) 127209.

\bibitem{Fang13a}
A.~F. Fang, G.~Xu, T.~Dong, P.~Zheng, N.~L. Wang, Structural phase transition
  in \protect{IrTe$_2$}: a combined study of optical spectroscopy and band
  structure calculations, Sci. Rep. 3 (2013) 1153.

\bibitem{Yang12a}
J.~J. Yang, Y.~J. Choi, Y.~S. Oh, A.~Hogan, Y.~Horibe, K.~Kim, B.~I. Min, S.-W.
  Cheong, Charge-orbital density wave and superconductivity in the strong
  spin-orbit coupled \protect{IrTe$_2$:Pd}, Phys.\ Rev.\ Lett. 108 (2012)
  116402.

\bibitem{Kohsaka12a}
Y.~Kohsaka, T.~Hanaguri, M.~Azuma, M.~Takano, J.~C. Davis, H.~Takagi,
  Visualization of the emergence of the pseudogap state and the evolution to
  superconductivity in a lightly hole-doped {M}ott insulator, Nat. Phys. 8
  (2012) 534--538.

\bibitem{Naito16a}
M.~Naito, Y.~Krockenberger, A.~Ikeda, H.~Yamamoto, Reassessment of the
  electronic state, magnetism, and superconductivity in high-\protect{T$_c$}
  cuprates with the \protect{Nd$_2$CuO$_4$} structure, Physica C 523 (2016)
  28--54.

\bibitem{Dagan07a}
Y.~Dagan, R.~L. Greene, Hole superconductivity in the electron-doped
  superconductor \protect{Pr$_{2-x}$Ce$_x$CuO$_4$}, Phys.\ Rev.\ Lett. 94
  (2007) 187003.

\bibitem{Masino17a}
M.~Masino, N.~Castagnetti, A.~Girlando, Phenomenology of the neutral-ionic
  valence instability in mixed stack charge-transfer crystals, Crystals 7
  (2017) 108, and references therein.

\end{thebibliography}
\end{document}